\documentclass[a4paper, 12pt, twoside]{book}
\usepackage{a4}
\usepackage{amsfonts}
\usepackage{amssymb}
\usepackage{epsfig}
\usepackage{comment}
\usepackage{graphicx}
\usepackage{amsmath}

\usepackage{axodraw}

\newcommand{\be}{\begin{equation}}
\newcommand{\ee}{\end{equation}}
\newcommand{\bea}{\begin{eqnarray}}
\newcommand{\eea}{\end{eqnarray}}
\newcommand{\mbb}{\mathbb}
\newcommand{\ti}{\times}
\newcommand{\half}{\frac{1}{2}}

\newcommand{\mc}{\mathcal}

\newcommand{\exd}{{\rm d}}
\newcommand{\efold}{$e$\,-fold}

\begin{document}

\pagestyle{headings}

\thispagestyle{empty}
\newcommand{\HRule}{\rule{\linewidth}{1mm}}
\setlength{\parindent}{1cm}
\setlength{\parskip}{1mm}
\vspace*{\stretch{1}}
\noindent
\HRule
\begin{center}
{\Huge String Loop Moduli Stabilisation and Cosmology in IIB Flux Compactifications} \\[5mm]
\end{center}
\HRule
\vskip 2cm
\begin{center} \Large Michele Cicoli \end{center}
\begin{center} \Large
 St John's College,
   Cambridge

\vskip 4cm
\epsfig{file=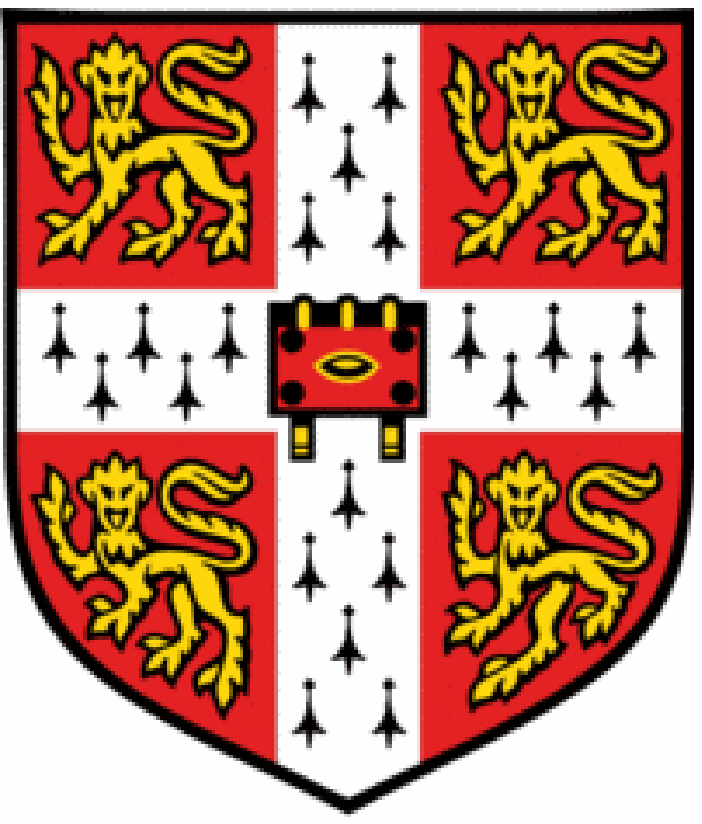, width=3.3cm, height=4cm}
\end{center}

\vspace*{\stretch{2}}
\begin{center}\Large \textsc {Dissertation submitted for the degree of
     Doctor of Philosophy \\ Cambridge University 2009}\end{center}

\newenvironment{dedication}   {\newpage \thispagestyle{empty}
\vspace*{\stretch{1}} \begin{center} \em}   {\end{center}
\vspace*{\stretch{3}} \clearpage}
\begin{dedication}\begin{flushright}{\it \Large{}}\end{flushright}\end{dedication}

\newenvironment{PreQuote}   {\newpage \thispagestyle{empty}
\vspace*{\stretch{1}} \begin{center} \em}   {\end{center}
\vspace*{\stretch{2}} \clearpage}
\begin{PreQuote}
Adde ergo scientiae caritatem, \newline

et utilis erit scientia;\newline

non per se, sed per caritatem. \phantom{g}
\newline
\newline
Saint Augustine\end{PreQuote}

\begin{dedication}\begin{flushright}{\it \Large{}}\end{flushright}\end{dedication}

\newpage
\thispagestyle{empty}
\vspace*{\stretch{0.8}}
\noindent
\begin{center}
{\Large Acknowledgements}
\end{center}
I would like to thank in particular my supervisor Fernando Quevedo
for introducing me to a completely new research field, for his
kind patience, continuous support, and contagious enthusiasm. On
the subjects contained in this thesis I have learned many things
from my collaborators Lilia Anguelova, Cliff Burgess, Vincenzo
Cal\`{o}, Joseph Conlon and Fernando Quevedo. I am also grateful
to Andres Collinucci, Christoph Mayrhofer and Maximilian Kreuzer
for collaboration on an unrelated project not described in this
thesis. Within Cambridge I have had stimulating discussions on
various topics with many people including Shehu Abdussalam, Ben
Allanach, Heng-Yu Chen, Daniel Cremades, Matthew Dolan, Mathieu
Ehrhardt, Chun-Hay Kom, Sven Krippendorf, Manuel Losi, Anshuman
Maharana, Ricardo Monteiro, Nelson Nunes, Miguel Paulos, Jorge
Santos, Aninda Sinha, Julian Sonner, Kerim Suruliz, and Hung Ling
Yan. Through conversations at conferences and email correspondence
I have also learnt many things from many physicists, and I am
indebted to them for their time and careful explanations. An
incomplete list includes Mina Aganagic, Nima Arkani-Hamed, Markus
Berg, Massimo Bianchi, Ralph Blumenhagen, Philip Candelas, Xenia
de la Ossa, Frederik Denef, Michael Douglas, Gian Francesco
Giudice, James Gray, Michael Haack, Arthur Hebecker, Luis Ibanez,
Costas Kounnas, David Lyth, Liam McAllister, Erik Plauschinn,
Michael Ratz, Emanuel Scheidegger, Gary Shiu, Michele Trapletti,
Angel Uranga, Licia Verde, and Alexander Westphal.

I am grateful to the Isaac Newton Trust for a European Research
Studentship, to St John's College, Cambridge for a Horne
Scholarship and to the Engineering and Physical Sciences Research
Council for financial support towards the cost of academic fees. I
would also like to thank both the Department of Applied
Mathematics and Theoretical Physics in Cambridge and St John's
College, Cambridge for contributions towards travel expenses.
\vspace*{\stretch{1}}

\newpage
\thispagestyle{empty}
\begin{center}
\Large{Declaration}
\end{center}
This dissertation is the result of my own work and includes nothing
which is the outcome of work done in collaboration except where
specifically indicated in the text. No part of this thesis has
previously been
submitted for a degree or other qualification at this or any other university.

\bigskip

\bigskip

\bigskip
This thesis
is based on the research presented in the following papers:
\begin{enumerate}

\item \textbf{Systematics of String Loop Corrections in Type IIB
Calabi-Yau Flux Compactifications}, Michele Cicoli, Joseph Conlon
and Fernando Quevedo, \textbf{JHEP01 (2008) 052}, arXiv:0708.1873
[hep-th].

\medskip
\item \textbf{General Analysis of LARGE Volume Scenarios with String
Loop Moduli Stabilisation}, Michele Cicoli, Joseph Conlon,
Fernando Quevedo, \textbf{JHEP10 (2008) 105}, arXiv:0805.1029
[hep-th].

\medskip
\item \textbf{Fibre Inflation: Observable Gravity Waves from IIB
String Compactifications}, Michele Cicoli, Cliff Burgess, Fernando
Quevedo, \textbf{JCAP03 (2009) 013}, arXiv:0808.0691 [hep-th].

\medskip
\item \textbf{LARGE Volume
String Compactifications at Finite Temperature}, Lilia Anguelova,
Vincenzo Cal\`{o}, Michele Cicoli, \textbf{arXiv:0904.0051
[hep-th]}.
\end{enumerate}

These papers are references \cite{Cicoli1, Cicoli2, Cicoli3,
Cicoli4} in the bibliography.

\vspace{2cm}

\begin{flushright}
MICHELE CICOLI

Cambridge, 5th June 2009
\end{flushright}

\newpage
\thispagestyle{empty}
\begin{center}
\Large{Summary} \\
\end{center}

String Theory is at present the most promising candidate for a
consistent theory of quantum gravity which is also able to unify
all the interactions and matter. However string compactifications
are plagued by the presence of moduli, which are massless
particles that would mediate unobserved long-range forces. It is
therefore crucial to generate a potential for those particles,
given that also the gauge and Yukawa couplings of the low energy
effective theory are determined by the vacuum expectation values
of the moduli.

The introductory chapter is a review of type IIB Calabi-Yau flux
compactifications where both the axio-dilaton and the complex
structure moduli are stabilised by turning on background fluxes.
However in order to fix the K\"{a}hler moduli one needs to
consider perturbative and non-perturbative corrections beyond the
leading order approximations. After presenting a survey of all the
existing solutions to this problem, I focus on the LARGE Volume
Scenario since it requires no fine-tuning of the fluxes and
provides a natural solution of the hierarchy problem. I then
derive the topological conditions on an arbitrary Calabi-Yau to
obtain such a construction. This result is illustrated with
explicit computations for various compactification manifolds,
showing that, in the absence of string loop corrections, the
potential always exhibits the presence of flat directions. Hence I
perform a systematic study of the behaviour of $g_s$ corrections
for general type IIB compactifications, and show how they play a
crucial r\^{o}le to achieve full K\"{a}hler moduli stabilisation
in the LARGE Volume Scenario.

In the next part of the thesis, I examine the possible
cosmological implication of these scenarios. After calculating the
moduli mass spectrum and couplings, I notice how, in the case of
K3-fibrations, string loop corrections give rise naturally to an
inflationary model which yields observable gravity waves. The
prediction for the tensor-to-scalar ratio $r\sim 5\cdot 10^{-3}$
is within reach of future cosmological observations. A further
chapter studies the finite-temperature behaviour of the LARGE
Volume Scenario. I compute the maximal temperature $T_{max}$ above
which the internal space decompactifies, as well as the
temperature $T_*$ that is reached after the decay of the heaviest
moduli. The requirement $T_*<T_{max}$ rules out a significant
region of parameter space, favouring values of the volume which
lead to TeV-scale supersymmetry instead of standard gauge coupling
unification. I then show that unwanted relics cannot be diluted by
the decay of the small moduli, nor by a low-energy period of
thermal inflation in the closed string moduli sector.

I finally conclude outlining the prospects for future work.

\begin{dedication}\begin{flushright}{\it \Large{}}\end{flushright}\end{dedication}

\tableofcontents

\part{Introduction}

\chapter{String Theory and the Real World}
\label{Intro} \linespread{1.3}

Scalar masses play a crucial r\^{o}le in fundamental physics
since, due to their high sensitivity to ultra-violet physics, they
give us invaluable insights into the structure of the underlying
theory describing Nature at higher energies. As we shall see later
on, the main two examples of such scalar particles are the Higgs
boson and the inflaton. They represent exceptions to the usual
decoupling of energy scales which allows us to understand the laws
at the basis of our Universe. In fact, the history of Science has
shown that it is possible to do physics since each energy scale
can be understood on its own terms without having to unveil all
the mysteries of our Universe at once. The reason is that the
properties of a particular scale do not have a strong dependence
on the details of the physics of the other scales.

The most famous example is the decoupling of quantum effects at
large distances with the consequent possibility to have a picture
of our world based on pure classical physics. Moreover, within
quantum theories, it is not necessary to master all the details of
nuclear physics in order to be able to understand the spectra and
the chemistry of atoms. In turn, the properties of nuclei can be
described without any need to explore the deep structure of
quantum chromo-dynamics, which is the theory describing the
behaviour of their constituents, quarks and gluons.

The concept of decoupling of scales in physics is closely tied up
with the notion of \textit{naturalness}, according to which, the
necessity of excessive fine-tuning of a particular parameter in
order to match the experimental data, is a clear sign of new
physics. This is the reason why the study of scalar masses is
important. In fact, their ultra-violet sensitivity together with
the requirement of avoiding fine-tuning, can guide us in the
search of new physics beyond our present knowledge.

The current view of our Universe is based on two theories with
strong experimental evidence:
\begin{itemize}
\item Standard Model of particle physics,

\item Standard Cosmological Model.
\end{itemize}
Let us now briefly review the main features and shortcomings of
each theory.

\section{Beyond standard particle physics}

\subsection{The Standard Model}

The Standard Model of particle physics is a four dimensional
relativistic quantum field theory based on the gauge symmetry
$SU(3)_c\times SU(2)_L\times U(1)_Y$. The theory is well defined
at the quantum mechanical level since it is unitary and
renormalisable. Strong interactions emerge from requiring local
invariance of the Lagrangian under the $SU(3)_c$ symmetry, whereas
the electro-weak forces are related to the $SU(2)_L\times U(1)_Y$
group. All the elementary particles are assumed to be point-like
according to present experimental evidence up to energies of about
1 TeV. The fundamental constituents of matter are three families
of quarks and leptons described in terms of left-handed Weyl
fermions which transform under the Standard Model group as
follows:
\begin{center}
\begin{tabular}{c||c|c|c}
  & $SU(3)_c$ & $SU(2)_L$ & $U(1)_Y$ \\
  \hline\hline
  $Q_L$ & $\textbf{3}$ & $\textbf{2}$ & 1/6 \\
  $U$ & $\bar{\textbf{3}}$ & $\textbf{1}$ & 1/3 \\
  $D$ & $\bar{\textbf{3}}$ & $\textbf{1}$ & -2/3 \\
  $L$ & $\textbf{1}$ & $\textbf{2}$ & -1/2 \\
  $E$ & $\textbf{1}$ & $\textbf{1}$ & 1
\end{tabular}\\
\vspace{0.3cm}{{\bf Table {1}:} Matter representations of the
Standard Model gauge group.}
\end{center}

Moreover a crucial property of these fermions is their
\textit{chirality} (parity-violation), given that the $SU(2)_L$
interactions act only on left-handed particles.

In order to break the electro-weak symmetry $SU(2)_L\times U(1)_Y$
down to the electromagnetic $U(1)$, the model contains a Higgs
sector, given by a complex scalar $\phi$ with quantum numbers
($\textbf{1}$, $\textbf{2}$, -1/2). Its vacuum expectation value
(VEV) $\langle\phi\rangle$ is determined by a potential of the
form:
\begin{equation}
V(\phi)=-m^2\phi^*\phi+\lambda(\phi^*\phi)^2,
\end{equation}
and fixes the scale of spontaneous electro-weak symmetry breaking:
\begin{equation}
M_W\simeq\langle\phi\rangle\simeq\frac{m}{\sqrt{\lambda}}\simeq
10^2\textrm{\ GeV}.
\end{equation}
This is the so-called Higgs mechanism through which the weak
bosons $Z$ and $W^{\pm}$ acquire a mass while leaving the photon
massless. Given that chirality forbids writing a Dirac mass term
for matter fermions, they get a mass via their coupling to the
Higgs multiplet through Yukawa-like interactions of the form $Q_L
U\phi$, $Q_L D\bar{\phi}$ and $LE\phi$. Hence the scale of fermion
masses is related to the scale of electro-weak symmetry breaking.

All the presently known high energy phenomena are described with a
remarkable success by the Standard Model of elementary particles
and fundamental interactions \cite{PDG}, although there are a
number of theoretical and phenomenological issues that the
Standard Model fails to address adequately:
\begin{itemize}
\item \textit{Hierarchy problem}. It consists of two issues:
\begin{enumerate}
\item \textit{Technical hierarchy problem}. It is associated with the absence
of a symmetry protecting the Higgs mass from getting large loop
corrections. Then power counting would suggest that the natural
value of quantum corrections to the Higgs mass term is of the
order $M_P^2\phi^2$, which would then push the electro-weak scale
up to the natural cut-off at the Planck scale.

\item \textit{Gauge hierarchy problem}. It is associated with explaining
the origin of the electro-weak scale, while a more fundamental
embedding theory is typically defined at the Planck scale, which
is $10^{15}$ times larger than the electro-weak scale.
\end{enumerate}
\item \textit{Electro-weak symmetry breaking}. In the Standard Model,
the Higgs sector is not constrained by any symmetry principles,
and it must be put into the theory by hand. Hence this issue needs
an explanation.

\item \textit{Gauge coupling unification}. The gauge couplings undergo
renormalisation group evolution in such a way that they tend to
meet at a point at a high scale. However, precise measurements of
the low energy values of the gauge couplings demonstrated that the
Standard Model cannot describe gauge coupling unification
accurately enough to imply it is more than an accident.

\item \textit{Family structure and fermion masses}.  The Standard Model does not
explain the existence of three families and can only parameterise
the strongly hierarchical values of the fermion masses. Moreover,
the Standard Model does not contain massive neutrinos.

\item \textit{Why questions}. In the Standard Model there is no
explanation for the number of dimensions which is assumed to be
four. In addition, the gauge group $SU(3)_c\times SU(2)_L\times
U(1)_Y$ is put in by hand.

\item \textit{Gravity}. The Standard Model does not contain
gravity. It still produces theoretical predictions in excellent
agreement with experimental results, given that gravitational
effects are negligible at the low scales where particle physics
experiments are performed.

Gravitational interactions are described by the classical theory
of General Relativity in perfect agreement with experiments
\cite{testGR}. Interactions are encoded in the space-time metric
$g_{\mu\nu}$ via the principle of diffeomorphism (or coordinate
reparameterisation) invariance of the physics. This leads to the
famous Einstein-Hilbert action of the form:
\begin{equation}
S_{EH}=\frac{M_P}{2}\int d^4 x \sqrt{-g}R.
\end{equation}
Since the interaction contains an explicit dimensionful coupling,
it is difficult to make sense of the theory at the quantum level.
The theory presents loss of unitarity at loop level and is
non-renormalisable. Hence it cannot be quantised in the usual
fashion and is not well defined in the ultra-violet.

The modern viewpoint is that Einstein theory should be regarded as
an effective field theory, which is a good approximation at
energies below $M_P$ (or some other cutoff scale at which four
dimensional classical Einstein theory ceases to be valid). There
should be an underlying, quantum mechanically well defined, theory
which exists for all ranges of energy, and reduces to General
Relativity at low energies, below the cutoff scale. Such a theory
would be called an ultra-violet completion of Einstein gravity.

\item \textit{Cosmological challenges}. In the Standard Model there is
no viable mechanism to explain the baryon asymmetry of the
Universe. In addition, there is no candidate particle for cold
dark matter and for the inflaton. Finally, probably the most
difficult problem the Standard Model has when trying to connect
with the gravitational sector, is the absence of the expected
scale of the cosmological constant.
\end{itemize}
Therefore, the Standard Model must be extended and its foundations
strengthened. Let us therefore review some proposals for physics
beyond the Standard Model \cite{Ibanez}.

\subsection{Supersymmetry}

Theories with low energy supersymmetry have emerged as the
strongest candidate for new physics beyond the Standard Model at
the TeV scale \cite{SUSY}. In supersymmetric models, each particle
has a superpartner which differs in spin by $1/2$ and is related
to the original particle by a supersymmetry transformation which
requires them to have equal mass.

Supersymmetry commutes with gauge symmetry. So in trying to build
a minimal extension of the Standard Model with low energy
supersymmetry, known as the Minimal Supersymmetric Standard Model
(MSSM), the simplest possibility is to add superpartners for all
observed particles: fermionic superpartners (gauginos) for gauge
bosons to promote them to vector multiplets; bosonic superpartners
(squarks and sleptons) for quarks and leptons to promote them to
chiral multiplets; and fermionic superpartners (Higgsinos) for the
Higgs scalars (due to anomaly cancellation a second Higgs chiral
multiplet must be added). Interactions are dictated by gauge
symmetry and supersymmetry. The main successes of the MSSM are:
\begin{itemize}
\item \textit{Hierarchy problem}.
\begin{enumerate}
\item \textit{Technical hierarchy problem}. It is solved by the fact that
since supersymmetry relates the scalar and the fermionic sectors,
the chiral symmetries which protect the masses of the fermions
also protect the masses of the scalars from quadratic divergences.
More precisely, the mass of a chiral fermion is forced to be zero
by chirality, so the mass of a scalar like the Higgs is protected
against getting large $\mathcal{O}(M_P)$ corrections.
Diagrammatically, any corrections to the Higgs mass due to
fermions in the theory are cancelled against corrections to the
Higgs mass due to their bosonic superpartners. There is a
non-renormalisation theorem of certain couplings in the Lagrangian
which guarantees this to any order in perturbation theory.

\item \textit{Gauge hierarchy problem}. It is mitigated by breaking the electro-weak
symmetry radiatively through logarithmic running, which explains
the large number $\sim 10^{15}$.
\end{enumerate}
\item \textit{Radiative electro-weak symmetry breaking}. The `Mexican
hat' potential with a minimum away from $\phi=0$ is derived rather
than assumed by starting with plausible boundary conditions at a
high scale and then running them down to the electro-weak scale,
where the $m^2$ parameter runs to negative values, driven by the
large top quark Yukawa coupling.

\item \textit{Gauge coupling unification}. The extrapolation of the low
energy values of the gauge couplings using renormalisation group
equations and the MSSM particle content shows that the gauge
couplings unify at the scale $M_{GUT}\sim 10^{16}$ GeV.

\item \textit{Cold dark matter}. In supersymmetric theories, the lightest
superpartner can be stable, so providing a nice cold dark matter
candidate.
\end{itemize}
However, unbroken supersymmetry implies a mass degeneracy between
superpartners, a possibility which is clearly forbidden by
experiment. Hence exact supersymmetry is not viable, but we can
keep all its desirable good properties, if it is broken
\textit{softly}. From a phenomenological perspective, a general
set of soft supersymmetry breaking terms are introduced explicitly
in the effective Lagrangian at the electro-weak scale
\cite{LSoft}. These terms render the superpartners more massive
than Standard Model fields. The cancellation of loop contributions
to the Higgs is not exact anymore, but it is only logarithmically,
and not quadratically, dependent on $M_P$. In order to retain
$10^2$ GeV as a natural scale, the superpartner mass scale should
be around $1$ TeV or so.

Together these successes and considerations provide powerful
indirect evidence that TeV scale supersymmetry is indeed part of
the correct description of Nature. Thus, it is likely that direct
evidence of the existence of superpartners should be discovered at
the forthcoming Large Hadron Collider (LHC) at CERN, which will
soon start operation with a centre of mass energy of 14 TeV.

Nevertheless, the soft Lagrangian $\mathcal{L}^{MSSM}_{soft}$
introduces about 105 new free parameters, so making the theory
lose all its predictive power. For phenomenological analyses, they
are reduced to usually less than five, assuming that
$\mathcal{L}^{MSSM}_{soft}$ takes on a simplified form at a given
high scale. These parameters must be then considered as boundary
conditions at the `input scale' for the renormalisation group
evolution that runs them down to the electro-weak scale. In this
way, we are able to predict all the phenomenologically interesting
features of the MSSM, as the mass spectrum, mixing angle,
cross-sections and decay rates, as functions only of a few free
parameters.

This picture is justified by the fact that we would like to
understand the \textit{explicit} soft supersymmetry breaking,
encoded in the $\mathcal{L}^{MSSM}_{soft}$ parameters, as the
result of \textit{spontaneous} supersymmetry breaking in a more
fundamental theory, which should provide also a \textit{dynamical}
explanation of such a mechanism. It turns out that the only
phenomenologically viable way to break supersymmetry is in the
context of the so-called \textit{hidden} sector models. In this
framework, one assumes that the theory is divided into two sectors
with no direct renormalisable couplings between them:
\begin{itemize}
\item  The observable or visible sector, which contains the MSSM
fields.

\item  The hidden sector, where spontaneous supersymmetry breaking
takes place triggered by a dynamical mechanism (such as gaugino
condensation, for example).
\end{itemize}
Within this framework, supersymmetry breaking is communicated from
the hidden to the observable sector via interactions involving a
third set of messenger fields. The result is the effective soft
supersymmetry breaking Lagrangian $\mathcal{L}^{MSSM}_{soft}$ in
the visible sector. The main known supersymmetry breaking
mediation scenarios are:
\begin{enumerate}
\item \textit{Gravity Mediated Supersymmetry Breaking}: the soft
parameters arise due to Planck-suppressed non-renormalisable
couplings which vanish as $M_P\to\infty$ \cite{SUGRAmediation};

\item \textit{Gauge Mediated Supersymmetry Breaking}: the soft
parameters arise from loop diagrams involving new messengers
fields with Standard Model quantum numbers \cite{GaugeMediation};

\item \textit{Anomaly Mediated Supersymmetry Breaking}: the supersymmetry
breaking is mediated by the supergravity auxiliary field which
couples to the visible sector fields at loop level due to the
super-conformal Weyl anomaly \cite{AMSB}.
\end{enumerate}
As we have seen, supersymmetry has several appealing features, but
still it does not answer any of the fundamental `why questions' at
the basis of the structure of the Standard Model. Furthermore,
even though supersymmetry is responsible for gauge couplings
unification, the MSSM does not treat the three non-gravitational
interactions in a unified manner, and, on top of that, it keeps
not containing gravity.

\subsection{Grand Unified Theories}

As we have already mentioned, the Standard Model gauge couplings
run with scale towards a roughly unified value at a scale around
$10^{16}$ GeV. This unification becomes very precise in the MSSM
by the inclusion of supersymmetric partners. Hence we have a clear
suggestion that the different low-energy interactions may be
unified at high energies.

Therefore a lot of attention has been put on Grand Unified
Theories (GUT) including the Standard Model group as low-energy
remnant of a larger gauge group \cite{Ross}. This group,
$G_{GUT}$, is usually taken to be $SU(5)$, $SO(10)$, or $E_6$, and
so unifies all low-energy gauge interactions into a unique kind.
The GUT group is broken spontaneously by a Higgs mechanism at a
high scale $M_{GUT}\sim 10^{16}$ GeV. This idea leads to a partial
explanation of the fermion family gauge quantum numbers, since the
different fermions are also unified into a smaller number of
representations of $G_{GUT}$. For example, a Standard Model family
fits into a $\textbf{10}+\bar{\textbf{5}}$ representation of
$SU(5)$, or within an irreducible $\textbf{16}$ representation of
$SO(10)$.

A disadvantage is that the breaking $G_{GUT}\to SU(3)_c\times
SU(2)_L\times U(1)_Y$ requires a complicated Higgs sector. In
minimal $SU(5)$ theories, the GUT-Higgs belongs to a
24-dimensional representation, and $SO(10)$ is even more involved.

An additional interesting feature of GUT theories is that extra
gauge interactions in $G_{GUT}$ mediate baryon number violating
processes of proton decay, which are suppressed by inverse powers
of $M_{GUT}$. The rough proton lifetime in these models is around
$10^{32}$ years, which is close to the experimental lower bounds.
In fact, some models like minimal $SU(5)$ are already
experimentally ruled out because they predict a too fast proton
decay.

Besides these nice features, it is fair to say that grand unified
theories do not address the fundamental problem of gravity at the
quantum level, or the relation between gravity and the other
interactions.

\subsection{Supergravity}

The first attempt to construct a theory that incorporates both
gravity and all the non-gravitational interactions within the same
description, was supergravity \cite{SUGRA}. Supersymmetry is a
global space-time symmetry and, in complete analogy with gauge
theories, one can try to make it local. This, in turn, implies
that the four-momentum operator, which generates global space-time
translations, is also promoted to a gauge generator. Given that
local translations are equivalent to coordinate
reparameterisations, the resulting theory contains General
Relativity.

A very important feature of supergravity theories is the presence
of a new supermultiplet, called the gravity multiplet, including a
spin-$2$ graviton $G_{\mu\nu}$ and its spin-$3/2$ superpartner
$\psi_{\alpha}^{\mu}$, the gravitino, which is the gauge field of
local supersymmetry. The supergravity Lagrangian is basically
obtained from the global supersymmetric one by adding the Einstein
term for the graviton, a kinetic term for the gravitino, and
coupling the graviton to the stress-energy tensor and the
gravitino to the supercurrent of the supersymmetric theory.

Supergravity can have interesting applications to phenomenology,
in particular, for the realisation of the gravity mediated
supersymmetry breaking scenario. In fact, spontaneous breaking of
local supersymmetry becomes, in the limit of energies much below
$M_P$, explicit breaking of global supersymmetry by soft terms.
Typically, local supersymmetry is broken in the hidden sector,
which is decoupled from the MSSM except by gravitational
interactions, at a scale of the order $M_{hidden}\sim 10^{11}$
GeV. Then transmission of supersymmetry breaking to the visible
sector is manifest at a lower scale, $M_{hidden}^2/M_P$, of around
$1$ TeV, which is the right superpartner mass scale.

We point out that, as we have already mentioned in the section
dedicated to supersymmetry, there is another mechanism of
supersymmetry breaking, called anomaly mediation, which is also
formulated within the supergravity framework.

Supergravity is a nice and inspiring idea, which attempts to
incorporate gravity. However, it does not provide an ultra-violet
completion of Einstein gravity, since it is neither finite nor
renormalisable.

\subsection{Extra Dimensions}

Many scenarios beyond the Standard Model propose a space-time with
more than four dimensions, the additional ones being unobservable
because they are compact and of very small size. We briefly
mention two ideas, which differ by whether ordinary Standard Model
matter is able to propagate in the new dimensions or not.

\subsubsection{Kaluza-Klein theories}

Kaluza and Klein independently \cite{Kaluza, Klein} proposed the
beautiful idea of unifying gravity and gauge interactions thanks
to the presence of tiny compactified extra dimensions. In fact,
they noticed that a purely gravitational theory in an arbitrary
$(4+D)$-dimensional space-time, upon compactification to four
dimensions, gives rise to the four-dimensional metric tensor plus
gauge bosons associated to a gauge group, which is the isometry
group of the compactification manifold.

In the simplest five-dimensional example with just one extra
dimension, the space-time takes the form $\mathbb{R}^{3,1}\times
S^1$, and is endowed with a five-dimensional metric $g_{MN}$,
$M,N=0,...,4$. From the viewpoint of the low-energy
four-dimensional theory (at energies much lower than the
compactification scale $M_{KK}=1/R$, with $R$ the radius of the
circle), the five-dimensional metric decomposes as ($\mu$,
$\nu=0,...,3$):
\begin{equation}
g_{MN}\longrightarrow \left\{
\begin{array}{c}
g_{\mu \nu }\textrm{, \ \ \ \ \ \ \ \ \ \ \ \ \ 4D graviton} \\
g_{\mu 4}\textrm{, \ \ \ \ \ \ \ \ 4D gauge boson} \\
g_{44}\textrm{, \ \ 4D scalar (modulus)}
\end{array}
\right.
\end{equation}
We then see that, in the resulting four-dimensional theory, we
obtain the metric tensor, a massless vector boson and a massless
scalar. Moreover, the diffeomorphism invariance in the fifth
dimension implies the four-dimensional $U(1)$ gauge invariance
associated with the vector boson $g_{\mu 4}$.

Despite its beauty, the Kaluza-Klein idea is very difficult to use
for phenomenology since it is not easy to construct manifolds with
the isometry group of the Standard Model. Moreover, it is
generically difficult to obtain four-dimensional chiral fermions
in this setup \cite{Witteno}. On top of that, although the idea
involves gravity, it still suffers from quantum inconsistencies.
Thus it does not provide an ultra-violet completion of General
Relativity.

\subsubsection{Brane-world scenario}

Another fascinating idea concerning the existence of extra
dimensions, is based on the assumption that only gravity can
propagate in the extra dimensions \cite{braneworld, add}. On the
other hand, Standard Model fields are assumed to be trapped in our
four-dimensional world. More precisely, the Standard Model is said
to live on a `brane' (generalisation of a membrane embedded in a
higher-dimensional space-time), while gravity propagates in the
`bulk' of space-time.

Given that gravity has been tested only down to $0.1$ mm without
observing any significant deviation from the four-dimensional
Newton's law \cite{TestGravity}, the most attractive feature of
this scenario is that it allows for compactification manifolds
with a very large volume, at least of the order $\mc{V}\sim
(0.01\textrm{\ mm})^D$, for arbitrary $D$ extra dimensions. We
stress that such a large volume is forbidden in Kaluza-Klein
theories, since it implies light Kaluza-Klein excitations of
Standard Model fields, in conflict with experiment. On the other
hand, in the brane-world scenario, such fields do not propagate in
the bulk, hence they do not have Kaluza-Klein replicas.

The large volume of the compactification manifold introduces, in
turn, a possible new solution of the hierarchy problem. In fact,
denoting with $M_{4+D}$ the $(4+D)$-dimensional fundamental scale,
the four-dimensional Planck mass is given by:
\begin{equation}
M_P^2=\left(M_{4+D}\right)^2\mathcal{V},
\end{equation}
where $\mathcal{V}$ is the volume of the internal manifold
measured in units of $\left(M_{4+D}\right)^{-1}$. Then in certain
models, a volume very close to the present experimental bounds
together with a fundamental scale of the order $M_{4+D}\sim 1$
TeV, would give exactly $M_P\sim 10^{18}$ GeV as a derived
quantity.

A very interesting generalisation of this scenario consists in
adding a warp factor in front of the four-dimensional metric. In
the five-dimensional case, this corresponds to:
\begin{equation}
ds^2 = W(y) g_{\mu\nu} dx^\mu dx^\nu + dy^2,
\end{equation}
where $y$ is the fifth dimension and $W(y)$ is the warp factor.
Considering branes located at fixed points in the $y$-direction,
they feel different scales in their metrics due to different
values of $W(y)$. Due to the exponential behaviour of the warp
factor, the scales change very fast, allowing for the possibility
of a small fundamental scale even for small volumes of the
compactification manifold \cite{RandallSundrum1}. In addition, a
further five-dimensional version of the brane-world scenario
proposes infinitely large extra dimensions in AdS space
\cite{RandallSundrum2}. In this case, it is still possible to
evade the present experimental bounds on the variation of gravity,
since warping localises gravity `close' to the Standard Model
brane.

Again, it is fair to emphasise that, despite its theoretical and
phenomenological appeal, this setup does not provide an
ultra-violet completion of General Relativity, since gravity is
treated classically.

\section{Beyond standard cosmology}

\subsection{The Standard Cosmological Model}

The Standard Cosmological Model is formulated within the framework
of the classical theory of Einstein gravity \cite{CosmoReview,
liddle}. The cosmological evolution of our Universe is determined
by considering a perfect fluid in a homogeneous and isotropic
space-time. This assumption constraints the form of the metric,
which is forced to take the famous Friedmann-Robertson-Walker
(FRW) form:
\begin{equation}
ds^2 = -dt^2 + a^2(t) \left[\frac{dr^2}{1-kr^2} +
r^2\left(d\theta^2+\sin^2\theta d\phi^2\right)\right], \label{frw}
\end{equation}
where the scale factor $a(t)$ is an unknown function which
measures the evolution of the Universe in cosmological time, while
$k$ is a discrete parameter, $k=-1, 0, 1$, that determines the
curvature of the spatial sections at fixed $t$, corresponding to
an open, flat or closed Universe, respectively. The perfect fluid
is described by an equation of state of the form $p = w\rho $,
where $\rho$ and $p$ are its energy density and pressure,
respectively, whereas the parameter $w$ is a constant describing
the nature of the energy source dominating the energy density of
the Universe.

The most important implication of the FRW ansatz (\ref{frw})
together with the positivity of the energy density $\rho$, is that
the space-time has an initial singularity at $t=0$, the big-bang,
from which the Universe started its expansion. The presence of the
singularity is a clear sign that we are using General Relativity
in a region beyond its regime of validity, where quantum effects
start playing a crucial r\^{o}le.

Plugging the FRW ansatz for the metric (\ref{frw}) into the
Einstein's equations, one obtains the so-called Friedmann's
equations:
\begin{eqnarray}
H^2 &\equiv& \left(\frac{\dot{a}}{a}\right)^ 2= \frac{8\pi G_N}{3} \rho -  \frac{k}{a^2}, \\
\frac{\ddot a}{a} & = & -\frac{4\pi G_N}{3} \left(\rho+ 3p\right),
\end{eqnarray}
which, at fixed $k$, are two differential equations for $a(t)$ and
$\rho(t)$. Usually the second Friedmann's equation is traded for
the derived energy conservation equation, which reads:
\begin{equation}
\dot{\rho}= -3H \left(\rho + p \right)=-3 \left(1+w\right)H \rho.
\label{conservation}
\end{equation}
Without solving these equations, but just looking at their form,
we can make several interesting observations:
\begin{itemize}
\item If $(\rho+ 3p)>0$, which is satisfied for many
physical cases, the Universe expands decelerating, as can be seen
from the second Friedmann's equation which would imply $\ddot
a<0$;

\item For $k=-1,0$ the first equation tells us that, for $\rho>0$,
the Universe keeps expanding forever, whereas for $k=1$ there can
be a value of $a$, for which the curvature term compensates the
energy density term, and $\dot a=0$. After this time, $a$
decreases and the Universe re-collapses.

\item It is not possible to have a closed Universe ($k=1$) that
re-collapses, if it is always accelerating. In fact, the second
Friedmann's equation, which is independent of $k$, for $(\rho
+3p)<0$, implies $\ddot{a}>0$. This is the case, for instance, of
vacuum domination ($w=-1$).
\end{itemize}
Let us now present in table 2, the solutions of Friedmann's
equations in the simplest case $k=0$ in which the curvature term
vanishes:
\begin{center}
\begin{tabular}{c||c|c|c}
  & $w$ & Energy density & Scale factor \\
  \hline\hline
  Matter &  $w=0$  & $\rho\sim a^{-3} $ & $ a(t)\sim t^{2/3} $ \\
  Radiation & $ w=\frac{1}{3}$ & $ \rho\sim a^{-4} $ & $a(t)\sim t^{1/2}$ \\
  Vacuum ($\Lambda$) & $ w=-1$ & $ \rho\sim \Lambda/(8\pi G_N)$ & $a(t)\sim \exp(\sqrt{\Lambda/3} t)$
\end{tabular}\\
\vspace{0.3cm}{{\bf Table {2}:} Solutions of the Friedmann's
equations in the case of a flat Universe ($k=0$).}
\end{center}

The value of $k$ can be determined experimentally by measuring the
parameter $\Omega$, defined as the ratio of the energy density of
our Universe and a critical density:
\begin{equation}
\Omega\equiv \frac{\rho}{\rho_c}\textrm{ \ \ \ with \ \ \ }\rho_c
\equiv \frac{3 H^2}{8\pi G_N}.
\end{equation}
In fact, with this definition, the first Friedmann's equation can
be rewritten as:
\begin{equation}
\Omega\ = 1 + \frac{k}{H^2 a^2},
\end{equation}
with a clear connection between the curvature of the spatial
sections and the departure from critical density. Then a flat
Universe ($k=0$) corresponds to $\Omega=1$, whereas an open
($k=-1)$ and closed ($k=1$) one corresponds to $\Omega < 1$ and
$\Omega>1$, respectively. In the case of multiple contributions to
the energy density of the Universe, we will have $\Omega=\sum_i
\Omega_i$.

Based on all these considerations, the Standard Cosmological Model
considers the early Universe as an expanding gas of particles in
thermal equilibrium, which underwent a cosmological evolution
through the following main steps \cite{CosmoReview, liddle}:
\begin{itemize}
\item Approximately 13 billion years ago, the Universe began expanding
from an almost inconceivably hot and dense state, until reaching
the present cold and sparse state after a long process of
continuous expansion and cooling.

\item The early Universe was a plasma of matter and radiation,
characterised by processes of creation and annihilation of
particles and anti-particle with the same interaction rate.
However, when the Universe became so cold to prevent the
production of certain kinds of particles, these species dropped
out of thermal equilibrium and are said to have frozen out. At
this point, a still poorly-understood process, called
\textit{baryogenesis}, should explain the observed asymmetry
between matter and antimatter.

\item When the Universe cooled down, the quarks bound together to
form hadrons. Subsequently, free protons and neutrons combined
together to give rise to various light elements such as $^2$H,
$^3$He, $^4$He, and $^7$Li. This process is called \textit{Big
Bang Nucleosynthesis} (BBN) \cite{BBN}, and took place during the
first 10 minutes or so. At this stage, the Universe was opaque to
light due to the very efficient absorption of photons by the large
number of free electrons.

\item Around a few thousand Kelvin, the Universe was cold enough
for free nuclei and electrons to begin to combine into atoms. This
process, called \textit{recombination}, occurred about $10^5$
years after the big-bang. Due to the formation of atoms, the
Universe became transparent and the light released at this time is
perceived today (after red-shifting by the Universe's expansion)
as the \textit{cosmic microwave background} (CMB). In addition, by
this time, dark matter had already begun to collapse into halos.

\item At approximately the same time when the Universe became transparent,
it also changed from being radiation dominated ($w=1/3$) to matter
dominated ($w=0$). After this, galaxies and stars began to form
when the baryonic gas and dust collapsed to the centre of the
pre-existing dark matter halos. The formation of these large scale
structures is probably due to the quantum fluctuations of the
early Universe, leading to our present time.
\end{itemize}
The main experimental successes of the Standard Cosmological Model
are the following ones:
\begin{itemize}
\item \textit{Hubble's law}. In 1929, after the observation of
the red-shift of the light emitted from distant galaxies, Hubble
formulated his famous law which consists in a linear relation
between the velocity and the distance of a galaxy with respect to
the Earth. This was the first empirical observation in favour of
the standard picture of an expanding Universe in continuous
deceleration.

\item \textit{Big-Bang Nucleosynthesis}. The relative abundance
of light elements predicted by Big-Bang Nucleosynthesis \cite{BBN}
matches the observational data with a stunning precision. The
percentages are approximately $75\%$ H, almost $24\%$ $^3$He, and
other light elements such as $^2$H and $^4$He, with small
fractions of a percent.

\item \textit{Cosmic Microwave Background}. In 1965,
Penzias and Wilson, in order to find the origin of an excess of
noise in their antenna, discovered by chance the cosmic microwave
background predicted by the Standard Cosmological Model. The CMB
presents a black-body spectrum at 2.7 Kelvin, with an extremely
uniform temperature in all the directions of our Universe.
However, in 1991 the COBE satellite observed small temperature
fluctuations in the CMB spectrum, signalling density fluctuations
in the sky corresponding to the presence of large scale
structures. Subsequent experiments, like BOOMERANG
\cite{BOOMERANG}, MAXIMA \cite{MAXIMA} and WMAP \cite{WMAP5},
provided very strong evidence for a flat Universe and measured
several fundamental parameters of FRW cosmology, such as $\Omega$,
$\Omega_{baryon}$, and the cosmological constant. New refined
measures of the properties of the CMB are going to be performed by
the PLANCK satellite which has been launched in May 2009.
\end{itemize}
Despite its observational and theoretical successes, the Standard
Cosmological Model is plagued by several severe problems, which we
list below:
\begin{itemize}
\item \textit{Initial singularity}. The initial singularity is a clear
sign of the breakdown of General Relativity. This is certainly the
main conceptual problem in cosmology, and it does not allow us to
answer the fundamental question of whether our Universe had a
beginning or not.

\item \textit{Flatness problem}. The recent results on the CMB
provide evidence for $\Omega\sim 1$. However, an extremely small
departure from $\Omega=1$ in the early Universe, would result in a
huge deviation from flatness today. Therefore, the flatness
problem is essentially a fine-tuning problem.

\item \textit{Horizon problem}. At leading order, the observed spectrum
of the CMB radiation coming from points in the space that do not
appear to be in causal contact with each other, is identical.
Therefore, the isotropy of the Universe turns out to be an
intriguing puzzle.

\item \textit{Origin of CMB anisotropies}. The Standard
Cosmological Model does not contain any explanation for the origin
of the observed CMB anisotropies, which are expected to be
produced from physics of the early Universe.

\item \textit{Baryogenesis}. The Standard
Cosmological Model does also not explain the observed excess of
matter over antimatter. It is known that baryogenesis can take
place only if the three Sakharov's conditions \cite{Sakharov}
(out-of-equilibrium decays, baryon number and CP violation) are
satisfied. However, this cannot be derived within the Standard
Model of particle physics.

\item \textit{Composition of our Universe}. Present observations of
the different contributions to the energy density of our Universe,
show that ordinary matter composed of protons, neutrons, and
electrons, (comprising gas, dust, stars, planets, people, etc)
accounts just for the $5\%$ of the total energy density
\cite{OmegaBaryon}. This leads to the dramatic consideration that
the nature of the remaining $75\%$ is presently unknown.

More precisely, there should be other two contributions to the
total energy density of our Universe:
\begin{itemize}

\item Cold Dark Matter: $\sim 25\%$

This is the so-called `missing mass' of the Universe
\cite{DarkMattero}. It comprises the dark matter halos that
surround galaxies and galaxy clusters, and should play an
important r\^{o}le in the explanation for the large scale
structure formation. It is said to be `cold' because it is
non-relativistic during the era of structure formation. The
Standard Model of particle physics does not contain any viable
dark matter candidate, which should, instead, correspond to a new
\textit{weakly interacting massive particle} or an axion.

\item Dark Energy: $\sim 70\%$

Recent results from the study of high red-shifted supernovae
\cite{supernovae} and the CMB \cite{WMAP5}, have independently
discovered that the expansion of the Universe appears to be
accelerating at present. This is due to some kind of `antigravity'
effect, called \textit{dark energy}, which provides $(\rho+3p)<0$,
and causes the Universe to accelerate. This acceleration could be
caused either by a time varying scalar field, named
`quintessence', or by the old Einstein's idea of an effective
\textit{cosmological constant}, whose present value is $\Lambda =
10^{-120} M_P^4 = (10^{-3}$ eV$)^4$. The fact that $\Lambda$ is
extremely close to zero but not vanishing, definitely requires an
explanation.
\end{itemize}
\end{itemize}

\subsection{Inflation}

The most compelling solution to the flatness and horizon problems
is achieved by requiring that in the first $10^{-34}$ seconds or
so, the Universe underwent a brief period of exponentially fast
expansion, known as \textit{inflation} \cite{inflation}. This
period would smooth out the Universe's original lumpiness and
leave it with the homogeneity and isotropy we see today, so
explain why some regions could be in causal contact with each
other.

Without any doubt, the most beautiful feature of inflation is that
it can render quantum effects visible in the sky. In fact, thanks
to inflation, quantum mechanical fluctuations during this process
could be imprinted on the Universe as density fluctuations
\cite{CMB}, which later seeded the formation of large scale
structures that are observed today as temperature fluctuations in
the CMB \cite{WMAP5}.

Moreover, most of the Grand Unified Theories beyond the Standard
Model predict the existence of topological defects, like
monopoles, whose presence would over-close the Universe. However,
inflation solves also this problem, since the exponential
expansion would dilute any unwanted relic.

This new cosmological paradigm has recently received a lot of
attention due to the striking agreement of its basic predictions
(a scale invariant, Gaussian and adiabatic CMB spectrum) with the
the measurements of the properties of the cosmic microwave
background radiation \cite{WMAP5}.

The simplest inflationary model is realised within an effective
field theory below $M_P$. It consists of a scalar field $\varphi$
with a potential $V(\varphi)$, whose value provides an effective
cosmological constant corresponding to the case $w=-1$, which
causes the scale factor $a(t)$ to increase exponentially.

However, there are still several unanswered questions regarding
inflation which can be answered only embedding it into an
underlying consistent theory of quantum gravity:
\begin{itemize}
\item What is the scalar field, called \textit{inflaton}, driving inflation?

\item Is the inflaton just driving inflation or it is also
generating the primordial quantum fluctuations? In the first case,
who is responsible for the density perturbations?

\item Is it possible to derive the inflaton's potential from first
principles?

\item Can we have control over the initial conditions of
inflation?

\item The best present model of inflation relies on the \textit{slow roll} of a scalar
field down a very shallow potential. This implies a very small
mass of the inflaton: $M_{inf}<H$, where $H$ is the Hubble
parameter during inflation. Given the notorious ultra-violet
sensitivity of scalar masses, how is it possible to keep the
inflaton's mass low protecting it from getting large quantum
corrections?

\item The ultra-violet sensitivity of inflation gets
even worse when one looks at possible models that would generate
observable tensor modes, since they require the inflaton to travel
a trans-Planckian distance in field space. Thus, in this case, the
contribution of Planck-suppressed operators cannot be neglected
anymore.
\end{itemize}
Finally it is worth mentioning that, although inflation is by far
the most compelling explanation for the observed properties of the
early Universe, it has nothing got to say about the resolution of
the initial singularity or the nature of dark energy. Moreover,
its status is the subject of some debate, and most authors find
that some degree of fine-tuning appears to be necessary. Thus, it
is worth keeping in mind the possibility of an entirely different
solution to the Big Bang problems (like pre-Big Bang \cite{pbb} or
ekpyrotic/cyclic scenarios \cite{ekpyrotic}).

\section{String Theory}

String theory is at present the most promising candidate for a
consistent theory of quantum gravity, which is also able to
incorporate all the four known interactions and matter in a
beautiful unified framework \cite{ScherkSchwarz, GreenSchwarz,
GSW, PolchinskiBook}. In this sense, it differs from all the
proposed ideas for physics beyond the Standard Model, since it
addresses precisely the toughest of all issues: the quantisation
of gravity. In fact, string theory provides an ultra-violet
completion of General Relativity, which is finite order by order
in perturbation theory.

Both four dimensional Einstein gravity and Standard Model-like
theories with chiral fermions can be obtained as a low-energy
effective theory for energies below a typical scale, the string
scale $M_s$. Furthermore, all the previous proposals for physics
beyond the Standard Model can be embedded into string theory,
which represents also the natural framework where inflation should
be derived.

As a theory of quantum gravity, string theory should be able to
provide an answer to most, if not all, of the fundamental
questions beyond the Standard Cosmological Model. On the other
hand, as a theory underlying gauge interactions, it has the
potential to solve crucial problems in particle physics, like the
number of families, the gauge group and the observed couplings,
the origin of chirality, the physics of supersymmetry breaking,
etc.

Furthermore, string theory has the beautiful feature that there
are no fundamental constants: it is the dynamics of the theory
that selects a particular vacuum state which, in turn, determines
all the masses and couplings.

String theory has an extremely rich structure, from the
mathematical, theoretical and phenomenological points of view. In
fact, the attempt to try to understand it in depth, has led to
many profound results, like the discovery of mirror symmetry
\cite{CLS, GreenePlesser}, an exact microscopic calculation of the
Bekenstein-Hawking black hole entropy \cite{BHinST}, a smooth
description of space-time topology change \cite{topchange1,
topchange2, topchange3, topchange4} and the AdS/CFT correspondence
\cite{AdSCFT}.

Despite these successes, it has to be said that, at present, the
theory still lacks any experimental evidence and a decisive
low-energy test of string theory does not seem possible.

\subsection{Perturbative strings and dualities}

String theory can be introduced as the theory describing tiny
one-dimensional objects that, moving through $D$-dimensional
space-time, sweep out two-dimensional `world sheets' $\Sigma$
which may be viewed as thickened Feynman diagrams. The
two-dimensional field theories living on such Riemann surfaces
define different perturbative string theories whose building
blocks are free two-dimensional bosons and fermions, $X_i(z)$ and
$\psi_i(z)$, corresponding to the bosonic and fermionic
coordinates of the string in $D$-dimensional space-time.
Combinations of these two-dimensional fields give rise to a finite
number of massless space-time fields, plus an infinite tower of
stringy excitations with arbitrary high masses and spins. The
level spacing of the excited states is governed by the string
scale, $M_s$, which gives also the size of the string.
Schematically, the most generic massless operators are ($\mu$,
$\nu=1,...,D$):
\begin{equation}
\left\{
\begin{array}{c}
\eta^{\mu \nu }\bar{\partial}X_{\mu}\partial X_{\nu}\longleftrightarrow\phi
\textrm{, \ \ \ \ \ \ dilaton} \\
\bar{\partial}X_{\mu}\partial X_{\nu}\longleftrightarrow
g_{\mu \nu}\textrm{, \ \ \ \ \ \ graviton} \\
\bar{\partial}X_{\mu}\partial X^a
\longleftrightarrow A_{\mu}^a\textrm{, \ \ gauge boson} \\
\bar{\partial}X^a\partial X^b\longleftrightarrow
\Phi^{ab}\textrm{, \ \ \ \ Higgs field}
\end{array}
\right.
\end{equation}
Moreover, two dimensional supersymmetry implies the existence of
also space-time fermions (`electrons' and `quarks'), which are
obtained by taking anti-periodic boundary conditions of the
$\psi_i(z)$ on $\Sigma$. Given that from the two-dimensional point
of view, the nature of all these operators is very similar, string
theory provides a beautiful unified description of all
interactions and matter. Moreover, due to the presence in the
massless spectrum of a spin-2 particle \cite{ScherkSchwarz}, the
graviton, we realise that the first `prediction' of string theory
is the existence of gravity.

The fact that string theory contains both General Relativity and
non-abelian gauge theories (plus stringy corrections strongly
suppressed by inverse powers of $M_s$), can de inferred by looking
at the perturbative effective action:
\begin{eqnarray}
S_{eff}^{(D)}(A_{\mu },g_{\mu \nu },...) &=&\sum_{\Sigma _{\gamma
}}e^{-\phi \chi (\Sigma _{\gamma })}\int_{M(\Sigma _{\gamma
})}\int d\psi d\chi ...e^{\int d^{2}z\mathcal{L}_{2D}(\psi ,\chi
,...,A_{\mu },g_{\mu \nu },...)} \nonumber
\\
&=&\int d^{D}x\sqrt{-g}e^{-2\phi }\left[ R+\textrm{Tr}F_{\mu \nu
}F^{\mu \nu }+... \right] +\mathcal{O}(M_{s}^{-1}), \label{EFAC}
\end{eqnarray}
where $A_{\mu },g_{\mu \nu },...$ are space-time fields providing
the background in which the string moves. The weighted sum in the
expression (\ref{EFAC}) is over all the possible two-dimensional
world-sheets $\Sigma_{\gamma}$, and corresponds to the usual loop
expansion of quantum field theory. The field $\phi$ is the dilaton
and its VEV sets the coupling constant for the perturbation
series, $g_s=\langle e^{\phi}\rangle$, whereas the Euler number
$\chi (\Sigma_{\gamma})$ is the analog of the loop-counting
parameter. It is interesting to notice that there is only one
`diagram' at any given order in string perturbation theory, which
comes along with integrals over all the possible inequivalent
shapes of the Riemann surface, corresponding to the momentum
integrations in ordinary quantum field theory. This string loop
expansion does not contain any ultra-violet divergence, so making
the theory a perfect candidate for a consistent treatment of
quantum gravity.

The two-dimensional field theory is characterised by
\textit{conformal invariance}, which is needed for its consistency
and requires the space-time to have $D=10$ dimensions. In
addition, the conformal symmetry, via Ward identities, implies
general coordinate and gauge invariance in space-time. This is the
reason why we could obtain an effective action of the form
(\ref{EFAC}).

By combining in various ways the building blocks of the
world-sheet field theories, one obtains five ten-dimensional
string theories which have completely different spectra, number of
supersymmetries and gauge symmetries at the \textit{perturbative
level}:
\begin{center}
\begin{tabular}{c||c|c}
  & Gauge Group & Supersymmetry \\
  \hline\hline
  Type IIA & $U(1)$ & non-chiral $N=2$ \\
  Type IIB & - & chiral $N=2$ \\
  Heterotic & $E_8\times E_8$ & chiral $N=1$ \\
  Heterotic' & $SO(32)$ & chiral $N=1$ \\
  Type I (open) & $SO(32)$ & chiral $N=1$
\end{tabular}\\
\vspace{0.3cm}{{\bf Table {3}:} Five perturbatively different
ten-dimensional string theories.}
\end{center}

At this point, the natural question to ask is: If one of these
five different theories in $D=10$ would be the fundamental theory,
what is then the r\^{o}le of the others? The answer to this
question relies on non-perturbative effects and the notion of
`duality' \cite{duality}, which open up a completely new
perspectives on the very nature of string theory.

Duality is a map between solitonic (non-perturbative, non-local,
`magnetic') degrees of freedom, and elementary (perturbative,
local, `electric') degrees of freedom. Typically, duality
transformations exchange weak and strong-coupling physics acting
on coupling constants like $g\to 1/g$.

Let us illustrate this concept in the simple example of
four-dimensional $N = 2$ $SU(2)$ gauge theory \cite{YMduality,
YMduality2}. The moduli space $\mathcal{M}$ is spanned by the VEV
of a Higgs field $\phi$, which determines also the holomorphic
effective gauge coupling $g(\phi)$. Moving around in $\mathcal{M}$
will change the vacuum state, and so also the value of $g(\phi)$
and the mass spectrum.

In the weak coupling semi-classical region near
$\langle\phi\rangle\to\infty$, the effective gauge coupling
becomes arbitrarily small, $g(\phi)\to 0$, and the perturbative
definition of the gauge theory is arbitrarily good. The
non-perturbative effects are strongly suppressed and the solitonic
magnetic monopoles become so heavy that they effectively decouple.
However, when we move in $\mathcal{M}$ towards the strong coupling
region $g(\phi)\to \infty$, the original perturbative definition
does not make sense, since the contribution of non-perturbative
instantons cannot be neglected anymore. The crucial observation is
that the inverse of $g(\phi)$, $g_D=1/g$, yields another
expression for the `dual' gauge coupling that is well defined near
$1/g\to 0$. Indeed, in this region, the infinite instanton series
for the dual coupling $g_D$ converges very well.

We realise that, in different regions of the moduli space, the
\textit{same} theory can be described by different perturbative
approximations in terms of different weakly coupled local degrees
of freedom. In fact, at weak coupling, there is a perturbative
formulation based on the $SU(2)$ gauge symmetry, while at strong
coupling, we have a $U(1)$ perturbative description with some
extra massless matter fields (`electrons'), which in the original
variables, correspond to some of the solitonic magnetic monopoles
that become light at strong coupling.

The analogs of these magnetic monopoles in string theory are given
by `$p$-branes' \cite{PolchinskiBook}. These are
$(p+1)$-dimensional extended objects, with $p=0,1,...,9$ space and
one time dimensions, that can wrap around $p$-dimensional cycles
$\gamma_p$ of a compactification manifold. In the singular regions
of the moduli space where such cycles shrink to zero size,
corresponding to the strong coupling regime of the field theory
living on the brane (since $g^2=1/\textrm{Vol}(\gamma_p)$), a
$p$-brane wrapped around $\gamma_p$, will give a massless state in
four dimensions \cite{topchange3}. On the other hand, in the limit
when $\gamma_p$ becomes very large, these states become
arbitrarily heavy and eventually decouple. In this case, the
relevant objects dual to certain solitonic states are special
kinds of $p$-branes, called `$Dp$-branes' \cite{DbraneDiscovery}.

In view of the previous remarks, we now point out that the five
perturbatively different string theories turned out to be all
connected by various dualities \cite{dualityWeb, dualityReview}.
Hence they represent just different approximations of a unique
underlying theory, called `M-theory' \cite{M1, M2}, whose deep
structure has not been unveiled yet.

\subsection{Compactification of extra dimensions}

As we have seen, string theory requires the existence of extra
space-time dimensions, and so the study of string
compactifications is a crucial issue to make contact with our real
world. We shall, therefore, assume that the space-time manifold is
not simply $\mathbb{R}^{1,9}$, but $\mathbb{R}^{1,3}\times Y_6$,
where $Y_6$ is some compact six-dimensional manifold.

It has to be said that there is no reason why a ten dimensional
theory wants at all to compactify down to $D=4$, since many
choices of space-time background vacua of the form
$\mathbb{R}^{10-n}\times Y_n$ appear to be on equal footing. It is
very likely that only cosmology will be able to give an answer to
this fundamental question.

The compactification is usually demanded to yield a supersymmetric
low energy effective theory mainly for two reasons:
\begin{itemize}
\item The supersymmetry breaking scale is supposed to be low;

\item Supersymmetry simplifies the calculations
(for example the holomorphicity of the superpotential and its
non-renormalisation).
\end{itemize}
The requirement of obtaining $N=1$ supersymmetry in $D=4$, forces
the internal manifold $Y_6$ to be a very complicated `Calabi-Yau'
space \cite{GreeneReview, HubschBook}. In reality, there is a
large number of choices for the compactification manifold, since
the number of Calabi-Yau spaces is perhaps of the order $10^4$.
Given that the particle content and forces in four dimensions are
determined by the topology of the extra dimensions, each of these
spaces leads to a different four dimensional physics. Then the
conceptual problem is that there does not seem to be a good way of
choosing between different vacua. At the moment, the best solution
is based only on a selection principle \cite{landscape0,
landscape1, landscape2, landscape3}.

Moreover, string compactifications are characterised by the
ubiquitous presence of moduli, which parameterise the shape and
the size of the extra dimensions. For a typical compactification
manifold, the number of these parameters can easily approach the
order of several hundreds. These moduli are massless uncharged
scalar particles which would mediate unobserved long-range fifth
forces due to their effective gravitational couplings to all
ordinary particles. Hence it is of primary importance to develop a
potential for these particles, and so to give them a mass. This
problem goes under the name of \textit{moduli stabilisation}. This
issue is also crucial to understand all the main features of the
low-energy effective field theory, since both the gauge and the
Yukawa couplings depend on the vacuum expectation value (VEV) of
the moduli. There is also a cosmological constraint over the
moduli masses, $m_{mod}\geq 10$ TeV, so that they decay before
baryogenesis evading the cosmological moduli problem \cite{CMP1,
CMP2, CMP3}.

The other fundamental task to make contact between string theory
and the real world, is to find a string compactification that
reproduces the Standard Model of particle physics. This means that
one should be able to reproduce the correct scales, gauge group,
chiral spectrum and Yukawa couplings. This problem is often called
\textit{model building}. String phenomenologists are also very
interested in building a supersymmetric extension of the Standard
Model and would like to embed these constructions in a framework
where gauge coupling unification is obtained as well.

As we have seen in the previous section, there are five different
perturbative string theories, all existing in ten dimensions and
all related by strong-weak coupling dualities. Therefore in order
to study string compactifications, one has first to choose which
ten-dimensional string theory wants to consider.

Historically the string theory that attracted most of the
attention was the heterotic $E_8\times E_8$ because it was the
most promising for phenomenology: upon Calabi-Yau compactification
to four dimensions it gives rise to $N=1$ chiral models with an
observable sector, coming from the first $E_8$, which contains the
Standard Model symmetry and several families of matter fields
\cite{heteroticMatter}. The second $E_8$ gives rise to a hidden
sector, which was proposed to break supersymmetry, in the attempts
of supersymmetric model building prior to string theory. Then the
Calabi-Yau breaks $E_8\to E_6$ which contains the Standard Model
gauge group or GUT's generalisations like $SU(5)$ or $SO(10)$
($SU(3)\times SU(2)\times U(1)\subset SU(5)\subset SO(10)\subset
E_6\subset E_8$) \cite{hetmodelbuilding}.

However the moduli stabilisation issue has always been a great
problem for heterotic strings. Even though there are some
solutions for simple examples mainly based on gaugino condensation
\cite{heteroticmodstab, DRSW, bqq, hetModStab}, a general deep
understanding of this issue is still lacking. Moreover, despite
all the promising features of these compactifications, no one has
managed yet to derive exactly the Standard Model from the
heterotic string. There are however some Standard Model-like
constructions which usually come along with exotic matter and
anomalous extra $U(1)$'s \cite{Zprime}. Heterotic string
phenomenology is still today an active area of research.

The other four string theories seemed much less interesting; for
example, it seemed to be impossible to obtain the Standard Model
out of type II string theories due to the absence of non-Abelian
gauge symmetries in their low-energy limit. On top of that, there
was also a no-go theorem preventing the turning on of background
fluxes \cite{nogo}, which were interesting candidate energy
sources to stabilise the moduli. Hence the other main problem of
type II theories was also their vacuum degeneracy.

However the discovery of $D$-branes by Polchinski in 1995
\cite{DbraneDiscovery}, opened up the possibility, not only to
understand the intricate web of dualities of M-theory, but it also
allowed to completely change our view of type II compactifications
for the following reasons:
\begin{itemize}
\item $D$-branes provide a new origin of non-Abelian gauge
symmetries and chirality. Gauge and matter fields are open strings
whose end-points are constrained to move on the brane
\cite{openStrings, IntersectingDbranes};

\item $D$-branes represent exceptions to the existing no-go theorem \cite{nogo},
so allowing to turn on background fluxes. This is a key-point to
be able to solve the moduli stabilisation problem \cite{gkp};

\item $D$-branes allow for a stringy realisation of the
`brane-world scenario' with large extra dimensions
\cite{braneworld}. The Standard Model lives on a particular
$D$-brane whereas the closed string sector, including gravity and
the dilaton, probes all the extra dimensions;

\item $D$-branes provide also new cosmologically
interesting degrees of freedom \cite{DDinflation, BBbarInfl}.
\end{itemize}
The reason why internal fluxes play a key-r\^{o}le to fix the
moduli in type II theories is that their back-reaction on the
Calabi-Yau geometry just renders the compactification manifold
conformally Calabi-Yau. On the contrary, the use of fluxes to
freeze out the moduli, in the heterotic case, is still poorly
understood since their back-reaction destroys the Calabi-Yau
background. Hence it is very difficult to have control over the
internal geometry. This is the main reason why moduli
stabilisation is better understood in the context of type II
theories which have, therefore, recently received so much interest
worldwide.

\subsection{String phenomenology}

Historically type II string phenomenologists focused their studies
either on local (brane) or global (bulk) properties of their
constructions with the general belief that the two issues were
almost independent. Hence each problem has been studied separately
assuming that a solution of the other would be found in an
independent way. This approach was, in a certain sense, justified
also by the intrinsic difficulty to solve each problem. Let us
summarise the main features of these two different approaches in
string phenomenology:
\begin{itemize}
\item \textit{Global models}. One insists in having a complete
compact Calabi-Yau compactification with, for example, full moduli
stabilisation, Ramond-Ramond tadpole cancellation and Freed-Witten
anomaly cancellation, being consistent at the global level. This
is often also called a \textit{top-down} approach.

\item \textit{Local models}. One considers local sets of lower
dimensional $Dp$-branes, which are localised in some area of the
Calabi-Yau and reproduce Standard Model physics. One does not then
care about global aspects of the compactification and assumes that
eventually the configuration may be embedded inside a fully
consistent global model. This is often called a \textit{bottom-up}
approach.
\end{itemize}
The latter \textit{bottom-up} approach is not available in
heterotic or type I models, since the Standard Model fields live
in the bulk six dimensions of the Calabi-Yau. In principle, a
globally consistent compactification is more satisfactory. On the
other hand, local configurations of $Dp$-branes may be more
efficient in trying to identify promising string vacua,
independent of the details of the global theory. The main examples
of \textit{local} constructions of the Standard Model via
$D$-branes are:
\begin{itemize}
\item Intersecting $D6$-branes in type IIA, where non-zero angles
between the branes give rise to chirality
\cite{IntersectingDbranes}. The Standard Model lives on $D$-branes
and gravity in the bulk, realising the brane world scenario;

\item Magnetised $D7$-branes in type IIB. This construction is
T-dual to the previous one. Open strings stretched between
differently magnetised (via turning on a non-zero field strength
on the brane world-volume) $D7$-branes give a chiral spectrum
\cite{blum};

\item $D3$-branes at singularities in type IIB \cite{DbranesAtSingularit,
Dbranesatsingluarities, Dbranesatsingluarities2, CMQ}. This case
can be considered as a limiting class of magnetised branes
wrapping cycles which are collapsed at a Calabi-Yau singularity.
Realistic models with $N=1,0$ supersymmetry have been found with
the small string scale $M_s\sim 10^{12}$ GeV or even $M_s\sim 1$
TeV. We mention that also in compactifications of M-theory on
$G_2$ holonomy manifolds, in order to have chiral fermions, the
matter needs to be at singular points \cite{G2}.
\end{itemize}
Some of the main issues of string compactifications that have to
be studied globally (bulk properties) and locally (brane
properties) are the following:
\begin{center}
\begin{tabular}{c|c}
Local (brane) issues & Global (bulk) issues \\ \hline\hline
Standard Model gauge group & Moduli stabilisation \\
Yukawa couplings & Supersymmetry breaking \\
Chiral spectrum & Soft terms \\
Gauge coupling unification & Cosmological constant \\
Mixing angles & Inflation \\
Hierarchies? & Hierarchies? \\
Proton stability &  \\
Baryogenesis &  \\
Reheating &
\end{tabular}\\
\vspace{0.3cm}{{\bf Table {4}:} Global and local aspects of string
compactifications.}
\end{center}
We wrote down the issue `hierarchies' on both sides with a
question mark, to stress the fact that it is not clear how to
solve this issue. In fact, in Nature there are several hierarchies
which demand an explanation (the only fully understood is the QCD
scale!):
\begin{itemize}
\item Planck scale: $M_P=2.4\cdot 10^{18}$ GeV,

\item GUT/Inflation scale: $M_{GUT}\sim 10^{16}$ GeV,

\item QCD axion (decay constant) scale\footnote{Assuming a solution of the strong CP problem via a
Peccei-Quinn axion \cite{PecceiQuinn1, PecceiQuinn2}.}:
$10^9\textrm{ \ GeV} \leq f_a \leq 10^{12}$ GeV,

\item Soft terms (masses of superpartners): $M_{soft}\sim 1$ TeV,

\item Weak scale: $M_W\sim 10^2$ GeV,

\item QCD scale (masses of hadrons): $\Lambda_{QCD}\sim 200$ MeV,

\item Neutrino mass scale: $0.05\textrm{\ eV} \leq m_{\nu} \leq 0.3$
eV,

\item Cosmological constant: $\Lambda\sim 10^{-120} M_P^4\sim\left(10^{-3}\textrm{\ eV}\right)^4$.
\end{itemize}
In this thesis, we shall consider the solution of the $M_P$ versus
$M_W$ hierarchy problem as a global aspect of the string
compactification \cite{LVS}. In fact, we shall make use of the
presence of exponentially large extra dimensions. The main scales
of a string compactification in terms of the internal volume
$\mc{V}$, take the following form \cite{LVS, LVS2, LVSGUT}:
\begin{eqnarray}
M_s &\sim& \frac{M_P}{\sqrt{\mc{V}}},\textrm{ \ \ \ \ \ \ \ }
M_{GUT} \sim \mathcal{V}^{1/6} M_s\sim  \frac{M_P}{\mathcal{V}^{1/3}}, \nonumber \\
M_{KK}&\sim&
\frac{M_s}{R}\sim\frac{M_s}{\mc{V}^{1/6}}\sim\frac{M_P}{\mc{V}^{2/3}},
\textrm{ \ \ \ \ \ }M_{soft}\sim m_{3/2} \sim
\frac{M_P}{\mathcal{V}}. \nonumber
\end{eqnarray}
Therefore a volume of the order $\mc{V}\sim 10^{15}$, gives rise
to TeV-scale supersymmetry ($M_{soft}\sim m_{3/2} \sim 1$ TeV),
even though the standard picture of gauge coupling unification
would be destroyed by a too low GUT scale: $M_{GUT}\sim 10^{13}$
GeV. The string scale would be intermediate \cite{Msintermediate},
$M_s\sim 10^{10}$ GeV, and the Kaluza-Klein scale a bit lower:
$M_{KK}\sim 10^8$ GeV.

In this thesis we shall focus mainly on the problem of moduli
stabilisation in type IIB string theory and the application of its
solution to cosmology (for moduli fixing in type IIA see
\cite{IIAmodst}). In fact, all the global aspects of string
compactifications depend on the solution to moduli stabilisation.
The final goal is to find solutions of all these problems via
building local explicit realistic models, and embedding them in a
global model independent framework which solves all the bulk
issues listed above.

\subsection{The type IIB LARGE Volume Scenario}

Taking the low-energy limit of type IIB string theory below $M_s$,
one obtains ten-dimensional type IIB supergravity, which has $N=2$
supersymmetries and whose field content is given by the massless
degrees of freedom of the string. We then compactify six of the
ten dimensions and perform the further low-energy limit below the
compactification scale $M_{KK}$, keeping only the zero-modes of
the Kaluza-Klein tower associated to each of the ten-dimensional
states. Zero modes correspond to zero eigenvalues of a particular
differential operator, which, for example, in the case of scalar
fields, is the ordinary Laplacian $\nabla^2$. We end up in
four-dimensional type IIB supergravity with $N=8$ supersymmetries.
Choosing the internal space to be a Calabi-Yau three-fold, the
number of supersymmetries is reduced from $N=8$ to $N=2$. We
finally end up with a $N=1$ theory by taking an orientifold
projection of the Calabi-Yau three-fold.

The moduli present in this low-energy effective field theory can
be classified according to their nature of closed or open string
modes. Examples of closed string moduli are:
\begin{enumerate}
\item The axio-dilaton whose VEV gives the string coupling,

\item Complex structure moduli parameterising the shape of the extra
dimensions,

\item K\"{a}hler moduli parameterising the size of the extra dimensions.
\end{enumerate}
Examples of open string moduli are:
\begin{enumerate}
\item $D3$-brane position moduli,

\item $D7$-brane deformation moduli,

\item Wilson lines.
\end{enumerate}
Given that we are focusing now on global properties of string
compactifications, we shall ignore open string moduli since their
presence and features depend on the particular local model one
decides to consider.

In the absence of internal fluxes, the superpotential is
vanishing. Hence we obtain a completely flat potential for all the
moduli. This is the reason why these particles are called moduli.

Historically there was a no-go theorem \cite{nogo} claiming that
it was impossible to have non-zero fluxes due to the tadpole
cancellation condition, given that the fluxes were the only
semi-positive definite contributions. However the discovery of
branes led to the possibility to overcome this no-go theorem (due
to the presence of local sources with negative $D3$-brane charge,
like $D7$-branes and $O3$-planes), and allow for non-vanishing
background fluxes \cite{gkp, drs}, which, in turn, generate a
semi-classical\footnote{Because the fluxes are quantised.}
superpotential.

Hence both the dilaton and the complex structure moduli can be
stabilised at tree level by turning on background fluxes. By
appropriate fine-tuning of the internal fluxes, one can always fix
the dilaton such that the string coupling is in the perturbative
regime. Given that there is no constraint on the choice of fluxes,
except for the tadpole cancellation condition, one is free to vary
these fluxes, generating an enormous number of different vacua,
which form the famous string landscape \cite{fluxlandscape0,
fluxlandscape1, fluxlandscape2, fluxlandscape3}.

Semi-classically, the potential for the K\"{a}hler moduli is still
flat due to the \emph{no-scale structure} \cite{noscale}, which is
the typical feature of these models. The tree-level flatness of
the potential for the K\"{a}hler moduli implies that, to study
K\"{a}hler moduli stabilisation, we should keep all possible
perturbative and non-perturbative corrections, while the dilaton
and the complex structure moduli can be integrated out at
tree-level \cite{IntOut}.

The respective order of magnitude of the perturbative versus the
non-perturbative corrections to the scalar potential is set by the
flux-dependent tree-level superpotential $W_0$. For natural values
of $W_0\sim \mathcal{O}(1)$, the perturbative corrections are
leading with respect to the non-perturbative ones \cite{bb}.

However historically, the first example of K\"{a}hler moduli
stabilisation \cite{kklt} considered only non-perturbative
corrections to the superpotential, and so it has been necessary to
fine-tune $W_0$ extremely small.

A more promising K\"{a}hler moduli stabilisation mechanism should
work without fine-tuning $W_0$. Therefore it has been put some
effort on trying to perform a pure perturbative stabilisation of
the K\"{a}hler moduli \cite{bhk2, hg, para}. However, also this
attempt turned out to require fine-tuning in the complex structure
sector.

The last possibility to fix the K\"{a}hler moduli naturally seems
to be by setting $W_0\sim \mathcal{O}(1)$, and then requiring
non-perturbative corrections to compete with the perturbative
ones, which include $\alpha'$ effects \cite{bbhl} and quantum
string loops (see \cite{bhk, bhp} and chapter 5 of this thesis).
This procedure was first investigated in \cite{LVS} neglecting
$g_s$ corrections, and led to the discovery of string
compactifications with exponentially large volume. This is the
main topic of this thesis and we will refer to those constructions
as LARGE Volume Scenarios (LVS)\footnote{The capitalisation of
LARGE is a reminder that the volume is exponentially large, and
not just large enough to trust the supergravity limit.}.

In Part II of this thesis, we shall give a general analysis of LVS
presenting the topological conditions on an arbitrary Calabi-Yau
three-fold under which the scalar potential admits a minimum at
exponentially large volume. Supersymmetry is broken through the
minimisation procedure thanks to non-vanishing background fluxes.

The two main conditions are the negativity of the Euler number of
the Calabi-Yau manifold and the presence of at least one del Pezzo
4-cycle (blow-up mode) resolving a point-like singularity. Then,
all the blow-up modes can be fixed at values large with respect to
the string scale by the interplay of non-perturbative and
$\alpha'$ corrections, which stabilise also the overall volume
mode exponentially large. All the other K\"{a}hler moduli, such as
those corresponding to fibrations, are not fixed by these effects
but by the inclusion of string loop corrections, which are always
dominant with respect to the non-perturbative ones (if they are
present).

The main advantages of LVS are the following:
\begin{enumerate}
\item There is a general analysis on the topological conditions
that an arbitrary Calabi-Yau has to satisfy to give rise to these
models (see Part II of this thesis);

\item All the possible perturbative and non-perturbative
corrections play a crucial r\^{o}le, which is largely understood
(see Part II of this thesis and \cite{LVSAalok});

\item The tree-level superpotential $W_0$ is not fine-tuned;

\item The existence of an exponentially large volume,
$\mathcal{V}\gg 1$, makes the effective field theory treatment
robust;

\item The exponentially large volume allows for a natural
explanation of many hierarchies we observe in Nature, which come
as different powers of $\mathcal{V}$ (examples are the weak
\cite{LVS}, the QCD axionic \cite{joe} and the right-handed
neutrino mass scale \cite{neutrino});

\item There is a two-fold possibility to make contact with
experiments via either particle phenomenology or cosmology:

a) Particle Phenomenology:
\begin{itemize}
\item Computation of moduli mass spectroscopy (see Part III of this thesis and
\cite{LVS2});

\item Study of supersymmetry breaking mediation mechanisms (with gravity
mediation that is leading with respect to anomaly mediation)
\cite{LVS2};

\item Computation of soft terms at $M_s$ with flavour
universality at leading order \cite{SoftSUSY, joemirror};

\item Running of the soft supersymmetry breaking terms down at $M_{EW}$
generating sample particle spectra for the LHC \cite{LVSatLHC}.
\end{itemize}
b) Cosmology:
\begin{itemize}
\item Derivation of a natural model of inflation, called \textit{Fibre Inflation},
with the prediction of observable gravity waves in the case of K3
fibrations by using string loop corrections (see Part III of this
thesis);

\item Some moduli, like the overall volume mode or large fibration
moduli, could form part of dark matter (see \cite{CQ} and Part III
of this thesis);

\item The 511 KeV line coming from the centre of our galaxy could
be interpreted as due to the decay of the overall volume mode to
$e^+ e^-$. The decay of $\mathcal{V}$, and possible large
fibration moduli, to photons should also give rise to clear new
monochromatic lines (see \cite{CQ} and Part III of this thesis);

\item Finite-temperature corrections to the scalar potential set an
upper bound on the temperature of our Universe, in order not to
fall into a decompactification limit. This tends to prefer larger
values of $\mathcal{V}$ of the order $\mathcal{V}\sim 10^{15}
l_s^6$ (see Part III of this thesis).
\end{itemize}
\end{enumerate}
The previous considerations would indicate a preferred value for
the overall volume of the order $\mathcal{V}\sim 10^{15} l_s^6$.
In fact, in this case, the mass scales of LVS would look
like\footnote{Assuming non-perturbative stabilisation of the cycle
supporting the MSSM.}:
\begin{itemize}
\item Planck scale: $M_P=2.4\cdot 10^{18}$ GeV,

\item Majorana scale for right handed neutrinos:
$M_{\nu_R}\sim\frac{M_P}{\mathcal{V}^{1/3}}\sim 10^{14}$ GeV,

\item String scale: $M_s\sim \frac{M_P}{\mathcal{V}}\sim 10^{10}$ GeV,

\item QCD axionic scale: $f_a\sim M_s\sim 10^{10}$ GeV

\item Kaluza-Klein scale: $M_{KK}\sim
\frac{M_P}{\mathcal{V}^{2/3}}\sim 10^8$ GeV,

\item Blow-up modes: $m_{\tau_s}\sim
m_{3/2}\ln\left(M_P/m_{3/2}\right)\sim 10^6$ GeV,

\item Gravitino mass: $m_{3/2}\sim \frac{M_P}{\mathcal{V}}\sim 10^4$
GeV

\item Complex structure moduli: $m_{U}\sim m_{3/2}\sim 10^4$ GeV,

\item Soft supersymmetry breaking terms: $M_{soft}\sim
\frac{m_{3/2}}{\ln\left(M_P/m_{3/2}\right)}\sim 10^3$ GeV,

\item Volume modulus: $m_{\tau_{big}}\sim
\frac{M_P}{\mathcal{V}^{3/2}}\sim 1$ MeV,

\item Large fibration moduli:
$m_{\tau_{fib}}\sim\frac{M_P}{\mathcal{V}^{5/3}}\sim 10$ KeV.
\end{itemize}
However, if the volume is set such that $\mathcal{V}\sim 10^{15}
l_s^6$, some shortcomings of these models are the following:
\begin{enumerate}
\item \textit{Tension between phenomenology and cosmology}, given that in
the inflationary scenario mentioned above, it is possible to match
the COBE normalisation for density fluctuations only if
$\mathcal{V}\sim 10^4 l_s^6$, but in this case we do not get
TeV-scale supersymmetry anymore. A possible solution to this
problem has been proposed in \cite{cklq}, assuming that the
overall volume mode is at values of the order $\mathcal{V}\sim
10^4 l_s^6$ during inflation but then, when the slow roll
conditions are not satisfied anymore, it rolls down towards larger
values of the order $\mathcal{V}\sim 10^{15} l_s^6$. However this
is a fine-tuned scenario. Another possibility would be to improve
the \emph{Fibre Inflation} model (see Part III of this thesis) by
having the inflaton that just drives inflation but does not
generate the primordial density fluctuations, which, on the
contrary, would be generated by another field that plays the
r\^{o}le of a curvaton \cite{curvaton}. In this case, it is likely
that it would be possible to obtain inflation and set, at the same
time, $\mathcal{V}\sim 10^{15} l_s^6$. However, tensor modes would
not be observable anymore, even though possible large
non-gaussianities could be produced \cite{NGcurvaton}.

\item \textit{No gauge coupling unification}, given that the string scale in
this case is intermediate, and so ruins the standard picture of
the running of all three non-gravitational coupling constants
which merge around $M_{GUT}\sim 10^{16}$ GeV. It has to be said
that, in general, in all the five perturbative string theories, it
is extremely difficult to derive an explicit string model which is
able to reproduce the standard picture of gauge coupling
unification.

\item \textit{Cosmological Moduli Problem} for the overall volume mode, and
possible lighter modes, given that these particles would storage
energy just after the end of inflation, and then, since they are
very light, would not decay before Big-Bang nucleosynthesis (BBN)
or would still be present in our Universe nowadays \cite{CQ}.
Therefore they could either destroy the good predictions of
standard BBN or overclose the Universe. There are two main
solutions to this problem: dilution of these moduli by the entropy
released by the non-thermal decay of a heavier modulus which is
dominating the energy density of the Universe \cite{Acharya,
Vafa}, or dilution by a low-energy period of thermal inflation
\cite{TI}. In Part III of this thesis, we shall show that the
first possibility can never happen in LVS with two K\"{a}hler
moduli; in addition, the study of the thermal potential for closed
string moduli revealed that in order to study if it can give rise
to any period of inflation, one has to go beyond the effective
field theory. In addition the possibility to have thermal
inflation in the open string moduli sector has still to be
studied.
\end{enumerate}

\subsubsection{Interplay between global and local issues}

Now that we have found and described a global framework such as
the LVS, which is theoretically robust, model-independent and very
promising to make contact with experiments, the next step is to
try to embed local brane constructions in this scenario. The
original plan was just to pick from the market one of the best
intersecting brane realisations of the MSSM and embed it in LVS by
wrapping these branes around some 4-cycles. For example the cycle
supporting the MSSM should be a small blow-up cycle, $\tau_s$, so
that the corresponding gauge coupling, $g^2=1/\tau_s$, is not
ridiculously small.

However, the authors of \cite{blumenhagen} pointed out that the
plan of first stabilising the moduli without any concern about the
local construction, and then embedding an intersecting brane
realisation of the MSSM, is definitely too na\"{i}ve. In fact,
they discovered another possible source of problems which is the
tension between moduli stabilisation and chirality. More
precisely, they noticed that the cycle supporting the MSSM is
likely not to get any non-perturbative correction, in order not to
give large VEVs to ordinary particles that would break any MSSM
gauge symmetry at the string scale where this stabilisation would
take place.

Hence they raised the problem of the stabilisation of the MSSM
cycle. They proposed to use $D$-terms, without, however, ever
managing to fix the MSSM cycle at values larger than the string
scale where it is possible to trust the effective field theory. In
Part II of this thesis, we shall propose to use string loop
corrections to solve this problem in a way very similar to the
case of K3 fibrations. In a fully realistic model, the $D$-term
contribution to the potential should also be included and the
combined $F$- and $D$-term potential studied. Usually the $D$-term
will include, besides the Fayet-Iliopoulos term depending on the
moduli, also the charged matter fields. Minimising the $D$-term
will generically fix one of the Standard Model singlets to
essentially cancel the Fayet-Iliopoulos term. Thus we can foresee
a scenario in which the MSSM cycle is fixed by string loop
corrections, whereas the $D$-term fixes, not the size of that
cycle, but instead the VEV of a Standard Model singlet as a
function of the moduli.

As far as supersymmetry breaking is concerned, the gaugino mass
terms depend just on the $F$-term of the MSSM cycle. If this cycle
is stabilised non-perturbatively, there is a $\ln(M_P/m_{3/2})$
suppression of the soft terms with respect to the gravitino mass
\cite{suppression0, suppression1, suppression2, suppression}. On
the contrary, it may be likely that perturbative stabilisation, as
in this case, gives $\mc{O}(m_{3/2})$ soft terms rather than
suppressed ones.

\subsection{From strings to cosmology}

As we have already mentioned, string theory was born about forty
years ago, but it still lacks an experimental evidence. However,
we are now in an exciting time for string phenomenology mainly for
two reasons. First that robust, consistent and well-defined
methods of moduli stabilisation, like the LVS, have been found.
Secondly, the forthcoming years are expected to be characterised
by a new set of experimental data coming from two crucial
experiments for fundamental physics: the PLANCK satellite, which
has been launched by the European Space Agency in May 2009, and
the Large Hadron Collider at CERN in Geneva which will start
operation soon.

It is, therefore, time to try to make contact of string theory
with our real world with the intention of deriving anything which
could look like a prediction from the theory. In this thesis we
shall focus mainly on cosmological implications of LARGE Volume
Scenarios which will be thoroughly discussed in Part III. The main
reasons are two. The first one is that the phenomenological
implications of these scenarios have already been largely studied
with the interesting computation of supersymmetric particle
spectra at the TeV scale \cite{LVSatLHC}. The second and most
important reason is that cosmology is definitely more promising
than phenomenology in order to find a way to test string theory.
In fact, the LHC will be able to probe energies which are $13$
orders of magnitude lower than the Planck scale, whereas the
observations of the cosmic microwave background, will give us
information about inflation, whose scale could be even just two
orders of magnitude lower than $M_P$. In addition, as we shall
explain in chapter 7, inflation is intrinsically ultra-violet
dependent, which means that it is strongly coupled to the
underlying quantum theory of gravity.

\begin{center}
---------------------------------------------------------
\end{center}

In summary, this thesis is organised as follows. Part I represents
a broad introduction to type IIB flux compactifications, paying
particular attention to the problem of K\"{a}hler moduli
stabilisation. Then Part II of this thesis will present a general
analysis of the LARGE Volume Scenario for arbitrary Calabi-Yau
manifolds, strengthened by a systematic study of the behaviour of
string loop corrections for type IIB compactifications. In Part
III of this thesis we shall apply the results of Part II to
cosmology, discussing a natural model of inflation which yields
observable gravity waves, and the finite-temperature behaviour of
the LARGE Volume Scenario.

The work contained in this thesis is based on the papers
\cite{Cicoli1} (chapter 3 and 5), \cite{Cicoli2} (chapter 4 and 6,
and appendix A), \cite{Cicoli3} (chapter 8 and appendix B), and
\cite{Cicoli4} (chapter 9 and appendix C). As indicated in the
preface, I am grateful to my collaborators Lilia Anguelova, Cliff
Burgess, Vincenzo Cal\`{o}, Joseph Conlon and Fernando Quevedo.

I have a final note on references. I have tried to cite relevant
work where appropriate, but it is inevitable that there are
lapses. As this is a thesis rather than a review article, I have
focused primarily on my own work and the results most directly
relevant to it, with the consequence that I have failed to cite
many interesting and important articles. I apologise in advance to
the authors of these papers.

\chapter{Type IIB Flux Compactifications}
\label{TypeIIBFluxCompactifications} \linespread{1.3}

In this chapter we shall review the fundamental concepts of type
IIB flux compactifications. In particular, we shall derive the
four-dimensional $N=1$ effective action for Calabi-Yau
orientifolds including background fluxes, $D3$/$D7$-branes and
$O3$/$O7$-planes \cite{louis0, louis1, louis2, louis3}. Useful
review papers extending this discussion are \cite{GreeneReview,
gkp, CHSW, CandelasDeLaOssa, SilversteinReview, Grana, dk,
germans}.

As we have explained in chapter 1, type IIB string theory seems to
be, at present, the most promising way to connect string theory
with particle physics and cosmology, due to the presence, within
its framework, of viable solutions both to stabilise the moduli
and to build local Standard Model-like constructions.

Assuming the space-time to be a product of the form
$\mathbb{R}^{3,1} \times X$, where $X$ is a Calabi-Yau three-fold,
one obtains an $N=2$ theory in four dimensions. Then taking
Calabi-Yau orientifolds, the number of supersymmetries can be
reduced from $N=2$ to $N=1$. The Standard Model, or any of its
possible generalisations, lives on a stack of space-time filling
$D$-branes in the bulk, so realising the so-called `brane-world
scenario'. The $N = 1$ supersymmetry can then be spontaneously
broken by additionally turning on background fluxes in the
orientifold bulk, which render the compactification manifold
conformally Calabi-Yau. The internal fluxes also generate a
potential that freezes all the scalar fields except the K\"{a}hler
moduli. Hence additional perturbative and non-perturbative effects
have to be employed in order to fix all the moduli and construct a
ground state. This aspect is particular important if one attempts
to construct de-Sitter vacua with a small cosmological constant.

We focus here on the computation of the four-dimensional effective
action, $S_{eff}$, which can be reconstructed, without any
approximation, from the calculation of string scattering
amplitudes \cite{scattering, scattering2}. However, this approach
is extremely complicated, because, in order to compute the
scattering amplitudes, one has to perturb around a string vacuum
where the underlying conformal field theory correlation functions
are known. Therefore, we shall take a less ambitious approach,
which is valid only in the region where the volume of $X$ is much
bigger than the string length, and is based on the traditional
Kaluza-Klein reduction of the ten-dimensional action, consisting
of the type II bulk supergravity action plus the Dirac-Born-Infeld
and Chern-Simons actions governing the dynamics of $D$-branes.

More precisely, in section 2.1 we describe the basic topological
properties of Calabi-Yau manifolds. In section 2.2 we then discuss
Calabi-Yau compactifications of type IIB string theory that lead
to $N=2$ theories in four dimensions. In section 2.3 we obtain
$N=1$ compactifications focusing on Calabi-Yau orientifolds, and
we show how, in the absence of internal fluxes, the potential for
all the moduli is exactly vanishing. In section 2.4, we then
discuss background fluxes, which can be turned on due to the
presence of local sources like $D$-branes, whose effective action
is presented in section 2.5. Finally, in section 2.6 we show how
it is possible to fix some moduli via the flux-generated
potential.

\section{Calabi-Yau manifolds}

Before presenting the effective action for type IIB
compactifications, let us briefly review the basic topological
properties of Calabi-Yau manifolds \cite{HubschBook}, paying
particular attention to their moduli spaces and to their
characteristic mirror symmetry.

\subsection{Basic features}

Given that we start with a supersymmetric ten-dimensional theory
and we want to perform a compactification to four dimensions
assuming a space-time of the form $\mathbb{R}^{3,1} \times X$, the
first question to ask is how to work out the number of
supersymmetries of the effective four-dimensional theory. This can
be done by looking at the decomposition of the spinor
representation of the ten-dimensional Lorentz group, and then
counting the number of singlets under the structure group of the
compactification manifolds. These singlets correspond to
covariantly constant Killing spinors, and their number gives the
number of supersymmetries in the effective four-dimensional
theory.

For a general compactification manifold $X$, the ten-dimensional
Lorentz group $SO(1,9)$ decomposes into:
\begin{equation}
SO(1,9)\longrightarrow SO(1,3)\times SO(6).
\end{equation}
The corresponding decomposition of the spinor representation
$\textbf{16}\in SO(1,9)$ looks like:
\begin{equation}
\textbf{16}\longrightarrow
(\textbf{2},\textbf{4})\oplus(\bar{\textbf{2}},\bar{\textbf{4}}),
\end{equation}
where $\textbf{4}$ and $\bar{\textbf{4}}$ are Weyl spinors of
$SO(6)$ conjugate to each other, while $\textbf{2}$ and
$\bar{\textbf{2}}$ are the usual Weyl spinors of $SO(1,3)$
transforming under $SL(2,\mathbb{C})$. Due to the absence of any
singlet under $SO(6)$, this compactification is
non-supersymmetric.

In order to preserve some supersymmetry, one has to choose a
particular compactification manifold $X$ with a reduced structure
group $SU(3)\subset SO(6)\cong SU(4)$. This implies the following
further decomposition of the $\textbf{4}\in SO(6)\cong SU(4)$
under $SU(3)$:
\begin{equation}
\textbf{4}\longrightarrow \textbf{3}\oplus\textbf{1}.
\end{equation}
We see that now we have obtained a singlet, which is a nowhere
vanishing and globally well defined invariant spinor, that we
shall call $\eta$. Requiring that $\eta$ is also covariantly
constant with respect to the Levi-Civita connection, we obtain a
Killing spinor. The geometrical meaning of this requirement is
that $SU(3)$ is also the holonomy group of $X$. Hence starting
from a $N=1$ theory in $D=10$ dimensions, and then compactifying
on a six-dimensional manifold with $SU(3)$ holonomy, one obtains a
$N=1$ theory in four dimensions. Now that we understood how to
count the number of supersymmetries in four dimensions, it is
straightforward to realise that a compactification manifold with
$SU(2)$ holonomy gives $N=2$ supersymmetries in four dimensions.
This is because the $\textbf{4}\in SO(6)$ decomposes under $SU(2)$
as:
\begin{equation}
\textbf{4}\longrightarrow
\textbf{2}\oplus\textbf{1}\oplus\textbf{1},
\end{equation}
giving rise to two covariantly constant spinors. Similarly
compactifications on a flat torus $T^6$, would lead to $N=4$
supersymmetries in four dimensions.

Given that we shall be interested in compactification manifolds
that preserve the minimal amount of supersymmetry in four
dimensions, we focus on the case of manifolds $X$ with $SU(3)$
holonomy group. Moreover, it turns out that these spaces are
Ricci-flat K\"{a}hler manifolds, corresponding to the famous case
of Calabi-Yau manifolds. Writing the complex dimension of the
Calabi-Yau as $\textrm{dim}_{\mathbb{C}}(X)=n$, there is an
enormous number of examples for $n\geq 3$ with $SU(n)$ holonomy,
while there is just one example for $n=1$, the torus $T^2$ with
$SU(1)$ holonomy, and one example for $n=2$, the K3 complex
surface with $SU(2)$ holonomy.

The fact that $X$ is K\"{a}hler and Ricci-flat is equivalent to
the presence of a globally defined $(1,1)$-form $J$ and a complex
holomorphic $(3,0)$-form $\Omega$, which are both closed. In our
case, these two forms can be built from the Killing spinor $\eta$
as:
\begin{equation}
\eta^{\dagger}_{\pm}\gamma^{mn}\eta_{\pm}=\pm
\frac{i}{2}J^{mn},\textrm{ \ \
}\eta^{\dagger}_-\gamma^{mnp}\eta_+=\frac{i}{2}\Omega^{mnp},\textrm{
\ \ }\eta^{\dagger}_+\gamma^{mnp}\eta_-=\pm
\frac{i}{2}\bar{\Omega}^{mnp},
\end{equation}
where $\eta_{\pm}$ denotes the two chiralities of the spinor
normalised as $\eta^{\dagger}_{\pm}\eta_{\pm}=1/2$. In addition,
the $\gamma^{m_1...m_p}=\gamma^{[m_1}\gamma^{m_2}...\gamma^{m_p]}$
are anti-symmetrised products of six-dimensional
$\gamma$-matrices. Using appropriate Fierz identities, one can
show that with this normalisation for the spinors, $J$ and
$\Omega$ are not independent but satisfy:
\begin{equation}
J\wedge J\wedge J=\frac{3i}{4}\Omega\wedge\bar{\Omega},\textrm{ \
\ \ }J\wedge\Omega=0.
\end{equation}
For a fixed metric and complex structure, $J$ is a closed
$(1,1)$-form while $\Omega$ is a closed $(3,0)$-form, ensuring
that the compactification manifold with $SU(3)$ holonomy structure
is indeed a Calabi-Yau three-fold.

In order now to work out the particle spectrum in four dimensions,
one has to study the splitting of the ten-dimensional equations of
motion in a compactified space-time background of the form
$\mathbb{R}^{3,1} \times X$. In the simplest case of a scalar
field $\phi$, one obtains:
\begin{equation}
\Delta_{10}\phi=\left(\Delta_4+\Delta_6\right)\phi=
\left(\Delta_4+m^2\right)\phi=0, \label{laplacian}
\end{equation}
where the second equation assumed that $\phi$ is an eigenfunction
of the internal, six-dimensional Laplace operator
$\Delta_6\phi=m^2\phi$. This result implies that the massless
modes of the $D = 4$ theory correspond to harmonic forms on $X$,
which are defined as the zero modes of $\Delta_6$. More in
general, not just the zero modes of the scalar fields, but also
all the zero modes arising from the other massless ten-dimensional
fields correspond to harmonic forms on $X$.

The beautiful property of Calabi-Yau three-folds, which allows us
to derive the main features of the low-energy effective field
theory from the topology of the compactification manifold, is that
the harmonic forms on $X$ are in one-to-one correspondence with
the elements of the Dolbeault cohomology groups $H_{p,q}(X)$. The
elements of the $H_{p,q}(X)$ are defined as the set of
\textit{closed} $(p,q)$-forms quotiented out by the set of
\textit{exact} $(p,q)$-forms, where $(p,q)$ is denoting the number
of holomorphic and anti-holomorphic differentials of the harmonic
forms. Important quantities for the effective field theory, are
the Hodge numbers $h_{p,q} = \textrm{dim}\left(H_{p,q}(X)\right)$,
which, for a Calabi-Yau three-fold, satisfy:
\begin{eqnarray*}
h_{0,1} &=&h_{0,2}=h_{1,3}=h_{2,3}=h_{3,2}
=h_{3,1}=h_{2,0}=h_{1,0}=0, \\
h_{0,0} &=&h_{0,3}=h_{3,3}=h_{3,0}=1, \textrm{ \ \ \
}h_{1,2}=h_{2,1},\textrm{ \ \ \ }h_{1,1}=h_{2,2}.
\end{eqnarray*}
We stress that the only non-trivial Hodge numbers are $h_{1,1}$
and $h_{1,2}$. A more conventional way to list the Hodge numbers,
is to arrange them in the so-called Hodge diamond, since it
renders all the possible symmetries manifest:
\begin{equation}
\begin{array}{ccccccc}
&  &  & h_{0,0} &  &  &  \\
&  & h_{1,0} &  & h_{0,1} &  &  \\
& h_{2,0} &  & h_{1,1} &  & h_{0,2} &  \\
h_{3,0} &  & h_{2,1} &  & h_{1,2} &  & h_{0,3} \\
& h_{3,1} &  & h_{2,2} &  & h_{1,3} &  \\
&  & h_{3,2} &  & h_{2,3} &  &  \\
&  &  & h_{3,3} &  &  &
\end{array}
=
\begin{array}{ccccccc}
&  &  & 1 &  &  &  \\
&  & 0 &  & 0 &  &  \\
& 0 &  & h_{1,1} &  & 0 &  \\
1 &  & h_{1,2} &  & h_{1,2} &  & 1 \\
& 0 &  & h_{1,1} &  & 0 &  \\
&  & 0 &  & 0 &  &  \\
&  &  & 1 &  &  &
\end{array}
\label{diamond}
\end{equation}
This Hodge diamond has three symmetries which are the complex
conjugation (reflection about the central vertical axis), the
Hodge-$\ast$ duality, also called Poincar\'{e} duality (reflection
about the central horizontal axis), and the mirror symmetry
\cite{GreenePlesser} (reflection about the central diagonal axis)
which we shall discuss in subsection 2.1.3.

We point out that there are particular deformations of the
Calabi-Yau metric which do not disturb the Calabi-Yau condition.
These deformations of the metric correspond to scalar fields in
the low energy effective action, which are called moduli. They can
be viewed as the coordinates of the geometrical moduli space of
the Calabi-Yau manifold \cite{modspace1, modspace2},
parameterising the shape and size of the three-fold. Let us
briefly review some properties of this moduli space in the next
subsection.

\subsection{Moduli space}

As we have seen in the previous section, a Calabi-Yau three-fold
is a Ricci-flat K\"{a}hler manifold with $SU(3)$ holonomy. This
implies that its metric $g_{m\bar{n}}$, $m,\bar{n}=1,...,3$,
satisfies $R_{m\bar{n}}(g)=0$. We shall now focus on deformations
of the metric $\delta g$ that preserve Ricci-flatness:
$R_{m\bar{n}}(g+\delta g)=0$. Given that we are not interested in
changes of coordinates, we need to eliminate them by fixing the
diffeomorphism invariance. This can be done through the following
gauge-fixing choice:
\begin{equation}
\nabla (\delta g)=0 \label{cond}
\end{equation}
In the case of K\"{a}hler manifolds, it turns out from
(\ref{cond}) that the conditions on $\delta g_{m\bar{n}}$ and
$\delta g_{mn}$ decouple and can be studied separately:
\begin{itemize}
\item $\delta g_{m\bar{n}}$: In this case (\ref{cond}) becomes:
\begin{equation}
\Delta \delta g_{m\bar{n}}=0,
\end{equation}
implying that $\delta g_{m\bar{n}}$ has to be a harmonic
$(1,1)$-form. Therefore it can be expanded in a basis of
$H_{1,1}(X)$, formed by harmonic $(1,1)$-forms $\hat{D}_i$:
\begin{equation}
\delta g_{m\bar{n}}=it^i(\hat{D}_i)_{m\bar{n}},\textrm{ \ \ \
}i=1,...,h_{1,1}, \label{11forms}
\end{equation}
where the $t^i$ are called K\"{a}hler moduli since these
deformations correspond to cohomologically non-trivial changes of
the K\"{a}hler form, defined as:
\begin{equation}
J=-i g_{m\bar{n}}dy^m\wedge d\bar{y}^{\bar{n}}.
\end{equation}
In order to make sure that the $t^i$ are such that the new metric
$g+\delta g$ is still positive definite, we impose:
\begin{equation}
\int_C J>0,\textrm{ \ \ }\int_S J\wedge J>0,\textrm{ \ \ }\int_X
J\wedge J \wedge J>0, \label{cono}
\end{equation}
for all complex curves $C$ and surfaces $S$ on the Calabi-Yau $X$.
The conditions (\ref{cono}) imply that the subset of
$\mathbb{R}^{h_{1,1}}$ spanned by the parameters $t^i$ is a cone,
called the K\"{a}hler cone, or the K\"{a}hler moduli space.
However, this moduli space is usually complexified, due to the
fact that the K\"{a}hler-form $J$ is complexified as $J_c = J +
iB_2$, where $B_2$ is the Neveu-Schwarz/Neveu-Schwarz two-form of
type II string theories. This, in turn, introduces complex
K\"{a}hler deformations $v^i$ which arise as the expansion of
$J_c$:
\begin{equation}
J_c = J + iB_2=v^i\hat{D}_i,\textrm{ \ \ \ }\hat{D}_i\in
H_{1,1}(X).
\end{equation}
The complex moduli space spanned by the coordinates $v^i$, denoted
as $\mathcal{M}_{h_{1,1}}^{K}$, is a special K\"{a}hler manifold,
which means that it admits a K\"{a}hler metric whose classical
K\"{a}hler potential is entirely determined by a holomorphic
prepotential $F(v)$:
\begin{equation}
K_K=-\ln\left(k_{ijk}t^i t^j t^k\right),\textrm{ \ \ \
}F(v)=k_{ijk}v^i v^j v^k, \label{prepot}
\end{equation}
where $k_{ijk}$ are topological intersection numbers:
\begin{equation}
k_{ijk}=\int_X \hat{D}_i\wedge\hat{D}_j\wedge\hat{D}_k.
\end{equation}

\item $\delta g_{m n}$: In this case (\ref{cond}) becomes:
\begin{equation}
\Delta \delta g_{m n}=0,
\end{equation}
implying that $\delta g_{m n}$ is a harmonic $(2,0)$-form.
However, $\delta g_{m n}$ cannot be expanded in a basis of
$(2,0)$-forms since $h_{2,0}=0$ for a Calabi-Yau. Hence we put
$H_{2,0}(X)$ in one-to-one correspondence to $H_{1,2}(X)$ via the
fundamental holomorphic $(3,0)$-form $\Omega$ in the following
way:
\begin{equation}
\delta g_{mn}=\frac{i}{\left\Vert\Omega\right\Vert^2}\bar{U}^a
\left(\bar{\chi}_a\right)_{m\bar{p}\bar{q}}\Omega^{\bar{p}\bar{q}}_n,
\textrm{ \ \ \ }a=1,...,h_{1,2}, \label{12forms}
\end{equation}
where $\bar{\chi}_{\alpha}$ denotes a basis of $H_{1,2}(X)$ and we
abbreviate $\left\Vert \Omega \right\Vert ^{2}\equiv \Omega
_{mnp}\bar{\Omega}^{mnp}/3!$. The complex scalar parameters $U^a$
are called complex structure moduli. The reason is that for the
new metric to be K\"{a}hler, there must be a coordinate system in
which it has only mixed components (since for a K\"{a}hler
manifold $g_{m n}=0$). Then, given that holomorphic coordinate
transformations do not change the index structure, it is clear
that $\delta g_{m n}$ can only be removed by a non-holomorphic
coordinate transformation. Thus $\delta g_{mn}$ corresponds to a
deformation of the complex structure.

As in the case of the K\"{a}hler moduli, the parameters $U^a$ span
a subset of $\mathbb{C}^{h_{1,2}}$ called the complex structure
moduli space $\mathcal{M}_{h_{1,2}}^{cs}$. This space is also a
special K\"{a}hler manifold, with a tree-level K\"{a}hler
potential given by \cite{modspace2}:
\begin{equation}
g_{a\bar{b}}=\frac{\partial^2 K_{cs}}{\partial U^a\partial
\bar{U}^{\bar{b}}},\textrm{ \ \ }K_{cs}=-\ln\left(-i\int_X
\Omega\wedge\bar{\Omega}\right)=-\ln\left[i\left(\bar{Z}^a
\mathcal{F}_a-Z^a\bar{\mathcal{F}}_a\right)\right]. \label{cs}
\end{equation}
The second form of $K_{cs}$ is obtained from the expansion of
$\Omega$:
\begin{equation}
\Omega(U)=Z^a(U)\alpha_a-\mathcal{F}_a(U)\beta^a,
\end{equation}
where $(\alpha_a,\beta^a)$ is a real, symplectic basis of $H_3(X)$
satisfying:
\begin{equation}
\int_X \alpha_a \wedge \beta^b=\delta_a^b,\textrm{ \ \ \ }\int_X
\alpha_a\wedge\alpha_b=0=\int_X \beta^a\wedge\beta^b.
\label{basis}
\end{equation}
Both $Z^a(U)$ and $\mathcal{F}_a(U)$ are holomorphic functions of
the $U$-moduli and furthermore $\mathcal{F}_a(U) =
\partial_{Z^a}\mathcal{F}(Z(U))$ is the derivative of a holomorphic
prepotential $\mathcal{F}(Z(U)))$.
\end{itemize}
At tree-level, the total moduli space $\mathcal{M}$ factorises and
takes the form of a direct product:
\begin{equation}
\mathcal{M}=\mathcal{M}_{h_{1,2}}^{cs}\times\mathcal{M}_{h_{1,1}}^{K}.
\label{moduliSpaces}
\end{equation}
We finally stress that these metric deformations which give rise
to moduli, are then seen in the four dimensional effective theory
as massless scalar fields. Giving them a mass via the generation
of a scalar potential for these fields, corresponds to fixing the
size and the shape of the Calabi-Yau three-fold.

\subsection{Mirror symmetry}

As we have seen in section 2.1.1, the Hodge diamond
(\ref{diamond}) is introduced because it highlights all the
symmetries of Calabi-Yau manifolds. Two of them, Poincar\'{e}
duality and complex conjugation, have been rigorously established
whereas the last one, mirror symmetry \cite{GreenePlesser}, has
been established only on a subspace of Calabi-Yau manifolds.
Therefore, strictly speaking, mirror symmetry is still a
conjecture about a not yet rigorously defined space of Calabi-Yau
three-folds, which was deduced from scatter plots of the Hodge
numbers of known Calabi-Yaus \cite{CLS}.

In terms of the Hodge diamond (\ref{diamond}), mirror symmetry
corresponds to a reflection about the diagonal axis, or, in other
words, the third cohomology $H_3 = H_{3,0}\oplus H_{2,1}\oplus
H_{1,2}\oplus H_{0,3}$ is interchanged with the even cohomologies
$H_{even} = H_{0,0}\oplus H_{1,1}\oplus H_{2,2}\oplus H_{3,3}$.
This means that for `every' Calabi-Yau $X$ there exists a mirror
manifold $\tilde{X}$ with reversed Hodge numbers:
\be
h_{1,1}(X)=h_{1,2}(\tilde{X}),\textrm{ \ \ \
}h_{1,2}(X)=h_{1,1}(\tilde{X}).
\ee
Therefore mirror symmetry exchanges also K\"{a}hler with complex
structure moduli, as well as their complexified moduli spaces:
\begin{equation}
\mathcal{M}_{h_{1,2}}^{cs}(X)\equiv\mathcal{M}_{h_{1,1}}^{K}(\tilde{X}),
\textrm{ \ \ \ }\mathcal{M}_{h_{1,1}}^{K}(X)
\equiv\mathcal{M}_{h_{1,2}}^{cs}(\tilde{X}).
\end{equation}
This, in turn, means that the underlying prepotentials are
identical:
\be
\mc{F}(X)\equiv F(\tilde{X}),\textrm{ \ \ \
}F(X)\equiv\mc{F}(\tilde{X}).
\ee
When one considers type II string theories compactified on
Calabi-Yau three-folds, mirror symmetry manifests itself has the
famous $T$-duality, which relates type IIA with type IIB in a
mirror symmetric background. In other words, the following
equivalence holds:
\begin{equation}
\textrm{IIA in background \ }\mathbb{R}^{3,1}\times
X\equiv\textrm{IIB in background \
}\mathbb{R}^{3,1}\times\tilde{X}.
\end{equation}
We immediately realise that mirror symmetry is enormously useful
to compute the effective action of string compactifications. In
fact, if one focuses on type IIB compactifications, as we shall do
in the next sections, the properties of the dual type IIA
compactification can be easily deduced considering the appropriate
mirror Calabi-Yau three-fold.

\section{$N=2$ type IIB compactifications}

In this section, we shall present Calabi-Yau compactifications of
type IIB string theory \cite{KKred1, KKred2, KKred3}. Since the
ten dimensional theory has $N=2$ supersymmetries (or 32
supercharges), these compactifications will preserve $N=2$
supersymmetries (or 8 supercharges) in $D=4$. Thus the low energy
effective action is an $N=2$ supergravity coupled to vector-,
hyper- and tensor multiplets.

\subsection{The spectrum}

We start by compactifying type IIB string theory on a Calabi-Yau
three-fold $X$, so that the background is of the form
$\mathbb{R}^{3,1} \times X$, and the ten-dimensional metric is
block-diagonal:
\begin{equation}
ds^2=g_{\mu\nu}dx^{\mu}dx^{\nu}+g_{m\bar{n}}dy^m
d\bar{y}^{\bar{n}},
\end{equation}
where $g_{\mu\nu}$, with $\mu,\nu=0,...,3$, is the Minkowski
metric, while $g_{m\bar{n}}$ is the Calabi-Yau metric.

The second step is to consider the low-energy limit below $M_s$,
focusing only on the massless bosonic degrees of freedom, which
are (with the hats denoting ten-dimensional fields) \cite{GSW,
PolchinskiBook}:
\begin{itemize}
\item Neveu-Schwarz/Neveu-Schwarz (NS-NS) sector: the metric $\hat{g}$,
a two-form $\hat{B}_2$, and the dilaton $\hat{\phi}$;

\item Ramond/Ramond (RR) sector: a four-form $\hat{C}_4$, a two-form $\hat{C}_2$,
and the axionic zero-form $\hat{C}_0$.
\end{itemize}
Using the notation of differential forms, the type IIB ten
dimensional effective action in Einstein frame takes the form
\cite{PolchinskiBook}:
\begin{eqnarray}
\label{10DLa} S_{IIB}^{(10D)} &=&-\int \left(
\frac{1}{2}\hat{R}\ast \mathbf{1}+\frac{1}{4}d \hat{\phi}\wedge
\ast d\hat{\phi}+\frac{1}{4}e^{-\hat{\phi}}\hat{H}
_{3}\wedge \ast\hat{H}_{3}\right)   \\
&&-\frac{1}{4}\int \left( e^{2\hat{\phi}}d\hat{C}_0\wedge \ast
d\hat{C}_0+e^{\hat{\phi}}\hat{F}_{3}\wedge \ast
\hat{F}_{3}+\frac{1}{2}\hat{F}_{5}\wedge \ast \hat{F}_{5}\right)
-\frac{1}{4}\int \hat{C}_{4}\wedge \hat{H}_{3}\wedge \hat{F}_{3},
\nonumber
\end{eqnarray}
where $\ast$ denotes the Hodge-$\ast$ operator and the field
strengths are defined as:
\begin{eqnarray}
\hat{H}_3&=&d\hat{B}_2,\textrm{ \ \ \ }\hat{F}_3=d\hat{C}_2
-\hat{C}_0d\hat{B}_2
\nonumber \\
\hat{F}_5&=&d\hat{C}_4-\frac{1}{2} d\hat{B}_2\wedge\hat{C}_2
+\frac{1}{2}\hat{B}_2\wedge d\hat{C}_2. \label{espansione}
\end{eqnarray}
The five-form field strength $\hat{F}_5$ additionally satisfies
the self-duality condition $\hat{F}_5 =\ast\hat{F}_5$, which is
imposed by hand at the level of the equations of motion.

According to the traditional mechanism of Kaluza-Klein reduction,
the four-dimensional spectrum is obtained by expanding all
ten-dimensional fields into eigenforms forms on $X$, and then
keeping only the zero modes. The reduction of the ten-dimensional
metric yields a four-dimensional metric tensor $g_{\mu\nu}$, a
one-form $V^0$, which is usually called `gravi-photon', and then
$h_{1,1}$ K\"{a}hler and $h_{1,2}$ complex structure moduli, as
already obtained in (\ref{11forms}) and (\ref{12forms}),
respectively. All the other type IIB ten dimensional bosonic
fields, appearing in the Lagrangian (\ref{10DLa}), are similarly
expanded in terms of harmonic forms on $X$ according to (the
absence of hats is denoting four-dimensional fields):
\begin{itemize}
\item NS-NS sector:
\begin{equation}
\label{expansione1} \hat{\phi}=\phi(x),\textrm{ \ \ \
}\hat{B}_2=B_2(x)+b^i(x)\hat{D}_i,\textrm{ \ \ }i=1,...,h_{1,1},
\end{equation}
where the $\hat{D}_i$ are a basis of harmonic $(1,1)$-forms of
$H_{1,1}(X)$. The four-dimensional fields $\phi(x)$ and $b^i(x)$
are scalars, whereas $B_2(x)$ is a two-form.
\item RR sector:
\begin{eqnarray}
\label{expansione} \hat{C_0}&=&C_0(x),\textrm{ \ \
\ }\hat{C}_2=C_2(x)+c^i(x)\hat{D}_i,\textrm{ \ \ }i=1,...,h_{1,1} \\
\hat{C}_4&=&Q_2^i(x)\wedge\hat{D}_i+V^a(x)\wedge\alpha_a
-\tilde{V}_a(x)\wedge\beta^a+\rho_i(x)\tilde{D}^i,\textrm{ \ \
}a=1,...,h_{1,2}, \nonumber
\end{eqnarray}
where the $\tilde{D}_i$ are a basis of harmonic $(2,2)$-forms of
$H_{2,2}(X)$, dual to the $(1,1)$-forms $\hat{D}_i$, while
$\left(\alpha_a,\beta^a\right)$ is the symplectic basis of
$H_3(X)$ introduced in (\ref{basis}). The four-dimensional fields
$C_0(x)$, $c^i(x)$ and $\rho^i(x)$ are scalars, $V^a(x)$ and
$\tilde{V}_a(x)$ are one-forms, whereas $C_2(x)$ and $Q^i_2(x)$
are two-forms.
\end{itemize}

The self-duality condition of $\hat{F}_5$ eliminates half of the
degrees of freedom in $\hat{C}_4$, and one conventionally chooses
to eliminate $Q_i^2$ and $\tilde{V}_a$ in favour of $\rho_i$ and
$V^a$. Altogether these fields assemble into $N=2$ multiplets
which are summarised in table 5.
\begin{center}
\begin{tabular}{c||c|c}
  & number & bosonic field components \\
  \hline\hline
  gravity multiplet & 1 & $\left(g_{\mu\nu},V^0\right)$ \\
  vector multiplet & $h_{1,2}$ & $\left(V^a,U^a\right)$ \\
  hypermultiplet & $h_{1,1}$ & $\left(t^i,b^i,c^i,\rho_i\right)$ \\
  double-tensor multiplet & 1 & $\left(B_2,C_2,\phi,C_0\right)$
\end{tabular}\\
\vspace{0.3cm}{{\bf Table {5}:} $N=2$ multiplets of type IIB
Calabi-Yau compactifications.}
\end{center}
We finally note that the two antisymmetric tensors $B_2$ and $C_2$
can be dualised to scalar fields, so that the double-tensor
multiplet can be treated as an extra hypermultiplet.

\subsection{Tree-level effective action}

The tree-level four dimensional low energy effective action can be
expressed in the standard $N=2$ supergravity form
\cite{WessBagger} by inserting (\ref{espansione}),
(\ref{expansione1}) and (\ref{expansione}) into the action
(\ref{10DLa}). Then integrating over the Calabi-Yau manifold and
performing an appropriate Weyl rescaling, one ends up with:
\begin{eqnarray}
\label{4DL} S_{IIB}^{(4D)} &=&-\int \frac{1}{2}R\ast
\mathbf{1}-\frac{1}{4} \textrm{Re}(\mc{M}_{ab})F^a\wedge
F^b-\frac{1}{4}
\textrm{Im}(\mc{M}_{ab})F^a\wedge\ast F^b \nonumber \\
&&+g_{a\bar{b}}dU^a\wedge \ast
d\bar{U}^{\bar{b}}+h_{IJ}dq^I\wedge\ast dq^J,
\end{eqnarray}
where $F^a=dV^a$, and the $\mc{M}(U)$ are gauge kinetic functions
which can be expressed in terms of the holomorphic prepotential
$\mc{F}(U)$. In addition, $g_{a\bar{b}}$ is the special K\"{a}hler
metric introduced in (\ref{cs}), whereas $h_{IJ}$ is the metric on
the space of the $4(h_{1,1}+1)$ moduli, collectively denoted as
$q^I$, which are the scalar components of the hypermultiplets
present in the theory.

The total moduli space of the $N=2$ theory, at tree-level,
factorises in the product of the complex structure moduli space
$\mc{M}^{cs}_{h_{1,2}}$, which is a special K\"{a}hler manifold
spanned by the scalars $U^a$ in the vector multiplets, and the
space of all the other moduli $\mc{M}^Q_{2(h_{1,1}+1)}$, which is
a quaternionic manifold spanned by the scalars $q^I$ in the
hypermultiplets:
\begin{equation}
\mc{M}=\mc{M}^{cs}_{h_{1,2}}\times \mc{M}^Q_{2(h_{1,1}+1)}.
\label{quaternionic}
\end{equation}
The $N=2$ moduli space $\mc{M}$ of (\ref{quaternionic}) contains
the geometrical moduli space (\ref{moduliSpaces}) as a
submanifold, and it is entirely determined by the two
prepotentials $\mc{F}(U)$ and $F(v)$, both of which are exactly
known due to mirror symmetry.

\section{$N=1$ type IIB compactifications}

The phenomenologically interesting compactifications are those
preserving $N=1$ supersymmetry, while, as we have seen, Calabi-Yau
compactifications of ten dimensional type IIB string theory lead
to $N=2$ supersymmetries. However, this $N=2$ can be further
broken to $N=1$ by introducing an appropriate orientifold
projection \cite{orientifold, orientifold2}. Type IIB
compactifications on Calabi-Yau orientifolds \cite{gkp, louis0,
louis1, louis2, louis3} are typically characterised by the
presence of non-trivial background fluxes and localised sources
like $D$-branes. In this section, we shall mainly concentrate on
the truncation of the $N=2$ four-dimensional spectrum due to the
orientifolding and the consequent modification of the supergravity
bulk effective action, ignoring the study of internal fluxes and
localised sources, which will be analysed in detail in section 2.4
and 2.5, respectively.

\subsection{Orientifold projection}

A four dimensional $N=1$ type IIB orientifold is obtained from an
$N=2$ Calabi-Yau compactification by gauging a discrete symmetry
of the form \cite{orientifold, orientifold2}:
\begin{equation}
(-1)^{\epsilon F_L}\Omega_p\sigma, \label{Oproj}
\end{equation}
with $\epsilon= 0, 1$. Employing common notation, $\Omega_p$
denotes world-sheet parity, which gives an orientation reversal of
the string world-sheet, $F_L$ is the left moving fermion number
and $\sigma:X\to X$ is an isometric and holomorphic involution of
the Calabi-Yau $X$. We stress that $\sigma$ is an `internal'
symmetry, in the sense that it acts solely on $X$, but leaves the
$D=4$ Minkowskian space-time untouched. In addition, the action of
$\sigma$ on the fundamental forms of the Calabi-Yau, $J$ and
$\Omega$, is given by the pull-back $\sigma^*$ which satisfies:
\begin{equation}
\sigma^{*}J=J\textrm{ \ \ \ and \ \ \
}\sigma^{*}\Omega=(-1)^{\epsilon}\Omega.
\end{equation}
Depending on the value of $\epsilon$, there are two classes of
models to consider:
\begin{enumerate}
\item $\epsilon = 0$: theories with $O5$/$O9$ orientifold
planes, in which the fixed point set of $\sigma$ is either one or
three complex dimensional;

\item $\epsilon = 1$: theories with $O3$/$O7$-planes, with $\sigma$ leaving
invariant zero or two complex dimensional submanifolds of $X$.
\end{enumerate}
From now on, we shall focus on the case $\epsilon=1$ corresponding
to the presence of $O3$/$O7$-planes, and compute the
four-dimensional $N=1$ spectrum and the low-energy effective
action, which can be expressed in terms of geometrical and
topological quantities of the orientifold.

Before doing that, let us mention that harmonic $(p,q)$-forms are
either even or odd eigenstates of $\sigma^*$ due to the fact that
$\sigma$ is a holomorphic involution. Therefore, the cohomology
groups $H_{p,q}$ split into two eigenspaces under the action of
$\sigma^*$:
\begin{equation}
H_{p,q}=H_{p,q}^+\oplus H_{p,q}^-,
\end{equation}
where $H_{p,q}^+$ has dimension $h_{p,q}^+$ and denotes the even
eigenspace of $\sigma^*$, while $H_{p,q}^-$ has dimension
$h_{p,q}^-$ and denotes the odd eigenspace of $\sigma^*$.
Moreover, the constraints on the Hodge numbers can be found as
follows:
\begin{itemize}
\item $h_{1,1}^{\pm}=h_{2,2}^{\pm}$ since
the Hodge $\ast$-operator commutes with $\sigma^*$, because
$\sigma$ preserves the orientation and the metric of the
Calabi-Yau manifold;

\item $h_{2,1}^{\pm}=h_{1,2}^{\pm}$ due to the
holomorphicity of $\sigma$;

\item $h_{3,0}^+=h_{0,3}^+=0$ and $h_{3,0}^-=h_{0,3}^-=1$
due to the property $\sigma^*\Omega=-\Omega$;

\item $h_{0,0}^+=h_{3,3}^+=1$ and $h_{0,0}^-=h_{3,3}^-=0$
since the volume-form, which is proportional to
$\Omega\wedge\bar{\Omega}$, is invariant under $\sigma^*$.
\end{itemize}

\subsection{The spectrum}

The orientifold projection (\ref{Oproj}) truncates the $N=2$
spectrum and reassembles the surviving fields in $N=1$ multiplets
\cite{louis0, orientifold2}. In fact, the four-dimensional
spectrum is found by using the Kaluza-Klein expansion given in
(\ref{11forms}), (\ref{12forms}), (\ref{expansione1}) and
(\ref{expansione}), but keeping only the fields which are left
invariant by the orientifold action.

Since $\sigma$ is a holomorphic isometry, it leaves both the
metric and the complex structure, and thus also the K\"{a}hler
form $J$, invariant. As a consequence only $h_{1,1}^+$ K\"{a}hler
deformations $t^{i_+}$ remain in the spectrum arising from:
\begin{equation}
J=t^{i_+}(x)\hat{D}_{i_+},\textrm{ \ \ \ }i_+=1,...,h_{1,1}^+,
\end{equation}
where $\hat{D}_{i_+}$ denotes a basis of $H_{1,1}^+$. Similarly,
from (\ref{12forms}), one sees that the invariance of the metric
together with $\sigma^* \Omega=-\Omega$, implies that the complex
structure deformations kept in the spectrum, correspond to
elements in $H_{1,2}^-$. Hence the expansion (\ref{12forms}) is
replaced by:
\begin{equation}
\delta g_{mn}=\frac{i}{\left\Vert\Omega\right\Vert^2}\bar{U}^{a_-}
\left(\bar{\chi}_{a_-}\right)_{m\bar{p}\bar{q}}
\Omega^{\bar{p}\bar{q}}_n, \textrm{ \ \ \ }a_-=1,...,h_{1,2}^-,
\end{equation}
where $\bar{\chi}_{a_-}$ denotes a basis of $H_{1,2}^-$.

The behaviour of all the other ten-dimensional type IIB bosonic
fields under the world sheet parity transformation $\Omega_p$ and
the `space-time fermion number' operator in the left moving sector
$(-1)^{F_L}$, are summarised in table 6.
\begin{center}
\begin{tabular}{c||c|c|c}
  & $\Omega_p$ & $(-1)^{F_L}$ & $(-1)^{F_L}\Omega_p$ \\
  \hline\hline
  $\hat{g}$ & + & + & + \\
  $\hat{\phi}$ & + & + & + \\
  $\hat{B}_2$ & - & + & - \\
  $\hat{C}_0$ & - & - & + \\
  $\hat{C}_2$ & + & - & - \\
  $\hat{C}_4$ & - & - & +
\end{tabular}\\
\vspace{0.3cm}{{\bf Table {6}:} Transformation properties under
$\Omega_p$ and $(-1)^{F_L}$ of the $D=10$ type IIB bosonic
fields.}
\end{center}
The results of table 6 imply that the invariant states have to
obey:
\begin{eqnarray}
\sigma^*\hat{g}&=&\hat{g},\textrm{ \ \
}\sigma^*\hat{\phi}=\hat{\phi},\textrm{ \ \
}\sigma^*\hat{B}_2=-\hat{B}_2,
\nonumber \\
\sigma^*\hat{C}_0&=&\hat{C}_0,\textrm{ \ \
}\sigma^*\hat{C}_2=-\hat{C}_2,\textrm{ \ \
}\sigma^*\hat{C}_4=\hat{C}_4, \label{sPhi}
\end{eqnarray}
and so the expansions (\ref{expansione1}) and (\ref{expansione})
are replaced by:
\begin{itemize}
\item NS-NS sector:
\begin{equation}
\label{newExpansion1} \hat{\phi}=\phi(x),\textrm{ \ \ \
}\hat{B}_2=b^{i_-}(x)\hat{D}_{i_-},\textrm{ \ \
}i_-=1,...,h_{1,1}^-, \nonumber
\end{equation}
where $\hat{D}_{i_-}$ is a basis of $H_{1,1}^-$. We see
immediately that $\phi$ remains in the spectrum, while in the
expansion of $\hat{B}_2$ only odd elements survive. In addition,
note that the $D=4$ two-form $B_2$ present in the $N=2$
compactification (see (\ref{expansione1})) has been projected out,
and in the expansion of $\hat{B}_2$ only scalar fields appear.

\item RR sector:
\begin{eqnarray}
\label{newExpansion} \hat{C_0}&=&C_0(x),\textrm{ \ \
\ }\hat{C}_2=c^{i_-}(x)\hat{D}_{i_-},\textrm{ \ \ }i_-=1,...,h_{1,1}^- \\
\hat{C}_4&=&Q_2^{i_+}(x)\wedge\hat{D}_{i_+}
+V^{a_+}(x)\wedge\alpha_{a_+}  \\
&&-\tilde{V}_{a_+}(x)\wedge\beta^{a_+}+\rho_{i_+}(x)\tilde{D}^{i_+},\textrm{
\ \ }a_+=1,...,h_{1,2}^+, \nonumber
\end{eqnarray}
where $\tilde{D}^{i_+}$ is a basis of $H_{2,2}^+$ which is dual to
$\hat{D}_{i_+}$, and $\left(\alpha_{a_+},\beta^{a_+}\right)$ is a
real, symplectic basis of $H_3^+=H_{1,2}^+\oplus H_{2,1}^+$. We
see immediately that $C_0$ remains in the spectrum, and in the
expansion $\hat{C}_2$ only odd elements survive, while for
$\hat{C}_4$ only even elements are kept. In addition, note that
the $D=4$ two-form $C_2$ present in the $N=2$ compactification
(see (\ref{expansione})) has been projected out, and in the
expansion of $\hat{C}_2$ only scalar fields appear.
\end{itemize}
The self-duality condition of $\hat{F}_5$ eliminates half of the
degrees of freedom in $\hat{C}_4$, and one conventionally chooses
to eliminate $Q^{i_+}_2$ and $\tilde{V}_{a_+}$ in favour of
$\rho_{i_+}$ and $V^{a_+}$. Altogether these fields assemble into
$N=1$ multiplets as summarised in table 7 \cite{louis0}. As we
have already mentioned, we can replace $h_{1,1}^+$ of the chiral
multiplets by linear multiplets.

\begin{center}
\begin{tabular}{c||c|c}
  & number & field components \\
  \hline\hline
  gravity multiplet & 1 & $g_{\mu\nu}$ \\
  vector multiplet & $h_{1,2}^+$ & $V^{a_+}$ \\
  chiral multiplet & $h_{1,2}^-$ & $U^{a_-}$ \\
  chiral multiplet & $1$ & $\left(\phi,C_0\right)$ \\
  chiral multiplet & $h_{1,1}^-$ & $\left(b^{i_-},c^{i_-}\right)$ \\
  chiral-linear multiplet & $h_{1,1}^+$ & $\left(t^{i_+},\rho_{i_+}\right)$
\end{tabular}\\
\vspace{0.3cm}{{\bf Table {7}:} $N=1$ spectrum of
$O3/O7$-orientifold compactifications.}
\end{center}
Compared to the $N=2$ spectrum of the Calabi-Yau compactification
given in table 5, we see that the gravi-photon $V^0$ `left' the
gravitational multiplet, while the $h_{1,2}$ $N=2$ vector
multiplets decomposed into $h_{1,2}^+$ $N=1$ vector multiplets
plus $h_{1,2}^-$ chiral multiplets. Furthermore, the $h_{1,1}+1$
hypermultiplets lost half of their physical degrees of freedom and
got reduced into $h_{1,1}+1$ chiral multiplets.

\subsection{Tree-level effective action}
\label{TreeLevelEffectiveAction}

The low energy effective action for orientifold compactifications,
can be obtained from the $N=2$ action (\ref{4DL}) by imposing the
truncation discussed before. The resulting $N=1$ low-energy
four-dimensional action can then be displayed in the standard
supergravity form \cite{WessBagger}, where it is expressed in
terms of a K\"{a}hler potential, $K$, a holomorphic
superpotential, $W$, and a holomorphic gauge-kinetic function,
$f_{ab}$, where the indices $a,b$ run over the various vector
multiplets:
\begin{equation}
S^{(4D)}=-\int
\frac{1}{2}R\ast\textbf{1}+K_{I\bar{J}}D\Phi^I\wedge\ast
D\bar{\Phi}^{\bar{J}}+\frac{1}{2}\textrm{Re}(f_{ab})F^a\wedge\ast
F^b +\frac{1}{2}\textrm{Im}(f_{ab})F^a\wedge F^b+V. \label{eff4D}
\end{equation}
Here $F^a=dV^a$ and the $\Phi^I$ collectively denote all complex
scalars in the theory. The $N=1$ $F$-term supergravity scalar
potential is given in terms of $K$ and $W$ (in four-dimensional
Planck units) by:
\begin{equation}
V_F=e^K\left(K^{I\bar{J}}D_I W D_{\bar{J}}\bar{W}-3|W|^2\right),
\label{bo}
\end{equation}
where $K^{I\bar{J}}$ is the inverse of the K\"{a}hler metric
$K_{I\bar{J}} = \partial_I \partial_{\bar{J}} K(\Phi,\bar{\Phi})$
and the definition of the K\"{a}hler covariant derivative $D_I W$
is:
\begin{equation}
\left\{
\begin{array}{l}
D_I W=\partial_{I}W+W \partial_{I} K,
\\
D_{\bar{J}}\bar{W}=\partial_{\bar{J}} \bar{W}+\bar{W}
\partial_{\bar{J}} K.
\end{array}
\right.
\end{equation}
On the other hand, the $D$-term scalar potential reads (denoting
with $T_a$ the gauge group generators):
\begin{equation}
V_D=\frac{1}{2}\left(\textrm{Re}f\right)^{-1\,ab}D_a D_b,\textrm{
\ \ \ }D_{a}=\left[ K_{I}+\frac{W_{I}}{W}\right] \left(
T_{a}\right) _{IJ}\Phi _{J}.
\end{equation}
Exactly as in $N=2$, the variables which appear naturally in the
Kaluza-Klein reduction are not necessarily the right variables to
put the effective action into the form (\ref{eff4D}). Instead one
again has to find the correct complex structure on the space of
scalar fields. It turns out that the complex structure
deformations $U$ are good K\"{a}hler coordinates since they are
the coordinates of a special K\"{a}hler manifold already in $N=2$.
For the remaining fields the definition of the K\"{a}hler
coordinates is not so obvious. For $O3/O7$-planes one finds:
\begin{enumerate}
\item Axio-dilaton: $S=e^{-\varphi}-i C_0$\footnote{Sometimes the
alternative definition $\tau\equiv iS=C_0+ie^{-\phi}$ is used.};

\item Two-form moduli: $G^{i_-}=c^{i_-}-i S b^{i_-}$,
$i_-=1,...,h_{1,1}^-(X)$;

\item Complex structure moduli: $U^{a_-}$,
$a_-=1,...,h_{1,2}^-(X)$;

\item K\"{a}hler moduli ($i_+=1,...,h_{1,1}^+(X)$) \cite{bbhl, louis0}:
\begin{equation}
T_{i_+}=\tau_{i_+}-\frac{1}{2(S+\bar{S})}k_{i_+ j_-
k_-}G^{j_-}\left(G-\bar{G}\right)^{k_-}+i\rho_{i_+},
\label{fullModuli}
\end{equation}
where $\tau_{i_+}$ is an implicit function of the $t_{i_+}$ and:
\begin{equation}
k_{i_+ j_- k_-}=\int_X \hat{D}_{i_+}\wedge \hat{D}_{j_-}\wedge
\hat{D}_{k_-}.
\end{equation}
\end{enumerate}
In what follows, we shall focus on orientifold projections such
that $h_{1,1}^{-} = 0 \Rightarrow h_{1,1}^{+} = h_{1,1}$. Hence
all the two-form moduli $b^{i_-}$ and $c^{i_-}$ are projected out.
In addition, the expression (\ref{fullModuli}), simplifies to
(where we have redefined $\rho_i\equiv b_i$ for later
convenience):
\begin{equation}
\label{fulloModuli} T_i=\tau_i+ib_{i},\textrm{ \ \
}i=1,...,h_{1,1}(X).
\end{equation}
The real part of the K\"{a}hler moduli $T_i$ can be related to the
initial quantities $t^i$ as follows. Expanding the K\"{a}hler form
$J$ in a basis $\{ \hat{D}_{i} \}$ of $H^{1,1}(X,\mathbb{Z})$ as:
\begin{equation}
J = \sum_{i=1}^{h_{1,1}} t^{i} \hat{D}_{i},
\end{equation}
the Calabi-Yau volume $\mathcal{V}$, measured with an Einstein
frame metric $g^{\scriptscriptstyle E}_{\mu
\nu}=e^{-\varphi/2}g^s_{\mu \nu}$ and in units of
$l_{s}=2\pi\sqrt{\alpha'}$, is given by:
\begin{equation}
  \mathcal{V} = \frac{1}{6}\int\limits_{X}J\wedge J\wedge
  J = \frac{1}{6} k_{ijk}t^{i}t^{j}t^{k}, \label{IlVolume}
\end{equation}
where $t^{i}$ are 2-cycle volumes and $k_{ijk}$ are the triple
intersection numbers of $X$:
\begin{equation}
  k_{ijk}=\int\limits_{X} \hat{D}_i \wedge \hat{D}_j \wedge
  \hat{D}_k.
\end{equation}
Then $\tau_{i}$ turns out to be the Einstein-frame volume (in
units of $l_s$) of the divisor $D_{i}\in H_{4}(X,\mathbb{Z})$,
which is the Poincar\'{e} dual to $\hat{D}_{i}$. Its axionic
partner $b_{i}$ is the component of the RR 4-form $C_{4}$ along
this cycle: $\int_{D_{i}} C_{4} = b_{i}$. The 4-cycle volumes
$\tau_{i}$ are related to the 2-cycle volumes $t^{i}$ by:
\begin{equation}
 \tau _{i}=\frac{\partial \mathcal{V}}{\partial
 t^{i}}=\frac{1}{2}\int\limits_{X}\hat{D}_{i}\wedge J\wedge
 J=\frac{1}{2} k_{ijk}t^{j}t^{k}, \label{defOfTau}
\end{equation}
Introducing now the following notation:
\begin{equation}
A_{ij}=\frac{\partial \tau _{i}}{\partial
t^{j}}=\int\limits_{X}\hat{D}_{i}\wedge \hat{D}_{j}\wedge
J=k_{ijk}t^{k}, \label{aa}
\end{equation}
some useful relations that we will use subsequently are:
\begin{equation}
\left\{
\begin{array}{c}
t^{i}\tau _{i}=3\mathcal{V}, \\
A_{ij}t^{j}=2\tau _{i}, \\
A_{ij}t^{i}t^{j}=6\mathcal{V}.
\end{array}
\right.   \label{useful}
\end{equation}

To leading order in the string-loop and $\alpha'$ expansions, the
resulting low-energy tree-level K\"{a}hler potential $K_{tree}$
has the form:
\begin{equation}
\frac{K_{tree}}{M_P^2}=-2\ln\left[ \mathcal{V}(T+\bar{T})\right]
-\ln \left(S+\bar{S}\right) -\ln \left( -i\int\limits_{X}\Omega(U)
\wedge \bar{\Omega}(\bar{U})\right), \label{eqtree}
\end{equation}
where $\Omega $ is the Calabi-Yau holomorphic (3,0)-form that
depends implicitly on the $U$-moduli, whereas the internal volume
$\mathcal{V}$ depends implicitly on the real part of the
$T$-moduli. The K\"{a}hler potential (\ref{eqtree}) is again block
diagonal, so that complex structure deformations $U$ do not mix
with the other scalars. Thus, the moduli space has the form:
\begin{equation}
\mc{M}=\mc{M}^{cs}_{h_{1,2}^-}\times\mc{M}^K_{h_{1,1}+1},
\label{factorisedmodulispace}
\end{equation}
where each factor is a K\"{a}hler manifold and
$\mc{M}^{cs}_{h_{1,2}^-}$ is even a special K\"{a}hler manifold.

We point out that $K_{tree}$ can be expressed as function of $T_i$
by solving the equations (\ref{defOfTau}) for the $t^i$ as
functions of the $\tau_i = \frac12(T_i +\overline{T}_i)$, and
substituting the result into (\ref{eqtree}), using
(\ref{IlVolume}) to evaluate $\mathcal{V}$. We point out that this
inversion cannot be done for a general Calabi-Yau, but the $t^i$
can be explicitly expressed as functions of the $\tau_i$ only in
some simple specific cases.

However, in the dual picture, where instead of the scalars
$\rho_{i_+}$, in the expansion (\ref{newExpansion}), one keeps the
two-forms $Q^{i_+}_2$, the K\"{a}hler deformations $t^{i_+}$ are
the lowest components of linear multiplets containing as bosonic
components $\left(t^{i_+},Q^{i_+}_2\right)$. In this case one can
give explicitly the metric for the linear multiplets, and the
somewhat involved definition of $T_{i_+}$ in (\ref{fullModuli})
can be understood as the superspace relation which expresses the
dualisation between chiral and linear multiplets \cite{louis0,
Binetruy}.

In the absence of internal fluxes, the tree-level superpotential
is vanishing and no $D$-terms are induced. Thus, as can be seen
from the expression (\ref{bo}), no scalar potential is generated
for any of the moduli, which, therefore, correspond to exactly
flat directions. This is the reason why these fields are called
moduli. However, given that the values of $S$ and the complex
structure moduli $U$, can become fixed once background fluxes are
turned on, we shall now discuss in depth the r\^{o}le played by
internal fluxes in section 2.4.

\section{Background fluxes}

As we have seen in section \ref{TreeLevelEffectiveAction}, all the
four dimensional scalar components of the chiral multiplets, which
arise in type IIB Calabi-Yau orientifold compactifications with
$O3/O7$ planes, have a completely flat potential. This is due to
the fact that the tree-level superpotential is vanishing. We shall
see in this section that $W_{tree}$ can indeed be generated by
turning on background fluxes \cite{gkp, louis0, fluxesonCY}. In
turn, a scalar potential is generated for the axio-dilaton and the
complex structure moduli, which therefore get stabilised. However,
we shall see that, due to the no-scale structure, the K\"{a}hler
moduli cannot be fixed by the internal fluxes, but by the
interplay of several corrections beyond the leading order
approximations, which will be studied in chapter
\ref{CapKaehlerModuliStabilisation}.

\subsection{Type IIB fluxes}

The definition of the flux of an arbitrary $p$-form field strength
$F_p$ through a $p$-cycle $\gamma^i_p$ in $X$, is nothing but the
straightforward generalisation of the electromagnetic flux:
\begin{equation}
\int_{\gamma_p^i\in X}F_p=n_i\neq 0.
\end{equation}
A more geometrical meaning of the concept of fluxes, can be
understood by expanding $F_p$ in terms of harmonic forms
$\omega^i_p$ with:
\begin{equation}
F_p=<F_p,\omega^i_p>\omega^i_p,\textrm{ \ \ \ }\omega_p\in H_p(X),
\end{equation}
where with $<F_p,\omega^i_p>$ we have denoted the projection of
$F_p$ along the space generated by the base element $\omega_i^p$.
From Poincar\'{e} duality, it turns out that:
\begin{equation}
<F_p,\omega^i_p>=\int_{\gamma_p^i\in X}F_p,\textrm{ \ \
}\Longrightarrow\textrm{ \ \ }F_p=n_i\omega^i_p.
\end{equation}
where the $p$-cycles $\gamma^i_p$ are Poincar\'{e} dual to the
$p$-forms $\omega_p^i$. The equation of motion and the Bianchi
identity, $dF_p=0=d^{\dagger}F_p$, require the background fluxes
$n_i$ to be constant. Moreover, these constants have to be
integers, since in string theory they are quantised due to a Dirac
quantisation condition of the form \cite{gkp}:
\begin{equation}
\frac{1}{2\pi\alpha'}\int_{\gamma_p^i\in X}F_p=n_i\in\mathbb{Z}.
\end{equation}
However in the low energy/large volume approximation we are
considering here, they appear as continuous parameters which
deform the low energy supergravity. The presence of background
fluxes leads to two crucial consequences. The first one is that
they provide a source of potential energy that partially lifts the
vacuum degeneracy of string theory, while the second important
application is that background fluxes generically break the
residual $N=1$ supersymmetry spontaneously.

In the particular case of type IIB Calabi-Yau compactifications,
one can turn on internal fluxes of the RR field strength $F_3$ and
the NS-NS field strength $H_3$ through $3$-cycles of the
Calabi-Yau $X$:
\begin{eqnarray}
\int_{\gamma_3^a\in X}F_3&=&m^{RR,a}\in\mathbb{Z},\textrm{ \ \ \ \
}\int_{\gamma_3^b\in X}F_3=n^{RR}_b\in\mathbb{Z}, \nonumber \\
\int_{\gamma_3^a\in X}H_3&=&m^{NS,a}\in\mathbb{Z},\textrm{ \ \ \ \
}\int_{\gamma_3^b\in X}H_3=n^{NS}_b\in\mathbb{Z}.
\label{integrations}
\end{eqnarray}
where the $2h_{1,2}$ RR flux parameters $(m^{RR,a},n^{RR}_b)$ and
the $2h_{1,2}$ NS-NS flux parameters $(m^{NS,a},n^{NS}_b)$ are the
coefficients of the expansion of $F_3$ and $H_3$ into the
symplectic base $(\alpha_a,\beta^b)$:
\begin{equation}
F_3=m^{RR,a}\alpha_a+n^{RR}_b\beta^b,\textrm{ \ \ \
}H_3=m^{NS,a}\alpha_a+n^{NS}_b\beta^b,\textrm{ \ \ \ }
a,b=1,...,h_{1,2}.
\end{equation}
However, it turns out that the relevant three-form flux $G_3$ is
defined as \cite{gkp}:
\begin{equation}
G_3\equiv F_3-iS H_3=\left(m^{RR,a}-iS
m^{NS,a}\right)\alpha_a+\left(n^{RR}_b-iS n^{NS}_b\right)\beta^b.
\label{G3}
\end{equation}
The effect of the these non-vanishing background fluxes is the
following. The electric fluxes $(n^{RR}_b,n^{NS}_b)$ gauge a
translational isometry of the quaternionic manifold $\mc{M}_Q$
acting on the dual scalars of the two space-time two-forms $B_2$
and $C_2$, which we shall denote as $q^{1,2}$. Hence the ordinary
derivatives are replaced by covariant derivatives:
\begin{equation}
\partial_{\mu}q^{1,2}\longrightarrow
D_{\mu}q^{1,2}=\partial_{\mu}q^{1,2}+\left(n^{RR,NS}_b
A_{\mu}^b\right) q^{1,2}.
\end{equation}
On the other hand, the magnetic fluxes $(m^{RR,a},m^{NS,a})$ are
related to the mass parameters of $B_2$ and $C_2$ \cite{KKred3}.
The total induced scalar potential (but no $D$-term) reads
\cite{TaylorVafa}:
\begin{equation}
V(U,S)=-\left(\bar{n}-\bar{\mc{M}}\cdot \bar{m}
\right)_a\left(\textrm{Im}\mc{M}\right)^{-1\,ab}\left(n-\mc{M}\cdot
m\right)_b, \label{fluxpotential}
\end{equation}
where $\mc{M}(U)$ is the gauge kinetic matrix appearing in
(\ref{4DL}). It is crucial to notice that $V$ depends on the
axio-dilaton $S$, as it is clear from the definition of $G_3$
(\ref{G3}), but it depends also on the complex structure moduli
$U$, since they give the volume of the $3$-cycles over which the
integrations in (\ref{integrations}) are performed. Therefore both
$S$ and $U$-moduli can be stabilised by turning on background
$3$-form fluxes. The flux-generated tree-level potential
(\ref{fluxpotential}) can be derived from a superpotential that
takes the famous Gukov-Vafa-Witten form \cite{gvw}:
\begin{equation}
W_{tree}(S, U)\sim \int\limits_{X}G_{3}\wedge \Omega \,.
\label{Wtree}
\end{equation}
After performing the orientifold projection to obtain an $N=1$
effective field theory, the expression (\ref{G3}) for $G_{3} =
F_{3}-iS H_{3}$, projects down to:
\begin{equation}
G_3=m^{a_-}\alpha_{a_-}+n_{a_-}\beta^{a_-},\textrm{ \ \ \ }
a_-=1,...,h_{1,2}^-,
\end{equation}
with $2\left(h_{1,2}^- +1\right)$ complex flux parameters:
\begin{equation}
m^{a_-}=m^{RR,a_-}-iS m^{NS,a_-},\textrm{ \ \ \
}n_{a_-}=n^{RR}_{a_-}-iS n^{NS}_{a_-}.
\end{equation}
Finally, let us mention that the backreaction of background fluxes
on the internal geometry causes warping, and so it renders $X$
conformally Calabi-Yau:
\begin{equation}
ds^2=e^{2A(y)}g_{\mu\nu}dx^{\mu}dx^{\nu}+e^{-2A(y)}g_{m\bar{n}}dy^m
d\bar{y}^{\bar{n}},
\end{equation}
where the warp factor $A(y)$ can depend only on the internal
coordinates $y^m$, in order not to break four-dimensional
Poincar\'{e} invariance. However, from now on, we shall ignore the
effect of any warp factor since it is negligible in the case of
exponentially large volume which we will be interested in. Here we
just recall that the main application of warping is the solution
of the hierarchy problem thanks to the red-shifting effect at the
end of a warped throat where chiral matter is localised
\cite{warping, fluxeshierarchy}. However, in our case, the
solution of the hierarchy problem is due to the presence of an
exponentially large volume of the extra dimensions.

\subsection{Tadpole cancellation}

Historically there was a no-go theorem \cite{nogo} claiming that
it was impossible to have non-zero fluxes due to the tadpole
cancellation condition, given that the fluxes were the only
semi-positive definite contribution to the $\hat{C}_4$ tadpole.
However the discovery of localised sources, such as $D$-branes
\cite{DbraneDiscovery}, led to the possibility to overcome this
no-go theorem and allow for non-vanishing background fluxes
$G_3=F_3-iSH_3$.

In fact, the fluxes turned out not to be the only source for the
$\hat{C}_4$ tadpole, since it was realised that several local
sources, like $D3$-, $\overline{D3}$-branes, wrapped $D7$-branes,
orientifold planes, or gauge fields on $D7$-branes (however, we
shall not concern ourselves with gauge fields here), can carry
$D3$-brane charge \cite{gkp}.

The requirement that the $\hat{C}_4$ tadpole must be cancelled,
means that the total internal $D3$-brane charge has to vanish.
This condition, in turn, guarantees the absence of anomalies in
the low-energy four dimensional theory, and reads:
\begin{equation}
N_{D3}-N_{\bar{D3}}+\frac{1}{(2\pi)^4\alpha'^2} \int H_3 \wedge
F_3 = \frac{\chi(Z)}{24},
\end{equation}
where $\chi(Z)$ collects the contribution to $D3$-brane charge
from orientifold planes and $D7$ branes. In the F-theory
interpretation \cite{Ftheory}, $\chi(Z)$ is the Euler number of
the corresponding four-fold.

Due to the great importance of $D$-branes in order to be able to
turn on background fluxes (as well as to obtain non-Abelian gauge
theories on the world-volume of the brane and stringy realisations
of the brane-world scenario), we shall now discuss the effective
action of these localised sources in section \ref{BraneSection}.
We just mention here that the number of $D$-branes that can be
introduced, is not arbitrary but is constrained by the requirement
of cancelling the $\hat{C}_4$ tadpole. Hence the number of
$D3$-branes that must be introduced varies with the discrete flux
choices. If we wish to avoid the need to include $D3$-branes, we
can always take the fluxes to saturate the tadpole condition.

\section{$D$-brane effective action} \label{BraneSection}

A very important ingredient of Calabi-Yau flux compactifications
is the presence of $Dp$-branes which have to fill the
four-dimensional space-time in order not to break Poincar\'{e}
invariance. Each space-time filling $Dp$-brane comes along with a
$U(1)$ gauge theory that lives on its world volume. More
generally, a stack of $N$ $Dp$-branes gives rise to a non-Abelian
$U(N)$ gauge theory \cite{Myers}.

The dynamics of a $D$-brane is governed by the Dirac-Born-Infeld
action $S_{DBI}$ together with a Chern-Simons action $S_{CS}$. For
a generic $Dp$-brane, they look like \cite{PolchinskiBook}:
\begin{eqnarray}
S_{DBI}&=&-T_p\int_{\mc{W}}d^{p+1}\xi
e^{-\hat{\phi}}\sqrt{-\textrm{det}\left[\varphi^*(g+B_2)+2\pi\alpha'
F_2\right]}, \nonumber \\
S_{CS}&=&\mu_p\int_{\mc{W}}\varphi^* \left(C_{p+1}e^{B_2}\right)
e^{2\pi\alpha' F_2}, \label{braneAction}
\end{eqnarray}
where $T_p$ is the tension, $\mu_p$ is the RR-charge of the
$Dp$-brane, and $F_2$ is the gauge field strength. In this case,
the integrals in (\ref{braneAction}) are taken over the
$(p+1)$-dimensional world-volume $\mc{W}$ of the $Dp$-brane, which
is embedded in the ten dimensional space-time manifold
$\mathbb{R}^{9,1}$ via the map $\varphi:\mc{W}\hookrightarrow
\mathbb{R}^{9,1}$. In order to preserve $N=1$ supersymmetry, the
$Dp$-brane has to satisfy a BPS condition, which amounts to the
fact that the tension $T_p$ must be equal to the RR-charge: $T_p =
\mu_p$.

Adding both $S_{DBI}$ and $S_{CS}$ to the ten-dimensional type IIB
bulk action and performing a Kaluza-Klein reduction, one derives
again a $D=4$ low energy effective action, which can be written in
the standard $N=1$ supergravity form (\ref{eff4D}). The tree-level
K\"{a}hler potential coincides with the $K_{tree}$ of the
orientifolds given in (\ref{eqtree}), but with a new definition of
the chiral K\"{a}hler coordinates, which depends on the fact that
we are considering $D3$- or $D7$-branes \cite{louis0, louis1,
louis2, louis3}. Let us start with $D3$-branes and then turn to
the case of $D7$-branes.

\subsection{$D3$-branes}

A space-time filling $D3$-brane is a point in the Calabi-Yau
orientifold whose position is parameterised by three gauge neutral
moduli $\zeta^i$, $i = 1, 2, 3$, which are the scalar components
of chiral matter multiplets living on the four-dimensional world
volume of the $D3$-brane and transforming in the adjoint
representation of $U(N)$ (in the case of a stack of $N$
$D3$-branes).

In this case, the definition of the $N=1$ chiral K\"{a}hler
coordinates (\ref{fullModuli}) is replaced by (in a small
$\zeta^i$ expansion) \cite{louis2}:
\begin{eqnarray}
T_{i_+}&=&\tau_{i_+} -\frac{1}{2(S+\bar{S})}k_{i_+ j_-
k_-}G^{j_-}\left(G-\bar{G}\right)^{k_-} \\
&&+i\left[\rho_{i_+}+\frac{\mu_3 l_s^4}{4\pi^2}
\left(\hat{D}_{i_+}\right)_{i\bar{j}}\textrm{Tr}
\zeta^i\left(\bar{\zeta}^{\bar{j}}-\frac{i}{2}\bar{U}^{a_-}
\left(\bar{\chi}_{a_-}\right)^{\bar{j}}_l
\zeta^l\right)+h.c.\right], \nonumber
\label{fullModuliWithD3Branes}
\end{eqnarray}
with $\hat{D}_{i_+}\in H_{1,1}^+$, $\chi_{a_-}\in H_{1,2}^-$ and
$\tau_i$ still given by (\ref{defOfTau}). Using the modified
definition of $T_{i^+}$, one again has to solve for $t^{i_+}$ in
terms of the chiral variables $S,T,G$ and $U$. Inserted into
$\mc{V}$, then results in a $K_{tree}(S, T, G, U, \zeta)$. We see
from (\ref{fullModuliWithD3Branes}) that the complex structure
moduli $U$ couple to the matter fields $\zeta^i$, and thus the
moduli space no longer is a direct product. The K\"{a}hler
potential is still the sum of two terms but both terms now depend
on $U$.

As an explicit example consider the situation $h_{1,1}^+=1$ and
$h_{1,1}^-=0$. In this case (\ref{fullModuliWithD3Branes}) can be
solved for $t$, and one obtains as the K\"{a}hler potential:
\begin{equation}
K= -3\ln\left[T+\bar{T}+\hat{D}_{i\bar{j}}\textrm{Tr}
\zeta^i\left(\bar{\zeta}^{\bar{j}}-\bar{U}^{a_-}
\left(\chi_{a_-}\right)^{\bar{j}}_l
\zeta^l\right)+h.c.\right]-\ln\left(S+\bar{S}\right)+K_{cs}(U).
\end{equation}
Using the Kaluza-Klein reduction, one also determines the gauge
kinetic function of the non-Abelian gauge theory of the $D3$-brane
to be $f\sim S$. Finally the superpotential is found to be:
\begin{equation}
W=\frac{1}{3}Y_{ijk}\textrm{Tr}\zeta^i \zeta^j \zeta^k,\textrm{ \
\ \ }Y_{ijk}=\Omega_{ijk}(U).
\end{equation}

\subsection{$D7$-branes}

A $D7$-brane is an extended object with seven space and one time
dimensions. As we have already stressed, it has to fill the
ordinary four-dimensional space time, while the remaining spatial
four dimensions have to wrap around an internal 4-cycle $\Sigma_4$
inside the Calabi-Yau orientifold (notice that the four cycle
$\Sigma_4$ includes both the cycle the $D7$-brane wraps and its
image with respect to the orientifold involution $\sigma$).
Therefore the world-volume of a $D7$-brane is of the form
$\mc{W}=\mathbb{R}^{3,1}\times \Sigma_4$. The possible
deformations of the internal $4$-cycle $\Sigma_4$ are again
parameterised by $D7$-moduli $\zeta^i(x)$, which are the scalar
components of chiral matter multiplets.

In addition, the eight-dimensional world-volume gauge field gives
rise to a four-dimensional $U(1)$ gauge field $A_{\mu}(x)$ and
several Wilson line moduli $a_{\alpha}(x)$. In this case, the the
appropriate chiral K\"{a}hler coordinates are found to be (in the
limit of small $D7$-brane fluctuations $\zeta^i$) \cite{louis1}:
\begin{eqnarray}
\mc{S}&=& i S-\mu_7\mathcal{L}_{i\bar{j}}
\zeta^i\bar{\zeta}^{\bar{j}}, \textrm{ \ \ }
S=e^{-\phi}-iC_0,\textrm{ \ \ }G^{i_-}=c^{i_-}-i S b^{i_-},
\nonumber \\
T_{i_+}&=&\tau_{i_+}-\frac{1}{2(S+\bar{S})}k_{i_+ j_-
k_-}G^{j_-}\left(G-\bar{G}\right)^{k_-} \\
&&+i\left(\rho_{i_+}-\frac{1}{2}k_{i_+ j_-
k_-}c^{j_-}b^{k^-}+\frac{\mu_7 l_s^4}{2\pi^2}
\mc{C}_{i_+}^{\alpha\bar{\beta}}a_{\alpha}\bar{a}_{\bar{\beta}}\right),
\nonumber \label{fullModuliWithD7Branes}
\end{eqnarray}
where $\mc{L}_{i\bar{j}}\mc{C}_{a_+}^{\alpha\bar{\beta}}$ are
intersection numbers on the $4$-cycle $\Sigma_4$ defined in
\cite{louis1}. In terms of these K\"{a}hler coordinates, the
K\"{a}hler potential for the low energy effective supergravity
action is given by:
\begin{equation}
K_{tree}=-2\ln\mc{V}-\ln\left[-i\left(\mc{S}-\bar{\mc{S}}\right)-2i\mu_7\mc{L}_{i\bar{j}}
\zeta^i\bar{\zeta}^{\bar{j}}\right]+K_{cs}(U),
\label{KaehlerPotentialWithD7Branes}
\end{equation}
where $K_{tree}(\mc{S},G,T,\zeta,a)$ is obtained by solving
(\ref{fullModuliWithD7Branes}) for $t^{i_+}$, exactly as before.
For the holomorphic gauge coupling function one finds $f\sim
\tilde{T}$, where $\tilde{T}$ includes the K\"{a}hler modulus
$\tilde{\tau}$ which parameterises the volume of the $4$-cycle
$\Sigma_4$ wrapped by the $D7$-brane.

Moreover, the presence of a $D7$-brane generates a non-vanishing
$D$-term of the form \cite{louis3}:
\begin{equation}
D=\frac{\mu_7 l_s^2}{\mc{V}}\int_{\Sigma_4}J\wedge B_2.
\end{equation}
However, by appropriately adjusting $B_2=b^{i_-}\hat{D}_{i_-}$,
this $D$-term can always be made to vanish, which just corresponds
to the BPS-condition for the $D7$-brane. Finally, one can consider
to turn on fluxes on the $D7$-brane. This requires that the
integral $\int_{\gamma_2}F_2$ is non-vanishing, where $F_2$ is the
`internal' field strength of the $D7$-gauge boson. These fluxes
generate additional contributions to the $D$-term and also a
superpotential \cite{louis1}.

\section{Flux-stabilisation}

In order to illustrate the r\^{o}le played by background fluxes to
stabilise the axio-dilaton and the complex structure moduli, let
us focus on orientifold projections such that
$h_{1,1}^{-}=0\Rightarrow h_{1,1}^{+}=h_{1,1}$, and with vanishing
open string moduli.

As we have seen in section 2.4, the background fluxes $G_{3} =
F_{3}-iS H_{3}$ generate a tree-level superpotential (but no
$D$-term) that takes the famous Gukov-Vafa-Witten form
\cite{gvw}\footnote{The prefactor in (\ref{Wtree}) is due to
careful dimensional reduction, as can be seen in Appendix A of
\cite{LVS2}. However, the authors of \cite{LVS2} define the
Einstein metric via
$g^s_{\mu\nu}=e^{(\phi-\langle\phi\rangle)/2}g^{\scriptscriptstyle
E}_{\mu\nu}$, so that it coincides with the string frame metric in
the physical vacuum. On the contrary, we opt for the more
traditional definition
$g^s_{\mu\nu}=e^{\phi/2}g^{\scriptscriptstyle E}_{\mu\nu}$, which
implies no factor of $g_s$ in the prefactor of $W$.}:
\begin{equation}
W_{tree}(S, U)=\frac{M_P^3}{\sqrt{4
\pi}}\int\limits_{X}G_{3}\wedge \Omega \,. \label{Wtree}
\end{equation}
Notice that the K\"{a}hler moduli $T_i$ do not appear in
$W_{tree}$ and so remain precisely massless at leading
semiclassical order. In order to understand this important issue,
let us write down explicitly the form that the scalar potential
acquires once background fluxes are turned on:
\begin{equation}\label{scalarpotentialfactorised}
V=e^{K}\left\{
K^{S\bar{S}}D_{S}WD_{\bar{S}}\bar{W}+K^{U\bar{U}}D_{U}WD_{\bar{U}}\bar{W}
+K^{i\bar{j}}D_{i}WD_{\bar{j}}\bar{W}-3\left\vert W\right\vert
^{2}\right\},
\end{equation}
where:
\begin{eqnarray}
D_{i}W &=&\frac{\partial W}{\partial T_{i}}+W\frac{\partial
K}{\partial T_{i}
}\equiv W_{i}+WK_{i}, \nonumber \\
D_{\bar{j}}\bar{W} &=&\frac{\partial \bar{W}}{\partial
\bar{T}_{\bar{j}}}+ \bar{W}\frac{\partial K}{\partial
\bar{T}_{\bar{j}}}\equiv \bar{W}_{\bar{j}}+ \bar{W}K_{\bar{j}}.
\nonumber
\end{eqnarray}
The form of the scalar potential given in
(\ref{scalarpotentialfactorised}) has used the factorisation of
the moduli space (\ref{factorisedmodulispace}): in general this
will be lifted by quantum corrections. As $W_{tree}$ is
independent of the K\"{a}hler moduli, this reduces to:
\begin{equation}
V=e^{K}\left\{
K^{S\bar{S}}D_{S}WD_{\bar{S}}\bar{W}+K^{U\bar{U}}D_{U}WD_{\bar{U}}\bar{W}+\left(
K^{i\bar{j}}K_{i}K_{\bar{j}}-3\right) \left\vert W\right\vert
^{2}\right\} \label{eqq}.
\end{equation}
Notice that $\mathcal{V}$ is a homogeneous function of degree
$3/2$ in the $\tau_i$'s, and so also ensures $K_{tree}$ satisfies
$K_{tree}(\lambda \tau_i) \equiv K_{tree}(\tau_i) - 3\ln\lambda$
as an identity for all $\lambda$ and $\tau_i$. It follows from
this that $K_{tree}$ satisfies the \textit{no-scale} identity
\cite{noscale}:
\begin{equation}
\left(\frac{\partial^{2} K_{tree}}{\partial T_{i} \partial \bar{
T_{\bar{j}}}}\right)^{-1}\frac{\partial K_{tree}}{\partial
T_{i}}\frac{\partial K_{tree}}{\partial \bar{T_{\bar{j}}}}=3,
\label{noscaleidentity}
\end{equation}
which implies the existence of a classical no-scale structure of
the potential for the K\"{a}hler moduli, since the last term of
(\ref{eqq}) vanishes:
\begin{equation}
V=e^{K}\left\{
K^{S\bar{S}}D_{S}WD_{\bar{S}}\bar{W}+K^{U\bar{U}}D_{U}WD_{\bar{U}}\bar{W}\right\}
\geq 0.
\end{equation}
As the scalar potential is positive semi-definite, it is possible
to fix supersymmetrically the dilaton and the complex structure
moduli at tree level by demanding $D_{S}W=0=D_{U}W$. Usually,
these fields are integrated out setting them equal to their vacuum
expectation values, but sometimes we will keep their dependence
manifest. However, since they are stabilised at tree level, even
though they will couple to quantum corrections, these will only
lead to subleading corrections to their VEVs, so it is safe just
to integrate them out \cite{IntOut}. From now on, we will set:
\begin{equation}
W_{tree}=\frac{M_P^3}{\sqrt{4\pi}}W_{0}\equiv
\frac{M_P^3}{\sqrt{4\pi}}\left\langle \int\limits_{X}G_{3}\wedge
\Omega \right\rangle,
\end{equation}
and:
\begin{equation}
K_{tree}= -2\ln \mathcal{V} - \ln\left(\frac{2}{g_{s}}\right) +
K_{cs},
\end{equation}
with:
\be
  e^{-K_{cs}} = \left\langle -i\int\limits_{X}\Omega \wedge
  \bar{\Omega}\right\rangle\,. \label{K0expr}
\ee A useful property of $K_{0}=-2\ln \mathcal{V}$ is:
\begin{equation}
K_{i}^{0}\equiv \frac{\partial K_{0}}{\partial \tau _{i}}=-
\frac{t_{i}}{\mathcal{V}},
\end{equation}
where, for later convenience, we have expressed the derivatives of
the K\"{a}hler potential in terms of derivatives with respect to
$\tau = \hbox{Re}(T)$, rather than derivatives with respect to
$T$. In addition, the general form of the K\"{a}hler metric is:
\begin{equation}
K_{ij}^{0}\equiv \frac{\partial ^{2}K_{0}}{\partial \tau
_{i}\partial \tau
_{j}}=\frac{1}{2}\frac{t_{i}t_{j}}{\mathcal{V}^{2}}-\frac{
A^{ij}}{\mathcal{V}},
\end{equation}
and its inverse looks like:
\begin{equation}
K_{0}^{ij}\equiv \left( \frac{\partial ^{2}K_{0}}{\partial \tau
_{i}\partial \tau _{j}}\right) ^{-1}=\tau _{i}\tau
_{j}-\mathcal{V} A_{ij}.
\end{equation}
From the previous relations it is also possible to show that:
\begin{equation}
K_{0}^{ij}K_{i}^{0}=-\tau _{j},  \label{p1}
\end{equation}
and the more important no-scale structure result:
\begin{equation}
K_{0}^{ij}K_{i}^{0}K_{j}^{0}=3.  \label{no scale}
\end{equation}
The classical flatness of the potential for the K\"{a}hler moduli,
$V \equiv 0$, implies that to study K\"{a}hler moduli
stabilisation we should keep all possible quantum corrections,
which will be studied in detailed in chapter
\ref{CapKaehlerModuliStabilisation}. We shall show that they can
indeed generate a scalar potential with stabilised $T$-moduli.

We finally stress that the fluxes may, but need not, break the
remaining four-dimensional $N=1$ supersymmetry
\cite{fluxesbreakSUSY, fluxesbreakSUSY2, fluxesbreakSUSY3},
corresponding to whether or not the resulting scalar potential for
the K\"{a}hler moduli is minimised where $F_i=D_i W=\partial_i W +
W
\partial_i K$ vanishes at the minimum. More precisely, for the superpotential (\ref{Wtree}), one finds
unbroken supersymmetry, that is $F_i=0$, for $G_3\in H_{2,1}^-$
\cite{gkp}. For $G_3\in H_{0,3}^-$, one finds a broken
supersymmetry $F_i\neq 0$ in Minkowski space, that is with $V=0$.
On the other hand, for $G_3\in H_{3,0}^-\oplus H_{1,2}^-$, one
obtains only unstable vacuum solutions.

\chapter{K\"{a}hler Moduli Stabilisation}
\label{CapKaehlerModuliStabilisation} \linespread{1.3}

\section{Corrections to the leading approximation} \label{22}

As seen in chapter 2, at tree level we can stabilise only the
dilaton and the complex structure moduli but not the K\"{a}hler
moduli. The only possibility to get mass for the $T$-moduli is
thus through corrections at sub-leading order in $\alpha'$ and
$g_s$ (string loops). We stress that this is not required for the
$S$ and $U$ moduli, whose potential is dominated by the leading
order contribution.

It is known that in $N=1$ four-dimensional supergravity, the
K\"{a}hler potential receives corrections at every order in
perturbation theory, while the superpotential receives
non-perturbative corrections only, due to the non-renormalisation
theorem \cite{beq}. The corrections will therefore take the
general form:
\begin{equation}
\left\{
\begin{array}{l}
K=K_{tree}+K_{p}+K_{np}, \\
W=W_{tree}+W_{np},
\end{array}
\right.
\end{equation}
and the hope is to stabilise the K\"{a}hler moduli through these
corrections to the tree-level approximation. In this section we
shall review the behaviour of the non-perturbative and $\alpha'$
corrections and then study the $g_{s}$ corrections in chapter 5 of
this thesis. Let us now review briefly all these kinds of
corrections.

\subsection{Non-perturbative corrections}

Since the superpotential receives no contributions at any finite
order in $\alpha'$ and $g_s$, its first corrections arise
non-perturbatively. These can be generated either by Euclidean
$D3$-brane ($ED3$) instantons \cite{ED3} wrapping 4-cycles in the
extra dimensions, or by gaugino condensation in the supersymmetric
gauge theories located on $D7$-branes that also wrap internal
4-cycles \cite{GauginoCondensation}. The superpotential that both
kinds of effects generate is of the form:
\begin{equation}
  W=\frac{M_P^3}{\sqrt{4
\pi}}\left(W_{0}+\sum\limits_{i}A_{i}e^{-a_{i}T_{i}}\right),
 \label{yuj}
\end{equation}
where the sum is over the 4-cycles generating nonperturbative
contributions to $W$, and as before $W_{0}$ is independent of
$T_i$. The coefficients $A_{i}$ correspond to threshold effects
and can depend on $U$ and $D3$-brane position moduli. The
constants $a_i$ in the exponential are given by $a_{i}=2\pi $ for
$ED3$-branes \cite{ED3}, or $a_{i}=2\pi /N$ for gaugino
condensation in an $SU(N)$ gauge theory
\cite{GauginoCondensation}. There may additionally be higher
instanton effects in (\ref{yuj}), but these can be neglected so
long as each $\tau_{i}$ is stabilised such that $a_{i}\tau_{i}\gg
1$. From (\ref{bo}), the presence of such a superpotential
generates a scalar potential for $T_i$, of the form (up to a
numerical prefactor and powers of $g_s$ and $M_P$):
\begin{eqnarray}
 \delta V_{(np)} &=& e^{K_0}  K_{0}^{j\bar\imath} \Bigl[
 a_{j}A_{j} \, a_{i} \bar{A}_{i} e^{-\left(
 a_{j}T_{j} + a_{i}\overline{T}_{i}\right) } \nonumber\\
 && \qquad\qquad  -\left(
 a_{j}A_{j}e^{-a_{j}T_{j}}\overline{W}
 \partial_{\bar\imath} K_{0}+a_{i}\bar{A}_{i} e^{-a_{i}\overline{T}_{i}} W
 \partial_j K_{0}\right) \Bigr].  \label{scalarWnp}
\end{eqnarray}

\subsection{Leading $\alpha'$ corrections}

Unlike the superpotential, the K\"{a}hler potential receives
corrections order-by-order in both the $\alpha'$ and string-loop
expansions:
\begin{equation}
K_{p}=\delta K_{(\alpha')}+\delta K_{(g_{s})}.
\end{equation}
On top of that, there are also non-perturbative effects $K_{np}$
which can come from either world-sheet or brane instantons and are
subdominant compared to the perturbative corrections to the
K\"{a}hler potential (see for instance \cite{bqq, kaplu}).
Therefore, we shall neglect them in the following.

In the effective supergravity description the $\alpha'$
corrections correspond to higher derivative terms. The leading
$\alpha'$ contribution comes from the the ten dimensional
$\mathcal{O}(\alpha'^{3})$ $\mathcal{R}^{4}$ term \cite{R4}, and
it leads to a K\"{a}hler potential for the K\"{a}hler moduli of
the form \cite{bbhl}:
\begin{equation}
  \frac{K}{M_P^2}= -2\ln \left( \mathcal{V}+\frac{\xi }{2g_{s}^{3/2}}\right)
  = -2\ln\mathcal{V}-\frac{\xi}{g_s^{3/2}\mathcal{V}}+\mathcal{O}\left(
  1/\mathcal{V}^{2}\right),
 \label{eq}
\end{equation}
with the constant $\xi$ given by:
\begin{equation}
\xi =-\frac{\chi (X)\zeta (3)}{2(2\pi )^{3}}.
\end{equation}
Here $\chi (X)=2\left(h_{1,1}-h_{2,1}\right)$ is the Euler number
of the Calabi-Yau $X$, and the relevant value for the Riemann zeta
function is $\zeta (3)\equiv \sum_{k=1}^{\infty }1/k^{3}\simeq
1.2$. We stress the point that the $\alpha' $ expansion is an
expansion in inverse volume and thus can be controlled only at
large volume. This is important, as very little is known about
higher $\alpha' $ corrections, the exact form of which are not
known even in the maximally supersymmetric flat ten-dimensional
IIB theory. From now on we focus only on situations in which the
volume can be stabilised at $\mathcal{V}\gg 1$ in order to have
theoretical control over the perturbative expansion in the
low-energy effective field theory.

Denoting for convenience $\hat{\xi}\equiv \xi/g_s^{3/2}$,
(\ref{eq}) implies to leading order the following contribution to
$V$ (again, up to a prefactor containing powers of $g_s$ and
$M_P$):
\begin{equation}
 \delta V_{(\alpha')} = 3e^{K_0}\hat{\xi}\frac{\left( \hat{\xi}^{2}+7\hat{\xi}
 \mathcal{V}+\mathcal{V}^{2}\right) }{\left( \mathcal{V}-\hat{\xi}
 \right) \left( 2\mathcal{V}+\hat{\xi} \right)^{2}} W_0^2
 \simeq \frac{3\xi W_0^2}{4 g_s^{3/2}\mathcal{V}^3}.
 \label{scalaralpha'}
\end{equation}

\subsection{String loop corrections}

In this section, we shall ignore corrections from string loops.
However, we shall present a detailed analysis of their behaviour
for general type IIB compactifications in chapter
\ref{SystematicsOfStringLoopCorrections}. Here we just briefly
mention that historically these quantum corrections have always
been the less understood, and so they have been neglected in most
mechanisms of K\"{a}hler moduli stabilisation. In section 3.2 we
shall present a detailed survey of all the main K\"{a}hler moduli
stabilisation mechanisms available in the literature, and we will
show that in most cases the neglecting of $g_s$ corrections is not
theoretically justified. In fact, their inclusion could
destabilise the moduli, or, as we shall see in chapter 6 for the
case of LARGE Volume Scenarios, string loop corrections could
generate a completely new set of vacua.

\subsection{Scalar potential}

Considering only the contributions $\delta V_{(np)}$ and $\delta
V_{(\alpha')}$, the large volume limit of the total scalar
potential can be obtained by combining (\ref{scalarWnp}) and
(\ref{scalaralpha'}), and takes the form:
\begin{eqnarray}
\label{scalar}
 V &=& e^{K_0} \left\{
 K_{0}^{j\bar\imath} \Bigl[
 a_{j}A_{j} \, a_{i} \bar{A}_{i} e^{-\left(
 a_{j}T_{j} + a_{i}\overline{T}_{i}\right) }
  \phantom{\frac12}\right. \\
 && \qquad\qquad  - \left. \left(
 a_{j}A_{j}e^{-a_{j}T_{j}}\overline{W}
 \partial_{\bar\imath} K_{0}+a_{i}\bar{A}_{i} e^{-a_{i}\overline{T}_{i}} W
 \partial_j K_{0}\right) \Bigr]
 + \frac{3 \,\hat\xi}{4 \mathcal{V}} \, |W_0|^2 \right\}.
 \nonumber
\end{eqnarray}

\section{Survey of moduli stabilisation mechanisms}
\label{SurveyOfModuliStabilisationMechanisms}

We have seen that the no-scale structure will be broken by several
contributions which will lead to the following general form of the
scalar potential for the K\"{a}hler moduli:
\begin{equation}
V=\delta V_{(np)}+\delta V_{(\alpha' )}+\delta V_{(g_{s})}+
V_{local}+V_{D}, \label{general}
\end{equation}
where $\delta V_{(np)}+\delta V_{(\alpha ' )}$ is given by
(\ref{scalar}), $\delta V_{(g_{s})}$ is the perturbative
contribution from string loop corrections, $V_{local}$ is the
potential generated by extra local sources, and $V_{D}$ is the
usual $D$-term scalar potential for $N=1$ supergravity:
\begin{equation}
V_{D}=\frac{1}{2}\left( \left( \hbox{Re}f\right) ^{-1}\right)
^{ab}D_{a}D_{b},\textrm{ \ \ \ }D_{a}=\left[
K_{i}+\frac{W_{i}}{W}\right] \left( T_{a}\right) _{ij}\varphi
_{j}. \label{D scalar}
\end{equation}
We now review the main K\"{a}hler moduli stabilisation mechanisms
proposed in the literature in order to illustrate the importance
of having a deeper understanding of the string loop corrections.
From the expression (\ref{scalar}) we realise that:
\begin{equation}
\delta V_{(np)}\sim e^{K}\left( W_{np}^{2}+W_{0}W_{np}\right)
,\textrm{ \ \ \ \ \ \ \ \ \ \ }\delta V_{(p)}\sim
e^{K}W_{0}^{2}K_{p}, \label{scaling}
\end{equation}
where in general we have:
\begin{equation}
\delta V_{(p)}=\delta V_{(\alpha ' )}+\delta V_{(g_{s})},
\end{equation}
for the full perturbative contributions to the scalar potential.
Let us explore the possible scenarios which emerge by varying
$W_0$. As stressed in subsection 3.1.2, we can trust the use of
solely the leading perturbative corrections to the scalar
potential only when the overall volume is stabilised at large
values $\mathcal{V}\gg 1$. The first systematic study of the
strength of perturbative and non-perturbative corrections to the
scalar potential has been performed in \cite{bb}. Neglecting
$\delta V_{(g_{s})}$, $V_{local}$ and $V_{D}$, \cite{bb} studied
the behaviour of the minima of the scalar potential when one
varies $\left\vert W_{0}\right\vert $. Their results are
summarised in the following table:

\bigskip

\begin{tabular}{|c|c|c|}
\hline 1) $\left\vert W_{0}\right\vert \sim \left\vert
W_{np}\right\vert \ll 1$ &
2) $\left\vert W_{np}\right\vert <\left\vert W_{0}\right\vert <1$ & 3) $%
\left\vert W_{np}\right\vert \ll \left\vert W_{0}\right\vert \simeq \mathcal{%
O}(1)$ \\ \hline $\left\vert \delta V_{(\alpha ' )}\right\vert \ll
\left\vert \delta V_{(np)}\right\vert $ & $\left\vert \delta
V_{(np)}\right\vert \simeq \left\vert \delta V_{(\alpha '
)}\right\vert $ & $\left\vert
\delta V_{(np)}\right\vert \ll \left\vert \delta V_{(\alpha ' )}\right\vert $ \\
\hline
\end{tabular}


\begin{enumerate}
\item $\left\vert W_{0}\right\vert \sim \left\vert W_{np}\right\vert \ll 1$
\ $\Longrightarrow $ $\left\vert \delta V_{(\alpha ' )}\right\vert
/\left\vert \delta V_{(np)}\right\vert \sim $\ $\left\vert \delta
K_{(\alpha ' )}\right\vert \sim 1/\mathcal{V}\ll 1$ \
$\Longleftrightarrow $ \ $\left\vert \delta V_{(\alpha '
)}\right\vert \ll \left\vert \delta V_{(np)}\right\vert $

This case is the well-known KKLT scenario \cite{kklt}. All
K\"{a}hler moduli are stabilised by non-perturbative corrections
at an AdS supersymmetric minimum with $D_{T}W=0$. A shortcoming of
this model is that $W_{0}$ must be tuned very small in order to
stabilise at large volume and neglect $\alpha '$ or other
perturbative corrections. KKLT gave the following fit for the
one-parameter case:
\begin{equation}
W_{0}=-10^{-4},\textrm{ \ }A=1,\textrm{ \ }a\simeq 2\pi
/60\textrm{ \ \ } \Longrightarrow \textrm{ \ \ }\left\langle \tau
\right\rangle \simeq 113\Longleftrightarrow \mc{V}\simeq 1.2\cdot
10^{2}.
\end{equation}
In addition to $\vert W_0 \vert \ll 1$, a large rank gauge group
(as in $SU(60)$ above) is also necessary to get $a\tau \gg 1$.
This is a bit inelegant but a lower rank of the gauge group would
imply a much worse fine tuning of $W_{0}$. The authors also
proposed a mechanism to uplift the solution to dS, by adding a
positive potential generated by the tension of
$\overline{D3}$-branes. This represents an explicit breaking
within four-dimensional supergravity. Remaining within a
supersymmetric effective theory\footnote{\cite{bkq} proposed using
$D$-term uplifting to keep manifest supersymmetry. However, since
$D$-terms can be shown in general to be proportional to $F$-terms,
this mechanism can work only for non-supersymmetric AdS vacua, as
in the LARGE Volume Scenario which will be presented later on.},
\cite{ss} instead proposed $F$-term uplifting using metastable
supersymmetry breaking vacua. Also \cite{cnp} pointed out that the
KKLT procedure in two steps (first the minimisation of the $S$ and
$U$ moduli at tree level, and then the $T$ moduli fixed
non-perturbatively) can miss important contributions such as a dS
minimum without the need to add any up-lifting term.

We finally notice that this mechanism also relies on the
assumption that $W_{np}$ depends explicitly on each K\"{a}hler
modulus. In the fluxless case, this assumption is very strong as
only arithmetic genus 1 cycles \cite{witten} would get stringy
instanton contributions and $D7$-brane deformation moduli would
remain unfixed. The presence of the corresponding extra fermionic
zero modes can prevent gaugino condensation and in general could
also destroy instanton contributions for non-rigid arithmetic
genus 1 cycles. However by turning on fluxes, the $D7$ moduli
should be frozen and the arithmetic genus 1 condition can be
relaxed \cite{GauginoCondensation, relax0, relax1, relax2,
relax3}. Therefore it is possible that also non-rigid cycles admit
nonperturbative effects.

\item $\left\vert W_{np}\right\vert <\left\vert W_{0}\right\vert
<1\Longrightarrow $ $\left\vert \delta V_{(\alpha ' )}\right\vert
/\left\vert \delta V_{(np)}\right\vert \sim \left\vert \delta
K_{(\alpha ' )}\right\vert /\left\vert W_{np}\right\vert
\left\vert W_{0}\right\vert \sim 1$ \ $ \Longleftrightarrow $ \
$\left\vert \delta V_{(np)}\right\vert \simeq \left\vert \delta
V_{(\alpha ' )}\right\vert $

\cite{bb} pointed out that there is an upper bound on the $|W_0|$
in order to find a KKLT minimum $\left\vert W_{0}\right\vert \leq
W_{\max }$. $W_{\max }$ is the value of $\left\vert
W_{0}\right\vert $ for which the leading $\alpha ' $ corrections
start becoming important and compete with the non-perturbative
ones to find a minimum. This minimum will be non-supersymmetric as
we can infer from looking at (\ref{scalar}) which implies that
$V\sim \mathcal{O}(1/\mathcal{V}^{3})$ at the minimum, while
$-3e^{K}\left\vert W\right\vert ^{2}\sim
\mathcal{O}(1/\mathcal{V}^{2})$. Now since the scalar potential is
a continuous function of $\left\vert W_{0}\right\vert $,
increasing $\left\vert W_{0}\right\vert $ from $\left\vert
W_{0}\right\vert =W_{\max }-\varepsilon $, where we have an AdS
supersymmetric minimum, to $\left\vert W_{0}\right\vert =W_{\max
}+\varepsilon $, will still lead to an AdS minimum which is now
non-supersymmetric. Subsequently, when $\left\vert
W_{0}\right\vert $ is further increased, the $\alpha'$ corrections
become more and more important, and the minimum rises to
Minkowski, then de Sitter and finally disappears. The
disappearance corresponds to the $\alpha'$ corrections completely
dominating the non-perturbative ones and the scalar potential is
just given by the last term in (\ref{scalar}) that has clearly a
runaway behaviour without a minimum.

Unfortunately there is no clear example in the literature that
realises this situation for $\mathcal{V}\gg 1$. In their analysis,
\cite{bb} considered the possibility of getting a Minkowski
minimum for the quintic Calabi-Yau
$\mathbb{C}P_{[1,1,1,1,1]}^{4}(5)$ ($\chi =-200$) \cite{quintic},
giving the following fit:
\begin{eqnarray}
W_{0} &=&-1.7,\textrm{ \ }A=1,\textrm{ \ }a=2\pi /10,\textrm{ \
}\xi =0.4,\textrm{\ }
\hbox{Re}\left( S\right) =1  \nonumber \\
&\Longrightarrow &\textrm{ \ \ }\left\langle \tau \right\rangle
\simeq 5\Longleftrightarrow \mathcal{V}\simeq 2.
\end{eqnarray}

We note that this example, in reality, belongs to the third case since $%
\left\vert W_{0}\right\vert \simeq \mathcal{O}(1)$ where we
claimed that no minimum should exist. That is true only for
$\mathcal{V}\gg 1$, but in this case $\mathcal{V}\simeq 2$ and
higher $\alpha'$ corrections cannot be neglected anymore. Moreover
with $g_{s}\simeq 1$ the string loop expansion is uncontrolled.

\item $\left\vert W_{np}\right\vert \ll \left\vert W_{0}\right\vert \sim
\mathcal{O}(1)$ \ $\Longrightarrow $ \ $\left\vert \delta
V_{(\alpha ' )}\right\vert /\left\vert \delta V_{(np)}\right\vert
\sim \left\vert \delta K_{(\alpha ' )}\right\vert /\left\vert
W_{np}\right\vert \gg 1$ \ $ \Longleftrightarrow $ \ \ $\left\vert
\delta V_{(\alpha ' )}\right\vert \gg \left\vert \delta
V_{(np)}\right\vert $

This is the more natural situation when $\left\vert
W_{0}\right\vert \sim \mathcal{O}(1)$. In this case, if we ignore
the non-perturbative corrections and keep only the $\alpha ' $
ones, no minimum is present. However there are still $\delta
V_{(g_{s})}$, $V_{local}$ and $V_{D}$. Thus, let us see two
possible scenarios:

\begin{enumerate}
\item $\delta V_{(np)}$ neglected, $\delta V_{(\alpha')}+ V_{local}$ considered

Bobkov \cite{bobkov} considered F-theory compactifications on an
elliptically-fibered Calabi-Yau four-fold $Z$ with a warped
Calabi-Yau three-fold $M$ that admits a conifold singularity at
the base of the fibration. Following the procedure proposed by
Saltman and Silverstein \cite{ss2} for flux compactifications on
products of Riemann surfaces, he added $n_{D7}$ additional pairs
of $D7/ \overline{D7}$-branes and $n_{7}$ extra pairs of $(p,q)$
$7/\overline{7}$-branes wrapped around the 4-cycles in $M$ placed
at the loci where the fiber $T^{2}$ degenerates. These extra local
sources generate positive tension and an anomalous negative
$D3$-brane tension contribution to $V_{local}$ which, in units of
$\left( \alpha ' \right) ^{3}$, reads:
\begin{eqnarray}
V &=&-\chi \left( 2\pi \right) ^{13}N_{flux}^{2}\left(
\frac{g_{s}^{4}}{ \mathcal{V}_{s}^{3}}\right) -N_{7} \left( \frac{
g_{s}^{3}}{\mathcal{V}_{s}^{2}}\right)   \nonumber \\
&&+n_{7}\left( \frac{g_{s}^{2}}{\mathcal{V}_{s}^{4/3}}\right)
+n_{D7}\left( \frac{g_{s}^{3}}{\mathcal{V}_{s}^{4/3}}\right) ,
\end{eqnarray}
where ${\mathcal{V}_{s}}$ is the string frame volume and
$N_{7}=\left( n_{D7}^{3}+n_{7}^{3}\right)$ is an effective
parameter given in terms of triple intersections of branes. By
varying the various parameters, this is argued to give a
discretuum of large-volume non-supersymmetric AdS, Minkowski and
metastable dS vacua for Calabi-Yau three-folds with $h_{1,1}=1$
(this implies $\chi <0$). The fit proposed is for the dS solution:
\begin{eqnarray}
\left\vert W_{0}\right\vert  &\simeq &\left( 2\pi \right)
^{2}N_{flux}>1, \textrm{ }\chi =-4,\textrm{ }N_{flux}=3,\textrm{
}n_{7}=1,\textrm{ }n_{D7}=73,
\nonumber \\
\textrm{ }g_{s} &\simeq &5\cdot 10^{-3}\textrm{ \ }\Longrightarrow
\textrm{ \ \ } \mathcal{V}\simeq 3\cdot 10^{4}.
\end{eqnarray}

The integer parameters are tuned to obtain a pretty small $g_{s}$
so that the effect of string loop corrections can be safely
neglected. In this scenario, in which supersymmetry is broken at
the Kaluza-Klein scale, the stabilisation procedure depends on
local issues, while we would prefer to have a more general
framework where we could maintain global control.

\item $\delta V_{(np)}$ neglected,
$\delta V_{(\alpha ' )}+\delta V_{(g_{s})}$ considered

Berg, Haack and K\"{o}rs \cite{bhk2}, following their exact
calculation of the loop corrections for the $N=1$ toroidal
orientifold $T^{6}/(\mathbb{Z}_{2}\times \mathbb{Z}_{2})$
\cite{bhk}, analysed if these corrections could compete with the
$\alpha '$ ones to generate a minimum for $V$. By treating the
three toroidal K\"{a}hler moduli in $T^{6}=T^{2}\times T^{2}\times
T^{2}$ on an equal footing, they reduce the problem to a
1-dimensional one. The schematic form of the scalar potential for
the volume modulus is found to be: \be V =\delta V_{(\alpha ')
}+\delta V_{(g_{s})}\sim \frac{\xi \vert W_0 \vert^2}{\mc{V}^3} +
\frac{\delta}{\mc{V}^{10/3}}. \ee It turns out that $\delta > 0$,
and so as $\xi \sim - \chi$, they need a positive Euler number
$\chi >0$ in order to find a minimum, while the $T^6/(\mbb{Z}_2
\ti \mbb{Z}_2)$ toroidal example has a negative Euler number. They
instead consider the $N=1$ toroidal orientifold $T^{6}/\mbb{Z}_6'$
that satisfies the condition $\chi
>0$. A non-supersymmetric AdS minimum is now present but as the
loop corrections are naturally subleading with respect to the
$\alpha ' $\ ones, they must fine tune the complex structure
moduli to get large volume. They find:
\begin{eqnarray}
\left\vert W_{0}\right\vert  &\sim &\mathcal{O}(1),\textrm{ \
}\hbox{Re}
(U)\simeq 650,\textrm{\ }\hbox{Re}\left( S\right) =10  \nonumber \\
\textrm{ \ } &\Longrightarrow &\textrm{ \ \ }\left\langle \tau
\right\rangle \simeq 10^{2}\Longleftrightarrow \mathcal{V}\simeq
10^{3}.
\end{eqnarray}
The fine-tuning comes from assuming the complex structure moduli
are stabilised at large values. A similar scenario has been
studied also by von Gersdorff and Hebecker \cite{hg}. In addition,
Parameswaran and Westphal \cite{para} studied the possibility to
have a consistent $D$-term uplifting to de Sitter in this
scenario.
\end{enumerate}

\item We have assumed above that when $\left\vert
W_{0}\right\vert \sim \mathcal{O}(1)$ perturbative corrections
always dominate non-perturbative ones, which can therefore be
neglected. But is this naturally always the case? In order to
answer this question, let us now consider scenarios in which
$\delta V_{(np)}$ and $\delta V_{(\alpha')}$ compete while
$\left\vert W_{0}\right\vert \sim \mathcal{O}(1)$.

%
\begin{enumerate}
\item[(a)] \bigskip $\delta V_{(g_{s})}$ neglected,
$\delta V_{(np)}+\delta V_{(\alpha ' )}$ considered $\
\Longrightarrow $ \ large volume

This situation was studied by Westphal \cite{Westphal} following
the work of Balasubramanian and Berglund, finding a dS minimum at
large volume for the quintic. However this result extends to other
Calabi-Yau three-folds with just one K\"{a}hler modulus. He
presents the following fit:
\begin{eqnarray}
W_{0} &=&-1.7,\textrm{ \ }A=1,\textrm{ \ }a=2\pi /100,\textrm{ \
}\xi =79.8,\textrm{
\ }\hbox{Re}\left( S\right) =1  \nonumber \\
\textrm{ \ } &\Longrightarrow &\textrm{ \ \ }\left\langle \tau
\right\rangle \simeq 52\Longleftrightarrow \mathcal{V}\simeq 376.
\end{eqnarray}

The non-perturbative corrections are rendered important by using a
large-rank gauge group $SU(100)$ for gaugino condensation. This is
not fine-tuned but is contrived. The loop corrections, which may
be important, are not considered here.

\bigskip

\item[(b)] $\delta V_{(g_{s})}$ neglected, $\delta
V_{(np)}+\delta V_{(\alpha ' )}$ considered $\ \Longrightarrow $ \
exponentially large volume
\end{enumerate}

This situation is very appealing since it provides a positive
answer to our basic question. Balasubramanian, Berglund, Conlon
and Quevedo \cite{LVS} developed these scenarios which now go
under the name of Large Volume Scenarios, which is a bit
misleading as large volume is always necessary to trust a
solution. They should be more correctly called LARGE Volume (or
WEAK coupling \cite{WEAK}) Scenarios because the volume is
exponentially large. In this framework, both non-perturbative and
$\alpha'$ corrections compete naturally to get a
non-supersymmetric AdS minimum of the scalar potential at
exponentially large volume. This is possible by considering more
than one K\"{a}hler modulus and taking a well-defined large volume
limit. For one modulus models, the work of \cite{bb} and
\cite{Westphal} shows that with the rank of the gauge group
$SU(N)$ in the natural range $N\simeq 1\div 10$, it is impossible
to have a minimum.

However, if we have more generally $h_{1,1}>1$, this turns out to
be possible. The simplest example of such models has been found
for the Calabi-Yau described by the degree 18 hypersurface
embedded in the complex weighted projective space
$\mathbb{C}P_{[1,1,1,6,9]}^{4}$ \cite{2ParamMirror, ddf}. The
overall volume in terms of 2-cycle volumes is given by:
\begin{equation}
\mathcal{V}=\frac{1}{6}\left(
3t_{1}^{2}t_{5}+18t_{1}t_{5}^{2}+36t_{5}^{3}\right),
\end{equation}
and the 4-cycle volumes take the form:
\begin{equation}
\tau _{4}=\frac{t_1^{2}}{2}\textrm{, \ \ \ \ \ }\tau
_{5}=\frac{\left( t_{1}+6t_{5}\right) ^{2}}{2},  \label{LVS}
\end{equation}
for which it is straightforward to see that:
\begin{equation}
\mathcal{V}=\frac{1}{9\sqrt{2}}\left( \tau _{5}^{3/2}-\tau
_{4}^{3/2}\right) .  \label{volume}
\end{equation}
The reason why $\tau_{4}$ and $\tau_{5}$ are considered instead of
$\tau_{1}$ and $\tau_{5}$, is that these are the only 4-cycles
which get instanton contributions to $W$ when fluxes are turned
off \cite{ddf}. As we will explain in chapter 4, in order to get
LARGE Volume Scenarios, we require that $W_{np}$ depends only on
blow-up modes which resolve point-like singularities, as
$\tau_{4}$ in this case. Such cycles are always rigid cycles and
thus naturally admit nonperturbative effects. If we now take the
large volume limit in the following way:
\begin{equation}
\left\{
\begin{array}{c}
\tau _{4}\textrm{ small}, \\
\textrm{\ }\tau _{5}\gg 1,
\end{array}
\right.   \label{assumption}
\end{equation}
the scalar potential looks like:
\begin{equation}
V=\delta V_{(np)}+\delta V_{(\alpha ' )}\sim \frac{\lambda
\sqrt{\tau _{4}} e^{-2a_{4}\tau _{4}}}{\mathcal{V}}-\frac{\mu \tau
_{4}e^{-a_{4}\tau _{4}}}{ \mathcal{V}^{2}}+\frac{\nu
}{\mathcal{V}^{3}},\textrm{ \ }\lambda ,\textrm{ }\mu ,\textrm{
}\nu \textrm{ constants,}  \label{minimumy}
\end{equation}
with a non-supersymmetric AdS minimum located at:
\begin{equation}
\tau _{4}\sim \left( 4\xi \right) ^{2/3}\textrm{ \ \ and \ \
}\mathcal{V}\sim \frac{\xi ^{1/3}\left\vert W_{0}\right\vert
}{a_{4}A_{4}}e^{a_{4}\tau _{4}}. \label{minimum}
\end{equation}

The result that we have found, confirms the consistency of our
initial assumption (\ref{assumption}) in taking the large volume
limit. Inserting in (\ref{minimum}) the correct parameter
dependence and with the following natural choice of parameters, we
find:
\begin{eqnarray}
W_{0} &=&1,\textrm{ \ }A_{4}=1,\textrm{ \ }a_{4}=2\pi /7,\textrm{
\ }\xi =1.31,
\textrm{\ }\hbox{Re}\left( S\right) =10  \nonumber \\
\textrm{ \ } &\Longrightarrow &\textrm{ \ \ }\left\langle \tau
_{4}\right\rangle \simeq 41\Longleftrightarrow \mathcal{V}\simeq
3.75\cdot 10^{15}. \label{exp}
\end{eqnarray}
Therefore $\tau _{4}$ is stabilised small whereas $\tau _{5}\gg
1$, and the volume can be approximated as:
\begin{equation}
\mathcal{V}\sim \tau _{5}^{3/2},
\end{equation}
and:
\begin{equation}
\tau _{4}\sim t_{1}^{2},\ \ \ \ \ \tau _{5}\sim t_{5}^{2}.
\end{equation}

Looking at (\ref{minimum}) we can realise why in this case we are
able to make $\delta V_{(np)}$ compete naturally with $\delta
V_{(\alpha ' )}$. In fact, in general $\delta V_{(\alpha ' )}\sim
1/\mathcal{V}^{3}$ and $\delta V_{(np)}\sim e^{-a_{4}\tau
_{4}}/\mathcal{V}^{2}$, but (\ref{minimum}) implies $\delta
V_{(np)}\sim 1/\mathcal{V}^{3}\sim \delta V_{(\alpha ' )}$. The
non-perturbative corrections in the big modulus $\tau _{5}$ will
be, as usual, subleading. An attractive phenomenological feature
of these models is that they provide a method of generating
hierarchies. In fact the result (\ref{exp}), for $M_{P}= 2.4 \cdot
10^{18}$ GeV, produces an intermediate string scale
\cite{Msintermediate}:
\begin{equation}
M_{s}\simeq \frac{M_{P}}{\sqrt{\mathcal{V}}}\sim 10^{11} \hbox{
GeV}, \label{string}
\end{equation}
and this can naturally give rise to the weak scale through
TeV-scale supersymmetry:
\begin{equation}
M_{soft}\sim m_{3/2}=e^{K/2}\left\vert W\right\vert \sim
\frac{M_{P}}{ \mathcal{V}}\sim 30\hbox{ TeV}.
\end{equation}
Changing the underlying parameters, one could easily find a volume
${\cal V} \sim 10^4$ in string units, as it is needed for $M_s
\sim M_{GUT}$, and much larger volumes of the order $\mc{V}\sim
10^{30}$ in the extreme case of TeV strings ($M_s\sim 1$ TeV). In
addition, the large volume allows massive string states to be
consistently integrated out and makes the effective field theory
description of the compactification more robust. We also stress
that explicitly obtaining exponentially large volume in string
theory, with all the geometric moduli stabilised, goes much
farther than the original large extra dimensions proposals
\cite{add} where the volume was simply assumed to be large. Since
the parameters (like $\xi$) appearing in the scalar potential
(\ref{minimumy}) are related to the topology of the underlying
Calabi-Yau space, the choices required for the existence of a
minimum at exponentially large volume imply conditions on this
underlying topology. These conditions will be derived in chapter 4
for an arbitrary Calabi-Yau.

However, this setup ignores further perturbative corrections as
the $g_{s}$\ ones. It is thus crucial to check if they do not
destroy the picture. In view of the known calculations of string
scattering amplitudes for toroidal orientifolds, Berg, Haack and
Pajer \cite{bhp} conjectured the string loop corrections to the
scalar potential to take the form:
\begin{equation}
\delta V_{(g_{s})}\sim \frac{\mathcal{C}_{4} ^{2}W_{0}^{2}}{
\hbox{Re}\left( S\right) ^{2}\mathcal{V}^{3}\sqrt{\tau
_{4}}}+O(\mathcal{V} ^{-10/3}).
\end{equation}
These corrections turn out to be subleading with respect to the
scalar potential (\ref{minimumy}) even if one tries to fine tune
the coefficients $\mathcal{C}_{4}$ pretty large,
$\mathcal{C}_{4}\simeq 20\div 40$. We therefore conclude that the
LARGE Volume Scenario is safe.
\end{enumerate}

This survey of moduli stabilisation mechanisms has shown that a
deeper understanding of string loop corrections to the K\"{a}hler
potential in Calabi-Yau backgrounds is highly desirable. In KKLT
stabilisation, the magnitude of the perturbative corrections is
what determines the regime of validity of the stabilisation
method. In all other methods of stabilisation, perturbative
corrections enter crucially into the stabilisation procedure, and
so not only $\alpha'$ but also $g_{s}$ corrections should be taken
into account.

These loop corrections are neglected in the cases (3a), (4a) and
(4b), but we learnt from the case (3b) that they can change the
vacuum structure of the system studied. However in this situation
a significant amount of fine tuning was needed to make them
compete with the $\alpha'$ corrections to produce a minimum at
large volume. In case (4b), the loop corrections did not
substantially affect the vacuum structure unless they were
fine-tuned large. Therefore one would tend to conclude that these
string loop corrections will in general be subdominant, and so
that it is safe to neglect them.

While this may be true for models with relatively few moduli, we
will see in chapter 6 that loop corrections can still play a very
important r\^{o}le in moduli stabilisation, in particular lifting
flat directions in LARGE Volume Scenarios. In this case the fact
that they are subdominant will turn out to be a good property of
these corrections since they can lift flat directions without
destroying the minimum already found in the other directions of
the K\"{a}hler moduli space. Moreover, these flat directions whose
potential is generated at subleading loop level, are perfect
candidates for inflaton fields, as we shall see in chapter 8.

\part{LARGE Volume Scenario}

\chapter{General Analysis and Calabi-Yau examples}
\label{GeneralAnalysis} \linespread{1.3}

As we have seen in chapter 3, there is a large variety of
K\"{a}hler moduli stabilisation mechanisms proposed in the
literature, which differ just from the particular corrections
beyond the leading order approximation which are taken into
account. The common feature of all these schemes is that $g_{s}$
has to be stabilised smaller than one in order to trust
perturbation theory, and the overall volume of the Calabi-Yau
$\mathcal{V}$ has to be fixed large, since the $\alpha'$
corrections to $K$ are an expansion in powers of
$\mathcal{V}^{-1}$, and we have presently control only over the
leading order term in this expansion.

However, the main problem is that no model is taking into account
all the possible quantum corrections. This is consistent since in
each case a large amount of fine tuning in the fluxes is allowed.
On the contrary, we would like to perform the K\"{a}hler moduli
stabilisation in a natural way by considering all the possible
sources of quantum corrections.

In reality the LARGE Volume Scenario (LVS), developed by
Balasubramanian, Berglund, Conlon and Quevedo in \cite{LVS} for
compactifications on an orientifold of the Calabi-Yau three-fold
$\mathbb{C}P^{4}_{[1,1,1,6,9]}(18)$ is an explicit example of
K\"{a}hler moduli stabilisation without fine-tuning. Moreover,
this model is able to generate most of the hierarchies we observe
in Nature, which come as different powers of the overall volume.
However, the minimum is found by making $W_{np}$ compete with
$\delta K_{(\alpha')}$ whereas $\delta K_{(g_{s})}$ is neglected.
There is also no rigorous systematic study of this model for
compactifications on an arbitrary Calabi-Yau. Due to the good
features of the LVS, Part II of this thesis focuses on this moduli
stabilisation mechanism, trying to improve it via the analysis of
its two theoretical shortcomings mentioned above. The two main
results of Part II of this thesis are:
\begin{enumerate}
\item A general analysis of the LVS for an arbitrary Calabi-Yau without string loop
corrections;
\item A detailed study of the behaviour of the string loop corrections
which are then added to the LVS, obtaining a full final picture.
\end{enumerate}
Point (1) is presented in section 4.1 of chapter 4, where we state
the LARGE Volume Claim, whose proof is given in appendix A, that
lists the topological conditions on an arbitrary Calabi-Yau to
find a non-supersymmetric AdS minimum of the scalar potential at
exponentially large volume. This is done including $\alpha'$
corrections to $K$ and non-perturbative corrections to $W$ but
neglecting the string loop corrections to the K\"{a}hler
potential. We will illustrate our general results in section 4.2
by applying them to some examples of Calabi-Yau three-folds that
are constructed as hypersurfaces in complex weighted projected
spaces. From this analysis, it turns out that a necessary and
sufficient condition for LARGE volume is the presence of blow-up
modes resolving point-like singularities. At this stage, where
$g_s$ corrections are neglected, it would also seem that the
Calabi-Yaus which have a fibration structure cannot present the
interesting phenomenological properties of the LVS.

The study of the string loop corrections mentioned in point (2)
will be performed in chapter 5, and, as we will explain in chapter
6, via the inclusion of the string loop corrections to the
K\"{a}hler potential, also K3 fibrations can present an
exponentially large volume minimum, provided a blow-up mode
exists.

In Part II of this thesis, we therefore managed to cure the two
shortcomings of the LVS, both finding a general analysis and
showing how all corrections, $\alpha'$, loop and non-perturbative,
play a crucial r\^{o}le in stabilising the K\"{a}hler moduli of a
generic Calabi-Yau three-fold without the need of doing any fine
tuning.

This chapter is organised as follows. In section 4.1 we state the
general conditions that have to be satisfied in order to have
exponentially large volume. The long detailed proof of this
general result is left to appendix A. Section 4.2 illustrates then
our general results for several Calabi-Yaus, including both Swiss
cheese models where all K\"{a}hler moduli other than the overall
volume are blow-ups, and also fibration Calabi-Yaus, such as K3
fibrations.

We shall not discuss obtaining be Sitter vacua in this chapter.
For a recent analysis of the conditions for de Sitter vacua from
supergravity, see \cite{marta1}.

\section{General analysis for the large volume limit}

We now investigate the topological conditions on an arbitrary
Calabi-Yau three-fold under which the scalar potential
(\ref{scalar}) admits an AdS non-supersymmetric minimum at
exponentially large volume deepening the analysis performed in
\cite{LVS}. We will refer to those constructions as LARGE Volume
Scenarios (LVS).

\begin{quotation}
\textbf{LARGE Volume Claim:} \textit{Let $X$ be a Calabi-Yau
three-fold and let the large volume limit be taken in the
following way:
\begin{equation}
\left\{
\begin{array}{c}
\tau _{j}\textrm{ remains small, }\forall j=1,...,N_{small}, \\
\mathcal{V}\rightarrow \infty \textrm{ \ \ for \ }\tau
_{j}\rightarrow \infty ,\textrm{\ }\forall
j=N_{small}+1,..,h_{1,1}(X),
\end{array}
\right. \label{limit}
\end{equation}
within type IIB $N=1$ four dimensional supergravity where the
K\"{a}hler potential and the superpotential in Einstein frame take
the form:
\begin{equation}
\left\{
\begin{array}{l}
K=K_{cs}-2\ln \left( \mathcal{V}+\hat{\xi}\right), \\
W= W_{0}+\sum\limits_{j=1}^{N_{small}}A_{j}e^{-a_{j}T_{j}}.
\end{array}
\right.  \label{explicit}
\end{equation}
Then the scalar potential admits a set $H$ of AdS
non-supersymmetric minima at exponentially large volume located at
$\mathcal{V}\sim e^{a_{j}\tau _{j}}$ $\forall j=1,...,N_{small}$
if and only if $h_{2,1}(X)>h_{1,1}(X)>1$, i.e. $\xi>0$ and
$\tau_{j}$ is a local blow-up mode resolving a given point-like
singularity $\forall j=1,...,N_{small}$. In this case:}
\begin{equation*}
\left\{
\begin{array}{c}
\text{if }h_{1,1}(X)=N_{small}+1\text{, }H=\left\{ \text{a
point}\right\} ,
\\
\text{if }h_{1,1}(X)>N_{small}+1\text{, }H=\left\{
(h_{1,1}(X)-N_{small}-1) \text{ flat directions}\right\} .
\end{array}
\right.
\end{equation*}
\end{quotation}

The proof of the previous Claim is presented in appendix A where
we show also that $\tau_{j}$ is the only blow-up mode resolving a
point-like singularity if and only if $K^{-1}_{jj}\sim
\mathcal{V}\sqrt{\tau_{j}}$. On the contrary when the same
singularity is resolved by several independent blow-ups, say
$\tau_{1}$ and $\tau_{2}$, then $K^{-1}_{11}\sim
\mathcal{V}h^{(1)}_{1/2}(\tau_{1},\tau_{2})$ and $K^{-1}_{22}\sim
\mathcal{V}h^{(2)}_{1/2}(\tau_{1},\tau_{2})$ with $h^{(j)}_{1/2}$
homogeneous function of degree 1/2 such that
$\frac{\partial^{2}h^{(j)}}{\partial\tau_{1}\partial\tau_{2}}\neq
0$ $\forall j=1,2$.

Let us now explain schematically the global picture of LVS for
arbitrary Calabi-Yau manifolds according to the LARGE Volume
Claim:

\begin{enumerate}
\item{}
The Euler number of the Calabi Yau manifold must be negative. More
precisely: $h_{12}> h_{11}>1$. This means that the coefficient
$\hat\xi$ must be positive in order to guarantee that in a
particular direction the potential goes to zero at infinity from
below \cite{LVS}. This is a both sufficient and necessary
condition.

\item{}
The Calabi-Yau manifold must have at least one blow-up mode
corresponding to a 4-cycle modulus that resolves a point-like
singularity. The associated modulus must have an induced
non-perturbative superpotential.
 This is usually guaranteed since these cycles
are rigid cycles of arithmetic genus one, which is precisely the
condition needed for the existence of
 non-perturbative
superpotentials  in the flux-less case \cite{witten}.

\item{}
This 4-cycle, together with other blow-up modes possibly present,
are fixed small by the interplay of non-perturbative and $\alpha'$
corrections, which stabilise also the overall volume mode. Here
small means larger than the string scale but not exponentially
large unlike the volume.

\item{}
All the other 4-cycles, such as those corresponding to fibrations,
cannot be stabilised small even though they may have induced
non-perturbative effects. They are sent large making their
non-perturbative corrections negligible.

\item{}
At this stage, non blow-up K\"{a}hler moduli, except the overall
volume mode, remain unfixed giving rise to essentially flat
directions.

\item{}
It turns out then that in order to freeze these moduli, it is
crucial to study string loop corrections as the leading term in a
$g_{s}$ expansion will be dominant over any potential
 non-perturbative correction.

\end{enumerate}
Notice that these are conditions to find exponentially large
volume minima and our results hold for generic $\mc{O}(1)$ values
of $W_0$. There may exist other minima which do not have
exponentially large volume for which our results do not have
anything to say. For example, $\vert W_0 \vert \ll 1$ may give
rise to KKLT-like minima.

Summarising, if there are $N_{small}$ blow-up modes and
$L=(h_{11}-N_{small}-1)$ modes which do not blow-up point-like
singularities nor correspond to the overall modulus, then our
results state that all the  $N_{small}$ can be fixed at values
large with respect to the string scale but not exponentially
large, the overall volume is exponentially large and the other $L$
K\"{a}hler moduli are not fixed by these effects.

In reality, the directions corresponding to the non blow-up modes,
if they have non-perturbative effects, will be lifted by these
tiny exponential terms, which however we neglect at this level of
approximation. The reason is that, as we will see in the next
sections, those directions will be lifted by the inclusion of
string loop corrections which are always dominant with respect to
the non-perturbative ones.

We would also like to stress that the previous general picture
shows how we need non-perturbative effects only in the blow-up
modes to get an exponentially large volume minimum. As blow-up
modes correspond to rigid exceptional divisors, the corresponding
non-perturbative corrections will be generally present even in the
fluxless case \cite{witten}. They can arise from either gaugino
condensation of the gauge theory living on the stack of branes
wrapping that $4$-cycle or from Euclidean $D3$-brane instantons.
On the contrary, it is not clear if all the other cycles can
indeed get non-perturbative corrections to $W$, but this is not
necessary to obtain LARGE Volume.

\section{Particular examples}
\label{examples}

Let us illustrate these results in a few explicit examples. At
this stage we ignore string loop corrections but as we will show
in section 6.1 and 6.2, these can in some cases actually be
important and change the configuration of the system studied.

\subsection{The single-hole Swiss cheese:
$\mathbb{C}P_{[1,1,1,6,9]}^{4}(18)$} \label{Swisscheese}

The original example of LVS of \cite{LVS} is given by the degree
18 hypersurface embedded in the complex weighted projective space
$\mathbb{C}P^4_{[1,1,1,6,9]}$ \cite{2ParamMirror, ddf}. The
overall volume in terms of 2-cycle volumes is given by:
\begin{equation}
\mathcal{V}=\frac{1}{6}\left(
3t_{1}^{2}t_{5}+18t_{1}t_{5}^{2}+36t_{5}^{3}\right).
\end{equation}
The divisor volumes take the form $\tau _{4}=\frac{t_{1}^{2}}{2}$,
$\tau _{5}=\frac{\left( t_{1}+6t_{5}\right) ^{2}}{2}$, from which
it is immediate to see that:
\begin{equation}
\mathcal{V}=\frac{1}{9\sqrt{2}}\left( \tau _{5}^{3/2}-\tau
_{4}^{3/2}\right) .  \label{volume form}
\end{equation}
$\xi$ is positive since $h_{1,1}<h_{2,1}$ and the limit
(\ref{limit}) can be correctly performed with $\tau_{5}\rightarrow
\infty $ and $\tau_{4}$ remaining small. Thus $N_{small}=1$ and we
have to check if this case satisfies the condition of the LARGE
Volume Claim which is $K^{-1}_{44}\simeq
\mathcal{V}\sqrt{\tau_{4}}$. This is indeed satisfied as it can be
seen either by direct calculation or by noticing that $\tau_{4}$
is a local blow-up. Omitting numerical factors, the scalar
potential takes the form:
\begin{equation}
V\simeq\frac{\sqrt{\tau_{4}}e^{-2a_{4}\tau_{4}}}{\mathcal{V}}
-\frac{W_{0}\tau_{4} e^{-2a_{4}\tau_{4}}}{\mathcal{V}^{2}}
+\frac{\hat{\xi}W_{0}^{2}}{\mathcal{V}^{3}}. \label{mio22}
\end{equation}
As the $\mathbb{C}P_{[1,1,1,6,9]}^{4}$ example is a particular
case of the LARGE Volume Claim, we conclude that the scalar
potential (\ref{mio22}) will admit an AdS minimum at exponentially
large volume with $(h_{1,1}-N_{small}-1)=0$ flat directions. This
is consistent with the original calculation in \cite{LVS}, which
shows that the minimum is located at:
\begin{equation}
\langle\tau_{4}\rangle\simeq (4\hat{\xi})^{2/3},\textit{ \ \ \
}\langle\mathcal{V}\rangle\simeq\frac{\hat{\xi}^{1/3}W_{0}}
{a_{4}A_{4}}e^{a_{4}\langle\tau_{4}\rangle}.
\end{equation}

\subsection{The multiple-hole Swiss cheese:
$\mathcal{F}_{11}$ and $\mathbb{C}P_{[1,3,3,3,5]}^{4}(15)$}
\label{Swisscheese2}

It is straightforward to realise that the LARGE Volume Claim can
be used to generalise the previous case by adding several blow-up
modes resolving point-like singularities that will be stabilised
small. In this case the overall volume looks like:
\begin{equation}
\mathcal{V}=\alpha \left( \tau
_{b}^{3/2}-\sum\limits_{i=1}^{N_{small}}\lambda _{i}\tau
_{i}^{3/2}\right), \label{cheese}
\end{equation}
where $\alpha $ and $\lambda_{i}$ are positive model-dependent
parameters and the Calabi-Yau manifold presents a typical ``Swiss
cheese'' shape. An explicit example is the Fano three-fold
$\mathcal{F}_{11}$ described in \cite{ddf}, which is topologically
a $\mathbb{Z}_{2}$ quotient of a $CY_{3}$ with Hodge numbers
$h_{1,1}=3$, $h_{2,1}=111$. The total volume of the
$\mathcal{F}_{11}$ reads:
\begin{equation}
\mathcal{V}=\frac{t_{1}^{2}t_{2}}{2}+\frac{t_{1}t_{2}^{2}}{2}+\frac{t_{2}^{3}
}{6}+\frac{t_{1}^{2}t_{3}}{2}
+2t_{1}t_{2}t_{3}+t_{2}^{2}t_{3}+t_{1}t_{3}^{2}+2t_{2}t_{3}^{2}+\frac{
2t_{3}^{3}}{3},
\end{equation}
and the 4-cycle moduli are given by:
\begin{equation}
\tau _{1}=\frac{t_{2}}{2}\left( 2t_{1}+t_{2}+4t_{3}\right)
,\textrm{ \ \ \ } \tau _{2}=\frac{t_{1}^{2}}{2},\textrm{ \ \ \ \
}\tau _{3}=t_{3}\left( t_{1}+t_{3}\right).
\end{equation}
It is then possible to express $\mathcal{V}$ in terms of the
$\tau$-moduli as:
\begin{equation}
\mathcal{V}=\frac{1}{3\sqrt{2}}\left( 2\left( \tau _{1}+\tau
_{2}+2\tau _{3}\right) ^{3/2}-\left( \tau _{2}+2\tau _{3}\right)
^{3/2}-\tau _{2}^{3/2}\right).
\end{equation}
The resemblance with the general ``Swiss cheese'' picture
(\ref{cheese}) is now manifest. Two further Calabi-Yau
realisations of this Swiss-Cheese structure have been presented in
\cite{blumenhagen}. They are the $h_{1,1}=3$ degree 15
hypersurface embedded in $\mathbb{C}P_{[1,3,3,3,5]}^{4}$ and the
$h_{1,1}=5$ degree 30 hypersurface in
$\mathbb{C}P_{[1,1,3,10,15]}^{4}$.

More generally, in \cite{blum} it was proved that examples of
Swiss-cheese Calabi-Yau three-folds with $h_{1,1}=n+2$, $0\leq
n\leq 8$, can be obtained by starting from elliptically fibred
Calabi-Yau manifolds over a del Pezzo $dP_n$ base\footnote{A del
Pezzo $dP_n$ surface is obtained by blowing-up $\mathbb{C}P^2$ (or
$\mathbb{C}P^1\times\mathbb{C}P^1$ ) on $0\leq n\leq 8$ points.},
and then performing particular flop transitions that flop away all
$n$ $\mathbb{C}P^1$-cycles in the base.

In this case, assuming that all the small cycles get
non-perturbative effects, the $4$-cycle $\tau_{b}$, controlling
the overall size of the Calabi-Yau, is stabilised exponentially
large:
\begin{equation}
\langle\mathcal{V}\rangle\sim \alpha \langle\tau_b\rangle^{3/2}
\sim W_0 e^{a_{i}\langle\tau_{i}\rangle},\textit{ \ }\forall
i=1,...,N_{small},
\end{equation}
with no orthogonal flat directions. The various 4-cycles,
$\tau_{i}$, controlling the size of the `holes' of the
Swiss-cheese, get fixed at small values $\tau_{i} \sim
\mathcal{O}(10)$, $\forall i=1,...,N_{small}$. However, in
\cite{collinucci} it was discovered that the Swiss-cheese
structure of the volume is not enough to guarantee that all the
rigid `small' cycles $\tau_i$ can indeed be stabilised small. In
fact, a further condition is that each rigid `small' cycle
$\tau_i$ must be del Pezzo. In \cite{collinucci}, there are three
examples of Swiss-cheese Calabi-Yau three-folds with $h_{1,1}=4$
where just one $4$-cycle has the topology $\mathbb{C}P^2$ (and so
it is $dP_0$). It is likely that in order to achieve full moduli
stabilisation when the topological condition, that all rigid
$4$-cycles be del Pezzo, is not satisfied, one needs to include
$g_s$ corrections.

We note that string loop corrections can be crucial also when one
imposes the phenomenological condition that the $4$-cycles
supporting chiral matter do not get non-perturbative effects
\cite{blumenhagen}.\footnote{Also $D$-terms could play a
significant r\^{o}le as pointed out still in \cite{blumenhagen}.}
In fact, if one wraps an Euclidean $D3$-brane instanton around a
rigid $4$-cycle which is also wrapped by $D7$-branes supporting
chiral matter, the $D7$-branes and the instanton will chirally
intersect, thereby forcing the insertion of charged superfields
next to the non-perturbative superpotential. At this point, the
phenomenological requirement that no Standard Model field gets a
non-vanishing VEV at the string scale, sets the prefactor of the
non-perturbative effect equal to zero.

Let us briefly review the geometric data of the resolution of the
$\mathbb{C}P_{[1,3,3,3,5]}^{4}(15)$ manifold, which has been used
by the authors of \cite{blumenhagen} to illustrate the tension
between moduli stabilisation and chirality mentioned above. As we
shall see, this turns out to be an interesting case in which loop
corrections may potentially stabilise the Standard Model cycle
that does not admit non-perturbative superpotential contributions.
This Calabi-Yau has $h_{1,1}=3$ and $h_{1,2}=75$. The K\"{a}hler
form $J$ can be expanded as $J=\sum_{i=1}^{3}t^{i}\hat{D}_{i}$ in
a base $\{\hat{D}_{i}\}_{i=1}^{3}$ of
$H_{1,1}(\mathbb{C}P_{[1,3,3,3,5]}^{4}(15),\mathbb{Z})$ where the
only non-vanishing intersection numbers look like:
\begin{center}
\begin{tabular}{|c|c|c|c|c|c|c|}
\hline $k_{111}$ & $k_{222}$ & $k_{333}$ & $k_{112}$ & $k_{122}$ &
$k_{223}$ & $ k_{233}$ \\ \hline $9$ & $-40$ & $-40$ & $-15$ &
$25$ & $-5$ & $15$ \\ \hline
\end{tabular}
\end{center}
Using (\ref{defOfTau}), the volumes of the divisors $D_{1}$,
$D_{2}$ and $D_{3}$ are given by:
\begin{eqnarray}
\tau _{1} &=&\frac{1}{2}\left( 3t_{1}-5t_{2}\right) ^{2},\text{ \
\ }\tau _{2}=\frac{5}{6}\left[ \left( 3t_{3}-t_{2}\right)
^{2}-\left(
5t_{2}-3t_{1}\right) ^{2}\right] ,  \notag \\
\tau _{3} &=&-\frac{5}{2}\left( t_{2}-4t_{3}\right) \left(
t_{2}-2t_{3}\right),
\end{eqnarray}
and the overall volume in terms of the $\tau$-moduli reads:
\begin{equation}
\mathcal{V}=\sqrt{\frac{2}{45}}\left[\left(5\tau_{1}+3\tau_{2}+\tau_{3}\right)^{3/2}
-\frac{1}{3}\left(5\tau_{1}+3\tau_{2}\right)^{3/2}-\frac{\sqrt{5}}{3}\tau_{1}^{3/2}\right].
\label{volOne}
\end{equation}
Looking at (\ref{volOne}) we realise that the form of the volume
becomes simpler if we introduce the following diagonal basis:
\begin{equation}
\tau_{a}=5\tau_{1}+3\tau_{2}+\tau_{3},\textrm{ \ \
}\tau_{b}=5\tau_{1}+3\tau_{2},\textrm{ \ \ }\tau_{c}=\tau_{1},
\label{modify}
\end{equation}
in which the only non-vanishing intersection numbers are:
\begin{center}
\begin{tabular}{|c|c|c|}
\hline $k_{aaa}$ & $k_{bbb}$ & $k_{ccc}$ \\ \hline $5$ & $45$ &
$9$ \\ \hline
\end{tabular}
\end{center}
and in the diagonal basis the total volume becomes:
\begin{equation}
\mathcal{V}=\sqrt{\frac{2}{45}}\left(\tau_{a}^{3/2}-\frac{1}{3}\tau_{b}^{3/2}
-\frac{\sqrt{5}}{3}\tau_{c}^{3/2}\right). \label{NuovoVolume}
\end{equation}
A Euclidean $D3$-brane instanton wraps the rigid $4$-cycle
$D_{E3}=\frac{1}{3}\left(D_{b}+D_{c}\right)$, giving a
non-perturbative superpotential term
$W_{np}=e^{-\frac{2\pi}{3}\left(\tau_{b}+\tau_{c}\right)}$. There
are also two stacks of D7-branes wrapping the rigid four cycles
$D_{D7A}=\frac{1}{3}\left(D_{b}-2D_{c}\right)$ and $D_{D7B}=D_{c}$
with line bundles
$\mathcal{L}_{A}=\frac{1}{3}\left(2D_{b}+5D_{c}\right)$ and
$\mathcal{L}_{B}=\mathcal{O}$. This choice guarantees that there
are no chiral zero modes on the $D7$-$E3$ intersections. The
``Standard Model'' is part of the $U(N_{A})$ gauge group on the
stack $A$ of $D7$-branes, with SM matter obtained from the
intersections $AA'$ and $AB$ where the prime denotes the
orientifold image.

Neglecting the $D$-term part of the scalar potential we obtain:
\begin{equation}
V=\frac{\lambda_{1}\left(\sqrt{5\tau_{b}}+\sqrt{\tau_{c}}\right)
e^{-\frac{4\pi}{3}\left(\tau_{b}+\tau_{c}\right)}}{\mathcal{V}}
-\frac{\lambda_{2}\left(\tau_{b}+\tau_{c}\right)
e^{-\frac{2\pi}{3}\left(\tau_{b}+\tau_{c}\right)}}{\mathcal{V}^{2}}
+\frac{\lambda_{3}}{\mathcal{V}^{3}}, \label{mio}
\end{equation}
where $\lambda_{i}>0$, $\forall i=1,2,3$ are unimportant numerical
factors. Now to make the study of the scalar potential (\ref{mio})
simpler, we perform the change of coordinates $\tau _{b}=2\tau
_{E3}+\tau _{SM}, \tau _{c}=\tau _{E3}-\tau _{SM}$, bringing
(\ref{mio}) to the form:
\begin{equation}
V=\frac{\lambda_{1}\left(\sqrt{5\left(2\tau_{E3}+\tau_{SM}\right)}
+\sqrt{\tau_{E3}-\tau_{SM}}\right)
e^{-4\pi\tau_{E3}}}{\mathcal{V}} -\frac{3\lambda_{2}\tau_{E3}
e^{-2\pi\tau_{E3}}}{\mathcal{V}^{2}}
+\frac{\lambda_{3}}{\mathcal{V}^{3}}. \label{mio2}
\end{equation}
The scalar potential (\ref{mio2}) then has a critical point at
$\tau_{E3}=2\tau_{SM}$. However, this is not a minimum of the full
scalar potential but is actually a saddle point along $\tau_{SM}$
at fixed $\tau_{E3}$ and $\mathcal{V}$. In subsection \ref{SM} we
will show how string loop corrections may give rise to a stable
LVS even though no non-perturbative corrections in $\tau_{SM}$ are
included (see \cite{blumenhagen} for a discussion of freezing
$\tau_{SM}$ by including $D$-terms with (\ref{mio2})).

\subsection{2-Parameter K3 fibration: $\mathbb{C}P_{[1,1,2,2,6]}^{4}(12)$}
\label{2modK3noLoop}

Our next example is a fibration Calabi-Yau, the degree 12
hypersurface embedded in $\mathbb{C}P_{[1,1,2,2,6]}^{4}$
\cite{Candelas}. The overall volume in terms of $2$-cycle volumes
is given by:
\begin{equation}
\mathcal{V}=t_{1}t_{2}^{2}+\frac{2}{3}t_{2}^{3}. \label{linearity}
\end{equation}
The 4-cycle volumes take the form $\tau _{1}=t_{2}^{2}$, $\tau
_{2}=2t_{2}\left( t_{1}+t_{2}\right)$, yielding:
\begin{equation}
\mathcal{V}=\frac{1}{2}\sqrt{\tau _{1}}\left( \tau
_{2}-\frac{2}{3}\tau _{1}\right).  \label{Vol11226}
\end{equation}
It is possible to invert the relations
$\tau_i=\partial{\mathcal{V}}/\partial t_i$  to produce:
\begin{equation}
t_{2}=\sqrt{\tau _{1}}\textrm{, \ \ \ \ \ }t_{1}=\frac{\tau
_{2}-2\tau _{1}}{2 \sqrt{\tau _{1}}}. \label{ecco}
\end{equation}
The Euler characteristic of the Calabi-Yau is negative and the
limit (\ref{limit}) can be performed only with
$\tau_{2}\rightarrow \infty $ and keeping $\tau_{1}$ small. This
corresponds to $t_{1}\rightarrow \infty $ and $t_{2}$ small. In
this limit the volume becomes:
\begin{equation}
\mathcal{V}=\frac{1}{2}\sqrt{\tau _{1}}\tau _{2}\simeq
t_{1}t_{2}^{2}\simeq t_{1}\tau_{1}. \label{lavel}
\end{equation}
Thus $N_{small}=1$ again and we need to check the condition of the
LARGE Volume Claim: $K^{-1}_{11}\simeq
\mathcal{V}\sqrt{\tau_{1}}$. However this is clearly not
satisfied, as $\tau_{1}$ is a fibration over the base $t_{1}$.

This is therefore a situation where no exponentially large volume
minimum is present, as can be confirmed by the explicit
calculation below.

\subsubsection{Explicit Calculation} \label{11226Sec}

Here we verify that the $\mathbb{C}P^4_{[1,1,2,2,6]}$ model does
not give a realisation of the LVS. We take the large volume limit
in the following way:
\begin{equation}
\left\{
\begin{array}{c}
\tau _{1}\textrm{ small}, \\
\textrm{\ }\tau _{2}\gg 1,
\end{array}
\right.  \label{ooooo}
\end{equation}which, after the axion minimisation ($W_{0}>0$), gives a
scalar potential of the form:
\begin{eqnarray}
V &=&\delta V_{(np)}+\delta
V_{(\alpha')}=\frac{4}{\mathcal{V}^{2}}\left[ a_{1}A_{1}^{2}\tau
_{1}\left( a_{1}\tau _{1}+1\right) e^{-2a_{1}\tau
_{1}}-a_{1}A_{1}\tau _{1}W_{0}e^{-a_{1}\tau _{1}}\right]  \nonumber \\
&&+\frac{3}{4}\frac{\xi }{\mathcal{V}^{3}}\left(
W_{0}^{2}+A_{1}^{2}e^{-2a_{1}\tau _{1}}-2A_{1}W_{0}e^{-a_{1}\tau
_{1}}\right).  \label{111}
\end{eqnarray}
We set $A_{1}=1$ and recall that to neglect higher order instanton
corrections we need $a_{1}\tau _{1}\gg 1$. (\ref{111}) becomes:
\begin{equation}
V =\frac{4}{\mathcal{V}^{2}}\left[ \left( a_{1}\tau
_{1}e^{-a_{1}\tau _{1}}-W_{0}\right) a_{1}\tau _{1}e^{-a_{1}\tau
_{1}}\right]+\frac{3}{4}\frac{\xi }{\mathcal{V}^{3}}\left[
W_{0}^{2}+\left( e^{-a_{1}\tau _{1}}-2W_{0}\right) e^{-a_{1}\tau
_{1}}\right]. \label{1111}
\end{equation}
The previous expression (\ref{1111}) can be rewritten as:
\begin{eqnarray}
V &=&\frac{e^{-2a_{1}\tau _{1}}}{\mathcal{V}^{2}}\left(
4a_{1}^{2}\tau _{1}^{2}+\frac{3}{4}\frac{\xi }{\mathcal{V}}\right)
-\frac{ 2W_{0}e^{-a_{1}\tau _{1}}}{\mathcal{V}^{2}}\left(
2a_{1}\tau _{1}+\frac{3}{4} \frac{\xi }{\mathcal{V}}\right)
+\frac{3}{4}\frac{\xi }{\mathcal{V}^{3}}
W_{0}^{2}  \nonumber \\
&&\underset{\mathcal{V}\gg 1}{\sim
}\frac{4}{\mathcal{V}^{2}}\left[ \left( a_{1}\tau
_{1}e^{-a_{1}\tau _{1}}-W_{0}\right) a_{1}\tau _{1}e^{-a_{1}\tau
_{1}}\right] +\frac{3}{4}\frac{\xi }{\mathcal{V}^{3}}W_{0}^{2}.
\label{1122}
\end{eqnarray}
Assuming a natural value $W_{0}\sim \mathcal{O}(1)$, then
(\ref{1122}) simplifies to:
\begin{equation}
V=-\frac{4}{\mathcal{V}^{2}}W_{0}a_{1}\tau _{1}e^{-a_{1}\tau
_{1}}+\frac{3}{4 }\frac{\xi }{\mathcal{V}^{3}}W_{0}^{2}.
\label{oooo}
\end{equation}
Extremising this scalar potential, we get:
\begin{equation}
\frac{\partial V}{\partial \tau _{1}}=\frac{4}{\mathcal{V}^{2}}
W_{0}a_{1}e^{-a_{1}\tau _{1}}\left( a_{1}\tau _{1}-1\right) =0,
\end{equation}
whose only possible solution for $W_{0}\neq 0$ is $a_{1}\tau
_{1}=1$, which is not in the controlled regime of parameter space.
However, when $W_{0}=0$, (\ref{1122}) gives:
\begin{equation}
V=\left( 4a_{1}^{2}\tau _{1}^{2}+\frac{3}{4}\frac{\xi
}{\mathcal{V}}\right) \frac{e^{-2a_{1}\tau
_{1}}}{\mathcal{V}^{2}}\underset{\mathcal{V}\gg 1}{\sim
}4a_{1}^{2}\tau _{1}^{2}\frac{e^{-2a_{1}\tau
_{1}}}{\mathcal{V}^{2}},
\end{equation}
and the first derivative with respect to $\tau _{1}$ is:
\begin{equation}
\frac{\partial V}{\partial \tau _{1}}=8a_{1}^{2}\tau _{1}\frac{
e^{-2a_{1}\tau _{1}}}{\mathcal{V}^{2}}\left( a_{1}\tau
_{1}-1\right) ,
\end{equation}
which also has no minimum. Thus we have shown that for $W_{0}\sim
\mathcal{O}(1)$, the $\mathbb{C}P^{4}_{[1,1,2,2,6]}$ model has no
exponentially large volume minimum. The last hope is to find a
minimum fine tuning $W_{0}\ll 1$. In this case taking the
derivatives of (\ref{1122}), one obtains:
\begin{equation}
\frac{\partial V}{\partial \tau _{1}}=\frac{4}{\mathcal{V}^{2}}
a_{1}e^{-2a_{1}\tau _{1}}\left( a_{1}\tau _{1}-1\right) \left(
W_{0}e^{a_{1}\tau _{1}}-2a_{1}\tau _{1}\right) =0,
\end{equation}
whose only possible solution is:
\begin{equation}
2a_{1}\left\langle \tau _{1}\right\rangle
=W_{0}e^{a_{1}\left\langle \tau _{1}\right\rangle},  \label{ede}
\end{equation}
but then fixing $\tau_{1}$, the scalar potential (\ref{1122})
along the volume direction looks like:
\begin{equation}
V\sim
\frac{W_{0}^{2}}{\mathcal{V}^{2}}\left(-1+\frac{3}{4}\frac{\xi
}{\mathcal{V}}\right)\sim -\frac{W_{0}^{2}}{\mathcal{V}^{2}}.
\label{efe}
\end{equation}
The potential (\ref{efe}) has no LARGE Volume minimum, and so we
conclude that the $\mathbb{C}P^{4}_{[1,1,2,2,6]}$ model does not
admit an exponentially large volume minimum for any value of
$W_{0}$.

It is still of course possible to fix the moduli using other
stabilisation schemes - for example KKLT. However, in this case
there will not be a large hierarchy between the two K\"{a}hler
moduli, with instead $\tau_{1}\lesssim \tau_{2}$, and the volume
can never be exponentially large.

\subsection{3-Parameter K3 fibration}
\label{3modK3noLoop}

In the previous subsections \ref{Swisscheese}, \ref{Swisscheese2}
and \ref{2modK3noLoop}, we have presented three examples which
illustrate two of the three possible situations which the general
analysis determines. We now illustrate the case when an
exponentially large volume minimum can be found, but with flat
directions still present. We will then explain how these can be
lifted using string loop corrections.

This example concerns Calabi-Yau three-folds which are single K3
Fibrations with three K\"{a}hler moduli. We start off with the
following expression for the overall volume in terms of the three
moduli:
\begin{equation}
\mathcal{V}=\alpha \left[ \sqrt{\tau _{1}}(\tau _{2}-\beta \tau
_{1})-\gamma \tau _{3}^{3/2}\right] ,  \label{hhh}
\end{equation}
where $\alpha $, $\beta $, $\gamma $ are positive model-dependent
constants. While we do not have any explicit realisation of such
kind of Calabi-Yau manifold, eq. (\ref{hhh}) is simply the
expression for the $\mathbb{C}P^{4}_{[1,1,2,2,6]}$ case
(\ref{Vol11226}), augmented by the inclusion of a blow-up mode
$\tau_{3}$. We also assume that $h_{2,1}(X)>h_{1,1}(X)=3$, thus
satisfying the other condition of the LARGE Volume Claim. There
are then two ways to perform the limit (\ref{limit}) without
obtaining an internal volume that is formally negative:

\begin{enumerate}
\item
\begin{equation}
\left\{
\begin{array}{l}
\tau _{i}\rightarrow \infty ,\textrm{ \ }\forall i=1,2\textrm{
with the
constraint }\tau _{1}<\tau _{2}, \\
\tau _{3}\textrm{\ remains small.}
\end{array}
\right.
\end{equation}
This case keeps both cycles associated with the fibration large,
while the blow-up cycle remains small. Given that
$\tau_{1}\rightarrow\infty$, this situation resembles the "Swiss
cheese" picture:
\begin{equation}
\mathcal{V}=\alpha \underset{\tau
_{big}^{3/2}}{[\underbrace{\sqrt{\tau _{1}} (\tau _{2}-\beta \tau
_{1})}}-\gamma \tau _{3}^{3/2}],
\end{equation}
and due to this analogy with the $\mathbb{C}P_{[1,1,1,6,9]}^{4}$
model, the condition $K^{-1}_{33}\simeq\mathcal{V}\sqrt{\tau_{3}}$
is verified. Thus we can apply the LARGE Volume Claim which states
that the scalar potential will have an AdS exponentially large
volume set of minima together with $(h_{1,1}-N_{small}-1)=1$ flat
directions. In the following section we shall confirm this with an
explicit calculation.

\item
\begin{equation}
\left\{
\begin{array}{l}
\tau _{2}\rightarrow \infty, \\
\tau _{1}\textrm{ and }\tau _{3}\textrm{\ remain small.}
\end{array}
\right.
\end{equation}
In this case $N_{small}=2$ and according to the LARGE Volume Claim
there will be an exponentially large volume minimum of the scalar
potential if and only if both $\tau_{1}$ and $\tau_{3}$ is a
blow-up mode. As we show in the next Section
\ref{3modK3noLoopCalc},
$K^{-1}_{33}\sim\mathcal{V}\sqrt{\tau_{3}}$, as is suggested by
the volume form (\ref{hhh}). However $K^{-1}_{11}\sim
\tau_{1}^{2}$, as could be guessed from the fact that the overall
volume (\ref{linearity}) in terms of the 2-cycles moduli is linear
in $t_{1}$. Hence $\tau_{1}$ is not a blow-up but a fibration
modulus that does not give rise to LVS.

\end{enumerate}

Before confirming these statements in the next subsection with
explicit calculations, we point out that more general examples of
this kind of LVS have been discovered in \cite{blum}. These
authors noticed that starting from an elliptically fibred
Calabi-Yau over a $dP_n$ base, and then flopping away only $r<n$
(instead of all $n$) of the $\mathbb{C}P^1$-cycles in the base,
one obtains another elliptically fibred Calabi-Yau (instead of a
Swiss-cheese one), whose volume looks like:
\begin{equation}
 \mathcal{V}=\textrm{Vol}\left(X_{n-r}\right)-\sum_{i=1}^{r}\lambda_{i}\tau_{i}^{3/2},
 \textrm{ \ }\lambda_{i} > 0\textrm{ \ }\forall
 i=1,...,r,
\end{equation}
where $X_{n-r}$ is the resulting elliptical fibration over a
$dP_{n-r}$ base. It is natural to expect that the scalar potential
for these examples has an AdS minimum at exponentially large
volume, together with $( h_{1,1} - N_{small} - 1)=n-r$ flat
directions that will be lifted by $g_s$ corrections.

\subsubsection{Explicit Calculation} \label{3modK3noLoopCalc}

We focus on the case in which:
\begin{equation}
\mathcal{V}=\alpha \left[ \sqrt{\tau _{1}}(\tau _{2}-\beta \tau
_{1})-\gamma \tau _{3}^{3/2}\right] ,
\end{equation}
where $\alpha $, $\beta $, $\gamma $ are positive model-dependent
constants and the K\"{a}hler potential and the superpotential take
the form (defining $\hat{\xi}\equiv \xi g_{s}^{-3/2}$):
\begin{equation}
K=K_{0}+\delta K_{(\alpha')}=-2\ln \left(
\mathcal{V}+\frac{\hat{\xi}}{2}\right) , \label{3}
\end{equation}
\begin{equation}
W=W_{0}+A_{1}e^{-a_{1}T_{1}}+A_{2}e^{-a_{2}T_{2}}+A_{3}e^{-a_{3}T_{3}}.
\label{sp}
\end{equation}
In the large volume limit the K\"{a}hler matrix and its inverse
look like:
\begin{equation}
K_{ij}^{0}=\frac{1}{4\mathcal{V}^{2}}\left(
\begin{array}{ccc}
\frac{\mathcal{V}^{2}}{\tau _{1}^{2}}+2\alpha ^{2}\beta ^{2}\tau
_{1} & \frac{\alpha ^{2}}{\sqrt{\tau _{1}}}\left( \gamma \tau
_{3}^{3/2}-2\beta \tau _{1}^{3/2}\right) & \frac{3\alpha \gamma
}{2}\frac{\sqrt{\tau _{3}}}{
\tau _{1}}\left( 2\alpha \beta \tau _{1}^{3/2}-\mathcal{V}\right) \\
\frac{\alpha ^{2}}{\sqrt{\tau _{1}}}\left( \gamma \tau
_{3}^{3/2}-2\beta \tau _{1}^{3/2}\right) & 2\alpha ^{2}\tau _{1} &
-3\alpha ^{2}\gamma \sqrt{
\tau _{1}}\sqrt{\tau _{3}} \\
\frac{3\alpha \gamma }{2}\frac{\sqrt{\tau _{3}}}{\tau _{1}}\left(
2\alpha \beta \tau _{1}^{3/2}-\mathcal{V}\right) & -3\alpha
^{2}\gamma \sqrt{\tau _{1}}\sqrt{\tau _{3}} & \frac{3\alpha \gamma
}{2}\frac{\mathcal{V}}{\sqrt{ \tau _{3}}}
\end{array}
\right), \label{LaDiretta}
\end{equation}
and:
\begin{equation}
K_{0}^{ij}=4\left(
\begin{array}{ccc}
\tau _{1}^{2} & \beta \tau _{1}^{2}+\gamma \sqrt{\tau _{1}}\tau
_{3}^{3/2} &
\tau _{1}\tau _{3} \\
\beta \tau _{1}^{2}+\gamma \sqrt{\tau _{1}}\tau _{3}^{3/2} &
\frac{\mathcal{V }^{2}}{2\alpha ^{2}\tau _{1}}+\beta ^{2}\tau
_{1}^{2} & \tau _{3}\left(
\frac{\mathcal{V}}{\alpha \sqrt{\tau _{1}}}+\beta \tau _{1}\right) \\
\tau _{1}\tau _{3} & \tau _{3}\left( \frac{\mathcal{V}}{\alpha
\sqrt{\tau _{1}}}+\beta \tau _{1}\right) & \frac{2}{3\alpha \gamma
}\mathcal{V}\sqrt{ \tau _{3}}
\end{array}
\right).  \label{Kinverse}
\end{equation}
Both of the ways outlined above to take the large volume limit
have $\tau _{2}\gg 1$ and so the superpotential (\ref{sp}) can be
simplified as follows:
\begin{equation}
W\simeq W_{0}+A_{1}e^{-a_{1}T_{1}}+A_{3}e^{-a_{3}T_{3}}.
\label{spot}
\end{equation}
The scalar potential takes its general form (\ref{scalar}). We
recall that in the large volume limit, the $\alpha'$ leading
contribution to the scalar potential becomes:
\begin{equation} \delta V_{\left( \alpha'\right) }=3e^{K}\hat{\xi}\frac{\left( \hat{\xi}
^{2}+7\hat{\xi} \mathcal{V}+\mathcal{V}^{2}\right) }{\left(
\mathcal{V}-\hat{\xi} \right) \left( 2 \mathcal{V}+\hat{\xi}
\right) ^{2}}\left\vert W\right\vert ^{2}\underset{\mathcal{V
}\rightarrow \infty }{\longrightarrow }\frac{3}{4}\frac{\hat{\xi}
}{\mathcal{V}^{3} }\left\vert W\right\vert ^{2}.
\end{equation}
Adding this to the non-perturbative part, we are left with:
\begin{eqnarray}
V &=&\frac{1}{\mathcal{V}^{2}}\left[
K_{0}^{ij}a_{i}A_{i}a_{j}\bar{A} _{j}e^{-\left(
a_{i}T_{i}+a_{j}\bar{T}_{j}\right) }+2a_{i}A_{i}\tau
_{i}e^{-a_{i}T_{i}}\bar{W}\right.  \nonumber \\
&&\left. +2a_{j}\bar{A}_{j}\tau _{j}e^{-a_{j}\bar{T}_{j}}W\right]
+\frac{ 3}{4}\frac{\hat{\xi} }{\mathcal{V}^{3}}\left\vert
W\right\vert ^{2}. \label{v1v}
\end{eqnarray}

We shall focus on the case $W_{0}\sim \mathcal{O}(1)$, (since a
tuned small value of $W_0$ cannot give rise to large volume)
 and
for $a_{1}\tau_{1}\gg 1$, $a_{3}\tau_{3}\gg 1$, after extremising
the axion directions, the scalar potential simplifies to (with
$\lambda\equiv 8/(3\alpha\gamma)$, $\nu\equiv 3\hat{\xi}/4$ and
$A_{1}=A_{3}=1$):
\begin{equation}
V =\frac{\lambda a_{3}^{2}}{\mathcal{V}}\sqrt{\tau
_{3}}e^{-2a_{3}\tau _{3}}-\frac{4}{\mathcal{V}^{2}}W_{0}a_{1}\tau
_{1}e^{-a_{1}\tau _{1}}-\frac{4}{\mathcal{V}^{2}}W_{0}a_{3}\tau
_{3}e^{-a_{3}\tau _{3}}+\frac{\nu }{\mathcal{V}^{3}}W_{0}^{2}.
\label{ygfdo}
\end{equation}
The scalar potential $V$ depends on $\mathcal{V}$, $\tau_{1}$ and
$\tau_{3}$: $V=V(\mathcal{V},\tau _{1},\tau _{3})$, with the
dependence on $\tau_{2}$ implicit in the internal volume
$\mathcal{V}$. The large volume limit can be taken in the two ways
(1 and 2) outlined in section \ref{3modK3noLoop}. The difference
between these two cases is that in limit 1
$\tau_{1}\rightarrow\infty$ whereas in limit 2 $\tau_{1}$ remains
small. Let us now study these two different cases in detail.

\bigskip

\textbf{1) }$\tau_{1}\rightarrow\infty$\textbf{ \
}$\Leftrightarrow \tau_{3}\ll \tau_{1}<\tau_{2}$

In this case, the superpotential (\ref{spot}) obtains
non-perturbative corrections only in $\tau_{3}$:
\begin{equation}
W\simeq W_{0}+A_{3}e^{-a_{3}T_{3}}. \label{ew}
\end{equation}
Since the $A_{1}$ term is not present in (\ref{ew}), we will be
unable to stabilise the corresponding K\"{a}hler modulus
$\tau_{1}$, thereby giving rise to an exactly flat direction. In
this case the scalar potential (\ref{ygfdo}) further reduces to:
\begin{equation}
V =\frac{\lambda a_{3}^{2}}{\mathcal{V}}\sqrt{\tau
_{3}}e^{-2a_{3}\tau _{3}}-\frac{4}{\mathcal{V}^{2}}W_{0}a_{3}\tau
_{3}e^{-a_{3}\tau _{3}}+\frac{\nu }{\mathcal{V}^{3}}W_{0}^{2}.
\label{41}
\end{equation}
and $V$ depends only on $\mathcal{V}$ and $\tau_{3}$: $V=V(V,$\
$\tau_{3})$. The potential (\ref{41}) has the same form as the
scalar potential found in section 3.2 of \cite{LVS} where the
$\mathbb{C}P_{[ 1,1,1,6,9]}^{4}$ case was first discussed.
Following the
 same reasoning, we
look for possible minima of the scalar potential (\ref{41}) by
working out the two minimisation conditions:
\begin{equation}
\frac{\partial V}{\partial \mathcal{V}}=0\textrm{ \ \
}\Leftrightarrow \textrm{ \ \ }\left(\lambda a_{3}^{2}\sqrt{\tau
_{3}}e^{-2a_{3}\tau _{3}}\right) \textrm{\
}\mathcal{V}^{2}-\left(8W_{0}a_{3}\tau _{3}e^{-a_{3}\tau
_{3}}\right) \mathcal{V}+3 \nu W_{0}^{2}=0, \label{III}
\end{equation}
\begin{equation}
\frac{\partial V}{\partial \tau _{3}}=0\textrm{ \ \
}\Leftrightarrow \textrm{ \ \ }\frac{\lambda a_{3}}{2\sqrt{\tau
_{3}}}\mathcal{V}\textrm{\
}e^{-a_{3}\tau_{3}}\left(1-4a_{3}\tau_{3}\right)+ 4W_{0}\left(
a_{3}\tau_{3}-1\right)=0. \label{IV}
\end{equation}
Equation (\ref{III}) admits a solution of the form:
\begin{equation}
\frac{\lambda a_{3}}{4
W_{0}\sqrt{\tau_{3}}}\mathcal{V}e^{-a_{3}\tau_{3}}=1\pm
\sqrt{1-\frac{3\lambda\nu}{16\tau_{3}^{3/2}}}, \label{uno}
\end{equation}
whereas in the approximation $a_{3}\tau_{3}\gg 1$, (\ref{IV})
becomes:
\begin{equation}
\frac{\lambda a_{3}}{2\sqrt{\tau
_{3}}}\mathcal{V}e^{-a_{3}\tau_{3}}= W_{0}. \label{due}
\end{equation}
Combining (\ref{uno}) and (\ref{due}), we find $\frac{1}{2}=1\pm
\sqrt{1-\frac{3\lambda\nu}{16\tau_{3}^{3/2}}}$, whose solution is
given by:
\begin{equation}
\langle\tau_{3}\rangle=\frac{1}{g_{s}}\left(\frac{\xi}
{2\alpha\gamma}\right)^{2/3}\sim \frac{1}{g_{s}}. \label{x}
\end{equation}
On the contrary, from (\ref{due}) we work out:
\begin{equation}
\langle\mathcal{V}\rangle=\frac{3(\alpha\gamma)^{2/3} W_{0}}{4
a_{3}\sqrt{g_{s}}}\left(\frac{\xi}{2}\right)^{1/3}
e^{\frac{a_{3}}{g_{s}}\left(\frac{\xi}
{2\alpha\gamma}\right)^{2/3}}\sim \frac{W_{0}}{
a_{3}\sqrt{g_{s}}}e^{\frac{a_{3}}{g_{s}}}. \label{duet}
\end{equation}
There is therefore an exponentially large volume minimum. Setting
$\alpha=\gamma=1$, $\xi=2$, $g_{s}=0.1$, $a_{3}=\pi$ and
$W_{0}=1$, we finally obtain $\langle\tau_{3}\rangle=10$ and
$\langle\mathcal{V}\rangle=3.324\cdot10^{13}$. However there is
still the presence of an exactly flat direction which can be
better appreciated after the following change of coordinates:
\begin{equation}
(\tau_{1},\tau_{2})\textrm{ \ \ }\longrightarrow\textrm{ \ \ }(
\mathcal{V},\Omega):\textrm{\ \ \ \ \ \ \ \ \ \ \ \ } \left\{
\begin{array}{l}
\mathcal{V}\simeq
\alpha\left[\sqrt{\tau_{1}}\left(\tau_{2}-\beta\tau_{1}\right)\right] \\
\Omega=\alpha\left[\sqrt{\tau_{1}}\left(\tau
_{2}+\beta\tau_{1}\right)\right]
\end{array}
\right.  \label{43}
\end{equation}

From (\ref{duet}) and (\ref{43}) we see that the stabilisation of
$\mathcal{V}$ and $\tau_{3}$ does not depend on $\Omega$ at all,
implying that $\Omega$ is a flat direction. We plot in Figure 4.1
the behaviour of this scalar potential where the flat direction is
manifest: $\tau_{3}$ has been already fixed as
$\langle\tau_{3}\rangle=10$, $\mathcal{V}$ is plotted on the
\textit{x}-axis and $\Omega$ on the \textit{y}-axis.

\begin{figure}[ht]
\begin{center}
\epsfig{file=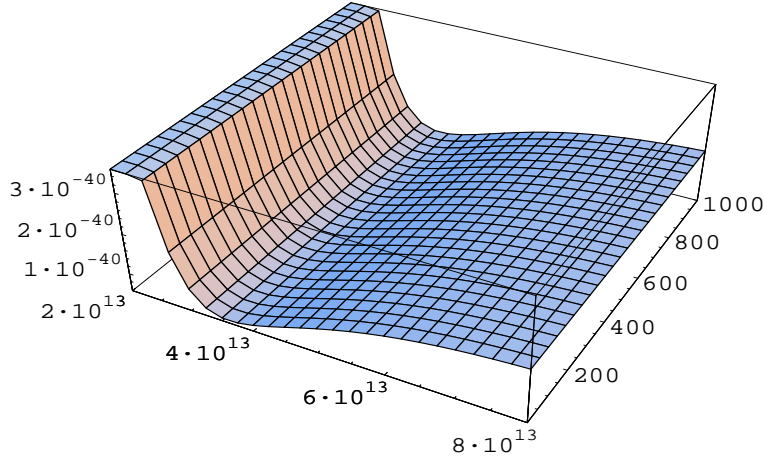, height=70mm,width=100mm} \caption{`Sofa'
potential with the presence of a flat direction.}
\end{center}
\end{figure}

\newpage

\textbf{2) $\tau_{1}$ small}

In this case the large volume limit is taken keeping $\tau_{1}$
small and the scalar potential takes the general form
(\ref{ygfdo}). The minimisation equation with respect to
$\tau_{1}$ reads
\begin{equation}
\frac{\partial V}{\partial \tau_{1}}=0\textrm{\ \ \ \
}\Leftrightarrow \textrm{ \ \ }\frac{4}{\mathcal{V}^{2}}W_{0}
a_{1}e^{-a_{1}\tau_{1}}\left(a_{1}\tau _{1}-1\right)=0, \label{II}
\end{equation}
which implies $a_{1}\tau_{1}=1$ and so we cannot neglect higher
instanton corrections.

There is therefore no trustable minimum for the $\tau_1$ field. We
may however think about a situation in which the system still has
an exponentially large internal volume, with $\tau_{3}$ and
$\mathcal{V}$ sitting at their minimum
$\langle\mathcal{V}\rangle\sim e^{a_{3}\langle\tau_{3}\rangle}$,
while $\tau_{1}$ plays the r\^{o}le of a quintessence field
rolling in a region at large $\tau_{1} \gg 1$ away from
$a_{1}\tau_{1}=1$. The quintessence scale would be set by the
$e^{-a_1 \tau_1}$ exponent. Setting $a_{1}\tau_{1}\gg 1$ it is
easy to see that this is possible. However, the values of $a_1$
and $\tau_1$ need to be tuned to get a realistically small mass
for $\tau_1$ and even if this is done the fifth force problems of
quintessence fields would seem to be unavoidable.

Finally, let us summarise in the table below the results found
without string loop corrections to $K$.
\begin{gather*}
1)\textrm{\ \ \ }\tau _{1}\rightarrow\infty\textrm{\ \ \ \ \ \
}\left\{
\begin{array}{l}
W_{0}\textrm{\ \ small \ \ \ \ \ No LVS,} \\
W_{0}\sim\mathcal{O}(1)\textrm{\ \ \ \ LVS + exactly flat
direction }\bot \mathcal{V},
\end{array}
\right. \\
2)\textrm{\ }\tau _{1}\textrm{ \ small \ }\left\{
\begin{array}{l}
W_{0}\textrm{\ \ small \ \ \ \ \ No LVS,} \\
W_{0}\sim\mathcal{O}(1)\textrm{\ and}\left\{
\begin{array}{l}
a_{1}\tau_{1}\gtrsim a_{3}\langle\tau_{3}\rangle\textrm{\ \ LVS +
almost
flat direction }\bot \textrm{\ }\mathcal{V}, \\
a_{1}\tau_{1}<a_{3}\langle\tau_{3}\rangle\textrm{\ \ No LVS.}
\end{array}
\right.
\end{array}
\right.
\end{gather*}

\chapter{Systematics of String Loop Corrections}
\label{SystematicsOfStringLoopCorrections}

In this chapter we shall study the behaviour of string loop
corrections to the K\"{a}hler potential for general type IIB
Calabi-Yau compactifications. We observe that there is an easier
and a harder part to compute the form of loop corrections. The
easier part involves the parametric scaling of moduli that control
the loop expansion: these are the dilaton, which controls the
string coupling, and the K\"{a}hler moduli, which controls the
gauge coupling on $D7$-branes. The harder part involves the actual
coefficients of the loop expansion, which depend on the complex
structure moduli and would require a explicit string computation.
This chapter focuses entirely on the `easier' part; however as the
K\"{a}hler moduli are unstabilised at tree-level, such knowledge
is very important for moduli stabilisation.

Recently, Berg, Haack and Pajer (BHP) \cite{bhp} gave arguments
for the general functional dependence of the leading order loop
corrections to $K$ on the K\"{a}hler moduli. By comparing with the
toroidal orientifold calculations and the standard transformations
required to go from the string frame, where string amplitudes are
computed, to the physical Einstein frame that enters the
supergravity action, they conjectured the parametric form of the
leading corrections for general Calabi-Yau compactifications as a
function of the K\"{a}hler moduli. As mentioned above, it is this
dependence (on the K\"{a}hler moduli) that is more relevant for
moduli stabilisation, as the dilaton and complex structure moduli
are usually stabilised directly from the fluxes and it is only the
K\"{a}hler moduli that need quantum corrections to the scalar
potential to be stabilised.

In this chapter we contribute to put on firmer grounds the leading
order string loop corrections to $K$ conjectured by BHP. The
results found do not apply only on LVS but are general features of
IIB flux compactifications. We provide a low-energy interpretation
of this conjecture in terms of the one-loop renormalisation of the
kinetic terms of the K\"{a}hler moduli. We check the consistency
of this interpretation explicitly in various examples. We then
prove that for arbitrary Calabi-Yaus, the leading contribution of
these corrections to the scalar potential is always vanishing, as
long as the corrections are homogeneous functions of degree $-2$
in the $2$-cycle volumes, which includes the BHP proposal. We call
this result `extended no-scale structure'.

We then use the Coleman-Weinberg potential to motivate this
cancellation from the viewpoint of low-energy field theory. We
show how the non-contribution of the leading order string loop
correction is no longer an accident but it is just a manifestation
of the underlying supersymmetry with equal number of bosons and
fermions, despite being spontaneously broken.

This `extended no-scale structure' is crucial to establish the
robustness of the LVS. In fact, if it were not present, the
leading string loop correction to $K$ would be dominant over the
$\alpha'$ corrections. On the contrary the first non-vanishing
one-loop contributions to the scalar potential, for which we give
a simple formula in terms of the tree-level K\"{a}hler metric and
the correction to the K\"{a}hler potential, are subdominant with
respect to the $\alpha'$ corrections. Therefore it is safe to
neglect the $g_{s}$ corrections in the
$\mathbb{C}P^{4}_{[1,1,1,6,9]}$ model, but according to our
general analysis for LVS, they are important when more complicated
Calabi-Yau manifolds are taken into account.

In fact, in chapter 6 we shall use the results found in this
chapter to include the string loop corrections in the study of LVS
for different classes of Calabi-Yau manifolds. In general, these
corrections will turn out to be very important for fixing all the
K\"{a}hler moduli.

Sections 5.1 and 5.2 are the main parts of this chapter, in which
we study in detail the proposed form of the string loop
corrections to the K\"{a}hler potential, their interpretation in
terms of the Coleman-Weinberg potential and examples of different
Calabi-Yau manifolds where these corrections are relevant. In
particular the `extended no-scale structure' of section 5.2 is
crucial to establish the robustness of the exponentially large
volume scenario. In chapter 6 we will use our results to study
moduli stabilisation in different classes of Calabi-Yau manifolds.

\section{General analysis of the string loop corrections}
\label{sec3}

\subsection{String loop corrections}

Our discussion of the form of the scalar potential in IIB flux
compactifications has still to include the string loop corrections
$\delta K_{(g_{s})}$. These have been computed in full detail only
for unfluxed toroidal orientifolds in \cite{bhk}. Subsequently the
same collaboration in \cite{bhp} made an educated guess for the
behaviour of these loop corrections for general smooth Calabi-Yau
three-folds by trying to understand how the toroidal calculation
would generalise to the Calabi-Yau case. To be self-contained, we
therefore briefly review the main aspects of the toroidal
orientifold calculation of \cite{bhk}.

\medskip

\subsubsection{Exact calculation: $N=2$ $K3\times
T^{2}$ and $N=1$ $T^{6}/(\mathbb{Z}_{2}\times \mathbb{Z} _{2})$}
\label{sec3.1.1}

\medskip

The string loop corrections to $N=1$ supersymmetric\textit{\
}$T^{6}/( \mathbb{Z}_{2}\times \mathbb{Z}_{2})$ orientifold
compactifications with $D3$- and $D7$-branes follow by
generalising the result for $N=2$ supersymmetric $K3\times T^{2}$
orientifolds. Therefore we start by outlining the result in the
second case.

The one-loop corrections to the K\"{a}hler potential from Klein
bottle, annulus and M\"{o}bius strip diagrams are derived by
integrating the one-loop correction to the tree level K\"{a}hler
metric. These corrections are given by 2-point functions and to
derive the corrections $\delta K_{(g_{s})}$ it is sufficient to
compute just one of these correlators and integrate, since all
corrections to the K\"{a}hler metric come from the same $\delta
K_{(g_{s})}$. From \cite{bhk} the one-loop correction to the
2-point function of the complex structure modulus $U$ of $T^{2}$
is given by, dropping numerical factors:
\begin{equation}
\label{correl} \left\langle V_{U}V_{\bar{U}}\right\rangle \sim
-\left( p_{1}\cdot p_{2}\right)
g_{s}^{2}\alpha'^{-4}V_{4}\frac{vol(T^{2})_{s}}{\left( U+
\bar{U}\right) ^{2}}\mathcal{E}_{2}(A_{i},U),
\end{equation}%
where $V_{4}$ is the regulated volume of the four dimensional
space-time, $ vol(T^{2})_{s}$ denotes the volume of $T^{2}$\ in
string frame and $A_{i}$ are open string moduli. The coefficient
$\mathcal{E}_{2}(A_{i},U)$ is a linear combination of
non-holomorphic Eisenstein series $E_{2}(A,U)$ given by:
\begin{equation}
E_{2}(A,U)=\sum_{(n,m)\neq
(0,0)}\frac{\hbox{Re}(U)^{2}}{\left\vert
n+mU\right\vert ^{4}}\exp \left[ 2\pi i\frac{A(n+m\bar{U})+\bar{A}(n+mU)}{U+%
\bar{U}}\right] .  \label{eisenstein}
\end{equation}%
The result (\ref{correl}) is converted to Einstein frame through a
Weyl rescaling:
\begin{equation}
\left\langle V_{U}V_{\bar{U}}\right\rangle _{E}=\left\langle V_{U}V_{\bar{U}%
}\right\rangle _{s}\frac{e^{2\varphi }}{vol(K3\times T^{2})_{s}},
\end{equation}%
giving:
\begin{equation}
\left\langle V_{U}V_{\bar{U}}\right\rangle \sim -\left( p_{1}\cdot
p_{2}\right) g_{s}^{2}\alpha'^{-4}V_{4}\frac{e^{2\varphi }}{\left( U+%
\bar{U}\right) ^{2}}\frac{\mathcal{E}_{2}(A_{i},U)}{vol(K3)_{s}}.
\end{equation}%
Writing the volume of the $K3$ hypersurface in Einstein frame,
$vol(K3)_{s}=e^{\varphi }vol(K3)_{E},$ produces the final result:
\begin{equation}
\left\langle V_{U}V_{\bar{U}}\right\rangle \sim -\left( p_{1}\cdot
p_{2}\right) g_{s}^{2}\alpha'^{-4}V_{4}\frac{e^{\varphi }}{\left( U+%
\bar{U}\right) ^{2}}\frac{\mathcal{E}_{2}(A_{i},U)}{vol(K3)_{E}}.
\label{U}
\end{equation}%
Now noticing that:
\begin{equation}
\partial _{U}\partial _{\bar{U}}E_{2}(A,U)\sim -\frac{E_{2}(A,U)}{\left( U+%
\bar{U}\right) ^{2}},
\end{equation}%
we can read off from (\ref{U}) the 1-loop correction to the
kinetic term for the field $U$, and using $vol(K3)_{E}=\tau $, the
1-loop correction to the K\"{a}hler potential becomes:
\begin{equation}
\delta K_{(g_{s})}= c \frac{\mathcal{E}_{2}(A_{i},U)}{
\hbox{Re}\left( S\right) \tau},  \label{UU}
\end{equation}%
where a full analysis determines the constant of proportionality
$c$ to be $c=-1/(128\pi ^{4})$\footnote{The constant $c$ given
here differs from the one calculated in \cite{bhk} only by a
factor of $(-\pi^{2})$ due to different conventions. In fact, in
\cite{bhk} the correction (\ref{UU}) takes the form $\delta
K_{(g_{s})}= -\frac{c}{8}
\frac{\mathcal{E}_{2}(A_{i},U)}{\hbox{Im}( S)\hbox{Im}(T)}$ with
$\hbox{Im}(S)\equiv \frac{e^{-\varphi}}{\sqrt{8}\pi}$ and
$\hbox{Im}(T)\equiv \frac{\tau}{\sqrt{8}\pi}$.}. This procedure
can be generalized to evaluate the loop corrections in the $N=1$
supersymmetric\textit{\ }$T^{6}/( \mathbb{Z}_{2}\times
\mathbb{Z}_{2})$ case, obtaining:
\begin{equation}
\delta K_{(g_{s})}=\delta K_{(g_{s})}^{KK}+\delta K_{(g_{s})}^{W},
\end{equation}%
where $\delta K_{(g_{s})}^{KK}$ comes from the exchange between
$D7$ and $D3$-branes of closed strings which carry Kaluza-Klein
momentum, and gives (for vanishing open string scalars):
\begin{equation}
\delta K_{(g_{s})}^{KK}= -\frac{1}{128\pi^{4}}\sum\limits_{i=1}^{3}%
\frac{\mathcal{E}_{i}^{KK}(U,\bar{U})}{\hbox{Re}\left( S\right)
\tau _{i}}. \label{KK}
\end{equation}
The other correction $\delta K_{(g_{s})}^{W}$ can again be
interpreted in the closed string channel as coming from the
exchange of winding strings between intersecting stacks of
$D7$-branes. These contributions are present in the $N=1$ case but
not in the $N=2$ case. They take the form:
\begin{equation}
\delta K_{(g_{s})}^{W}=-\frac{1}{128\pi^{4}}\sum\limits_{i\neq
j\neq k=1}^{3}\frac{\mathcal{E}_{i}^{W}(U,\bar{U})}{\tau _{j}\tau
_{k}}. \label{W}
\end{equation}%

\medskip
\subsubsection{Generalisation to Calabi-Yau three-folds}
\medskip

The previous calculation teaches us that, regardless of the
particular background under consideration, a Weyl rescaling will
always be necessary to convert to four-dimensional Einstein frame.
This implies the 2-point function should always be suppressed by
the overall volume:
\begin{equation}
\left\langle V_{U}V_{\bar{U}}\right\rangle _{s}\sim g(U,T,S)%
\Longleftrightarrow \left\langle V_{U}V_{\bar{%
U}}\right\rangle _{E}\sim g(U,T,S)\frac{e^{\varphi
/2}}{\mathcal{V}_{E}}.
\end{equation}
This allowed \cite{bhp} to conjecture the parametric form of the
loop corrections even for Calabi-Yau cases. $g(U, T, S)$
originates from KK modes as $m_{KK}^{-2}$ and so should scale as a
2-cycle volume $t$. Conversion to Einstein frame then leads to:
\begin{equation}
\delta K_{(g_{s})}^{KK}\sim
\sum\limits_{i=1}^{h_{1,1}}g(U)\frac{\left(
a_{l}t^{l}\right) e^{\varphi }}{\mathcal{V}}=\sum\limits_{i=1}^{h_{1,1}}%
\frac{\mathcal{C}_{i}^{KK}(U,\bar{U})\left( a_{il}t^{l}\right) }{\hbox{Re}%
\left( S\right) \mathcal{V}},  \label{UUU}
\end{equation}%
where $a_{l}t^{l}$ is a linear combination of the basis 2-cycle volumes $%
t_{l}$. A similar line of argument for the winding corrections
(where the function $g(U, T, S)$ goes as $m_{W}^{-2} \sim t^{-1}$)
gives:
\begin{equation}
\delta K_{(g_{s})}^{W}\sim \sum\limits_{i=1}^{h_{1,1}}\frac{\mathcal{C}%
_{i}^{W}(U,\bar{U})}{\left( a_{il}t^{l}\right) \mathcal{V}}.
\label{UUUU}
\end{equation}%

Notice that $\mathcal{C}^{KK}_i $ and $\mathcal{C}^W_i$ are
unknown functions of the complex structure moduli and therefore
this mechanism is only useful to fix the leading order dependence
on K\"{a}hler moduli. This is similar to the K\"{a}hler potential
for matter fields whose dependence on K\"{a}hler moduli can be
extracted by scaling arguments \cite{ccq}, while the complex
structure dependence is unknown. Fortunately it is the K\"{a}hler
moduli dependence that is more relevant in both cases due to the
fact that complex structure moduli are naturally fixed by fluxes
at tree-level. On the other hand, the K\"{a}hler moduli need
quantum corrections to be stabilised and are usually more relevant
for supersymmetry breaking.

We now turn to trying to understand the loop corrections from a
low-energy point of view.

\subsection{Low energy approach}

The low energy physics is described by a four dimensional
supergravity action. We ask here whether it is possible to
understand the form of the loop corrections in terms of the
properties of the low energy theory, without relying on a full
string theory computation.

We first ask what one could reasonably hope to understand. The
form of equations (\ref{KK}) and (\ref{W}) show a very complicated
dependence on the complex structure moduli, and a very simple
dependence on the dilaton and K\"ahler moduli. The dependence on
the complex structure moduli is associated with an Eisenstein
series originating from the structure of the torus, and so we
cannot expect to reproduce this without a full string computation.
On the other hand the dilaton and K\"ahler moduli appear with a
very simple scaling behaviour. This we may hope to be able to
understand using low-energy arguments, and to be able to
conjecture the generalisation to the Calabi-Yau case.

There is one paper in the literature that has already tried to do
that. In an interesting article \cite{hg}, von Gersdorff and
Hebecker considered models with one K\"{a}hler modulus $\tau $,
such that $\mathcal{V}=\tau ^{3/2}=R^{6}$ $\Longleftrightarrow
\tau =R^{4}$, and argued for the form of $\delta K_{(g_{s})}^{KK}$
using the Peccei-Quinn symmetry, scaling arguments and the
assumption that the loop corrections arise simply from the
propagation of ten dimensional free fields in the compact space
and therefore do not depend on $M_{s}$. This led to the proposal:
\begin{equation}
 \delta K_{(g_{s})}^{KK}\simeq \tau
^{-2}.  \label{von Gersdorff}
\end{equation}
However, at the level of the K\"{a}hler potential (but not the
scalar potential) this result disagrees with the outcome of the
exact toroidal calculation (\ref{KK}). It seems on the contrary to
reproduce the corrections due to the exchange of winding strings
(\ref{W}), but as $m_W > M_s > m_{KK}$ we do not expect to see
such corrections at low energy. In reality, $\ \delta
K_{(g_{s})}^{KK}$ should contain all contributions to the 1-loop
corrections to the kinetic term of $\tau$. From the reduction of
the DBI action we know that $\tau $ couples to the field theory on
the stack of $D7$-branes wrapping the 4-cycle whose volume is
given by $\tau $. It therefore does not seem that the string loop
corrections will come from the propagation of free fields as
$\tau$ will interact with the corresponding gauge theory on the
brane. In fact the reduced DBI action contains a term which looks
like:
\begin{equation}
\delta S_{DBI}\supset \int d^{4}x\sqrt{-g^{(4)}}\tau F^{\mu \nu
}F_{\mu \nu },
\end{equation}%
and when $\tau $\ gets a non-vanishing VEV, expanding around this
VEV in the following way:
\begin{equation}
\tau =\left\langle \tau \right\rangle +\tau ' ,
\end{equation}
we obtain:
\begin{equation}
\delta S_{DBI}\supset \int d^{4}x\sqrt{-g^{(4)}}\left(
\left\langle \tau \right\rangle F^{\mu \nu }F_{\mu \nu }+\tau '
F^{\mu \nu }F_{\mu \nu }\right) .  \label{DBI}
\end{equation}

%
\begin{figure}[ht]
\begin{center}
\epsfig{file=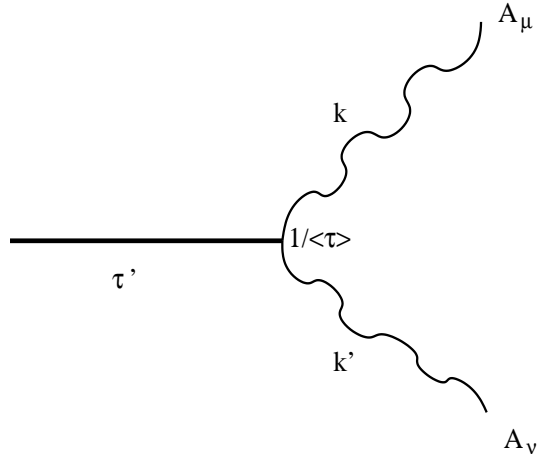, height=60mm,width=70mm} \caption{
 Coupling of the K\"ahler modulus with the gauge fields on the brane.}
\end{center}
\end{figure}

From the first term in (\ref{DBI}) we can readily read off the
coupling constant of the gauge group on the brane:
\begin{equation}
g^{2}=\frac{1}{M_{s}^{4}\tau },  \label{coupling}
\end{equation}%
where we have added $M_{s}^{4}$\ to render it correctly
dimensionless. On the other hand, the second term in (\ref{DBI})
will give rise to an interaction vertex of the type shown in
Figure 5.1 that will affect the 1-loop renormalisation of the
$\tau$ kinetic term.

In any ordinary quantum field theory, generic scalar fields
$\varphi $ get 1-loop quantum corrections to their kinetic terms
(wavefunction renormalisation) of the form:
\begin{equation}
\int d^{4}x\sqrt{-g^{(4)}}\frac{1}{2}\left( 1+A\right) \partial
_{\mu }\varphi
\partial ^{\mu }\varphi,  \label{hj}
\end{equation}%
where $A$ is given by $A\simeq \frac{g^{2}}{16\pi ^{2}}$, with $g$
the coupling constant of the gauge interaction this scalar couples
to.

$\tau$ is a modulus and not a gauge-charged field. Nonetheless, we
still expect loop corrections to generate corrections to the
moduli kinetic terms. We expect to be able to write the kinetic
terms as:
\begin{equation}
K_{i \bar{j}} = K_{i \bar{j}, tree} + \delta K_{i \bar{j},
1-loop}.
\end{equation}
We also expect the loop correction to the kinetic term to always
involve a suppression by the coupling that controls the loop
expansion. This is the analogue to the correction in (\ref{hj})
depending on the gauge coupling constant, which controls the loop
expansion of ordinary field theory. For a brane wrapping a cycle
$\tau$, the value of $\tau$ is the gauge coupling for branes
wrapping the cycle, and we expect loop corrections involving those
branes to involve a suppression, relative to tree-level terms, by
a factor of $\tau$ (see \cite{joe} for related arguments).

This is not a rigorous derivation, but we consider this a
reasonable assumption. We will find that it gives the correct
scaling of the loop correction for the toroidal case where the
correction has been computed explicitly, and that, while it has a
different origin, it agrees with the BHP conjecture for the
parametric form of loop corrections in the Calabi-Yau case. The
loop corrections to the K\"ahler potential $K$ should then be such
as to generate corrections to the kinetic terms for $\tau$ that
are suppressed by a factor of $g^2$ for the gauge theory on branes
wrapping the cycle $\tau$. The K\"{a}hler potential upon double
differentiation yields the kinetic terms in the four dimensional
Einstein frame Lagrangian:
\begin{equation}
S_{\textrm{\textit{Einstein}}}\supset \int d^{4}x\sqrt{-g^{(4)}}\left( \frac{%
\partial ^{2}\left( K_{tree}\right) }{\partial \tau ^{2}}+\frac{\partial
^{2}\left( \delta K_{(g_{s})}^{KK}\right) }{\partial \tau
^{2}}\right) \left( \partial \tau \right) ^{2},
\end{equation}%
and the general canonical redefinition of the scalar fields:
\begin{equation}
\tau \longrightarrow \varphi =\varphi (\tau ),
\end{equation}%
will produce a result similar to (\ref{hj}), which implies:
\begin{equation}
\frac{\partial ^{2}\left( K_{tree}\right) }{\partial \tau ^{2}}%
\longrightarrow \frac{1}{2},\textrm{ \ \ \ \ }\frac{\partial
^{2}\left( \delta
K_{(g_{s})}^{KK}\right) }{\partial \tau ^{2}}\longrightarrow \frac{1}{2}%
A\sim \frac{1}{2}\frac{g^{2}}{16\pi ^{2}},
\end{equation}%
and thus:
\begin{equation}
\frac{\partial ^{2}\left( \delta K_{(g_{s})}^{KK}\right) }{\partial \tau ^{2}%
}\sim \frac{g^{2}}{16\pi ^{2}}\frac{\partial ^{2}\left(
K_{tree}\right) }{\partial \tau ^{2}}.  \label{dds}
\end{equation}%
Using  equation (\ref{coupling}) we then guess for the scaling
behavior of the string loop corrections to the K\"{a}hler
potential:
\begin{equation}
\frac{\partial ^{2}\left( \delta K_{(g_{s})}^{KK}\right) }{\partial \tau ^{2}%
}\sim \frac{f(\hbox{Re}(S))}{16\pi ^{2}}\frac{1}{\tau
}\frac{\partial ^{2}\left( K_{tree}\right) }{\partial \tau ^{2}},
\label{fundamentale}
\end{equation}%
where we have introduced an unknown function of the dilaton
$f(\hbox{Re}(S))$ representing an integration constant\footnote{In
general there should be also an unknown function of the complex
structure and open string moduli but we dropped it since, as we
stated at the beginning of this section, its full determination
would require an exact string calculation.}. However we may be
able to use similar reasoning to determine $f(\hbox{Re}(S))$. The
same correction $\delta K_{(g_{s})}^{KK}$, upon double
differentiation with respect to the dilaton, has to give rise to
the 1-loop quantum correction to the corresponding dilaton kinetic
term. We also recall that $S$ couples to all field theories on
$D3$-branes as the relative gauge kinetic function is the dilaton
itself. Using the same argument as above we end up with the
further guess for $\delta K_{(g_{s})}^{KK}$:
\begin{equation}
\frac{\partial ^{2}\left( \delta K_{(g_{s})}^{KK}\right) }{\partial \hbox{Re}%
(S)^{2}}\sim \frac{h(\tau )}{16\pi
^{2}}\frac{1}{\hbox{Re}(S)}\frac{\partial ^{2}\left(
K_{tree}\right) }{\partial \hbox{Re}(S)^{2}}\simeq \frac{h(\tau
)}{16\pi ^{2}}\frac{1}{\hbox{Re}(S)^{3}},  \label{jkj}
\end{equation}%
where $h(\tau )$ is again an unknown function which parameterises
the dependence on the K\"{a}hler modulus. Integrating (\ref{jkj})
twice, we obtain:
\begin{equation}
\delta K_{(g_{s})}^{KK}\sim \frac{h(\tau )}{16\pi ^{2}}\frac{1}{\hbox{Re}(S)}%
,  \label{dilatonic}
\end{equation}%
where $h(\tau )$ can be worked out from (\ref{fundamentale}):
\begin{equation}
\frac{\partial ^{2}\left( h(\tau )\right) }{\partial \tau ^{2}}\sim \frac{1}{%
\tau }\frac{\partial ^{2}\left( K_{tree}\right) }{\partial \tau
^{2}}. \label{fundamental}
\end{equation}%

We now apply the above methods to several Calabi-Yau cases,
comparing to either the exact results or the conjecture of
equation (\ref{UUU}).

\subsubsection{Case 1: $N=1$ $T^{6}/(%
\mathbb{Z}
_{2}\times
\mathbb{Z}
_{2})$} \label{sec3.2.1}

We first consider the case of toroidal compactifications, for
which the loop corrections have been explicitly computed
\cite{bhk}. In that case the volume can be expressed as (ignoring
the 48 twisted K\"{a}hler moduli obtained by blowing up orbifold
singularities):
\begin{equation}
\mathcal{V}=\sqrt{\tau _{1}\tau _{2}\tau _{3}},  \label{torus}
\end{equation}%
and so (\ref{fundamental})\ takes the form:
\begin{equation}
\frac{\partial ^{2}\left( \delta K_{(g_{s})}^{KK}\right)
}{\partial \tau _{i}^{2}}\sim \frac{f(\hbox{Re}(S))}{16\pi
^{2}}\frac{1}{\tau _{i}^{3}}\textrm{ \ \ \ \ \ }\forall i=1,2,3.
\end{equation}%
Upon integration we get:
\begin{equation}
\delta K_{(g_{s})}^{KK}\sim \frac{1}{16\pi
^{2}}\frac{f(\hbox{Re}(S))}{\tau _{i}}\textrm{ \ \ \ \ \ }\forall
i=1,2,3.
\end{equation}%
Now combining this result with the analysis for the dilatonic
dependence of the string loop corrections, we obtain:
\begin{equation}
\delta K_{(g_{s})}^{KK}\sim \frac{1}{16\pi
^{2}}\sum_{i=1}^{3}\frac{1}{\hbox{Re}(S)\tau _{i}},
\end{equation}%
which reproduces the scaling behaviour of the result (\ref{KK})\
found from string scattering amplitudes.

\subsubsection{Case 2: $%
\mathbb{C}
P_{[1,1,1,6,9]}^{4}(18)$} \label{sec3.2.2}

We next consider loop corrections to the K\"{a}hler potential for
an orientifold of the Calabi-Yau $\mathbb{C}
P_{[1,1,1,6,9]}^{4}(18)$. We will compare the form of (\ref{UUU})
to that arising from our method (\ref{fundamental}) to work out
the behaviour of $\delta K_{(g_{s})}^{KK}$,
finding again a matching.\footnote{We note that the topology of $%
\mathbb{C} P_{[1,1,1,6,9]}^{4}(18)$ does not allow to have $\delta
K_{(g_{s})}^{W}\neq 0$ \cite{curio}.}
In the large volume limit we can write the volume 
as follows:
\begin{equation}
\mathcal{V}=\frac{1}{9\sqrt{2}}\left( \tau _{5}^{3/2}-\tau
_{4}^{3/2}\right) \simeq \tau _{5}^{3/2},
\end{equation}%
and (\ref{UUU}) becomes:
\begin{equation}
\delta K_{(g_{s})}^{KK}\sim \frac{\mathcal{C}_{4}^{KK}\sqrt{\tau _{4}}}{%
\hbox{Re}\left( S\right)
\mathcal{V}}+\frac{\mathcal{C}_{5}^{KK}\sqrt{\tau
_{5}}}{\hbox{Re}\left( S\right) \mathcal{V}}\simeq \frac{\mathcal{C}_{4}^{KK}%
\sqrt{\tau _{4}}}{\hbox{Re}\left( S\right) \mathcal{V}}+\frac{\mathcal{C}%
_{5}^{KK}}{\hbox{Re}\left( S\right) \tau _{5}}.  \label{hjkl}
\end{equation}%
From the tree-level K\"{a}hler matrix we read:
\begin{equation}
\frac{\partial ^{2}\left( K_{tree}\right) }{\partial \tau
_{4}^{2}}\simeq \frac{1}{\sqrt{\tau _{4}}\mathcal{V}},\textrm{ \ \
\ \ }\frac{\partial ^{2}\left( K_{tree}\right) }{\partial \tau
_{5}^{2}}\simeq \frac{1}{\tau _{5}^{2}}.
\end{equation}%
Requiring loop corrections to be suppressed by a factor of
$g^2_{c}$ for the field-theory on the brane gives:
\begin{equation}
\left\{
\begin{array}{c}
\frac{\partial ^{2}\left( \delta K_{(g_{s})}^{KK}\right)
}{\partial \tau _{4}^{2}}\sim \frac{1}{16\pi
^{2}}\frac{1}{\hbox{Re}(S)}\frac{1}{\tau
_{4}^{3/2}\mathcal{V}} \\
\frac{\partial ^{2}\left( \delta K_{(g_{s})}^{KK}\right)
}{\partial \tau _{5}^{2}}\sim
\frac{1}{16\pi ^{2}}\frac{1}{\hbox{Re}(S)}\frac{1}{\tau _{5}^{3}}%
\end{array}%
\right.
\end{equation}%
which, upon double integration, matches exactly the scaling
behaviour of the result (\ref{hjkl}).

\subsubsection{Case 3: $%
\mathbb{C}
P_{[1,1,2,2,6]}^{4}(12)$} \label{sec3.2.3}

\bigskip
As another example we study the expected form of loop corrections
for the case of the Calabi-Yau manifold $\mathbb{C}
P_{[1,1,2,2,6]}^{4}(12)$, defined by the degree 12 hypersurface
embedding. This Calabi-Yau is a K3 fibration and has $(h^{1,1},
h^{2,1})=(2,128)$ with $\chi =-252$. Including only the complex
structure deformations that survive the mirror map, the defining
equation is:
\begin{equation}
z_{1}^{12}+z_{2}^{12}+z_{3}^{6}+z_{4}^{6}+z_{5}^{2}-12\psi
z_{1}z_{2}z_{3}z_{4}z_{5}-2\phi z_{1}^{6}z_{2}^{6}=0.
\end{equation}%
In terms of 2-cycle volumes the overall volume takes the form:
\begin{equation}
\mathcal{V}=t_{1}t_{2}^{2}+\frac{2}{3}t_{2}^{3},
\end{equation}%
giving relations between the 2- and 4-cycle volumes: \bea
\label{tayo} \tau _{1}=t_{2}^{2}, & \qquad &
\tau_{2}=2t_{2}\left(t_{1}+t_{2}\right),
\nonumber \\
t_{2}=\sqrt{\tau _{1}}, & \qquad & t_{1}=\frac{\tau _{2}-2\tau
_{1}}{2\sqrt{\tau _{1}}}, \eea allowing us to write:
\begin{equation}
\mathcal{V}=\frac{1}{2}\sqrt{\tau _{1}}\left( \tau
_{2}-\frac{2}{3}\tau _{1}\right) .  \label{vol11226}
\end{equation}%

Let us now investigate what the arguments above suggest for the
form of the
string loop corrections for the $%
\mathbb{C}
P_{[1,1,2,2,6]}^{4}$ model should look like. Applying (\ref{UUU})
and (\ref{UUUU}) for the one-loop correction to $K$, we find:
\begin{equation}
\delta K_{(g_{s})}^{KK}\sim
\frac{\mathcal{C}_{1}^{KK}}{\hbox{Re}\left(
S\right) \mathcal{V}}\frac{\tau _{2}-2\tau _{1}}{2\sqrt{\tau _{1}}}+\frac{%
\mathcal{C}_{2}^{KK}\sqrt{\tau _{1}}}{\hbox{Re}\left( S\right)
\mathcal{V}}, \label{ggg}
\end{equation}%
along with:
\begin{equation}
\delta K_{(g_{s})}^{W}\sim \frac{\mathcal{C}_{1}^{W}}{\mathcal{V}}\frac{2%
\sqrt{\tau _{1}}}{\tau _{2}-2\tau _{1}}+\frac{\mathcal{C}_{2}^{W}}{\mathcal{V%
}\sqrt{\tau _{1}}}.  \label{gggg}
\end{equation}%
The arguments summarised in the relation (\ref{fundamental})
reproduce exactly the behaviour of these corrections. The
tree-level K\"{a}hler metric reads:
\begin{equation}
\frac{\partial ^{2}\left( K_{tree}\right) }{\partial \tau _{1}^{2}}=\frac{1}{%
\tau _{1}^{2}}+\frac{2}{9}\frac{\tau _{1}}{\mathcal{V}^{2}},\textrm{ \ \ \ \ }%
\frac{\partial ^{2}\left( K_{tree}\right) }{\partial \tau _{2}^{2}}=\frac{1}{%
2}\frac{\tau _{1}}{\mathcal{V}^{2}}. \label{matrixElem}
\end{equation}%
Given that we are interested simply in the scaling behaviour of
these corrections, we notice that either in the case $\tau_{1}
\lesssim \tau_{2}$ such that:
\begin{equation}
\mathcal{V}=\frac{1}{2}\sqrt{\tau _{1}}\left( \tau
_{2}-\frac{2}{3}\tau _{1}\right) \simeq \tau _{1}^{3/2}\simeq \tau
_{2}^{3/2},  \label{ooo}
\end{equation}%
or in the large volume limit $\tau_{1} \ll \tau_{2}$ where:
\begin{equation}
\mathcal{V} \simeq \sqrt{\tau _{1}}\tau _{2},
\end{equation}
the matrix elements (\ref{matrixElem}) take the form:
\begin{equation}
\frac{\partial ^{2}\left( K_{tree}\right) }{\partial \tau
_{1}^{2}}\sim
\frac{1}{\tau _{1}^{2}},\ \ \ \ \frac{\partial ^{2}\left( K_{tree}\right) }{%
\partial \tau _{2}^{2}}\sim \frac{1}{\tau _{2}^{2}}.
\end{equation}%
We can now see that our method (\ref{fundamental})\ yields:

\begin{equation}
\left\{
\begin{array}{c}
\frac{\partial ^{2}\left( \delta K_{(g_{s})}^{KK}\right)
}{\partial \tau
_{1}^{2}}\sim \frac{1}{16\pi ^{2}}\frac{1}{\hbox{Re}(S)\tau _{1}}\frac{%
\partial ^{2}\left( K_{tree}\right) }{\partial \tau _{1}^{2}}%
\Longleftrightarrow \delta K_{(g_{s},\tau _{1})}^{KK}\sim \frac{1}{\hbox{Re}%
(S)\tau _{1}} \\
\frac{\partial ^{2}\left( \delta K_{(g_{s})}^{KK}\right)
}{\partial \tau _{2}^{2}}\sim \frac{1}{16\pi
^{2}}\frac{1}{\hbox{Re}(S)\tau _{2}}\frac{\partial ^{2}\left(
K_{tree}\right) }{\partial \tau _{2}^{2}}\Longleftrightarrow
\delta
K_{(g_{s},\tau _{2})}^{KK}\sim \frac{1}{\hbox{Re}(S)\tau _{2}}%
\end{array}%
\right.
\end{equation}%
which, both in the case $\tau_{1} \lesssim \tau_{2}$ and $\tau_{1}
\ll \tau_{2}$, matches the scaling behaviour of (\ref{ggg}):
\begin{equation}
\delta K_{(g_{s})}^{KK}\sim
\frac{\mathcal{C}_{1}^{KK}}{\hbox{Re}\left(
S\right) \mathcal{V}}\frac{\tau _{2}-2\tau _{1}}{2\sqrt{\tau _{1}}}+\frac{%
\mathcal{C}_{2}^{KK}\sqrt{\tau _{1}}}{\hbox{Re}\left( S\right) \mathcal{V}}%
\sim \frac{\mathcal{C}_{1}^{KK}}{\hbox{Re}\left( S\right) \tau _{1}}+\frac{%
\mathcal{C}_{2}^{KK}}{\hbox{Re}\left( S\right) \tau _{2}}.
\end{equation}

\section{Extended no-scale structure}
\label{sec4}

The examples in the previous section give support to the notion
that loop corrections to the K\"{a}hler potential can be
understood by requiring that the loop-corrected kinetic terms for
a modulus $\tau$ are suppressed by a factor of $g^2$ for the gauge
group on branes wrapping the $\tau$ cycle. We repeat again that
these arguments only apply to moduli that control loop factors.

While not proven, we now assume the validity of this parametric
form of the corrections and move on to study the effect of such
corrections in the scalar potential. We shall show that the
leading contribution to the scalar potential is null, due to a
cancellation in the expression for the scalar potential.
 We shall see that this cancellation holds so long as $\delta
K_{(g_{s})}^{KK}$ is an homogeneous function of degree $n=-2$ in
the 2-cycle volumes. We call this ``extended no scale structure",
as the cancellation in the scalar potential that is characteristic
of no-scale models extends to one further order, so that compared
to a naive expectation the scalar potential is only non-vanishing
at sub-sub-leading order. Let us state clearly the ``extended
no-scale structure" result:

\begin{quotation}
\textit{Let $X$ be a Calabi-Yau three-fold and consider type IIB
$N=1$ four dimensional supergravity where the K\"{a}hler potential
and the superpotential in the Einstein frame take the form:
\begin{equation}
\left\{
\begin{array}{l}
K=K_{tree}+\delta K,\\
W=W_{0}.
\end{array}%
\right.
\end{equation}%
If and only if the loop correction $\delta K$ to $K$ is a
homogeneous function in the 2-cycles volumes of degree $n=-2$,
then at leading order:}
\begin{equation}
\delta V_{(g_{s})}=0.
\end{equation}
\end{quotation}
We shall provide now a rigorous proof of the previous claim. We
are interested only in the perturbative part of the scalar
potential. We therefore focus on:
\begin{equation}
\delta V_{(g_{s})}=\left( K^{ij}\partial _{i}K\partial _{j}K-3\right) \frac{%
\left\vert W\right\vert ^{2}}{\mathcal{V}^{2}},  \label{pp}
\end{equation}%
where $K=-2\ln \left( \mathcal{V}\right) +\delta K_{(g_{s})}$. We
focus on $\delta K$ coming from $g_s$ (rather than $\alpha'$)
corrections. We require the inverse of the quantum corrected
K\"{a}hler matrix, which can be found using the Neumann series.
Introducing an expansion parameter $\varepsilon $, and writing
$K_{tree}$ as $K_{0}$, we define:
\begin{equation}
\mathcal{K}_{0}=\left\{ \frac{\partial ^{2}K_{0}}{\partial \tau
_{i}\partial
\tau _{j}}\right\} _{i,j=1,...,h_{1,1}},\textrm{ \ \ \ \ \ }\delta \mathcal{K}%
=\left\{ \frac{\partial ^{2}\left( \delta K_{(g_{s})}\right)
}{\partial \tau _{i}\partial \tau _{j}}\right\}
_{i,j=1,...,h_{1,1}}
\end{equation}%
and have:
\begin{equation}
K^{ij}=\left( \mathcal{K}_{0}+\varepsilon \delta
\mathcal{K}\right)
^{ij}=\left( \mathcal{K}_{0}\left( \mathbf{1}+\varepsilon \mathcal{K}%
_{0}^{-1}\delta \mathcal{K}\right) \right) ^{ij}=\left( \mathbf{1}%
+\varepsilon \mathcal{K}_{0}^{-1}\delta \mathcal{K}\right)
^{il}K_{0}^{lj}.
\end{equation}%
Now use the Neumann series:
\begin{equation}
\left( \mathbf{1}+\varepsilon \mathcal{K}_{0}^{-1}\delta
\mathcal{K}\right) ^{il}=\delta _{l}^{i}-\varepsilon
K_{0}^{im}\delta K_{ml}+\varepsilon ^{2}K_{0}^{im}\delta
K_{mp}K_{0}^{pq}\delta K_{ql}+\mathcal{O}(\varepsilon ^{3}),
\end{equation}%
to find:
\begin{equation}
K^{ij}=K_{0}^{ij}-\varepsilon K_{0}^{im}\delta
K_{ml}K_{0}^{lj}+\varepsilon
^{2}K_{0}^{im}\delta K_{mp}K_{0}^{pq}\delta K_{ql}K_{0}^{lj}+\mathcal{O}%
(\varepsilon ^{3}).  \label{ppp}
\end{equation}%
Substituting (\ref{ppp}) back in (\ref{pp}), we obtain:
\begin{equation}
\delta V_{(g_{s})}=V_{0}+\varepsilon \delta V_{1}+\varepsilon
^{2}\delta V_{2}+\mathcal{O}(\varepsilon ^{3}), \label{expansion
of V}
\end{equation}%
where $V_{0}=\left( K_{0}^{ij}K_{i}^{0}K_{j}^{0}-3\right)
\frac{\left\vert W\right\vert ^{2}}{\mathcal{V}^{2}}=0$ due to
(\ref{no scale}) is the usual no-scale structure and:
\begin{equation}
\left\{
\begin{array}{c}
\delta V_{1}=\left( 2K_{0}^{ij}K_{i}^{0}\delta
K_{j}-K_{0}^{im}\delta
K_{ml}K_{0}^{lj}K_{i}^{0}K_{j}^{0}\right) \frac{\left\vert W\right\vert ^{2}%
}{\mathcal{V}^{2}} \\
\delta V_{2}=\left( K_{0}^{ij}\delta K_{i}\delta
K_{j}-2K_{0}^{im}\delta
K_{ml}K_{0}^{lj}K_{i}^{0}\delta K_{j}\right.  \\
\left. +K_{0}^{im}\delta K_{mp}K_{0}^{pq}\delta
K_{ql}K_{0}^{lj}K_{i}^{0}K_{j}^{0}\right) \frac{\left\vert W\right\vert ^{2}%
}{\mathcal{V}^{2}}.%
\end{array}%
\right.   \label{pppp}
\end{equation}%
We caution the reader that (\ref{expansion of V}) is not a loop
expansion of the scalar potential but rather an expansion of the
scalar potential arising from the 1-loop quantum corrected
K\"{a}hler metric. The statement of extended no-scale structure is
that $\delta V_{1}$ will vanish, while
 $\delta V_{2}$ will be non-vanishing.
Recalling (\ref{p1}), $\delta V_{1}$\ simplifies to:
\begin{equation}
\delta V_{1}=-\left( 2\tau _{j}\frac{\partial \left( \delta K\right) }{%
\partial \tau _{j}}+\tau _{m}\tau _{l}\frac{\partial ^{2}\left( \delta
K\right) }{\partial \tau _{m}\partial \tau _{l}}\right)
\frac{\left\vert W\right\vert ^{2}}{\mathcal{V}^{2}}.
\end{equation}%
Let us make a change of coordinates and work with the 2-cycle
volumes instead of the 4-cycles. Using the second of the relations
(\ref{useful}), we deduce:
\begin{equation}
2\tau _{j}\frac{\partial }{\partial \tau _{j}}=t_{l}\frac{\partial }{%
\partial t_{l}},
\end{equation}%
and:
\begin{equation}
\tau _{m}\tau _{l}\frac{\partial ^{2}}{\partial \tau _{m}\partial \tau _{l}}=%
\frac{1}{4}t_{i}t_{k}\frac{\partial ^{2}}{\partial t_{i}\partial t_{k}}+%
\frac{1}{4}A_{li}t_{i}t_{k}\frac{\partial \left( A^{lp}\right)
}{\partial t_{k}}\frac{\partial }{\partial t_{p}}.
\end{equation}%
From the definition (\ref{aa}) of $A_{li}$, we notice that
$A_{li}$ is an homogeneous function of degree $n=1$ $\forall l,i$.
Inverting the matrix, we still get homogeneous matrix elements but
now of degree $n=-1$. Finally the Euler theorem for homogeneous
functions, tells us that:
\begin{equation}
t_{k}\frac{\partial \left( A^{lp}\right) }{\partial t_{k}}=\left(
-1\right) A^{lp},
\end{equation}%
which gives:
\begin{equation}
\tau _{m}\tau _{l}\frac{\partial ^{2}}{\partial \tau _{m}\partial \tau _{l}}=%
\frac{1}{4}t_{i}t_{k}\frac{\partial ^{2}}{\partial t_{i}\partial t_{k}}-%
\frac{1}{4}t_{p}\frac{\partial }{\partial t_{p}},
\end{equation}%
and, in turn:
\begin{equation}
\delta V_{1}=-\frac{1}{4}\left( 3t_{l}\frac{\partial \left( \delta
K\right)
}{\partial t_{l}}+t_{i}t_{k}\frac{\partial ^{2}\left( \delta K\right) }{%
\partial t_{i}\partial t_{k}}\right) \frac{\left\vert W\right\vert ^{2}}{%
\mathcal{V}^{2}}.
\end{equation}
The form of equation (\ref{UUU}) suggests that for arbitrary
Calabi-Yaus the string loop corrections to $K$ will be homogeneous
functions of the $2$-cycle volumes, and in particular that the
leading correction will be of degree $-2$ in $2$-cycle volumes.
Therefore if the degree of $\delta K$ is $n $, the Euler theorem
tells us that:
\begin{equation}
\delta V_{1}=-\frac{\left\vert W\right\vert ^{2}}{\mathcal{V}^{2}}\frac{1}{4}%
\left( 3n+n(n-1)\right) \delta K=-\frac{\left\vert W\right\vert ^{2}}{%
\mathcal{V}^{2}}\frac{1}{4}n(n+2)\delta K.  \label{ghj}
\end{equation}%
It follows then, as we claimed above, that only $n=-2$ implies
$\delta V_{1}=0$. In particular, from the conjectures (\ref{UUU})
and (\ref{UUUU}), we see that:
\begin{equation}
\left\{
\begin{array}{c}
n=-2\textrm{ \ for \ }\delta K_{(g_{s})}^{KK}, \\
n=-4\textrm{ \ for \ }\delta K_{(g_{s})}^{W},
\end{array}%
\right.
\end{equation}%
and so:
\begin{equation}
\left\{
\begin{array}{c}
\delta V_{(g_{s}),1}^{KK}=0, \\
\delta V_{(g_{s}),1}^{W}=-2\delta K_{(g_{s})}^{W}\frac{\left\vert
W\right\vert ^{2}}{\mathcal{V}^{2}}.%
\end{array}%
\right.   \label{finaly}
\end{equation}

\subsection{General formula for the effective scalar potential}
\label{sec4.1}

\bigskip Let us now work out the general formula for the effective scalar
potential evaluating also the first non-vanishing contribution of
$\delta K_{(g_{s})}^{KK}$, that is the $\varepsilon ^{2}$ terms
(\ref{pppp})\ in $V$:
\begin{eqnarray}
\delta V_{2} &=&\left( K_{0}^{ij}\delta K_{i}\delta
K_{j}-2K_{0}^{im}\delta
K_{ml}K_{0}^{lj}K_{i}^{0}\delta K_{j}\right.   \nonumber \\
&&\left. +K_{0}^{im}\delta K_{mp}K_{0}^{pq}\delta
K_{ql}K_{0}^{lj}K_{i}^{0}K_{j}^{0}\right) \frac{\left\vert W\right\vert ^{2}%
}{\mathcal{V}^{2}}.
\end{eqnarray}%
Using (\ref{p1}), $\delta V_{2}$\ simplifies to:
\begin{equation}
\delta V_{2}=\left( K_{0}^{ij}\delta K_{i}\delta K_{j}+2\tau
_{m}\delta K_{ml}K_{0}^{lj}\delta K_{j}+\tau _{m}\tau _{q}\delta
K_{ml}K_{0}^{lp}\delta K_{pq}\right) \frac{\left\vert W\right\vert
^{2}}{\mathcal{V}^{2}}. \label{zz}
\end{equation}%
We now stick to the case where $\delta K_{(g_{s})}^{KK}$ is given
by the conjecture (\ref{UUU}). Considering just the contribution
from one modulus (as the contributions from different terms are
independent), and dropping the dilatonic dependence, we have:
\begin{equation}
\delta K\rightarrow \delta K_{(g_{s}),\tau _{a}}^{KK}\sim \frac{\mathcal{C}%
_{a}^{KK}t_{a}}{\mathcal{V}}.  \label{uhu}
\end{equation}%
From (\ref{uhu}) we notice that:
\begin{eqnarray}
\delta K_{m} &=&A^{mj}\frac{\partial \left( \delta K\right) }{\partial t^{j}}%
=\mathcal{C}_{a}^{KK}A^{mj}\left( -\frac{t_{a}}{\mathcal{V}^{2}}\frac{%
\partial \left( \mathcal{V}\right) }{\partial t^{j}}+\frac{\delta _{aj}}{%
\mathcal{V}}\right)   \label{zzz} \\
&=&\mathcal{C}_{a}^{KK}\left( -\frac{1}{2}\frac{t_{a}t_{m}}{\mathcal{V}^{2}}+%
\frac{A^{am}}{\mathcal{V}}\right)
=-\mathcal{C}_{a}^{KK}K_{am}^{0},
\end{eqnarray}%
thus:
\begin{equation}
K_{0}^{ij}\delta K_{j}=-\mathcal{C}_{a}^{KK}K_{0}^{ij}K_{aj}^{0}=-\mathcal{C}%
_{a}^{KK}\delta _{ai}.
\end{equation}%
With this consideration (\ref{zz}) becomes:
\begin{equation}
\delta V_{2}=\left( -\mathcal{C}_{a}^{KK}\delta K_{a}-2\mathcal{C}%
_{a}^{KK}\tau _{m}\delta K_{ma}+\tau _{m}\tau _{q}\delta
K_{ml}K_{0}^{lp}\delta K_{pq}\right) \frac{\left\vert W\right\vert ^{2}}{%
\mathcal{V}^{2}}.
\end{equation}%
We need now to evaluate:
\begin{equation}
\tau _{m}\delta K_{ml}=\frac{1}{2}t_{p}\frac{\partial }{\partial t_{p}}%
\left( A^{li}\frac{\partial \left( \delta K\right) }{\partial t_{i}}\right) =%
\frac{1}{2}t_{p}\frac{\partial }{\partial t_{p}}\left( \delta
K_{l}\right) =-2\delta K_{l},
\end{equation}%
that yields:
\begin{eqnarray}
\delta V_{2} &=&\left( -\mathcal{C}_{a}^{KK}\delta K_{a}+4\mathcal{C}%
_{a}^{KK}\delta K_{a}+4\delta K_{l}K_{0}^{lp}\delta K_{p}\right) \frac{%
\left\vert W\right\vert ^{2}}{\mathcal{V}^{2}}= \\
&=&\left( -\mathcal{C}_{a}^{KK}\delta
K_{a}+4\mathcal{C}_{a}^{KK}\delta
K_{a}-4\mathcal{C}_{a}^{KK}\delta K_{a}\right) \frac{\left\vert
W\right\vert
^{2}}{\mathcal{V}^{2}} \\
&=&-\mathcal{C}_{a}^{KK}\delta K_{a}\frac{\left\vert W\right\vert ^{2}}{%
\mathcal{V}^{2}}.
\end{eqnarray}%
With the help of the relation (\ref{zzz}) and replacing the
dilatonic dependence, we can write the previous expression in
terms of the tree-level K\"{a}hler metric:
\begin{equation}
\delta V_{2}=\frac{\left( \mathcal{C}_{a}^{KK}\right) ^{2}}{\hbox{Re}(S)^{2}}%
K_{aa}^{0}\frac{\left\vert W\right\vert ^{2}}{\mathcal{V}^{2}}.
\label{zzzz}
\end{equation}%
Putting together (\ref{finaly}) and (\ref{zzzz}), we can now
write the quantum correction to the scalar potential at leading
order at 1 loop for general Calabi-Yaus, in terms of the cycles
$i$ wrapped by the branes and the quantum corrections to the
K\"ahler potential:
\begin{equation}
\frame{$\delta V_{\left( g_{s}\right)
}^{1-loop}=\sum\limits_{i}\left( \frac{\left( \mathcal{C}%
_{i}^{KK}\right) ^{2}}{\hbox{Re}(S)^{2}}K_{ii}^{0}-2\delta
K_{(g_{s}),\tau _{i}}^{W}\right)
\frac{W_{0}^{2}}{\mathcal{V}^{2}}$}.  \label{1 loop}
\end{equation}
We emphasise that this formula assumes the validity of the BHP
conjecture, and only focuses on corrections of this nature.
Moreover, we considered branes wrapped only around the basis
4-cycles. If this were not the case, we should replace the first
term $K^0_{i\bar\imath}$ with the more general combination
$K^0_{i\bar\imath}\to a_{ik} a_{ij} K^0_{k\bar\jmath}$.

Finally we point out that, due to the extended no-scale structure,
in the presence of non-perturbative contributions to the
superpotential, it is also important to check that the leading
quantum corrections to the general scalar potential (\ref{scalar})
are indeed given by (\ref{1 loop}) and the contribution to the
non-perturbative part of the scalar potential generated by string
loop corrections
\begin{equation}
\delta V_{(np)}=\left(2 K_{0}^{ij}W _{i}\delta K_{(g_{s}),j} W +
\delta
K_{(g_{s})}^{ij}W _{i}W _{j} \right) \frac{%
\left\vert W\right\vert ^{2}}{\mathcal{V}^{2}},  \label{dVnp}
\end{equation}
is irrelevant. A quick calculation shows that this is indeed the
case.\footnote{We shall not discuss the effects of higher loop
contributions to the scalar potential. We expect that these will
be suppressed compared to the one-loop contribution by additional
loop factors of $(16 \pi^2)$, and so will not compete with the
terms considered in (\ref{1 loop}).}

\subsection{Field theory interpretation}

We now interpret the above arguments and in particular the
existence of the extended no-scale structure in light of the
Coleman-Weinberg potential \cite{cw}.\footnote{For a previous
attempt at matching string effective actions onto the
Coleman-Weinberg potential, see \cite{cwmatching}.}
 We will see that this gives
a quantitative explanation for the cancellation that is present.
The Coleman-Weinberg potential is given in supergravity by (e.g.
see \cite{fkz, ckn}):
\begin{equation}
\delta V_{1-loop}=\frac{1}{64\pi ^{2}}\left[ \Lambda ^{4}
STr\left( M^{0}\right) \ln \left( \frac{\Lambda ^{2}}{\mu
^{2}}\right) +2\Lambda ^{2} STr\left( M^{2}\right) + STr \left(
M^{4}\ln \left( \frac{M^{2}}{\Lambda ^{2}}\right) \right) \right]
, \label{Coleman}
\end{equation}%
where $\mu $ is a scale parameter, $\Lambda$ the cut-off scale
and:
\begin{equation}
 STr \left( M^{n}\right) \equiv \sum_{i}\left( -1\right) ^{2j_{i}}\left(
2j_{i}+1\right) m_{i}^{n},
\end{equation}%
is the supertrace, written in terms of the the spin of the
different particles $j_{i}$ and the field-dependent mass
eigenvalues $m_{i}$.

The form of (\ref{Coleman}) gives a field theory interpretation to
the scalar potential found in section \ref{sec4.1}. Let us try and
match the 1-loop expression with the potential (\ref{Coleman})
interpreting the various terms in the Coleman-Weinberg potential
as different terms in the $\epsilon$ expansion in (\ref{expansion
of V}). We first notice that in any spontaneously broken
supergravity theory, $STr\left( M^{0}\right) =0$, as the number of
bosonic and fermionic degrees of freedom must be equal. The
leading term in (\ref{Coleman}) is therefore null.

We recall that due to the extended no-scale structure the
coefficient of the $\mc{O}(\epsilon)$ term in (\ref{expansion of
V}) is also vanishing. Our comparison should therefore involve the
leading non-zero terms in both cases. In the following paragraphs,
we will re-analyse the three examples studied in section
\ref{sec3} and show how we always get a matching. This gives a
nice physical understanding of this cancellation at leading order
in $\delta V^{KK}_{(g_{s}),1-loop}$ which is due just to
supersymmetry: the cancellation must take place if the resulting
1-loop potential is to match onto the Coleman-Weinberg form.
Supersymmetry causes the vanishing of the first term in
(\ref{Coleman}) and we notice, for each example, that the second
term in (\ref{Coleman}) scales as the $\mc{O}(\epsilon^2)$ term in
(\ref{expansion of V}), therefore, in order to match the two
results, the $\mc{O}(\epsilon)$ term in (\ref{expansion of V})
also has to be zero. This is, in fact, what the extended no-scale
structure guarantees.

We note here that both with the use of the supergravity expression
for the Coleman-Weinberg formula and for the earlier discussions
of section 6.1, supersymmetry has played a crucial r\^{o}le. In
the Coleman-Weinberg formula, the presence of low-energy
supersymmetry is used to evaluate the supertraces and to relate
these to the gravitino mass. In the discussion of kinetic terms,
the fact that the corrections are written as corrections to the
K\"ahler metric automatically implies that the structure of
low-energy supersymmetry is respected.

\subsubsection{Case 1: $N=1$ $T^{6}/(%
\mathbb{Z}
_{2}\times
\mathbb{Z}
_{2})$} \label{sec4.2.1}

The case of the $N=1$ toroidal orientifold background was studied
in sections \ref{sec3.1.1} and \ref{sec3.2.1}. We here treat all
three moduli on equal footing, reducing the volume form
(\ref{torus}) to the one-modulus case:
\begin{equation}
\mathcal{V}=\tau ^{3/2} = \left(\frac{T+\bar{T}}{2}\right)^{3/2}.
\end{equation}
We therefore take:
\begin{equation}
\left\langle \tau _{1}\right\rangle \simeq \left\langle \tau
_{2}\right\rangle \simeq \left\langle \tau _{3}\right\rangle .
\end{equation}%
We write out very explicitly the correction to the scalar
potential due to the correction to the K\"ahler potential as
computed in \cite{bhk}. We focus only on the K\"ahler moduli
dependence. The tree level K\"ahler potential is:
$$
K = -3 \ln (T + \bar{T})
$$
and the loop-corrected K\"ahler potential has the form:
$$
K = - 3 \ln (T + \bar{T}) + \frac{\epsilon}{(T + \bar{T})}.
$$
The scalar potential is:
$$
V = M_P^4 e^K \left( K^{i \bar{j}} \partial_i K \partial_{\bar{j}}
K - 3 \right) \vert W \vert^2.
$$
Evaluated, this gives: \bea
V & = & \frac{M_P^4}{(T + \bar{T})^3} \left( 0 + \frac{0 \ti \mc{O}(\epsilon)}
{T + \bar{T}} + \frac{\mc{O}(\epsilon^2)}{(T + \bar{T})^2} \right) \nonumber \\
& = & \frac{M_P^4 \epsilon^2}{(T + \bar{T})^5} \sim \frac{M_P^4
\epsilon^2}{\mc{V}^{10/3}}. \label{potsimp} \eea The cancellation
of the $\mc{O}(T + \bar{T})^{-3}$ term in (\ref{potsimp}) is due
to the original no-scale structure. The cancellation of the
$\mc{O}(T + \bar{T})^{-4}$ term in (\ref{potsimp}) is due to the
extended no-scale structure that is satisfied by the loop
corrected K\"ahler potential, giving a leading contribution at
$\mc{O}(T + \bar{T})^{-5}$. This gives the behaviour of the
leading contribution to the scalar potential, which we want to
compare with the Coleman-Weinberg expression.

To compare with (\ref{Coleman}) we recall that in supergravity the
supertrace is proportional to the gravitino mass:
\begin{equation}
STr\left( M^{2}\right) \simeq m_{3/2}^{2}.
\end{equation}%
The dependence of the gravitino mass on the volume is always given
by:
\begin{equation}
m_{3/2}^{2}=e^{K}W_{0}^{2}\simeq \frac{1}{\mathcal{V}^{2}}\textrm{ }%
\Longrightarrow \textrm{ } STr\left( M^{2}\right) \simeq \frac{1}{\mathcal{V}%
^{2}}.  \label{STr}
\end{equation}%
We must also understand the scaling behaviour of the cut-off
$\Lambda $. $\Lambda$ should be identified with the energy scale
above which the four-dimensional effective field theory breaks
down. This is the compactification scale at which many new KK
states appear, and so is given by:
\begin{equation}
\Lambda =m_{KK}\simeq \frac{M_{s}}{R}=\frac{M_{s}}{\tau ^{1/4}}=\frac{1}{%
\tau ^{1/4}}\frac{1}{\sqrt{\mathcal{V}}}M_{P}=\frac{M_{P}}{\mathcal{V}^{2/3}}%
.
\end{equation}%
In units of the Planck mass, (\ref{Coleman}) therefore scales as:
\begin{eqnarray}
\label{fqx} \delta V_{1-loop} &\simeq &0\cdot \Lambda ^{4}+\Lambda
^{2}STr\left( M^{2}\right) +STr\left( M^{4}\ln \left(
\frac{M^{2}}{\Lambda ^{2}}\right)
\right) \simeq   \nonumber \\
&\simeq &0\cdot \frac{1}{\mathcal{V}^{8/3}}+\frac{1}{\mathcal{V}^{10/3}}+%
\frac{1}{\mathcal{V}^{4}},
\end{eqnarray}%
in agreement with (\ref{potsimp}).

\subsubsection{Case 2: $%
\mathbb{C}
P_{[1,1,1,6,9]}^{4}(18)$} \label{sec4.2.2}

This case, studied in section \ref{sec3.2.2}, is more involved, as
it includes two K\"{a}hler moduli, the large modulus
 $\tau _{b}\simeq \mathcal{V}^{2/3}$ and the small modulus $\tau _{s}$. The
effective potential gets contributions from loop corrections for
both moduli and in these two cases, (\ref{pp}) takes the form (the
dilaton is considered fixed and its dependence is reabsorbed in
$\mathcal{C}^{KK}_{b}$ and $\mathcal{C}^{KK}_{s}$):

\begin{enumerate}
\item Big modulus%
\begin{eqnarray}
\delta V_{\left( g_{s}\right) ,1-loop}^{KK} &=&\left( 0\cdot \frac{\mathcal{C}%
^{KK}_{b}}{\tau _{b}}+\frac{\alpha _{2,b}\left(
\mathcal{C}^{KK}_{b}\right) ^{2}}{\tau
_{b}^{2}}+\frac{\alpha _{3,b}\left( \mathcal{C}^{KK}_{b}\right) ^{3}}{\tau _{b}^{3}%
}+\mathcal{O}\left( \frac{\partial ^{4}K_{0}}{\partial \tau
_{b}^{4}}\right)
\right) \frac{W_{0}^{2}}{\mathcal{V}^{2}}\textrm{ }  \nonumber \\
&\simeq &\left( 0\cdot \frac{\mathcal{C}^{KK}_{b}}{\mathcal{V}^{8/3}}+\frac{%
\alpha _{2,b}\left( \mathcal{C}^{KK}_{b}\right) ^{2}}{\mathcal{V}^{10/3}}+\frac{%
\alpha _{3,b}\left( \mathcal{C}^{KK}_{b}\right)
^{3}}{\mathcal{V}^{4}}\right) W_{0}^{2}.  \label{big1}
\end{eqnarray}

\item Small modulus%
\begin{equation}
\delta V_{\left( g_{s}\right) ,1-loop}^{KK}=\left( 0\cdot
\mathcal{C}^{KK}_{s} \frac{\sqrt{\tau
_{s}}}{\mathcal{V}^{3}}+\frac{\alpha _{2,s}\left( \mathcal{C}_{s}
^{KK}\right) ^{2}}{\mathcal{V}^{3}\sqrt{\tau _{s}}}+\frac{\alpha
_{3,s}\left(\mathcal{C}^{KK}_{s}\right) ^{3}}{\mathcal{V}^{3}\tau _{s}^{3/2}}+\mathcal{O}%
\left( \mathcal{V}^{-2}\frac{\partial ^{4}K_{0}}{\partial \tau
_{s}^{4}}\right) \right) W_{0}^{2}.  \label{small1}
\end{equation}
\end{enumerate}

In the Coleman-Weinberg potential, the supertrace has the same
scaling $\sim \mc{V}^{-2}$ as in (\ref{STr}), but there now exist
different values of the cut-off $\Lambda$ for the field theories
living on branes wrapping the big and small 4-cycles:

\medskip

\begin{equation}
\left\{
\begin{array}{c}
\Lambda _{b}=m_{KK,b}\simeq \frac{1}{\tau _{b}^{1/4}}\frac{1}{\sqrt{\mathcal{%
V}}}M_{P}=\frac{M_{P}}{\mathcal{V}^{2/3}}, \\
\Lambda _{s}=m_{KK,s}\simeq \frac{1}{\tau _{s}^{1/4}}\frac{1}{\sqrt{\mathcal{%
V}}}M_{P}.%
\end{array}%
\right.   \label{KK mass scale}
\end{equation}

%
\begin{figure}[ht]
\begin{center}
\epsfig{file=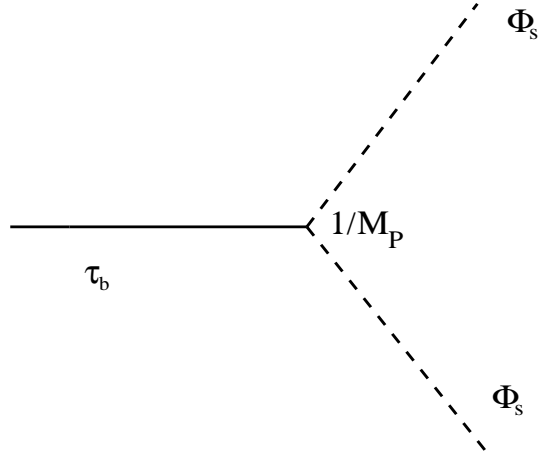, height=60mm,width=70mm} \caption{
 Coupling of the big modulus KK modes to a generic field
  $\Phi_s$ living on the brane wrapping the small 4-cycle.}
\label{diag2}
\end{center}
\end{figure}

The existence of two cut-off scales requires some explanation. At
first glance, as $\Lambda _{b}<\Lambda _{s}$ and the KK modes of
the big K\"{a}hler modulus couple to the field theory on the brane
wrapping the small 4-cycle, one might think that there is just one
value of the cut-off $\Lambda$, which is given by $\Lambda_{b}$ =
$m_{KK,b}$. This corresponds to the mass scale of the lowest
Kaluza-Klein mode present in the theory. For a field theory living
on a brane wrapping the large cycle, this represent the mass scale
of Kaluza-Klein replicas of the gauge bosons and matter fields of
the theory. However, we do not think this is the correct
interpretation for a field theory living on the small cycle. The
bulk Kaluza-Klein modes are indeed lighter than those associated
with the small cycle itself.

%
\begin{figure}[ht]
\begin{center}
\epsfig{file=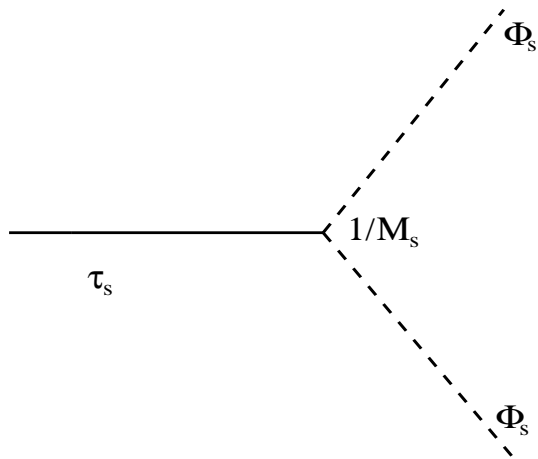, height=60mm,width=70mm} \caption{
 Coupling of the small modulus KK modes to a generic field
  $\Phi_s$ living on the brane wrapping the small 4-cycle.}
\label{setup}
\end{center}
\end{figure}

However it is also the case that the bulk modes couple extremely
weakly to this field theory compared to the local modes. The bulk
modes only couple gravitationally to this field theory, whereas
the local modes couple at the string scale \cite{CQ}. In the case
that the volume is extremely large, this difference is
significant. For a field theory on the small cycle, the cutoff
should be the scale at which KK replicas of the quarks and gluons
appear, rather than the scale at which new very weakly coupled
bulk modes are present. As the local modes are far more strongly
coupled, it is these modes that determine the scale of the UV
cutoff. This is illustrated in Figure 5.2 and 5.3.\footnote{Notice
that the cut-off dependence of the $STr(M^2)$ term could
potentially be dangerous for the stability of the magnitude of
soft terms computed for this model in references \cite{SoftSUSY,
LVSatLHC}. With our analysis here it is easy to see that the
contribution of this term to the scalar potential and then to the
structure of soft breaking terms is suppressed by inverse powers
of the volume and is therefore harmless.}

\bigskip

We now move on to make the matching
of (\ref{big1}) and (\ref{small1}) with the Coleman-Weinberg potential (\ref%
{Coleman}). For the big modulus, we find:
\begin{eqnarray}
\delta V_{1-loop} &\simeq &0\cdot \Lambda _{b}^{4}+\Lambda
_{b}^{2}STr\left(
M^{2}\right) +STr\left( M^{4}\ln \left( \frac{M^{2}}{\Lambda _{b}^{2}}%
\right) \right) \simeq   \nonumber \\
&\simeq &0\cdot \frac{1}{\mathcal{V}^{8/3}}+\frac{1}{\mathcal{V}^{10/3}}+%
\frac{1}{\mathcal{V}^{4}},
\end{eqnarray}%
which yields again a scaling matching that of (\ref{big1}). For
the small modulus we obtain, proceeding as in the previous case:
\begin{eqnarray}
\delta V_{1-loop} &\simeq &0\cdot \Lambda _{s}^{4}+\Lambda
_{s}^{2}STr\left(
M^{2}\right) +STr\left( M^{4}\ln \left( \frac{M^{2}}{\Lambda _{s}^{2}}%
\right) \right) \simeq   \nonumber \\
&\simeq &0\cdot \frac{1}{\tau _{s}}\frac{1}{\mathcal{V}^{2}}+\frac{1}{\sqrt{%
\tau _{s}}}\frac{1}{\mathcal{V}^{3}}+\frac{1}{\mathcal{V}^{4}},
\label{jklm}
\end{eqnarray}%
where we have a matching only of the second term of (\ref{jklm})
with the second term of (\ref{small1}). This is indeed the term
which we expect to match, given that is the first non-vanishing
leading contribution to the effective scalar potential at 1-loop.
There is no reason the first terms need to match as they have
vanishing coefficients.

As an aside, we finally note that the third term in (\ref{small1})
can also match with the Coleman-Weinberg effective potential,
although we should not try to match this with the third term in
(\ref{Coleman}) but with a subleading term in the expansion of the
second term in (\ref{Coleman}). This is due to the fact that we do
not have full control on the expression for the Kaluza-Klein scale
(\ref{KK mass scale}). In the presence of fluxes, this is more
reasonably given by (for example see the discussion in appendix D
of \cite{bhp}):
\begin{eqnarray}
\Lambda _{s} &=&m_{KK,s}\simeq \frac{1}{\tau _{s}^{1/4}}\frac{M_{P}}{\sqrt{%
\mathcal{V}}}\left( 1+\frac{1}{\tau _{s}}+...\right)
=\frac{1}{\tau
_{s}^{1/4}}\frac{M_{P}}{\sqrt{\mathcal{V}}}+\frac{1}{\tau _{s}^{5/4}}\frac{%
M_{P}}{\sqrt{\mathcal{V}}}+...  \nonumber \\
&\Longrightarrow &\Lambda _{s}^{2}\simeq \frac{1}{\tau
_{s}^{1/2}}\frac{M_{P}^{2}%
}{\mathcal{V}}+\frac{2}{\tau
_{s}^{3/2}}\frac{M_{P}^{2}}{\mathcal{V}}.
\end{eqnarray}%
This, in turn, produces:
\begin{equation}
\Lambda _{s}^{2}STr\left( M^{2}\right) \simeq \frac{1}{\tau _{s}^{1/2}}\frac{%
1}{\mathcal{V}^{3}}+\frac{2}{\tau
_{s}^{3/2}}\frac{1}{\mathcal{V}^{3}}. \label{bnbn}
\end{equation}%
In this case the second term in (\ref{bnbn}) reproduces the
scaling behaviour of the third term in (\ref{small1}).

\subsubsection{Case 3: $%
\mathbb{C}
P_{[1,1,2,2,6]}^{4}(12)$}

In section \ref{sec3.2.3} we have seen that there are two regimes
where the case of the K3 fibration with two K\"{a}hler moduli can
be studied. When the VEVs of the two moduli are of the same order
of magnitude, they can be treated on equal footing and the volume
form (\ref{vol11226}) reduces to the classical one parameter
example which, as we have just seen in section \ref{sec4.2.1},
gives also the scaling behaviour of the toroidal orientifold case.
We do not need therefore to repeat the same analysis and we
automatically know that the scaling of our general result for the
effective scalar potential at 1-loop matches exactly the
Coleman-Weinberg formula also in this case.

The second situation when $\tau_{2} \gg \tau_{1}$ is more
interesting. The relations (\ref{tayo}) tell us that the large
volume limit $\tau_{2} \gg \tau_{1}$ is equivalent to $t_{1} \gg
t_{2}$ and thus they reduce to:
\begin{equation}
\tau _{1}=t_{2}^{2}\textrm{, \ \ \ \ \ }\tau _{2} \simeq
2t_{2}t_{1}, \qquad \mc{V} \simeq \half \sqrt{\tau_1} \tau_2
\simeq t_1 t_2^{2}. \label{stay}
\end{equation}
The KK scale of the compactification is then set by the large
2-cycle $t_1$: \be m_{KK} \sim \frac{M_s}{\sqrt{t_1}} \sim
\frac{M_P}{t_1 t_2}, \ee while in the large volume limit the
gravitino mass is: \be m_{3/2} \sim \frac{M_P}{\mc{V}} \sim
\frac{M_P}{t_1 t_2^2}. \ee The bulk KK scale is therefore
comparable to that of the gravitino mass, and it is not clear that
this limit can be described in the language of four-dimensional
supergravity. Let us nonetheless explore the consequences of using
the same analysis as in the previous sections. The evaluation of
(\ref{expansion of V}) gives (reabsorbing the VEV of the dilaton
in $\mathcal{C}^{KK}_{1}$ and $\mathcal{C}^{KK}_{2}$):

\begin{enumerate}
\item Small modulus $\tau_{1}$%
\begin{eqnarray}
\delta V_{\left( g_{s}\right) ,1-loop}^{KK}&\simeq&\left( 0\cdot
\frac{\mathcal{C}^{KK}_{1}}{\tau_{1}\mathcal{V}^{2}}+\frac{\alpha
_{2,1}\left( \mathcal{C} ^{KK}_{1}\right)
^{2}}{\tau_{1}^{2}\mathcal{V}^{2}}+\frac{\alpha _{3,1}\left(
\mathcal{C}^{KK}_{1}\right) ^{3}}{\tau_{1}^{3}\mathcal{V}^{2}}
\right) W_{0}^{2}.  \label{SMALL}
\end{eqnarray}

\item Big modulus $\tau_{2}$%
\begin{equation}
\delta V_{\left( g_{s}\right) ,1-loop}^{KK} \simeq \left( 0\cdot
\mc{C}^{KK}_{2}\frac{\sqrt{\tau_{1}}}{\mathcal{V}^{3}}+ \alpha
_{2,2}\left( \mathcal{C}^{KK}_{2}\right)
^{2}\frac{\tau_{1}}{\mathcal{V}^{4}}+ \alpha _{3,2}\left(
\mathcal{C}^{KK}_{2}\right)
^{3}\frac{\tau_{1}^{3/2}}{\mathcal{V}^{5}}\right) W_{0}^{2}.
\label{BIG}
\end{equation}
\end{enumerate}
Let us now derive the two different values of the cut-off
$\Lambda$ for the field theories living on branes wrapping the big
and small 4-cycles. We realise that the Kaluza-Klein radii for the
two field theories on $\tau_{1}$ and $\tau_{2}$ are given by:
\begin{equation}
\left\{
\begin{array}{c}
R _{1}\simeq \sqrt{t_{2}}, \\
R_{2}\simeq \sqrt{t_{1}},
\end{array}%
\right.   \label{KK radii}
\end{equation}
and consequently:
\begin{equation}
\left\{
\begin{array}{c}
\Lambda _{1}=m_{KK,1}\simeq \frac{M_{s}}{\sqrt{t_{2}}} \simeq
\frac{1}{\tau_{1}^{1/4}\sqrt{\mathcal{V}}} M_{P},\\
\Lambda _{2}=m_{KK,2}\simeq \frac{M_{s}}{\sqrt{t_{1}}} \simeq
\frac{\sqrt{\tau_{1}}}{\mathcal{V}} M_{P}.
\end{array}%
\right.   \label{KKMassScales}
\end{equation}
We note that $m_{KK,2}$ coincides with the scale of the lightest
KK modes $m_{KK}$. If we try to match the result (\ref{SMALL}) for
the small cycle with the corresponding Coleman-Weinberg potential
for the field theory on $\tau_{1}$:
\begin{eqnarray}
\delta V_{1-loop} &\simeq &0\cdot \Lambda _{1}^{4}+\Lambda
_{1}^{2} \hbox{STr}\left(
M^{2}\right) + \hbox{STr}\left( M^{4}\ln \left( \frac{M^{2}}{\Lambda _{1}^{2}}%
\right) \right) \simeq   \nonumber \\
&\simeq &0\cdot \frac{1}{\tau_{1}^2\mathcal{V}^{2}}+\frac{1}{\sqrt{\tau_{1}}\mathcal{V}^{3}}+%
\frac{1}{\mathcal{V}^{4}},
\end{eqnarray}%
we do not find any agreement. This is not surprising since
effective field theory arguments only make sense when:
\begin{equation}
\delta V_{\left( g_{s}\right) ,1-loop}^{KK} \ll m_{KK}^{4},
\label{condition}
\end{equation}
but this condition is not satisfied in our case. In fact, using
the mass of the lowest KK mode present in the theory, we have:
\begin{equation}
m_{KK}^4 = m_{KK,2}^{4}\simeq\ \frac{\tau_{1}^2}{\mathcal{V}^{4}}
\ll \frac{1}{\tau_{1}^{2}\mathcal{V}^{2}}\simeq\delta V_{\left(
g_{s}\right) ,1-loop}^{KK}.  \label{mkk24}
\end{equation}
Energy densities couple universally through gravity, and so this
implies an excitation of Kaluza-Klein modes, taking us
 beyond the regime of validity of effective field theory.
Thus in this limit the use of the four-dimensional supergravity
action with loop corrections to compute the effective potential
does not seem trustworthy, as it gives an energy density much
larger than $m_{KK}^4$.

For the field theory on the large cycle $\tau_{2}$ the
Coleman-Weinberg potential gives:
\begin{eqnarray}
\delta V_{1-loop} &\simeq &0\cdot \Lambda _{2}^{4}+\Lambda
_{2}^{2}STr\left( M^{2}\right) +STr\left( M^{4}\ln \left(
\frac{M^{2}}{\Lambda _{2}^{2}}
\right) \right) \simeq   \nonumber \\
&\simeq &0\cdot
\frac{\tau_{1}^2}{\mathcal{V}^{4}}+\frac{\tau_{1}}{\mathcal{V}^{4}}+
\frac{1}{\mathcal{V}^{4}}. \label{fff}
\end{eqnarray}
In this case the energy density given by the loop corrections
(\ref{BIG}) is (marginally) less than $m_{KK}^4 \simeq\
\tau_{1}^2\mathcal{V}^{-4}$, being smaller by a factor of
$\tau_1$. Equation (\ref{fff}) then matches the result (\ref{BIG})
at leading order.

Again, we also note as an aside that if we expand the KK scale as
in in section \ref{sec4.2.2}, then we obtain:
\begin{eqnarray}
\Lambda _{2} &=&m_{KK,2}\simeq
\frac{\sqrt{\tau_{1}}}{\mathcal{V}}\left( 1+\frac{1}{\tau
_{2}}+...\right)M_{P} \simeq
\left(\frac{\sqrt{\tau_{1}}}{\mathcal{V}}+\frac{\tau_{1}}{\mathcal{V}^{2}}\right)M_{P}
\nonumber \\
&\Longrightarrow &\Lambda _{2}^{2}\simeq
\left(\frac{\tau_{1}}{\mathcal{V}^{2}}+\frac{\tau_{1}^{3/2}}{\mathcal{V}^{3}}\right)M_{P}^{2}.
\end{eqnarray}
This, in turn, produces:
\begin{equation}
\Lambda _{2}^{2} STr\left( M^{2}\right) \simeq
\frac{\tau_{1}}{\mathcal{V}^{4}}+\frac{\tau
_{1}^{3/2}}{\mathcal{V}^{5}}. \label{bnbnyu}
\end{equation}
In this case the second term in (\ref{bnbnyu}) also reproduces the
scaling behaviour of the third term in (\ref{BIG}).

\begin{center}
---------------------------------------------------------
\end{center}

The purpose of this chapter has been to study, as far as possible,
the form of loop corrections to the K\"ahler potential for general
Calabi-Yau compactifications and their effect on the scalar
potential. The aim has been to extract the parametric dependence
on the moduli that control the loop expansion. We have contributed
to put the proposed form of leading order string loop corrections
on firmer grounds in the sense that they agree with the low-energy
effective action behaviour. In particular, it is reassuring that
the Coleman-Weinberg formula for the scalar potential fits well
with that arising from the BHP conjecture for the corrections to
the K\"ahler potential. Furthermore, the non-contribution of the
leading order string loop correction is no longer an accident but
it is just a manifestation of the underlying supersymmetry with
equal number of bosons and fermions, despite being spontaneously
broken.

These results are important for K\"{a}hler moduli stabilisation.
In particular, even though the string loop corrections to the
K\"{a}hler potential are subdominant with respect to the leading
order $\alpha'$ contribution, they can be more important than
non-perturbative superpotential corrections to stabilise non
blow-up moduli. The general picture is that all corrections -
$\alpha'$, loop and non-perturbative - play a r\^{o}le in a
generic Calabi-Yau compactification. We will discuss these matters
in more detail in chapter 6.

\chapter{String Loop Moduli Stabilisation}
\label{StringLoopModuliStabilisation} \linespread{1.3}

Given that, in the case of the LARGE Volume Scenario (LVS), the
$\alpha'$ corrections can play a significant r\^{o}le in moduli
stabilisation in the phenomenologically relevant regime of large
volume and weak coupling, it is natural to wonder whether $g_s$
corrections may also have a significant effect. This is the main
topic of this chapter. At first sight this seems unavoidable, as
at large volume the corrections to the K\"{a}hler potential
induced by string loops are parametrically larger than those
induced by $\alpha'$ corrections \cite{bhk}. However, as we have
seen in chapter 5, the scalar potential exhibits an extended
no-scale structure, and the loop corrections contribute to the
scalar potential at a level subleading to their contribution to
the K\"{a}hler potential, and subleading to the $\alpha'$
corrections \cite{bhk, hg}. \cite{bhp} studied the effect of loop
corrections on the $\mathbb{C}P^4_{[1,1,1,6,9]}$ large volume
model and found it only gave minor corrections to the moduli
stabilisation and sub-sub-leading corrections to the soft term
computation. It is then natural to ask whether loop corrections to
the scalar potential can give a qualitative, rather than only
quantitative, change to moduli stabilisation.

We study this question in this chapter and find that the answer is
affirmative. The LARGE Volume Scenario stabilises the overall
volume at an exponentially large value using $\alpha'$ and $g_s$
corrections. Most previous work has focused on `Swiss-Cheese'
Calabi-Yaus, where one cycle controls the overall volume (`size of
the cheese') and the remaining moduli control the volume of
blow-up cycles (`holes in the cheese'). However for Calabi-Yaus
with a fibration structure - the torus is the simplest example -
multiple moduli enter into the overall volume. For the overall
volume to be made large in a homogeneous fashion, several moduli
must become large. In these cases, while the existence of at least
one blow-up mode is still necessary, loop corrections turn out to
be necessary in order to realise the LVS and obtain a stable
minimum at exponentially large volume. The loop corrections lift
directions transverse to the overall breathing mode and stabilise
these.

More precisely, the general picture is that in order to find LVS
we need at least one blow-up mode which resolves point-like
singularities. Calabi-Yau orientifolds should generically have
this property, and so many compactifications should present a
non-supersymmetric minimum at exponentially large volume. All
these blow-up modes are stabilised small whereas the overall
volume is fixed exponentially large due to $\alpha'$ and
non-perturbative corrections as in the
$\mathbb{C}P^{4}_{[1,1,1,6,9]}$ model. However the potential for
all the non blow-up moduli, except the overall volume, still
remains flat, but it will be naturally lifted by string loop
corrections. This claim is illustrated with a detailed explicit
calculation for a K3 fibration with 3 K\"{a}hler moduli. In Part
III of this thesis we shall present some interesting cosmological
implications of this LVS for K3-fibered Calabi-Yau manifolds.

We will also show that the string loop corrections may play an
important r\^{o}le in addressing the problem stressed in
\cite{blumenhagen}. The authors there argued that the $4$-cycle on
which the Standard Model lives, cannot get non-perturbative
corrections since their prefactor is proportional to the VEV of
Standard Model fields which, at this stage, is required to vanish.
However, due to the constraints of the Standard Model gauge
couplings, this cycle must still be stabilised at a relatively
small size.

This problem may be cured through having at least two blow-up
modes and then adding $g_{s}$ corrections. The loop corrections
have the ability to stabilise the Standard Model cycle, while the
`transverse' cycle is stabilised non-perturbatively as usual. This
possible solution is discussed for the example
$\mathbb{C}P^{4}_{[1,3,3,3,5]}(15)$, studied in detail in
subsection 6.2.2 (following the discussion of the same model
performed in subsection 4.2.2 without the inclusion of string loop
corrections). We will see that the inclusion of the $g_{s}$
corrections can freeze the $4$-cycle supporting the Standard Model
small producing a minimum of the full scalar potential at
exponentially large volume.

This chapter is organised as follows. Section 6.1 describes the
r\^{o}le played by string loop corrections in the LVS for an
arbitrary Calabi-Yau compactification, while in section 6.2 we
illustrate these general remarks by reconsidering all the
Calabi-Yau examples studied in section 4.2, showing how each of
them is modified by the inclusion of $g_s$ effects. Finally in
section 6.3, we discuss the prospects for possible
phenomenological and cosmological applications of our results.

\section{LARGE Volume and string loop corrections}

The results obtained in chapter 5 are very important for
K\"{a}hler moduli stabilisation. The general picture for LVS which
we presented in chapter 4, was neglecting the effect of string
loop corrections to the scalar potential. However just looking at
the K\"{a}hler potential we have seen that, in terms of powers of
the K\"{a}hler moduli, the leading order $\alpha'$ correction
(\ref{eq}) scales as $\delta K_{(\alpha')}\sim
\frac{1}{\mathcal{V}}$, whereas from (\ref{UUU}), the scaling
behaviour of the Kaluza-Klein loop correction is $\delta
K_{(g_{s})}^{KK}\sim \frac{\sqrt{\tau}}{\mathcal{V}}$. Naively it
seems incorrect to neglect $\delta K_{(g_{s})}^{KK}$ while
including the effects of $\delta K_{(\alpha')}$. However, as
discussed in chapter 5, due to the extended no-scale structure, at
the level of the scalar potential the $\alpha'$ corrections
dominate over the $g_s$ corrections. This allows loop corrections
to be neglected compared to $\alpha'$ corrections for the
stabilisation of the volume.

However in our general analysis presented in chapter 4, we saw
that for fibration models the inclusion of $\alpha'$ corrections
still left almost flat directions corresponding to non blow-up
moduli orthogonal to the overall volume. Loop corrections to the
scalar potential are much more important than non-perturbative
superpotential corrections, and we realise that they can play a
crucial r\^{o}le in stabilising these non blow-up moduli
transverse to the overall volume.

Thus we conclude that the extended no-scale structure renders the
LVS robust not only because it allows $\delta V_{(g_{s})}$ to be
neglected when stabilising the volume, but also because it ensures
that when $\delta V_{(g_{s})}$ is introduced to lift the remaining
flat directions, even though it will reintroduce a dependence in
$V$ on $\mathcal{V}$ and blow-up moduli, it will not destroy the
minimum already found but will give just a small perturbation
around it.

The general picture is that all corrections - $\alpha'$, loop and
non-perturbative - play a r\^{o}le in a generic Calabi-Yau
compactification. We can summarise our general analysis for the
LVS as:

\begin{enumerate}

\item{}
In order to stabilise all the K\"{a}hler moduli at exponentially
large volume one needs at least one 4-cycle which is a blow-up
mode resolving a point-like singularity (a del Pezzo complex
surface).

\item{}
 This 4-cycle,
together with other blow-up modes possibly present, are fixed
small by the interplay of non-perturbative and $\alpha'$
corrections, which stabilise also the overall volume mode.

\item{}
 The
$g_{s}$ corrections are subleading and so can be safely neglected.

\item{}
All the other 4-cycles, as those corresponding to fibrations, even
though they have non-perturbative effects, cannot be stabilised
small. Thus they are sent large so making their non-perturbative
corrections negligible.

\item{}
These moduli, which are large and transverse to the overall
volume,
 can then be frozen by $g_{s}$ corrections,
which dominate over the (tiny) non-perturbative ones.

\end{enumerate}

In general $\delta V_{(g_{s})}$ only lifts the flat directions
associated to non blow-up moduli transverse to the overall volume.
One could wonder whether they indeed yield a real minimum for such
moduli as opposed to a runaway direction. We do not address this
problem in general terms here and so in principle this looks like
a model-dependent issue. However, as the overall volume is
stabilised, the internal moduli space is compact. Therefore these
non blow-up moduli cannot run-away to infinity and so we expect
that loop corrections will induce a minimum for the potential. In
fact, one example in the next section will illustrate this idea
explicitly.

\section{Moduli stabilisation via string loop corrections}
\label{6}

We will now see in detail how the inclusion of string loop
corrections can affect the results found in the previous examples
which, neglecting $g_{s}$ corrections, can be summarised as:
\begin{enumerate}
\item  $\mathbb{C}P_{[1,1,1,6,9]}^{4}(18)\rightarrow$
LVS without flat directions.

\item  3-parameter K3 fibration with
$\tau_{1}$ `small' and $a_{1}\tau_{1}>a_{3}\langle\tau_{3}\rangle
\rightarrow$ LVS with an almost flat direction.

\item  3-parameter K3 fibration with $\tau_{3}\ll\tau_{1}<\tau_{2}
\rightarrow$ LVS with one flat direction.

\item  $\mathbb{C}P_{[1,3,3,3,5]}^{4}(15)\rightarrow$ LVS with a tachyonic
direction.

\item  $\mathbb{C}P_{[1,1,2,2,6]}^{4}(12)$ and 3-parameter K3 fibration with
$\tau_{1}$ `small' and $a_{1}\tau_{1}<a_{3}\tau_{3}\rightarrow$ No
LVS.
\end{enumerate}
We shall find that the inclusion of loop corrections modifies the
previous picture as follows:
\begin{enumerate}
\item  $\mathbb{C}P_{[1,1,1,6,9]}^{4}(18)\rightarrow$
Not affected by $\delta V_{(g_{s})}$.

\item  3-parameter K3 fibration with
$\tau_{1}$ `small' and $a_{1}\tau_{1}>a_{3}\langle\tau_{3}\rangle
\rightarrow \delta V_{(g_{s})}$ turns the almost flat direction
into a stabilised one $\Longrightarrow$ LVS without flat
directions\footnote{Notice that this case is the same as case 3
below but in a different region of moduli space. This means that
for this model the LARGE Volume Scenario can be realised with the
fibration modulus $\tau_1$ both `small' (but still hierarchically
larger than the blow-up mode $\tau_3$) and large.}.

\item  3-parameter K3 fibration with $\tau _{3}\ll\tau _{1}<\tau
_{2}\rightarrow \delta V_{(g_{s})}$ lifts the flat direction
$\Longrightarrow$ LVS without flat directions.

\item  $\mathbb{C}P_{[1,3,3,3,5]}^{4}(15)\rightarrow \delta V_{(g_{s})}$
stabilises the tachyonic direction $\Longrightarrow$ LVS without
flat directions.

\item  $\mathbb{C}P_{[1,1,2,2,6]}^{4}(12)$ and 3-parameter K3 fibration with
$\tau_{1}$ `small' and $a_{1}\tau_{1}<a_{3}\tau_{3}\rightarrow$
Not affected by $\delta V_{(g_{s})}$ - still no LVS.
\end{enumerate}

The $\mathbb{C}P_{[1,1,2,2,6]}^{4}$ case can never give large
volume due to the fibration 4-cycle $\tau_{1}$ which is impossible
to stabilise small. However in the example of the 3-parameter K3
fibration, LARGE Volume can be achieved by including a third
K\"{a}hler modulus which is a local blow-up and then sending
$\tau_{1}$ large. We shall use the expression (\ref{1 loop}) for
the form of the leading order string loop corrections to the
scalar potential.

\subsection{The single-hole Swiss cheese:
$\mathbb{C}P_{[1,1,1,6,9]}^{4}(18)$}

The influence of the $g_{s}$ corrections in the
$\mathbb{C}P_{[1,1,1,6,9]}^{4}$ case has been studied in detail in
\cite{bhp}. The authors showed that the loop corrections are
subleading and so can be neglected, as we claimed above. The loop
corrected K\"{a}hler potential looks like\footnote{We note that in
this case, as argued by Curio and Spillner \cite{curio}, $\delta
K_{(g_{s})}^{W}$ is absent, because in
$\mathbb{C}P_{[1,1,1,6,9]}^{4}(18)$ there is no intersection of
the divisors that give rise to nonperturbative superpotentials if
wrapped by $D7$-branes.}:
\begin{eqnarray}
K &=&K_{tree}+\delta K_{(\alpha ^{\prime })}+\delta K_{(g_{s},\tau
_{5})}^{KK}+\delta K_{(g_{s},\tau _{4})}^{KK}  \nonumber \\
&=&-2\ln \mathcal{V}-\frac{\xi }{\mathcal{V}g_{s}^{3/2}}+\frac{
g_{s}C_{5}^{KK}\sqrt{\tau
_{5}}}{\mathcal{V}}+\frac{g_{s}C_{4}^{KK}\sqrt{\tau
_{4}}}{\mathcal{V}}, \label{Gret}
\end{eqnarray}
but due to the ``extended no scale structure'', we obtain for the
scalar potential:
\begin{eqnarray}
V &=&\delta V_{(np)}+\delta V_{(\alpha ^{\prime })}+\delta
V_{(g_{s},\tau
_{5})}^{KK}+\delta V_{(g_{s},\tau _{4})}^{KK} \\
&=&\frac{\lambda _{1}\sqrt{\tau _{4}}e^{-2a_{4}\tau
_{4}}}{\mathcal{V}}- \frac{\lambda _{2}W_{0}\tau _{4}e^{-a_{4}\tau
_{4}}}{\mathcal{V}^{2}}+\frac{ 3\xi
W_{0}^{2}}{4\mathcal{V}^{3}g_{s}^{3/2}}+\frac{g_{s}^{2}(C_{5}^{KK})^{2}
}{\mathcal{V}^{3}\sqrt{\tau
_{5}}}+\frac{g_{s}^{2}(C_{4}^{KK})^{2}}{\mathcal{V}^{3}\sqrt{\tau
_{4}}}. \nonumber
\label{Grett}
\end{eqnarray}
Without taking the loop corrections into account, we have found a
minimum located at $\mathcal{V}\sim
e^{a_{4}\tau_{4}}\Leftrightarrow a_{4}\tau_{4}\sim
\ln\mathcal{V}$. Therefore the various terms in (\ref{Grett})
scale as:
\begin{eqnarray}
V &=&\delta V_{(np)}+\delta V_{(\alpha ^{\prime })}+\delta
V_{(g_{s},\tau
_{5})}^{KK}+\delta V_{(g_{s},\tau _{4})}^{KK}  \nonumber \\
&\sim&\frac{\sqrt{\ln\mathcal{V}}}{\mathcal{V}^{3}}-
\frac{\ln\mathcal{V}}{\mathcal{V}^{3}}+\frac{1}{\mathcal{V}^{3}}
+\frac{1}{\mathcal{V}^{10/3}}+\frac{1}{\mathcal{V}^{3}\sqrt{\ln\mathcal{V}}},
\label{Gretta}
\end{eqnarray}
and it is straightforward to realise that at exponentially large
volume the last two terms in (\ref{Gretta}) are suppressed with
respect to the first three ones.

\subsection{The multiple-hole Swiss cheese:
$\mathbb{C}P_{[1,3,3,3,5]}^{4}(15)$} \label{SM}

In section \ref{Swisscheese2} we have seen that if the
non-perturbative corrections in the SM cycle $\tau_{SM}$ are
absent, the $F$-term scalar potential (\ref{mio2}) for the
$\mathbb{C}P_{[1,3,3,3,5]}^{4}(15)$ Calabi-Yau does not present a
LVS with all the K\"{a}hler moduli stabilised. Following the same
procedure as in \cite{joe}, we shall now illustrate how the
$g_{s}$ corrections can turn the maximum in the $\tau_{SM}$
direction into a minimum without destroying the exponentially
large volume minimum
$\mathcal{V}\sim\sqrt{\tau_{E3}}e^{2\pi\tau_{E3}}$.

To derive the conjectured scaling behaviour of the loop
corrections, we use the formula (\ref{1 loop}) setting
$\mathcal{C}_{i}^{KK}=\hbox{Re}(S)$ $\forall i$ and $W_{0}=1$. Two
stacks of $D7$-branes wrap the $\tau_{SM}$ and $\tau_{c}$ cycle
respectively and both will give rise to Kaluza-Klein $g_{s}$
corrections. From (\ref{1 loop}), we estimate the first kind of
corrections by writing the overall volume (\ref{NuovoVolume}) in
the $(\tau_{a},\tau_{SM},\tau_{c})$ basis and computing the
relevant elements of the direct K\"{a}hler metric. We find:
\begin{equation}
\mathcal{V}=\sqrt{\frac{2}{45}}\left(\tau_{a}^{3/2}-\frac{1}{3}\left(3\tau_{SM}+2\tau_{c}\right)^{3/2}
-\frac{\sqrt{5}}{3}\tau_{c}^{3/2}\right), \label{NuovoVolume2}
\end{equation}
along with:
\begin{equation}
\frac{\partial^{2}K_{tree}}{\partial\tau_{SM}^{2}}\simeq
\frac{3}{\sqrt{10}}\frac{1}{\mathcal{V}\sqrt{3\tau_{SM}+2\tau_{c}}},
\end{equation}
and:
\begin{equation}
\frac{\partial^{2}K_{tree}}{\partial\tau_{c}^{2}}\simeq
\frac{2\sqrt{2}}{3\sqrt{5}}\left(\frac{\sqrt{5}}{4\sqrt{\tau_{c}}}
+\frac{1}{\sqrt{3\tau_{SM}+2\tau_{c}}}\right)\frac{1}{\mathcal{V}},
\label{az2}
\end{equation}
where in the large volume limit we have approximated the volume as
$\mathcal{V}\simeq\sqrt{\frac{2}{45}}\tau_{a}^{3/2}$. Thus the
Kaluza-Klein loop corrections to (\ref{mio2}) look like:
\begin{equation}
\delta V_{(g_{s})}^{KK}\simeq \left(\frac{5}{\sqrt{\tau_{c}}}
+\frac{13\sqrt{5}}{\sqrt{3\tau_{SM}+2\tau_{c}}}\right)
\frac{1}{15\sqrt{2}\mathcal{V}^{3}}. \label{azs2}
\end{equation}
Writing (\ref{azs2}) back in terms of $\tau_{SM}$ and
$\tau_{E3}=\tau_{c}+\tau_{SM}$, we obtain:
\begin{equation}
\delta V_{(g_{s})}^{KK}\simeq
\left(\frac{5}{\sqrt{\tau_{E3}-\tau_{SM}}}
+\frac{13\sqrt{5}}{\sqrt{2\tau_{E3}+\tau_{SM}}}\right)
\frac{1}{15\sqrt{2}\mathcal{V}^{3}}. \label{imp1}
\end{equation}
Due to the particulary simple form of the volume
(\ref{NuovoVolume2}), it is very sensible to expect that the
winding corrections will scale like the Kaluza-Klein ones
(\ref{imp1}). Therefore adding (\ref{imp1}) to (\ref{mio2}) we end
up with:
\begin{eqnarray}
V+\delta V_{(g_{s})} &=&\frac{\lambda _{1}\left( \sqrt{5\left(
2\tau _{E3}+\tau _{SM}\right) }+\sqrt{\tau _{E3}-\tau
_{SM}}\right) e^{-4\pi \tau _{E3}}}{\mathcal{V}}-\frac{3\lambda
_{2}\tau _{E3}e^{-2\pi \tau _{E3}}}{
\mathcal{V}^{2}}  \notag \\
&&+\frac{\lambda _{3}}{\mathcal{V}^{3}}+\left( \frac{\lambda
_{4}}{\sqrt{ \tau _{E3}-\tau _{SM}}}+\frac{\lambda
_{5}}{\sqrt{2\tau _{E3}+\tau _{SM}}} \right)
\frac{1}{\mathcal{V}^{3}}.  \label{mio20}
\end{eqnarray}
We notice that the string loop corrections are suppressed with
respect to the $\alpha'$ ones by a factor of $1/\sqrt{\tau_{E3}}$
and so do not affect the large volume minimum
$\mathcal{V}\sim\sqrt{\tau_{E3}}e^{2\pi\tau_{E3}}$ given that we
require $\tau_{E3}\gg 1$ to neglect higher order instanton
contributions. On the contrary $\delta V_{(g_{s})}$ can become
important to fix the SM direction when $\tau_{SM}$ gets small. In
fact, the maximum in that direction is now accompanied by a
minimum, as illustrated in Figure 6.1.
\begin{figure}[ht]
\begin{center}
\epsfig{file=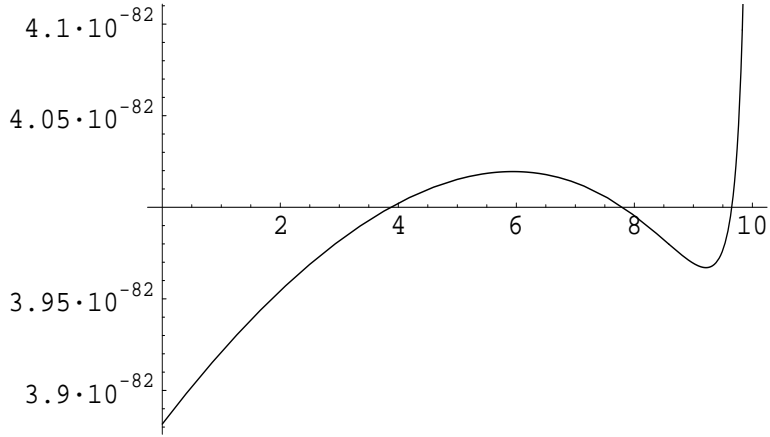, height=70mm,width=100mm}
\caption{$\tau_{SM}$ fixed by string loop corrections. The
numerical values used are $\lambda_{1}=\lambda_{2}=1$,
$\lambda_{3}=50$, $\lambda_{4}=\lambda_{5}=5$ and then we have
fixed $\tau_{E3}=10$ and $\mathcal{V}=\sqrt{10}e^{20\pi}$. }
\end{center}
\end{figure}

Thus we have shown that $g_{s}$ corrections can indeed freeze the
SM direction so giving rise to a LVS without any tachyonic
direction. The physics of this stabilisation is simply that if a
$D7$-brane wraps a 4-cycle, then loop corrections induced by the
brane will become large as the cycle size collapses. This repels
the modulus from collapsing and induces a minimum of the the
potential.

This example is illustrative in nature and shows how a cycle,
which is required to be small and which does not admit
nonperturbative effects, can potentially be stabilised by loop
corrections. In a fully realistic model, the $D$-term contribution
to the potential should also be included and the combined $F$- and
$D$-term potential studied. Usually the $D$-term will include,
besides the Fayet-Iliopoulos term depending on the moduli, also
the charged matter fields. Minimising the $D$-term will
generically fix one of the Standard Model singlets to essentially
cancel the Fayet-Iliopoulos term. Thus we can foresee a scenario
in which the Standard Model cycle is fixed by string loop
corrections, whereas the $D$-term fixes not the size of that cycle
but instead the VEV of a Standard Model singlet as a function of
the moduli. The axion corresponding to the SM cycle will not be
fixed by stringy instanton effects, but by standard QCD
nonperturbative effects \cite{joe} ($D$-term moduli stabilisation
has also been discussed in \cite{Dstabilisation}).

In this way we address the challenge of \cite{blumenhagen}. The
form of the $D$-term however depends on the model and in
particular on the details of the charged matter content and
whether or not they acquire VEVs. We therefore do not try and
specify this, but note that it will be necessary to include it in
a realistic model. However, we just notice that in string
compactification Standard Model-like constructions usually come
with extra anomalous $U(1)$'s. Then if the SM singlet fixed by
$D$-terms, were charged only under $U(1)_L$, this scenario would
be very attractive to break lepton number and so generate neutrino
masses. A perfect candidate for such a SM singlet would be a right
handed neutrino that corresponds to an open string going from two
of the $U(1)$ branes of the `Madrid model' with four intersecting
branes \cite{MadridModel}.

\subsection{2-Parameter K3 fibration:
$\mathbb{C}P_{[1,1,2,2,6]}^{4}(12)$}

One could wonder whether including the string loop corrections in
the case of the K3 fibration with two K\"{a}hler moduli treated in
section \ref{2modK3noLoop}, could generate an exponentially large
volume minimum which was absent when only non-perturbative and the
$\alpha'$ corrections are included. In reality, the answer is
negative as these further perturbative corrections produce a
contribution $\delta V_{(g_{s})}^{KK}+ \delta V_{(g_{s})}^{W}$ to
the scalar potential (\ref{oooo}), which is subdominant and cannot
help to stabilise the moduli. In fact, in the large volume limit
(\ref{ooooo}) and for $W_{0}\sim \mathcal{O}(1)$, the full
corrected scalar potential, now takes the form:
\begin{eqnarray}
V &=&\delta V_{(np)}+\delta V_{(\alpha' )}+\delta
V_{(g_{s},\tau_{1})}^{KK} +\delta V_{(g_{s},\tau_{2})}^{KK}+\delta
V_{(g_{s},\tau_{1})}^{W}
+\delta V_{(g_{s},\tau_{2})}^{W}\simeq \nonumber \\
&\simeq &-\frac{4}{\mathcal{V}^{2}}W_{0}a_{1}\tau
_{1}e^{-a_{1}\tau _{1}}+ \frac{3}{4}\frac{\xi \hbox{Re}\left(
S\right)^{3/2}}{\mathcal{V}^{3}}W_{0}^{2}\nonumber \\
&&+\frac{W_{0}^{2}}{\mathcal{V}^{2}}\left( \frac{\left(\mathcal{C}
_{1}^{KK}\right) ^{2}}{\hbox{Re}\left( S\right)
^{2}}\frac{1}{\tau_{1}^{2}}+ \frac{\left(
\mathcal{C}_{2}^{KK}\right) ^{2}}{\hbox{Re}\left( S\right)
^{2}}\frac{1}{2}\frac{\tau
_{1}}{\mathcal{V}^{2}}-2\mathcal{C}_{1}^{W}\frac{\tau
_{1}}{\mathcal{V}^{2}}-\frac{2\mathcal{C}_{2}^{W}}{\mathcal{V}\sqrt{\tau_{1}
}}\right) \\
&\simeq &-\frac{4}{\mathcal{V}^{2}}W_{0}a_{1}\tau
_{1}e^{-a_{1}\tau _{1}}+ \frac{3}{4}\frac{\xi\hbox{Re}\left(
S\right)^{3/2}}{\mathcal{V}^{3}}W_{0}^{2}+\frac{W_{0}^{2}}{\mathcal{V}
^{2}}\left( \frac{\left( \mathcal{C}_{1}^{KK}\right)
^{2}}{\hbox{Re}\left( S\right) ^{2}}\frac{1}{\tau
_{1}^{2}}-\frac{2\mathcal{C}_{2}^{W}}{\mathcal{V}\sqrt{\tau
_{1}}}\right).  \nonumber
\label{freg}
\end{eqnarray}
First of all we have to check that the minimum in the volume is
exponentially large. Therefore let us take the derivative:
\begin{equation}
\frac{4\mathcal{V}^{4}}{W_{0}^{2}}\frac{\partial V}{\partial
\mathcal{V}} =\left( \frac{32}{W_{0}}a_{1}\tau _{1}e^{-a_{1}\tau
_{1}}-\frac{\left( \mathcal{C}_{1}^{KK}\right)
^{2}}{\hbox{Re}\left( S\right) ^{2}}\frac{8}{ \tau
_{1}^{2}}\right) \mathcal{V}+\left(
\frac{24\mathcal{C}_{2}^{W}}{\sqrt{\tau _{1}}}-9\xi
\hbox{Re}\left( S\right) ^{3/2}\right) =0,
\end{equation}
whose solution is:
\begin{equation}
\left\langle \mathcal{V}\right\rangle
=\frac{3}{8}\frac{\hbox{Re}\left( S\right) ^{2}\left\langle \tau
_{1}\right\rangle ^{3/2}W_{0}\left( 8\mathcal{
C}_{2}^{W}-3\sqrt{\left\langle \tau _{1}\right\rangle
}\xi\hbox{Re}\left( S\right) ^{3/2} \right) }{\left( \left(
\mathcal{C}_{1}^{KK}\right) ^{2}W_{0}e^{a_{1}\left\langle \tau
_{1}\right\rangle }-4a_{1}\hbox{Re}\left( S\right)
^{2}\left\langle \tau _{1}\right\rangle ^{3}\right)
}e^{a_{1}\left\langle \tau _{1}\right\rangle }. \label{ttt}
\end{equation}
From (\ref{ttt}) we realise that in order to have an exponentially
large volume, we need to fine tune $\left(
\mathcal{C}_{1}^{KK}\right) ^{2}\sim e^{-a_{1}\tau_{1}}\ll 1$. We
assume that this is possible and so the denominator of (\ref{ttt})
scales as:
\begin{equation}
W_{0}-4a_{1}\hbox{Re}(S)^{2}\langle\tau_{1}\rangle^{3}\simeq
-4a_{1}\hbox{Re}(S)^{2}\langle\tau_{1}\rangle^{3},
\end{equation}
given that we are working in a regime where
$W_{0}\sim\mathcal{O}(1)$, $\hbox{Re}(S)\simeq 10$ and
$a_{1}\tau_{1}\gg 1$. Finally the VEV of the volume reads:
\begin{equation}
\left\langle \mathcal{V}\right\rangle \simeq
\frac{3}{8}\frac{W_{0}\left( 8
\mathcal{C}_{2}^{W}-3\sqrt{\left\langle \tau _{1}\right\rangle
}\xi \hbox{Re}(S)^{3/2}\right) }{-4a_{1}\left\langle \tau
_{1}\right\rangle ^{3/2}}e^{a_{1}\left\langle \tau
_{1}\right\rangle },  \label{t3}
\end{equation}
with $\mathcal{C}_{2}^{W}$ chosen such that:
\begin{equation}
\left( 1-\frac{3\sqrt{\left\langle \tau _{1}\right\rangle
}\xi\hbox{Re}(S)^{3/2}}{8\mathcal{C} _{2}^{W}}\right) <0,
\label{t2}
\end{equation}
to have a positive result. Now we neglect the
$\left(\mathcal{C}_{1}^{KK}\right) ^{2}$ term in $V$ (\ref{freg})
when we perform the derivative with respect to $\tau_{1}$ and we
obtain:
\begin{equation}
\frac{\mathcal{V}^{2}}{W_{0}}\frac{\partial V}{\partial \tau _{1}}
=4a_{1}e^{-a_{1}\left\langle \tau _{1}\right\rangle }\left(
a_{1}\left\langle \tau _{1}\right\rangle -1\right)
+\frac{W_{0}\mathcal{C}_{2}^{W}}{\left\langle
\mathcal{V}\right\rangle \left\langle \tau _{1}\right\rangle
^{3/2}}=0.  \label{t4}
\end{equation}
Substituting back (\ref{t3}), (\ref{t4}) becomes:
\begin{equation}
a_{1}\left\langle \tau _{1}\right\rangle =1+\frac{1}{3\left(
1-\frac{3\sqrt{ \left\langle \tau _{1}\right\rangle
}\xi\hbox{Re}(S)^{3/2}}{8\mathcal{C}_{2}^{W}}\right) },
\end{equation}
but (\ref{t2}) forces us to get $a_{1}\left\langle \tau
_{1}\right\rangle <1$, clearly in disagreement with our starting
point when we ignored higher order instanton corrections. Hence we
conclude that the inclusion of the string loop corrections does
not help to stabilise the moduli at exponentially large volume
since they render this attempt even worse.

\subsection{3-Parameter K3 fibration}
\label{bene}

The results of the study of the K3 fibration with three K\"{a}hler
moduli are summarised in the table at the end of section 4.2.4. We
will now try to address the problem left unsolved in that section.
Without loop corrections it was possible to find an exponentially
large volume in this class of models but there was still a flat
direction left, which we named $\Omega$. Let us see now how this
direction is lifted. We shall work in the regime
$W_{0}\sim\mathcal{O}(1)$ where the perturbative corrections are
important. We start off wrapping stacks of $D7$-branes around all
the $4$-cycle $\tau_{1}$, $\tau_{2}$ and $\tau_{3}$. We
immediately notice that the Kaluza-Klein loop correction to $V$ in
$\tau_{3}$ takes the form:
\begin{equation}
\delta V^{KK}_{(g_{s}),\tau _{3}}=
\frac{g_{s}^{2}(\mathcal{C}_{3}^{KK})^{2}}
{\sqrt{\tau_{3}}\mathcal{V}^{3}}, \label{eq3}
\end{equation}
and so does not depend on $\Omega$ and is subdominant to the
$\alpha'$ correction. Thus we will confidently neglect it. More
precisely, it could modify the exact locus of the minimum but not
the main feature of the model, that is the presence of an
exponentially large volume. Let us now focus on the region:
$\tau_{3}\ll\tau_{1}<\tau_{2}$. We recall the form of the scalar
potential and the K\"{a}hler potential without loop corrections:
\begin{equation}
V =\frac{16 a_{3}^{2}}{3\mathcal{V}}\sqrt{\tau_{3}}e^{-2a_{3}\tau
_{3}}-\frac{4}{\mathcal{V}^{2}}a_{3}\tau_{3}e^{-a_{3}\tau
_{3}}+\frac{3}{2g_{s}^{3/2}\mathcal{V}^{3}}, \label{41new}
\end{equation}
\begin{equation}
K=K_{tree}+\delta K_{(\alpha')}\underset{\mathcal{V}\gg 1}{\simeq
}-2\ln \mathcal{V}-\frac{2}{g_{s}^{3/2}\mathcal{V}}.
\end{equation}
We study now the possible corrections to $V$ coming from
$\tau_{1}$ and $\tau_{2}$ according to the general 1-loop formula
(\ref{1 loop}). We realise that the form of the volume (\ref{hhh})
implies that in this base of the K\"{a}hler cone, the blow-up mode
$\tau_{3}$ has only its triple self-intersection number
non-vanishing and so it does not intersect with any other cycle.
This is a typical feature of a blow-up mode which resolves a
point-like singularity: due to the fact that this exceptional
divisor is a \textit{local} effect, it is always possible to find
a suitable basis where it does not intersect with any other
divisor. Now we have seen that some string loop corrections come
from the exchange of closed winding strings at the intersection of
stacks of $D7$-branes. Hence the topological absence of these
intersections, implies an absence of these corrections. At the
end, the only relevant loop corrections are:
\begin{equation}
\delta V_{(g_{s})}=\delta V^{KK}_{(g_{s}), \tau_{1}}+\delta
V^{KK}_{(g_{s}), \tau_{2}}+\delta V^{W}_{(g_{s}),
\tau_{1}\tau_{2}},
\end{equation}
which look like:
\begin{eqnarray}
\delta V_{(g_{s}),\tau _{1}}^{KK} &=&g_{s}^{2}\left(
C_{1}^{KK}\right) ^{2}\left( \frac{1}{\tau _{1}^{2}}+\frac{2\beta
^{2}}{P}\right) \frac{
W_{0}^{2}}{\mathcal{V}^{2}}, \notag \\
\delta V_{(g_{s}),\tau _{2}}^{KK} &=&g_{s}^{2}\left(
C_{2}^{KK}\right) ^{2}
\frac{2}{P}\frac{W_{0}^{2}}{\mathcal{V}^{2}}, \label{LOOP} \\
\delta V_{(g_{s}),\tau _{1}\tau _{2}}^{W}
&=&-2C_{12}^{W}\frac{W_{0}^{2}}{ \mathcal{V}^{3}t_{\ast }}, \notag
\end{eqnarray}
where the 2-cycle $t_{*}$ is the intersection locus of the two
4-cycles whose volume is given by $\tau_{1}$ and $\tau_{2}$. In
order to work out the form of $t_{*}$, we need to write down the
volume of the K3 fibration (\ref{hhh}) in terms of 2-cycle moduli:
\begin{equation}
\mathcal{V}=(\lambda_{1}t_{1}+\lambda_{2}t_{2})t_{2}^{2}+\lambda_{3}t_{3}^{3}.
\label{volumeet}
\end{equation}
Then:
\begin{equation}
\left\{
\begin{array}{c}
\tau _{1}=\frac{\partial \mathcal{V}}{\partial t_{1}}=t_{2}\left(
\lambda
_{1}t_{2}\right) , \\
\tau _{2}=\frac{\partial \mathcal{V}}{\partial t_{2}}=t_{2}\left(
2\lambda _{1}t_{1}+3\lambda _{2}t_{2}\right),
\end{array}
\right. \label{taus}
\end{equation}
and so $t_{*}=t_{2}=\sqrt{\frac{\tau_{1}}{\lambda_{1}}}$.
Therefore the $g_{s}$ corrections to the scalar potential
(\ref{LOOP}) take the general form:
\begin{equation}
\delta V_{(g_{s})}= \left(\frac{A}{\tau_{1}^{2}}
+\frac{B}{\mathcal{V}\sqrt{\tau_{1}}} +\frac{C\tau_{1}}
{\mathcal{V}^{2}}\right)\frac{W_{0}^{2}}{\mathcal{V}^{2}},
\label{74}
\end{equation}
where:
\begin{equation}
\left\{
\begin{array}{c}
A=g_{s}^{2}\left( C_{1}^{KK}\right) ^{2}>0, \\
B=-2C_{12}^{W}\sqrt{\lambda _{1}}\equiv -\frac{C_{12}^{W}}{\alpha }, \\
C=2\alpha ^{2}g_{s}^{2}\left[ \left( C_{1}^{KK}\right) ^{2}\beta
^{2}+\left( C_{2}^{KK}\right) ^{2}\right] >0.
\end{array}
\right.
\end{equation}
Notice that due to the ``extended no-scale structure'' which
causes the vanishing of the leading Kaluza-Klein correction to
$V$, we know the sign of the coefficients $A$ and $C$ because the
parameters are squared (see (\ref{1 loop}) but we do not have any
control over the sign of $B$. It is now convenient to take
advantage of the field redefinition (\ref{43}) and recast the loop
corrections (\ref{LOOP}) in terms of $\mathcal{V}$ and $\Omega$.
Inverting the relation (\ref{43}), we get:
\begin{equation}
\tau_{1}=\left(\frac{\Omega-\mathcal{V}}{2\alpha\beta}\right)^{2/3},\text{
\ \ \
}\tau_{2}=\left(\frac{\beta}{4\alpha^{2}}\right)^{1/3}\frac{\left(\Omega+\mathcal{V}\right)}
{\left(\Omega-\mathcal{V}\right)^{1/3}}. \label{CambiaCoord}
\end{equation}
Substituting these results back in (\ref{LOOP}) we can find the
relevant dependence of the scalar potential on $\Omega$:
\begin{equation}
\delta
V_{(g_{s})}=\frac{d_{1}\Omega^{2}+d_{2}\Omega\mathcal{V}+d_{3}\mathcal{V}^{2}}
{\left(\Omega-\mathcal{V}\right)^{4/3}\mathcal{V}^{4}}, \label{mo}
\end{equation}
where:
\begin{eqnarray}
d_{1} &=&g_{s}^{2}\left( \frac{2\alpha ^{4}}{\beta ^{2}}\right)
^{1/3}\left[ \left( C_{1}^{KK}\right) ^{2}\beta ^{2}+\left(
C_{2}^{KK}\right) ^{2}\right]
W_{0}^{2}, \label{1} \\
d_{2} &=&-\left( \frac{2}{\alpha ^{2}\beta ^{2}}\right)
^{1/3}\left\{ \beta C_{12}^{W}+2\alpha ^{2}g_{s}^{2}\left[ \left(
C_{1}^{KK}\right) ^{2}\beta
^{2}+\left( C_{2}^{KK}\right) ^{2}\right] \right\} W_{0}^{2}, \label{2} \\
d_{3} &=&\left( \frac{2}{\alpha ^{2}\beta ^{2}}\right)
^{1/3}\left\{ \beta C_{12}^{W}+\alpha ^{2}g_{s}^{2}\left[ 3\left(
C_{1}^{KK}\right) ^{2}\beta ^{2}+\left( C_{2}^{KK}\right)
^{2}\right] \right\} W_{0}^{2}. \label{3}
\end{eqnarray}
For generic values of $d_{1}$, $d_{2}$ and $d_{3}$ we expect to
lift the flat direction $\Omega$. Consistency requirements imply
that any meaningful minimum must lie within the K\"{a}hler cone so
that no 2-cycle or 4-cycle shrinks to zero and the overall volume
is always positive. Let us work out the boundaries of the
K\"{a}hler moduli space in terms of $\mathcal{V}$ and $\Omega$ and
then look for a minimum in the $\Omega$ direction. Given that we
are sending both $\tau_{1}$ and $\tau_{2}$ large while keeping
$\tau_{3}$ small we can approximate the volume
(\ref{hhh})-(\ref{volumeet}) as follows:
\begin{equation}
\mathcal{V}\simeq\alpha\sqrt{\tau _{1}}(\tau _{2}-\beta \tau
_{1})=(\lambda_{1}t_{1}+\lambda_{2}t_{2})t_{2}^{2},
\label{volumes}
\end{equation}
where $\lambda_{1}=\frac{1}{4\alpha^{2}}>0$ and
$\lambda_{2}=\frac{\beta}{4\alpha^{2}}>0$. Then looking at
(\ref{taus}) and (\ref{volumes}) it is clear that when $t_{1}$ and
$t_{2}$ are positive then also $\mathcal{V}>0$ and $\tau_{i}>0$
$\forall i=1,2$. Hence the boundaries of the K\"{a}hler cone are
where one of the 2-cycle moduli $t_{1,2} \to 0$. The expression of
the 2-cycles in terms of $\mathcal{V}$ and $\Omega$ reads:
\begin{equation}
t_{1}=\left(\frac{2\mathcal{V}-\Omega}{\lambda_{1}}\right)\left(\frac{\lambda_{2}}
{\Omega-\mathcal{V}}\right)^{2/3},\text{ \ \ \
}t_{2}=\left(\frac{\Omega-\mathcal{V}}{\lambda_{2}}\right)^{1/3},
\label{ts}
\end{equation}
and so we realise that the K\"{a}hler cone is given by
$\mathcal{V}<\Omega<2\mathcal{V}$. In fact, looking at
(\ref{CambiaCoord}) and (\ref{ts}) we obtain:
\begin{equation*}
\left\{
\begin{array}{c}
\Omega \rightarrow \mathcal{V}^{+}\Longleftrightarrow \tau
_{1}\rightarrow 0\Longleftrightarrow \tau _{2}\rightarrow \infty
\Longleftrightarrow
t_{1}\rightarrow \infty \Longleftrightarrow t_{2}\rightarrow 0, \\
\Omega \rightarrow \left( 2\mathcal{V}\right)
^{-}\Longleftrightarrow \tau _{1}\rightarrow \lambda _{1}\left(
\frac{\mathcal{V}}{\lambda _{2}}\right) ^{2/3}\Longleftrightarrow
\tau _{2}\rightarrow 3\lambda _{2}^{1/3}\mathcal{V}
^{2/3}\Longleftrightarrow t_{1}\rightarrow 0\Longleftrightarrow
t_{2}\rightarrow \left( \frac{\mathcal{V}}{\lambda _{2}}\right)
^{1/3}.
\end{array}
\right.
\end{equation*}
We look now for possible minima along the $\Omega$ direction
considering the volume already fixed:
\begin{equation}
\frac{\partial(\delta V_{(g_{s})})}{\partial\Omega}=\frac{2 d_{1}
\Omega
\left(\Omega-3\mathcal{V}\right)-\mathcal{V}\left(d_{2}\left(\Omega+3\mathcal{V}\right)+4
d_{3}\mathcal{V} \right)
}{3\left(\Omega-\mathcal{V}\right)^{7/3}\mathcal{V}^{4}}=0.
\label{der}
\end{equation}
Equation (\ref{der}) admits a solution of the form
$\langle\Omega\rangle=\kappa\langle\mathcal{V}\rangle$ where:
\begin{equation}
\kappa=\frac{6 d_1 + d_2 + \sqrt{36 d_1^2+36 d_1 d_2 + d_2^2 + 32
d_1 d_3 }}{4 d_{1}}. \label{kappa}
\end{equation}
A consistent minimum within the walls of the K\"{a}hler cone
requires a choice of $d_{1}$, $d_{2}$ and $d_{3}$ such that
$1<\kappa<2$. In section 4.2.4 we have set the parameters
$\alpha=\gamma=1$, $\xi=2$, $g_{s}=0.1$, $a_{3}=\pi$ and
$W_{0}=1$, and then obtained $\langle\tau_{3}\rangle=10$ and
$\langle\mathcal{V}\rangle\simeq 3.324\cdot10^{13}$ from
(\ref{x})-(\ref{duet}). We now keep the same choice of parameters
and set also $\beta=1/2$, $C_{1}^{KK}=1$,
$C_{2}^{KK}=C_{12}^{W}=10$. It follows that $d_{1}=2.005$,
$d_2=-14.01$ and $d_3=12.015$ which gives $\kappa\simeq 1.004$
correctly in the required regime. Then the minimum for $\Omega$
shown in Figure 6.2, is located at
$\langle\Omega\rangle=\kappa\cdot 3.324\cdot 10^{13}\simeq
3.337\cdot 10^{13}$. We stress that we have stabilised $\Omega$
without fine tuning any parameter.
\begin{figure}[ht]
\begin{center}
\epsfig{file=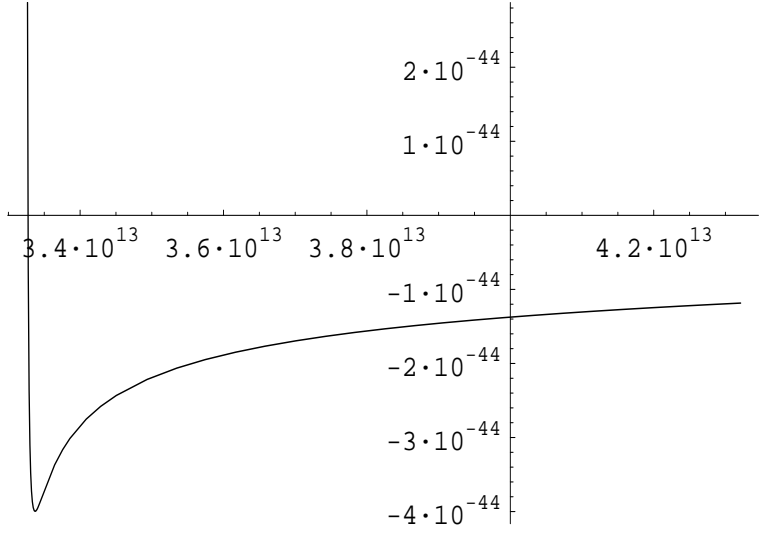, height=70mm,width=100mm} \caption{$V$
versus $\Omega$ at $\mathcal{V}$ and $\tau_{3}$ fixed.}
\end{center}
\end{figure}

We could also contemplate the case where we do not have
$D7$-branes wrapping one of the 4-cycles $\tau_{1}$ and
$\tau_{2}$. In this case there is no correction coming from the
exchange of winding strings because we have just one stack of
$D7$-branes with no intersection. Only the Kaluza-Klein
corrections would survive.

\begin{enumerate}
\item no $D7$-brane wrapping the $\tau _{1}$ cycle

In this case the 1-loop correction looks like:
\begin{equation}
\delta V_{(g_{s})}=\delta V^{KK}_{(g_{s}),
\tau_{2}}=2\alpha^{2}g_{s}^{2}\left( C_{2}^{KK}\right) ^{2}
\frac{W_{0}^{2}\tau_{1}}{\mathcal{V}^{4}}=
d\frac{\left(\Omega-\mathcal{V}\right)^{2/3}}{\mathcal{V}^{4}},
\label{xy}
\end{equation}
where $d=\left(\alpha^{2}\sqrt{2}/\beta\right)^{2/3}
g_{s}^{2}\left(C_{2}^{KK}\right)^{2}W_{0}^{2}$. However (\ref{xy})
has no minimum in $\Omega$ regardless of the value of $d$.

\item no $D7$-brane wrapping the $\tau _{2}$ cycle:

\begin{equation}
\delta V_{(g_{s})}=\delta V^{KK}_{(g_{s}),
\tau_{1}}=g_{s}^{2}\left( C_{1}^{KK}\right) ^{2}\left(
\frac{1}{\tau
_{1}^{2}}+\frac{2(\alpha\beta)^{2}\tau_{1}}{\mathcal{V}^{2}}\right)
\frac{ W_{0}^{2}}{\mathcal{V}^{2}}=\frac{\mu_{1}\Omega^{2}
+\mu_{2}\Omega\mathcal{V}+\mu_{3}\mathcal{V}^{2}}
{\left(\Omega-\mathcal{V}\right)^{4/3}\mathcal{V}^{4}}, \label{mm}
\end{equation}
where:
\begin{equation}
\mu _{1}=\delta ,\text{ \ \ \ \ }\mu _{2}=-2\delta ,\text{ \ \ \ \
}\mu _{3}=3\delta ,\text{ \ \ \ \ \ }\delta =2^{1/3}\left( \alpha
\beta \right) ^{4/3}g_{s}^{2}\left( C_{1}^{KK}\right)
^{2}W_{0}^{2}.
\end{equation}
Since the potential (\ref{mm}) has the same form of (\ref{mo}), we
can follow the same line of reasoning as above and conclude that
this case admits a minimum located at
$\langle\Omega\rangle=\kappa\langle\mathcal{V}\rangle$ if
$1<\kappa<2$, where $\kappa$ now is given by (\ref{kappa}) with
the replacement $d_i\leftrightarrow \mu_i$ $\forall i=1,2,3$.
\end{enumerate}

We consider now the matrix of second derivatives in the
$(\tau_1,\tau_2)$ space $M_{ij}=V_{ij}$ and using the known
expression for the K\"{a}hler metric $K_{ij}$ we construct the
matrix $K^{-1}_{ik}M_{kj}$. The two eigenvalues of this matrix
correspond to the mass-squared of the canonically normalised
particles corresponding to the volume modulus and the originally
flat direction $\Omega$:
$$m^2_{\mathcal{V}}\ \sim\ 1/\mathcal{V}^3, \ \ \ m^2_{\Omega}\ \sim\ 1/\mathcal{V}^{10/3}.$$

Finally we found also minima for which $\tau_2\gg \tau_1>\tau_3$,
where the inclusion of the string loop corrections turns the
almost flat direction, which we had found before, into a
stabilised one with an actual minimum at exponentially large
volume. This is an interesting configuration because of the
following observation: since $t_2=\sqrt{\tau_1}$ and
$t_1=(\tau_2-2\tau_1)/2\sqrt{\tau_1}$, we can see that $\tau_2\gg
\tau_1$ would imply that $t_1\gg t_2$ and we would effectively
have a very anisotropic compactification with the 2-cycle much
bigger than its dual 4-cycle. This could then lead to a
realisation of the supersymmetric 2 large extra dimensions
scenario \cite{add, cliff}, in which the extra dimension could be
as large as a fraction of a millimetre. This would correspond to
looking for solutions $\langle\Omega\rangle
=(1+\epsilon)\langle\mathcal{V}\rangle$ with $\epsilon\to 0$.

However, we shall postpone the detailed discussion of this class
of minima to chapter 8, where we shall show that, in this case,
the inclusion of string loop corrections allows to build a very
promising inflationary model that predicts observable gravity
waves without fine-tuning.

\section{Potential applications}
\label{7}

Part II of this thesis has studied the general conditions needed
to find exponentially large volume in type IIB compactifications.
The necessary and sufficient conditions are simple to state:
negative Euler number, more than one K\"{a}hler moduli with at
least one of them being a blow-up mode resolving a point-like
singularity.

We have also uncovered the important r\^{o}le played by $g_{s}$
corrections in moduli stabilisation.
 This has allowed us to find new
classes of LVS with a fibration structure in which not only the
volume but the fibre moduli  are exponentially large whereas the
blow-up modes are stabilised at the usual small values. Therefore
in general all of $\alpha'$, non-perturbative and $g_s$
corrections, may be important to stabilise the different classes
of K\"{a}hler moduli.

Here we briefly discuss some of the applications. First, our
results do not appear to change significantly the standard
phenomenology of LVS explored in \cite{SoftSUSY, LVSatLHC}, where
we imagine the Standard Model localised on $D7$-branes wrapping a
small 4-cycle. The reason is that the volume modulus is still the
main source of supersymmetry breaking leading to an approximate
no-scale structure, which can be argued in general terms
\cite{joemirror}. As the Standard Model is localised around a
blow-up cycle, the effects from other exponentially large moduli
will be suppressed. However, it may be interesting to explore the
potential implications of hidden sectors localised on those
cycles. Also, in the multiple-hole Swiss cheese case where the
Standard Model cycle is stabilised by perturbative rather than
non-perturbative effects, the general structure of soft terms will
not change significantly, again since the main source of
supersymmetry breaking is the volume modulus. The only difference
could be the absence of the small hierarchy between the scale of
the soft terms $M_{soft}$ and the gravitino mass $m_{3/2}$, since
if the SM cycle is not stabilised non-perturbatively, then the
suppression of $M_{soft}$ with respect to $m_{3/2}$ by
$\ln(M_P/m_{3/2})$ \cite{suppression0, suppression1, suppression2,
suppression} is probably not present, but it would be interesting
to study this case in further detail.

A potentially more interesting application is to cosmology. The
cosmological implications of LVS have been explored in
\cite{kahler, TwoFieldKMI, aalok} only for Swiss cheese
compactifications.
 Small moduli were found to be good candidates for
inflatons as long as $h_{11}>2$ without the need to fine tune.
However a difficulty with this is that loop corrections are
expected to modify this result if there is a $D7$-brane wrapping
the inflaton cycle, while if there is no such brane then it is
difficult to reheat the Standard Model brane since there is no
direct coupling of the inflaton to Standard Model fields.

The volume modulus is not suitable for inflation as $m_{\mc{V}}
\sim H$ and so it suffers directly from the $\eta$ problem.
However for K3 fibration models, there is the transverse field
$\Omega$ which is stabilised by the loop corrections. As the loop
corrections are parametrically weaker than the $\alpha'$
corrections which stabilise the volume, $\Omega$ is parametrically
lighter than the volume modulus and thus the Hubble scale. In fact
 $m_{\Omega}\sim
\mathcal{V}^{-(3/2+\alpha)}$, with $\alpha=1/6$. It follows that
the slow-roll $\eta$ parameter is:
\begin{equation}
\eta\sim M_P^2\,  \frac{m_{\Omega}^{2}}{H^2}\sim
\frac{1}{\mathcal{V}^{1/3}}\ll 1.
\end{equation}

Therefore such fibration models seem promising for string theory
realisations of modular inflation, as at large volume the mass
scale induced by loop corrections is parametrically smaller than
the Hubble scale. A detailed study of the potential for large
values of the field, away from the minimum, will be required in
order to see if this is a viable model of inflation, including the
value of density perturbations and the potential for reheating. A
very interesting model of large field inflation with the
prediction of observable gravity waves, where the inflaton is a
fibration modulus, as discussed above, will be presented and
described in detail in chapter 8.

The fact that the spectrum of moduli fields includes further
candidate light fields, besides the volume modulus, is a new
source for the cosmological moduli problem. In the LARGE Volume
Scenario this problem is already present as long as the string
scale is smaller than $10^{13}$ GeV, since the volume modulus
would be lighter than $10$ TeV, and coupling with gravitational
strength interactions it would overclose the universe or decay so
late to ruin nucleosynthesis. Given a solution to this problem -
such as a late period of inflation or the dilution due to entropy
released by the non-thermal decay of a heavier particle, as will
be discussed in chapter 9 - the corresponding modulus becomes a
dark matter candidate. With an intermediate string scale and
TeV-scale supersymmetry, the volume modulus has a mass $m \sim 1$
MeV. The additional light moduli fields are also potential dark
matter candidates and have masses $m \sim 10$ KeV. Furthermore,
they can decay into photons with a clean monochromatic line
similar to the volume modulus. A proper analysis of their
couplings to photons along the lines of \cite{CQ} should be made
in order to see if this effect could be eventually detected.

It is worth pointing out that the multiple-hole Swiss cheese case
provides an explicit example of K\"{a}hler moduli inflation, in
which at least three K\"{a}hler moduli were needed (but no
explicit example was provided in \cite{kahler}). Also this is a
good example to explore the issue about stabilisation of the
Standard Model cycle that has to be small (and then a blow-up
mode) but without the presence of a non-perturbative
superpotential which is not desired if the corresponding axion is
the QCD axion \cite{joe} and if $D$-terms could induce a breaking
of the Standard Model group \cite{blumenhagen}. Our results
indicate that it is actually possible to achieve this.

We would finally like to emphasise that this is only a first
attempt to investigate the relevance of loop corrections in the
LVS and much work remains to be done. In particular, although we
have used a well motivated volume dependence of the leading
quantum corrections to the K\"{a}hler potential, explicit
calculations are still lacking. While we believe that given the
general importance of loop corrections, it is important to study
their effects even with incomplete knowledge of their form,
further information about these corrections for general
Calabi-Yaus is very desirable.

\part{Applications to Cosmology}

\chapter{String Cosmology}
\label{StringCosmology} \linespread{1.3}

In this chapter we shall provide a brief review of string
cosmology \cite{GloriousHistory} to prepare the stage for the
description of possible cosmological implications of the LARGE
Volume Scenario which will be presented in chapters 8 and 9. We
start by recalling in section 7.1 the basics of inflation, and
then in section 7.2 we outline the main reasons for believing that
string theory and inflation, or cosmology more in general, are
tightly related. In the same section, we also describe the main
challenges for inflationary model building in string theory.

\section{Inflation}

As we have seen in chapter 1, a very early period in the history
of our Universe where the scale factor $a(t)$ increases
exponentially, named inflation \cite{inflation}, can provide a
solution to the flatness, horizon and monopole problems of the
Standard Cosmological Model, and, at the same time, it can explain
the origin of the CMB anisotropies, so providing a beautiful
mechanism for large scale structure formation \cite{CMB}. In this
section, we provide a very brief more detailed introduction to
inflation \cite{liddle}, focusing on the case in which the
exponential expansion is driven by a scalar field $\varphi$, whose
flat potential $V(\varphi)$ provides an effective cosmological
constant. In this case, the Friedmann's equation takes the form
(in the case of a flat Universe):
\begin{equation}
H^2= \frac{8\pi G_N}{3} \left( V+ \frac{\dot\varphi^2}{2} \right).
{\label{finfl}}
\end{equation}
Now when the following condition is satisfied:
\begin{equation}
\epsilon \equiv \frac{M_P^2}{2} \left(\frac{V'}{V}\right)^2\ll 1,
\end{equation}
corresponding to the case in which the potential energy dominates
over the kinetic energy, the Friedmann's equation (\ref{finfl})
simplifies to:
\begin{equation}
H^2\simeq \frac{8\pi G_N}{3} V. {\label{finflo}}
\end{equation}
For a very flat potential that behaves as an effective
cosmological constant, $V\sim \Lambda>0$, (\ref{finflo}) admits an
inflationary solution of the form $a\sim e^{Ht}\sim
e^{\sqrt{\Lambda/3}}$ (for $8\pi G_N=1$). On the other hand, the
equation of motion for the inflaton reads:
\begin{equation}
\ddot\varphi - 3 H\dot\varphi = -\frac{dV}{d\varphi},
\label{inflaton}
\end{equation}
and one can guarantee that the inflationary period lasts for some
time if the friction term $- 3 H\dot\varphi$ dominates the LHS of
(\ref{inflaton}), so forcing the inflaton to roll \textit{slowly}
on the potential $V(\varphi)$. This can take place if the
following condition is satisfied:
\begin{equation}
\eta \equiv M_P^2 \frac{V''}{V} \ll 1. \label{definizionedieta}
\end{equation}
The conditions $\epsilon\ll 1$ and $\eta\ll 1$ are named `slow
roll conditions' and, if satisfied, guarantee the presence of an
inflationary expansion. The amount of inflation is measured in
terms of the number of e-foldings $N_e$ defined as:
\begin{equation}
N_e(t)\equiv \int_{t_{in}}^{t_{end}} H(t') dt'=
\int_{\varphi_{in}}^{\varphi_{end}} \frac{H}{\dot\varphi} d\varphi
= \frac{1}{M_P^2} \int_{\varphi_{end}}^{\varphi_{in}} \frac{V}{V'}
d\varphi,
\end{equation}
where $\varphi_{end}$ is defined as the point in field space where
the slow-roll conditions cease to be valid
($\epsilon(\varphi_{end})\sim 1$). The exact number of e-foldings
which are needed to solve the horizon problem, depends both on the
inflationary scale and the reheating temperature. However in most
models of inflation, one needs at least $N_e\geq 50$.

The observed CMB temperature fluctuations $\delta T/T$
\cite{WMAP5}, corresponding to density perturbations $\delta
\rho/\rho$, are generated by the quantum fluctuations of the
inflaton which give rise to scalar and tensor perturbations to the
metric \cite{CMB} with a primordial spectrum of the form:
\begin{equation}
\mathcal{P}_{\mathcal {R}}(k)\propto k^{n_s-1},\textrm{ \ \ \ and
\ \ \ }\mathcal{P}_{grav}(k)\propto k^{n_{grav}}.
\end{equation}
In the previous expression, the constants $n_s$ and $n_{grav}$ are
the `scalar spectral index' and the `gravitational spectral
index', respectively. They represent two important CMB observables
which have to be supplemented with the amplitude of both kinds of
primordial perturbations. The COBE normalisation of the scalar
density perturbations reads:
\begin{equation}
\delta_H = \frac{2}{5} {\mathcal P}_{\mathcal R}^{1/2} =
\frac{1}{5 \pi \sqrt{3}}{V^{3/2}\over M_P^3\,V'}= 1.91\cdot
10^{-5}, \label{deltaH}
\end{equation}
whereas the last CMB observable $r$ is defined as the ratio of the
amplitude of tensor fluctuations to the amplitude of scalar
fluctuations.

All these four CMB observables, $n_s$, $n_{grav}$, $\delta_H$ and
$r$, are only sensitive to essentially three numbers in any
slow-roll inflationary model: the inflationary Hubble scale,
$H_{inf}$, and the two small slow-roll parameters, $\epsilon$ and
$\eta$, evaluated at `horizon exit', that is at 50 or 60
e-foldings before the end of inflation when the relevant scales
left the horizon and got frozen. In principle these three
parameters should provide one relationship amongst the four CMB
observables, but at the moment it is not possible to exploit the
full power of this prediction since tensor fluctuations have not
been detected yet. In the meantime, one may instead constrain the
inflationary scale $H_{inf}$ by demanding to match the COBE
normalisation for the density fluctuations (\ref{deltaH}); then
one can work out the value of the spectral index $n_s$ and the
tensor-to-scalar ratio $r$ in terms of $\epsilon$ and $\eta$ as:
\be \label{rnsslowrollo}
    n_s = 1 + 2\eta - 6\epsilon \qquad \hbox{and} \qquad
    r =16 \, \epsilon,
\ee
showing that for slow rolling ($\eta,\epsilon \ll 1$), the
spectrum is almost scale invariant ($n\sim 1$), in very good
agreement with experimental observations. Finally, varying
$\epsilon$ and $\eta$ leads to predictions which fill regions of
the observable $(r-n_s)$-plane. The hope is to use experimental
data to be able to distinguish between the predictions of
different broad classes of models whose main examples can be
classified as \cite{liddle}:
\begin{enumerate}
\item \textit{Large-Field Models}, for which $\epsilon$ and $\eta$
vary inversely with the value of the inflaton field: $\propto
(M_P/\varphi)^p$, for some $p > 0$;

\item  \textit{Small-Field Models}, for which $\epsilon$ and $\eta$
are proportional to a positive power of the value of the inflaton
field: $\propto (\varphi/M_P)^p$, for some $p > 0$. Most of the
small-field models sweep out a preferred region of the
$(r-n_s)$-plane corresponding to $n_s<1$;

\item \textit{Hybrid Models}, for which field evolution at the end of
inflation involves at least a two-dimensional field space, and for
which the slow-roll parameters depend on parameters in the
potential which govern the couplings between these fields. Most of
the hybrid models sweep out a preferred region of the
$(r-n_s)$-plane corresponding to $n_s > 1$.
\end{enumerate}

\section{String theory and inflation}

The first reason why one would be tempted to find a connection
between string theory and cosmology is the fact that the basic
questions in cosmology, like the resolution of space-like
singularities, the determination of initial conditions, and the
nature of dark energy, can be answered only within a fundamental
theory of quantum gravity.

However, in order to understand the origin of the special initial
conditions (homogeneity, isotropy, flatness, and a spectrum of
primordial density fluctuations) which have to be chosen to start
the Hot Big Bang model, we need to invoke a very early epoch of
inflationary expansion \cite{inflation}. Then, according to its
intrinsic nature, inflation screens from observational view much
of the physics of the ultra-violet completion of the effective
field theory framework where it is formulated. Indeed, inflation
explains the \textit{absence} of observable relics, such as
monopoles from possible extensions of the Standard Model.

Therefore, one may wonder whether string theory, as a candidate
ultra-violet completion of particle physics and gravity, should
play much of a r\^{o}le in cosmology at all. However, there are
several reasons for believing that inflation itself, although
modelled within the framework of effective field theory, is
closely connected to challenging issues in quantum gravity, and
hence it is the simplest application of string theory to cosmology
to motivate.
\begin{itemize}
\item Inflation could easily involve energy scales which are so high
that they could plausibly directly probe string-related physics. A
natural energy scale for inflation is around the GUT scale, and an
observation of primordial tensor fluctuations in the CMB
\cite{Verde, rBounds} would establish beyond doubt that the CMB
fluctuations were generated at such energies. No planned
terrestrial experiment can reach a fraction of this energy. Hence,
unless the string scale is extremely low (so that the LHC can tell
us something about string theory), signals of string theory will
be seen, if they are seen at all, in the sky.

\item Specific models of inflation
often depend on the existence of very shallow scalar potentials
\cite{LindeReview} which give rise to extremely light scalar
masses, $M_{\phi} \leq H$, where $H$ is the Hubble scale at the
epoch of interest. But scalar masses are famously difficult to
keep from getting large contributions when the short-distance
sector of the theory is integrated out. Therefore the vast
majority of inflationary models rely for their phenomenological
success on properties which are notoriously sensitive to
microscopic details; this is usually summarised by saying that
`inflation is ultra-violet sensitive'. This issue is tightly
related to the concept of naturalness. In fact, we would like to
understand if the choices made in order to obtain acceptable
values for $H_{inf}$, $\epsilon$ and $\eta$ are inordinately
sensitive to short-distance effects, and so they must be
finely-tuned in order to achieve sufficient inflation (this is
usually called the `$\eta$ problem' \cite{etaproblem}).

\item Many field-theoretic
cosmological models, as all the examples of large-field inflation
that predict observable gravity waves, not only require very
shallow potentials, but also that this flatness is present over a
trans-Planckian region in field space. In this case, generic
string or Planck-scale corrections become significant.

\item Inflation provides a robust
field-theoretic mechanism that addresses many cosmological
problems but the data does not yet pin down anything close to a
precise model. The further requirement that each inflationary
model should be sensibly embedded into microscopic physics, could
strongly restrict the number of viable models.

\item The experimental measurements of the properties of
the CMB \cite{WMAP5} will soon be made more precise due to the
recent launch of the PLANCK satellite and the plan of new
experiments for the not too far future, such as EPIC, BPol or
CMBPol \cite{Verde, rBounds}.

\item It is possible to understand the mechanism of reheating
only embedding it in a fundamental theory, like string theory,
which would allow us to know what are the relevant degrees of
freedom at inflationary energies. In fact, in studying how, at the
end of inflation, the energy associated with inflation gets
converted into observable heat, one has to make sure that too much
energy is not lost into any unobserved degrees of freedom
\cite{warpedreheat}.

\item The main reason why inflation has been introduced is to
explain the unusual initial conditions of Hot Big Bang cosmology.
However, inflation itself can require special initial conditions
for some kinds of inflationary models. Therefore, for such models,
the full microscopic theory is required in order to understand the
origin of these initial conditions.

\item The study of cosmological implications of string theory,
like inflation, can shed some light into the better understanding
of the theory itself. We already have the experience with the
study of black hole backgrounds in string theory which has led to
some of the main successes of the theory, namely the explicit
calculation of the black hole entropy \cite{BHinST} and the
identification of the AdS/CFT correspondence \cite{AdSCFT}.

\item Inflationary models derived from string theory could be
characterised by the presence of purely stringy signals that
cannot arise in any low-energy effective quantum field theory.
Important examples of these phenomena which do not fully decouple
at low energies, are cosmic strings \cite{cosmicStrings}. They are
topological defects that get formed at the end of brane/anti-brane
inflation \cite{DDinflation, BBbarInfl} when the brane and
anti-brane annihilate.

\item Inflation could be a very effective way to test,
or at least to constrain, string theory thanks to the experimental
observation of signals which are generic in string-derived
effective Lagrangians, but highly unnatural from a conventional
field-theory point of view. As an example, in many, but
\textit{not} all, string inflationary models, the primordial
tensor signal is very small \cite{BMcA, SmallGW}. Hence, an
observation of primordial tensors would eliminate the great
majority of presently-known string inflationary models. In
addition, one might hope that in string theory not all of the
three-dimensional inflationary parameter space ($H_{inf}$,
$\epsilon$ and $\eta$) of the four-dimensional field theories, is
generated by varying the underlying parameters of the string
models through all of their allowed values. This would be an
attractive possibility if it were true, since it might permit a
definitive test of string-based inflation by observations.

\item Inflation could also be crucial to realise which vacuum
solution of the string landscape \cite{landscape0, landscape1,
landscape2, landscape3} is actually, if any, describing our real
world by the requirement of giving a reasonable description of
cosmology, combined with all the possible phenomenological
constraints coming from particle physics. It is very likely also
that cosmology will play a major r\^{o}le in addressing the
central question for string theorists related to understanding why
the Universe should end up being described by a particular
solution rather than by the many other possible solutions.
\end{itemize}

\subsection{Challenges for string inflation}

The first step in string cosmology is to specify a consistent
string compactification, including the total dimensionality, the
geometry and topology, the locations of any $D$-branes,
orientifold planes, and other localised sources, and the amount of
flux turned on through each cycle. Such a configuration would
uniquely specify a four-dimensional classical Lagrangian, and our
knowledge of this theory would be limited only by the accuracy of
the dimensional reduction, for example, by $\alpha'$ and $g_s$
corrections, or by backreaction effects from the localised
sources. Then this low-energy effective Lagrangian should be
capable of producing inflation that is consistent with current
observations.

However, this turns out to be a surprisingly difficult problem.
The main difficulty is that string compactifications invariably
involve more than one scalar field, and so are very complicated.
The four-dimensional potential depends, in general, on all the
moduli of the compactification, which parameterise the geometry of
the internal space. For a typical Calabi-Yau manifold, the number
of these moduli can easily reach the order of some hundreds
\cite{HubschBook}. Computing the full potential as a function of
all these fields is a formidable task, and even if one could
succeed, it would then be necessary to search through the
high-dimensional field space for a path along which the resulting
potential is sufficiently flat for inflation.

An essential point is that it does \textit{not} suffice to hold
fixed, by hand, all fields but one, and then find a path along
which the potential for that single field, $\phi_1$, is flat. One
reason is that the full potential will typically have a steep
downhill direction coinciding with one or more of these other
fields, $\phi_2,...,\phi_n$. If the steepest such direction is
more steep than the desired, and nearly flat, inflaton direction,
the full system will evolve by rolling downhill in this steepest
direction rather than along the putative inflaton direction
$\phi_1$. Thus, one must actually arrange that the potential has
positive curvature in the directions of $\phi_2,...,\phi_n$.

Once this has been achieved, one can search for special features
of the string-derived Lagrangian that might provide characteristic
signatures of the model. Let us now explain two general and
important obstacles in string inflation model-building: the
cosmological moduli problem \cite{CMP1, CMP2, CMP3} and the $\eta$
problem \cite{etaproblem}.

\subsubsection{The Cosmological Moduli Problem}

Any field $\chi$ with $0<m_{\chi}^2<\frac{3}{2}H^2$ will undergo
quantum fluctuations during inflation. These fluctuations carry
the field away from its minimum and hence lead to storage of
energy in $\chi$. After inflation, the field $\chi$ behaves like
non-relativistic matter and therefore its energy density decreases
with temperature as $\rho\sim T^{-3}$, whereas radiation decreases
faster: $\rho_r\sim T^{-4}$. Hence these fields quickly dominate
the energy density of the Universe, $\rho/\rho_r \sim 1/T$, as the
Universe cools down.

This causes a serious cosmological problem because if the scalar
field $\chi$ happens to be quite light and couples only
gravitationally (as moduli do), the $\chi$ particles will not have
decayed by the present day, and will overclose the Universe. On
the other hand, if $\chi$ is somewhat heavier, it will have
decayed during or after Big Bang Nucleosynthesis, spoiling the
delicate predictions of the light element abundances (such as
those of $^4$He and D nuclei). To avoid this `cosmological moduli
problem' \cite{CMP1, CMP2, CMP3}, one must therefore arrange that
$m_{\chi} > 10$ TeV, and preferably that $m_{\chi}^2 \gg H^2$.

At present there is no compelling solution to this problem. The
two major options are the dilution of these unwanted relics by a
late period of low-energy inflation caused by thermal effects
\cite{TI}, or by the entropy released by the non-thermal decay of
a heavier particle \cite{Acharya, Vafa}, which could be another
modulus not suffering from the cosmological moduli problem.

\subsubsection{The $\eta$ problem}

As we have already explained, inflation takes place whenever the
slow-roll conditions, $\epsilon\ll 1$ and $\eta\ll 1$, are
satisfied. Recalling the definition of the slow roll parameter
$\eta$ (\ref{definizionedieta}), and the fact that $V=3 H^2 M_P^2$
and $m_{\varphi}^2= V''$, we have:
\begin{equation}
\eta = \frac{m_{\varphi}^2}{3 H^2}.
\end{equation}
Thus, the slow-roll condition $\eta\ll 1$ can be turned into
$m_{\varphi}^2 \ll H^2$, and the inflaton mass being of order $H$
is equivalent to having $\eta \sim 1$.

The so-called `$\eta$ problem' \cite{etaproblem} then arises from
the fact that, in string theory, the inflaton is nothing but a
carefully-chosen modulus whose potential will generically be
affected by any mechanism of moduli stabilisation. In fact, the
natural expectation is that, whatever mechanism lifts the flat
directions of all the other moduli, will also lift the inflaton's
flat direction, implying $m_{\varphi} \sim H$, and so $\eta\sim
1$.

More precisely, the origin of the $\eta$ problem is the presence
of Planck-suppressed corrections which take the general form:
\begin{equation}
\Delta V = \frac{\mc{O}_4}{M_P^2}\varphi^2,
\end{equation}
for some operator $\mc{O}_4$ of dimension four, and lead to mass
terms:
\begin{equation}
\Delta m_{\varphi}^2 \propto \frac{\langle \mc{O}_4
\rangle}{M_P^2}.
\end{equation}
Then, if $\langle \mc{O}_4 \rangle \sim V=3 H^2 M_P^2$, these
contributions lead to $\Delta \eta \sim 1$.

In the context of models with a low-energy supergravity
description, the fact that one is often obliged to consider
moduli-stabilising energies of order the inflationary energy, can
be seen by looking at the $F$-term scalar potential:
\begin{equation}
\label{equ:vf} V_F = e^{K/M_P^2}\left(K^{i\bar{j}}D_i W
D_{\bar{j}}\overline{W} -\frac{3|W|^2}{M_P^2}  \right).
\end{equation}
Expanding $K$ around $\phi=\langle\phi\rangle+\varphi$ for small
$\varphi$, as:
\begin{equation}
K=K(\langle\phi\rangle)+\frac{\partial^2
K}{\partial\phi\partial\bar{\phi}}(\langle\phi\rangle)\varphi\bar{\varphi}+...,
\end{equation}
we find:
\begin{equation}
V_F =
e^{K(\langle\phi\rangle)/M_P^2}\left(1+\frac{1}{M_P^2}\frac{\partial^2
K}{\partial\phi\partial\bar{\phi}}(\langle\phi\rangle)
\varphi\bar{\varphi}+...\right)\left(K^{i\bar{j}}D_i
W D_{\bar{j}}\overline{W} -\frac{3|W|^2}{M_P^2}  \right).
\label{equ:vfe}
\end{equation}
Noticing that the canonically-normalised inflaton $\varphi_c$
obeys:
\begin{equation}
\partial\varphi_c\partial\bar{\varphi_c}\simeq
\frac{\partial^2
K}{\partial\phi\partial\bar{\phi}}(\langle\phi\rangle)\partial\varphi\partial\bar{\varphi},
\end{equation}
we see that the contribution to the mass term of $\varphi_c$ is
$\Delta m_{\varphi_c}^2 \simeq V_F(\langle\phi\rangle)/M_P^2 = 3
H^2$, so that $\Delta \eta \simeq 1$.

In some cases, there is a concrete framework for computing these
Planck-suppressed contributions to the potential. In $N=1$
supergravity, these corrections more often appear in the
K\"{a}hler potential than in the superpotential, because the
latter is protected by holomorphicity. String loop (see \cite{bhk,
bhp} and chapter 5 of this thesis) and $\alpha'$ \cite{bbhl}
corrections to the K\"{a}hler potential then lead to the relevant
Planck-suppressed operators.

In the absence of such information, appeals to fine-tuning are
necessary: one argues that in some restricted subset of possible
models, the net correction might be accidentally small.

Another proposed solution to the $\eta$ problem is to include
protective symmetries, like the axionic shift symmetry in type IIB
\cite{shiftInflation} . Although well-motivated, this approach has
been surprisingly difficult to achieve in explicit models, because
the desired symmetries do not always survive quantum corrections.

As we shall see later on, the no-scale property of the K\"{a}hler
potential in type IIB compactifications, helps to evade the $\eta$
problem, which, however, as we shall see in chapter 8, can find a
definite solution only via the discovery of the \textit{extended
no-scale structure}, which we described in chapter 5. This further
property of the string loop corrections to the K\"{a}hler
potential, will allow us to satisfy the slow-roll conditions
without any fine-tuning for the case where the inflaton is a
particular K\"{a}hler modulus (a K3 fiber modulus, namely).

However, at present, the $\eta$ problem is one of the most serious
constraints limiting our ability to construct explicit models of
string inflation.

\section{Models of string inflation}

As we have already pointed out, the fact that string
compactifications are generically characterised by the ubiquitous
presence of several massless scalars with effective gravitational
couplings, is one of the main problems that one faces in trying to
connect string theory with our real world. However, once a solid
mechanism for moduli stabilisation is found, it is actually
possible to turn this apparent hurdle into a virtue. In fact, all
the moduli of string compactifications then emerge as natural good
candidates for inflaton fields.

Of course, each direction of the whole scalar potential has to be
studied in detail, but experience during the last decade has shown
that flat directions suitable for inflation can indeed be found.
Because proposing a model involves identifying a scalar field as
an inflaton candidate, we can classify all the string inflationary
models in two big groups according to the origin of the inflaton
field \cite{Openclosedinflatons0, Openclosedinflatons1}.
\begin{itemize}
\item \textit{Open string models} are those in which the inflaton
is a scalar field arising from open strings ending on a $D$-brane.
Usually this scalar parameterises transverse motion of the
$D$-brane, and hence governs the location of the $D$-brane in the
compactification. In M-theory, there is a closely-related
alternative in which the inflaton corresponds to the position of
an $M5$-brane \cite{M5inflation}. Particular examples of open
string models are:
\begin{enumerate}
\item $D$-brane inflation \cite{DDinflation},

\item Slow roll warped brane/anti-brane inflation \cite{BBbarInfl},

\item DBI warped $D$-brane inflation \cite{DBI1, DBI2},

\item M-theory inflation \cite{M5inflation}.
\end{enumerate}

\item \textit{Closed string models} are those in which the inflaton
is a closed string modulus. These models are rather promising
since there is a well-defined choice of background for which a
subset of these moduli enjoy nearly flat potentials. Particular
examples of closed string models are:
\begin{enumerate}
\item K\"{a}hler moduli inflation \cite{kahler, KModInfl},

\item Axion inflation \cite{Nflation},

\item Wilson line moduli inflation \cite{WLInfl},

\item Volume modulus inflation \cite{cklq, volume}.
\end{enumerate}
\end{itemize}
Each of these models of string inflation gives a particular
prediction for the inflationary observables, which, in turn,
translates into a different dot in the $(r,n_s)$-plane. Given that
observational advances will certainly zoom in on a tiny fraction
of the present $(r,n_s)$-space, we hope to be able to rule out
some models rather soon. This would represent the first contact of
string theory with experimental testability.

Finally, we always have to keep in mind that inflation is not the
only paradigm describing the early Universe, since different
approaches, like the pre big-bang \cite{pbb} or the
ekpyrotic/cyclic scenario \cite{ekpyrotic}, have been put forward
as alternatives to inflation.

\subsection{Naturalness}

As far as the solution of the $\eta$ problem is concerned, it
seems that models of string inflation based on closed string modes
\cite{kahler, KModInfl} tend to be more promising that the ones
based on open moduli \cite{DDinflation, BBbarInfl}. In fact, in
the latter case, the slow-roll conditions can be satisfied only
via fine-tuning.

Let us see this in the illustrative example of the
$D3$/$\overline{D3}$-brane inflationary model \cite{BBbarInfl}. In
this case, the inflaton field has a geometrical interpretation as
the distance, in the extra dimensions, of the colliding worlds,
described by the brane and anti-brane respectively. Moreover this
scenario has a stringy mechanism to end inflation by the
appearance of an open string tachyon at a critical distance, so
providing a string theory realisation of hybrid inflation.
Furthermore, the annihilation of the branes comes with the
creation of cosmic strings which could be observable in the sky.
The fact that fine-tuning is needed in order to solve the
$\eta$-problem, in effective supergravity, can be seen as coming
from the K\"{a}hler potential which takes the form:
\begin{equation}
K=-3\ln\left[\left(T+\bar{T}\right)-\varphi\bar{\varphi}\right],
\end{equation}
where the distance between the $D3$ and $\overline{D3}$-branes is
related to $\varphi$ and $\left(T+\bar{T}\right)$ is related to
the volume of the compactification. Setting
$\left(T+\bar{T}\right)$ at its VEV, given by $\langle\left(
T+\bar{T}\right)\rangle$, there is a standard K\"{a}hler potential
$K=\varphi_c\bar{\varphi_c}$ for the canonically normalised
distance field:
\begin{equation}
\varphi_c=\frac{\varphi\sqrt{3}}{\sqrt{\langle\left(
T+\bar{T}\right)\rangle}},
\end{equation}
which comes from the expansion near the minimum of the volume:
\begin{equation}
K=-3\ln\left[\left(T+\bar{T}\right)-\varphi\bar{\varphi}\right]\simeq
-3\ln\left[\langle\left(
T+\bar{T}\right)\rangle\right]+\varphi_c\bar{\varphi_c}+...\,.
\end{equation}
The inflaton potential for the field $\varphi_c$, has a form $e^K
U$, where $U=|DW|^2-3|W|^2$. With
$e^K=e^{\varphi_c\bar{\varphi_c}}$, the $\eta$-problem is due to
the $e^K$ part of the potential. Because of this term, the second
derivative of the potential, is of order $H^2$, instead of $\sim
10^{-2}H^2$, as required by the flatness of the spectrum of
inflationary perturbations. It seems that this problem can be
cured only via fine-tuning.

On the other hand, in the case of moduli inflation \cite{kahler,
KModInfl}, the no-scale property of the scalar potential renders
inflation natural. In fact, in the case \cite{kahler}, where the
inflaton is a modulus measuring the volume of a blow-up cycle of
characteristic size $L$, the scalar potential takes the schematic
form:
\begin{equation}
\label{NaturalInflation} V(\varphi) = V_0  - A
\left(\frac{\varphi}{\mc{V}}\right)^{4/3} e^{- a
\mc{V}^{2/3}(\varphi/M_P)^{4/3}},
\end{equation}
where $V_0$, $A$ and $a$ are constants, and the canonically
normalised inflaton is given by $\varphi/M_P \sim
L^3/\mc{V}^{1/2}$. This potential has a slow roll provided that
$\mc{V}^{1/2}\varphi \gg M_P$, but the point is that this is
\textit{generic} to the domain of validity of the effective theory
because $\mc{V}^{1/2}\varphi\gg M_P$ corresponds to the condition
$L^3 \gg 1$ (in string units), which is a prerequisite for
trusting the effective field theory.

However, as we shall see in chapter 8, it turns out that string
loop corrections to the scalar potential (\ref{NaturalInflation})
destroy the flatness of the inflaton's direction. Starting from
the promising results of this scenario, still in chapter 8 of this
thesis, we shall derive a natural model of K\"{a}hler moduli
inflation where the inflaton is not a blow-up mode anymore, but a
fibration modulus. In this way, thanks to the extended no-scale
structure described in chapter 5, string loop corrections do not
destroy the flatness of the inflaton's potential anymore, but
represent the main effect that yields such a potential.

\subsection{Bounds on field ranges}

In all of the models of `large-field' inflation, the distance in
field space travelled by the inflaton, is large compared to the
Planck mass. As pointed out by Lyth \cite{Lyth}, and as we will
explain more in detail in section 8.1.1, this is a necessary
condition for the production of a detectably-large primordial
tensor signal. To realise any such large-field model in string
theory, one would need to find a trajectory in field space that is
large in Planck units, and along which the effective action is
suitable for inflation. This has proved to be very difficult. One
way to see this problem in the case of models with a low-energy
supergravity description, is to write the possible corrections to
the K\"{a}hler potential:
\begin{equation}
K=K_{cl}(\varphi,\bar{\varphi}) + M_P^2 \sum_{i}
{c_i}\left({\varphi\bar{\varphi}\over{M_P^2}}\right)^{1+i},
\end{equation}
where the dimensionless coefficients $c_i$ may be true constants
or may depend on other fields in the system. Unless the $c_i$ are
all very small, this series is badly divergent for $\varphi \gg
M_P$, and so over trans-Planckian distances, the metric on moduli
space is poorly-described by the classical metric on moduli space
derived from $K_{cl}$.

In certain specific contexts in string theory, we can compute the
field ranges in regimes appropriate to candidate inflation models.
An example of this is the field range for a $D3$-brane in a warped
throat region \cite{BBbarInfl}. It has been shown \cite{BMcA} that
for any sort of warped throat arising from a cone over an Einstein
manifold, the field range in Planck units is small. Similar bounds
are easily derived for $D3$-branes moving in toroidal
compactifications. This result implies that $D3$-brane inflation
in Calabi-Yau throats, or in most tori, cannot give rise to an
observably-large primordial tensor signal.

In closed string models, the field ranges correspond to distances
in the space of geometric moduli, not distances in the
compactification itself. An example of an infinite direction in
the moduli space is the decompactification direction. In the case
of low-energy supergravity, the K\"{a}hler potential depends on
the total volume $\mc{V}$ as:
\begin{equation}
K= - 2 M_P^2 \ln\mc{V},
\end{equation}
so that $ R=M_P \sqrt{2} \ln\mc{V}$ has a canonical kinetic term.
The range of $R$ between any fixed $\mc{V} \sim \mc{V}_0$ and the
limiting point $\mc{V} \to \infty$ is arbitrarily large. This
would seem to be a promising set-up for large-field inflation, but
it remains difficult to find a suitable inflaton potential along
this direction if one tries to avoid fine-tuning\footnote{The
authors of \cite{Nflation} managed to obtain a model of
large-field inflation by combining the displacements of $N \gg 1$
string axions into an effective displacement of a collective
field. However, it is not clear if this scenario is stable against
radiative corrections.}. As we shall see in chapter 8, instead of
considering the overall volume, but a specific modulus measuring
the volume of a K3 fiber, it is indeed possible to achieve a
stringy model of large-field inflation without the need to
fine-tune any parameter.

However, most closed string models involve potentials that are
flat over very small ranges of the canonical inflaton, and so they
predict a lower tensor signal, but they have the advantage that
physics associated with Planckian displacements plays no r\^{o}le.

\chapter{Fibre Inflation}
\label{FibreInflation}

\section{Preliminary considerations}

As we have seen in chapter 7, much progress has been made over the
past few years towards the goal of finding cosmological inflation
amongst the controlled solutions of string theory
\cite{GloriousHistory}. Part of the motivation for so doing has
been the hope that observable predictions might emerge that are
robust to all (or many) realisations of inflation in string
theory, but not generic to inflationary models as a whole. The
amplitude of primordial gravity waves has recently emerged as a
possible observable of this kind \cite{BMcA, SmallGW}, with
unobservably small predictions being a feature of most of the
known string-inflation proposals.

The prediction arises because the tensor amplitude is related (see
subsection \ref{lythsection}) to the distance traversed in field
space by the inflaton during inflation, and this turns out to have
upper limits in extant models, despite there being a wide
variation in the nature of the candidate inflaton fields
considered: including brane separation \cite{BBbarInfl}; the real
and imaginary parts of K\"{a}hler moduli \cite{kahler, KModInfl};
Wilson lines \cite{WLInfl}; the volume \cite{cklq, volume} and so
on. Furthermore, the same prediction appears also to be shared by
some of the leading proposed alternatives to inflation, such as
the cyclic/ekpyrotic models.

Since the observational constraints on primordial tensor
fluctuations are about to improve considerably --- with
sensitivity reaching down to $r \simeq 0.001$ (for $r = T/S$ the
ratio of amplitudes of primordial tensor and scalar fluctuations)
\cite{Verde, rBounds} --- it is important to identify precisely
how fatal to string theory would be the observation of primordial
gravity waves at this level. This has launched a search amongst
theorists either to prove a no-go theorem for observable $r$ from
string theory, or to derive explicit string-inflationary scenarios
that can produce observably large values of $r$. Silverstein and
Westphal \cite{SW} have taken the first steps along these lines,
proposing the use of monodromies in a particular class of IIA
string compactifications. In such models the inflaton field
corresponds to the position of a wrapped $D4$-brane that can move
over a potentially infinite range, thereby giving rise to
observably large tensor perturbations.

In this chapter we provide a concrete example of large field
inflation in the context of moduli stabilisation within the well
studied IIB string compactifications. Working within such a
framework allows us to use the well-understood properties of
low-energy four dimensional supergravity, with the additional
control this implies over the domain of validity of the
inflationary calculations.

More generally, we believe the inflationary model we propose to be
the simplest member of a broad new family of inflationary
constructions within the rich class of IIB stabilisations known as
the LARGE Volume Scenario (LVS) \cite{LVS}. Most useful for
inflationary purposes is the classification, within the LVS
framework, of the order in the $\alpha'$ and string-loop
expansions that governs the stabilisation of the various K\"ahler
moduli for general IIB Calabi-Yau compactifications. In
particular, in chapter 6, it was found that for K3-fibred
Calabi-Yaus, LVS moduli stabilisation only fixes the overall
volume and blow-up modes if string loop corrections to $K$ are
ignored. The fibre modulus --- call it $\Omega$, say -- then
remains with a flat potential that is only lifted once string loop
corrections are also included. Consequently $\Omega$ has a flatter
potential than does the overall volume modulus, making it
systematically lighter, and so also an attractive candidate for an
inflaton. Our proposal here is the first example of the family of
inflationary models which exploits this flatness mechanism, and
which we call {\it Fibre Inflation}.

This class of models is also attractive from the point of view of
obtaining large primordial tensor fluctuations. This is because
the relatively flat potential for $\Omega$ allows it to traverse a
relatively large distance in field space compared with other
K\"ahler moduli. In this chapter we use these LVS results to
explicitly derive the inflaton potential in this scenario, where
the range of field values is large enough to easily give rise to
60 \efold ings of slow-roll inflation.

Unlike most string-inflation models (but similar to K\"ahler
modulus inflation (KMI) \cite{kahler}) slow roll is ensured by
large field values rather than tuning amongst parameters in the
potential. Most interestingly, within the inflationary regime all
unknown potential parameters appear only in the normalisation of
the inflaton potential and not in its shape. Consequently,
predictions for the slow-roll parameters (and for observables
determined by them) are completely determined by the number of
\efold ings, $N_e$, between horizon exit and inflation's end.
Elimination of $N_e$ then implies the slow-roll parameters are
related by $\varepsilon \simeq \frac32 \, \eta^2$, implying a
similar relation between $r$ and the scalar spectral tilt: $r
\simeq 6(n_s - 1)^2$. [By contrast, the corresponding predictions
for KMI are $\varepsilon \simeq 0$ and so $r \simeq 0$, leaving
$n_s \simeq 1 + 2\eta$.]

Since the value of $N_e$ depends somewhat on the post-inflationary
reheat history, the precise values of $r$ and $n_s$ are more model
dependent, with larger $N_e$ implying smaller $r$. In a simple
reheat model (described in more detail below) $N_e$ is correlated
with the reheat temperature, $T_{rh}$, and the inflationary scale,
$M_{inf}$, through the relationship:
\begin{equation}
 N_e \simeq 62 + \ln\left(\frac{M_{inf}}{10^{16}
 \textrm{GeV}}\right) - \frac{\left(1-3w\right)}{3\left(1+w\right)}
 \ln\left(\frac{M_{inf}}{T_{rh}}\right) \,,
\end{equation}
where $w = p/\rho$ parameterises the equation of state during
reheating. Numerically, choosing $M_{inf}\simeq 10^{16}$ GeV and
$T_{rh}\simeq 10^{9}$ GeV (respectively chosen to provide
observably large primordial scalar fluctuations, and to solve the
gravitino problem), we find that $N_e \simeq 58$, and so $n_s
\simeq 0.970$ and $r \simeq 0.005$. Tensor perturbations this
large would be difficult to see, but would be within reach of
future cosmological observations like EPIC, BPol or CMBPol
\cite{Verde, rBounds}.

Our preliminary investigations reveal several features likely to
be common to the broader class of Fibre Inflation models. On one
hand, as already mentioned, slow roll is ultimately controlled by
the large values of the moduli rather than on the detailed tuning
of parameters in the scalar potential. On the other hand, large
volumes imply low string scales, $M_s$, and this drives down the
inflationary scale $M_{inf}$. This is interesting because it may
lead to inflation even at low string scales but could be  a a
problem inasmuch as it makes it more difficult to obtain large
enough scalar fluctuations to account for the primordial
fluctuations seen in the CMB. (It also underlies the well-known
tension between TeV scale supersymmetry and the scale of inflation
\cite{cklq, kl}.) This suggests studying alternative methods to
generate density fluctuations within these models,\footnote{We
thank Toni Riotto for numerous discussions of this point.}, to
allow lower inflationary scales to co-exist with observably large
primordial fluctuations. Although fluctuations generated in this
way would not produce large tensor modes, they might be testable
through their predictions for non-gaussianities.

The biggest concern for Fibre Inflation and K\"{a}hler Modulus
Inflation is whether higher-loop contributions to the potential
might destabilise slow roll. In KMI this problem arises already at
one loop, and leads to the requirement that no branes wrap the
inflationary cycle (from which the dangerous contributions arise).
Fibre Inflation models do not have the same problems, and this is
likely to simplify greatly the ultimate reheating picture in these
models. They may yet have similar troubles once contributions from
blow-up modes or higher loops can be estimated,\footnote{We thank
Markus Berg for conversations about this.} but we find that
current best estimates for these corrections are not a problem.

Finally, it is relatively simple in these models to obtain large
hierarchies amongst the size of the moduli, in a way that leads to
some dimensions becoming larger than others (rather than making
the extra dimensions into a frothy Swiss cheese). This potentially
opens up the possibility of `sculpting' the extra dimensions, by
having some grow relatively slowly compared to others as the
observed four dimensions become exponentially large.

After a short digression, next, summarising why large $r$ has
proven difficult to obtain in past constructions, and a brief
review of KMI, the remainder of this chapter is devoted to
explaining Fibre Inflation, and why it is possible to obtain in it
$r \simeq 0.005$. After describing the case of a 3-parameter K3
fibration whose volume takes a simplified form with respect to the
more general case studied in chapter 4 and 6, section 8.2 focuses
on the inflationary potential to which these K3-fibered
Calabi-Yaus give rise. Finally, in our section 8.3 we conclude
with a brief summary and discussion of our results.

\subsection{The Lyth bound}
\label{lythsection}

What is so hard about obtaining observably large primordial tensor
fluctuations in string constructions? In 1996 David Lyth
\cite{Lyth} derived a general correlation between the ratio $r$
and the range of values through which the (canonically normalized)
inflaton field, $\varphi$, rolls in single-field slow-roll models:
\be
    r = 16 \, \varepsilon = \frac{8}{N_{\rm eff}^2}
    \left(\frac{\Delta\varphi}{M_P}\right)^2 \,,
\ee
where $\varepsilon = \frac12 (V'/V)^2$ is the standard first
slow-roll parameter, and:
\be
    N_{\rm eff} = \int_{t_{\rm he}}^{t_{\rm end}}
    \left( \frac{\xi}{r} \right)^{1/2}
    H \exd t \,.
\ee
Here $\xi(t) = 8(\dot\varphi/H M_P)^2$ is the quantity whose value
at horizon exit gives the observed tensor/scalar ratio, $r =
\xi(t=t_{\rm he})$, $H(t) = \dot a/a$ is (as usual) the Hubble
parameter, and the integral runs over the $N_e \gtrsim 50$ \efold
ings between horizon exit and inflation's end. Notice in
particular that $N_{\rm eff} = N_e$ if $\xi$ is a constant. Lyth's
observation was that the validity of the slow roll and
measurements of the scalar spectral index, $n_s-1$, constrain
$N_{\rm eff} \gtrsim 50$, and so $r \gtrsim 0.01$ requires the
inflaton to roll through a trans-Planckian range, $\Delta \varphi
\gtrsim M_P$.

This observation has proven useful because the inflaton usually
has some sort of a geometrical interpretation when inflationary
models are embedded into string theory, and this allows the
calculation of its maximum range of variation. For instance, as we
have already mentioned in chapter 7, suppose inflation occurs due
to the motion of the position, $x$, of a brane within 6 extra
dimensions, each of which has length $L$. Then expressing the
geometric upper limit $\Delta x < L$ in terms of the canonically
normalised inflaton field, $\varphi = M_s^2 x$, gives $\Delta
\varphi/M_P < M_s^2 L/M_P$, where $M_s$ is of order the string
scale. But $L$ is not itself independent of $M_s$ and the four
dimensional Planck constant, $M_P$. For instance, in the absence
of warping one often has $M_P^2 \simeq M_s^8 L^6$, which allows
one to write $\Delta \varphi/M_P < (M_s/M_P)^{2/3}$. Finally,
consistency of calculations performed in terms of a
(higher-dimensional) field theory generally require the hierarchy,
$1/L \ll M_s$ which implies $M_s/M_P \simeq (M_s L)^{-3} \ll 1$,
showing that $\Delta \varphi/M_P \ll 1$.

More careful estimates of brane motion within an extra-dimensional
throat, with the condition that it geometrically cannot move
further than the throat itself is long, lead to similar
constraints \cite{BMcA}. It is considerations such as these that
show (on a case-by-case basis) for each of the extant
string-inflation constructions that the distance moved by the
inflaton is too small to allow $r \gtrsim 0.01$. However, in the
absence of a no-go theorem, there is strong motivation to find
stringy examples which evade these kinds of constraints, and allow
the inflaton to undergo large excursions.

\subsection{K\"{a}hler modulus inflation}

In this subsection, we briefly review the mechanism of K\"{a}hler
moduli inflation \cite{kahler}, since many of the features of the
model presented here draw on this example. The starting point for
this model is a Swiss cheese Calabi-Yau manifold, which must have
at least \textit{two} blow-up modes ($N_{small} \ge 2$ and so
$h_{1,1} \ge 3$), such as is true, for instance, for the
$\mathbb{C} P_{[1,3,3,3,5]}$ model \cite{blumenhagen} described in
chapters 4 and 6.

Assuming the minimal three K\"ahler moduli of this form, our
interest is in that part of moduli space for which these satisfy
$\tau_b \gg \tau \gg \tau_s$, where $\tau$ and $\tau_s$ are the
blow-up modes while $\tau_b$ controls the overall volume. As a
first approximation neglect string loop corrections as well as
exponentials of the large moduli $\tau_b$ and $\tau$ in $V$. Then
one finds that $\tau_b$ and $\tau_s$ can both be stabilised with
$\mathcal{V} \sim e^{a_s \tau_s}$ and $\tau_s \gg 1$.

Fixing these to their stabilised values, but now considering the
subdominant dependence on $\tau$, the potential for the remaining
modulus takes the form:
\be
 V= A \frac{\sqrt{\tau}\,e^{-2a\tau}}{\mc{V}}-B\frac{\tau
 e^{-a\tau}}{\mc{V}^2}+C\frac{\hat\xi}{\mc{V}^3} \,,
\ee
where the volume $\mc{V}$ should be regarded as being fixed.
Varying $\tau$ with $\mc{V}$ fixed (this is the reason why
$h_{11}\geq 3$ is needed), the potential for large $\tau$ is
dominated for the last two terms, which is naturally very flat due
to its exponential form.

The above potential gives rise to slow-roll inflation,
\textit{without} the need for fine-tuning parameters in the
potential. For the canonically normalised inflaton: \be \varphi =
\sqrt{4 \lambda / (3\mc{V})} \; \tau^{3/4}, \ee
 the above potential becomes:
\be
 V\simeq V_0 -\beta\left(\frac{\varphi}{\mc{V}}\right)^{4/3}\,
 e^{-a'\mc{V}^{2/3}\varphi^{4/3}} \,,
\ee
with $V_0 \sim \hat\xi/\mc{V}^3$. This is very similar to textbook
models of large-field inflation \cite{liddle}, although with the
search for observably large tensor modes in mind, one must also
keep in mind an important difference. This is because although in
both cases slow roll requires a large argument for the exponential
in the potential, this is accomplished differently in the two
cases. In the textbook examples the argument of the exponential is
typically given by $\varphi/M_P$, and so slow roll requires
$\varphi \gtrsim M_P$. In the present case, however, slow roll is
typically accomplished for small $\varphi$, due to the large
factor of $\mc{V}^{2/3}$ in the exponent. In this crucial way,
what we have is actually a small-field model of inflation.

\subsubsection{Naturalness}

It is remarkable that this is one of the only string-inflation
models that does not suffer from the $\eta$ problem, inasmuch as
slow roll does not require a delicate adjustment amongst the
parameters in the scalar potential.
However, one worries that the extreme flatness of the potential
might be affected by sub-leading corrections not yet included in
the scalar potential, such as string-loop corrections to the
K\"{a}hler potential.

Although a definitive analysis requires performing a string loop
calculation, some conclusion may be drawn using the conjectured
modulus dependence \cite{bhp} discussed in chapter 5. In fact,
examination of the previous formulae shows that dangerous
contributions can arise if $D7$-branes wrap the inflationary
cycle, since in this case string-loop corrections take the form:
\be
 \delta V_{1-loop}\sim \frac{1}{\sqrt{\tau} \; \mc{V}^3}
 \sim \frac{1}{\varphi^{2/3} \mc{V}^{10/3}} \,.
\ee
This is dangerous because it gives a contribution to the slow-roll
parameter, $\eta = M_P^2 V''/V$, of the form $\delta \eta \sim
M_P^2 \delta V''/V_0 \sim \varphi^{-8/3} \mc{V}^{-1/3}
\hat\xi^{-1}$, which for the typical values of interest, $\varphi
\sim \mc{V}^{-1/2} \ll 1$, may be large.

One way out of this particular problem is simply not to wrap
$D7$-branes about the inflationary cycle. In this case the
remaining loop corrections discussed above do not destroy the slow
roll. (Although it is not yet possible to quantitatively
characterise the contributions of higher loops, see appendix B.2
for a related discussion of some of the issues.) Of course, if
ordinary Standard Model degrees of freedom reside on a $D7$, not
wrapping $D7$-branes on the inflationary cycle is likely to
complicate the eventual reheating mechanism because it acts to
decouple the inflaton from the observable sector. However we do
not regard this particular objection as being too worrisome, since
a proper study of reheating in these (and most other models) of
string inflation remains a long way off \cite{warpedreheat}.

\section{Fibre Inflation}

We now return to our main line of argument, and describe the
simplest K3-Fibration inflationary model. We regard this model as
being a representative of a larger class of constructions (Fibre
Inflation), which rely on choosing the inflaton to be one of those
K\"ahler moduli whose potential is first generated at the
string-loop level.\footnote{Even though these moduli are also
K\"ahler moduli, their behaviour is very different from the volume
and in particular the blowing-up modes that drive K\"ahler moduli
inflation. In this sense the previous scenario might be more
properly called `blow-up inflation' to differentiate it from the
later `volume' inflation \cite{volume} and `fibre' inflation
developed here.}

\subsection{The simplest K3 fibration Calabi-Yau} \label{examples}

To describe the model we first require an explicit example of a
Calabi-Yau compactification which has a modulus that is not
stabilised by nonperturbative corrections to $W$ together with the
leading $\alpha'$ corrections to $K$. The simplest such examples
are given by Calabi Yaus which have a K3 fibration structure such
as those described in chapters 4 and 6. However, in this case, we
shall focus on a simplified version of these constructions, which
represents the simplest possible K3 fibration and can be thought
of as a particular case of the more general Calabi-Yaus described
in chapters 4 and 6. Therefore we briefly describe here the main
features of this compactification manifold.

For our present purposes, the simplest K3 fibered Calabi-Yau can
be regarded as one whose volume is linear in one of the 2-cycle
sizes, $t_j$ \cite{oguiso}. That is, when there is a $j$ such that
the only non-vanishing coefficients are $k_{jlm}$ and $k_{klm}$
for $k,l,m \neq j$, then the Calabi-Yau manifold is a K3 fibration
having a $\mbb{C}P^1$ base of size $t_j$, and a K3 fibre of size
$\tau_j$. The simplest such K3 fibration has two K\"ahler moduli,
with $\mc{V}= \tilde t_1 \tilde t_2^{\,2} + \frac{2}{3}\tilde
t_2^{\,3}$. This becomes $\mc{V} = \frac12 \sqrt{\tilde\tau_1}
\left( \tilde\tau_2 - \frac23 \tilde\tau_1 \right)$ when written
in terms of the 4-cycle volumes $\tilde\tau_1 = \tilde t_2^{\,2}$
and $\tilde\tau_2 = 2(\tilde t_1 + \tilde t_2) \tilde t_2$,
corresponding to the geometry $\mbb{C}P^4_{[1,1,2,2,6]}(12)$
\cite{Candelas}. For later convenience we prefer to follow a
slightly different basis of cycles in this geometry:
\begin{equation}
 \tau_1 = \tilde\tau_1, \textit{ \ \ \ \ \
 } \tau_2 = \tilde\tau_2 - \frac{2}{3}\tilde \tau_1,
\end{equation}
with a similar change in the 2-cycle basis, $\{ \tilde t_i \} \to
\{ t_i \}$. In terms of these the overall volume becomes:
\begin{equation}
 \mc{V} = t_1 t_2^2
 = \frac{1}{2}\sqrt{\tau_1} \; \tau_2\textit{ \ \ \
 }\Leftrightarrow\textit{ \ \ \ } \mc{V} = t_1\tau_1,
\end{equation}
where $t_1$ is the base and $\tau_1$ the K3 fiber.

For inflationary purposes we also require a third K\"ahler
modulus, which we can achieve by simply adding an extra blow-up
mode, as is required in any case to guarantee the existence of
controlled large volume solutions. We therefore begin by assuming
a compactification whose volume is given in terms of its three
K\"ahler moduli in the following way:
\begin{equation}
 \mathcal{V} = \lambda_{1}t_{1}
 t_{2}^{2} + \lambda_{3}t_{3}^{3}
 = \alpha \left( \sqrt{\tau _{1}}\tau _{2} - \gamma \tau _{3}^{3/2}\right)
 = t_1\tau_1-\alpha\gamma\tau_3^{3/2},  \label{hhhoi}
\end{equation}
where the constants $\alpha $ and $\gamma $ are given in terms of
the model-dependent numbers, $\lambda_i$, by $\alpha = \frac12 \,
\lambda_1^{-1/2}$ and $\gamma = (4\lambda_1/27 \lambda_3)^{1/2}$,
related to the two independent intersection numbers, $d_{122}$ and
$d_{333}$, by $\lambda_1 = \frac12 \, d_{122}$ and $\lambda_3 =
\frac16 \, d_{333}$. (Clearly, including more blow-up modes than
we have done here is straightforward.) Given that (\ref{hhhoi})
simply expresses the addition of the blow-up mode $\tau_3$, to the
geometry $\mathbb{C}P^{4}_{[1,1,2,2,6]}(12)$ described in chapter
4, we do not expect there to be any obstruction to the existence
of a Calabi-Yau manifold with these features.

We further assume that $h_{2,1}(X) > h_{1,1}(X) = 3$, thus
satisfying the other general LVS condition. Since we seek
stabilisation with $\mc{V}$ large and positive, we work in the
parameter regime:
\begin{equation} \label{V0hieroi}
 \mc{V}_0 := \alpha\sqrt{\tau_1}
 \; \tau_2 \gg \alpha\gamma\tau_3^{3/2} \gg 1 \,,
\end{equation}
with the constant $\gamma$ taken to be positive and order unity.
This limit keeps the volume of the Calabi-Yau large, while the
blow-up cycle remains comparatively small. Regarding the relative
size of $\tau_1$ and $\tau_2$, we consider two situations in what
follows: $\tau_2 \gtrsim \tau_1 \gg \tau_3$ and $\tau_2 \gg \tau_1
\gg \tau_3$. (We notice in passing that the second case
corresponds to $t_1 \sim \tau_2/\sqrt{\tau_1} \gg t_2 \sim
\sqrt{\tau_1} \gg t_3 \sim \sqrt{\tau_3}$, corresponding to
interesting geometries having the two dimensions of the base,
spanned by the cycle $t_1$, hierarchically larger than the other
four of the K3 fibre, spanned by $\tau_1$.) The similarity of eq.
(\ref{V0hieroi}) with the `Swiss cheese' Calabi-Yaus:
\begin{equation}
\mathcal{V}=\alpha \underset{\tau
_{big}^{3/2}}{(\underbrace{\sqrt{\tau _{1}}\tau _{2}}}-\gamma \tau
_{3}^{3/2}) \,,
\end{equation}
leads us to expect (and our calculations below confirm) that the
scalar potential has a minimum at exponentially large volume,
together with $( h_{1,1} - N_{small} - 1) = 1$ flat directions.

\subsubsection{The potential without string loops}
\label{3modK3noLoopCalc}

We start by considering the scalar potential computed using the
leading $\alpha'$ corrections to the K\"{a}hler potential, as well
as including nonperturbative corrections to the superpotential.
\begin{equation} \label{3}
 K = K_{0} + \delta K_{(\alpha')} = -2 \ln \left(
 \mathcal{V} + \frac{\hat{\xi}}{2} \right) \qquad \hbox{and}
 \qquad
 W = W_{0} + \sum_{k=1}^3 A_{k} e^{-a_{k}T_{k}} \,.
\end{equation}
Because our interest is in large volume $\mc{V}_0 \gg
\alpha\gamma\tau_3^{3/2}$, we may to first approximation neglect
the dependence of $T_{1,2}$ in $W$ and use instead:
\begin{equation} \label{spoi}
 W \simeq W_{0} + A_{3} e^{-a_{3} T_{3}} \,.
\end{equation}
In the large volume limit the K\"{a}hler metric and its inverse
become:
\begin{equation}
 K_{i\bar\jmath}^{0}=\frac{1}{4\tau_2^2}\left(
 \begin{array}{ccccc}
 \frac{\tau_2^2}{\tau _{1}^{2}} && \gamma\left(
 \frac{\tau_{3}}{\tau_{1}}\right)^{3/2} && -\frac{3 \gamma
 }{2}\frac{\sqrt{\tau _{3}}}{
 \tau _{1}^{3/2}}\tau_{2} \\
 \gamma\left(\frac{\tau_{3}}{\tau_{1}}\right)^{3/2} && 2 &&
 -3\gamma \frac{\sqrt{\tau_{3}}}{\sqrt{\tau_{1}}} \\
 -\frac{3\gamma }{2}\frac{\sqrt{\tau_{3}}}{\tau_{1}^{3/2}}\tau_{2} && -3\gamma
 \frac{\sqrt{\tau_{3}}}{\sqrt{\tau_{1}}} && \frac{3\alpha \gamma
 }{2}\frac{\tau_2^2}{\mathcal{V}\sqrt{\tau_{3}}}
 \end{array}
 \right), \label{LaDirettaoi}
\end{equation}
and:
\begin{equation}
 K_{0}^{\bar\imath j}=4\left(
 \begin{array}{ccccc}
 \tau _{1}^{2} && \gamma\sqrt{\tau_{1}}\tau_{3}^{3/2} &&
 \tau _{1}\tau _{3} \\
 \gamma\sqrt{\tau_{1}}\tau_{3}^{3/2} && \frac12
 \, \tau_2^2 &&
  \tau_2 \tau _{3} \\
 \tau _{1}\tau _{3} && \tau_2
 \tau_{3} && \frac{2}{3\alpha \gamma }\mathcal{V}\sqrt{ \tau _{3}}
 \end{array}
 \right) \,,  \label{Kinverseoi}
\end{equation}
where we systematically drop all terms that are suppressed
relative to those shown by factors of order
$\sqrt{\tau_3/\tau_2}$. In particular, here (and below), $\mc{V}$
now denotes $\mc{V}_0 = \alpha \sqrt{\tau_1}\tau_2$ rather than
the full volume, $\mc{V}_0 - \alpha \gamma \tau_3^{3/2}$.

We now use these expressions in (\ref{scalarWnp}), adding the
linearisation of $\delta V_{(\alpha')}$ in $\hat\xi$, eq.
(\ref{scalaralpha'}). The following identity (to the accuracy of
eqs. (\ref{LaDirettaoi}) and (\ref{Kinverseoi})) proves very
useful when doing so:
\be \label{oddKid}
    K_0^{3\bar 1} K^0_{\bar 1} + K_0^{3\bar 2} K^0_{\bar 2} +
    \hbox{c.c.} = -3\tau_3 \,.
\ee
The result may be explicitly minimised with respect to the $T_3$
axion direction, $b_3 = \hbox{Im} \, T_3$, with a minimum at $b_3
= 0$ if $W_0 < 0$ or at $b_3 = \pi/a_3$ if $W_0 > 0$. Once this is
done, the resulting scalar potential simplifies to:
\begin{equation}
 V = \frac{8 \, a_{3}^{2}A_3^2}{3\alpha\gamma}
 \left( \frac{\sqrt{\tau _{3}}}{\mathcal{V}}
 \right) e^{-2a_{3}\tau_3}
 -4 W_{0}a_{3} A_3 \left( \frac{\tau _{3}}{\mathcal{V}^{2}}
 \right) \, e^{-a_{3}\tau_3}
 +\frac{3 \, \hat\xi W_0^2}{4 \mathcal{V}^{3}} \,,
 \label{ygfdooi}
\end{equation}
where we take $W_0$ to be positive and neglect terms that are
subdominant relative to the ones displayed by inverse powers of
$\mc{V}$ without compensating powers of $e^{a_3\tau_3}$.

\begin{figure}[ht]
\begin{center}
\epsfig{file=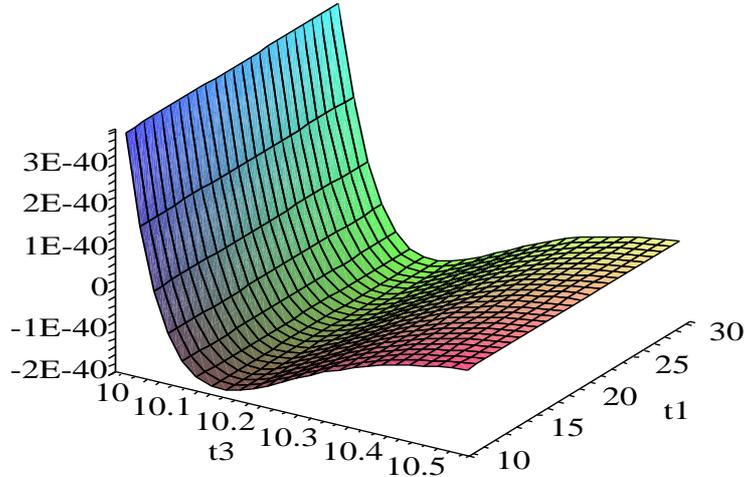, height=70mm,width=100mm} \caption{$V$
(arbitrary units) versus $\tau_1$ and $\tau_3$ for one of the
parameter sets discussed in the text, with $\mathcal{V}$ evaluated
at its minimum.} \label{Fig:sofa}
\end{center}
\end{figure}

Now comes the main point. Notice that by virtue of (\ref{oddKid}),
$V$ depends only on two of the three moduli on which it could have
depended: $V=V(\mc{V},$\ $\tau_{3})$. This occurs because we take
$a_1 \tau_{1}$ to be large enough to switch off its
non-perturbative dependence in $W$. This observation has two
consequences: First, it implies that within these approximations
there is one modulus
--- any combination (call it $\Omega$, say) of $\tau_1$ and
$\tau_2$ independent of $\mc{V}$ --- which describes a direction
along which $V$ is (so far) completely flat. This plays the
r\^{o}le of our inflaton in subsequent sections.

Second, the potential (\ref{ygfdooi}) completely stabilises the
combinations $\tau_3$ and $\mc{V}$ (and, in fact, has precisely
the same form as the scalar potential of the original
$\mathbb{C}P_{[ 1,1,1,6,9]}^{4}(18)$ LVS example of \cite{LVS}).
In particular, the only minimum satisfying $a_3\tau_3 \gg 1$ is
given explicitly by $\mc{V} = \langle \mc{V} \rangle$ and $\tau_3
= \langle \tau_3 \rangle$ with:
\begin{equation}
 \langle \tau _{3}\rangle = \left(
 \frac{\hat\xi}{2\, \alpha \gamma } \right) ^{2/3}
 \qquad \hbox{and} \qquad
 \langle
 \mathcal{V}\rangle = \left( \frac{ 3 \,\alpha \gamma }{4a_{3}A_3}
 \right) W_0 \, \sqrt{\langle \tau _{3}\rangle }
 \; e^{a_{3} \langle \tau_3
 \rangle }\text{\ .}  \label{xx}
\end{equation}
This is the minimum corresponding to exponentially large
volume\footnote{The two relations (\ref{xx}) do not take into
account the shift in the volume minimum due to the up-lifting
term, which are worked out explicitly in appendix B.1 (and
incorporated in our numerics).}.

The flat direction of the potential (\ref{ygfdooi}) is manifest in
Figure \ref{Fig:sofa}, which plots this scalar potential with
$\mc{V}$ fixed (using the LV parameter set discussed below), as a
function of $\tau_3$ (on the \textit{x}-axis) and $\Omega$
--- which represents any third field coordinate independent of
$\tau_3$ and $\mc{V}$ (such as $\tau_1$, for instance) --- (on the
\textit{y}-axis). In order properly to understand the potential
for $\Omega$, we must go beyond the approximations that underly
eq. (\ref{ygfdooi}), in order to lift this flat direction, such as
by including the leading string-loop contributions to the
potential.

\subsubsection*{Sample parameter sets}

In what follows it is useful to follow some concrete numerical
choices for the various underlying parameters. To this end we
track several sets of choices throughout this chapter, listed in
Table 8.1. One of these sets (call it `LV') gives very large
volumes, $\mc{V} \simeq 10^{13}$ (and so a string scale of order
$M_s \propto \mc{V}^{-1/2} \sim 10^{12}$ GeV), and is
representative of what the LARGE volume scenario likes to give for
simple choices of parameters. The others (`SV1' and `SV2', say)
instead have $\mc{V} \sim 10^3$ much smaller (and so with $M_s
\sim 10^{16}$ GeV). While all naturally provide an inflationary
regime, the LV choice has the disadvantage that the value of the
classical inflationary potential turns out too small to provide
observable primordial density fluctuations. The other choices are
chosen to remedy this problem, and to provide illustrations of
different inflationary parameter regimes. We regard all of these
choices as being merely illustrative, and have not attempted to
perform a systematic search through the allowed parameter space.

\begin{figure}[ht]
\begin{center}
\begin{tabular}{c||c|c|c}
  & LV & SV1 & SV2 \\
  \hline\hline
  $\lambda_1$ & 1 & 15 & 21/2 \\
  $\lambda_3$ & 1 & 1/6 & 1/6 \\
  $g_s$ & 0.1 & 0.3 & 0.3 \\
  $\xi$ & 0.409 & 0.934 & 0.755 \\
  $W_0$ & 1 & 100 & 100 \\
  $a_3$ & $\pi$ & $\pi/5$ & $\pi/4$ \\
  $A_3$ & 1 & 1 & 1 \\
  \hline
  $\alpha$ & 0.5 & 0.1291 & 0.1543 \\
  $\gamma$ & 0.385 & 3.651 & 3.055 \\
  $\langle\tau_3\rangle$ & 10.46 & 4.28 & 3.73 \\
  $\langle\mc{V}\rangle$ & $2.75 \cdot 10^{13}$ & 1709.55 & 1626.12
\end{tabular}\\
\vspace{0.3cm}{{\bf Table {8.1}:} Some model parameters (the
up-lifting to a Minkowski minimum has been taken into account).}
\end{center}
\end{figure}

\subsubsection{Inclusion of string loops} \label{bene}

We now specialise the string-loop corrections to the K3 fibration
of interest, using expression (\ref{1 loop}) and working in the
regime $W_{0}\gtrsim\mathcal{O}(1)$ where the perturbative
corrections are important.

Consider first the contribution coming from stacks of $D7$-branes
wrapping the blow-up cycle, $\tau_{3}$. The Kaluza-Klein loop
correction to $V$ coming from this wrapping takes the form:
\begin{equation}
 \delta V^{KK}_{(g_{s}),\tau _{3}}=
 \frac{g_{s}^{2}(\mathcal{C}_{3}^{KK})^{2}}
 {\sqrt{\tau_{3}} \; \mathcal{V}^{3}}, \label{eq3oi}
\end{equation}
which does not depend on $\Omega$, and is subdominant to the
$\alpha'$ correction. These features imply such a term may modify
the exact locus of the potential's minimum, but not the main
features of the model, such as the existence of the flat direction
in $\Omega$ and the minimisation of $\mc{V}$ at exponentially
large values.

Similarly, we have seen that the winding-mode contributions to
string-loop corrections arise from the exchange of closed winding
strings at the intersection of stacks of $D7$-branes. But the form
of the volume (\ref{hhhoi}) shows that the blow-up mode,
$\tau_{3}$, only has its triple self-intersection number
non-vanishing, and so does not intersect with any other cycle.
This is a typical feature of a blow-up mode which resolves a
point-like singularity: due to the fact that this exceptional
divisor is a \textit{local} effect, it is always possible to find
a suitable basis where it does not intersect with any other
divisor. Hence the topological absence of the required cycle
intersections implies an absence of the corresponding
winding-string corrections. In the end, only three types of loop
corrections turn out to arise:
\begin{equation}
 \delta V_{(g_{s})}=\delta V^{KK}_{(g_{s}), \tau_{1}}+\delta
 V^{KK}_{(g_{s}), \tau_{2}}+\delta V^{W}_{(g_{s}),
 \tau_{1}\tau_{2}},
\end{equation}
which have the form:
\begin{eqnarray}
 \delta V_{(g_{s}),\tau _{1}}^{KK} &=&g_{s}^{2} \frac{\left(
 C_{1}^{KK}\right) ^{2}}{\tau _{1}^{2}}
  \frac{W_{0}^{2}}{\mathcal{V}^{2}}, \notag \\
 \delta V_{(g_{s}),\tau _{2}}^{KK} &=& 2g_{s}^{2}\frac{\left( C_{2}^{KK}\right)^{2}}{\tau_2^2}
 \frac{W_{0}^{2}}{\mathcal{V}^{2}}, \label{LOOPoi} \\
 \delta V_{(g_{s}),\tau _{1}\tau _{2}}^{W}
 &=&- \left( \frac{2 \, C_{12}^{W}}{t_{\ast }} \right)
 \frac{W_{0}^{2}}{\mathcal{V}^{3}} \,.
 \notag
\end{eqnarray}
Here the 2-cycle $t_{*}$ denotes the intersection locus of the two
4-cycles whose volumes are given by $\tau_{1}$ and $\tau_{2}$. In
order to work out the form of $t_{*}$, we need the relations:
\begin{equation}
 \tau _{1} = \frac{\partial \mathcal{V}}{\partial t_{1}}
 = \left( \lambda_{1} t_{2}\right) t_{2}
 \qquad\hbox{and} \qquad
 \tau _{2} = \frac{\partial \mathcal{V}}{\partial t_{2}}
 =2t_{1}(\lambda _{1}t_{2}),
 \label{tausoi}
\end{equation}
and so $t_{*} =\lambda_1 t_{2} = \sqrt{\lambda_{1}\tau_1}$.
Therefore the $g_{s}$ corrections to the scalar potential
(\ref{LOOPoi}) take the general form:
\begin{equation}
 \delta V_{(g_{s})}= \left(\frac{A}{\tau_{1}^{2}}
 - \frac{B}{\mathcal{V}\sqrt{\tau_{1}}} +\frac{C\tau_{1}}
 {\mathcal{V}^{2}}\right)\frac{W_{0}^{2}}{\mathcal{V}^{2}},
 \label{74oi}
\end{equation}
where:
\begin{eqnarray}
 A &=& \left(g_{s} C_{1}^{KK}\right) ^{2}>0, \nonumber\\
 B &=& 2 \, C_{12}^{W}\lambda _{1}^{-1/2}
 = 4\alpha C_{12}^{W}, \label{GREAToi} \\
 C &=& 2\,\left(\alpha g_{s}C_{2}^{KK}\right)^{2}>0. \nonumber
\end{eqnarray}
Notice that $A$ and $C$ are both positive (and suppressed by
$g_s^2$) but the sign of $B$ is undetermined. The structure of
$\delta V_{(g_s)}$ makes it very convenient to use $\Omega \equiv
\tau_1$ as our parameter along the flat directions at fixed
$\mc{V}$ and $\tau_3$.

In this way, it is also easier to have a pictorial view of the
inflationary process since the K3 fiber modulus $\tau_1$ will turn
out to be mostly the inflaton. Inflation will correspond to an
initial situation, with the K3 fibre much larger than the base,
which will dynamically evolve to a final situation with the base
larger than the K3 fibre.

For generic values of $A$, $B$ and $C$ we expect the potential of
eq. (\ref{74oi}) to lift the flat direction and so to stabilise
$\Omega\equiv\tau_1$ at a minimum. Indeed, minimising $\delta
V_{(g_s)}$ with respect to $\tau_1$ at fixed $\mc{V}$ and $\tau_3$
gives:
\be
  \frac{1}{\tau_1^{3/2}} = \left( \frac{B}{8 A \mc{V}} \right)
  \left[ 1 + (\hbox{sign} \, B) \sqrt{1 + \frac{32 AC}{B^2}}
  \right] \label{tau1soln1} \,,
\ee
which, when $32 AC \ll B^2$, reduces to:
\be
  \tau_1 \simeq \left(-\frac{B \mc{V}}{2C} \right)^{2/3}
  \quad \hbox{if $B<0$}
  \qquad \hbox{or} \qquad
  \tau_1 \simeq \left(\frac{4A \mc{V}}{B} \right)^{2/3}
  \quad \hbox{if $B>0$} \,. \label{tau1soln2}
\ee

Any meaningful minimum must lie within the K\"{a}hler cone defined
by the conditions that no 2-cycle or 4-cycle shrink to zero and
that the overall volume be positive, and so we must check that
this is true of the above solution. Since we take $\tau_{1}$ and
$\tau_{2}$ both much larger than $\tau_{3}$, we may approximate
$\mc{V}$ by $\mathcal{V} \simeq \alpha\sqrt{\tau _{1}}\tau _{2}=
\lambda_{1}t_{1}t_{2}^{2}$ where $\lambda_{1}={1}/{4\alpha^{2}} >
0$, and this together with eq. (\ref{tausoi}) shows that positive
$t_{1}$ and $t_{2}$ suffices to ensure $\tau_1$, $\tau_2$ and
$\mathcal{V}$ are all positive. Consequently, the boundaries of
the K\"{a}hler cone arise where one of the 2-cycle moduli,
$t_{1,2}$, degenerates to zero. Since in terms of $\mathcal{V}$
and $\Omega \equiv \tau_1$ we have:
\begin{equation}
 t_1 = \frac{\mc{V}}{\tau_1} \,,
 \quad
 t_2 = \left( \frac{\tau_1}{\lambda_1} \right)^{1/2}
 \quad\hbox{and}\quad
 \tau_2 = 2\mc{V} \left( \frac{\lambda_1}{\tau_1} \right)^{1/2}\,,
 \label{tsoi}
\end{equation}
the K\"{a}hler cone is given by $0 < \tau_1 < \infty$. At its
boundaries we have:
\begin{equation*}
 \tau_{1}\rightarrow 0\Longleftrightarrow \tau _{2}
 \rightarrow \infty \Longleftrightarrow
 t_{1}\rightarrow \infty \Longleftrightarrow t_{2}\rightarrow 0,
\end{equation*}
while:
\begin{equation*}
 \tau _{1}\rightarrow \infty\Longleftrightarrow
 \tau _{2}\rightarrow 0
 \Longleftrightarrow t_{1}\rightarrow 0\Longleftrightarrow
 t_{2}\rightarrow \infty.
\end{equation*}
Comparing the solutions of eqs. (\ref{tau1soln2}) with the walls
of the K\"{a}hler cone shows that when $32AC \ll B^2$ we must
require either $C> 0$ (if $B<0$) or $A > 0$ (if $B>0$), a
condition that is always satisfied (see (\ref{GREAToi})).

In Table 8.1 we chose for numerical purposes several
representative parameter choices, and these choices are extended
to the loop-generated potential in Table 8.2. (The entries for
$\langle\tau_3\rangle$ and $\langle\mc{V}\rangle$ in this table
are simply carried over from Table 8.1 for ease of reference.) The
LV case shows that loop corrections can indeed stabilise the
remaining modulus, $\Omega \equiv \tau_1$, at hierarchically large
values, $\tau_2 \gg \tau_1 \gg \tau_3$ without requiring the
fine-tuning of parameters in the potential, while the SV examples
illustrate cases where $\tau_2 \gg \tau_1 \gtrsim \tau_3$
(although $e^{-a_1 \tau_1} \ll e^{-a_3 \tau_3}$).

\begin{figure}[ht]
\begin{center}
\begin{tabular}{c||c|c|c}
  & LV & SV1 & SV2 \\
  \hline\hline
  $C^{KK}_1$ & 0.1 & 0.15 & 0.18 \\
  $C^{KK}_2$ & 0.1 & 0.08 & 0.1 \\
  $C^W_{12}$ & 5 & 1 & 1.5 \\
  \hline
  $A$ & $10^{-4}$ & $2\cdot 10^{-3}$ & $2.9\cdot 10^{-3}$ \\
  $B$ & 10 & 0.52 & 0.93 \\
  $C$ & $5\cdot 10^{-5}$ & $1.9 \cdot 10^{-5}$ & $4.3 \cdot 10^{-5}$ \\
  $\langle\tau_3\rangle$ & 10.46 & 4.28 & 3.73 \\
  $\langle\tau_1\rangle$ & $1.07 \cdot 10^6$ & 8.96 & 7.5 \\
  $\langle\mc{V}\rangle$ & $2.75 \cdot 10^{13}$ & 1709.55 & 1626.12
\end{tabular}\\
\vspace{0.3cm}{{\bf Table {8.2}:} Loop-potential parameters.}
\end{center}
\end{figure}

\subsubsection{Canonical normalisation}

To discuss dynamics and masses requires the kinetic terms in
addition to the potential, which we now display in terms of the
variables $\mc{V}$ and $\Omega \equiv \tau_1$. Neglecting the
small blow-up cycle, $\tau_{3}$, the non canonical kinetic terms
for the large moduli $\tau_1$ and $\tau_2$ are given at leading
order by:
\begin{eqnarray}
 -\mathcal{L}_{kin} &=& K^0_{i\bar\jmath} \,
 \Bigl( \partial_{\mu } T_{i} \, \partial^{\mu }
 \overline{T}_{j} \Bigr)
 = \frac14 \, \frac{\partial^2 K_0}{\partial \tau_i \partial \tau_j}
 \; \Bigl( \partial_\mu \tau_i \, \partial^\mu \tau_j +
 \partial_\mu b_i \, \partial^\mu b_j \Bigr) \\
 &=& \frac{\partial_{\mu } \tau_{1} \partial^{\mu }
 \tau_{1}}{4\tau_{1}^{2}}
 + \frac{\partial_{\mu } \tau_{2} \partial^{\mu }
 \tau_{2} }{2 \tau_2^2} \;
 + \cdots \,, \nonumber\label{Lkinoi}
\end{eqnarray}
where the ellipses denote both higher-order terms in $\sqrt{\tau_3
/ \tau_{1,2}}$, as well as axion kinetic terms. Trading $\tau_2$
for $\mc{V}$ with eq. (\ref{tsoi}), the previous expression
becomes:
\begin{equation}
 -\mathcal{L}_{kin} =
 \frac{3}{8\tau_{1}^{2}} \; \partial_{\mu } \tau_{1}
 \partial^{\mu } \tau_{1}
 - \frac{1}{2\tau_1 \mc{V}} \; \partial_{\mu } \tau_{1}
 \partial^{\mu } \mc{V} + \frac{1}{2 \mc{V}^2} \;
 \partial_{\mu }\mc{V} \partial^{\mu }
 \mc{V} + \cdots \,. \label{Lkin2oi}
\end{equation}
Notice that the kinetic terms in this sector can be made field
independent by redefining $\vartheta_1 = \ln \tau_1$ and
$\vartheta_v = \ln \mc{V}$, showing that this part of the target
space is flat (within the approximations used). The canonically
normalised fields satisfy $-{\cal L}_{kin} = \frac12[(\partial
\varphi_1)^2 + (\partial \varphi_2)^2 ]$, and so may be read off
from the above to be given by:
\be
 \left( \begin{array}{c}
  \partial_\mu \tau_1/\tau_1 \\ \partial_\mu \mc{V}/\mc{V} \\
 \end{array} \right) =
 M \cdot
 \left( \begin{array}{c}
  \partial_\mu \varphi_1 \\
  \partial_\mu \varphi_2 \\
 \end{array} \right) \,,
\ee
where the condition:
\be
    M^{\scriptscriptstyle T} \cdot
 \left( \begin{array}{rrr}
  \frac34 && -\frac12 \\
  -\frac12 && 1 \\
 \end{array} \right) \cdot M
 = I\,,
\ee
implies $M^2 = \left( \begin{array}{cc} 2 & 1 \\ 1 & \frac32
\end{array} \right)$, and so if $M = \left( \begin{array}{cc} a & b \\ b &
c \end{array} \right)$ then $a_\pm = \sqrt{2-b_\pm^2}$, $c_\pm =
\sqrt{\frac32 - b_\pm^2}$ and $b_\pm^2 = 2/\left(7\pm 4\sqrt
2\right)$ (so explicitly $a_+ \simeq 1.357$, $b_+ \simeq 0.398$,
$c_+ \simeq 1.158$ and $a_- \simeq 0.715$, $b_- \simeq 1.220$,
$c_- \simeq 0.105$).

Finally, we may use these results to estimate the mass of the
propagation eigenstates, $\varphi_{1,2}$, obtained at the
potential's minimum. Before diagonalising the kinetic terms, but
writing $\vartheta_v = \ln \mc{V}$ and $\vartheta_1 = \ln \tau_1$,
we find that the derivatives of the potential at its minimum scale
as $\partial^2 V/\partial \vartheta_v^2 \sim \hat\xi/\mc{V}^3$ ---
since it is dominated by contributions from $\delta V_{(\alpha')}$
and $\delta V_{(np)}$ --- while $\partial^2 V/\partial
\vartheta_1^2 \sim \partial^2 V/\partial \vartheta_v \partial
\vartheta_1 \sim 1/\mc{V}^{10/3}$ --- since these are dominated by
$\delta V_{(g_s)}$. These properties remain true for the physical
mass eigenvalues after diagonalising the kinetic terms, since this
mixing changes the form of the eigenvectors but not the leading
scaling of the eigenvalues at large $\mc{V}$. This confirms the
qualitative expectation that the $\Omega \equiv \tau_1$ direction
is systematically lighter than $\mc{V}$ in the large-$\mc{V}$
limit.

\subsection{Inflationary potential} \label{7}

Having established the existence of a consistent LVS minimum of
the potential for all fields, we now explore the inflationary
possibilities that can arise when some of these fields are
displaced from these minima. Since the potential for
$\Omega\equiv\tau_1$ is systematically flat in the absence of
string loop corrections, it is primarily this field that we
displace in the hopes of finding it to be a good candidate for a
slow-roll inflaton.

In the approximation that string-loop effects are completely
turned off, we have seen that the leading large-$\mc{V}$ potential
stabilising both $\mc{V}$ and $\tau_3$ is completely flat in the
$\Omega\equiv\tau_1$ direction. We therefore perform our initial
inflationary analysis within an approximation where both
$\mathcal{V}$ and $\tau_{3}$ remain fixed at their respective
$\tau_1$-independent minima while $\tau_1$ rolls towards its
minimum from initially larger values. In this approximation the
important evolution involves only the single field $\tau_1$,
making it very simple to calculate. This single-field approach
should be an excellent approximation for large enough $\mc{V}$,
and we return below to the issue of whether or not $\mc{V}$ can be
chosen large enough to call this approximation into question.

\subsubsection{The single-field inflaton}

As before, we choose $\Omega\equiv\tau_1$ as the coordinate along
the inflationary direction. When $\tau_3 = \langle \tau_3 \rangle$
and $\mc{V} = \langle \mc{V} \rangle$ are fixed at their
$\tau_1$-independent minima, so that $\partial_{\mu}
\tau_3=\partial_{\mu}\mc{V}=0$, (\ref{Lkin2oi}) shows that the
relevant dynamics reduces to:
\be
 {\cal L}_{inf} = - \frac38 \left( \frac{\partial_\mu \tau_1
 \partial^\mu \tau_1 }{\tau_1^2} \right)\,
 - V_{inf}(\tau_1) \,, \label{LKINoi}
\ee
with scalar potential given by:
\begin{equation}
 V_{inf} = V_0 + \left(\frac{A}{\tau_{1}^{2}}
 -\frac{B}{\mathcal{V}\sqrt{\tau_{1}}} +\frac{C\tau_{1}}
 {\mathcal{V}^{2}}\right)\frac{W_{0}^{2}}{\mathcal{V}^{2}} \,.
 \label{inflpotoi}
\end{equation}
Notice that (\ref{LKINoi}) does not depend on the intersection
numbers, $\lambda_1$ and $\lambda_3$, implying that tuning these
cannot help with the search of a canonical normalisation more
suitable for an inflationary roll. The $\tau_1$-independent
constant, $V_0$, of eq. (\ref{inflpotoi}) consists of:
\begin{equation}
 V_0 =\frac{8 \,a_{3}^{2}A_3^2\sqrt{\langle \tau_{3}
 \rangle}}{3\alpha\gamma \langle \mathcal{V} \rangle}
 \; e^{-2a_{3} \langle \tau_{3} \rangle} -
 \frac{4 W_{0}a_{3}A_3 \langle \tau_{3} \rangle}{
 \langle \mathcal{V} \rangle^{2}}
 \; e^{-a_{3} \langle \tau_{3} \rangle}
 +\frac{3 \,\hat\xi W_0^2}{4 \langle \mathcal{V}
 \rangle^{3}}
 + \delta V_{up},
 \label{V0potoi}
\end{equation}
where $\delta V_{up}$ is an up-lifting potential, such as might be
produced by the tension of an $\overline{D3}$ brane in a warped
region somewhere in the extra dimensions: $\delta V_{up} \sim
\delta_{up}/\mc{V}^{4/3}$. For the present purposes, what is
important about this term is that it does not depend at all on
$\tau_1$ once $\mc{V}$ is fixed. We imagine $\delta_{up}$ to be
tuned to ensure the complete vanishing of $V$ (or a tiny positive
value) at the minimum, with $\delta_{up} \sim 1/\langle\mc{V}
\rangle^{5/3}$ required to cancel the non-perturbative and
$\alpha'$-correction parts of the potential (which together scale
like $\langle \mc{V} \rangle^{-3}$ at their minimum). In addition
a second adjustment ($\delta_{up}\to\delta_{up}+\mu_{up}$) of
order $\mu_{up}/\langle \mc{V} \rangle^{4/3}  = - \delta
V_{(g_{s})}(\langle \mathcal{V}\rangle, \langle \tau_1\rangle)$ is
required to cancel the loop-generated part of $V$, for which $V_0
\sim \mathcal{O}\left(1/\langle \mathcal{V}\rangle^{10/3}\right)$.

The canonical inflaton is therefore given by:
\begin{equation}
 \varphi =\frac{\sqrt{3}}{2} \, \ln \tau_1 \,,
 \qquad \hbox{and so} \qquad
 \tau_1 = e^{ \kappa \varphi }
 \quad\hbox{with} \quad
 \kappa = \frac{2}{\sqrt3} \,. \label{cambiooi}
\end{equation}
In terms of this field the walls of the K\"{a}hler cone are
located at:
\begin{equation}
 0 < \tau_1 < \infty \Longleftrightarrow
 -\infty < \varphi < +\infty,
\end{equation}
implying that any inflationary dynamics can in principle take
place over an \textit{infinite} range in field space. The
potential (\ref{inflpotoi}) becomes:
\begin{eqnarray}
 V_{inf} &=& V_0 + \frac{W_{0}^{2}}{\mathcal{V}^{2}}
 \left(A \, e^{-2\kappa \varphi}
 -\frac{B}{\mathcal{V}} \, e^{-\kappa\varphi/2}
 +\frac{C} {\mathcal{V}^{2}} \, e^{\kappa \varphi}\right)
 \nonumber\\
 &=& \frac{1}{\left\langle \mathcal{V}
 \right\rangle^{10/3}}\left(\mathcal{C}_0
 \, e^{\kappa \hat{\varphi}}
 - \mathcal{C}_1 \, e^{-\kappa \hat{\varphi}/2}
 + \mathcal{C}_2 \, e^{-2\kappa \hat{\varphi}}
 + \mathcal{C}_{up} \right) \,,
 \label{VVVoi}
\end{eqnarray}
where we shift $\varphi = \langle \varphi \rangle + \hat\varphi$
by its vacuum value, (\ref{tau1soln2}), and adjust $V_0 =
\mathcal{C}_{up}/\langle \mc{V} \rangle^{10/3}$ to ensure
$V_{inf}(\langle \varphi \rangle) = 0$. Choosing, for
concreteness' sake, $32AC \ll B^2$ \footnote{Notice that this is a
natural choice since for $B>0$, $CA/B^2\sim g_s^4.$ } we have
$\langle \varphi \rangle = \frac{1}{\sqrt3} \ln \left( \zeta
\mc{V} \right)$, with $\zeta \simeq -B/2C$ if $B<0$ or $\zeta
\simeq 4A/B$ if $B>0$. With these choices the coefficients
$\mathcal{C}_i$ do not depend on $\langle \mc{V} \rangle$, being
given by:
\begin{equation}
 \mathcal{C}_0 = C W_0^2 \zeta^{2/3}, \quad
 \mathcal{C}_1 = B W_0^2 \zeta^{-1/3},\quad
 \mathcal{C}_2 = A W_0^2 \zeta^{-4/3}
 \quad\hbox{and} \quad
 \mathcal{C}_{up} = \mathcal{C}_1 - \mathcal{C}_0
 - \mathcal{C}_2. \label{Ci2}
\end{equation}
Notice that because $A$ and $C$ are both positive, we know that
$\mathcal{C}_0$ and $\mathcal{C}_2$ must also be. By contrast, not
knowing the sign of $C_{12}^{W}$ precludes having similar control
over the sign of $\mathcal{C}_1$. Table 8.3 gives the values for
these coefficients as computed using the parameter sets of the
previous tables.

\begin{figure}[ht]
\begin{center}
\begin{tabular}{c||c|c|c}
  & LV & SV1 & SV2 \\
  \hline\hline
  $\mathcal{C}_0$ & $5.8 \cdot 10^{-8}$ & 0.012 & 0.023 \\
  $\mathcal{C}_1$ & 292.4 & 20629.4 & 39786.9 \\
  $\mathcal{C}_2$ & 73.1 & 5157.35 & 9946.73 \\
  $\mathcal{C}_{up}$ & 219.3 & 1200.8 & 29840.2 \\
  $R = \mathcal{C}_0/\mathcal{C}_2$
  & $8\cdot 10^{-10}$ & $2.3\cdot 10^{-6}$
  & $2.3\cdot 10^{-6}$ \\
\end{tabular}\\
\vspace{0.3cm}{{\bf Table {8.3}:} Coefficients of the inflationary
potential for the various parameter sets discussed in the text.}
\end{center}
\end{figure}

\begin{figure}[ht]
\begin{center}
\epsfig{file=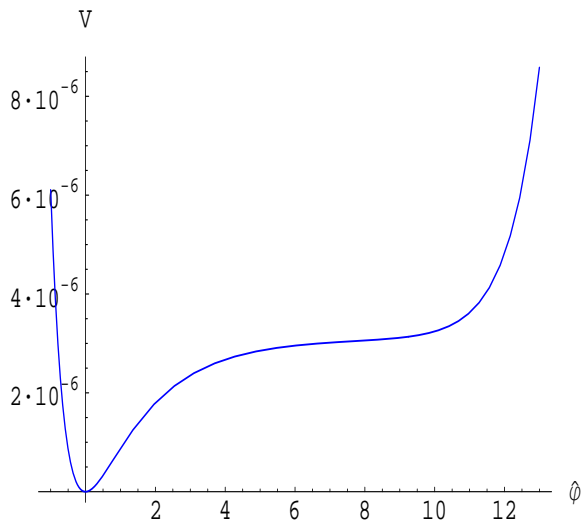, height=70mm,width=80mm}
\caption{$V$ (in
 arbitrary units) versus $\hat{\varphi}$, with $\mathcal{V}$ and
 $\tau_{3}$ fixed at their minima. The plot assumes the parameters
 used in the text (for which $\hat\varphi_{ip} \simeq 0.80$, $\hat\varphi_{end} = 1.0$,
 and $R\equiv\mathcal{C}_0/\mathcal{C}_2\sim 10^{-6}$).} \label{Fig:grafico3}
\end{center}
\end{figure}

Of particular interest is the case where both $A$ and $C$ are
small compared with $|B|$, as might be expected by their explicit
suppression by the factor $g_s^2$. For concreteness we focus in
what follows on the case $B>0$ (and so $\mathcal{C}_1>0$), for
which $\zeta \simeq 4A/B\ll 1$. This leads to two very useful
simplifications. First, it implies that $\mathcal{C}_0 /
\mathcal{C}_1 = \zeta C/B = 4AC/B^2$ and $R := \mathcal{C}_0 /
\mathcal{C}_2 = \zeta^2 C/A = 16 AC/B^2$ and so $\mathcal{C}_0$ is
systematically smaller than either $\mathcal{C}_1$ or
$\mathcal{C}_2$. This observation allows us to neglect completely
the $\mathcal{C}_0 \, e^{\kappa\hat\varphi}$ term of the potential
in the vicinity of the minimum and in most of the inflationary
region, as we shall see in what follows. Second, this limit
implies $\mathcal{C}_1/\mathcal{C}_2 = \zeta B/A = 4$, showing
that $\mathcal{C}_1$ and $\mathcal{C}_2$ are both positive, with a
fixed, order-unity ratio. This observation precludes using the
ratio of these parameters in the next section as a variable for
tuning the inflationary potential. These choices are visible in
Table 8.3, for which $A,C \ll B$, and so $\mathcal{C}_0$ is small
and $\mathcal{C}_1/\mathcal{C}_2 \simeq 4$. Figure
\ref{Fig:grafico3} plots the resulting scalar potential against
$\hat{\varphi}$.

\subsection{Inflationary slow roll}
\label{SlowRoll}

We next ask whether the scalar potential (\ref{VVVoi}) can support
a slow roll, working in the most natural limit identified above,
with $A,C \ll B$ and $B > 0$. As we have seen, this case also
implies $0 < \mathcal{C}_0 \ll \mathcal{C}_1 = 4 \mathcal{C}_2$,
leaving a potential well approximated by:
\begin{equation}
 V \simeq \frac{\mathcal{C}_{2}}{\left\langle
 \mathcal{V}\right\rangle ^{10/3}} \left[ (3 - R) - 4
 \left( 1 + \frac16 \, R \right) \,e^{-\kappa
 \hat{\varphi}/2} + \left( 1 + \frac23 \, R \right)
 \, e^{-2\kappa \hat{\varphi}} + R\text{ }
 e^{\kappa \hat{\varphi}}\right] \label{SCALAoi}
\end{equation}
which uses $\mathcal{C}_{up} \simeq \mathcal{C}_1 - \mathcal{C}_0
- \mathcal{C}_2$ and $\mathcal{C}_1/\mathcal{C}_2 \simeq 4$, and
works to linear order in:
\begin{equation}
 R := \frac{\mathcal{C}_{0} }{ \mathcal{C}_{2}}
 = 2 g_s^4 \left(\frac{C_1^{KK} C_{2}^{KK} }{C_{12}^W}
 \right)^2 \ll 1 \,.
\end{equation}
The normalisation of the potential may instead be traded for the
mass of the inflaton field at its minimum: $m_\varphi^2 = V''(0) =
4\, \left( 1 + \frac76 \, R \right) {\mathcal{C}_2} / {\left
\langle \mathcal{V} \right \rangle^{10/3}}$.

In practice the powers of $R$ can be neglected in all but the last
term in the potential, where it multiplies a positive exponential
which must eventually become important for sufficiently large
$\hat{\varphi}$. For smaller $\hat\varphi$, $R$ is completely
negligible and the potential is fully determined by its overall
normalisation. Furthermore, the range of $\hat\varphi$ for which
this is true becomes larger and larger the smaller $R$ is, and so
we start by neglecting $R$.

\begin{figure}[ht]
\begin{center}
\epsfig{file=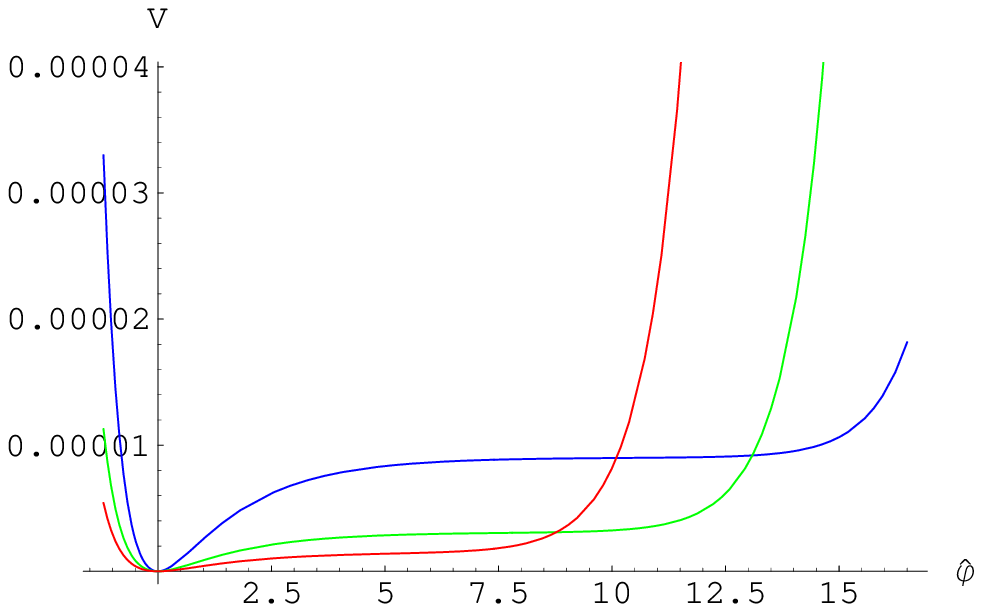, height=50mm,width=47mm}
\epsfig{file=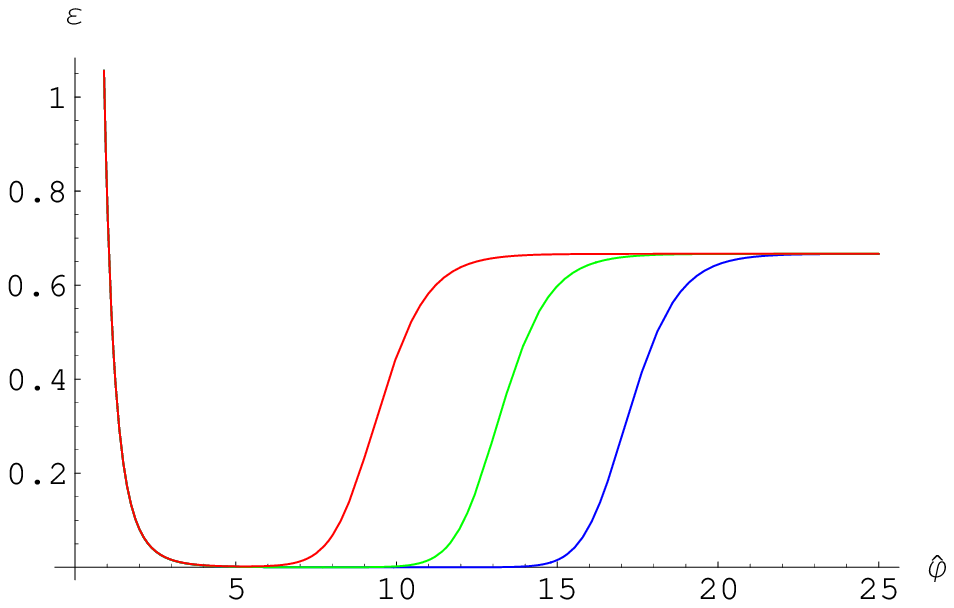, height=50mm,width=47mm}
\epsfig{file=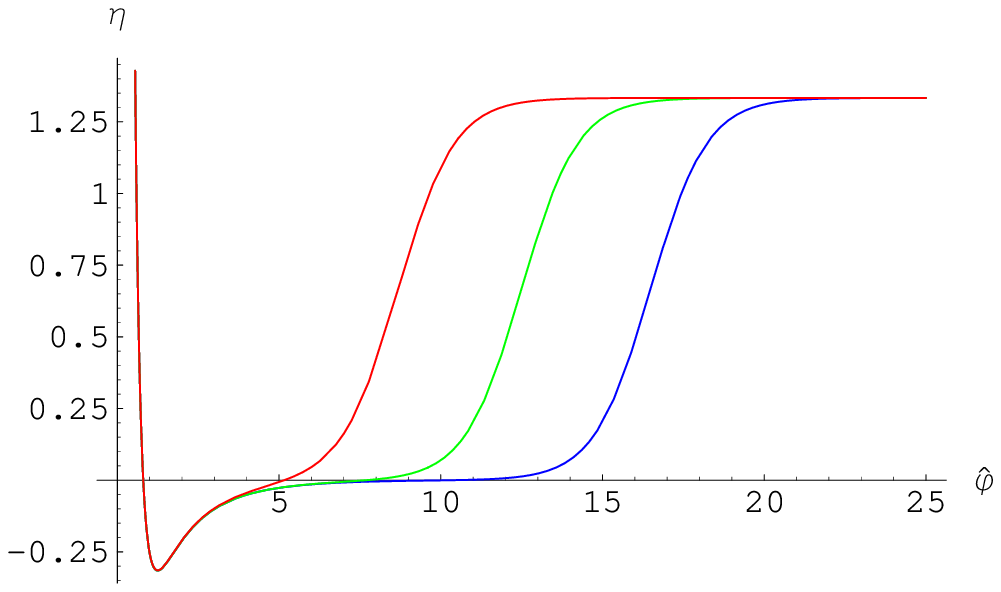, height=50mm,width=47mm} \caption{
 Plots of the potential and the slow-roll parameters $\varepsilon$
  and $\eta$ vs $\hat\varphi$ for $R=10^{-8}$ (blue curve), $R=10^{-6}$ (green
  curve), and $R=10^{-4}$ (red curve).}
   \label{NewFig}
\end{center}
\end{figure}

We seek inflationary rolling focusing on the situation in which
$\hat{\varphi}$ rolls down to its minimum (at $\hat\varphi = 0$)
from positive values. Defining, as usual, the slow-roll
parameters, $\varepsilon$ and $\eta$, by (recalling our use of
Planck units, $M_P = 1$):
\begin{eqnarray}
 \varepsilon  =\frac{1}{2V^{2}}
 \left( \frac{\partial V}{\partial
 \hat{\varphi} }\right)^{2}, \qquad
 \eta  = \frac{1}{V} \left( \frac{\partial^{2}V}{\partial
 \hat{\varphi}^{2}} \right),
\end{eqnarray}
we find (using $\kappa^2 = \frac43$ and keeping $R$ only when it
comes multiplied by $e^{\kappa \hat\varphi}$):
\begin{eqnarray}
 \varepsilon  &\simeq& \frac{8}{3} \left(
 \frac{ e^{-\kappa\hat{\varphi}/2}
 - e^{-2 \kappa\hat{\varphi}}
 + \frac12 \, R \, e^{\kappa \hat\varphi}}
 {3 - 4 \, e^{-\kappa\hat{\varphi}/2} +
 e^{-2\kappa \hat{\varphi}} + R \, e^{\kappa\hat\varphi}}
 \right)^{2}, \label{epsoi} \\
 \eta  &\simeq& -\frac{4}{3} \left( \frac{
 e^{-\kappa\hat{\varphi}/2}
 - 4 \, e^{-2\kappa\hat{\varphi}} - R \, e^{\kappa \hat\varphi}}
 {3 - 4 \, e^{-\kappa\hat{\varphi}/2}
 + e^{-2\kappa\hat{\varphi}}
 + R \, e^{\kappa \hat\varphi}} \right)\,.
 \label{etaoi}
\end{eqnarray}
Plots of these expressions are given in Figure \ref{NewFig}, which
show three qualitatively different regimes.

\newpage

\medskip\noindent{\em Slow-Roll Regime}

\medskip\noindent
Both slow roll parameters are naturally exponentially small in the
regime $R^{1/3} \ll e^{-\hat\varphi/2} \ll 1$. In this regime it
is the term $e^{-\kappa\hat\varphi/2}$ that dominates in
(\ref{SCALAoi}), and so the dynamics is effectively governed by
the approximate potential:
\begin{equation}
 V \simeq \frac{\mathcal{C}_2}{\left\langle \mathcal{V}
 \right\rangle^{10/3}}\left(3
 - 4 \, e^{-\kappa\hat{\varphi}/2}
 \right) \,. \label{SCALAREoi}
\end{equation}
This resembles a standard potential for large-field inflation,
which drives the field to evolve towards smaller
values\footnote{It would be interesting to see how our
inflationary mechanism fits in the general analysis of
supergravity conditions for inflation performed in \cite{marta0,
marta1, marta2}.}. The slow-roll parameters (\ref{epsoi}) and
(\ref{etaoi}) in this regime simplify to:
\begin{eqnarray}
 \varepsilon  &\simeq&
 \frac{8}{3 \left[3
 \, e^{\kappa\hat{\varphi}/2} -4\right]^{2}},
 \label{Epsoi} \\
 \eta &\simeq& -\frac{4}{3\left[
 3 \, e^{\kappa\hat{\varphi}/2} -4\right]} \,,
 \label{Etaoi}
\end{eqnarray}
and for all $\hat\varphi$ in this regime we have the interesting
relation:
\begin{equation} \label{epsvsetarelation}
 \varepsilon \simeq \frac{3 \,\eta^2}{2} \,.
\end{equation}

\medskip\noindent{\em Small-$\hat\varphi$ Regime}

\medskip\noindent
The slow-roll conditions break down once $\hat\varphi$ is small
enough that the two negative exponentials are comparative in size
to produce a zero in $\eta$. An inflection point occurs in this
regime, located where:
\begin{equation}
 \left( \frac{\partial^{2} V}{\partial \hat{\varphi}^{2}}
 \right)_{\hat\varphi_{ip}} \simeq \frac{4\mathcal{C}_2}{3 \langle
 \mathcal{V}\rangle^{10/3}}
 \left( -  e^{-\kappa\hat{\varphi}_{ip}/2}
 + 4 \, e^{-2\kappa \hat{\varphi}_{ip}}
 \right) = 0,
\end{equation}
and so:
\be
 \hat{\varphi}_{ip} = \frac{1}{\sqrt{3}} \ln \left(
 \frac{16 \, \mathcal{C}_2}{\mathcal{C}_1} \right)
 \simeq \frac{\ln 4 }{\sqrt{3}} \simeq 0.8004..\,.
\ee
As Figure \ref{NewFig} shows, to the left of this point
$\varepsilon$ grows quickly, while at the inflection point
$\hat{\varphi} = \hat{\varphi}_{ip}$, we have $\varepsilon_{ip} =
1.464$ and $\eta_{ip} = 0$. Just to the right of this, at $\hat
\varphi_{end} = 1$ we have $\varepsilon_{end} = 0.781$ and
$\eta_{end} = -0.256$, making this as good a point as any to end
inflation. (In what follows we verify numerically that our results
are not sensitive to precisely where we end inflation in this
regime.)

\newpage

\medskip\noindent{\em Large-$\hat\varphi$ Regime}

\medskip\noindent
Once $R \, e^{\kappa \hat\varphi} \gg 3$ the positive exponential
dominates the potential (\ref{SCALAoi}), which becomes
well-approximated by:
\begin{equation}
 V \simeq \frac{m_{\varphi }^{2}}{4} \, R
 \, e^{\kappa\hat{\varphi}} \,,
\label{SCALAreoi}
\end{equation}
and so the slow-roll parameters plateau at constant values: $\eta
\simeq 2\varepsilon \simeq \kappa^2 = \frac43$ (as is seen in
Figure \ref{NewFig}). This shows that the slow-roll conditions
also break down for $\kappa\hat{\varphi} \simeq \ln(1/R)$,
providing an upper limit to the distance over which the slow roll
occurs (and so also on the number of \efold ings, $N_e$).

An interesting feature of transition to this large-$\hat\varphi$
regime is the necessity for $\eta$ to change sign. This is
interesting because, as Figure \ref{NewFig} shows, $\varepsilon$
is still small where it does, and so there is a slow-roll region
for which $\eta \gg \varepsilon > 0$. This regime is unusual
because it allows $n_s > 1$ (see Figure \ref{Fig:nsComparison}),
unlike generic single-field inflationary models. In practice, in
what follows we choose horizon exit to occur for $\hat\varphi$
smaller than this, due to the current observational preference for
$n_s < 1$. A precise upper limit on $\hat\varphi$ this implies can
be defined as the inflection point where $\eta$ vanishes due to
the competition between the $e^{\kappa\hat{\varphi}}$ and
$e^{-\kappa\hat{\varphi}/2}$ terms of the potential. This occurs
when $e^{-\kappa\hat\varphi/2} \simeq R \, e^{\kappa
\hat\varphi}$, or $\hat{\varphi}(R) \simeq \hat{\varphi}_0(R) :=
-\ln(R)/\sqrt{3}$.

\begin{figure}[ht]
\begin{center}
\epsfig{file=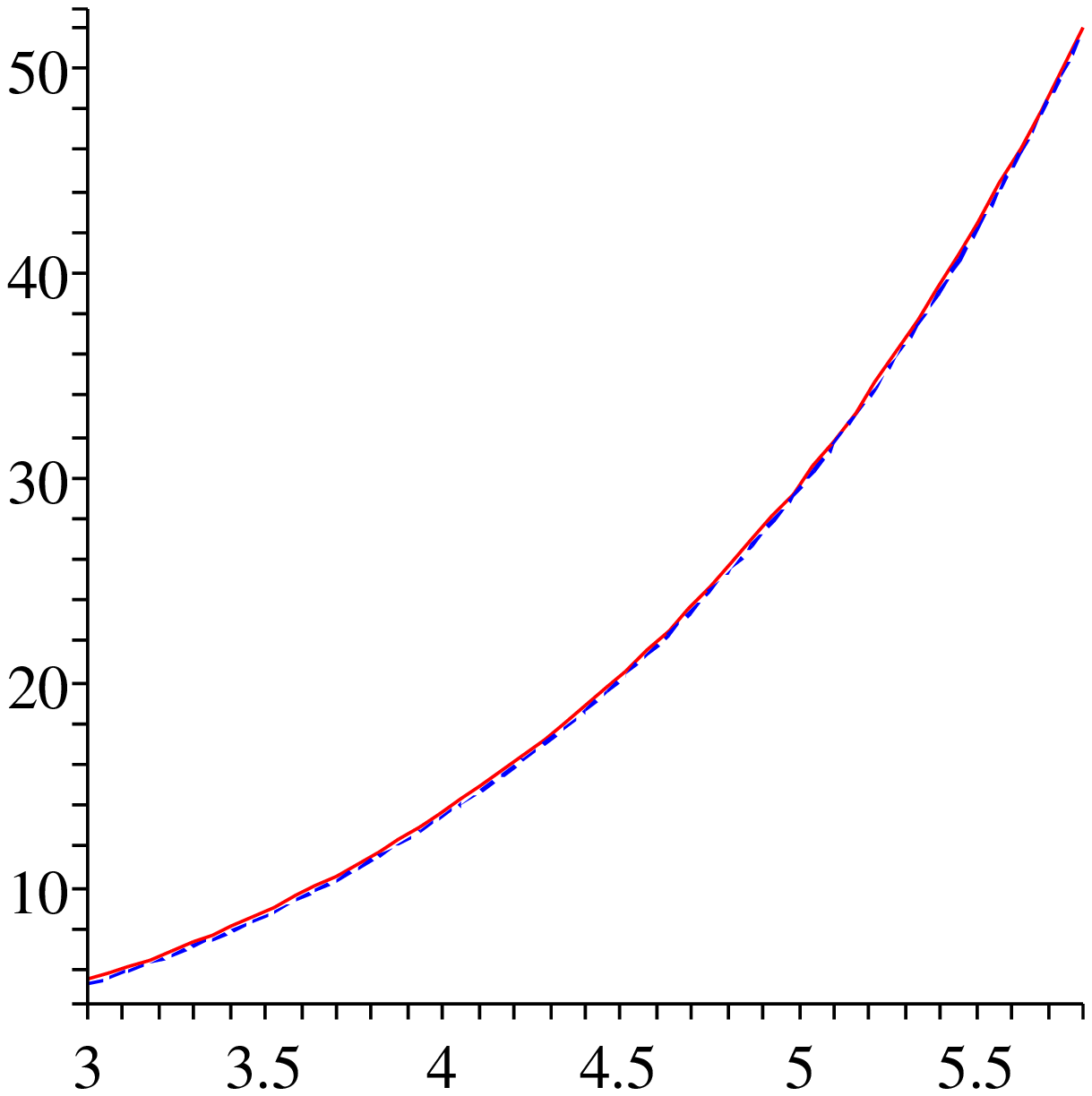, height=60mm,width=67mm}
\epsfig{file=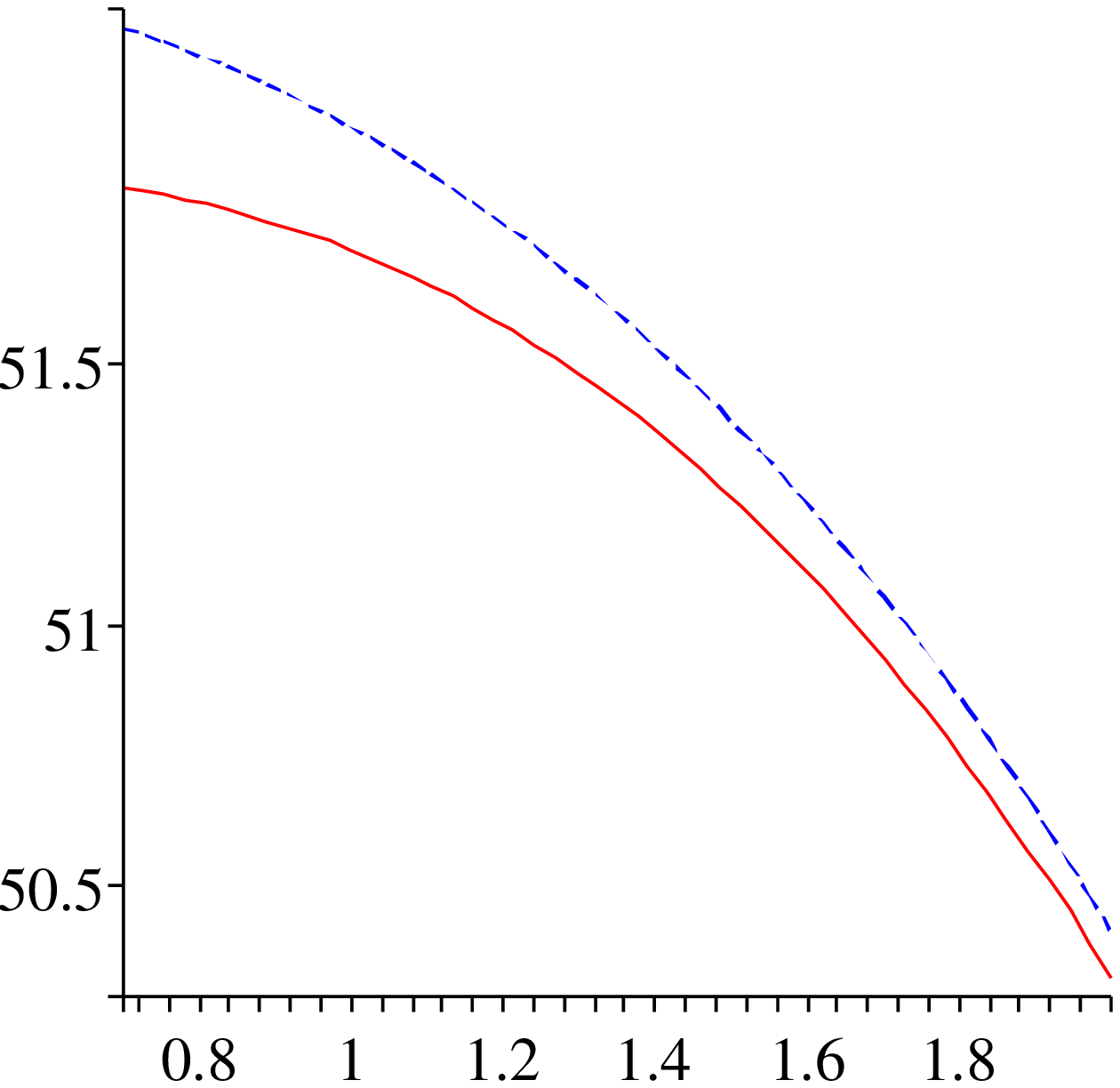, height=60mm,width=67mm} \caption{
 Plot of the number of $e$-foldings, $N_e$, vs $\hat\varphi_*$ (left)
 and $\hat\varphi_{end}$ (right) for $R=0$. The inflection point occurs at
  $\hat\varphi_{ip} \simeq 0.8$ and $\hat\varphi_{end} = 1$ in the left-hand
  plot. $\hat\varphi_* = 5.7$ in the right-hand plot. The solid (red) curves
  are computed using the full
  potential (\ref{SCALAoi}) while the dashed (blue) curves are computed
  using the approximate potential (\ref{SCALAREoi}).}
   \label{Fig:efolds}
\end{center}
\end{figure}

We may now ask whether the slow-roll regime is large enough to
allow 60 \efold ings of inflation. The number of \efold ings $N_e$
occurring during the slow-roll regime can be computed using the
approximate potential (\ref{SCALAREoi}), which gives:
\begin{equation}
 N_{e}=\int_{\hat{\varphi}_{end}}^{\hat{\varphi}_{\ast}}
 \frac{V}{V'} \; \exd \hat{\varphi}
 \simeq \frac{\sqrt{3}}{4} \int_{\hat{\varphi}_{end}}^{\hat{
 \varphi}_{\ast }} \left[3 \, e^{\kappa\hat{\varphi}/2}
 -4 \right] \, \exd\hat{\varphi}
 = \left[ \frac94 \, e^{\kappa\hat{\varphi}/2}
 - \sqrt3 \, \hat\varphi
 \right]_{\hat{\varphi}_{end}}^{\hat{\varphi}_{\ast }} \,,
 \label{Nefunc}
\end{equation}
where $e^{\kappa\hat{\varphi}_{end}} \simeq 16\, \mathcal{C}_2 /
\mathcal{C}_1 \simeq 4  \ll e^{\kappa\hat\varphi_*}$ represents
the onset of the small-$\hat\varphi$ regime, as described above,
and $\hat\varphi = \hat\varphi_*$ denotes the value of
$\hat\varphi$ at horizon exit. Figure \ref{Fig:efolds} shows how
the number of \efold ings depends on the assumed field value
during horizon exit, as well as the insensitivity of this result
to the assumed point where inflation ends. This shows that
interesting inflationary applications require $\hat\varphi$ to
roll through an interval of at least $\mathcal{O}(5)$.

\begin{figure}[ht]
\begin{center}
\epsfig{file=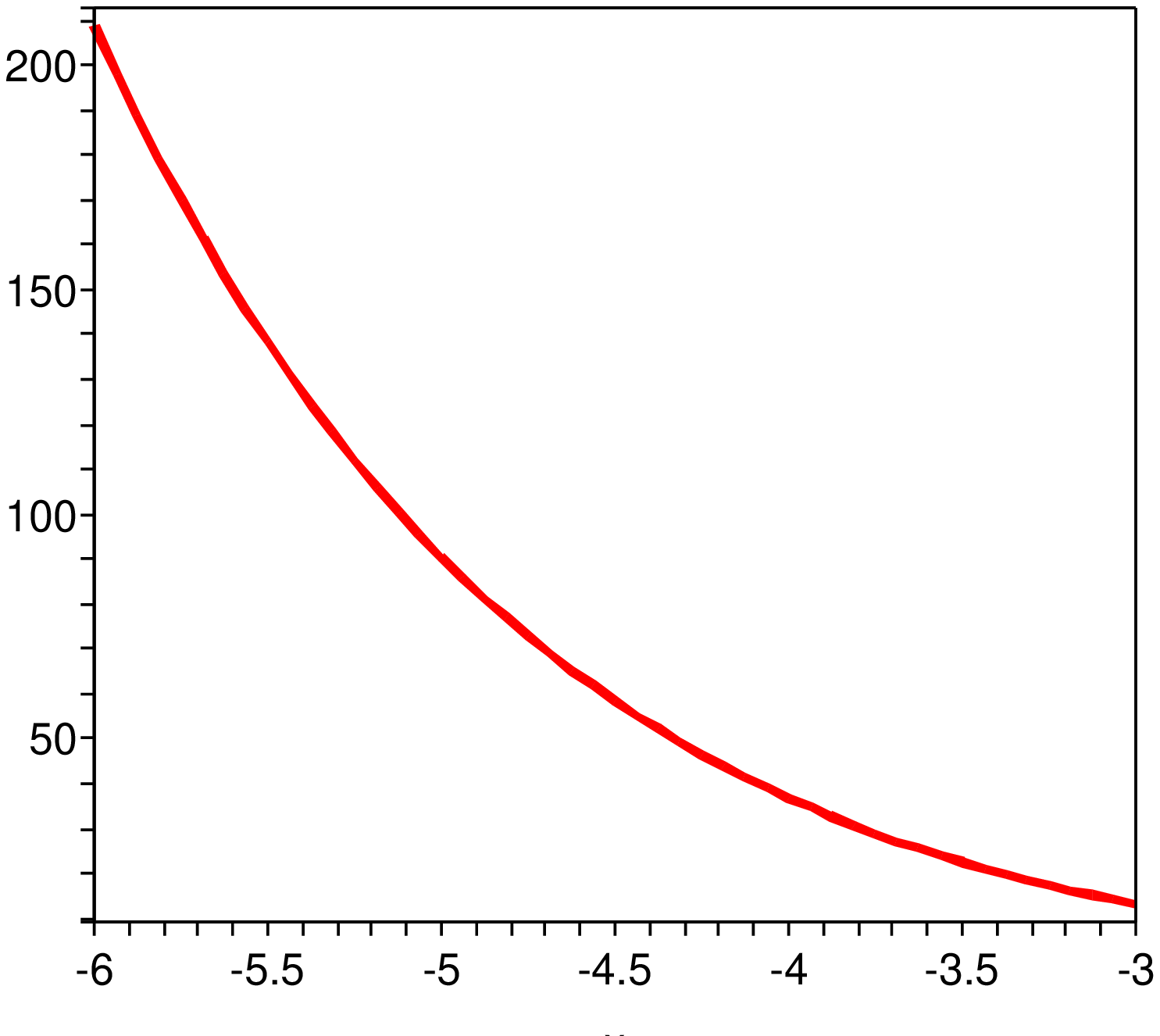, height=60mm,width=67mm} \caption{
 Plot of the maximum number of $e$-foldings, $N_e^{max}$, vs
 $x=\log_{10}R$, defined by the condition
 $\hat\varphi_* = \hat\varphi_0(R)$ as described in the text.
 The integration takes $\hat\varphi_{end} = 1$, and the
 curves are computed
  using the approximate potential (\ref{SCALAREoi}).}
   \label{Fig:Nemax}
\end{center}
\end{figure}

An estimate for the upper limit to $N_e$ that can be obtained as a
function of $R$ can be found by using $\hat\varphi_* =
\hat\varphi_0(R)$ in eq. (\ref{Nefunc}). This leads to:
\begin{equation}
 N_e^{max} \simeq \frac94 \left( R^{-1/3} - 2 \right)
 - \left[ \ln \left( \frac{1}{R} \right) - \ln 8 \right]
  \,,
 \label{Nemax}
\end{equation}
This result is plotted in Figure \ref{Fig:Nemax}, and shows that
more than 60 \efold ings of inflation requires $R \lesssim 3 \cdot
10^{-5}$.

The validity of the $\alpha'$ and $g_s$ expansions also set a
limit to how large $\hat{\varphi}_*$ can be taken, since the
exponential growth of $\delta V_{(g_s)}$ for large $\hat\varphi$
would eventually allow it to become larger than the lower-order
contributions, $\delta V_{(np)}+\delta V_{(\alpha')}$.
Microscopically this arises because $\hat{\varphi}\to\infty$
corresponds to $\tau_1\to\infty$ and $\tau_2\to 0$, leading to the
failure of the expansion of $\delta V^{KK}_{(g_s),\tau_2}$ in
inverse powers of $\tau_2$. However, as is argued in appendix B.2,
it is the slow-roll condition $\eta \ll 1$ that breaks down first
as $\hat\varphi$ increases, and so provides the most stringent
upper edge to the inflationary regime. For the two sample sets SV1
and SV2 given in the Tables, we obtain $R \simeq 2.3\cdot
10^{-6}$, and this gives $\hat{\varphi}_{max} \simeq 12.4$ (in
particular allowing more than 60 \efold ings of inflation).

\subsubsection{Observable footprints}

We now turn to the observable predictions of the model. These
divide naturally into two types: those predictions depending only
on the slow roll parameters, which are insensitive to the
underlying potential parameters; and those which also depend on
the normalization of the inflationary potential, and so depend on
more of the details of the underlying construction.

\subsubsection*{Model-independent predictions}

The most robust predictions are for those observables whose values
depend only on the slow roll parameters, such as the spectral
index and tensor-to-scalar ratio, which are given as functions of
the slow-roll parameters (evaluated at horizon exit) :
\be \label{rnsslowroll}
    n_s = 1 + 2\eta_* - 6\varepsilon_* \qquad \hbox{and} \qquad
    r =16 \, \varepsilon_* \,.
\ee
In general, as can be seen from (\ref{epsoi}) and (\ref{etaoi}),
the two quantities $\varepsilon_*$ and $\eta_*$ are functions of
two parameters, $\hat\varphi_*$ and $R$; hence $n_s = n_s( \hat
\varphi_*, R)$ and $r=r(\hat\varphi_*,R)$. However we have also
seen that having a significant number of \efold ings requires $R
\ll 1$, and so to a good approximation $n_s = n_s(\hat\varphi_*)$
and $r = r(\hat\varphi_*)$, unless $\hat\varphi_*$ is large enough
that $R e^{\kappa\hat\varphi_*}$ cannot be neglected.

\begin{figure}[ht]
\begin{center}
\epsfig{file=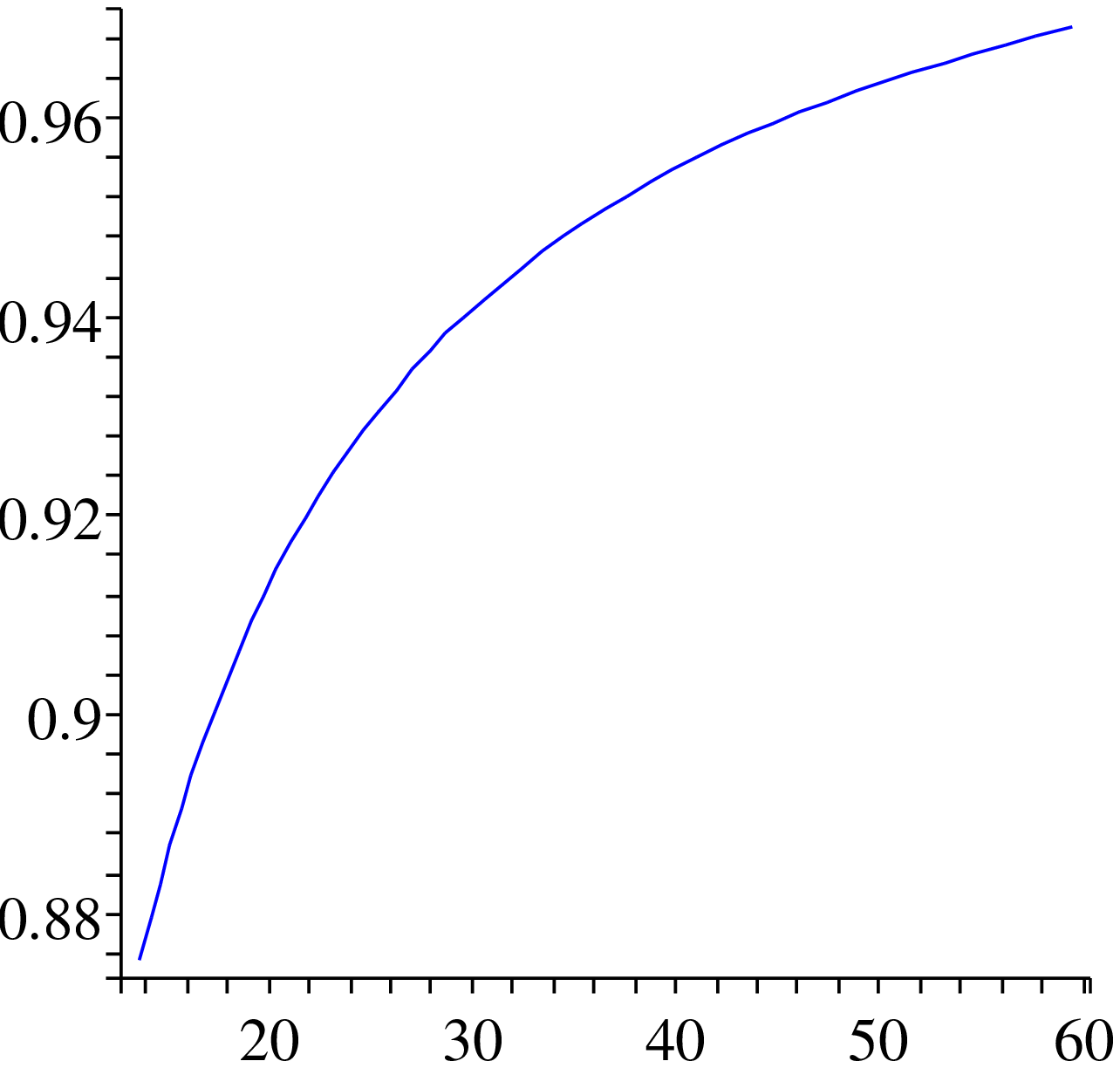, height=40mm,width=47mm}
\epsfig{file=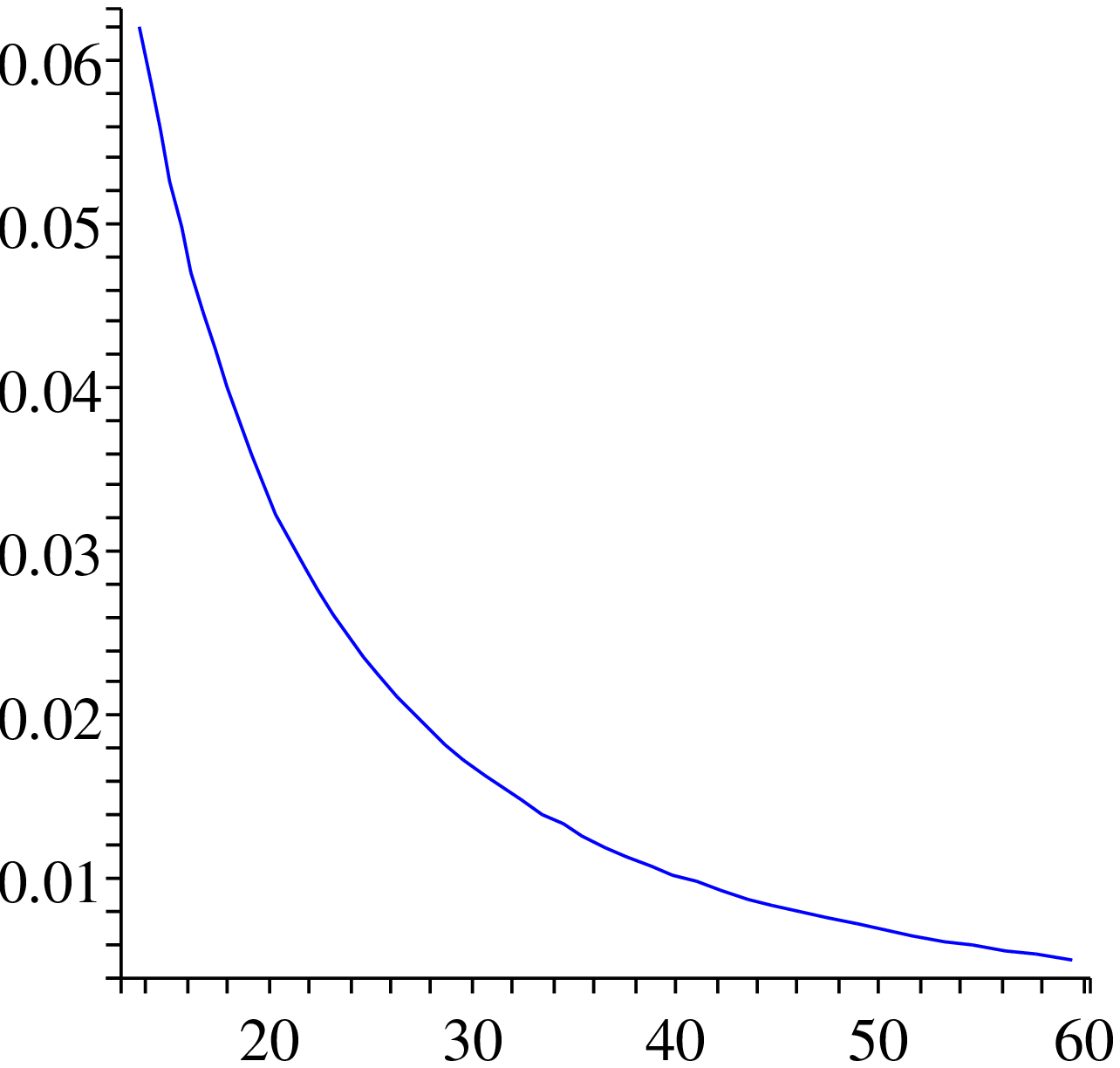, height=40mm,width=47mm}
\epsfig{file=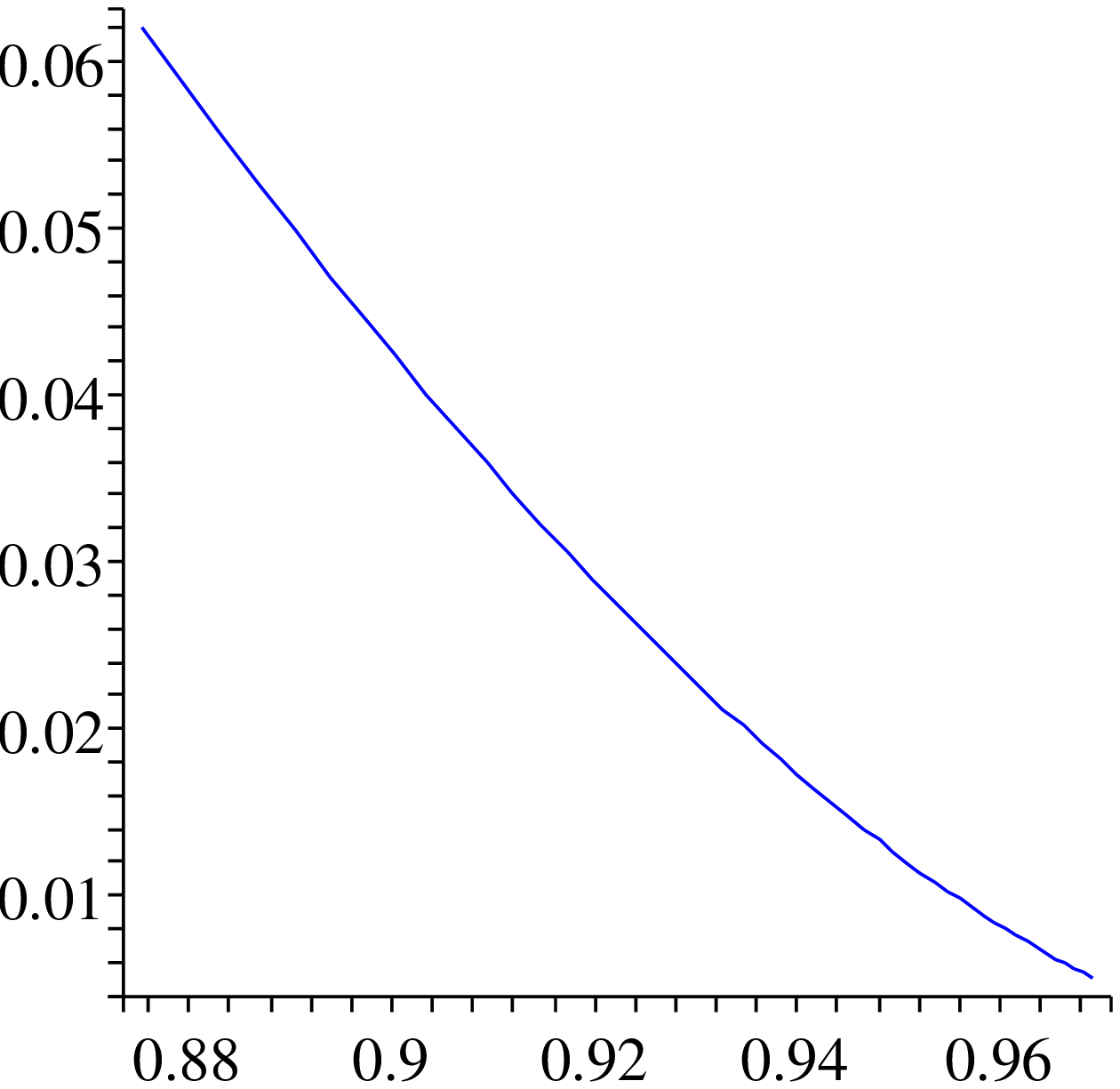, height=40mm,width=47mm}
 \caption{
 A plot of $n_s$ (left panel) and $r$ (center panel) vs the number
 of $e$-foldings, $N_e$. The right panel plots the correlation $r$
 vs $n_s$ that results when $N_e$ is eliminated, resembling simple
 single-field large-field models.}
   \label{Fig:randns}
\end{center}
\end{figure}

For small $R$ we find the robust correlation predicted amongst
$r$, $n_s$ and $N_e$, as described in the introduction. The
implied relation between $r$ and $n_s$ is most easily found by
using the relation $\varepsilon_* = \frac32 \, \eta_*^2$, eq.
(\ref{epsvsetarelation}) in (\ref{rnsslowroll}) and dropping
$\varepsilon_*$ relative to $\eta_*$ in $n_s -1$:
\begin{equation}
 r \simeq 6 (n_s - 1)^2 \,,
\end{equation}
showing that a smaller ratio of tensor-to-scalar perturbations,
$r$, correlates with larger $n_s$. Figure \ref{Fig:randns} plots
the predictions for $r$ and $n_s$ that are obtained in this way.

\begin{figure}[ht]
\begin{center}
\epsfig{file=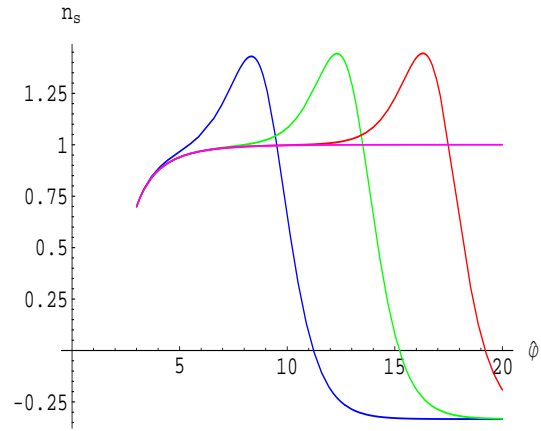, height=60mm,width=70mm} \caption{
 Plots of the spectral index $n_s$
 vs $\hat\varphi$ for $R=0$ (purple curve),
 $R=10^{-8}$ (red curve), $R=10^{-6}$ (green
 curve), and $R=10^{-4}$ (blue curve).}
   \label{Fig:nsComparison}
\end{center}
\end{figure}

Deviations from this correlation arise for large enough
$\varphi_*$, for which $N_e$ approaches the maximum number of
\efold ings possible, and this is illustrated in Figure
\ref{Fig:nsComparison}, which plots $n_s$ vs $\hat\varphi_*$ for
several choices of $R$. (Notice in particular the excursion to
values $n_s > 1$ shown in the figure for $\hat\varphi_* \simeq
\hat\varphi_0(R)$ when $R \ne 0$, as discussed above.) In the
extreme case where $\hat\varphi_* = \hat\varphi_0(R)$ we have
$\eta_* \simeq 0$ and $\varepsilon_* \simeq \frac23 \, R^{2/3}$,
leading to:
\begin{equation}
 r \simeq \frac{32}{3} \, R^{2/3}
 \qquad \hbox{and} \qquad
 n_s \simeq 1-4 \, R^{2/3} \,.
\end{equation}
Recall that $N_e^{max} \gtrsim 60$ implies $R \lesssim 3 \times
10^{-5}$, and in the extreme case $R \simeq 3 \cdot 10^{-5}$ the
above formulae lead to $r \simeq 0.01$ and $n_s \simeq 0.996$.
Should $r \simeq 0.01$ be observed and ascribed to this scenario,
the close proximity of horizon exit to the beginning of inflation
would likely imply other observable implications for the CMB,
along the lines of those discussed in \cite{tPsignals}.

\subsubsection*{Model-dependent predictions}

We next turn to those predictions which depend on the
normalisation, $V_0$, of the inflaton potential, and so depend
more sensitively on the parameters of the underlying supergravity.

\medskip \noindent {\it Number of $e$-foldings:}
The first model-dependent prediction is the number of \efold ings
itself, since this depends on the value $\hat\varphi_*$ taken by
the scalar field at horizon exit. Indeed we have already seen that
the constraint that there be enough distance between
$\hat\varphi_*$ and $\hat\varphi_{end}$ to allow many \efold ings
of inflation imposes upper limits on parameters such as $R$. The
strongest such limit turned out to be the requirement that $n_s$
be low enough to agree with observed values (see the discussion
surrounding eq. (\ref{Nemax})). For numerical comparison of our
benchmark parameter sets we formalise this by requiring
$\hat\varphi < \hat{\varphi}_{max}$, defined as the value for
which $n_s < 0.974$, since this is the $68\%$ C.L. observational
upper bound (for small $r$). Table 8.4 then lists the maximal
number of \efold ings that are possible given the constraint
$\hat\varphi_* < \hat\varphi_{max}$ for the models given in Tables
8.1 and 8.2.

\begin{figure}[ht]
\begin{center}
\begin{tabular}{c||c|c|c}
  & LV & SV1 & SV2 \\
  \hline\hline
  $\langle \varphi \rangle$ & 12.02 & 1.9 & 1.7 \\
  $\hat\varphi_{max}$ & 6.3 & 6.14 & 6.16 \\
  $N_e^{max}$ & 72 & 64 & 64 \\
  $A_{COBE}$ & $2.1 \cdot 10^{-45}$
             & $1.2 \cdot 10^{-7}$ & $2.8 \cdot 10^{-7}$ \\
  $\mathcal{R}_{cv}$ & 1201.6 & 29.7 & 12.2 \\
\end{tabular}\\
\vspace{0.3cm}{{\bf Table {8.4}:} Model parameters for the
inflationary potential. $N_e^{max}$ denotes the number of \efold
ings computed when rolling from $\hat\varphi_{max}$ to
$\hat\varphi = 1$. $A_{COBE}$ is calculated at $N_e\simeq 60$ and
we set $K_{cs}=3\ln g_s\simeq -3.6$.}
\end{center}
\end{figure}

But how many \efold ings of inflation are required is itself a
function of both the inflationary energy scale and the
post-inflationary thermal history. For instance, suppose the
inflaton energy density, $\rho_{inf}\sim M_{inf}^4=V_{end}$,
re-thermalises during a re-reheating epoch during which the
equation of state is $p=w\rho$, at the end of which the
temperature is $T_{rh}$, and after this the radiation-dominated
epoch lasts right down to the present epoch. With these
assumptions, $M_{inf}$, $T_{rh}$, $w$ and $N_{e}$ are related
by\footnote{We thank Daniel Baumann for identifying an error in
this formula in an earlier version.}:
\begin{equation}
 N_e\simeq 62+\ln\left(\frac{M_{inf}}{10^{16}
 GeV}\right)-\frac{\left(1-3w\right)}{3\left(1+w\right)}
 \ln\left(\frac{M_{inf}}{T_{rh}}\right). \label{cosmology}
\end{equation}
This formula is obtained by equating the product $aH$ at horizon
exit during inflation and horizon re-entry in the cosmologically
recent past, $a_{he}H_{he}=a_0 H_0$, and using the intervening
cosmic expansion to relate these two quantities to $N_e$, $T_{rh}$
and $M_{inf}$ \cite{liddle}. In particular it shows (if $w <
\frac13$) that lower reheat temperatures (for fixed $M_{inf}$)
require smaller $N_e$. For instance, if $M_{inf} \simeq 10^{16}$
GeV and $w=0$ then an extremely low reheat temperature, $T_{rh}
\simeq 1$ GeV, allows $N_e \simeq 50$.

\begin{figure}[ht]
\begin{center}
\begin{tabular}{|c|c|c|c|c|}
$T_{rh}$ (GeV) & $N_e$ & $n_s$ & $r$ \\
\hline \hline
$10^{10}$ & 57 & 0.9702 & 0.0057 \\
\hline
$5\cdot 10^{7}$ & 55 & 0.9690 & 0.0060 \\
\hline
$10^{5}$ & 53 & 0.9676 & 0.0064 \\
\hline
$5\cdot 10^{3}$ & 52 & 0.9669 & 0.0066 \\
\end{tabular} \\\smallskip
{\bf Table {8.5}:} Predictions for cosmological observables as a
function of $T_{rh}\leq 10^{10}$ GeV fixing $M_{inf}=5\cdot
10^{15}$ GeV (for $w=0$ and $R=2.3\cdot 10^{-6}$).
\end{center}
\end{figure}

Given that $M_{inf}$ is constrained by the requirement that
inflation generate the observed primordial scalar fluctuations
(see below), eq. (\ref{cosmology}) is most usefully read as giving
the post-inflationary reheat temperature that is required to have
modes satisfying $k = (aH)_*$ be the right size to be re-entering
the horizon at present. That is, given a measurement of $n_s$ one
can invert the prediction $n_s(N_e)$ to learn $N_e$, and so also
$r$ and the two slow roll parameters, $\varepsilon_*$ and
$\eta_*$. Then computing $M_{inf}$ from the amplitude of
primordial fluctuations allows (\ref{cosmology}) to give $T_{rh}$.
In particular, eq. (\ref{cosmology}) represents an obstruction to
using the cosmology (without assuming more complicated reheating)
if $N_e$ is too low, since the required $T_{rh}$ would be so low
as to be ruled out.

\newpage

\begin{figure}[hb]
\begin{center}
\epsfig{file=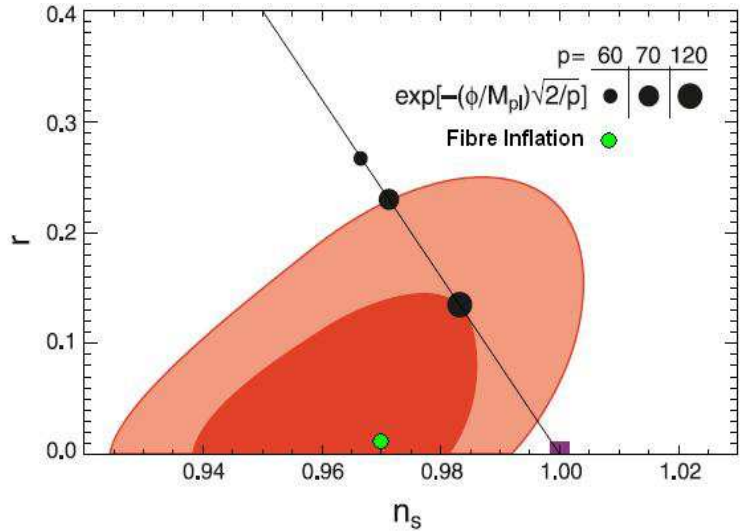, height=70mm,width=100mm}
\caption{The contours show the $68\%$ and $95\%$ C.L. derived from
WMAP+BAO+SN in the $(r-n_s)$ plane.} \label{Fig8}
\end{center}
\end{figure}

A few illustrative values are listed in Table 8.5, which assumes a
matter-dominated reheating epoch ($w=0$) and takes $M_{inf}=4
\times 10^{15}$ GeV, to compute $N_e\simeq 57$ and $T_{rh}$ as a
function of $n_s$ and $r$. These all show respectable reheat
temperatures, with $10^3 \, \hbox{GeV} < T_{rh} < 10^{10}$ GeV,
with the upper bound motivated by the requirement that gravitini
not be overproduced during reheating \cite{sarkar}. Furthermore,
as shown in Figure \ref{Fig8}, these values for $n_s$ and $r$ that
are predicted lie well within the observably allowed range.
Furthermore $r$ is large enough to allow detection by forthcoming
experiments such as EPIC, BPol or CMBPol \cite{Verde, rBounds}.

\medskip\noindent{\it Amplitude of Scalar Perturbations:}
It is not impressive to have relatively large values for the
tensor-to-scalar ratio, $r$, unless the amplitude of primordial
scalar perturbations are themselves observably large. Since this
depends on the size of Hubble scale at horizon exit, it is
sensitive to the constant $V_0 = m_\varphi^2/4 = \mathcal{C}_2/
\mc{V}^{10/3}$ that pre-multiplies the inflationary potential. The
condition that we reproduce the COBE normalisation for primordial
scalar density fluctuations, $\delta_{H} = 1.92 \cdot 10^{-5}$,
can be expressed as:
\begin{equation}
 A_{COBE}\equiv\left(\frac{g_s e^{K_{cs}}}{8\pi}\right)
 \left(\frac{V^{3/2}}{V'}\right)^{2} \simeq 2.7\cdot
 10^{-7}, \label{cobe}
\end{equation}
where the prefactor $\left(g_s e^{K_{cs}}/8\pi\right)$ is the
correct overall normalisation of the scalar potential obtained
from dimensional reduction \cite{LVS}.

As Table 8.4 shows, it is possible to obtain models with many
\efold ings and which satisfy the COBE normalisation condition,
but this clearly prefers relatively large values for $g_s$ and
$1/\mc{V}$, and so tends to prefer models whose volumes are not
inordinately large. It is then possible to evaluate the
inflationary scale as (setting $K_{cs}=3\ln g_s\simeq -3.6$):
\begin{equation}
 M_{inf}=V^{1/4}_{end}\simeq
 V_0^{1/4}M_P=\left(\frac{g_s\mathcal{C}_{2}}{8\pi}\right)^{1/4}\frac{e^{K_{cs}/4}}
 {\langle\mathcal{V}\rangle^{5/6}}M_P\sim 5\cdot 10^{15} \textrm{GeV},
\end{equation}
as can be deduced from Table 8.6 which summarises the different
inflationary scales obtained for the models SV1 and SV2 with
smaller values for the overall volume. These results were used
above in Table 8.5 to determine the correlation between
observables and reheat temperature.

\begin{figure}[ht]
\begin{center}
\begin{tabular}{c||c|c}
  & SV1 & SV2 \\
  \hline\hline
  $\mathcal{C}_2$ & 5157.35 & 9946.73 \\
  $\langle\mathcal{V}\rangle$ & 1709.55 & 1626.12 \\
  $N_e^{max}$ & 64 & 64 \\
  $M_{inf}$ & $5.5 \cdot 10^{15}$ & $6.8 \cdot 10^{15}$ \\
\end{tabular}\\
\vspace{0.3cm}{{\bf Table {8.6}:} Inflationary scales for models
with large $r$ (setting $K_{cs}=3\ln g_s\simeq -3.6$).}
\end{center}
\end{figure}

We have seen that although the Fibre Inflation mechanism can
naturally produce inflation with detectable tensor modes if the
moduli start at large enough values for $\hat\varphi$ ({\em i.e.}
high-fibre models), the generic such model ({\em e.g.} the LV
model of the Tables) predicts too small a Hubble scale during
inflation to have observable fluctuations. Such models may
nonetheless ultimately prove to be of interest, either by using
alternative mechanisms \cite{DiffMechs} to generate perturbations,
or as a way to generate a second, shorter and relatively late
epoch of inflation \cite{2ndInf} (as might be needed to eliminate
relics in the later universe).

\subsection{Two-field cosmological evolution}

Given that the resulting volumes, $\mathcal{V}\gtrsim 10^3$, are
not extremely large, one could wonder whether the approximations
made above are fully justified or not. We pause now to re-examine
in particular the assumption that $\mc{V}$ and $\tau_3$ remain
fixed at constant values while $\tau_1$ rolls during inflation. We
first identify the combination of parameters that controls this
approximation, and then re-analyse the slow roll with these fields
left free to move. This more careful treatment justifies our use
of the single-field approximation elsewhere.

\subsubsection{Inflaton back-reaction onto $\mc{V}$ and $\tau_3$}

Recall that the approximation that $\mc{V}$ and $\tau_3$ not move
is justified to the extent that the $\tau_1$-independent
stabilising forces of the potential $\delta V_{(\alpha')}$ remain
much stronger than the forces in $\delta V_{(g_s)}$ that try to
make $\mc{V}$ and $\tau_3$ also move. And this hierarchy of forces
seems guaranteed to hold because of the small factors of $g_s$ and
$1/\mc{V}^{1/3}$ that suppress the string-loop contribution
relative to the $\alpha'$ corrections. However we also see, from
(\ref{cobe}) and Table 8.4, that observably large primordial
fluctuations preclude taking $g_s e^{K_{cs}}/\mc{V}^{10/3}$ to be
too small -- at least when they are generated by the standard
mechanism. This implies a tension between the COBE normalisation
and the validity of our analysis at fixed $\mc{V}$, whose severity
we now try to estimate.

Since the crucial issue is the relative size of the forces on
$\mc{V}$ due to $\delta V_{(\alpha')}$, $\delta V_{(np)}$ and
$\delta V_{(g_s)}$, we first compare the derivatives of these
potentials. Keeping in mind that it is the variable $\vartheta_v
\sim \ln \mc{V}$ that satisfies the slow-roll condition, we see
that the relevant derivative to be compared is $\mc{V} \partial
/\partial \mc{V}$. Furthermore, since it is competition between
derivatives of $\delta V_{(np)}$ and $\delta V_{(\alpha')}$ in
eq.~(\ref{ygfdooi}) that determines $\mc{V}$ in the leading
approximation, it suffices to compare the string-loop potential
with only the $\alpha'$ corrections, say. We therefore ask when:
\be
 \left| \mc{V} \frac{\partial \delta V_{(g_s)}}{\partial \mc{V}}
 \right| \ll \left| \mc{V} \frac{\partial
 \delta V_{(\alpha')}}{\partial \mc{V}}
 \right| \,,
\ee
or when:
\begin{equation}
 \frac{10 \, \mathcal{C}_2}{\mc{V}^{10/3}}
 \ll \frac{9 \, \xi}{4g_{s}^{3/2}}
 \frac{W_{0}^{2}}{\mathcal{V}^{3}} \,,
\label{Valfa}
\end{equation}
where we take $3 \gg 4 \, e^{-\kappa\hat\varphi/2}$ during
inflation when simplifying the left-hand side. Grouping terms we
find the condition:
\begin{equation}
 \mathcal{R}_{cv} :=
 \left(\frac{9 \xi W_{0}^{2}}{40 g_{s}^{3/2}} \right)
 \frac{\mathcal{V}^{1/3}}{\mathcal{C}_2}
 \simeq \left( \frac{9 \xi \zeta^{4/3}}{40 g_{s}^{3/2}} \right)
 \frac{\mathcal{V}^{1/3}}{A}
 \gg 1 \,,
 \label{stab}
\end{equation}
which is clearly satisfied if we can choose $g_s$ and
$1/\mc{V}^{1/3}$ to be sufficiently small. The value for
$\mathcal{R}_{cv}$ predicted by the benchmark models of Tables 8.1
and 8.2 is given in Table 8.4. This Table shows that large
$\mathcal{R}_{cv}$ is much larger in large-$\mc{V}$ models, as
expected, with $\mathcal{R}_{cv} > 10^3$ in the LV model. By
contrast, $\mathcal{R}_{cv} \gtrsim 10$ for inflationary parameter
choices (SV1 and SV2) that satisfy the COBE normalisation.
Although these are large, the incredible finickiness of
inflationary constructions leads us, in the next section, to study
the multi-field problem where the volume modulus is free to roll
in addition to the inflaton. Be doing so we hope to widen the
parameter space of acceptable inflationary models.

\subsubsection{Relaxing the Single-Field Approximation}

\begin{figure}[ht]
\begin{center}
\epsfig{file=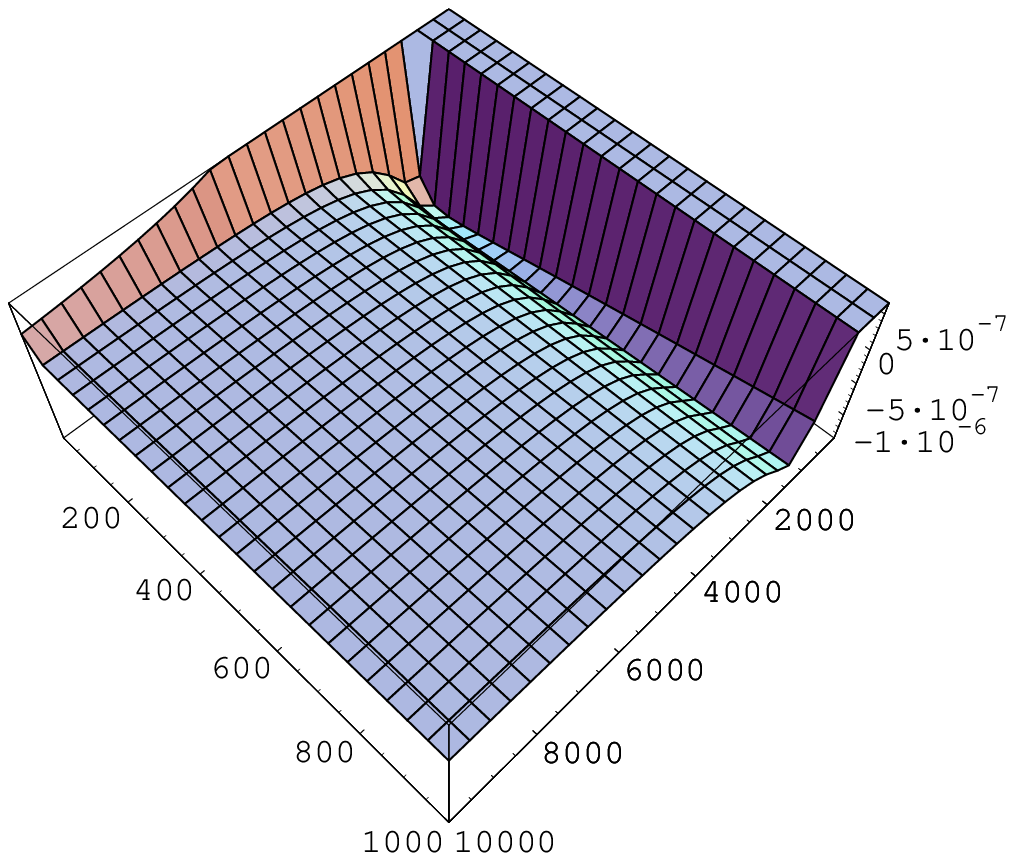, height=60mm,width=60mm}
\epsfig{file=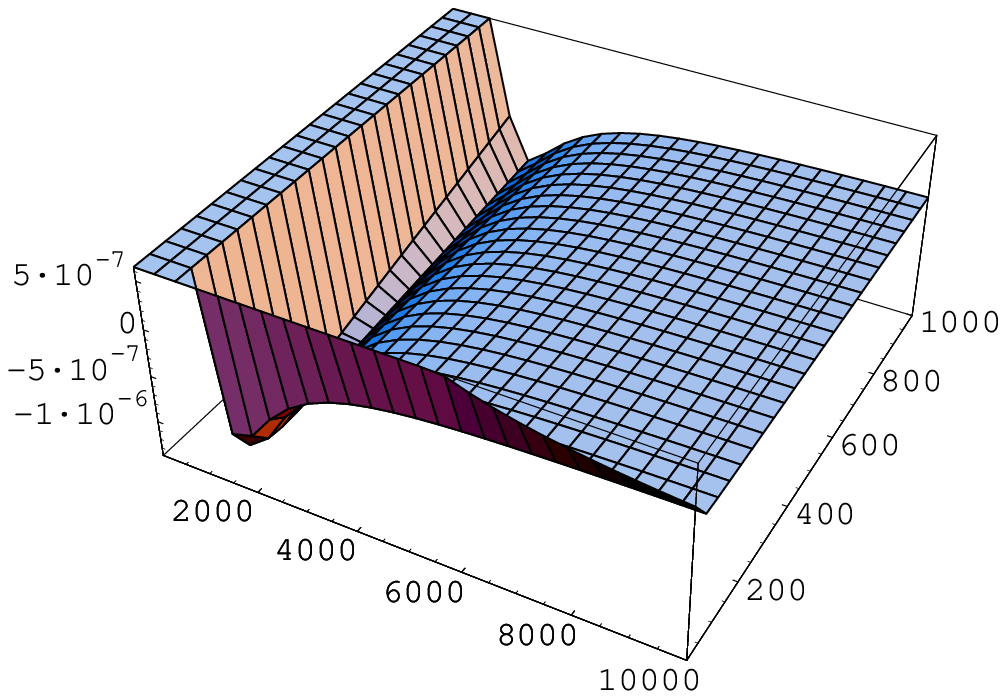, height=60mm,width=60mm} \caption{Two
  views of the
  inflationary trough representing the potential as a function of the
  volume and $\tau_1$ for $R=0$. The rolling is mostly in the $\tau_1$ direction
  (`north-west' direction in the left-hand figure and `south-west'
  direction in the
  right-hand figure).} \label{Fig:trough}
\end{center}
\end{figure}

In this section we redo the inflationary analysis without making
the single-field approximation. We start from the very general
scalar potential, whose form is displayed in Figure
\ref{Fig:trough}:
\begin{eqnarray}
V &=&\mu _{1}\frac{\sqrt{\tau _{3}}}{\mathcal{V}}e^{-2a_{3}\tau
_{3}}-\mu
_{2}W_{0}\frac{\tau _{3}e^{-a_{3}\tau _{3}}}{\mathcal{V}^{2}}+\mu _{3}\frac{%
W_{0}^{2}}{\mathcal{V}^{3}}+\frac{\delta _{up}}{\mathcal{V}^{4/3}}  \notag \\
&&+\frac{D}{\mathcal{V}^{3}\sqrt{\tau _{3}}}+\left( \frac{A}{\tau _{1}^{2}}-%
\frac{B}{\mathcal{V}\sqrt{\tau _{1}}}+\frac{C\tau _{1}}{\mathcal{V}^{2}}%
\right) \frac{W_{0}^{2}}{\mathcal{V}^{2}}\,.  \label{ilpot}
\end{eqnarray}
Here:
\begin{equation}
 \mu_{1}\equiv\frac{8a_{3}^{2}A_{3}^{2}}{3\alpha\gamma},\text{ \ \
 \ }\mu_{2}\equiv 4 a_{3}A_{3},\text{ \ \ \
 }\mu_{3}\equiv\frac{3\xi}{4 g_{s}^{3/2}}.
\end{equation}
Recall that the correction proportional to $D$ does not depend on
$\tau_1$ which is mostly the inflaton, but it can change the
numerical value obtained for $\tau_3$ and $\mc{V}$ at the minimum.
However, for $D=g_s^2\left(\mathcal{C}_3^{KK}\right)^2\ll 1$ this
modification is negligible. Thus we set $D=0$ from now on.

The result for $\mc{V}$ obtained by solving ${\partial
V}/{\partial \tau _{3}}=0$, in the limit $a_{3}\tau _{3}\gg 1$,
reads:
\begin{equation}
 \mathcal{V} =\frac{2\mu_{2}W_{0}}{\mu_{1}} \sqrt{\tau _{3}}
 \left( \frac{1- a_3 \tau_3}{1 - 4 a_3 \tau_3} \right)
 e^{a_{3}\tau_{3}}
 \simeq \frac{\mu_{2}W_{0}}{2\mu_{1}}\sqrt{\tau _{3}}
 \, e^{a_{3}\tau_{3}}.  \label{tautre}
\end{equation}
Solving (\ref{tautre}) for $\tau_3$, we obtain the result:
\be \label{app:tau3vsV}
    a_3 \tau_3 \simeq a_3 \tau_3 + \ln \left(\frac{\sqrt\tau_3}{2}
    \right)  := \ln \left(
    c \mc{V} \right) \,,
\ee
where $c = 2 a_3 A_3/(3\alpha \gamma \, W_0)$. Here the first
approximate equality neglects the slowly-varying logarithmic
factor, bearing in mind that in most of our applications we find
$\sqrt{\tau_3} \simeq 2$. Using this to eliminate $\tau_3$ in
(\ref{ilpot}) then gives the following approximate potential for
$\mc{V}$ and $\tau_1$:
\begin{equation}
 V=\left[-\mu_{4}(\ln
 \left(c\mathcal{V}\right))^{3/2}+\mu_{3}\right]
 \frac{W_{0}^{2}}{\mathcal{V}^{3}}+\frac{\delta
 _{up}}{\mathcal{V}^{4/3}}+\left( \frac{A}{\tau
 _{1}^{2}}-\frac{B}{\mathcal{V}\sqrt{\tau _{1}}}+\frac{C\tau
 _{1}}{\mathcal{V}^{2}}\right) \frac{W_{0}^{2}}{\mathcal{V}^{2}}, \label{PPpot1}
\end{equation}
where $\mu_{4} = \frac32 \alpha\gamma a_3^{-3/2}$.

Given that we set $\tau_3$ at its minimum, $\partial_{\mu}
\tau_3=0$, and so the non canonical kinetic terms look like
(\ref{Lkin2oi}). In order now to study inflation, we let the two
fields $\mathcal{V}$ and $\tau_1$ evolve according to the
cosmological evolution equations for non-canonically normalised
scalar fields:
\begin{equation}
 \left\{
 \begin{array}{c}
 \ddot{\varphi}^{i}+3H\dot{\varphi}^{i}
 +\Gamma_{jk}^{i}\dot{\varphi}^{j}\dot{
 \varphi}^{k}+g^{ij}\frac{\partial V}{\partial \varphi ^{j}}=0, \\
 H^{2}=\left( \frac{\dot{a}}{a}\right) ^{2}=\frac{1}{3}\left(
 \frac{1}{2} g_{ij}\dot{\varphi}^{i}\dot{\varphi}^{j}+V\right),
 \end{array}
 \right.
\end{equation}
where $\varphi_i$ represents the scalar fields ($\mathcal{V}$ and
$\tau_1$ in our case), $a$ is the scalar factor, and $\Gamma
_{jk}^{i}$ are the target space Christoffel symbols using the
metric $g_{ij}$ for the set of real scalar fields $\varphi^i$ such
that $\frac{\partial^2 K}{\partial \Phi^{I}
\partial_{\mu} \Phi^{*J}} \partial^{\mu}\Phi^I\partial\Phi^{*J}
=\frac{1}{2}g_{ij}\partial_{\mu}\varphi^i\partial^{\mu}\varphi^j$.

For numerical purposes it is more convenient to write down the
evolution of the fields as a function of the number $N_e$ of
\efold ings rather than time. Using:
\begin{equation}
 a(t)=e^{N_e},\textit{ \ \ \ \ \ \ \ \ \ \
 }\frac{d}{dt}=H\frac{d}{dN_e} ,
\end{equation}
we avoid having to solve for the scale factor, instead directly
obtaining $\mathcal{V}(N_e)$ and $\tau_1(N_e)$. The equations of
motion are (with $'$ denoting a derivative with respect to $N_e$):
\begin{eqnarray}
 \tau _{1}^{\prime \prime } &=&-\left( \mathcal{L}_{kin}+3\right)
 \left( \tau _{1}^{\prime }+2\tau _{1}^{2}\frac{V_{,
 \tau_{1}}}{V}+\tau _{1}\mathcal{V} \frac{V_{,\mathcal{V}}}{V}\right)
 +\frac{\tau _{1}^{\prime 2}}{\tau _{1}},
 \notag \\
 \mathcal{V}^{\prime \prime } &=&-\left( \mathcal{L}_{kin}+3\right)
 \left( \mathcal{V}^{\prime }+\tau _{1}\mathcal{V}\frac{V_{,
 \tau_{1}}}{V}+\frac{3
 \mathcal{V}^{2}}{2}\frac{V_{,\mathcal{V}}}{V}\right)
 +\frac{\mathcal{V} ^{\prime 2}}{\mathcal{V}},
\end{eqnarray}
We shall focus on the parameter case SV2, for which a numerical
analysis of the full potential gives:
\begin{equation}
 \left\langle \mathcal{V}\right\rangle =1413.26,\text{ \ \ \
 }\left\langle \tau _{1}\right\rangle =6.77325,\text{ \ \ \ }
 \delta_{up}=0.082.
\end{equation}
To evaluate the initial conditions, we fix $\tau_1\gg\left\langle
\tau _{1}\right\rangle$ and then we work out numerically the
minimum in the volume direction $\langle \mc{V} \rangle = \langle
\mc{V} \rangle( \tau_1)$.

Notice that, in general, in the case of unwarped up-lifting
$\frac{\delta_{up}}{\mc{V}^2}$, the volume direction develops a
run-away for large $\tau_1$, whereas the potential is well behaved
for the case with warped up-lifting $\frac{ \delta_{up}
}{\mc{V}^{4/3}}$. Thus we set the following initial conditions:
\begin{equation}
 \tau_1(0)=5000\text{ \ \ }\Rightarrow\text{ \ \
 }\mc{V}(0)\equiv\langle\mathcal{V}\rangle(\tau_1=5000)=1841.25,\text{
 \ \ \ }\tau_1'(0)=0,\text{ \ \ \ }\mc{V}'(0)=0.
\end{equation}
We need to check now that for this initial point we both get
enough \efold ings and the spectral index is within the allowed
range. In order to do this, we start by recalling the
generalisation of the slow-roll parameter $\varepsilon$ in the
two-field case:
\begin{equation}
 \varepsilon=-\frac{\left(V_{,\tau_1}\dot{\tau_1}+V_{,\mathcal{V}}
 \dot{\mathcal{V}}\right)^2}{4\mathcal{L}_{kin}V^2},
\end{equation}
and so it becomes a function of the number of \efold ings. In the
case SV2, $\varepsilon\ll 1$ for the first 65 \efold ings as it is
shown in Figure \ref{Fig1} below.

\begin{figure}[ht]
\begin{center}
\epsfig{file=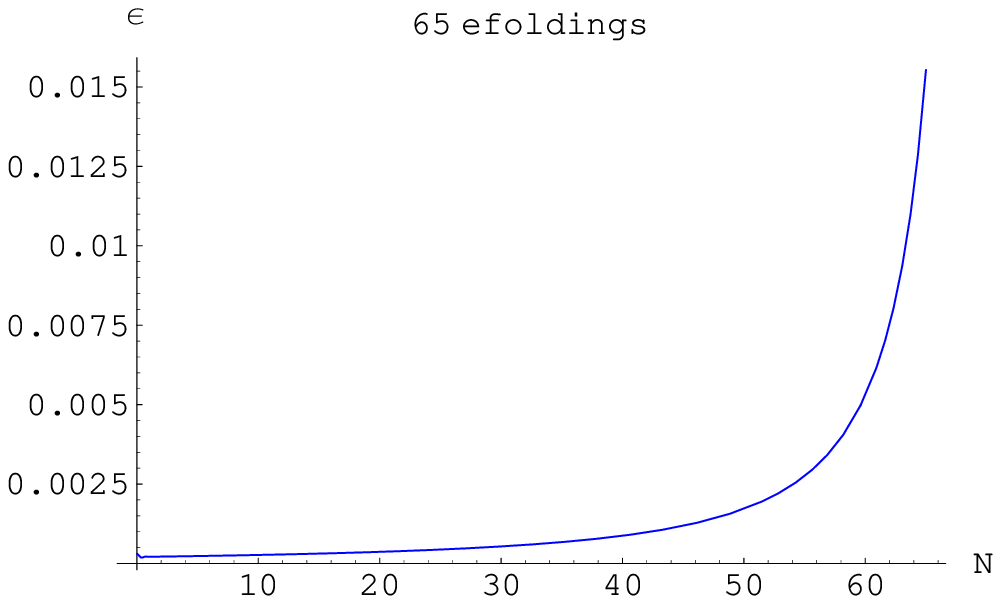, height=50mm,width=60mm}
\epsfig{file=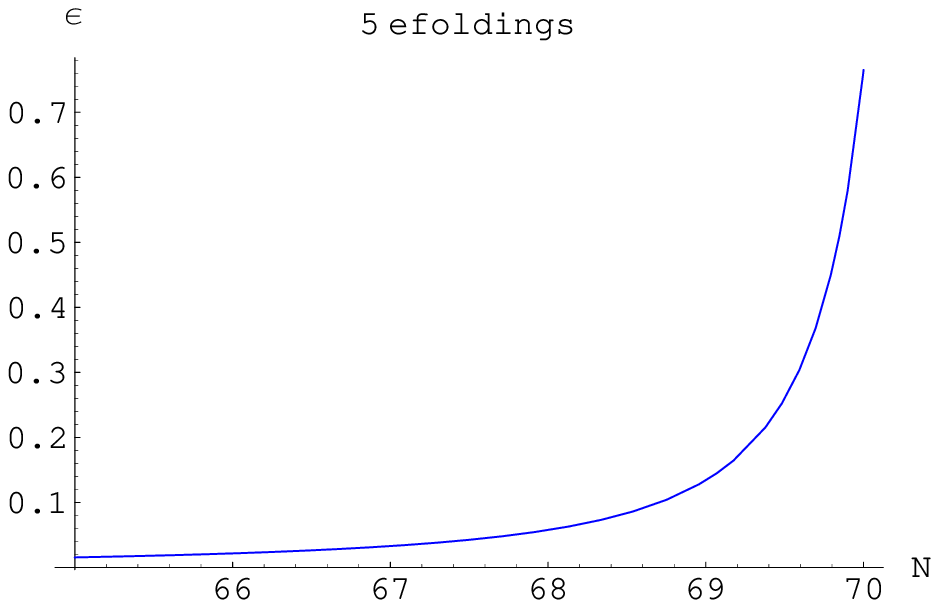, height=50mm,width=60mm}
\caption{$\varepsilon$ versus $N$ for (left) the first 65 \efold
ings of inflation and (right) the last 5 \efold ings.}
\label{Fig1}
\end{center}
\end{figure}

However $\varepsilon$ grows faster during the last 5 \efold ings
until it reaches the value $\varepsilon\simeq 0.765$ at $N=70$ at
which point the slow-roll approximation ceases to be valid and
inflation ends. This can be seen in Figure \ref{Fig1}. (From here
on we save $N_e$ to refer to the physical number of \efold ings,
and denote by $N$ the variable that parameterises the cosmological
evolution of the fields).

\begin{figure}[ht]
\begin{center}
\epsfig{file=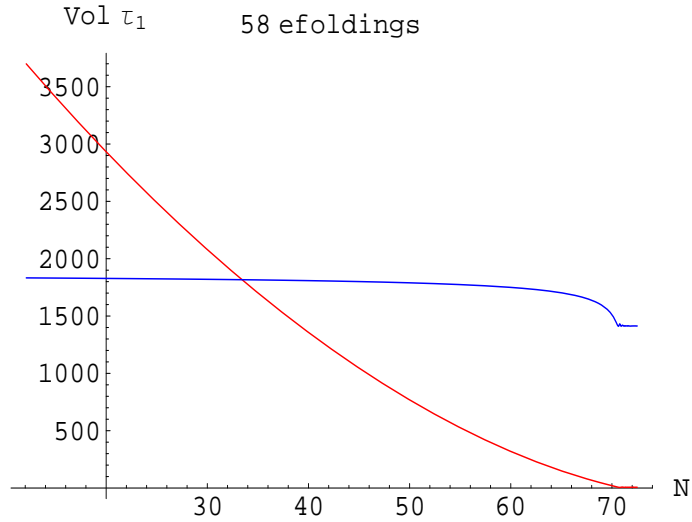, height=70mm,width=100mm}
\caption{$\tau_1$ (red curve) and $\mathcal{V}$ (blue curve)
versus $N$ for the last 58 \efold ings of inflation.} \label{Fig3}
\end{center}
\end{figure}

Therefore focusing on horizon exit at 58 \efold ings before the
end of inflation, we need to start at $N=12$. We also find
numerically that at horizon exit $\varepsilon(N=12)=0.0002844$
which corresponds to a tensor-to-scalar ration $r=4.6\cdot
10^{-3}$. Figure \ref{Fig3} shows the cosmological evolution of
the two fields during the last 58 \efold ings of inflation before
the fields start oscillating around the minimum. It it clear how
the motion is mostly along the $\tau_1$ direction, as expected.

\begin{figure}[ht]
\begin{center}
\epsfig{file=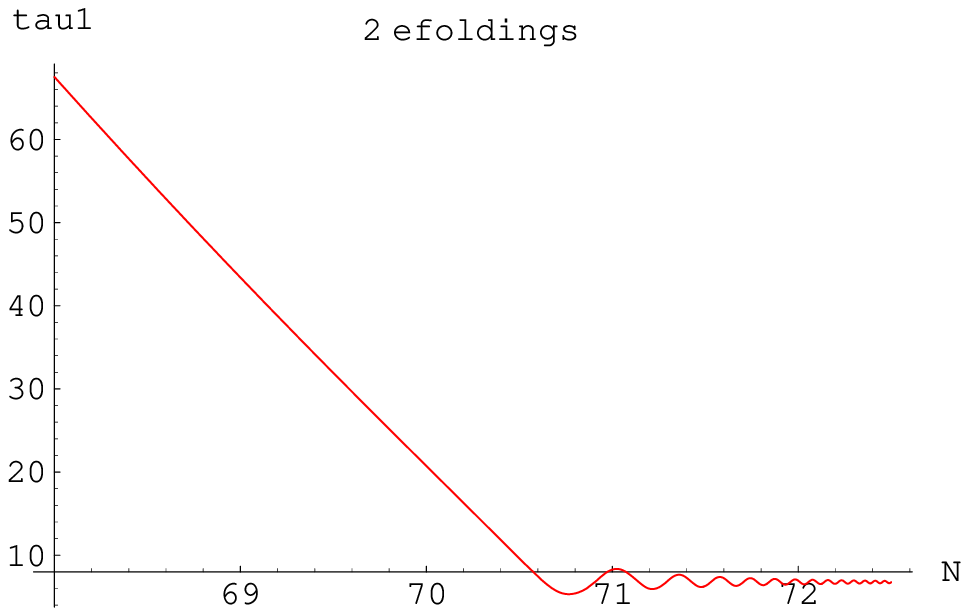, height=60mm,width=67mm}
\epsfig{file=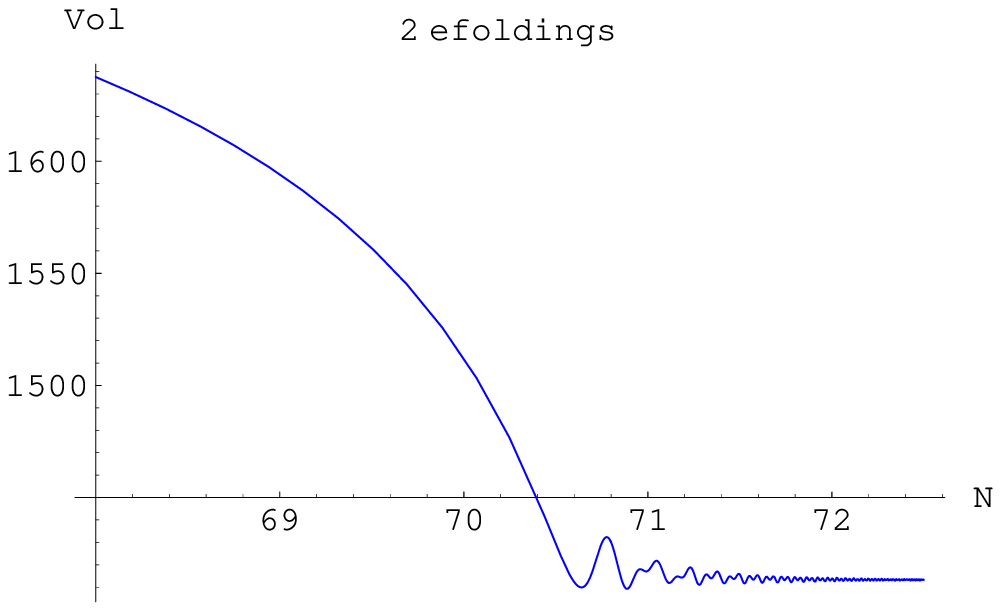, height=60mm,width=67mm} \caption{
 Plot of the $\tau_1$ (red curve on the left)
 and $\mc{V}$ (blue curve on the right) vs $N$ for the last 2 \efold ings of inflation.}
   \label{Fig4}
\end{center}
\end{figure}

Figure \ref{Fig4} gives a blow-up of the $\tau_1$ and $\mc{V}$
trajectory close to the minimum for the last 2 \efold ings of
inflation, where it is evident how the fields oscillate before
sitting at the minimum.

\begin{figure}[ht]
\begin{center}
\epsfig{file=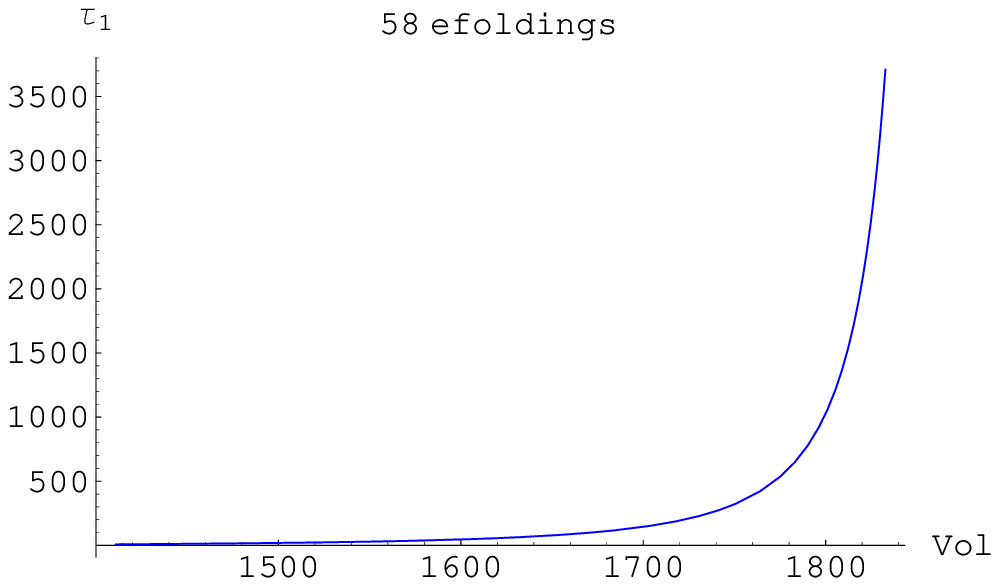, height=60mm,width=67mm}
\epsfig{file=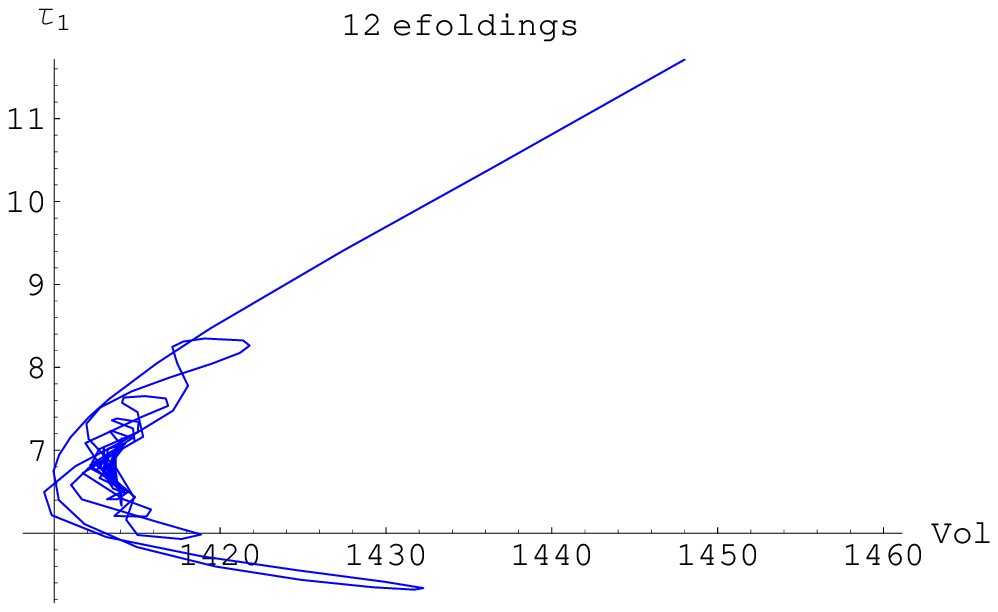, height=60mm,width=67mm}
\caption{Path of the inflation trajectory in the $\tau_1$-$\mc{V}$
space for the last 58 (left) and 12 (right) \efold ings of
inflation.}
   \label{Fig5}
\end{center}
\end{figure}

Finally, Figure \ref{Fig5} illustrates the path of the inflation
trajectory in the $\tau_1$-$\mc{V}$ space.

\begin{figure}[ht]
\begin{center}
\epsfig{file=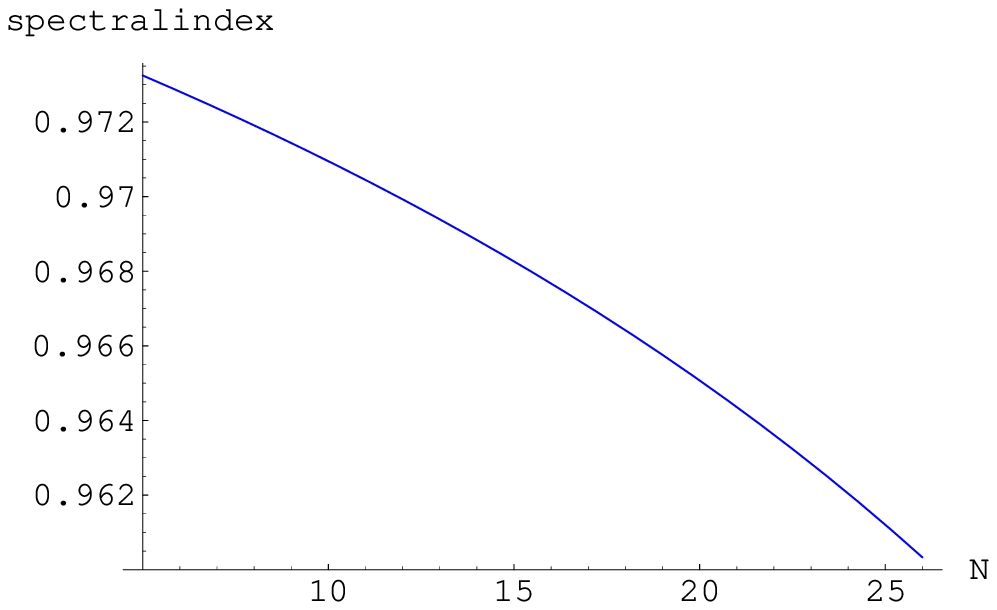, height=50mm,width=60mm}
\epsfig{file=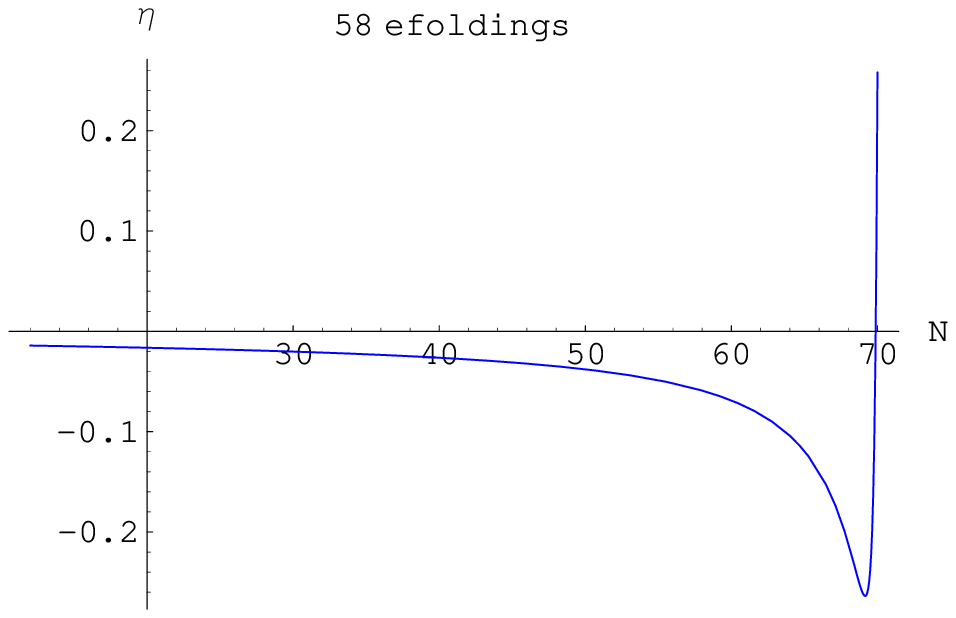, height=50mm,width=60mm} \caption{Left: $n_s$
versus $N$ between 65 and 44 \efold ings before the end of
inflation. Right: $\eta$ versus $N$ during the last 58 \efold ings
of inflation.} \label{Fig6}
\end{center}
\end{figure}

To consider the experimental predictions of Fibre Inflation we
need to make sure that the inflaton is able to generate the
correct amplitude of density fluctuations. After multiplying the
scalar potential (\ref{PPpot1}) by the proper normalisation factor
$g_s e^{K_{cs}}/(8\pi)$, the COBE normalisation on the power
spectrum of scalar density perturbations is given by:
\begin{equation}
 \sqrt{P}\equiv\frac{\sqrt{g_s}e^{K_{cs}/2}}{20\sqrt{3}\pi^{3/2}}\sqrt{\frac{V}{\varepsilon}}=2\cdot
 10^{-5},
\end{equation}
where both $V$ and $\varepsilon$ have to be evaluated at horizon
exit for $N=12$ corresponding to $N_e=58$. We find numerically
that the COBE normalisation is perfectly matched:
\begin{equation}
 \text{at \ }N=12\text{: \ }\tau_1=3710.5,\textit{ \ }
 \mc{V}=1832.74,\Rightarrow V=6.1\cdot
 10^{-7}\Rightarrow\sqrt{P}=2.15\cdot
 10^{-5}. \notag
\end{equation}
We need also to evaluate the spectral index which is defined as:
\begin{equation}
 n_s=1+\frac{d \ln{P(k)}}{d \ln{k}}\simeq 1+\frac{d \ln{P(N)}}{d
N},
\end{equation}
where the latter approximation follows from the fact that
$k=aH\simeq He^{H}$ at horizon exit, so $d\ln{k}\simeq dN$. In
Figure \ref{Fig6} we plot the spectral index versus $N$ around
horizon exit, namely between 65 and 44 \efold ings before the end
of inflation. It turns out that $n_s(N=12)=0.96993$, and so our
starting point is within the experimentally allowed region for the
spectral index.

We also checked that the second slow-roll parameter $\eta$,
obtainable from $\eta=(n_s+6\varepsilon-1)/2$, is always less than
unity during the last 58 \efold ings as shown in Figure
\ref{Fig6}. It is interesting to notice that $\eta$ vanishes very
close to the end of inflation for $N=69.88$. This is perfect
agreement with the presence of the inflection point previously
found in the fixed-volume approximation.

The inflationary scale evaluated at the end of inflation turns out
to be:
\begin{equation}
 M_{inf}=V^{1/4}_{end}M_P=V(N=70)^{1/4}M_P=5.2\cdot 10^{16} \textrm{GeV},
\end{equation}
and so, using (\ref{cosmology}) for $w=0$, we deduce that we can
obtain $N_e=58$ if $T_{rh}=2.27\cdot 10^{9}$ GeV which is
correctly below $10^{10}$ GeV to solve the gravitino problem.
Finally we conclude that we end up with the following experimental
predictions:
\begin{equation}
 n_s\simeq 0.970,\textit{ \ \ \ }r\simeq4.6\cdot 10^{-3} \,,
\end{equation}
in agreement with our earlier single-field results.

\newpage

\subsection{Naturalness}

Finally, we return to the issue of the stability of the
inflationary scenario presented here to various kinds of
perturbations, and argue that it is much more robust than are
generic inflationary mechanisms because of the control afforded by
the LARGE volume approximation.\footnote{We thank Liam McAllister
for several helpful conversations on this point.}

As we have seen in chapter 7, there are several reasons why
inflationary models are generically sensitive to perturbations of
various kinds, of which we list several, explaining in each case
why Fibre Inflation is not affected by these kinds of
perturbations.

\medskip\noindent{\it Dimension-six Operators and the
$\eta$ problem:}

\medskip\noindent A generic objection to the stability of an
inflationary scenario rests on the absence of symmetries
protecting scalar masses. This line of argument \cite{etaproblem}
grants that it is possible to arrange a regime where the scalar
potential is to a good approximation constant, $V = V_0$, chosen
to give the desired inflationary Hubble scale, $3 M_P^2 H^2 =
V_0$. It then asks whether there are dangerous higher-dimension
interactions in the effective theory that are small enough to
allow an effective field theory description, but large enough to
compete with the extraordinarily flat inflationary potential.

In particular, since $V_0$ is known (by assumption) not to be
precluded by symmetries of the problem, and since scalar masses
are notoriously difficult to rule out by symmetries, one worries
about the possibility of the following dimension-six combination
of the two:
\be
    {\cal L}_{\rm eff} = \frac{1}{M^2} \; V_0 \varphi^2 \,,
\ee
where $\varphi$ is the canonically normalised inflaton and $M$ is
a suitable heavy scale appropriate to any heavy modes that have
been integrated out. Such a term is dangerous, even with $M \simeq
M_P$, because it contributes an amount $V_0/M_P^2 \simeq H^2$ to
the inflaton mass, corresponding to $\delta \eta \simeq {\cal
O}(1)$. Supersymmetric versions of this argument use the specific
form $V_F = e^K \, U$, where $U$ is constructed from the
superpotential, and argue that inflation built on regions of
approximately constant $U$ get destabilised by generic $\delta K
\simeq \varphi^*\varphi$ corrections to the K\"ahler potential.

A related question that is specific to the large-field models
required for large tensor fluctuations asks what controls the
expansion of the effective theory in powers of $\varphi$ if fields
run over a range as large as $M_P$.

We believe that neither of these problems arise in the Fibre
Inflation models considered here. First, both arguments rely on
generic properties of an expansion in powers of $\varphi$, which
is strictly a valid approximation only for small excursions about
a fixed point in field space. As the previous paragraph points
out, such an expansion cannot be used for large field excursions
and one must instead identify a different small parameter with
which to control calculations. In the present instance this small
parameter is given both by powers of $1/\mc{V}$ and by powers of
$g_s$, since these control the underlying string perturbation
theory and low-energy approximations. In particular, as we find
explicitly in appendix B.1, in the supersymmetric context these
ensure that perturbations to $K$ have the form $\delta e^K \simeq
\delta (1/\mc{V}^2) \simeq -2\delta \mc{V}/\mc{V}^3$, and it is
the suppression by additional powers of the LARGE volume that
makes such corrections less dangerous than they would generically
be.

Normally when dangerous corrections are suppressed by a small
expansion parameter, the suppression can be traced to additional
symmetries that emerge in the limit that the parameter vanishes.
But for large-volume expansions, the corrections vanish strictly
in the de-compactification limit, $\mc{V} \to \infty$, which {\em
does} enjoy many symmetries (like higher-dimensional general
covariance) that are not evident in the effective
lower-dimensional theory. It would be worth understanding in more
detail whether the natural properties of the large-volume
expansion can be traced to these additional symmetries of higher
dimensions.

\medskip\noindent{\it Integrating out sub-Planckian modes:}

\medskip\noindent There is a more specific objection, related to the
above. Given the potential sensitivity of inflation to
higher-dimension operators, this objection asks why the inflaton
potential is not destabilised by integrating out the many heavy
particles that are likely to live above the inflationary scale,
$M_I \simeq V_0^{1/4}$ and below the Planck scale? (See, for
instance, \cite{tPsignals} for more specific variants of this
question.)

In particular, for string inflation models one worries about the
potential influence of virtual KK modes, since these must be
lighter than the string and four dimensional Planck scales, and
generically couple to any inflaton field. For Fibre Inflation this
question can be addressed fairly precisely, since virtual KK modes
are included in the string loop corrections that generate the
inflationary potential in the first place.

There are generically two ways through which loops can introduce
the KK scale into the low-energy theory. First, the lightest KK
masses enter as a cutoff for the virtual contribution of the very
light states that can be studied purely within the four
dimensional effective theory. These states contribute following
generic contributions to the inflaton potential:
\be \label{4Dloop}
    \delta V_{inf}^{4D} \simeq c_1 \hbox{STr} \, M^4 + c_2 m_{3/2}^2
    \hbox{STr} \, M^2 + \cdots \,,
\ee
where $c_1$ and $c_2$ are dimensionless constants, the gravitino
mass, $m_{3/2}$, measures the strength of supersymmetry breaking
in the low-energy theory, and the super-traces are over powers of
the generic four dimensional mass matrix, $M$, whose largest
elements are of order the KK scale, $M_{KK}$. In general
low-energy supersymmetry ensures $c_1 = 0$, making the second term
the leading contribution.

Now comes the important point. In the LARGE volume models of
interest, we know that $m_{3/2} \sim \mc{V}^{-1}$, and we know
that $M_{KK}$ is suppressed relative to the string scale by
$\mc{V}^{-1/6}$, and so in Planck units $M_{KK} \sim
\mc{V}^{-2/3}$. These together imply that $\delta V_{inf}^{4D}
\sim \mc{V}^{-10/3}$, in agreement with the volume-dependence of
the loop-generated inflationary potential discussed above.

But $\delta V_{inf}$ also potentially receives contributions from
scales larger than $M_{KK}$ and these cannot be described by the
four dimensional loop formula, eq. (\ref{4Dloop}). These must
instead be computed using the full higher-dimensional (string)
theory, potentially leading to the dangerous effective
interactions in the low-energy theory. This calculation is the one
that is explicitly performed for toroidal orientifolds in
\cite{bhk} and whose properties were estimated more generally in
\cite{bhp} and in chapter 5. Their conclusion is that such
effective contributions {\em do} arise in the effective four
dimensional theory, appearing there as contributions to the
low-energy K\"ahler potential. The contributions from open-string
loops wrapped on a cycle whose volume is $\tau$ have the generic
form:
\be
    \delta K \simeq \frac{1}{\mc{V}} \left[a_1 \sqrt\tau+ \frac{a_2}{\sqrt{\tau}}
    + \cdots \right] \,,
\ee
where it is $1/\tau$ that counts the loop expansion.

These two terms can potentially give contributions to the scalar
potential, and if so these would scale with $\mc{V}$ in the
following way:
\be
    \delta V^{\rm he}_{inf} \sim \frac{a_1}{\mc{V}^{8/3}} +
    \frac{a_2}{\mc{V}^{10/3}} + \cdots \,.
\ee
Notice that the first term is therefore potentially dangerous,
scaling as it does like $M_{KK}^4$. However, as we have seen in
chapter 5, a simple calculation shows that the contribution of a
term $\delta K \propto \tau^\omega/\mc{V}$ gives a contribution to
$V_F$ of the form $\delta V_F \propto \left(\omega - \frac12
\right) \mc{V}^{-8/3}$, implying that the leading correction to
$K$ happens to drop out of the scalar potential (although it does
contribute elsewhere in the action).

These calculations show how LARGE volume and four dimensional
supersymmetry can combine to keep the potentially dangerous loop
contributions of KK and string modes from destabilising the
inflaton potential. We regard the study of how broadly this
mechanism might apply elsewhere in string theory as being well
worthwhile.

\section{Discussion}

This year the PLANCK satellite is expected to start a new era of
CMB observations, and to be joined over the next few years by
other experiments aiming to measure the polarisation of the cosmic
microwave background and to search for gravitational waves. In
this chapter, we have presented a new class of explicit string
models, with moduli stabilisation, that both agrees with current
observations and can predict observable gravitational waves, most
probable not at PLANCK but at future experiments. Many of the
models' inflationary predictions are also very robust against
changes to the underlying string/supergravity parameters, and in
particular predict a definite correlation between the scalar
spectral index, $n_s$, and tensor-to-scalar ratio, $r$. It is also
encouraging that these models realise inflation in a comparatively
natural way, inasmuch as a slow roll does not rely on fine-tuning
parameters of the potential against one another.

\newpage

Other important features of the model are:
\begin{itemize}
\item The comparative flatness of the inflaton direction,
$\Omega$, is guaranteed by general features of the modulus
potential that underly the LARGE volume constructions. These
ultimately rely on the no-scale structure of the lowest-order
K\"{a}hler potential and the fact that the leading $\alpha'$
corrections depend on the K\"{a}hler moduli only through the
Calabi-Yau volume.
\item The usual $\eta$ problem of generic supergravity theories is
also avoided because of the special features of the no-scale
LARGE-volume structure. In particular, the expansion of the
generic $e^K = \mc{V}^{-2}$ factor of the $F$-term potential are
always punished by the additional powers of $1/\mc{V}$, which they
bring along: $\delta e^K = -2 \mc{V}^{-3} \delta \mc{V}$. This
result is explicitly derived in appendix B.1.
\item The exponential form of the inflationary potential is a
consequence of two things. First, the loop corrections to $K$ and
$V$ depended generically on powers of $\Omega$ and the volume. And
second, the leading-order K\"{a}hler potential gives a kinetic
term for $\Omega$ of the form $(\partial \ln \Omega)^2$, leading
to the canonically normalised quantity $\varphi$, with $\Omega=
e^{\kappa\hat\varphi}$, with $\kappa = 2/\sqrt3$. So we know the
potential can have a typical large-field inflationary form, $V =
K_1 - K_2 e^{-\kappa_1\hat\varphi} + K_3 e^{-\kappa_2 \hat\varphi}
+ \cdots $, without knowing any details about the loop
corrections.
\item The robustness of some of the predictions then follows
because the coefficients $K_{i}$ turn out to be proportional to
one another. They are proportional because of our freedom to shift
$\varphi$ so that $\hat\varphi = 0$ is the minimum of $V$, and our
choice to uplift this potential so that it vanishes at this
minimum. The two conditions $V(0) = V'(0) = 0$ impose two
conditions amongst the three coefficients $K_1$, $K_2$ and $K_3$
(where three terms in the potential are needed to have a minimum).
The remaining normalisation of the potential can then be expressed
without loss of generality in terms of the squared mass,
$m^2_\varphi = V''(0)$.
\item The exact range of the field $\hat\varphi$ depends only on
the ratio of two parameters ($B/A$) of the underlying
supergravity. This quantity is typically much greater than one due
to the string coupling dependence of this ratio, leading to
`high-fiber' models for which $\hat\varphi$ can naturally run
through trans-Planckian values. $B/A \gg 1$ also suffices to
ensure that the minimum $\langle \varphi \rangle$ lies inside the
K\"ahler cone. But the range of $\hat\varphi$ also cannot be too
large, since it depends only logarithmically on $B/A$. This
implies that $\hat\varphi$ at most rolls through a few Planck
scales, which can allow $50-60$ \efold ings, or even a bit more.
This makes the models potentially sensitive to details of the
modulus dynamics at horizon exit, along the lines of
\cite{tPsignals}, since this need not be deep in an inflationary
regime.
\item The COBE normalisation is the most constraining restriction
to the underlying string/supergravity parameters. In particular,
as usual, it forbids the volume from being very large because it
restricts the string scale to be of the order of the GUT scale.
This leads to the well known tension between the scale of
inflation and low-energy supersymmetry \cite{cklq}. Of course,
this conclusion assumes the standard production mechanism for
primordial density fluctuations, and it remains an interesting
open question whether alternative mechanisms might allow a broader
selection of inflationary models in this class. In particular,
this makes the development of a reheating mechanism particularly
pressing for this scenario.
\item The model is extremely predictive since the requirement of
generating the correct amplitude of scalar perturbations fixes the
inflationary scale of the order the GUT scale, which, in turn,
fixes the numbers of \efold ings. Lastly the number of \efold ings
is correlated with the cosmological observables and we end up with
the general prediction: $n_s\simeq 0.970$ and $r\simeq 0.005$. We
find examples with $r\simeq 0.01$ and $n_s \simeq 1$ also to be
possible, but only if horizon exit occurs very soon after the
onset of inflation.
\end{itemize}
For these reasons, even though the string-loop corrections to the
K\"{a}hler potential are not fully known for general Calabi-Yau
manifolds, because they come as inverse powers of K\"{a}hler
moduli and the dilaton, we believe the results we find here are
likely to be quite generic. Of course, it would in any case be
very interesting to have more explicit calculations of the loop
corrections to K\"{a}hler potentials in order to better understand
this scenario. Furthermore even though blow-up modes are very
common for Calabi-Yau manifolds, it would be useful to have
explicit examples of K3 fibration Calabi-Yau manifolds with the
required intersection numbers.

During Fibre Inflation an initially large K3 fiber modulus
$\tau_1$ shrinks, with the volume $\mc{V}=t_1\tau_1$ approximately
constant. Consequently, the value of the 2-cycle modulus $t_1$,
corresponding to the base of the fibration, must grow during
inflation. This forces us to check that $t_1$ is not too small at
the start of inflation, in particular not being too close to the
singular limit $t_1\to 0$ where perturbation theory breaks down.
We show in appendix B.2 that the inflationary region can start
sufficiently far away from this singular limit. The more
restrictive limit on the range of the inflationary regime is the
breakdown of the slow-roll conditions as $t_1$ gets smaller,
arising due to the growth of a positive exponentials in the
potential when expressed using canonical variables. One can
nonetheless show that natural choices of the underlying parameters
can guarantee that enough \efold ings of inflation are achieved
before reaching this region of field space.

It is worth emphasising that, independent of inflation and as
mentioned in subsection 8.2.1, we have also shown that our
scenario allows for the LARGE volume to be realised in such a way
that there is a hierarchy of scales in the K\"ahler moduli,
allowing the interesting possibility of having two dimensions much
larger than the rest and  making contact with the potential
phenomenological and cosmological implications of two large extra
dimensions scenarios \cite{add, cliff}.

We do not address the issues of initial conditions, which in our
case ask why the other fields start initially near their minimum,
and why inflationary modulus should start out high up a fiber. As
for K\"{a}hler modulus inflation, one argument is that {\em any}
initial modulus configuration must evolve towards its stabilised
value, and so if the last modulus to reach is minimum happens to
be a fibre modulus we expect this inflationary mechanism to be
naturally at work.

\chapter{Finite Temperature Effects}
\label{FiniteTemperatureEffects}

\section{Thermal effects in string compactifications}

String compactifications with stabilised moduli typically admit a
slightly de Sitter metastable vacuum that breaks supersymmetry
along with a supersymmetric minimum at infinity. In fact, the
exponentially large volume minimum of LVS is AdS with broken
supersymmetry, even before any uplifting. In contrast, in KKLT
constructions the AdS minimum is supersymmetric and the uplifting
term is the source of supersymmetry breaking. The two minima are
separated by a potential barrier $V_b$, whose order of magnitude
is very well approximated by the value of the potential at the AdS
vacuum before uplifting.

\begin{figure}[t]
\begin{center}
\scalebox{0.9}{\includegraphics{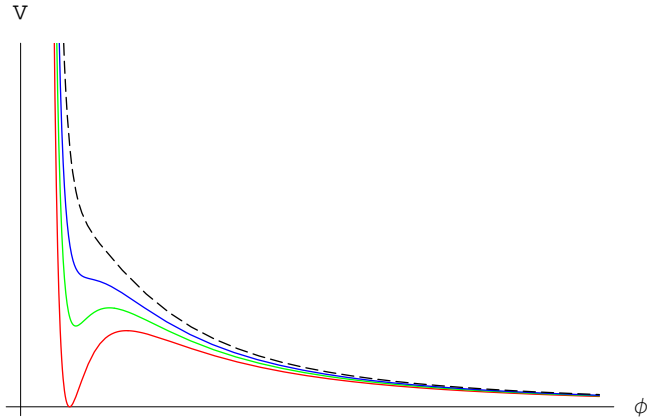}}
\end{center}
\vspace{-0.5cm} \caption{The effective potential $V$ versus the
volume modulus $\phi$ for a typical potential of KKLT or LARGE
Volume compactifications. The different curves show the effect of
various sources of energy that, if higher than the barrier of the
potential, can lead to a decompactification of the internal
space.} \label{Fig:decomp}
\end{figure}

As is well-known, the modulus related to the overall volume of the
Calabi-Yau couples to any possible source of energy due to the
Weyl rescaling of the metric needed to obtain a four dimensional
supergravity effective action in the Einstein frame. Thus, in the
presence of any source of energy greater then the height of the
potential barrier, the system will be driven to a dangerous
decompactification limit. For example, during inflation the energy
of the inflaton $\varphi$ could give an additional uplifting term
of the form $\Delta
V(\varphi,\mathcal{V})=V(\varphi)/\mathcal{V}^n$ for $n>0$, that
could cause a run-away to infinity \cite{linde}. Another source of
danger of decompactification is the following. After inflation the
inflaton decays to radiation and, as a result, a high-temperature
thermal plasma is formed. This gives rise to temperature-dependent
corrections to the moduli potential, which could again destabilise
the moduli and drive them to infinity, if the finite-temperature
potential has a run-away behaviour. The decompactification
temperature, at which the finite-temperature contribution starts
dominating over the $T=0$ potential, is very well approximated by
$T_{max}\sim V_b^{1/4}$ since $V_T\sim T^4$. Clearly, $T_{max}$
sets also an upper bound on the reheating temperature after
inflation. The discussion of this paragraph is schematically
illustrated in Figure \ref{Fig:decomp}.

On the other hand, if, instead of having a run-away behaviour, the
finite-temperature potential develops new minima, then there could
be various phase transitions, which might have played an important
r\^{o}le in the early Universe and could have observable
signatures today. The presence of minima at high $T$ could also
have implications regarding the question how natural it is for the
Universe to be in a metastable state at $T=0$. More precisely,
recent studies of various toy models \cite{craig, abel, fischler,
ART, papineau} have shown that, despite the presence of a
supersymmetric global minimum, it is thermodynamically preferable
for a system starting in a high $T$ minimum to end up at low
temperatures in a (long-lived) local metastable minimum with
broken supersymmetry. Similar arguments, if applicable for more
realistic systems, could be of great conceptual value given the
present accelerated expansion of the Universe.

For cosmological reasons then, it is of great importance to
understand the full structure of the finite temperature effective
potential. In this chapter we shall investigate this problem in
great detail for the type IIB LARGE Volume Scenario. Contrary to
the traditional thought that moduli cannot thermalise due to their
Planck-suppressed couplings to ordinary matter and radiation, we
show that in LVS some of the moduli {\it can} be in thermal
equilibrium with MSSM particles for temperatures well below the
Planck scale. The main reason is the presence of an additional
large scale in this context, namely the exponentially large
volume, which enters the various couplings and thus affects the
relevant interaction rates. The unexpected result, that some
moduli can thermalise, in principle opens up the possibility that
the finite temperature potential could develop new minima instead
of just having a run-away behaviour as, for example, in
\cite{BHLR}. However, we show that this is not the case since, for
temperatures below the Kaluza-Klein scale, the $T$-dependent
potential still has a run-away behaviour. Although it is
impossible to find exactly the decompactification temperature
$T_{max}$, as it is determined by a transcendental equation, we
are able to extract a rather precise analytic estimate for it. As
expected, we find that $T_{max}$ is controlled by the
supersymmetry breaking scale: $T_{max}^4\sim m_{3/2}^3 M_P$. This
expression gives also an upper bound on the temperature in the
early Universe. We show that this constraint can be translated
into a lower bound on the value of the Calabi-Yau volume, by
computing the temperature of the Universe $T_*$, just after the
heaviest moduli of LVS decay, and then imposing $T_*<T_{max}$.

Our lower bound implies that, for cosmological reasons, larger
values of the volume of the order $\mathcal{V}\sim 10^{15}l_s^6$,
which naturally lead to TeV-scale supersymmetry, are favoured over
smaller values of the order $\mathcal{V}\sim 10^{4}l_s^6$, which
lead to standard GUT theories. More precisely, what we mean by
this is the following. Upon writing the volume as $\mathcal{V}\sim
10^{x}$ and encoding the fluxes and the Calabi-Yau topology in the
definition of a parameter $c$, we are able to rule out a
significant portion of the $(x,c)$-parameter space that
corresponds to small $x$ (for example, for $c=1$ we obtain $x>6$).
This is rather intriguing, given that other cosmological
considerations seem to favour smaller values of the volume.
Indeed, the inflationary model presented in chapter 8 requires
$\mathcal{V}\sim 10^{4}l_s^6$, in order to generate the right
amount of density perturbations. Despite that, our lower bound on
${\cal V}$ does not represent an unsurmountable obstacle for the
realisation of inflation. The reason is that the Fibre Inflation
model can give rise to inflation even for large values of the
volume. Hence, a modification of it, such that the density
fluctuations are generated by a curvaton-like field different from
the inflaton, would be a perfectly viable model with large ${\cal
V}$. The large value of $\mathcal{V}$ would imply a low-energy
inflationary scale, and so, in turn, gravity waves would not be
observable. However, it is likely that both inflation and
TeV-scale supersymmetry could be achieved at the same time, with
also the generation of a relevant amount of non-gaussianities in
the CMB, which is a typical feature of curvaton models.

On the other hand, we pose a challenge for the solution of the
cosmological moduli problem, that the overall breathing mode of
LVS with $\mathcal{V}> 10^{10}l_s^6$ is afflicted by \cite{CQ}.
This is so, because we show that unwanted relics cannot be diluted
by the entropy released by the decay of the heaviest moduli of
LVS, nor by a low-energy period of thermal inflation. More
precisely, we show that the heaviest moduli of LVS decay before
they can begin to dominate the energy density of the Universe and,
also, that in order to study thermal inflation in the closed
string moduli sector, it is necessary to go beyond our low energy
EFT description.

The present chapter is organised as follows. In section
\ref{EffPotFT}, we recall the general form of the effective
potential at finite temperature and discuss in detail the issue of
thermal equilibrium in an expanding Universe. In section
\ref{Mmc}, we derive the masses and the couplings to visible
sector particles of the moduli and modulini in LVS. Using these
results, in section \ref{Sec:ModuliTherm} we investigate moduli
thermalisation and show that, generically, the moduli
corresponding to the small cycles can be in thermal equilibrium
with MSSM particles, due to their interaction with the gauge
bosons. In section \ref{FTCLVS}, we study the finite temperature
effective potential in LVS. We show that it has a runaway
behaviour and find the decompactification temperature $T_{max}$.
Furthermore, we establish a lower bound on the Calabi-Yau volume,
which follows from the constraint that the temperature of the
Universe just after the small moduli decay should not exceed
$T_{max}$. Finally, in section \ref{ConDis}, we summarise our
results and discuss some open issues, among which the question why
thermal inflation does not occur within our approximations.

\section{Effective potential at finite temperature}
\label{EffPotFT}

At nonzero temperature, the effective potential receives a
temperature-dependent contribution. The latter is determined by
the particle species that are in thermal equilibrium and, more
precisely, by their masses and couplings. In this section, we
review the general form of the finite temperature effective
potential and discuss in detail the establishment of thermal
equilibrium in an expanding Universe. In particular, we elaborate
on the relevant interactions at the microscopic level. This lays
the foundation for the explicit computation, in section
\ref{Sec:ExplicitTPot}, of the finite temperature effective
potential in LVS.

\subsection{General form of temperature corrections}
\label{Sec:GenTemp}

The general structure of the effective scalar potential is the
following one:
\begin{equation}
V_{TOT}=V_0+V_T,
\end{equation}
where $V_0$ is the $T=0$ potential and $V_T$ the thermal
correction. As discussed in chapter 3, $V_0$ has the general form:
\begin{equation}
V_0=\delta V_{(np)}+\delta V_{(\alpha')}+\delta V_{(g_s)},
\end{equation}
where the tree level part is null due to the no-scale structure
(recall that we are studying the scalar potential for the
K\"{a}hler moduli), $\delta V_{(np)}$ arises due to
non-perturbative effects, $\delta V_{(\alpha')}$ are $\alpha'$
corrections and the contribution $\delta V_{(g_s)}$ comes from
string loops and, as noticed in chapter 5, matches the
Coleman-Weinberg potential of the effective field theory. In
addition, $\delta V_{(g_s)}$ has an extended no-scale structure,
which is crucial for the robustness of LVS since it renders
$\delta V_{(g_s)}$ subleading with respect to $\delta V_{(np)}$
and $\delta V_{(\alpha')}$.

On the other hand, the finite temperature corrections $V_T$ have
the generic loop expansion:
\begin{equation}
V_{T}=V_T^{1-loop}+V_T^{2-loops}+... \,\, .
\end{equation}
The first term $V_T^{1-loop}$ is a 1-loop thermal correction
describing an ideal gas of non-interacting particles. It has been
derived for a renormalisable field theory in flat space in
\cite{DJ}, using the zero-temperature functional integral method
of \cite{Jackiw}, and reads
\begin{equation}
V_T^{1-loop}=\pm \frac{T^4}{2\pi^2}\int_0^{\infty}dx \,x^2
\ln\left(1\mp e^{-\sqrt{x^2+m(\varphi)^2/T^2}}\right), \label{klo}
\end{equation}
where the upper (lower) signs are for bosons (fermions) and $m$ is
the background field dependent mass parameter. At temperatures
much higher than the mass of the particles in the thermal bath,
$T\gg m(\varphi)$, the 1-loop finite temperature correction
(\ref{klo}) has the following expansion:
\begin{equation}
V_T^{1-loop}=-\frac{\pi^2 T^4}{90}\alpha +\frac{T^2
m(\varphi)^2}{24}+\mathcal{O}\left(T m(\varphi)^3\right),
\label{1-LOOP}
\end{equation}
where for bosons $\alpha=1$ and for fermions $\alpha=7/8$. The
generalisation of (\ref{1-LOOP}) to supergravity, coupled to an
arbitrary number of chiral superfields, takes the form \cite{BG1}:
\begin{equation}
V_T^{1-loop}=-\frac{\pi^2 T^4}{90} \left( g_B+ \frac{7}{8} g_F
\right)+\frac{T^2}{24}\left(Tr M_b^2+Tr M_f^2
\right)+\mathcal{O}\left(T M_b^3\right), \label{1-loop}
\end{equation}
where $g_B$ and $g_F$ are, respectively, the numbers of bosonic
and fermionic degrees of freedom and $M_b$ and $M_f$ are the
moduli-dependent bosonic and fermionic mass matrices of all the
particles forming the thermal plasma.

If the particles in the thermal bath interact among themselves, we
need to go beyond the ideal gas approximation. The effect of the
interactions is taken into account by evaluating higher thermal
loops. The high temperature expansion of the 2-loop contribution
looks like:
\begin{equation}
V_T^{2-loops}=\alpha_2 T^4\left(\sum_i f_i(g_i) \right)+\beta_2
T^2\left(Tr M_b^2+Tr M_f^2 \right)\left(\sum_i f_i(g_i)
\right)+... \,\, , \label{2-loop}
\end{equation}
where $\alpha_2$ and $\beta_2$ are known constants, $i$ runs over
all the interactions through which different species reach thermal
equilibrium, and the functions $f_i$ are determined by the
couplings $g_i$ and the number of bosonic and fermionic degrees of
freedom. For example, for gauge interactions $f(g)=const \times
g^2$, whereas for the scalar $\lambda \phi^4$ theory one has that
$f(\lambda)=const \times \lambda$.

Now, since we are interested in the moduli-dependence of the
finite temperature corrections to the scalar potential, we can
drop the first term on the RHS of (\ref{1-loop}) and focus only on
the $T^2$ term, which indeed inherits moduli-dependence from the
bosonic and fermionic mass matrices. However, notice that in
string theory the various couplings are generically functions of
the moduli. Thus, also the first term on the RHS of (\ref{2-loop})
depends on the moduli and, even though it is a 2-loop effect, it
could compete with the second term on the RHS of (\ref{1-loop}),
because it scales as $T^4$ whereas the latter one scales only as
$T^2$. This issue has to be addressed on a case by case basis, by
studying carefully what particles form the thermal bath.

\subsection{Thermal equilibrium}
\label{Sec:ThermEq}

In an expanding Universe, a particle species is in equilibrium
with the thermal bath if its interaction rate, $\Gamma$, with the
particles in that bath is larger than the expansion rate of the
Universe. The latter is given by $H\sim g_{*}^{1/2}T^2/M_P$,
during the radiation dominated epoch, with $g_{*}$ being the total
number of degrees of freedom. Thermal equilibrium can be
established and maintained by $2\leftrightarrow 2$ interactions,
like scattering or annihilation and the inverse pair production
processes, and also by $1\leftrightarrow 2$ processes, like decays
and inverse decays (single particle productions). Let us now
consider each of these two cases in detail.

\subsubsection{$2\leftrightarrow 2$ interactions}
\label{Par:2to2}

In this case the thermally averaged interaction rate can be
inferred on dimensional grounds by noticing that:
\begin{equation}
\langle\Gamma\rangle\sim\frac{1}{\langle t_c \rangle},
\end{equation}
where $\langle t_c \rangle$ is the mean time between two
collisions (interactions). Moreover:
\begin{equation}
t_c \sim \frac{1}{n\sigma v},
\end{equation}
where $n$ is the number density of the species, $\sigma$ is the
effective cross section and $v$ is the relative velocity between
the particles. Thus $\langle\Gamma\rangle\sim n\langle\sigma
v\rangle$. For relativistic particles, one has that $\langle
v\rangle\sim c$ ($\equiv 1$ in our units) and also $n \sim T^3$.
Therefore:
\begin{equation}
\langle\Gamma\rangle \sim \langle \sigma \rangle T^3 \, .
\end{equation}
The cross-section $\sigma$ has dimension of $(length)^2$ and for
$2\leftrightarrow 2$ processes its thermal average scales with the
temperature as:
\begin{enumerate}
\item{For renormalisable interactions:}
\begin{equation}
\langle \sigma \rangle \sim\alpha^2
\frac{T^2}{\left(T^2+M^2\right)^{2}}, \label{2to2}
\end{equation}
where $\alpha=g^2/(4\pi)$ ($g$ is the gauge coupling) and $M$ is
the mass of the particle mediating the interactions under
consideration.
\begin{itemize}
\item[a)] For long-range interactions $M=0$ and
(\ref{2to2}) reduces to:
\begin{equation}
\langle \sigma \rangle \sim\alpha^2 T^{-2}\text{ \
}\Rightarrow\text{ \ }\langle\Gamma\rangle\sim\alpha^2 T.
\label{UNOs}
\end{equation}
This is also the form that (\ref{2to2}) takes for short-range
interactions at energies $E>\!\!>M$.
\item[b)] For short-range interactions at scales lower than the
mass of the mediator, the coupling constant becomes dimensionful
and (\ref{2to2}) looks like:
\begin{equation}
\langle \sigma \rangle \sim\alpha^2\frac{T^{2}}{M^{4}}\text{ \
}\Rightarrow\text{ \
}\langle\Gamma\rangle\sim\alpha^2\frac{T^{5}}{M^4}. \label{3ef}
\end{equation}
\end{itemize}

\item{For processes including gravity:}
\begin{itemize}
\item[a)] Processes with two gravitational vertices:
\begin{equation}
\langle\sigma\rangle\sim d\frac{T^{2}}{M_P^4}\text{ \
}\Rightarrow\text{ \ }\langle\Gamma\rangle\sim
d\frac{T^{5}}{M_P^4}, \label{gravit}
\end{equation}
where $d$ is a dimensionless moduli-dependent constant.
\item[b)]{Processes with one renormalisable and one gravitational vertex:}
\begin{equation}
\langle\sigma\rangle\sim \sqrt{d}\frac{g^{2}}{M_P^2}\text{ \
}\Rightarrow\text{ \ }\langle\Gamma\rangle\sim \sqrt{d}\frac{g^2
T^{3}}{M_P^2}, \label{renorm-gravit}
\end{equation}
where $d$ is the same moduli-dependent constant as before.
\end{itemize}
\end{enumerate}

Let us now compare these interaction rates with the expansion rate
of the Universe, $H\sim g_{*}^{1/2}T^2/M_P$, in order to determine
at what temperatures various particle species reach or drop out of
thermal equilibrium, depending on the degree of efficiency of the
relevant interactions.
\begin{enumerate}
\item[1.a)]{Renormalisable interactions with massless
mediators:}
\begin{equation}
\langle\Gamma\rangle>H \text{ \ }\Leftrightarrow\text{ \ }\alpha^2
T>g_{*}^{1/2}T^2 M_P^{-1}\text{ \ }\Rightarrow\text{ \
}T<\alpha^2g_{*}^{-1/2} M_P.
\end{equation}
QCD processes, like the ones shown in Figure \ref{QCD}, are the
main examples of this kind of interactions. The same behaviour of
$\sigma$ is expected also for the other MSSM gauge groups for
energies above the EW symmetry breaking scale. Therefore, MSSM
particles form a thermal bath via strong interactions for
temperatures $T<\alpha_s^2g_{*}^{-1/2}M_P\sim 10^{15}$ GeV
\cite{en}.

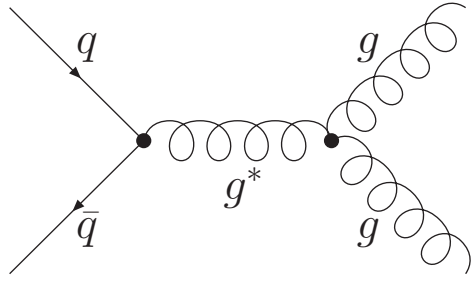
\begin{figure}
\begin{center}
\vspace{1.0cm}
  \begin{picture}(200,80) (0,0)
     \SetColor{Black}

    \ArrowLine(50,50)(0,0)
    \ArrowLine(0,100)(50,50)
    \Vertex(50,50){2.83}
    \Gluon(50,50)(120,50){7.5}{4}
    \Vertex(120,50){2.83}
    \Gluon(120,50)(170,100){7.5}{4}
    \Gluon(120,50)(170,0){7.5}{4}
    \Text(25, 10)[lb]{\Large{\Black{$\bar{q}$}}}
    \Text(25, 80)[lb]{\Large{\Black{$q$}}}
    \Text(80, 25)[lb]{\Large{\Black{$g^*$}}}
    \Text(130, 10)[lb]{\Large{\Black{$g$}}}
    \Text(130, 80)[lb]{\Large{\Black{$g$}}}
 \end{picture}
 \end{center}
 \caption{QCD scattering process $q\bar{q}\rightarrow g g$ through
which quarks and gluons reach thermal equilibrium.} \label{QCD}
\vspace{1.2cm}
\end{figure}

\item[1.b)]{Renormalisable interactions with massive mediators:}
\begin{equation}
\langle\Gamma\rangle>H \text{ \ }\Leftrightarrow\text{ \ }\alpha^2
\frac{T^5}{M^{4}}>g_{*}^{1/2}\frac{T^2}{M_P}\text{ \
}\Rightarrow\text{ \ }\left(\frac{g_{*}^{1/2}M^4}{\alpha^2
M_P}\right)^{1/3}<T<M.
\end{equation}
Examples of interactions with effective dimensionful couplings are
weak interactions below $M_{EW}$. In this case, the theory is well
described by the Fermi Lagrangian. An interaction between
electrons and neutrinos, like the one shown in Figure \ref{WEAK},
gives rise to a cross-section of the form of (\ref{3ef}):
\begin{equation}
\langle\sigma_w\rangle\sim\frac{\alpha_w^2}{M_Z^4}\langle p^2
\rangle \sim\frac{\alpha_w^2}{M_Z^4}T^2,
\end{equation}
where $\alpha_w$ is the weak fine structure constant and $p\sim
T$. Thus, neutrinos are coupled to the thermal bath if and only if
\begin{equation}
T>\left(\frac{g_{*}^{1/2}M_Z^4}{\alpha_w^2 M_P}\right)^{1/3}\sim
1\text{ \ MeV}.
\end{equation}

\begin{figure}
\begin{center}
  \begin{picture}(200,80) (0,0)
     \SetColor{Black}
    \ArrowLine(50,50)(0,0)
    \ArrowLine(0,100)(50,50)
    \Vertex(50,50){2.83}
    \DashLine(50,50)(120,50){7}
    \Vertex(120,50){2.83}
    \ArrowLine(170,100)(120,50)
    \ArrowLine(120,50)(170,0)
    \Text(25, 10)[lb]{\Large{\Black{$e^+$}}}
    \Text(25, 80)[lb]{\Large{\Black{$e^-$}}}
    \Text(80, 30)[lb]{\Large{\Black{$Z_0$}}}
    \Text(130, 10)[lb]{\Large{\Black{$\nu_e$}}}
    \Text(130, 80)[lb]{\Large{\Black{$\bar{\nu}_e$}}}
 \end{picture}
 \end{center}
\caption{Weak interaction between electrons and neutrinos through
which they reach thermal equilibrium.} \label{WEAK}
\end{figure}
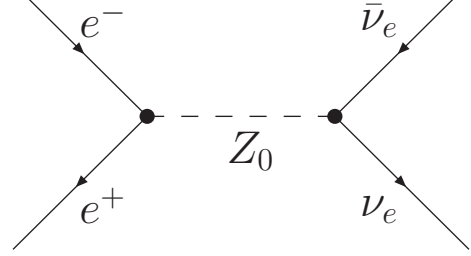

\item[2.]{Gravitational interactions:}
\begin{equation}
\hspace*{-1cm}\text{a) \ } \hspace*{1cm}\langle\Gamma\rangle>H
\text{ \ }\Leftrightarrow\text{ \ }d
\frac{T^5}{M_P^{4}}>g_{*}^{1/2}\frac{T^2}{M_P}\text{ \
}\Rightarrow\text{ \ }T>g_{*}^{1/6}\frac{M_P}{d^{1/3}}.
\label{casea}
\end{equation}
\begin{equation}
\hspace*{-0.6cm}\text{b) \ } \hspace*{1cm}\langle\Gamma\rangle>H
\text{ \ }\Leftrightarrow\text{ \ }\sqrt{d} \frac{g^2
T^3}{M_P^{2}}>g_{*}^{1/2}\frac{T^2}{M_P}\text{ \
}\Rightarrow\text{ \ }T>\frac{g_{*}^{1/2}M_P}{g^2 \sqrt{d}}.
\label{caseb}
\end{equation}
As before, case (a) refers to $2\leftrightarrow 2$ processes with
two gravitational vertices, whereas in case (b) one vertex is
gravitational and the other one is a renormalisable interaction. A
typical K\"{a}hler modulus of string compactifications generically
couples to the gauge bosons of the field theory, that lives on the
stack of branes wrapping the cycle whose volume is given by that
modulus. Scattering processes, annihilation and pair production
reactions, that arise due to that coupling, all have
cross-sections of the form (\ref{gravit}) and
(\ref{renorm-gravit}). For all the K\"{a}hler moduli in KKLT
constructions $d\sim\mathcal{O}(1)$ and so $\langle\Gamma\rangle$
is never greater than $H$ for temperatures below the Planck scale,
for both cases (a) and (b). Therefore, those moduli will never
thermalise through $2\leftrightarrow 2$ processes. However, we
shall see in section \ref{Sec:ModuliTherm} that the situation is
different for the small modulus in LVS, since in that case $d\sim
\mathcal{V}^2\gg 1$. A typical $2\leftrightarrow 2$ process of
type (b), with a modulus $\Phi$ and a non-abelian gauge boson $g$
going to two $g$'s, is shown in Figure \ref{ren-grav}. Here $\Phi$
denotes the canonically normalised field, which at leading order
in the large-volume expansion corresponds to the small modulus. We
will give the precise definition of $\Phi$ in section \ref{Mmc}.
\end{enumerate}

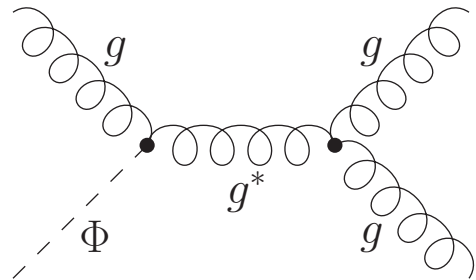
\begin{figure}
\begin{center}
\vspace{1.0cm}
  \begin{picture}(200,80) (0,0)
     \SetColor{Black}
    \DashLine(0,0)(50,50){6}
    \Gluon(0,100)(50,50){7.5}{4}
    \Vertex(50,50){2.83}
    \Gluon(50,50)(120,50){7.5}{4}
    \Vertex(120,50){2.83}
    \Gluon(120,50)(170,100){7.5}{4}
    \Gluon(120,50)(170,0){7.5}{4}
    \Text(25, 10)[lb]{\Large{\Black{$\Phi$}}}
    \Text(34, 80)[lb]{\Large{\Black{$g$}}}
     \Text(80, 25)[lb]{\Large{\Black{$g^*$}}}
     \Text(130, 10)[lb]{\Large{\Black{$g$}}}
      \Text(130, 80)[lb]{\Large{\Black{$g$}}}
   \end{picture}
   \end{center}
\caption{Scattering process $\Phi g\rightarrow g g$ through which
the modulus $\Phi$ and gluons can reach thermal equilibrium.}
\label{ren-grav}
\end{figure}

\subsubsection{$1\leftrightarrow 2$ interactions}
\label{Par:1to2}

In order to work out the temperature dependence of the interaction
rate for decay and inverse decay processes, recall that the rest
frame decay rate $\Gamma_{D}^{(R)}$ does not depend on the
temperature. For renormalisable interactions with massless
mediators or mediated by particles with mass $M$ at temperatures
$T>M$, it takes the form:
\begin{equation}
\Gamma_{D}^{(R)}\sim \alpha m, \label{decadim}
\end{equation}
where $m$ is the mass of the decaying particle and $\alpha\sim
g^2$, with $g$ being either a gauge or a Yukawa coupling. On the
other hand, for gravitational interactions or for renormalisable
interactions mediated by particles with mass $M$ at temperatures
$T<M$, we have ($M\equiv M_P$ in the case of gravity):
\begin{equation}
\Gamma_{D}^{(R)}\sim D\frac{m^3}{M^2}, \label{decadime}
\end{equation}
with $D$ a dimensionless constant (note that in the case of
gravity $D=\sqrt{d}$, where $d$ is the same moduli-dependent
constant as in the previous subsection on $2\leftrightarrow 2$
interactions).

Now, the decay rate that has to be compared with $H$ is not
$\Gamma_{D}^{(R)}$, but its thermal average
$\langle\Gamma_{D}\rangle$. In order to evaluate this quantity, we
need to switch to the `laboratory frame' where:
\begin{equation}
\Gamma_{D}=\Gamma_{D}^{(R)}\sqrt{1-v^2}=\Gamma_{D}^{(R)}\frac{m}{E}
\, ,
\end{equation}
and then take the thermal average:
\begin{equation}
\langle\Gamma_{D}\rangle=\Gamma_{D}^{(R)}\frac{m}{\langle
E\rangle} \, .
\end{equation}
In the relativistic regime, $T\gtrsim m$, the Lorentz factor
$\gamma=\langle E\rangle/m\sim T/m$, whereas in the
non-relativistic regime, $T\lesssim m$, $\gamma=\langle
E\rangle/m\sim 1$.

Notice that, by definition, in a thermal bath the decay rate of
the direct process is equal to the decay rate of the inverse
process. However, for $T<m$ the energy of the final states of the
decay process is of order $T$, which means that the final states
do not have enough energy to re-create the decaying particle. So
the rate for the inverse decay, $\Gamma_{ID}$, is
Boltzmann-suppressed: $\Gamma_{ID}\sim e^{-m/T}$. Hence, the
conclusion is that, for $T<m$, one can never have
$\Gamma_{D}=\Gamma_{ID}$ and thermal equilibrium will not be
attained. Let us now summarise the various decay and inverse decay
rates:

\begin{enumerate}
\item{Renormalisable interactions with massless mediators or mediated by particles with mass $M$ at $T>M$:}
\begin{equation}
\langle \Gamma _{D}\rangle \simeq \left\{
\begin{array}{c}
g^{2}\frac{m^{2}}{T}\text{, \ \ for }T\gtrsim m \\
g^{2}m\text{, \ \ for }T\lesssim m \,\, ,
\end{array}
\right.
\end{equation}
\begin{equation}
\langle \Gamma _{ID}\rangle \simeq \left\{
\begin{array}{c}
g^{2}\frac{m^{2}}{T}\text{, \ \ for }T\gtrsim m \\
g^{2}m\left( \frac{m}{T}\right) ^{3/2}e^{-m/T}\text{, \ \ for
}T\lesssim m \,\, .
\end{array}
\right.
\end{equation}
Therefore, particles will reach thermal equilibrium via decay and
inverse decay processes if and only if:
\begin{equation}
\langle\Gamma\rangle>H \text{ \ }\Leftrightarrow\text{ \ }g^2
\frac{m^2}{T}>g_{*}^{1/2}\frac{T^2}{M_P}\text{ \
}\Rightarrow\text{ \ }m<T<\left(\frac{g^2 m^2
M_P}{g_{*}^{1/2}}\right)^{1/3}.
\end{equation}

\item{Gravity or renormalisable interactions mediated by particles with mass $M$ at $T<M$:}
\begin{equation}
\langle \Gamma _{D}\rangle \simeq \left\{
\begin{array}{c}
D\frac{m^{4}}{M^2 T}\text{, \ \ for }T\gtrsim m \\
D\frac{m^3}{M^2}\text{, \ \ for }T\lesssim m \,\, ,
\end{array}
\right.
\end{equation}
\begin{equation}
\langle \Gamma _{ID}\rangle \simeq \left\{
\begin{array}{c}
D\frac{m^{4}}{M^2 T}\text{, \ \ for }T\gtrsim m \\
D\frac{m^3}{M^2}\left( \frac{m}{T}\right) ^{3/2}e^{-m/T}\text{, \
\ for }T\lesssim m
\end{array}
\right.
\end{equation}
with $M\equiv M_P$ in the case of gravity. Therefore, particles
will reach thermal equilibrium via decay and inverse decay
processes if and only if:
\begin{equation}
\langle\Gamma\rangle>H \text{ \ }\Leftrightarrow\text{ \
}D\frac{m^{4}}{M^2 T}>g_{*}^{1/2}\frac{T^2}{M_P}\text{ \
}\Rightarrow\text{ \ }1<\frac{T}{m}<\left(D\frac{m
M_P}{g_{*}^{1/2}M^2}\right)^{1/3}. \label{izz}
\end{equation}
In the case of gravitational interactions, (\ref{izz}) becomes:
\begin{equation}
1<\frac{T}{m}<\left(D\frac{m}{g_{*}^{1/2}M_P}\right)^{1/3}.
\label{uzz}
\end{equation}
In KKLT constructions, $D\sim\mathcal{O}(1)$ and $m\sim m_{3/2}$.
So (\ref{uzz}) can never be satisfied and hence moduli cannot
reach thermal equilibrium via decay and inverse decay processes.
However, we shall see in section \ref{Sec:ModuliTherm} that in LVS
one has $D\sim\mathcal{V}\gg 1$ and so $1\leftrightarrow 2$
processes could, in principle, play a r\^{o}le in maintaining
thermal equilibrium between moduli and ordinary MSSM particles.
\end{enumerate}

\section{Moduli masses and couplings}
\label{Mmc}

As we have seen in the previous section, the temperature, at which
a thermal bath is established or some particles drop out of
thermal equilibrium, depends on the masses and couplings of the
particles. To determine the latter, one needs to use canonically
normalised fields. In this section, we study the canonical
normalisation of the K\"{a}hler moduli kinetic terms and use the
results to compute the masses of those moduli and their couplings
to visible sector particles.

\subsection{Single-hole Swiss cheese}
\label{Sec:ModCoupl}

We start by focusing on the simplest Calabi-Yau realisation of
LVS, the `single-hole Swiss cheese' case described in chapter 4
(i.e., the degree 18 hypersurface embedded in
$\mathbb{C}P^4_{[1,1,1,6,9]}$). First of all, we shall review the
canonical normalisation derived in \cite{CQ}. In order to obtain
the Lagrangian in the vicinity of the zero temperature vacuum, one
expands the moduli fields around the $T=0$ minimum: \bea
\tau_b &=& \langle\tau_b\rangle +\delta \tau_b\, , \\
\tau_s &=& \langle\tau_s\rangle  +\delta \tau_s \, . \eea where
$\langle\tau_b\rangle$ and $\langle\tau_s\rangle$ denote the VEV
of $\tau_b$ and $\tau_s$. One then finds:
\be {\cal L} = K_{i \bar{j} } \partial_{\mu} (\delta \tau_i)
\partial^{\mu} (\delta \tau_j) -\langle V_0 \rangle - \frac{1}{2}
V_{i\bar{j}} \delta \tau_i \delta \tau_j + {\cal O}(\delta \tau^3)
\, , \label{Lago} \ee
where $i=b,s$ and $\langle V_0 \rangle$ denotes the value of the
zero temperature potential at the minimum. To find the canonically
normalised fields $\Phi$ and $\chi$, let us write $\delta\tau_b$
and $\delta\tau_s$ as:
\be \delta \tau_i=\frac{1}{\sqrt{2}}\left[(\vec{v}_{\Phi})_i\Phi+
(\vec{v}_{\chi})_i\chi\right]. \label{oiu} \ee
Then the conditions for the Lagrangian (\ref{Lago}) to take the
canonical form:
\be {\cal L}= \half \partial_{\mu} {\Phi}
\partial^{\mu} {\Phi}+\half
\partial_{\mu} {\chi} \partial^{\mu} {\chi}- \langle V_0 \rangle -
\half m_{\Phi}^2 \Phi^2 - \half m_{\chi}^2 \chi^2 \ee
are the following:
\begin{equation}
K_{i\bar{j}}(\vec{v}_{\alpha})_{i}(\vec{v}_{\beta})_j=
\delta_{\alpha\beta}\text{ \ \ and \ \
}\frac{1}{2}V_{i\bar{j}}(\vec{v}_{\alpha})_{i}(\vec{v}_{\beta})_j
=m_{\alpha}^2\delta_{\alpha\beta} \, . \label{gg}
\end{equation}
These relations are satisfied when $\vec{v}_{\Phi}$,
$\vec{v}_{\chi}$ (properly normalised according to the first of
(\ref{gg})) and $m_{\Phi}^2$, $m_{\chi}^2$ are, respectively, the
eigenvectors and the eigenvalues of the mass-squared matrix
$\left(M^2\right)_{ij}\equiv\frac{1}{2}\left(K^{-1}\right)_{i\bar{k}}V_{\bar{k}j}$.

Substituting the results of \cite{CQ} for $\vec{v}_{\Phi}$ and
$\vec{v}_{\chi}$ in (\ref{oiu}), we can write the original
K\"{a}hler moduli $\delta\tau_i$ as (for $a_s\tau_s\gg 1$):
\bea \delta\tau_b &=&
\left(\sqrt{6}\langle\tau_b\rangle^{1/4}\langle\tau_s\rangle^{3/4}\right)
\frac{\Phi}{\sqrt{2}}
+\left(\sqrt{\frac{4}{3}}\langle\tau_b\rangle\right)\frac{\chi}{\sqrt{2}}
\sim\mathcal{V}^{1/6}\Phi
+\mathcal{V}^{2/3}\chi \, , \label{big}\\
\delta\tau_s &=&
\left(\frac{2\sqrt{6}}{3}\langle\tau_b\rangle^{3/4}\langle\tau_s\rangle^{1/4}\right)
\frac{\Phi}{\sqrt{2}}
+\left(\frac{\sqrt{3}}{a_s}\right)\frac{\chi}{\sqrt{2}}
\sim\mathcal{V}^{1/2}\Phi+\mathcal{O}\left(1\right)\chi \, .
\label{small} \eea
As expected, these relations show that there is a mixing of the
original fields. Nevertheless, $\delta\tau_b$ is mostly $\chi$ and
$\delta\tau_s$ is mostly $\Phi$. On the other hand, the
mass-squareds are \cite{CQ}:
\bea m^2_{\Phi} &\simeq& Tr\left(M^{2}\right)
\simeq\left(\frac{g_s e^{K_{cs}}}{8\pi}\right)\frac{24\sqrt{2}\nu
a_s^2 \langle\tau_s\rangle^{1/2}}{\mathcal{V}^2}M_P^2
\sim \left(\frac{\ln{\mathcal{V}}}{\mathcal{V}}\right)^2 \!M_P^2 \label{canmass1} \\
m^2_{\chi} &\simeq&
\frac{Det\left(M^2\right)}{Tr\left(M^2\right)}\simeq
\left(\frac{g_s e^{K_{cs}}}{8\pi}\right)\frac{27 \nu}{4 a_s
\langle\tau_s\rangle
\mathcal{V}^3}M_P^2\sim\frac{M_P^2}{\mathcal{V}^3\ln{\mathcal{V}}}.
\label{canmass2} \eea
We can see that there is a large hierarchy of masses among the two
particles, with $\Phi$ being heavier than the gravitino mass
(recall that $m_{3/2} \sim M_P/ {\cal V}$)
and $\chi$ lighter by a factor of $\sqrt{\mc{V}}$.\\

Using the above results and assuming that the MSSM is built via
magnetised $D7$-branes wrapped around the small cycle, we can
compute the couplings of the K\"{a}hler moduli fields of the
$\mathbb{C} P^{4}_{[1,1,1,6,9]}$ model to visible gauge and matter
fields. This is achieved by expanding the kinetic and mass terms
of the MSSM particles around the moduli VEVs. The details are
provided in appendix C, where we focus on $T>M_{EW}$ since we are
interested in thermal corrections at high temperatures. This, in
particular, means that all fermions and gauge bosons are massless
and the mixing of the Higgsinos with the EW gauginos, that gives
neutralinos and charginos, is not present. We summarise the
results for the moduli couplings in Tables 9.1 and 9.2.

\begin{figure}[ht]
\begin{center}
\begin{tabular}{c||c|c|c|c}
  & Gauge bosons & Gauginos
 & Matter fermions
 & Higgsinos  \\
  \hline\hline
  $\chi$ & $\frac{1}{M_P \, \rm{ln} \mc{V}}$ & $\frac{1}{\mc{V} \textrm{ln}\mc{V}}$
  & No coupling & $\frac{1}{\mathcal{V}\ln\mathcal{V}}$
  \\ & & & & \vspace{-0.3cm}\\
  $\Phi$ & $ \frac{\sqrt{\mc{V}} } {M_P} $ & $\frac{1}{\mc{V}^{3/2} \rm{ln}\mc{V}} $
  & No coupling & $\frac{1}{\sqrt{\mathcal{V}}\ln\mathcal{V}}$ \\
\end{tabular} \\\smallskip
{\bf Table {9.1}:} $\mathbb{C} P^{4}_{[1,1,1,6,9]}$ case: moduli
couplings to spin $1$ and $1/2$ MSSM particles for $T>M_{EW}$.
\end{center}
\end{figure}

\begin{figure}[ht]
\begin{center}
\begin{tabular} {c|| c | c | c | c | c }
 & Higgs & Higgs-Fermions & SUSY scalars & $\chi^2$ & $\Phi^2$ \\
  \hline\hline
  $\chi$ & $\frac{M_P}{\mathcal{V}^2(\ln\mathcal{V})^2}$ & $\frac{1}{M_P\mathcal{V}^{1/3}}$
  & $\frac{M_P}{\mathcal{V}^2(\ln\mathcal{V})^2}$ & $\frac{M_P}{\mc{V}^3}$
  & $ \frac{M_P}{\mc{V}^2} $
  \\ & & & & \vspace{-0.3cm}\\
  $\Phi$ & $\frac{M_P}{\mathcal{V}^{5/2}(\ln\mathcal{V})^2}$ & $\frac{1}{M_P\mathcal{V}^{5/6}}$
  & $\frac{M_P}{\mathcal{V}^{5/2}(\ln\mathcal{V})^2}$ & $\frac{M_P}
  { \mc{V}^{5/2} }$  & $ \frac{M_P} {\mc{V}^{3/2} } $ \\
\end{tabular} \\\smallskip
{\bf Table {9.2}:} $\mathbb{C} P^{4}_{[1,1,1,6,9]}$ case: moduli
couplings to spin $0$ and $1/2$ MSSM particles and cubic
self-couplings for $T>M_{EW}$.
\end{center}
\end{figure}

\newpage

\subsection{Multiple-hole Swiss cheese}

Let us now consider the more general Swiss cheese Calabi-Yau
three-folds with more than one small modulus and with volume given
by (\ref{cheese}). In this case we find:
\bea \label{Lagrangiankin} {\cal L}_{kin} &=& \frac{3}{4 \langle
\tau_b \rangle^2} \partial_{\mu} (\delta \tau_b) \partial^{\mu}
(\delta \tau_b) + \frac{3}{8} \sum_i \frac{\lambda_i \epsilon_i}
{\langle \tau_b \rangle \langle \tau_i \rangle} \partial_{\mu} (\delta \tau_i)
\partial^{\mu} (\delta \tau_i) \nonumber \\
&-& \frac{9}{4} \sum_i \frac{\lambda_i \epsilon_i}{\langle \tau_b
\rangle^2} \partial_{\mu} (\delta \tau_b) \partial^{\mu} (\delta
\tau_i) + \frac{9}{4} \sum_{i<j} \frac{\lambda_i \lambda_j
\epsilon_i \epsilon_j}{\langle \tau_b \rangle^2} \partial_{\mu}
(\delta \tau_i) \partial^{\mu} (\delta \tau_j) \,\, , \eea
where $\epsilon_i \equiv \sqrt{\frac{\tau_i}{\tau_b}} \ll 1$ and
also we have kept only the leading contribution in each term (in
the limit $\tau_b\gg\tau_i \, \forall i$). Notice that the mixed
terms are subleading compared to the diagonal ones. So, to start
with, one can keep only the first line in (\ref{Lagrangiankin}).
Then at leading order the canonically normalised fields $\chi$ and
$\Phi_i$, $i=1,...,N_{small}$, are defined via:
\be \delta \tau_b = \sqrt{\frac{2}{3}} \langle \tau_b \rangle \chi
\sim {\cal O} \left( {\cal V}^{2/3} \right) \chi \, , \qquad
\delta \tau_i = \frac{2}{\sqrt{3 \lambda_i}} \langle \tau_b
\rangle^{3/4} \langle \tau_i \rangle^{1/4} \Phi_i \sim {\cal O}
\left( {\cal V}^{1/2} \right) \Phi_i \, . \ee
As was to be expected, this scaling with the volume agrees with
the behaviour of $\delta \tau_b$ and $\delta \tau_s$ in
(\ref{big}), (\ref{small}). Now, let us work out the volume
scaling of the subdominant mixing terms since it is important for
the computation of the various moduli couplings. Proceeding order
by order in a large-$\mathcal{V}$ expansion, we end up with:
\begin{gather}
\delta \tau _{b} \sim \mathcal{O}\left( \mathcal{V}^{2/3}\right)
\chi+\sum_i\mathcal{O}\left( \mathcal{V}
^{1/6}\right) \Phi_i \, ,  \label{SCbig11} \\
\delta \tau _{i} \sim \mathcal{O} \left( \mathcal{V}^{1/2}\right)
\Phi_i+\mathcal{O }\left( 1\right) \chi +\sum_{j\neq i
}\mathcal{O}\left( \mathcal{V}^{-1/2}\right) \Phi_{j} \, .
\label{SCSmall1}
\end{gather}
This shows that the mixing between the small moduli is strongly
suppressed by inverse powers of the overall volume, in accord with
the subleading behaviour of the last term in
(\ref{Lagrangiankin}). Furthermore, the fact that the leading
order volume-scaling of (\ref{SCbig11})-(\ref{SCSmall1}) is the
same as (\ref{big})-(\ref{small}), implies that all small moduli
behave in the same way as the only small modulus of the
$\mathbb{C} P^{4}_{[1,1,1,6,9]}$ model. Hence, if all the small
moduli are stabilised by non-perturbative effects, the moduli mass
spectrum in the general case will look like
(\ref{canmass1})-(\ref{canmass2}), with (\ref{canmass1}) valid for
all the small moduli. In addition, if we assume that all the
4-cycles corresponding to small moduli are wrapped by MSSM
$D7$-branes, the moduli couplings to matter fields are again given
by Tables 9.1 and 9.2, where now $\Phi$ stands for any small
modulus $\Phi_i$.

However, in general the situation may be more complicated. In
fact, the authors of \cite{blumenhagen} pointed out that 4-cycles
supporting MSSM chiral matter cannot always get non-perturbative
effects.\footnote{This is because an ED3 wrapped on the same cycle
will have, in general, chiral intersections with the MSSM branes.
Thus the instanton prefactor would be dependent on the VEVs of
MSSM fields which are set to zero for phenomenological reasons. In
the case of gaugino condensation, this non-perturbative effect
would be killed by the arising of chiral matter.} A possible way
to stabilise these 4-cycles is to use $g_s$ corrections as
proposed in chapter 6. In this case, the leading-order behaviour
of (\ref{canmass1}) should not change: $m_{\Phi_i}^2\sim
\frac{M_P^2}{\mathcal{V}^2}$\,.\footnote{It may be likely that
$m_{\Phi_i}^2$ depends on subleading powers of $(\ln\mathcal{V})$
due to the fact that the loop corrections are subdominant with
respect to the non-perturbative ones, but the main
$\mathcal{V}^{-2}$ dependence should persist.} However, the moduli
couplings to MSSM particles depend on the underlying brane set-up.
So let us consider the following main cases:
\begin{enumerate}
\item All the small 4-cycles are wrapped by MSSM D7 branes except
$\tau_{np}$ which is responsible for non-perturbative effects,
being wrapped by an ED3 brane. It follows that the MSSM couplings
of $\Phi_{np}$ are significantly suppressed compared to the MSSM
couplings of the other small cycles (still given by Tables 9.1 and
9.2). This is due to the mixing term in (\ref{SCSmall1}) being
highly suppressed by inverse powers of $\mathcal{V}$.

\item All the small 4-cycles are wrapped by MSSM D7 branes except
$\tau_{np}$ which is supporting a pure $SU(N)$ hidden sector that
gives rise to gaugino condensation. This implies that the coupling
of $\Phi_{np}$ to hidden sector gauge bosons will have the same
volume-scaling as the coupling of the other small moduli with
visible sector gauge bosons. However, the coupling of the MSSM
4-cycles with hidden sector gauge bosons will be highly
suppressed.

\item All the small 4-cycles $\tau_i$ support MSSM D7 branes which are also wrapped
around the 4-cycle responsible for non-perturbative effects
$\tau_{np}$, but they have chiral intersections only on the other
small cycles. In this case, the coupling of $\Phi_{np}$ to MSSM
particles would be the same as the other $\Phi_i$. However, if
$\tau_{np}$ supports an hidden sector that undergoes gaugino
condensation, the coupling of the MSSM 4-cycles with the gauge
bosons of this hidden sector would still be highly suppressed.
\end{enumerate}

\subsection{K3 fibration}
\label{K3CanNorm}

We turn now to the case of the simplest K3 fibration described in
chapter 8. We shall consider first the `LV' case, in which the
modulus related to the K3 divisor is fixed at a very large value,
and then the `SV' case, in which the overall volume is of the
order $\mathcal{V}\sim 10^{3}$ and the K3 fiber is small.

In order to compute the moduli mass spectroscopy and couplings, it
suffices to canonically normalise the fields just in the vicinity
of the vacuum. The non-canonical kinetic terms look like (with
$\varepsilon\equiv\sqrt{\langle\tau_s\rangle/\langle\tau_1\rangle}$):
\begin{align}
\mathcal{L}_{kin}& = \frac{1}{4\langle\tau_{1}\rangle^{2}}\partial
_{\mu }(\delta\tau _{1})\partial ^{\mu }(\delta\tau
_{1})+\frac{1}{2\langle\tau _{2}\rangle^{2}}\partial _{\mu
}(\delta\tau _{2})\partial ^{\mu }(\delta\tau
_{2})-\frac{3\gamma\varepsilon}{4\langle\tau_2\rangle\langle\tau_1\rangle}
\partial_{\mu }(\delta\tau _{1})\partial ^{\mu
}(\delta\tau_{s})\nonumber\\
& -\frac{3\gamma\varepsilon}{2\langle\tau_{2}\rangle^{2}}
\partial _{\mu }(\delta\tau_2)
\partial ^{\mu }(\delta\tau_s) +\frac{ \gamma
\varepsilon^{3}}{2\langle\tau_2\rangle^2}
\partial _{\mu }(\delta\tau_{1})\partial ^{\mu }
(\delta\tau_{2})+\frac{3\gamma\varepsilon
}{8\langle\tau_2\rangle\langle\tau_{s}\rangle}
\partial_{\mu}(\delta\tau_{s})\partial^{\mu}(\delta\tau_{s}). \label{LKinetic}
\end{align}

\medskip\noindent{\em Large K3 fiber}

\medskip\noindent
In the `LV' case where the K3 fiber is stabilised at large value,
$\varepsilon\ll 1$. Therefore at leading order in a large volume
expansion, where
$\langle\tau_2\rangle>\langle\tau_1\rangle\gg\langle\tau_s\rangle$,
all the cross-terms in (\ref{LKinetic}) are subdominant to the
diagonal ones, and so can be neglected:
\begin{equation}
\mathcal{L}_{kin} \simeq
\frac{1}{4\langle\tau_{1}\rangle^{2}}\partial_{\mu}(\delta\tau
_{1})\partial ^{\mu }(\delta\tau _{1})+\frac{1}{2\langle\tau
_{2}\rangle^{2}}\partial _{\mu }(\delta\tau _{2})\partial ^{\mu
}(\delta\tau _{2})+\frac{ 3\gamma\varepsilon
}{8\langle\tau_2\rangle\langle\tau_{s}\rangle}\partial _{\mu
}(\delta\tau _{s})\partial ^{\mu }(\delta\tau _{s}).
\label{LKinetica}
\end{equation}
Therefore, at leading order the canonical normalisation close to
the minimum becomes rather easy and reads:
\begin{eqnarray}
\delta \tau _{1} &=& \sqrt{2}\langle \tau _{1}\rangle \chi
_{1}\sim
\mathcal{O}\left( \mathcal{V}^{2/3}\right) \chi _{1}, \label{recup} \\
\delta \tau _{2} &=&\langle \tau
_{2}\rangle \chi _{2}\sim \mathcal{O}\left( \mathcal{V}^{2/3}\right) \chi _{2},  \label{K3big} \\
\delta \tau _{s} &=&\sqrt{\frac{4\langle \tau _{1}\rangle
^{1/2}\langle \tau _{2}\rangle \langle \tau
_{s}\rangle^{1/2}}{3\gamma }}\Phi \sim \mathcal{O}\left(
\mathcal{V}^{1/2}\right) \Phi . \label{K3small}
\end{eqnarray}
However, in order to derive all the moduli couplings, we need also
to work out the leading order volume-scaling of the subdominant
mixing terms in (\ref{K3big}) and (\ref{K3small}). This can be
done order by order in a large-$\mathcal{V}$ expansion and, after
some algebra, we obtain:
\begin{gather}
\delta \tau _{1}=\alpha _{1}\langle \tau _{1}\rangle \chi
_{1}+\alpha _{2} \frac{\sqrt{\langle \tau _{1}\rangle }}{\langle
\tau _{2}\rangle }\langle \tau _{s}\rangle ^{3/2}\chi _{2}+\alpha
_{3}\frac{\langle \tau _{1}\rangle ^{3/4}}{\sqrt{\langle \tau
_{2}\rangle }}\langle \tau _{s}\rangle ^{3/4}\Phi,  \label{K3big1} \\
\delta \tau _{2}=\alpha _{4}\frac{\sqrt{\langle \tau _{1}\rangle
}}{\langle \tau _{2}\rangle }\langle \tau _{s}\rangle ^{3/2}\chi
_{1}+\alpha _{5}\langle \tau _{2}\rangle \chi _{2}+\alpha
_{6}\frac{\sqrt{\langle \tau _{2}\rangle }}{ \langle \tau
_{1}\rangle ^{1/4}}\langle \tau _{s}\rangle
^{3/4}\Phi,  \label{K3big2} \\
\delta \tau _{s}=\alpha _{7}\frac{\langle \tau _{1}\rangle
}{\langle \tau _{2}\rangle }\langle \tau _{s}\rangle \chi
_{1}+\alpha _{8}\langle \tau _{s}\rangle \chi _{2}+\alpha
_{9}\langle \tau _{1}\rangle ^{1/4}\sqrt{ \langle \tau _{2}\rangle
}\langle \tau _{s}\rangle ^{1/4}\Phi, \label{K3Small}
\end{gather}
where the $\alpha_i$, $i=1,...,9$ are $\mathcal{O}(1)$
coefficients. The volume-scalings of (\ref{K3big1}),
(\ref{K3big2}) and (\ref{K3Small}) are the following:
\begin{gather}
\delta \tau _{1} \sim \mathcal{O}\left( \mathcal{V}^{2/3}\right)
\chi _{1}+\mathcal{O}\left( \mathcal{V}^{-1/3}\right) \chi
_{2}+\mathcal{O}\left( \mathcal{V}
^{1/6}\right) \Phi ,  \label{K3big11} \\
\delta \tau _{2}\sim \mathcal{O}\left( \mathcal{V}^{-1/3}\right)
\chi _{1}+\mathcal{O}\left( \mathcal{V}^{2/3}\right) \chi_2
+\mathcal{O}\left( \mathcal{V}^{1/6}\right)
\Phi ,  \label{K3big21} \\
\delta \tau _{s} \sim \mathcal{O }\left( 1\right) \chi
_{1}+\mathcal{O}\left( 1\right) \chi _{2}+\mathcal{O} \left(
\mathcal{V}^{1/2}\right) \Phi.  \label{K3Small1}
\end{gather}
This shows that, if we identify each of $\tau_1$ and $\tau_2$ with
the large modulus $\tau_b$ in the Swiss cheese case,
(\ref{K3big11}) and (\ref{K3big21}) have the same volume scaling
as (\ref{big}), as one might have expected. Moreover, the
similarity of (\ref{K3Small1}) and (\ref{small}) shows that also
the small moduli in the two cases behave in the same way.
Therefore, we can conclude that (\ref{canmass1}) is valid also for
the K3 fibration case under consideration:
\begin{equation}
m_{\Phi}\sim\left(\frac{\ln\mathcal{V}}{\mathcal{V}}\right)M_P.
\label{XX}
\end{equation}
On the other hand, we need to be more careful in the study of the
mass spectrum of the large moduli $\tau_1$ and $\tau_2$. We can
work out this `fine structure', at leading order in a
large-$\mathcal{V}$ expansion, first integrating out $\tau_{s}$
and then computing the eigenvalues of the matrix. The latter are
obtained by multiplying the inverse K\"{a}hler metric by the
Hessian of the potential both evaluated at the minimum. The
leading order behaviour of the determinant of this matrix is:
\begin{equation}
Det\left(K^{-1}d^2V\right)\sim
\frac{\tau_2^{4}\sqrt{\ln\mathcal{V}}}{\mathcal{V}^9},\text{ \ \
with \ \ }\mathcal{V}\sim\sqrt{\tau_1}\tau_2.
\end{equation}
Because $m_{\chi_2}^2\gg m_{\chi_1}^2$, we have at leading order
at large volume:
\bea m^2_{\chi_2} &\simeq& Tr\left(K^{-1}d^2V\right)
\sim \frac{\sqrt{\ln\mathcal{V}}}{\mathcal{V}^3}M_P^2 \label{Canmass1} \\
m^2_{\chi_1} &\simeq&
\frac{Det\left(K^{-1}d^2V\right)}{Tr\left(K^{-1}d^2V\right)}
\sim\frac{\tau_2^{4}}{\mathcal{V}^6}M_P^2\sim\frac{M_P^2}{\tau_1^3\tau_2^2}.
\label{Canmass2} \eea
Identifying $\tau_1$ with $\tau_2$, (\ref{Canmass2}) simplifies to
$m^2_{\chi_1} \sim \mathcal{V}^{-10/3}$, confirming the
qualitative expectation that the $\tau_1$ direction is
systematically lighter than $\mc{V}$ in the large-$\mc{V}$ limit.

Using the results of this section and assuming that the MSSM
branes are wrapped around the small cycle\footnote{We also ignore
the incompatibility between localising non-perturbative effects
and the MSSM on the same 4-cycle.}, it is easy to repeat the
computations of appendix C for the K3 fibration. Due to the fact
that the leading order $\mc{V}$-scaling of
(\ref{K3big11})-(\ref{K3Small1}) matches that of the single-hole
Swiss cheese model, we again find the same couplings as those
given in Tables 9.1 and 9.2, where now $\chi$ stands for any of
$\chi_1$ and $\chi_2$.

\medskip\noindent{\em Small K3 fiber}

\medskip\noindent
In the `SV' case where the K3 fiber is stabilised at small value,
$\varepsilon\simeq 1$. Therefore at leading order in a large
volume expansion, where
$\langle\tau_2\rangle\gg\langle\tau_1\rangle>\langle\tau_s\rangle$,
the first term in (\ref{LKinetic}) is dominating the whole kinetic
Lagrangian. Hence we conclude that, at leading order, the
canonical normalisation of $\delta\tau_1$ close to the $T=0$
minimum is again given by (\ref{recup}). However, now its volume
scaling reads:
\begin{equation}
\delta \tau _{1}\sim \mathcal{O}\left(1\right) \chi
_{1}+\left(\text{subleading \ mixing \ terms}\right).
\label{K3Big}
\end{equation}
To proceed order by order in a large volume expansion, note that
the third and the sixth term in (\ref{LKinetic}) are suppressed by
just one power of $\langle\tau_2\rangle$, whereas the second,
fourth and fifth term are suppressed by two powers of the large
modulus. Thus, we obtain the following leading order behaviour for
the canonical normalisation of the two remaining moduli:
\begin{gather}
\delta \tau _{2}\sim \mathcal{O}\left( \mathcal{V}\right) \chi
_{1}+\mathcal{O}\left( \mathcal{V}\right) \chi_2
+\mathcal{O}\left( \mathcal{V}\right)
\Phi ,  \label{K3BIG} \\
\delta \tau _{s} \sim \mathcal{O}\left( \mathcal{V}^{1/2}\right)
\chi _{1}+\mathcal{O}\left( \mathcal{V}^{1/2}\right)
\Phi+\text{subleading \ mixing \ terms}. \label{K3SMALL}
\end{gather}
Notice that the canonically normalised field $\chi_1$ corresponds
to the K3 divisor $\tau_1$, whereas $\Phi$ is a mixing of $\tau_1$
and the blow-up mode $\tau_s$. Finally $\chi_2$ is a combination
of all the three states, and so plays the role of the `large'
field. The moduli mass spectrum will still be given by (\ref{XX}),
(\ref{Canmass1}) and (\ref{Canmass2}). However now the volume
scaling of (\ref{Canmass2}) simplifies to $m^2_{\chi_1} \sim
\mathcal{V}^{-2}$, confirming the qualitative expectation that
$\chi_1$ is also a small field with a mass of the same order of
magnitude of $m_{\Phi}$.

The computation of the moduli couplings depends on the
localisation of the MSSM within the compact Calabi-Yau. Given that
As the scalar potential receives non-perturbative corrections in
the blow-up mode $\tau_s$, in order for the non-perturbative
contributions to be non-vanishing, the MSSM branes have to wrap
either the small K3 fiber $\tau_1$ or the 4-cycle given by the
formal sum $\tau_s+\tau_1$ with chiral intersections on $\tau_1$.
In both cases, we cannot immediately read off the moduli couplings
from the results of appendix C. This is due to the difference of
the leading order volume scaling of the canonical normalisation
between the `SV' case for the K3 fibration and the Swiss cheese
scenario.\footnote{We stress also that presently there is no
knowledge of the K\"{a}hler metric for chiral matter localised on
deformable cycles.}

However, as we shall see in the next section, in the Swiss cheese
case, the relevant interactions through which the small moduli can
thermalise, are with the gauge bosons. As we shall see in section
\ref{Sec:K3FibrModTherm}, these interactions will also be the ones
that are crucial for moduli thermalisation in the K3 fibration
case. Therefore, here we shall focus on them only. Following the
calculations in subsection \ref{Sec:ModCouplOrdPart} of appendix
C, we infer that if only $\tau_1$ is wrapped by MSSM branes, then
the coupling of $\chi_1$ with MSSM gauge bosons is of the order
$g\sim 1/M_P$ without any factor of the overall volume, while the
coupling of $\Phi$ with gauge bosons will be more suppressed by
inverse powers of $\mathcal{V}$. On the other hand, if both
$\tau_1$ and $\tau_s$ are wrapped by MSSM branes, then the
couplings of both small moduli with the gauge bosons are similar
to the ones in the Swiss cheese case:
$g\sim\sqrt{\mathcal{V}}/M_P$. Moreover, if gaugino condensation
is taking place in the pure $SU(N)$ theory supported on $\tau_s$,
then both $\chi_1$ and $\Phi$ couple to the hidden sector gauge
bosons with strength $g\sim\sqrt{\mathcal{V}}/M_P$.

We end this subsection by commenting on K3 fibrations with more
than one blow-up mode. In such a case, it is possible to localise
the MSSM on one of the small blow-up modes and the situation is
very similar to the one outlined for the multiple-hole Swiss
cheese. The only difference is the presence of the extra modulus
related to the K3 fiber, which will couple to the MSSM gauge
bosons with the same strength as the small modulus supporting the
MSSM. This is because of the particular form of the canonical
normalisation, which, for example in the case of two blow-up modes
$\tau_{s1}$ and $\tau_{s2}$, looks like (\ref{K3Big}) and
(\ref{K3BIG}) together with:
\begin{gather}
\delta \tau _{s1} \sim \mathcal{O}\left( \mathcal{V}^{1/2}\right)
\chi _{1}+\mathcal{O}\left( \mathcal{V}^{1/2}\right)
\Phi_1+\text{subleading \ mixing \ terms}, \label{K3SMALL2} \\
\delta \tau _{s2} \sim \mathcal{O}\left( \mathcal{V}^{1/2}\right)
\chi _{1}+\mathcal{O}\left( \mathcal{V}^{1/2}\right)
\Phi_2+\text{subleading \ mixing \ terms}. \label{K3SMALL3}
\end{gather}

\subsection{Modulini}
\label{Sec:Modulini}

In this subsection we shall concentrate on the supersymmetric
partners of the moduli, the modulini. More precisely, we will
consider the fermionic components of the chiral superfields, whose
scalar components are the K\"{a}hler moduli. The kinetic
Lagrangian for these modulini reads:
\begin{equation}
\mathcal{L}_{kin} = \frac{i}{4}\frac{\partial^2 K}{\partial
\tau_i\partial\tau_j} \delta \bar{\tilde{\tau}}_j\gamma^{\mu}
\partial_{\mu} (\delta \tilde{\tau}_i) \, , \label{LagModulini}
\end{equation}
where the K\"{a}hler metric is the same as the one that appears in
the kinetic terms of the K\"{a}hler moduli. Therefore, the
canonical normalisation of the modulini takes exactly the same
form as the canonical normalisation of the corresponding moduli.
For example, in the single-hole Swiss cheese case, we have:
\bea \delta\tilde{\tau}_b &=&
\left(\sqrt{6}\langle\tau_b\rangle^{1/4}\langle\tau_s\rangle^{3/4}\right)
\frac{\tilde{\Phi}}{\sqrt{2}}
+\left(\sqrt{\frac{4}{3}}\langle\tau_b\rangle\right)\frac{\tilde{\chi}}{\sqrt{2}}
\sim\mathcal{V}^{1/6}\tilde{\Phi}
+\mathcal{V}^{2/3}\tilde{\chi} \, , \label{Bbig}\\
\delta\tilde{\tau}_s &=&
\left(\frac{2\sqrt{6}}{3}\langle\tau_b\rangle^{3/4}\langle\tau_s\rangle^{1/4}\right)
\frac{\tilde{\Phi}}{\sqrt{2}}
+\left(\frac{\sqrt{3}}{a_s}\right)\frac{\tilde{\chi}}{\sqrt{2}}
\sim\mathcal{V}^{1/2}\tilde{\Phi}+\mathcal{O}\left(1\right)\tilde{\chi}.
\label{Ssmall} \eea
We focus now on the modulini mass spectrum. We recall that in LVS
the minimum is non-supersymmetric, and so the Goldstino is eaten
by the gravitino via the super-Higgs effect. The Goldstino is the
supersymmetric partner of the scalar field, which is responsible
for supersymmetry breaking. In our case this is the modulus
related to the overall volume of the Calabi-Yau, as can be checked
by studying the order of magnitude of the various $F$-terms.
Therefore, the volume modulino is the Goldstino. More precisely,
in the $\mathbb{C}P^4_{[1,1,1,6,9]}$ case, $\tilde{\chi}$ is eaten
by the gravitino, whereas the mass of $\tilde{\Phi}$ can be
derived as follows:
\be \label{TrMfer} m_{\tilde{\Phi}}^2={\rm Tr} M_f^2 = \langle e^G
K^{i \bar{j}} K^{l \bar{m}} (\nabla_i G_l +\frac{G_i G_l}{3} )
(\nabla_{\bar{j}} G_{\bar{m}} + \frac{G_{\bar{j}} G_{\bar{m}}}{3})
\rangle, \ee
where the function $G = K + \ln |W|^2$ is the supergravity
K\"{a}hler invariant potential, and $\nabla_i G_j = G_{ij} -
\Gamma^{l}_{ij}G_{l}$, with the connection $\Gamma^{l}_{ij} = K^{l
\bar{m}} \partial_i K_{j \bar{m}}$. Equation (\ref{TrMfer}) at
leading order in a large volume expansion, can be approximated as:
\be m_{\tilde{\Phi}}^2 \simeq \langle e^G | (K^{s \bar{s}}
(\nabla_s G_s + \frac{G_s G_s}{3}) |^2\rangle \label{TrMfer11169}
\ee
where $\nabla_s G_s \simeq G_{ss} - \Gamma^{s}_{ss} G_s $ and
$\Gamma^{s}_{ss} \simeq K^{s\bar{s}}
\partial_{s} K_{s\bar{s}}$. In the single-hole Swiss cheese case, for $a_s\tau_s\gg 1$, we obtain:
\begin{equation}
m_{\tilde{\Phi}}^{2}\simeq\langle \frac{g_s
e^{K_{cs}}M_P^2}{\pi}\left(36 a_{s}^{4}A_{s}^{2}\tau
_{s}e^{-2a_{s}\tau
_{s}}-\frac{6\sqrt{2}a_{s}^{2}A_{s}W_{0}}{\mathcal{V}} \sqrt{\tau
_{s}}e^{-a_{s}\tau _{s}}+\frac{
W_{0}^{2}}{2\mathcal{V}^{2}}\right) \rangle. \label{ModulinoMass}
\end{equation}
Evaluating (\ref{ModulinoMass}) at the minimum, we find that the
mass of the modulino $\tilde{\Phi}$ is of the same order of
magnitude as the mass of its supersymmetric partner $\Phi$:
\begin{equation}
m_{\tilde{\Phi}}^{2} \simeq \frac{a_{s}^{2}\langle \tau
_{s}\rangle ^{2}W_{0}^{2}}{ \mathcal{V} ^{2}}M_P^2\sim \left(
\frac{\ln \mathcal{V}}{ \mathcal{V} }\right) ^{2}M_{P}^{2}\sim
m_{\Phi }^{2}. \label{mtilde}
\end{equation}
Similarly, it can be checked that, in the general case of
multiple-hole Swiss cheese Calabi-Yaus and K3 fibrations, the
masses of the modulini also keep being of the same order of
magnitude as the masses of the corresponding supersymmetric
partners.

We now turn to the computation of the modulini couplings. In fact,
we are interested only in the modulino-gaugino-gauge boson
coupling since, as we shall see in section \ref{Sec:ModuliTherm},
this is the relevant interaction through which the modulini reach
thermal equilibrium with the MSSM thermal bath. This coupling can
be worked out by recalling that the small modulus $\tau_s$ couples
to gauge bosons $X$ as (see appendix \ref{Sec:ModCouplOrdPart}):
\begin{equation}
\mathcal{L}_{gauge}\sim\frac{\tau_s}{M_P}F_{\mu\nu}F^{\mu\nu} \, .
\label{tauFF}
\end{equation}
The supersymmetric completion of this interaction term contains
the following modulino-gaugino-gauge boson coupling:
\begin{equation}
\mathcal{L}\sim\frac{\tilde{\tau}_s}{M_P}\sigma^{\mu\nu}\lambda'
F_{\mu\nu} \, . \label{tildetaulambdaF}
\end{equation}
Now, expanding $\tilde{\tau}_s$ around its minimum and going to
the canonically normalised fields $G_{\mu\nu}$ and $\lambda$
defined as (see appendices \ref{Sec:ModCouplOrdPart} and
\ref{Sec:ModCouplSUSY}):
\begin{equation}
G_{\mu\nu}=\sqrt{\langle\tau_s\rangle}F_{\mu\nu} \, ,\text{ \ \
}\lambda=\sqrt{\langle\tau_s\rangle}\lambda' \, , \label{Redef}
\end{equation}
we obtain:
\begin{equation}
\mathcal{L}\sim \frac{\delta\tilde{\tau}_s}
{M_P\langle\tau_s\rangle}\sigma^{\mu\nu}\lambda G_{\mu\nu} \, .
\end{equation}
Hence, by means of (\ref{Ssmall}), we end up with the following
\textit{dimensionful} couplings:
\begin{eqnarray}
\mathcal{L}_{\tilde{\chi}\tilde{X}X}&\sim&\left(\frac{1}
{M_P\ln{\mathcal{V}}}\right)\tilde{\chi} \sigma^{\mu\nu}\lambda G_{\mu\nu} \, , \\
\mathcal{L}_{\tilde{\Phi}\tilde{X}X}&\sim&\left(\frac{\sqrt{\mathcal{V}}}
{M_P}\right)\tilde{\Phi} \sigma^{\mu\nu}\lambda G_{\mu\nu} \, .
\label{ImpModuliniCoupl}
\end{eqnarray}

\section{Study of moduli thermalisation}
\label{Sec:ModuliTherm}

Using the general discussion of section \ref{Sec:ThermEq} and the
explicit expressions for the moduli masses and couplings of
section \ref{Mmc}, we can now study in detail which particles form
the thermal bath. Consequently, we will be able to write down the
general form that the finite temperature corrections of section
\ref{Sec:GenTemp} take in the LVS.

We shall start by focusing on the simple geometry
$\mathbb{C}P^4_{[1,1,1,6,9]}(18)$, and then extend our analysis to
more general Swiss cheese and fibred Calabi-Yau manifolds. We will
show below that, unlike previous expectations in the literature,
the moduli corresponding to small cycles that support chiral
matter can reach thermal equilibrium with the matter fields.

\subsection{Single-hole Swiss cheese}
\label{Sec:11169ModTherm}

As we have seen in section \ref{Sec:ThermEq}, both
$2\leftrightarrow 2$ and $1\leftrightarrow 2$ processes can
establish and maintain thermal equilibrium. Let us now apply the
general conditions of section \ref{Sec:ThermEq} to our case.

As we have already pointed out, scattering and annihilation
processes involving strong interactions will establish thermal
equilibrium between MSSM particles for temperatures $T<\alpha_s^2
g_{*}^{-1/2}M_P\sim 10^{15}$ GeV. Let us now concentrate on the
moduli.

\newpage
\medskip\noindent{\em Small modulus $\Phi$}

\medskip\noindent
From section \ref{Sec:ModCoupl}, we know that the largest coupling
of the small canonical modulus $\Phi$ is with the non-abelian
gauge bosons denoted by $X$:
\begin{equation}
\mathcal{L}_{\Phi XX}=g_{\Phi XX}\Phi F_{\mu\nu}F^{\mu\nu},\text{
\ \ }g_{\Phi XX}\sim\frac{\sqrt{\cal V}}{M_P}\sim\frac{1}{M_s}.
\label{1oi}
\end{equation}
Therefore according to (\ref{casea}), scattering or annihilation
and pair production processes with two gravitational vertices like
$X+X\leftrightarrow\Phi+\Phi$, $X+\Phi\leftrightarrow X+\Phi$, or
$X+X\leftrightarrow X+X$, can establish thermal equilibrium
between $\Phi$ and $X$ for temperatures:
\begin{equation}
T>T_f^{(1)}\equiv g_{*}^{1/6}\frac{M_P}{\mathcal{V}^{2/3}},
\label{caseA}
\end{equation}
where $T_f^{(1)}$ denotes the freeze-out temperature of the
modulus. Taking the number of degrees of freedom $g_*$ to be
${\cal O}(100)$, as in the MSSM, we find that (\ref{caseA})
implies $T>5 \times 10^{8}$ GeV for ${\cal V} \sim 10^{15}$,
whereas $T>10^{16}$ GeV for ${\cal V} \sim 10^4$. In fact, for a
typically large volume (${\cal V}>10^{10}$) a more efficient $2
\leftrightarrow 2$ process is $X+X\leftrightarrow X+\Phi$ with one
gravitational and one renormalisable vertex with coupling constant
$g$. Indeed, according to (\ref{caseb}), such scattering processes
maintain thermal equilibrium for temperatures:
\begin{equation}
T>T_f^{(2)}\equiv\frac{g_{*}^{1/2}M_P}{g^2 \mathcal{V}}\sim
10^3\frac{M_P}{\mathcal{V}}\text{ \ for \ }g_{*}\sim 100\text{ \
and \ }g\sim 0.1 \, , \label{caseB}
\end{equation}
which for ${\cal V} \sim 10^{15}$ gives $T>10^6$ GeV while for
${\cal V} \sim 10^{4}$ it gives $T>10^{17}$ GeV.

Finally, let us investigate the role played by decay and inverse
decay processes of the form $\Phi\leftrightarrow X+X$. We recall
that such processes can, in principle, maintain thermal
equilibrium only for temperatures:
\begin{equation}
T>m_{\Phi}\sim \frac{\ln{\mathcal{V}}}{\mathcal{V}}M_P,
\label{mass}
\end{equation}
because the energy of the gauge bosons is given by $E_{X}\sim T$
and hence for $T<m_{\Phi}$ it is insufficient for the inverse
decay process to occur. However, for $T>m_{\Phi}$ the process $X +
X \to \Phi$ does take place and so one only needs to know the rate
of the decay $\Phi \to X + X$ in order to find out whether thermal
equilibrium is achieved. According to (\ref{uzz}) with $D\sim
g_{\Phi XX}^2/4\pi\sim\mathcal{V}/4\pi$, where we have also used
(\ref{1oi}), the condition for equilibrium is that:
\begin{equation}
T<T_{eq}\equiv\left(\frac{\mathcal{V}m_{\Phi}}{4\pi
g_{*}^{1/2}M_P}\right)^{1/3}m_{\Phi}\,\sim
\,\left(\frac{\ln{\mathcal{V}}}{4\pi
g_{*}^{1/2}}\right)^{1/3}m_{\Phi}\equiv \kappa m_{\Phi} \, .
\label{Uzz}
\end{equation}
Hence thermal equilibrium between $\Phi$ and $X$ can be maintained
by $1\leftrightarrow 2$ processes only if $\kappa>1$.\footnote{The
exact value of $\kappa$ can be worked out via a more detailed
calculation, very similar to the one that we will carry out in
section \ref{Sec:VolBound}. It turns out that this value differs
from the estimate in (\ref{Uzz}) just by a multiplicative factor
$c^{1/3}$ of $\mathcal{O}(1)$. More precisely, $c=18
(\pi\langle\tau_s\rangle)^{-3/2}e^{K_{\textrm{cs}}/2}W_0\sqrt{10g_s}$
and so, for natural values of all the parameters: $W_0=1$,
$g_s=0.1$, $\langle\tau_s\rangle=5$, $K_{\textrm{cs}}=3$, we
obtain $c^{1/3}=1.09$.}  However, estimating the total number of
degrees of freedom as $g_{*}\sim\mathcal{O}(100)$, and writing the
volume as $\mathcal{V}\sim 10^{x}$, we obtain that $\kappa>1$
$\Leftrightarrow$ $x>55$. Such a large value is unacceptable, as
it makes the string scale too small to be compatible with
observations. Therefore, we conclude that in LVS the small modulus
$\Phi$ never thermalises via decay and inverse decay processes.

The final picture is the following:
\begin{itemize}
\item For ${\cal V}$ of order $10^{15}$ ($10^{10}$), as in typical LVS,
from (\ref{caseB}) we deduce that the modulus $\Phi$ is in thermal
equilibrium with MSSM particles for temperatures $T
>T_f^{(2)}\simeq 10^6$ GeV ($T > T_f^{(2)}\simeq10^{11}$ GeV) due to
$X+X\leftrightarrow\Phi+X$ processes.

\item On the other hand,
for ${\cal V}<10^{10}$, as for LVS that allow gauge coupling
unification, the main processes that maintain thermal equilibrium
of the modulus $\Phi$ with MSSM particles are purely
gravitational: $X+X\leftrightarrow\Phi+\Phi$,
$\Phi+X\leftrightarrow\Phi+X$ or $X+X\leftrightarrow X+X$ and the
freeze-out temperature is given by (\ref{caseA}). For example for
${\cal V}\sim 10^{4}$ ($\Leftrightarrow$ $M_{s} \sim 10^{16}$
GeV), $\Phi$ is in thermal equilibrium for temperatures $T
>T_f^{(1)}\simeq 5\times 10^{15}$ GeV.
\end{itemize}
We stress that this is the first example in the literature of a
modulus that reaches thermal equilibrium with ordinary particles
for temperatures significantly less than $M_P$, and so completely
within the validity of the low energy effective theory. Note that
we did not focus on the interactions of $\Phi$ with other ordinary
and supersymmetric particles, since the corresponding couplings,
derived in appendix C, are not large enough to establish thermal
equilibrium.

Finally, let us also note that, once the modulus $\Phi$ drops out
of thermal equilibrium, it will decay before its energy density
can begin to dominate the energy density of the Universe, unlike
traditional expectations in the literature. We will show this in
more detail in section \ref{Sec:MaxTemp}.

\medskip\noindent{\em Large modulus $\chi$}

\medskip\noindent
As summarised in section \ref{Sec:ModCoupl}, the coupling of the
large modulus $\chi$ with gauge bosons is given by
\begin{equation}
\mathcal{L}_{\chi XX}=g_{\chi XX}\chi F_{\mu\nu}F^{\mu\nu},\text{
\ \ }g_{\chi XX }\sim\frac{1}{M_P\ln\mathcal{V}}. \label{10}
\end{equation}
Consequently purely gravitational $2\leftrightarrow 2$ processes
like $X+X\leftrightarrow\chi+\chi$, $X+\chi\leftrightarrow
X+\chi$, or $X+X \leftrightarrow X+X$, could establish thermal
equilibrium between $\chi$ and $X$ for temperatures:
\begin{equation}
T>T_{f}^{(1)}\equiv
g_{*}^{1/6}M_P\left(\ln\mathcal{V}\right)^{4/3}. \label{CaseA}
\end{equation}
On the other hand, scattering processes like $X+X\leftrightarrow
X+\chi$ with one gravitational and one renormalisable vertex with
coupling constant $g$, could maintain thermal equilibrium for
temperatures:
\begin{equation}
T>T_{f}^{(2)}\equiv\frac{g_{*}^{1/2}M_P}{g^2}\left(\ln\mathcal{V}\right)^2\sim
10^3M_P\left(\ln\mathcal{V}\right)^2,\text{ \ for \ }g_{*}\sim
100\text{ \ and \ }g\sim 0.1 \, . \label{CaseB}
\end{equation}
Clearly, both $T_{f}^{(1)}$ and $T_{f}^{(2)}$ are greater than
$M_P$ and so we conclude that $\chi$ can never thermalise via
$2\leftrightarrow 2$ processes. It is also immediate to notice
that thermal equilibrium cannot be maintained by $1\leftrightarrow
2$ processes, like $\chi\leftrightarrow X+X$, either. The reason
is that, as derived in \cite{CQ}, for typical LARGE values of the
volume $\mathcal{V}\sim 10^{10} - 10^{15}$, the lifetime of the
large modulus $\chi$ is greater than the age of the Universe.
Hence this modulus could contribute to dark matter and its decay
to photons or electrons could be one of the smoking-gun signal of
LVS.

Furthermore, as can be seen from section \ref{Sec:ModCoupl}, the
couplings of $\chi$ to other MSSM particles are even weaker than
its coupling to gauge bosons. So $\chi$ cannot thermalise via any
other kind of interaction. Finally, one can also verify that
thermal equilibrium between $\chi$ and $\Phi$ can never be
maintained via $1\leftrightarrow 2$ and $2\leftrightarrow 2$
processes involving only the moduli, which processes arise due to
the moduli triple self-couplings computed in appendix \ref{SELF}.
Therefore, $\chi$ behaves as a typical modulus studied in the
literature.

\subsection{Multiple-hole Swiss cheese}
\label{Sec:MultHoleModTherm}

We shall now extend the results of section \ref{Sec:11169ModTherm}
to the more general case of Calabi-Yau three-folds with one large
cycle and several small ones. We shall not focus on explicit
models since this is beyond the scope of this chapter, but we will
try to discuss qualitatively the generic behaviour of small moduli
in the case of `multiple-hole Swiss cheese' Calabi-Yau manifolds.

As we have seen in section \ref{Sec:ModCoupl}, the couplings with
MSSM particles of all the small cycles wrapped by MSSM branes have
the same volume scaling as the corresponding couplings of the
single small modulus in the $\mathbb{C}P^4_{[1,1,1,6,9]}$ case.
Moreover, in section \ref{Sec:11169ModTherm} we have learned that
$\Phi$ can thermalise via its interaction with gauge bosons.
Hence, we conclude that the same arguments as in section
\ref{Sec:11169ModTherm} can be applied for $h_{1,1}>2$ and so all
small cycles, that support MSSM chiral matter, reach thermal
equilibrium with the gauge bosons.

Note however that, as we already pointed out in section
\ref{Sec:ModCoupl}, the situation may be more complicated in
concrete phenomenological models due to the possibility that
non-perturbative effects may be incompatible with MSSM branes,
which are localised on the same 4-cycle \cite{blumenhagen}.
Whether or not such an incompatibility arises depends on the
particular features of the model one considers, including the
presence or absence of charged matter fields with non-vanishing
VEVs. As a consequence of these subtleties, the issue of moduli
thermalisation is highly dependent on the possible underlying
brane set-ups. To gain familiarity with the outcome, let us
explore in more detail several brane set-ups in the case of only
two small moduli. At the end we will comment on the generalisation
of these results to the case of arbitrary $h_{1,1}$.

We will focus on the case $h_{1,1}=3$ with two small moduli
$\tau_1$ and $\tau_2$, that give the volumes of the two rigid
divisors $\Gamma_1$ and $\Gamma_2$. The results of subsection
\ref{Sec:11169ModTherm} imply the following for the different
brane set-ups below:

\begin{enumerate}
\item If $\Gamma_1$ is wrapped by an $ED3$ instanton and $\Gamma_2$ is wrapped by MSSM branes:

\begin{itemize}
\item $\tau_1$ couples to MSSM gauge bosons with strength
$g\sim1/(\sqrt{\mathcal{V}}M_P)$ $\Rightarrow$ $\tau_1$ does not
thermalise.\footnote{The coupling $g\sim1/(\sqrt{\mathcal{V}}M_P)$
can be worked out by substituting the expression (\ref{SCSmall1})
in (\ref{FmunuFmunu}). As pointed out in point 1 at the end of
section 9.3.2, the weakness of this coupling is due to the mixing
term in (\ref{SCSmall1}) being highly suppressed by inverse powers
of $\mathcal{V}$.}
\item $\tau_2$ couples to MSSM gauge bosons with strength
$g\sim\sqrt{\mathcal{V}}/M_P$ $\Rightarrow$ $\tau_2$ thermalises.
\end{itemize}

\item If $\Gamma_1$ is wrapped by an $ED3$ instanton and $\Gamma_1+\Gamma_2$ is wrapped by MSSM
branes with chiral intersections on $\Gamma_2$:\footnote{We assume
that a single $D7$-brane is wrapping $\Gamma_2$ in order to get
chirality from the intersection with the MSSM branes. The same
assumption applies throughout this chapter everywhere we use the
expression `chiral intersections on some divisor'.}

\begin{itemize}
\item $\tau_1$ couples to MSSM gauge bosons with strength
$g\sim\sqrt{\mathcal{V}}/M_P$ $\Rightarrow$ $\tau_1$ thermalises.
\item $\tau_2$ couples to MSSM gauge bosons with strength
$g\sim\sqrt{\mathcal{V}}/M_P$ $\Rightarrow$ $\tau_2$ thermalises.
\end{itemize}

\item If $\Gamma_1$ is supporting a pure $SU(N)$ theory, that undergoes gaugino condensation,
and $\Gamma_2$ is wrapped by MSSM branes:

\begin{itemize}
\item $\tau_1$ couples to MSSM gauge bosons with strength
$g\sim 1/(\sqrt{\mathcal{V}}M_P)$ and to hidden sector gauge
bosons with strength $g\sim\sqrt{\mathcal{V}}/M_P$ $\Rightarrow$
$\tau_1$ thermalises via its interaction with hidden sector gauge
bosons.
\item $\tau_2$ couples to MSSM gauge bosons with strength
$g\sim\sqrt{\mathcal{V}}/M_P$ and to hidden sector gauge bosons
with strength $g\sim1/(\sqrt{\mathcal{V}}M_P)$ $\Rightarrow$
$\tau_2$ thermalises via its interaction with MSSM gauge bosons.
\end{itemize}
Hence in this case there are two separate thermal baths: one
contains $\tau_1$ and the hidden sector gauge bosons at
temperature $T_1$, whereas the other one is formed by $\tau_2$ and
the MSSM particles at temperature $T_2$. Generically, we would
expect that $T_1\neq T_2$ since the two thermal baths are not in
contact with each other.

\item If $\Gamma_1$ is supporting a pure $SU(N)$ theory, that undergoes gaugino condensation,
and $\Gamma_1+\Gamma_2$ is wrapped by MSSM branes with chiral
intersections on $\Gamma_2$:

\begin{itemize}
\item $\tau_1$ couples both to MSSM and hidden sector gauge bosons
with strength $g\sim\sqrt{\mathcal{V}}/M_P$ $\Rightarrow$ $\tau_1$
thermalises.
\item $\tau_2$ couples to MSSM gauge bosons with strength
$g\sim\sqrt{\mathcal{V}}/M_P$ and to hidden sector gauge bosons
with strength $g\sim 1/(\sqrt{\mathcal{V}}M_P)$ $\Rightarrow$
$\tau_2$ thermalises via its interaction with MSSM gauge bosons.
\end{itemize}
Unlike the previous case, now there is only one thermal bath,
which contains both $\tau_1$ and $\tau_2$ together with the MSSM
particles and the hidden sector gauge bosons, since in the present
case $\tau_1$ interacts strongly enough with the MSSM gauge
bosons.
\end{enumerate}
We can now extend these results to the general case with
$h_{1,1}>3$ by noticing that a small 4-cycle wrapped by MSSM
branes will always thermalise via its interaction with MSSM gauge
bosons. On the other hand, for a 4-cycle that is not wrapped by
MSSM branes there are the following two options. If it is wrapped
by an $ED3$ instanton, it will not thermalise. If instead it is
supporting gaugino condensation, it will reach thermal equilibrium
with the hidden sector gauge bosons.

\subsection{K3 fibration}
\label{Sec:K3FibrModTherm}

Let us now turn to the issue of moduli thermalisation for K3
fibrations. As we have seen in section \ref{K3CanNorm}, there is
an essential difference between the cases when the K3 fiber is
stabilised at a large and at a small value. Let us consider
separately each of these two situations.

\medskip\noindent{\em Large K3 fiber}

\medskip\noindent

As we have already stressed in section \ref{K3CanNorm}, in the
case `LV' where the K3 divisor is stabilised large, the small
modulus $\Phi$ plays exactly the same r\^{o}le as the small
modulus of the single-hole Swiss cheese case, whereas both
$\chi_1$ and $\chi_2$ behave as the single large modulus. Hence we
can repeat the same analysis as in section \ref{Sec:11169ModTherm}
and conclude that only $\Phi$ will reach thermal equilibrium with
the MSSM particles via its interaction with the gauge bosons.

\medskip\noindent{\em Small K3 fiber}

\medskip\noindent

The study of moduli thermalisation in the case of small K3 fiber
is more complicated. We shall first focus on Calabi-Yau
three-folds with just one blow-up mode and later on will infer the
general features of the situation with several blow-ups.

K3 fibrations with $h_{1,1}=3$ are characterised by two small
moduli: $\tau_1$ that gives the volume of the K3 divisor
$\Gamma_1$, and $\tau_s$ which is the volume of the rigid divisor
$\Gamma_s$. The canonically normalised fields $\chi_1$ and $\Phi$
are defined by (\ref{K3Big}) and (\ref{K3SMALL}). We recall that
one has to be careful about the possible incompatibility of MSSM
branes on $\Gamma_s$ with the non-perturbative effects that this
cycle supports. Hence, to avoid dealing with such subtleties,
below we will assume that the MSSM branes are not wrapping
$\Gamma_s$. Again, using the results of section
\ref{Sec:11169ModTherm}, we infer the following for the different
brane set-ups below:

\begin{enumerate}
\item If $\Gamma_s$ is wrapped by an $ED3$ instanton and $\Gamma_1$ is wrapped by MSSM branes:

\begin{itemize}
\item $\chi_1$ couples to MSSM gauge bosons with strength
$g\sim 1/M_P$ $\Rightarrow$ $\chi_1$ does not thermalise.
\item $\Phi$ couples to MSSM gauge bosons more weakly than $\chi_1$
$\Rightarrow$ $\Phi$ does not thermalise.
\end{itemize}

\item If $\Gamma_s$ is wrapped by an $ED3$ instanton and $\Gamma_s+\Gamma_1$ is wrapped by MSSM
branes with chiral intersections on $\Gamma_1$:

\begin{itemize}
\item $\chi_1$ couples to MSSM gauge bosons with strength
$g\sim\sqrt{\mathcal{V}}/M_P$ $\Rightarrow$ $\chi_1$ thermalises.
\item $\Phi$ couples to MSSM gauge bosons with strength
$g\sim\sqrt{\mathcal{V}}/M_P$ $\Rightarrow$ $\Phi$ thermalises.
\end{itemize}

\item If $\Gamma_s$ is supporting a pure $SU(N)$ theory, that undergoes gaugino condensation,
and $\Gamma_1$ is wrapped by MSSM branes:

\begin{itemize}
\item $\chi_1$ couples to MSSM gauge bosons with strength
$g\sim 1/M_P$ and to hidden sector gauge boson with strength
$g\sim\sqrt{\mathcal{V}}/M_P$ $\Rightarrow$ $\chi_1$ thermalises
via its interaction with hidden sector gauge bosons.
\item $\Phi$ couples to MSSM gauge bosons more weakly than $\chi_1$
and to hidden sector gauge bosons with strength
$g\sim\sqrt{\mathcal{V}}/M_P$ $\Rightarrow$ $\Phi$ thermalises via
its interaction with hidden sector gauge bosons.
\end{itemize}
In this case, two separate thermal baths are established: one
contains $\chi_1$, $\Phi$ and the hidden sector gauge bosons at
temperature $T_1$, whereas the other one is formed by the MSSM
particles at temperature $T_2$. Generically, we expect that
$T_1\neq T_2$ since the two thermal baths are not in contact with
each other.

\item If $\Gamma_s$ is supporting a pure $SU(N)$ theory, that undergoes gaugino condensation,
and $\Gamma_s+\Gamma_1$ is wrapped by MSSM branes with chiral
intersections on $\Gamma_1$:

\begin{itemize}
\item $\chi_1$ couples both to MSSM and hidden sector gauge bosons
with strength $g\sim\sqrt{\mathcal{V}}/M_P$ $\Rightarrow$ $\chi_1$
thermalises.
\item $\Phi$ couples both to MSSM and hidden sector gauge bosons
with strength $g\sim\sqrt{\mathcal{V}}/M_P$ $\Rightarrow$ $\Phi$
thermalises.
\end{itemize}
Now only one thermal bath is established containing $\chi_1$,
$\Phi$, the hidden sector gauge bosons and the MSSM particles,
since both moduli interact with equal strength with the gauge
bosons of the MSSM and of the hidden sector.
\end{enumerate}
It is interesting to notice that both moduli $\chi_1$ and $\Phi$
thermalise in all situations, except when the blow-up mode is
wrapped by an $ED3$ instanton only. In this particular case, no
modulus thermalises. It is trivial to generalise these conclusions
for more than one blow-up mode and the MSSM still localised on the
K3 fiber.

On the other hand, if the MSSM is localised on one of the rigid
divisors, then for the case of more than one blow-up mode one can
repeat the same general conclusions as at the end of subsection
\ref{Sec:MultHoleModTherm}, with in addition the fact that
$\chi_1$ will always thermalise as soon as one of the blow-up
modes thermalises. This is due to the leading order mixing between
$\Phi$ and any other small modulus, as can be seen explicitly in
(\ref{K3SMALL2}) and (\ref{K3SMALL3}).

\subsection{Modulini thermalisation}
\label{Sec:ModuliniTherm}

The study of modulini thermalisation is straightforward since, as
we have seen in section \ref{Sec:Modulini}, the canonical
normalisation for the modulini takes exactly the same form as the
canonical normalisation for the moduli. This implies that, after
supersymmetrisation, the small modulino-gaugino-gauge boson
coupling has the same strength as the small modulus-gauge
boson-gauge boson coupling. Given that this is the relevant
interaction for moduli thermalisation, we can repeat the same
considerations as those in sections \ref{Sec:11169ModTherm} and
\ref{Sec:K3FibrModTherm}, and conclude that the modulini
thermalise every time, when their supersymmetric partners reach
thermal equilibrium with the MSSM thermal bath. Note however that,
if for the moduli the relevant processes are $2\leftrightarrow 2$
interactions with gauge bosons, the crucial $2\leftrightarrow 2$
processes for the modulini are:
\begin{itemize}
\item $2\leftrightarrow 2$ processes with two gravitational vertices dominant for $\mathcal{V}<10^{10}$:
$\tilde{X}+\tilde{X}\leftrightarrow \tilde{\Phi}+\tilde{\Phi}$,
$X+X\leftrightarrow \tilde{\Phi}+\tilde{\Phi}$,
$\tilde{X}+\tilde{\Phi}\leftrightarrow \tilde{X}+\tilde{\Phi}$,
$X+\tilde{\Phi}\leftrightarrow X+\tilde{\Phi}$,
$\tilde{X}+\tilde{X}\leftrightarrow X+X$,
$\tilde{X}+X\leftrightarrow \tilde{X}+X$.
\item $2\leftrightarrow 2$ processes with one gravitational and one renormalisable
vertex dominant for $\mathcal{V}>10^{10}$:
$X+\tilde{\Phi}\leftrightarrow \tilde{X}+\tilde{X}$,
$\tilde{X}+\tilde{\Phi}\leftrightarrow X+X$.
\end{itemize}

\section{Finite temperature corrections in LVS}
\label{FTCLVS}

In this section we study the finite temperature effective
potential in LVS. We show that it has runaway behaviour at high
$T$ and compute the decompactification temperature $T_{max}$. We
also investigate the cosmological implications of the small
modulus decay. By imposing that the temperature just after its
decay (regardless of whether or not that decay leads to reheating)
be less than $T_{max}$, in order to avoid decompactification of
the internal space, we find important restrictions on the range of
values of the Calabi-Yau volume.

\subsection{Effective potential}
\label{Sec:ExplicitTPot}

We shall now derive the explicit form of the finite temperature
effective potential for LVS, following the analysis of moduli
thermalisation performed in section \ref{Sec:ModuliTherm}. We will
study in detail the behaviour of thermal corrections to the $T=0$
potential of the simple $\mathbb{C}P^4_{[1,1,1,6,9]}$ model, and
then realise that the single-hole Swiss cheese case already
incorporates all the key properties of the general LVS.

\medskip\noindent{\em Single-hole Swiss cheese}

\medskip\noindent

As we have seen in section \ref{Sec:11169ModTherm}, not only
ordinary MSSM particles thermalise via Yang-Mills interactions but
also the small modulus and modulino reach thermal equilibrium with
matter via their interactions with the gauge bosons. Therefore,
the general expression (\ref{1-loop}) for the 1-loop finite
temperature effective potential, takes the following form:
\begin{equation}
V_T^{1-loop}=-\frac{\pi^2 T^4}{90} \left( g_B+ \frac{7}{8} g_F
\right)+\frac{T^2}{24}\left(m_{\Phi}^2+m_{\tilde{\Phi}}^2+\sum_i
M_{soft,i}^2\right)+...\,\,. \label{ONE-loop}
\end{equation}
We recall that (\ref{ONE-loop}) is a high temperature expansion of
the general 1-loop integral (\ref{klo}), and so it is valid only
for $T\gg m_{\Phi},m_{\tilde{\Phi}},M_{soft,i}$. The general
moduli-dependent expression for the modulino mass-squared
$m_{\tilde{\Phi}}^2$ is given by (\ref{ModulinoMass}) without the
vacuum expectation value. On the other hand, in the limit
$\tau_b\gg\tau_s$, $m_{\Phi}^2$ can be estimated as follows:
\begin{equation}
m_{\Phi}^2\simeq{\rm Tr} M^2_b = \frac{ K^{ij}}{2}
\frac{\partial^2 V_0} {\partial \tau_i\partial\tau_j}\simeq
\frac{K^{ss}}{2}\frac{\partial^2 V_0} {\partial \tau_s^2}.
\label{TraceMb}
\end{equation}
For $a_s\tau_s\gg 1$, the previous expression (\ref{TraceMb}), at
leading order, becomes:
\begin{equation}
m_{\Phi}^{2}\simeq \frac{A_s a_s^3 g_s
e^{K_{cs}}M_P^2}{\pi}\left(72 A_s a_s \tau_s
e^{-2a_{s}\tau_{s}}-\frac{3
W_0\tau_s^{3/2}e^{-a_s\tau_s}}{\sqrt{2}\mathcal{V}} \right).
\label{GenModulusMass}
\end{equation}
It can be shown that the gaugino and scalar masses arising from
gravity mediated supersymmetry breaking\footnote{The contribution
from anomaly mediation is subleading with respect to gravity
mediation as shown in \cite{SoftSUSY}.} are always parametrically
smaller than $m_{\Phi}$ and $m_{\tilde{\Phi}}$, and so we shall
neglect them. Moreover we shall drop also the $\mathcal{O}(T^4)$
term in (\ref{ONE-loop}) since it has no moduli dependence.
Therefore, the relevant 1-loop finite-temperature effective
potential reads:
\begin{equation}
V_T^{1-loop}=\frac{T^2}{24}\left(m_{\Phi}^2+m_{\tilde{\Phi}}^2\right)+...\,\,,
\label{1LOOP}
\end{equation}
which using (\ref{ModulinoMass}) and (\ref{GenModulusMass}), takes
the form:
\begin{equation}
V_{T}^{1-loop}=\frac{T^{2}}{24}\left( \frac{g_s
e^{K_{cs}}M_{P}^{2}}{ \pi }\right) \left[ \lambda _{1}\tau
_{s}e^{-2a_{s}\tau _{s}}-\lambda _{2}\left(4+a_s\tau_s\right)
\frac{\sqrt{\tau _{s}} e^{-a_{s}\tau
_{s}}}{\mathcal{V}}+\frac{W_{0}^{2}}{2\mathcal{V}^{2}}\right]
+...\,\,, \label{Full1-loop}
\end{equation}
with:
\begin{equation}
\lambda _{1}\equiv 108 A_{s}^{2}a_{s}^{4},\text{ \ \ \ \ \ \
}\lambda _{2}\equiv 3 a_{s}^{2}A_{s}W_{0}/\sqrt{2}.
\end{equation}
Given that the leading contribution in (\ref{ONE-loop}), namely
the $\mathcal{O}(T^4)$ term, does not bring in any moduli
dependence, we need to go beyond the ideal gas approximation and
consider the effect of 2-loop thermal corrections, as the latter
could in principle compete with the terms in (\ref{Full1-loop}).
The high temperature expansion of the 2-loop contribution looks
like:
\begin{equation}
V_T^{2-loops}=T^4\left(\kappa_1 g_{MSSM}^2+\kappa_2 g_{\Phi XX}^2
m_{\Phi}^2+\kappa_3 g_{\tilde{\Phi}\tilde{X}X}^2
m_{\tilde{\Phi}}^2+...\right)+...\,\,, \label{TWO-loop}
\end{equation}
where the $\kappa$'s are $\mathcal{O}(1)$ coefficients and:
\begin{itemize}
\item the $\mathcal{O}(g_{MSSM}^2)$ contribution comes from two
loops involving MSSM particles;
\item the $\mathcal{O}(g_{\Phi XX}^2)$ contribution is due to two loop
diagrams with $\Phi$ and two gauge bosons;
\item the $\mathcal{O}(g_{\tilde{\Phi}\tilde{X}X}^2)$ contribution
comes from two loops involving the modulino $\tilde{\Phi}$, the
gaugino $\tilde{X}$ and the gauge boson $X$;
\item all the other two loop diagrams give
rise to subdominant contributions, and so they have been
neglected. Such diagrams are the ones with $\Phi$ or
$\tilde{\Phi}$ plus other MSSM particles, the self-interactions of
the moduli and of the modulini, and two loops involving both
$\Phi$ and $\tilde{\Phi}$. For example, the subleading
contribution originating from the two-loop vacuum diagram due to
the $\Phi^3$ self-interaction takes the form: $\delta
V_T^{2-loops}=\kappa_4 T^4 \frac{g_{\Phi^3}^2}{m_{\Phi}^2}\sim T^4
\frac{const}{{\cal V}(\ln {\cal V})^2}$.
\end{itemize}
Note that in (\ref{TWO-loop}) we have neglected the
$\mathcal{O}(T^2)$ term since it is subleading compared to both
the ${\cal O}(T^4)$ 2-loop term and the ${\cal O}(T^2)$ 1-loop
one. Now, the relevant gauge couplings in (\ref{TWO-loop}), have
the following moduli dependence:
\begin{itemize}
\item $g_{MSSM}^2=4\pi/\tau_s$ since we assume that the
MSSM is built via magnetised $D7$-branes wrapping the small cycle.
In the case of a supersymmetric $SU(N_c)$ gauge theory with $N_f$
matter multiplets, the coefficient $\kappa_1$ reads \cite{FTBook}:
\begin{equation}
\kappa_1=\frac{1}{64}\left(N_c^2-1\right)(N_c+3N_f)>0.
\label{kappa1}
\end{equation}
\item $g_{\Phi XX}^2\sim g_{\tilde{\Phi}\tilde{X}X}^2
\sim \frac{\sqrt{\mathcal{V}}}{M_P}$ as derived in
(\ref{ImpModuliniCoupl}) and (\ref{ImpModuliCoupl}).
\end{itemize}
Adding (\ref{1LOOP}) and (\ref{TWO-loop}) to the $T=0$ potential
$V_0$, we obtain the full finite temperature effective potential:
\begin{equation}
V_{TOT}=V_0+T^4\left(\kappa_1 g_{MSSM}^2+\kappa_2 g_{\Phi XX}^2
m_{\Phi}^2+\kappa_3 g_{\tilde{\Phi}\tilde{X}X}^2
m_{\tilde{\Phi}}^2\right)+\frac{T^2}{24}\left(m_{\Phi}^2
+m_{\tilde{\Phi}}^2\right)+...\,\,. \label{FULL}
\end{equation}
Despite the thermalisation of $\Phi$ and $\tilde{\Phi}$, which in
principle leads to a modification of $V_{TOT}$ compared to
previous expectations in the literature, we shall now show that
the thermal corrections due to $\Phi$ and $\tilde{\Phi}$ are, in
fact, negligible compared to the other contributions in
(\ref{FULL}), everywhere in the moduli space of these models. In
particular, the 2-loop MSSM effects dominate the
temperature-dependent term.\footnote{Note that this is consistent
with the results of \cite{AC} in the context of the O'KKLT model,
where it was also found that the T-dependent contribution of
moduli, that were assumed to be in thermal equilibrium, is
negligible compared to the dominant contribution of the rest of
the effective potential.}

Let us start by arguing that the $\mathcal{O}(T^4)$ corrections
arising from the modulus $\Phi$ and the modulino $\tilde{\Phi}$
are subleading compared to the 1-loop $\mathcal{O}(T^2)$ term.
Indeed, the relevant part of the effective scalar potential
(\ref{FULL}) may be rewritten as:
\begin{equation}
T^{4}\left( \kappa _{2}g_{\Phi XX}^{2}m_{\Phi }^{2}+\kappa
_{3}g_{\tilde{\Phi}\tilde{X}X}^{2}m_{\tilde{\Phi}}^{2}\right) \sim
T^{2}\left( m_{\Phi }^{2}+m_{\tilde{\Phi}}^{2}\right) \underset
{\left( \frac{T}{M_{s}}\right) ^{2}\ll
1}{\underbrace{T^{2}\frac{\mathcal{V} }{M_{P}^{2}}}},
\end{equation}
where the $<\!\!<$ inequality is due to the fact that our
effective field theory treatment makes sense only at energies
lower than the string scale $M_s$. Therefore, we can neglect the
effect of 2-loop thermal corrections involving $\Phi$ and
$\tilde{\Phi}$. So we see that, although the interactions of
$\Phi$ and $\tilde{\Phi}$ with gauge bosons and gauginos are
strong enough to make them thermalise, they are not sufficient to
produce thermal corrections large enough to affect the form of the
total effective potential. Let us also stress that this result is
valid everywhere in moduli space, i.e. for each value of
$m_{\Phi}^2$ and $m_{\tilde{\Phi}}^2$, not just in the region
around the zero-temperature minimum.

We now turn to the study of the general behaviour of the 1-loop
$\mathcal{O}(T^2)$ term arising from $\Phi$ and $\tilde{\Phi}$. We
shall show that it is always subdominant compared to the
zero-temperature potential, and so it can be safely neglected. In
fact, the sum of the $T=0$ potential and the 1-loop thermal
correction (\ref{Full1-loop}) can be written as (ignoring the
subleading loop corrections in $V_0$):
\begin{equation}
V_0+\frac{T^2}{24}\left(m_{\Phi}^2
+m_{\tilde{\Phi}}^2\right)=\frac{g_s e^{K_{cs}}M_{P}^{4}}{8\pi
}\left[ p_{1}A_{1}\sqrt{\tau _{s}} \frac{e^{-2a_{s}\tau
_{s}}}{\mathcal{V}}-p_{2}A_{2}\frac{\tau _{s}e^{-a_{s}\tau
_{s}}}{\mathcal{V}^{2}}+p_{3}A_{3}\frac{1}{\mathcal{V}^{3}}
\right] ,  \label{VSimple}
\end{equation}
with:
\begin{equation}
p_{1}=36 a_s^4 A_s^2,\text{ \ \ \ \ \ \
}p_{2}=4a_{s}A_{s}W_{0},\text{ \ \ \ \ \ }p_{3}=W_{0}^{2}/6,
\label{Numbers}
\end{equation}
and:
\begin{equation}
A_{1}\equiv \frac{2\sqrt{2}}{3 a_s^2}+\underset{\left(
\frac{T}{M_{KK}}\right) ^{2}\ll 1}{
\underbrace{\frac{T^{2}\mathcal{V}\sqrt{\tau
_{s}}}{M_{P}^{2}}}},\text{ \ }A_{2}\equiv
1+\frac{a_s^2}{4\sqrt{2}}\underset{\left( \frac{T}{M_{KK}}\right)
^{2}\ll 1}{\underbrace{\frac{T^{2}\mathcal{V}\sqrt{\tau
_{s}}}{M_{P}^{2}}}} \left( 1+\frac{4}{a_{s}\tau _{s}}\right)
,\text{ \ }A_{3}\equiv \frac{9\hat{\xi}}{2}+\underset{\left(
\frac{T}{M_{s}}\right) ^{2}\ll 1}{\underbrace{
\frac{T^{2}\mathcal{V}}{M_{P}^{2}}.}} \notag
\end{equation}
where the appearance of the Kaluza-Klein scale comes from the
assumption that the MSSM branes are wrapping the small cycle
$\tau_s$:
\begin{equation}
M_{KK}\sim\frac{M_s}{\tau_s^{1/4}}\simeq\frac{M_P}{\sqrt{\mathcal{V}}\tau_s^{1/4}}
\, . \label{Mkk}
\end{equation}
Therefore, we can see that the 1-loop $\mathcal{O}(T^2)$ thermal
corrections can never compete with $V_0$ for temperatures below
the compactification scale $M_{KK}<M_s$, where our low energy
effective field theory is trustworthy. Once again, we stress that
the previous considerations are valid in all the moduli space
(within our large volume approximations) and not just in the
vicinity of the $T=0$ minimum. We have seen that the only
finite-temperature contribution that can compete with $V_0$ is the
2-loop $T^4 g_{MSSM}^2$ term, and so we can only consider from now
on the following potential:
\begin{eqnarray}
\label{effPot}
V_{TOT} &=&V_{0}+4\pi \kappa _{1}\frac{T^{4}}{\tau _{s}} \\
&=&\left( \frac{g_{s}e^{K_{cs}}}{8\pi }\right) \left[
\frac{\lambda \sqrt{ \tau _{s}}e^{-2a_{s}\tau
_{s}}}{\mathcal{V}}-\frac{\mu \tau _{s}e^{-a_{s}\tau
_{s}}}{\mathcal{V}^{2}}+\frac{\nu }{\mathcal{V}^{3}}+\frac{ 4\pi
\tilde{\kappa}_{1}}{\tau _{s}}\left( \frac{T}{M_{P}}\right)
^{4}\right] M_{P}^{4},   \nonumber
\end{eqnarray}
valid for temperatures $T\gg M_{soft}$\footnote{For convenience,
here we have redefined $\tilde{\kappa}_1\equiv8 \pi\kappa_1
g_s^{-1} e^{-K_{cs}}$.}. We realise that the leading
moduli-dependent finite temperature contribution to the effective
potential comes from 2-loops instead of 1-loop. This, however,
does not mean that perturbation theory breaks down, since 1-loop
effects still dominate when one takes into account the moduli
independent $\mathcal{O}(T^4)$ piece that we dropped.

Now, from (\ref{effPot}) it is clear that the thermal correction
cannot induce any new $T$-dependent extremum of the effective
potential. Its presence only leads to destabilization of the $T=0$
minimum at a certain temperature, above which the potential has a
runaway behaviour. Therefore, we are led to the following
qualitative picture. Let us assume that at the end of inflation
the system is sitting at the $T=0$ minimum. Then, after reheating
the MSSM particles thermalise and the thermal correction $T^4
g_{MSSM}^2\sim T^4/\tau_s$ gets switched on. As a result, the
system starts running away along the $\tau_s$ direction only,
since $V_T$ does not depend on ${\cal V}$. However, as soon as
$\tau_s$ becomes significantly larger than its $T=0$ VEV, the two
exponential terms in (\ref{effPot}) become very suppressed with
respect to the $\mathcal{O}(\mathcal{V}^{-3})$ $\alpha'$
correction (the $\nu$ term). Hence, the potential develops a
run-away behaviour also along the $\mathcal{V}$-direction, thus
allowing the K\"{a}hler moduli to remain within the K\"{a}hler
cone.

In section \ref{Sec:DecTemp}, we shall compute the
decompactification temperature, at which the $T=0$ minimum gets
destabilised. Hence we shall focus on the region in the vicinity
of the zero-temperature minimum, where the regime of validity of
the expression (\ref{effPot}) takes the form:
\begin{equation}
M_{soft}\ll T \ll M_{KK}\textit{ \ \ \ }\Leftrightarrow\textit{ \
\ \
}\frac{1}{\mathcal{V}\ln\mathcal{V}}\ll\frac{T}{M_P}\ll\frac{1}
{\sqrt{\mathcal{V}}\tau_s^{1/4}}. \label{validity}
\end{equation}
In the typical LVS where $\mathcal{V}\sim 10^{14}$ allows low
energy supersymmetry, we get $M_{soft}\sim 10^{3}$ GeV and
$M_{KK}\sim 10^{11}$ GeV; thus, in that case, eq. (\ref{effPot})
makes sense only for energies $10^3$ GeV $\ll T\ll 10^{11}$ GeV.
On the other hand, for LVS that allow GUT string scenarios,
$\mathcal{V}\sim 10^{4}$, which implies $M_{soft}\sim 10^{13}$ GeV
and $M_{KK}\sim 10^{16}$ GeV; thus, in that case, (\ref{validity})
becomes $10^{13}$ GeV $\ll T\ll 10^{16}$ GeV.

\newpage
\medskip\noindent{\em General LARGE Volume Scenario}

\medskip\noindent

As we have seen in chapter 4, one of the conditions on an
arbitrary Calabi-Yau to obtain LVS, is the presence of a blow-up
mode resolving a point-like singularity (del Pezzo 4-cycle). The
moduli scaling of the scalar potential, at leading order and in
the presence of $N_{small}$ blow-up modes $\tau_{s_{i}}$,
$i=1,...,N_{small}$, is given by (neglecting loop corrections):
\begin{equation}
V_0=\left(\frac{g_s e^{K_{cs}}M_P^4}{8
\pi}\right)\left[\sum_{i=1}^{N_{small}}\left(\frac{\lambda
\sqrt{\tau_{s_{i}}} e^{-2 a_{s_{i}} \tau_{s_{i}}}}{{\cal V}} -
\frac{\mu \tau_{s_{i}} e^{-a_{s_{i}} \tau_{s_{i}}}}{{\cal
V}^2}\right) + \frac{\nu}{{\cal V}^3}\right]. \label{GenVsimple}
\end{equation}
All the other moduli which are neither the overall volume nor a
blow-up mode will appear in the scalar potential at subleading
order. Moreover, due to the topological nature of $\tau_{s,i}$,
$K^{-1}_{s_{i}s_{i}}\sim \mathcal{V}\sqrt{\tau_{s_{i}}}$ $\forall
i=1,...,N_{small}$ (see appendix A).

As derived in section \ref{Sec:ModCoupl}, these blow-up modes
correspond to the heaviest moduli and modulini, which play the
same role as $\Phi$ and $\tilde{\Phi}$ in the single-hole Swiss
cheese case. Hence the leading order behaviour of the
mass-squareds of the blow-up moduli $\tau_{s_{i}}$ and the
corresponding modulini $\tilde{\tau}_{s_{i}}$ are still given by
(\ref{GenModulusMass}) and (\ref{ModulinoMass}) $\forall
i=1,...,N_{small}$. Therefore we can repeat the same
considerations made in the previous paragraph and conclude that,
for a general LVS, the 1-loop $\mathcal{O}(T^2)$ thermal
corrections are always subdominant with respect to $V_0$ for
temperatures below the compactification scale\footnote{As we have
seen in section \ref{Sec:MultHoleModTherm}, if all the
$\tau_{s_{i}}$ are wrapped by $ED3$ instantons then they do not
thermalise. Only the moduli corresponding to 4-cycles wrapped by
MSSM branes would then thermalise but, since they are lighter than
the $ED3$ moduli, our argument is still valid. The same is true
for all the possible scenarios outlined for the K3 fibration case
in section \ref{Sec:K3FibrModTherm}.}. The only finite-temperature
contribution that can compete with $V_0$ is again the 2-loop $T^4
g_{MSSM}^2$ term.

\subsection{Decompactification temperature}
\label{Sec:DecTemp}

As we saw in the previous subsection, the finite temperature
corrections destabilise the large volume minimum of a general LVS.
In this subsection we will derive the decompactification
temperature $T_{max}$, that is the temperature above which the
full effective potential has no other minima than the one at
infinity.

Before performing a more precise calculation of $T_{max}$, let us
present a qualitative argument that gives a good intuition for its
magnitude. Let us denote by $V_b$ the height of the potential
barrier that separates the supersymmetric minimum at infinity from
the zero temperature supersymmetry breaking one. Now, in order for
the moduli to overcome the potential barrier and run away to
infinity, one needs to supply energy of at least the same order of
magnitude as $V_b$. In our case, the source of energy is provided
by the finite-temperature effects, which give a contribution to
the scalar potential of the order $V_T\sim T^4$. Hence a very good
estimate for the decompactification temperature is given by
$T_{max}\sim V_b^{1/4}$.

It is instructive to compare the implications of this estimate for
the KKLT and LVS cases. In the simplest KKLT models the potential
reads:
\begin{equation}
V_{KKLT}=\lambda_1\frac{e^{-2a\tau}}{\tau}-\lambda_2 W_0
\frac{e^{-a\tau}}{\tau^2},
\end{equation}
where $\lambda_1$ and $\lambda_2$ are constants of order unity.
The minimum is achieved by fine tuning the flux parameter $W_0\sim
\tau e^{-a\tau}$ and so the height of the barrier is given by:
\begin{equation}
V_b\sim \langle V_{KKLT}\rangle\sim
\frac{W_0^2}{\mathcal{V}^2}M_P^4\sim m_{3/2}^2 M_P^2,
\end{equation}
where we have used the fact that $\mathcal{V}=\tau^{3/2}$ and
$m_{3/2}=W_0 M_P/\cal V$. Therefore the decompactification
temperature becomes $T_{max}\sim\sqrt{m_{3/2}M_P}\sim 10^{10}$
GeV, as estimated in \cite{BHLR}.

In the case of LVS, the height of the barrier is lower and so we
expect a lower decompactification temperature $T_{max}$. Indeed,
to leading order the potential is given by:
\begin{equation}
V_{LVS}=\lambda_1\sqrt{\tau_s}\frac{e^{-2a_s\tau_s}}{\cal
V}-\lambda_2 W_0
\tau_s\frac{e^{-a_s\tau_s}}{\mathcal{V}^2}+\lambda_3
\frac{W_0^2}{\mathcal{V}^3}
\end{equation}
with $\lambda_1$, $\lambda_2$ and $\lambda_3$ being constants of
order one. The minimum is achieved for natural values of the flux
parameter $W_0\sim \mathcal{O}(1)$ and at exponentially large
values of the overall volume $\mathcal{V}\sim
W_0\sqrt{\tau_s}e^{a_s\tau_s}$. Hence the height of the barrier
can be estimated as:
\begin{equation}
V_b\sim \langle V_{LVS}\rangle\sim
\frac{W_0^2}{\mathcal{V}^3}M_P^4\sim m_{3/2}^3 M_P,
\end{equation}
which gives a decompactification temperature of the order:
\begin{equation}
T_{max}\sim\left(m_{3/2}^3
M_P\right)^{1/4}\sim\frac{M_P}{\mathcal{V}^{3/4}}.
\label{KeyResult}
\end{equation}

Let us now turn to a more precise computation. Without loss of
generality, we shall focus here on the effective potential
(\ref{effPot}), valid for the single-hole Swiss cheese case, and
look for its extrema. Given that the thermal contribution does not
depend on the volume, the derivative of the potential with respect
to $\mathcal{V}$ gives the same result as in the $T=0$ case:
\begin{equation}
\frac{\partial V_{TOT}}{\partial \mathcal{V}}=0\text{ \ \ \ \ }
\Longrightarrow \text{ \ \ \ \ }\mathcal{V}_*=\frac{\mu}{\lambda
}A(\tau_s)\sqrt{\tau_s}e^{a_s\tau_s}, \label{Vpmc}
\end{equation}
where\footnote{We discard the solution with the positive sign in
front of the square root in (\ref{A}) since, upon its substitution
one finds that the other extremum condition, $\partial
V_{TOT}/\partial\tau_s=0$, does not have any solution.}:
\begin{equation}
A(\tau_s)\equiv 1-\sqrt{1-\frac{3}{4}\left(
\frac{\langle\tau_s\rangle}{\tau _{s}}\right) ^{3/2}}, \label{A}
\end{equation}
and $\langle\tau_s\rangle \simeq \left(4 \lambda
\nu/\mu^2\right)^{2/3}$ is the $T=0$ VEV of $\tau_s$. Substituting
(\ref{Vpmc}) in the derivative of $V_{TOT}$ with respect to
$\tau_s$ and working in the limit $a_s \tau_s\gg 1$, in which one
can neglect higher order instanton corrections, we obtain:
\begin{equation}
\left. \frac{\partial V_{TOT}}{\partial \tau _{s}}\right\vert
_{\mathcal{V}= \mathcal{V}_*}=0\text{ \ \ }\Longrightarrow \text{
\ \ }4\pi \tilde{\kappa}_1 \frac{ \mu e^{3a_{s}\tau _{s}}}{\lambda
^{2}a_{s}\tau _{s}^{2}}\left( \frac{T}{M_{P}}\right)^{4}\!A(\tau
_{s})^{2}+2A(\tau _{s})-1=0.  \label{tausc}
\end{equation}
Notice that at zero temperature (\ref{tausc}) simplifies to
$A(\tau_s)=1/2$, which from (\ref{A}) correctly implies
$\tau_s=\langle\tau_s\rangle$. Now, since equation (\ref{tausc})
is transcendental, one cannot write down an analytical solution,
that gives the general relation between the location of the
$\tau_s$ extrema and the temperature. Nevertheless, we will see
shortly that it is actually possible to extract an analytic
estimate for the decompactification temperature. To understand
why, let us gain insight into the behaviour of the function on the
LHS of (\ref{tausc}) by plotting it and looking at its
intersections with the $\tau_s$-axis.

\begin{figure}[t]
\begin{center}
\scalebox{0.9}{\includegraphics{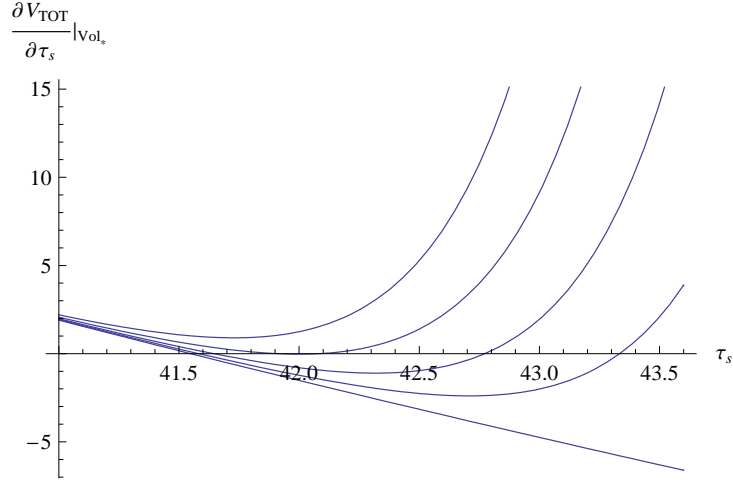}}
\end{center}
\vspace{-0.5cm} \caption{The LHS of eq. (\ref{tausc}) is plotted
versus $\tau_s$. The temperature increases from right to left. The
straight line represents the zero temperature case. The other
values of the temperature are $T/M_P= 0.8 \cdot 10^{-10}$, $1.0
\cdot 10^{-10}$, $1.2 \cdot 10^{-10}$, $1.4 \cdot 10^{-10}$. To
obtain the plots we used the following numerical values: $\xi
=1.31$, $A_s=1$, $W_0=1$, $a_s=\pi/4$, $e^{K_{cs}}= 8\pi /g_s$,
$g_s=0.1$, $N_c=5$, $N_f=7$. With these values one has that
$\langle\tau_{s} \rangle = 41.55$ and $\langle \mathcal{V} \rangle
= 7.02\cdot 10^{13}$, which implies that $T_{max}=1.58 \cdot
10^{-10} M_P\simeq 3.79 \cdot 10^{8}$ GeV according to
(\ref{final}). Note that the numerically found value of the
decompactification temperature is $T_{max,num} = 1.20 \cdot
10^{-10} M_P$.} \label{Fig:dVdtaus}
\end{figure}

We plot the LHS of equation (\ref{tausc}) on Figure
\ref{Fig:dVdtaus} for several values of the temperature; $T$
increases from right to left. From this figure it is easy to see
that the temperature-dependent correction to $V_{TOT}$ behaves
effectively as an up-lifting term. Namely, the finite-temperature
contribution lifts the potential, giving rise to a local maximum
(the right intersection with the $\tau_s$ axis) in addition to the
$T=0$ minimum (the left intersection). As the temperature
increases, the maximum increases as well and shifts towards
smaller values of $\tau_s$. On the other hand, the minimum remains
very close to the zero-temperature one at all temperatures.
Clearly, the decompactification temperature $T_{max}$ is reached
when the two extrema coincide. The key observation here is that
this happens in a small neighborhood of the $T=0$ minimum, located
at $\langle \tau_s \rangle \simeq \left(4 \lambda
\nu/\mu^2\right)^{2/3}$.

In view of the considerations of the previous paragraph, to find
an analytic estimate for $T_{max}$ we shall utilize the following
strategy. We will Taylor-expand the function $F(\tau_s)$, defined
by the LHS of equation (\ref{tausc}), to second order in a small
neighborhood of the point $\tau_s = \langle \tau_s \rangle$. Then
we will use the resulting quadratic function $f(\delta)$, where
$\delta \equiv \tau_s - \langle \tau_s \rangle$, as an
approximation of $F(\tau_s)$ in a larger neighborhood and will
look for the zeros of $f(\delta)$. Requiring that the two roots of
$f(\delta)$ coincide, will give us an estimate for the
decompactification temperature. Clearly, this procedure is not
exact. In particular, the function $F(\tau_s)$ is better
approximated by keeping higher orders in the Taylor expansion. In
our case, we have checked numerically that a really good
approximation is obtained by going to at least sixth order.
However, in doing so one again ends up with an equation that
cannot be solved analytically. So the key point is that the
systematic error introduced by the quadratic approximation is
rather small (we have checked that the analytical results obtained
by following the above procedure are in very good agreement with
the exact numerical values).

Now let us substitute $\tau_s = \langle\tau_s\rangle + \delta$ in
(\ref{tausc}) and read off the terms up to order $\delta^2$. The
result is:
\begin{equation}
a\,\delta^2 + b\,\delta +c =0, \label{eqcri}
\end{equation}
where the corresponding coefficients, in the limit
$a_s\langle\tau_s\rangle\gg 1$, take the form:
\begin{equation}
\left\{
\begin{array}{c}
a\simeq \frac{9}{2}\mathcal{T}a_{s}^{2}+\frac{171}{8}\lambda
^{2}a_{s}, \\
b\simeq 3\mathcal{T}a_{s} -9\lambda ^{2}a_{s}\langle \tau
_{s}\rangle,\text{ \ \ \ }c\simeq \mathcal{T},
\end{array}
\right.
\end{equation}
and we have set:
\begin{equation}
{\cal T} \equiv 4 \pi \tilde{\kappa}_1
\left(\frac{T}{M_P}\right)^4 \mu e^{3a_s \langle\tau_s\rangle}.
\end{equation}
Finally, to find the decompactification temperature, we require
that the two solutions $\delta_1$ and $\delta_2$ coincide:
\begin{equation}
\delta_1=\delta_2\text{ \ \ \ }\Longleftrightarrow\text{ \ \ \
}b^2-4a\, c =0,
\end{equation}
which, for $a_s \langle \tau_s \rangle \gg 1$, gives:
\begin{equation}
{\cal T}_{max} = 3 (\sqrt{2}-1)\lambda ^2 \langle\tau_s\rangle
\text{ \ \ \ }\Longleftrightarrow\text{ \ \ \ }T_{max}^4= \frac{3
(\sqrt{2}-1) \lambda^2 \langle\tau_s\rangle }{4 \pi
\tilde{\kappa}_1 \mu} e^{-3a_s \langle\tau_s\rangle} M_P^4.
\end{equation}
Notice that we can rewrite the decompactification temperature in
terms of $\mathcal{V}$ as:
\begin{equation}
T_{max}^4=\frac{3(\sqrt{2}-1)}{32 \pi}\frac{\mu^2}{\lambda
\tilde{\kappa}_1}\frac{\langle\tau_s\rangle^{5/2}}
{\mathcal{V}^{3}}M_P^4\text{ \ \ \ }\Longrightarrow\text{ \ \ \
}T_{max}\sim\left(m_{3/2}^3
M_P\right)^{1/4}\sim\frac{M_P}{\mathcal{V}^{3/4}}\, ,
\label{final}
\end{equation}
where we have used the relation between the $T=0$ VEV of the
volume and $\langle\tau_s\rangle$, which is given by (\ref{Vpmc})
with $\tau_s=\langle\tau_s\rangle$ and $A=1/2$. It is reassuring
that (\ref{final}) is of the same form as the result
(\ref{KeyResult}), obtained from the intuitive arguments based on
the height of the potential barrier.

\subsection{Small moduli cosmology}
\label{Sec:MaxTemp}

Clearly, the decompactification temperature (\ref{final}) sets an
upper bound on the temperature in the early Universe, in
particular on the reheating temperature, $T_{RH}^0$, at the end of
inflation. We will investigate now how this constraint affects the
moduli thermalisation picture studied in subsection
\ref{Sec:11169ModTherm}.\footnote{Similar considerations apply for
the more general multiple-hole Swiss cheese and K3 fibration
cases.}

Recall that there we derived the following:
\begin{itemize}
\item For small values of
the Volume ($\mathcal{V}<10^{10}$), the freeze-out temperature for
the small modulus $\Phi$ is given by (\ref{caseA}): $T_f^{SV}\sim
M_P \mathcal{V}^{-2/3}$.

\item For large values of
the Volume ($\mathcal{V}>10^{10}$), the freeze-out temperature for
$\Phi$ is given by (\ref{caseB}): $T_f^{LV}\sim 10^3 M_P
\mathcal{V}^{-1}$.
\end{itemize}
Note also that, in both cases, the condition
$T_f<T_{RH}^0<T_{max}$ has to be satisfied in order for the
modulus to reach equilibrium with the MSSM thermal bath. Now, for
small values of $\mathcal{V}$ we have that:
\begin{equation}
\frac{T_{max}}{T_f^{SV}}\sim\frac{\mathcal{V}^{2/3}}{\mathcal{V}^{3/4}}
=\mathcal{V}^{-1/12}<1,
\end{equation}
which implies that $\Phi$ actually never thermalises. On the other
hand, for large values of $\mathcal{V}$ we have that (writing
$\mathcal{V}\sim 10^x$):
\begin{equation}
\frac{T_{max}}{T_f^{LV}}\sim\frac{\mathcal{V}^{1/4}}{10^3}
=10^{x/4-3}>1\text{ \ \ }\Leftrightarrow\text{ \ \ }x>12.
\end{equation}
Hence, for $\mathcal{V}>10^{12}$, $\Phi$ can reach thermal
equilibrium with the MSSM plasma, as long as $T_{RH}^0$ is such
that $T_f^{LV}<T_{RH}^0<T_{max}$. Let us stress, however, that if
$T_{RH}^0<T_f^{LV}$ the modulus will never thermalise even though
$T_f^{LV}<T_{max}$. Note that, since the temperature $T_{RH}^0$
depends on the concrete realization of inflation and the details
of the initial reheating process, its determination is beyond the
scope of this chapter. So we will treat it as a free parameter,
satisfying only the constraint $T_{RH}^0<T_{max}$.

We would like now to study the cosmological history of $\Phi$
which, in our case, presents two possibilities:
\begin{enumerate}
\item The modulus $\Phi$ decays at the end of
inflation being the main responsible for initial reheating. We may
envisage two physically different situations where this could
happen: in one case, $\Phi$ is the inflaton and it decays at the
end of inflation. In the other case, $\Phi$ is not the inflaton,
but it starts oscillating around its VEV when the inflaton is
still driving inflation by rolling down its flat potential. In
this case, the decay of $\Phi$ occurs just after the slow-roll
conditions stop being satisfied and the inflaton reaches its VEV.

After $\Phi$ decays, its energy density is converted into
radiation. The decay products thermalise rapidly and re-heat the
Universe to a temperature $T_{RH}=T_{RH}^0$. The latter can be
computed by noticing that the $\Phi$ energy density
$\rho_{\Phi}\sim \Gamma_{\Phi\to XX}^2 M_P^2$ will be converted
into radiation energy density $\rho_{R}\sim g_{*}T^4$. Hence
$T_{RH}^0$ can be obtained by comparing $\Gamma_{\Phi\to XX}$ with
the value of $H$, given by the Friedmann equation for radiation
dominance:
\begin{gather}
\Gamma _{\Phi \rightarrow XX}\sim \frac{\ln \mathcal{V}}{16\pi
}\frac{ m_{\Phi }^{2}}{M_{P}}\simeq H\sim g_{\ast
}^{1/2}\frac{\left(
T_{RH}^{0}\right) ^{2}}{M_{P}}  \notag \\
\Leftrightarrow\text{ \ } T_{RH}^{0}\simeq \left( \frac{\ln
{\mathcal{V}}}{16\pi \sqrt{g_{\ast }}}\right) ^{1/2}m_{\Phi
}=\frac{\left( \ln {\mathcal{V}} \right) ^{3/2}}{4\sqrt{\pi
}g_{\ast }^{1/4}}\frac{M_{P}}{\mathcal{V}}. \label{222}
\end{gather}
In order for this picture to be compatible with the presence of a
decompactification temperature (\ref{final}), that sets the
maximal temperature of the Universe, we need to require that
$T_{RH}^0<T_{max}$. As we shall see in subsection
\ref{Sec:VolBound}, this requirement can be translated into a
constraint on the values that the internal volume can take.

\item The modulus $\Phi$ is not the main source of initial reheating,
which we suppose to be the inflaton. After the inflaton decays,
the Universe is re-heated to a temperature $T_{RH}^0$ and an epoch
of radiation dominance begins. The modulus $\Phi$ will only
thermalise if $\mathcal{V}>10^{12}$ and $T_f^{LV}<T_{RH}^0$.
However, $T_f^{LV}$ is rather close to $T_{max}$ and so, even when
$\Phi$ thermalises, it will drop out of equilibrium very quickly
at $T_f^{LV}$. Then, for general values of $\mathcal{V}$, the
modulus $\Phi$ will decay out of equilibrium at a temperature
$T_D<T_{RH}^0$. As we shall show below, this decay will occur
during radiation domination, since $T_D>T_{dom}$, with $T_{dom}$
being the temperature at which the modulus energy density would
dominate over the radiation energy density. So the temperature
$T_D$ at which $\Phi$ decays, is still given by (\ref{222}) upon
replacing $T_{RH}^0$ with $T_D$:
\begin{equation}
T_D\simeq
\frac{\left(\ln{\mathcal{V}}\right)^{3/2}}{4\sqrt{\pi}g_{*}^{1/4}}
\frac{M_P}{\mathcal{V}}. \label{DUE}
\end{equation}
Note that the above expression satisfies $T_D<T_f^{SV,LV}$, as
should be the case for consistency. Another important observation
is that (\ref{DUE}) is also the usual expression for the
temperature $T_{RH}$, to which the Universe is re-heated by the
decay of a particle releasing its energy to the thermal bath. In
other words, for us $T_{RH}=T_D$ since the modulus $\Phi$ decays
during radiation domination. On the contrary, if a modulus decays
when its energy density is dominating the energy density of the
Universe, then $T_D < T_{RH}$ and the decay produces an increase
in the entropy density $S$, which is determined by:
\begin{equation}
\Delta \equiv
\frac{S_{fin}}{S_{in}}\sim\left(\frac{T_{RH}}{T_D}\right)^3.
\label{Delta}
\end{equation}
As already mentioned, since for us $T_{RH}=T_D$, the decay of
$\Phi$ does not actually lead to reheating or, equivalently, to an
increase in the entropy density, given that from (\ref{Delta}) we
have $\Delta = 1$. As a consequence, $\Phi$ cannot dilute any
unwanted relics, like for example the large modulus $\chi$ which
suffers from the cosmological moduli problem.\footnote{This kind
of solution of the cosmological moduli problem, i.e. dilution via
saxion or modulus decay, is used both in \cite{Acharya} and in
\cite{Vafa}.}

To recapitulate: in the present case 2, we have the following
system of inequalities:
\begin{eqnarray}
\text{for }\mathcal{V} &<&10^{12}\text{: \ \ \
}T_{dom}<T_{D}<T_{RH}^{0}<T_{max}, \\
\text{for }\mathcal{V} &>&10^{12}\text{: \ \ \ }
T_{dom}<T_{D}<T_{f}^{LV}<T_{RH}^{0}<T_{max}.
\end{eqnarray}
As in case 1 above, the condition $T_D<T_{max}$ implies a
constraint on $\mathcal{V}$, that we will derive in section
\ref{Sec:VolBound}. We underline again that this condition is
necessary but not sufficient, since for us $T_{RH}^0$ is an
undetermined parameter. In concrete models, in which one could
compute $T_{RH}^0$, the condition $T_{RH}^0 < T_{max}$ might lead
to further restrictions.
\end{enumerate}

Let us now prove our claim above that, when the modulus $\Phi$ is
not responsible for the initial reheating (case 2), it will decay
before its energy density begins to dominate the energy density of
the Universe. $\Phi$ will start oscillating around its VEV when
$H\sim m_{\Phi}$ at a temperature $T_{osc}$ given by:
\begin{equation}
T_{osc}\sim g_{*}^{-1/4}\sqrt{m_{\Phi}M_P}.  \label{Tosc}
\end{equation}
The energy density $\rho_{\Phi}$, stored by $\Phi $, and the ratio
between $\rho_{\Phi}$ and the radiation energy density at
$T_{osc}$ read as follows:
\begin{equation}
\left. \rho _{\Phi }\right\vert _{T_{osc}}\sim m_{\Phi
}^{2}\langle \tau_s \rangle ^{2}\text{ \ \ }\Rightarrow \text{ \ \
}\left. \left( \frac{\rho _{\Phi }}{\rho _{r}}\right) \right\vert
_{T_{osc}}\sim\frac{m_{\Phi }^{2}\langle \tau_s \rangle
^{2}}{g_{\ast }T_{osc}^{4}}\sim \frac{\langle \tau_s \rangle
^{2}}{M_{P}^{2}}.
\end{equation}
By definition, the temperature $T_{dom}$, at which $\rho_{\Phi}$
becomes comparable to $\rho_r$ and hence $\Phi$ begins to dominate
the energy density of the Universe, is such that:
\begin{equation}
\left. \left( \frac{\rho _{\Phi }}{\rho _{r}}\right) \right\vert
_{T_{dom}}\sim 1.
\end{equation}
Now, given that $\rho_{\Phi}$ redshifts as $T^3$ whereas $\rho_r$
scales as $T^4$, we can relate $T_{dom}$ with $T_{osc}$:
\begin{equation}
T_{dom}\left. \left( \frac{\rho _{\Phi }}{\rho _{r}}\right)
\right\vert _{T_{dom}}\sim T_{osc}\left. \left( \frac{\rho _{\Phi
}}{\rho _{r}}\right) \right\vert _{T_{osc}}\text{ \ \
}\Leftrightarrow \text{ \ \ }T_{dom}\sim
g_*^{-1/4}\frac{\langle\tau_s\rangle^2}{M_P^2}\sqrt{m_{\Phi}M_P}.
\label{Tdom}
\end{equation}
We shall show now that $T_{dom}<T_D$ with $T_D$ being the decay
temperature during radiation dominance, which is obtained by
comparing $H$ with $\Gamma_{\Phi\to XX}$:
\begin{equation}
T_D\sim g_{*}^{-1/4}\sqrt{\Gamma_{\Phi\to XX}M_{P}}. \label{TD}
\end{equation}
The ratio of (\ref{TD}) and (\ref{Tdom}) gives:
\begin{equation}
\frac{T_{D}}{T_{dom}}\sim \frac{\sqrt{\Gamma _{\Phi \rightarrow
XX}}}{\sqrt{ m_{\Phi }}}\frac{M_{P}^{2}}{\langle \tau _{s}\rangle
^{2}}.
\end{equation}
Using that $\Gamma _{\Phi \rightarrow XX}\sim \mathcal{V}m_{\Phi
}^{3}M_{P}^{-2}$ and $\langle \tau _{s}\rangle \sim 10 M_s \sim
10M_{P}\mathcal{V}^{-1/2}$, the last relation becomes:
\begin{equation}
\frac{T_{D}}{T_{dom}}\sim \frac{\left( \ln \mathcal{V}\right)
\sqrt{\mathcal{ V}}}{100}>1\text{ \ for \ }\mathcal{V}>10^{2.5}.
\end{equation}
Hence, we conclude that $T_D>T_{dom}$ and, therefore, $\Phi$
decays before it can begin to dominate the energy density of the
Universe. The main consequence of this is that $\Phi$ cannot
dilute unwanted relics via its decay.

\subsection{Lower bound on the volume}
\label{Sec:VolBound}

As we saw in the previous subsection, there are two possible
scenarios for the cosmological evolution of the small modulus
$\Phi$. However, since the RHS of (\ref{222}) and (\ref{DUE})
coincide, in both cases the crucial quantity is the same, although
with a different physical meaning. Let us denote this quantity by
$T_*\sim (\Gamma_{\Phi}M_P)^{1/2}$. We shall impose that
$T_*<T_{max}$ and shall show below that from this requirement one
can derive a lower bound on the possible values of $\mathcal{V}$
in a general LVS. Before we begin, let us first recall that:
\begin{enumerate}
\item If $\Phi$ is responsible for the initial reheating via its
decay, then $T_*=T_{RH}^0$;

\item If $\Phi$ decays after the original reheating in a radiation
dominated era, then $T_*=T_D<T_{RH}^0$.
\end{enumerate}
Regardless of which of these two situations we consider, $T_*$ is
the temperature of the Universe after $\Phi$ decays. Then, in
order to prevent decompactification of the internal space, we need
to impose $T_*<T_{max}$. In general, this condition is necessary
but not sufficient because in case 2 one must ensure also that
$T_{RH}^0 < T_{max}$. This is a constraint that we cannot address
given that in this case $T_{RH}^0$ is an undetermined parameter
for us.

Let us now compute $T_*$ precisely. We start by using the exact
form of the decay rate $\Gamma_{\Phi\to XX}$:
\begin{equation}
\Gamma_{\Phi\to XX}=\frac{g_{\Phi XX}^2 m_{\Phi}^3}{64\pi M_P^2},
\label{gammaPhi}
\end{equation}
where:
\begin{equation}
g_{\Phi
XX}=\frac{2^{5/4}\sqrt{3}}{\langle\tau_s\rangle^{3/4}}\sqrt{\mathcal{V}}.
\end{equation}
The mass of $\Phi$ is given by:
\begin{equation}
m_{\Phi}=\sqrt{P}\frac{2 a_s\langle\tau_s\rangle
W_0}{\mathcal{V}}M_P,
\end{equation}
where we are denoting with $P$ the prefactor of the scalar
potential: $P\equiv g_s e^{K_{cs}}/(8\pi)$. From the minimisation
of the scalar potential we have that:
\begin{equation}
a_s\langle\tau_s\rangle=\ln\left(p\mathcal{V}\right)=\ln p+\ln
\mathcal{V},
\end{equation}
where:
\begin{equation}
p\equiv \frac{12\sqrt{2}a_s A_s}{W_0\sqrt{\tau_s}}\sim
\mathcal{O}(1)\text{ \ \ }\Rightarrow\text{ \ \
}a_s\langle\tau_s\rangle\simeq\ln \mathcal{V},
\end{equation}
and so:
\begin{equation}
m_{\Phi}=\sqrt{P}\frac{2W_0\ln \mathcal{V}}{\mathcal{V}}M_P.
\end{equation}
Therefore, the decay rate $\Gamma_{\Phi\to XX}$ turns out to be:
\begin{equation}
\Gamma_{\Phi\to XX}=P^{3/2}\frac{3W_0^3(\ln
\mathcal{V})^3}{\sqrt{2}\pi\langle\tau_s\rangle^{3/2}}\frac{M_P}{\mathcal{V}^2}.
\end{equation}
Finally, in order to obtain the total decay rate, we need to
multiply $\Gamma_{\Phi\to XX}$ by the total number of gauge bosons
for the MSSM $N_X=12$:
\begin{equation}
\Gamma_{\Phi\to XX}^{TOT}=P^{3/2}\frac{36W_0^3(\ln
\mathcal{V})^3}{\sqrt{2}\pi\langle\tau_s\rangle^{3/2}}\frac{M_P}{\mathcal{V}^2}.
\end{equation}
Now, we can find $T_*$ by setting $4\left(\Gamma_{\Phi\to
XX}^{TOT}\right)^2/3$ equal to $3 H^2$, with $H$ read off from the
Friedmann equation for radiation dominance:
\begin{equation}
T_*=\left(\frac{40}{\pi^2g_*}\right)^{1/4}\sqrt{\Gamma_{\Phi}^{TOT}M_P}=
P^{3/4}\frac{6}{\pi}\left(\frac{20}{g_*}\right)^{1/4}\frac{(W_0\ln
\mathcal{V})^{3/2}}{\langle\tau_s\rangle^{3/4}}\frac{M_P}{\mathcal{V}}.
\label{good}
\end{equation}
We are finally ready to explore the constraint $T_* < T_{max}$.
Recall that the maximal temperature is given by the
decompactification temperature (\ref{final}):
\begin{equation}
T_{max}=\left(\frac{P}{4\pi\kappa_1}\right)^{1/4}
\left[\frac{(\sqrt{2}-1)}{4\sqrt{2}}\right]^{1/4}
\frac{\sqrt{W_0}\langle\tau_s\rangle^{5/8}}
{\mathcal{V}^{3/4}}M_P.
\end{equation}
Let us now consider the ratio $T_{max}/T_*$ and impose that it is
larger than unity (using $g_*(MSSM)=228.75$):
\begin{equation}
R\equiv\frac{T_{max}}{T_*}=c\frac{\mathcal{V}^{1/4}}{(\ln
\mathcal{V})^{3/2}}\text{ \ \ with \ \ }c\equiv
J\left[\frac{(\sqrt{2}-1)g_*}{80\sqrt{2}}\right]^{1/4}
\frac{\pi\langle\tau_s\rangle^{11/8}}{6W_0}\simeq
\frac{\langle\tau_s\rangle^{11/8}}{2W_0}, \label{c}
\end{equation}
where we have defined:
\begin{equation}
J\equiv \left(4\pi\kappa_1
P^2\right)^{-1/4}=\frac{8.42}{\kappa_1^{1/4}}e^{-K_{cs}/2}\text{ \
\ for \ \ }g_s=0.1, \label{J}
\end{equation}
and in (\ref{c}) we have set $J=1$. In fact, from (\ref{kappa1}),
we find that in the case of SQCD with $N_c=3$ and $N_f=6$,
$\kappa_1=2.625$. However for the MSSM we expect a larger value of
$\kappa_1$ which we assume to be of the order $\kappa_1=10$. Then
for natural values of $K_{cs}$ like $K_{cs}=3$ \footnote{The
dependence of the K\"{a}hler potential on the complex structure
moduli can be worked out by computing the different periods of the
CY three-fold under consideration. As derived in \cite{Periods},
for the simplest example of a CY manifold with just one complex
structure modulus $U$ (the mirror of the quintic), for natural
values of $U$, $|U|\sim \mathcal{O}(1)\Rightarrow
K_{\textrm{cs}}\sim \mc{O}(1)$.}, from (\ref{J}), we find
$J=1.05$. Let us consider now the maximum and minimum values that
the parameter $c$ can take for natural values of
$\langle\tau_s\rangle$ and $W_0$:
\begin{eqnarray}
\left\{
\begin{array}{c}
\langle \tau _{s}\rangle _{max }=100 \\
W_{0,min }=0.01
\end{array}
\right.  &\Longrightarrow &c_{max }\simeq 10^{4}, \label{cmax}\\
\left\{
\begin{array}{c}
\langle \tau _{s}\rangle _{min }=2 \\
W_{0,max }=100
\end{array}
\right.  &\Longrightarrow &c_{min }\simeq 10^{-2}. \label{cmin}
\end{eqnarray}

\begin{figure}[ht]
\begin{center}
\begin{tabular}{l||l}
& $R>1$ $\Leftrightarrow$ $T_{max}>T_*$ \\
  \hline\hline
  $c=4$ & \hspace*{1.4cm}$\forall x$ \\
  $c=3$ & \hspace*{1cm}$x>2.1$ \\
  $c=2$ & \hspace*{1cm}$x>3.8$ \\
  $c=1$ & \hspace*{1cm}$x>5.9$ \\
  $c=0.5$ & \hspace*{1cm}$x>7.6$ \\
  $c=0.1$ & \hspace*{1cm}$x>11.3$ \\
  $c=0.05$ & \hspace*{1cm}$x>12.8$ \\
  $c=0.01$ & \hspace*{1cm}$x>16.1$ \\
\end{tabular} \\\smallskip
{\bf Table {9.3}:} Lower bounds on the physical volume as seen
  by the string $\mathcal{V}_s\sim 10^{x-3/2}$ for some benchmark scenarios.
\end{center}
\end{figure}

\begin{figure}[ht]
\begin{center}
\epsfig{file=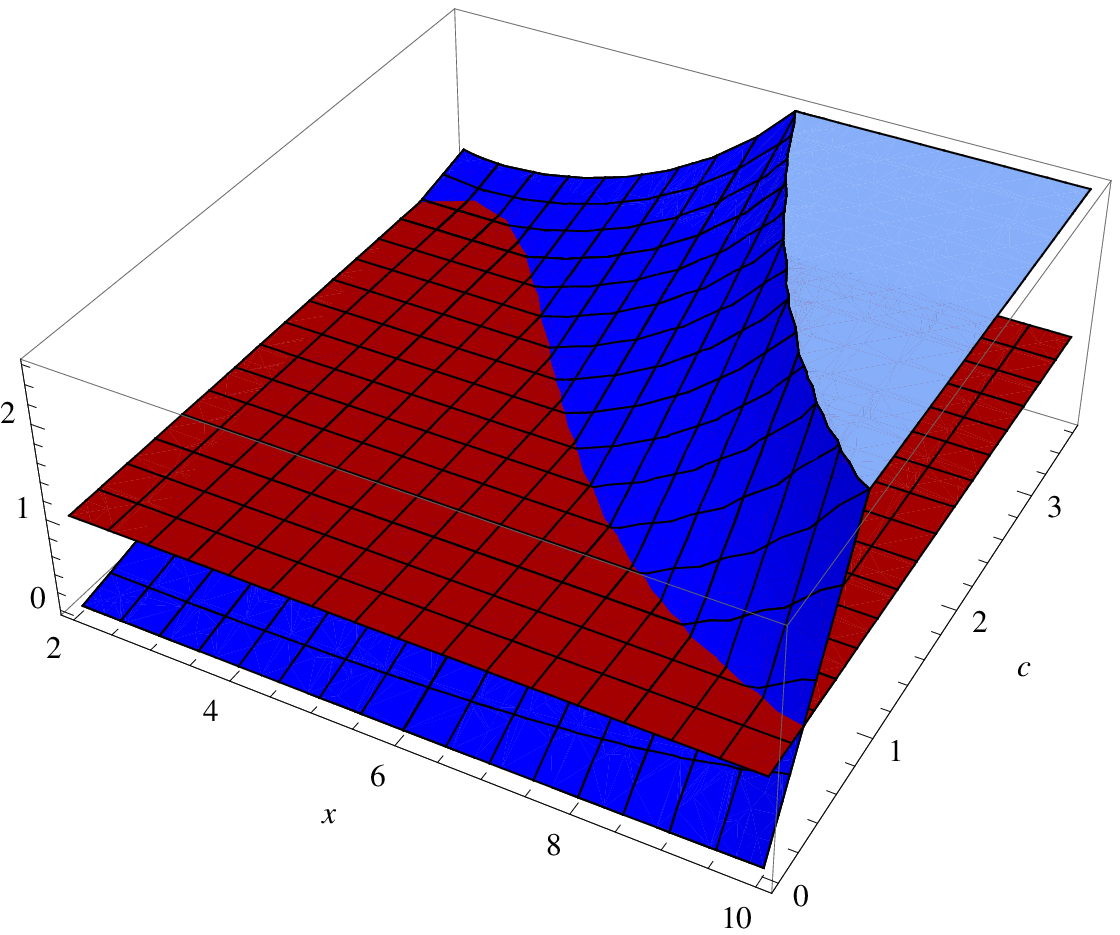, height=60mm,width=67mm}
\epsfig{file=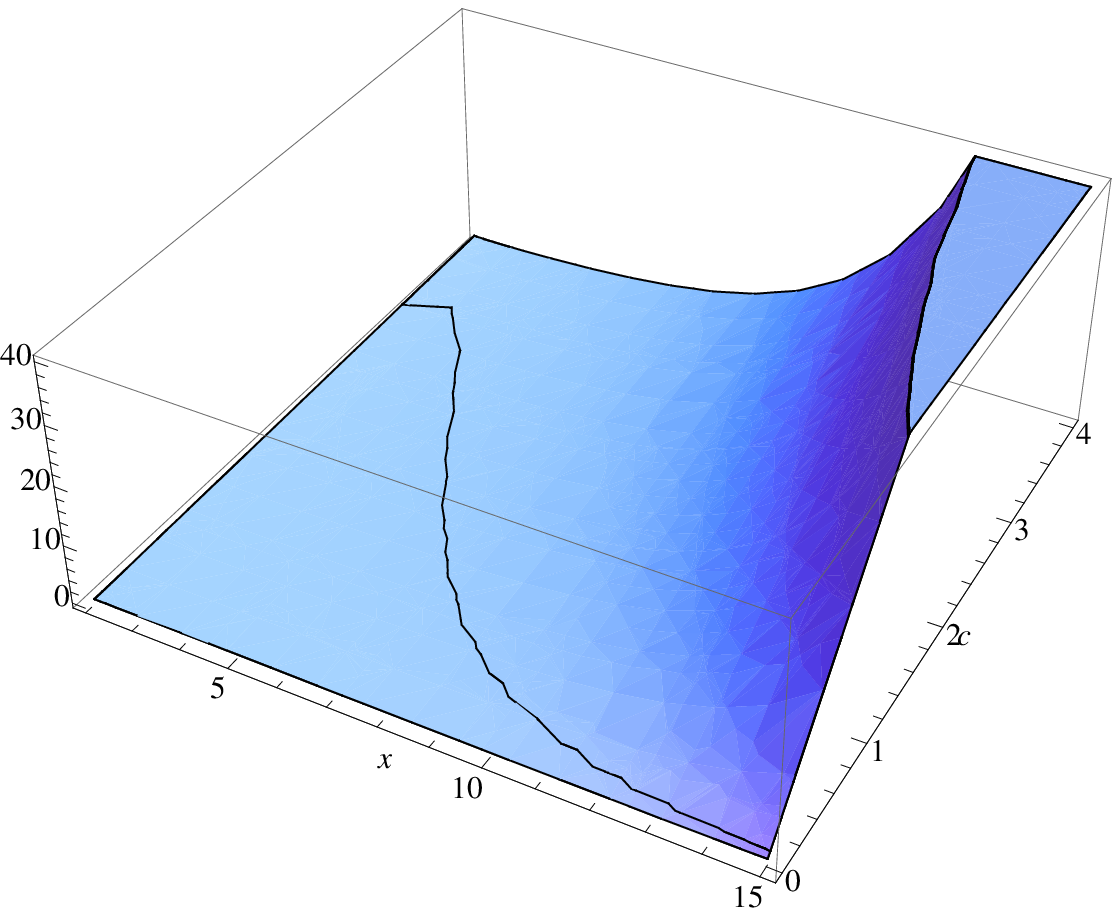, height=60mm,width=67mm}
\caption{Plots of the ratio $R\equiv T_{max}/T_*$ as a function of
$\mathcal{V}=10^x$ and the parameter $c_{min}<c<c_{max}$ as
defined in (\ref{c}), (\ref{cmax}) and (\ref{cmin}). In the left
plot, the red surface is the constant function $R=1$, whereas in
the right plot the black line denotes the curve in the
($x$,$c$)-plane for which $R=1$.}
   \label{VolConstraint1}
\end{center}
\end{figure}

\begin{figure}[t]
\begin{center}
\scalebox{0.9}{\includegraphics{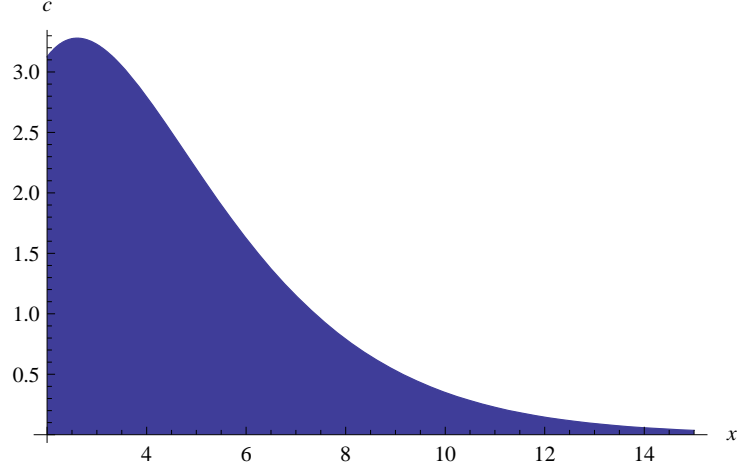}}
\end{center}
\vspace{-0.4cm} \caption{Plot of the $R=1$ curve in the
($x$,$c$)-plane. The shaded region represents the
phenomenologically forbidden area, in which the values of $x$ and
$c$ are such that $R<1$ $\Leftrightarrow$ $T_{max}<T_*$.}
\label{VolConstraint2}
\end{figure}

Now writing $\mathcal{V}\simeq 10^x$, $R$ becomes a function of
$x$ and $c$. Finally, we can make a 3D plot of $R$ with
$c_{min}<c<c_{max}$ and $2<x<15$, and see in which region $R>1$.
This is done in Figure \ref{VolConstraint1}. In order to
understand better what values of $\mathcal{V}$ are disfavoured, we
also plot in Figure \ref{VolConstraint2}, as the shaded region,
the region in the ($x$,$c$)-plane below the curve $R=1$, which
represents the phenomenologically forbidden area for which
$T_{max}<T_*$. We conclude that small values of the volume, which
would allow the standard picture of gauge coupling unification and
GUT theories, are disfavoured compared to larger values of
$\mathcal{V}$, that naturally lead to TeV-scale supersymmetry and
are thus desirable to solve the hierarchy problem. In Table 9.3,
we show explicitly how the lower bound on the volume, for some
benchmark scenarios, favours LVS with larger values of
$\mathcal{V}$.

From the definition (\ref{c}) of the parameter $c$, it is
interesting to notice that for values of $\langle\tau_s\rangle$
far from the edge of consistency of the supergravity approximation
$\langle\tau_s\rangle\sim \mc{O}(10)$, $c$ should be fairly large,
and hence the bound very weak, for natural values of $W_0\sim
\mc{O}(1)$, while $c$ should get smaller for larger values of
$W_0$ that lead to a stronger bound. In addition, it is reassuring
to notice that for typical values of $\mathcal{V}\sim 10^{15}$,
$T_{max}>T_*$ except for a tiny portion of the ($x$,$c$)-space. It
also important to recall that the physical value of the volume as
seen by the string is the one expressed in the string frame
$\mathcal{V}_s$, while we are working in the Einstein frame where
$\mathcal{V}_s=g_s^{3/2}\mathcal{V}_E$. Hence if we write
$\mathcal{V}_E\sim 10^x$, then we have that
$\mathcal{V}_s=10^{x-3/2}$, upon setting $g_s=0.1$.

\medskip\noindent{\em General LARGE Volume Scenario}

\medskip\noindent

Let us now generalise our lower bound on $\mathcal{V}$ to the four
cases studied in sections \ref{Sec:MultHoleModTherm} and
\ref{Sec:K3FibrModTherm} for the multiple-hole Swiss cheese and K3
fibration cases (focusing on the small K3 fiber scenario)
respectively.

First of all, we note that, since in all the cases the 4-cycle
supporting the MSSM is stabilised by string loop corrections, we
can estimate the actual height of the barrier seen by this modulus
as (see (\ref{74oi})):
\begin{equation}
V_b\sim\frac{W_0^2}{\mathcal{V}^3\sqrt{\tau}},
\end{equation}
where we are generically denoting any small cycle (either a
blow-up or a K3 fiber divisor) as $\tau$, given that the values of
the VEV of all these 4-cycles will have the same order of
magnitude. Then setting $V_b\sim T_{max}^4/\tau$, we obtain:
\begin{equation}
T_{max}^4\sim\frac{\sqrt{\tau}W_0^2}{\mathcal{V}^3}.
\label{newTmax}
\end{equation}
We notice that (\ref{newTmax}) is a bit lower than (\ref{final})
but the two expressions for $T_{max}$ share the same leading order
$\mathcal{V}$-dependence.

Let us now examine the 4 cases of section
\ref{Sec:MultHoleModTherm} in more detail, keeping the same
notation as in that subsection, and denoting as $\Phi$ the small
modulus of the single-hole Swiss cheese scenario studied above:
\begin{enumerate}
\item The relevant decay is the one of $\tau_2$ to MSSM gauge bosons.
The order of magnitude of the mass of $\tau_2$ is:
\begin{equation}
m_{\tau_2}^2\sim\frac{\left(\ln\mathcal{V}\right)^2 W_0^2
}{\mathcal{V}^2\tau^2}, \label{masstau2}
\end{equation}
and so $\tau_2$ is lighter than $\Phi$, and, in turn, $T_*$ will
be smaller. In fact, plugging (\ref{masstau2}) in
(\ref{gammaPhi}), we end up with (ignoring numerical prefactors):
\begin{equation}
T_*\sim\frac{\left(\ln\mathcal{V}\right)^{3/2}
W_0^{3/2}}{\mathcal{V}\tau^{9/4}}. \label{tstarnew}
\end{equation}
Hence we obtain:
\begin{equation}
R^{(1)}\equiv\frac{T_{max}}{T_*}=
c^{(1)}\frac{\mathcal{V}^{1/4}}{\left(\ln\mathcal{V}\right)^{3/2}}\text{
\ \ with \ \ }c^{(1)}\sim \frac{\tau^{19/8}}{W_0}.
\end{equation}
Comparing this result with (\ref{c}), we realise that $R^{(1)}\sim
R\,\tau$ and so the lower bound on $\mathcal{V}$ turns out to be
less stringent. The final results can still be read from Table 9.3
upon replacing $c$ with $c^{(1)}$.

\item The relevant decay is the one of $\tau_1$ to MSSM gauge
bosons since $m_{\tau_1}\sim m_{\Phi}$, and so $\tau_1$ is heavier
than $\tau_2$. Therefore $T_*$ will still be given by
(\ref{good}). Hence we obtain:
\begin{equation}
R^{(2)}\equiv\frac{T_{max}}{T_*}=
c^{(2)}\frac{\mathcal{V}^{1/4}}{\left(\ln\mathcal{V}\right)^{3/2}}\text{
\ \ with \ \ }c^{(2)}\sim \frac{\tau^{7/8}}{W_0}.
\end{equation}
Comparing this result with (\ref{c}), we realise that $R^{(2)}\sim
R\,\tau^{-1/2}$ and so the lower bound on $\mathcal{V}$ turns out
to be more stringent. The final results can still be read from
Table 9.3 upon replacing $c$ with $c^{(2)}$.

\item The relevant decay is the one of $\tau_1$ to hidden sector gauge
bosons. Hence we point out that the considerations of case 2 apply
also for this case.

\item The relevant decay is the one of $\tau_1$ to MSSM gauge
bosons, and so we can repeat the same considerations of case 2.
\end{enumerate}
The final picture is that for all cases the
$\mathcal{V}$-dependence of the ratio $T_{max}/T_*$ is the same as
in (\ref{c}). The only difference is a rescaling of the parameter
$c$. Thus we conclude that, as far as the lower bound on
$\mathcal{V}$ is concerned, the single-hole Swiss cheese case
shows all the qualitative features of a general LVS.

Finally, we mention that in the case of a K3 fibration with small
K3 fiber, cases 2, 3, and 4 of section \ref{Sec:K3FibrModTherm}
have the same behaviour as case 2 of the multiple-hole
Swiss-Cheese, so giving a more stringent lower bound on
$\mathcal{V}$. We should note though that this lower bound does
not apply to case 1 of subsection \ref{Sec:K3FibrModTherm}, since
both of the moduli have an $M_P$-suppressed, instead of
$M_s$-suppressed, coupling to MSSM gauge bosons. However, these
kinds of models tend to prefer larger values of $\mathcal{V}$ (due
to the fact that $a_s=2\pi$ for an $ED3$ instanton) which are not
affected by the lower bound that we derived.

\section{Discussion}
\label{ConDis}

In this chapter, we studied how finite-temperature corrections
affect the $T=0$ effective potential of type IIB LVS and what are
the subsequent cosmological implications in this context.

We showed that the small moduli and modulini can reach thermal
equilibrium with the MSSM particles. Despite that, we were able to
prove that their thermal contribution to the effective potential
is always subleading compared to the $T=0$ potential, for
temperatures below the Kaluza-Klein scale. As a result, the
leading temperature-dependent part of the effective potential is
due only to the MSSM thermal bath and it turns out to have runaway
behaviour at high $T$. We derived the decompactification
temperature $T_{max}$, above which the $T=0$ minimum is completely
erased and the volume of the internal space starts running towards
infinity. Clearly, in this class of IIB compactifications the
temperature $T_{max}$ represents the maximal allowed temperature
in the early Universe. Hence, in particular, it gives an upper
bound on the initial reheating temperature after inflation:
$T_{RH}^0<T_{max}$.\footnote{Note, however, that it may be
possible to relax this constraint to a certain degree by studying
the dynamical evolution of the moduli in presence of finite
temperature corrections, in the vein of the considerations of
\cite{barreiro} for the KKLT set-up.} The temperature $T_{RH}^0$
is highly dependent on the details of the concrete inflationary
model and re-heating process, and so in principle its
determination is beyond the scope of this chapter. Nevertheless,
we can compute the temperature of the Universe after the small
moduli decay. They are rather short-lived and their decay can
either be the main source of initial reheating (in which case the
temperature after their decay is exactly $T_{RH}^0$) or it can
occur during a radiation dominated epoch, after initial reheating
has already taken place. In both cases, the resulting temperature
of the Universe $T_*$ has to satisfy $T_* < T_{max}$\,, which
implies a lower bound on the allowed values of $\mathcal{V}$. We
were able to derive this bound and show that it rules out a large
range of smaller ${\cal V}$ values (which lead to standard GUT
theories), while favouring greater values of ${\cal V}$ (that lead
to TeV-scale supersymmetry). Note though, that the condition
$T_*<T_{max}$ is both necessary and sufficient in the case the
decay of the small moduli is the origin of initial reheating,
whereas it is just necessary but not sufficient in the case the
small moduli decay below $T_{RH}^0$.

Let us now discuss some of the possible applications of these
results, as well as directions for future work. As we have
emphasised throughout this chapter, there are two kinds of LVS,
depending on the magnitude of the value of the internal volume
$\mathcal{V}$. Their main cosmological characteristics are the
following:

\medskip\noindent{\em  \textbf{LV case}}

\medskip\noindent

In this case the volume is stabilised at large values of the order
$\mathcal{V}\sim 10^{15}$ which allows to solve the hierarchy
problem yielding TeV-scale supersymmetry naturally. Here are the
main cosmological features of these scenarios:

\begin{itemize}
\item The moduli spectrum includes a light field $\chi$ related to the overall volume.
This field is a source for the cosmological moduli problem (CMP)
as long as $M_s<10^{13}$ GeV, corresponding to
$\mathcal{V}>10^{10}$. There are two main possible solutions to
this CMP:
\begin{enumerate}
\item The light modulus $\chi$ gets diluted due to an increase in the entropy
that occurs when a short-lived modulus decays out of equilibrium
and while dominating the energy density of the Universe
\cite{Acharya, Vafa};

\item The volume modulus gets diluted due to a late period of low
energy inflation caused by thermal effects \cite{TI}.
\end{enumerate}
Assuming this problem is solved, the volume modulus becomes a dark
matter candidate (with a mass $m \sim 1$ MeV, if $\mathcal{V}\sim
10^{15}$) and its decay to $e^+ e^-$ could be one of the sources
that contribute to the observed $511$ KeV line, coming from the
centre of our galaxy.\footnote{However, recently it has been
discovered with the INTEGRAL spectrometer SPI \cite{INTEGRAL} that
the 511 KeV line emission appears to be asymmetric. This
distribution of positron annihilation resembles that of low mass
X-ray binaries, suggesting that these systems may be the dominant
origin of the positrons and so reducing the need for more exotic
explanations, such as the one presented in this paper.} The light
modulus $\chi$ can also decay into photons, producing a clean
monochromatic line that would represent a clear astrophysical
smoking-gun signal for these scenarios \cite{CQ}. We point out
that in the case of K3 fibrations, where the K3 fiber is
stabilised large, the spectrum of moduli fields includes an
additional light field. This field is also a potential dark matter
candidate with a mass $m \sim 10$ KeV, that could produce another
monochromatic line via its decay to photons.

\item At present, there are no known models of inflation in
LVS with intermediate scale $M_s$. However, the Fibre Inflation
model of chapter 8 can give rise to inflation for every value of
$\mathcal{V}$. The only condition, which fixes $\mathcal{V}\sim
10^3$, and so $M_s\sim M_{GUT}$, is the matching with the COBE
normalisation for the density fluctuations. Such a small value of
$\mathcal{V}$ is also necessary to have a very high inflationary
scale (close to the GUT scale) which, in turn, implies detectable
gravity waves. However, in principle it is possible that the
density perturbations could be produced by another scalar field
(not the inflaton), which is playing the r\^{o}le of a curvaton.
In such a case, one could be able to get inflation also for
$\mathcal{V}\sim 10^{15}$. In this way, both inflation and
TeV-scale supersymmetry would be achieved within the same model,
even though gravity waves would not be observable. It would be
interesting to investigate whether such scenarios are indeed
realisable.

\item As derived in section \ref{Sec:DecTemp}, if the volume is stabilised such that
$\mathcal{V} \sim 10^{15}$, the decompactification temperature is
rather low: $T_{max}\sim 10^{7}$ GeV.
\end{itemize}

\medskip\noindent{\em  \textbf{SV case}}

\medskip\noindent

In this case the volume is stabilised at smaller values of the
order $\mathcal{V}\sim 10^{4}$, which allows to reproduce the
standard picture of gauge coupling unification with $M_s\sim
M_{GUT}$. Here are the main cosmological features of these
scenarios:

\begin{itemize}
\item Given that in this case $\mathcal{V}<10^{13}$, all the
moduli have a mass $m>10$ TeV, and so they decay before Big-Bang
nucleosynthesis. Hence these scenarios are not plagued by any CMP.

\item As we have already pointed out in the LV case above, smaller
values of $\mathcal{V}$ more naturally give rise to inflationary
models, as the one presented in chapter 8. Here we observe that
the predictions for cosmological observables of Fiber Inflation
were sensitive to the allowed reheating temperature. Since for
$\mathcal{V}\sim 10^4$ GeV we have $T_{RH}^0<T_{max}\sim 10^{15}$
GeV and since in chapter 8 we considered already a more stringent
upper bound $T_{RH}^0<10^{10}$ GeV (in order to avoid thermal
gravitino overproduction), the presence of a maximal temperature
does not alter the predictions of that inflationary scenario.

\item Fixing the volume at small values of the order
$\mathcal{V} \sim 10^3$, the decompactification temperature turns
out to be extremely high: $T_{max}\sim 10^{15}$ GeV.
\end{itemize}

According to the discussion above, it would seem that cosmology
tends to prefer smaller values of $\mathcal{V}$. The reason is
that in the SV case there is no CMP and robust models of inflation
are known, whereas for $\mathcal{V}\sim 10^{15}$ the light modulus
suffers from the CMP and no model of inflation has been found yet.
Interestingly enough, the lower bound on $\mathcal{V}$ derived in
this chapter, suggests exactly the opposite. Namely, larger values
of $\mathcal{V}$ are favoured since, writing the volume as
$\mathcal{V}\sim 10^{x}$ and recalling the definition (\ref{c}) of
the parameter $c$, the constraint $T_*<T_{max}$ rules out a
relevant portion of the $(x,c)$-parameter space, that corresponds
to the SV case.

In view of this result, let us point out again that the LV case
has its advantages. For example, the decay of the light modulus
into $e^+ e^-$ could contribute to explain the origin of the $511$
KeV line. In addition, its decay to photons could produce a clean
smoking-gun signal of LVS. Furthermore, finding a realisation of
inflation, that is compatible with the LV case, is not necessarily
an unsurmountable problem. In that regard, let us note that the
authors of \cite{cklq} proposed a model, which relates the LV to
the SV case. More precisely, the inflaton is the volume modulus
and inflation takes place at a high scale for small values of
$\mathcal{V}$. Then after inflation the modulus ends up at a VEV
located at $\mathcal{V}\sim 10^{15}$, thus obtaining TeV scale
supersymmetry. However, as we have already mentioned above, it
could even be possible to realise inflation directly in the LV
case by appropriately modifying the Fibre Inflation scenario.

Now, even if inflation turns out not to be a problem for the LV
case, there is still the CMP due to the presence of the light
volume modulus. The results of this chapter pose a challenge for
the solution of this problem. Indeed, as we have shown in section
\ref{Sec:MaxTemp}, the CMP cannot be solved by diluting the volume
modulus via the entropy increase caused by the decay of the small
moduli. The reason is that the latter moduli decay before they can
begin to dominate the energy density of the Universe. So let us
now discuss in more detail the prospects of the other main
possible solution of the CMP in LVS, namely thermal inflation.

\newpage
\medskip\noindent{\em  \textbf{Thermal Inflation}}

\medskip\noindent

Thermal inflation has been studied in the literature from the
field theoretic point of view \cite{TI}. The basic idea is that a
field $\phi$, whose VEV is much larger than its mass (and so is
called `flaton') can be trapped by thermal corrections at a false
vacuum in the origin. At a certain temperature, its vacuum energy
density can start dominating over the radiation one, thus leading
to a short period of inflation. This period ends when the
temperature drops enough to destabilise the local minimum the
flaton was trapped in.

Since the flaton $\phi$ has to have a VEV $\langle\phi\rangle\gg
m_{\phi}$, it is assumed that the quartic piece in its potential
is absent. However in this way, the 1-loop thermal corrections
cannot trap the flaton in the origin because they go like:
\begin{equation}
V_T\sim T^2 m_{\phi}^2=T^2 \frac{d^2 V}{d \phi^2},
\end{equation}
and there is no quartic term in $V$ that would give rise to a term
like $T^2\phi^2$. Hence, it is usually assumed that there is an
interaction of the flaton with a very massive field, say a scalar
$\psi$, of the form $g\psi^2\phi^2$, where $g\sim 1$ so that
$\psi$ thermalises at a relatively low temperature. At this point,
a 1-loop thermal correction due to $\psi$ would give the required
term:
\begin{equation}
V_T\sim T^2 m_{\psi}^2=gT^2\phi^2.
\end{equation}
When $\phi$ gets a nonzero VEV, the interaction term
$g\psi^2\phi^2$ generates a mass term for $\psi$ of the order
$m_{\psi}\sim \langle\phi\rangle$. Hence, when $\phi$ is trapped
in the origin at high $T$, $\psi$ becomes very light. Close to the
origin, the potential looks like:
\begin{equation}
V=V_0+(g T^2-m_{\phi}^2)\phi^2+...\,,
\end{equation}
where $V_0$ is the height of the potential in the origin. A period
of thermal inflation takes place in the temperature window
$T_c<T<T_{in}$, where $T_{in}\sim V_0^{1/4}$ is the temperature at
which the flaton starts to dominate the energy density of the
Universe (beating the radiation energy density $\rho_r\sim T^4$)
and $T_c\sim m_{\phi}/g$ is the critical temperature at which the
flaton undergoes a phase transition rolling towards the $T=0$
minimum. The number of e-foldings of thermal inflation is given
by:
\begin{equation}
N_e\sim
\ln\left(\frac{T_{in}}{T_{c}}\right)\sim\ln\sqrt{\frac{\langle\phi\rangle}{m_{\phi}}}.
\end{equation}

Let us see how the above picture relates to the LVS. In the case
of $\mathcal{V}\sim 10^{15}$, the modulus $\tau_s$ has the right
mass scale and VEV to produce $N_e\sim 10$ e-foldings of
inflation, which would solve the CMP without affecting the density
perturbations generated during ordinary high-energy inflation.
However, in section \ref{Sec:ExplicitTPot} we derived the relevant
1-loop temperature corrections to the scalar potential and showed
that they are always subleading with respect to the $T=0$
potential, for temperatures below the Kaluza-Klein scale. Hence,
since thermal inflation requires the presence of new minima at
finite-temperature, we would be tempted to conclude that it does
not take place in the LVS. In fact, this was to be expected also
for the following reason. According to the field theoretic
arguments above, in order for thermal inflation to occur, it is
crucial that the flaton be coupled to a very massive field $\psi$.
However, in our model there is no particle, which is heavier than
the flaton candidate $\tau_s$. It is not so surprising, then, that
we are not finding thermal inflation.

Let us now discuss possible extensions of our model that could,
perhaps, allow for thermal inflation to occur, as well as the
various questions that they raise.

\begin{enumerate}
\item Since in our case $\tau_s$ is the candidate flaton field,
the necessary $\psi$ field would have a mass of the order
$m_{\psi}\sim\langle\tau_s\rangle M_s$, and so it is likely to be
a stringy mode. In such a case, it is not a priori clear how to
compute thermal corrections to $V_T$ due to the presence of $\psi$
in the thermal bath.

\item Even if we can compute $V_T$, it is not clear why these
corrections should trap $\tau_s$ at the origin. Note, however,
that this is not implausible, as the origin is a special point in
moduli space, where new states may become massless or the local
symmetry may get enhanced. Any such effect might turn out to play
an important r\^{o}le.

\item Even assuming that $V_T$ does trap $\tau_s$ in the origin,
one runs into another problem. Namely, the corresponding small
cycle shrinks below $M_s$ and so we cannot trust the low-energy
effective field theory (EFT). For a full description, we should go
to the EFT that applies close to the origin. The best known
examples of these are EFTs for blow-up fields at the actual
orbifold point. In addition, one should verify that $\mathcal{V}$
stays constant when the $\tau_s$ cycle shrinks to zero size.

\item When $\tau_s$ goes to zero, the field $\psi$ should
become massless, according to the comparison with the field
theoretic argument (if this comparison is valid). So possible
candidates for the role of the $\psi$ field could be winding
strings or $D1$-branes wrapping a 1-cycle of the collapsing
4-cycle.

\item If $\psi$ corresponds to a winding string, the interaction of the flaton
$\tau_s$ with $\psi$ cannot be seen in the EFT and it would be
very difficult to have a detailed treatment of this issue.

\item The field $\psi$ could also be a right handed neutrino,
or sneutrino, heavier than $\tau_s$. The crucial question would
still be if it would be possible to see $\psi$ in our EFT
description. In addition, one would need to write down $m_{\psi}$
as a function of $\tau_s$ and $\mathcal{V}$. It goes without
saying that this issue is highly dependent on the particular
mechanism for the generation of neutrino masses.

\item Besides the small modulus $\tau_s$, another possible flaton
candidate could be a localised matter field such as an open string
mode. However we notice that the main contribution to the scalar
potential of this field should come from $D$-terms, and that a
$D$-term potential usually gives rise to a mass of the same order
of the VEV. Hence it may be difficult to find an open string mode
with the typical behaviour of a flaton field.
\end{enumerate}
In general, all of the above open questions are rather difficult
to address. This poses a significant challenge for the derivation
of thermal inflation in LVS and the corresponding solution of the
CMP. However, let us note that the CMP could also be solved by
finding different models of low-energy inflation, which do not
rely on thermal effects.

\part{Conclusions and Outlook}

\chapter{Conclusions}
\label{seCON}

Let us conclude by summarising the results of this thesis and by
outlining the prospects for future work.

The thesis has been concerned with moduli stabilisation in IIB
string theory and its cosmological applications. The three
chapters of Part I were introductory. After motivating in chapter
1 the use of string theory as a framework for physics beyond the
Standard Model, in chapter 2 we focused on type IIB flux
compactifications and the study of the respective $N=1$ four
dimensional effective action in the case of Calabi-Yau
orientifolds. We then reviewed the use of background fluxes to
stabilise moduli, and in chapter 3 we presented a detailed survey
of all the main K\"{a}hler moduli mechanisms available in the
literature that use perturbative and non-perturbative corrections
beyond the leading order approximations.

Part II was concerned with developing a detailed understanding of
the very promising moduli stabilisation mechanism that goes under
the name of LARGE Volume Scenario. More precisely, in chapter 4,
we studied the topological conditions for general Calabi-Yaus to
get a non-supersymmetric exponentially large volume minimum of the
scalar potential in flux compactifications of IIB string theory.
We showed that negative Euler number and the existence of at least
one blow-up mode resolving point-like singularities are necessary
and sufficient conditions for moduli stabilisation with
exponentially large volumes.

In chapter 5, we then studied the behaviour of the string loop
corrections to the K\"{a}hler potential for general type IIB
compactifications. We gave a low energy interpretation for the
conjecture of Berg, Haack and Pajer for the form of the loop
corrections to the K\"{a}hler potential, checking the consistency
of this interpretation in several examples. We also showed that
for arbitrary Calabi-Yaus, the leading contribution of these
corrections to the scalar potential is always vanishing, giving an
``extended no-scale structure''. This result holds as long as the
corrections are homogeneous functions of degree $-2$ in the
2-cycle volumes. We used the Coleman-Weinberg potential to
motivate this cancellation from the viewpoint of low-energy field
theory. Finally we gave a simple formula for the 1-loop correction
to the scalar potential in terms of the tree-level K\"ahler metric
and the conjectured correction to the K\"ahler potential.

Chapter 6 then used these results in the study of K\"{a}hler
moduli stabilisation. The final picture is that, while the
combination of $\alpha'$ and nonperturbative corrections are
sufficient to stabilise blow-up modes and the overall volume,
quantum corrections are needed to stabilise other directions
transverse to the overall volume. This allows exponentially large
volume minima to be realised for fibration Calabi-Yaus, with the
various moduli of the fibration all being stabilised at
exponentially large values. String loop corrections may also play
an important r\^{o}le in stabilising 4-cycles which support chiral
matter and cannot enter directly into the non-perturbative
superpotential. We illustrated these ideas by studying the scalar
potential for various Calabi-Yau three-folds including K3
fibrations.

In Part III of this thesis, we discussed interesting cosmological
implications of the LARGE Volume Scenario. After the brief
introduction to string cosmology of chapter 7, in chapter 8 we
introduced a simple string model of inflation, in which the
inflaton field can take trans-Planckian values while driving a
period of slow-roll inflation. This leads naturally to a
realisation of large field inflation, inasmuch as the inflationary
epoch is well described by the single-field scalar potential $V =
V_0 \left( 3-4 e^{-\hat\varphi/\sqrt{3}}\right)$. Remarkably, for
a broad class of vacua all adjustable parameters enter only
through the overall coefficient $V_0$, and in particular do not
enter into the slow-roll parameters. Consequently these are
determined purely by the number of \efold ings, $N_e$, and so are
not independent: $\varepsilon \simeq \frac32 \eta^2$. This implies
similar relations among observables like the primordial
scalar-to-tensor amplitude, $r$, and the scalar spectral tilt,
$n_s$: $r \simeq 6(n_s - 1)^2$. $N_e$ is itself more
model-dependent since it depends partly on the post-inflationary
reheat history. In a simple reheating scenario, a reheating
temperature of $T_{rh}\simeq 10^{9}$ GeV gives $N_e\simeq 58$,
corresponding to $n_s\simeq 0.970$ and $r\simeq 0.005$, within
reach of future observations. The model is an example of a class
that arises naturally in the LARGE Volume Scenario, and takes
advantage of the generic existence there of K\"ahler moduli whose
dominant appearance in the scalar potential arises from string
loop corrections to the K\"{a}hler potential. The inflaton field
is a combination of K\"{a}hler moduli of a K3-fibered Calabi-Yau
manifold. We believe there are likely to be a great number of
models in this class -- `high-fibre models' -- in which the
inflaton starts off far enough up the fibre to produce observably
large primordial gravity waves.

Chapter 9 presented a detailed study of the finite-temperature
behaviour of the LARGE Volume Scenario. We showed that certain
moduli can thermalise at high temperatures. Despite that, their
contribution to the finite-temperature effective potential is
always negligible and the latter has a runaway behaviour. We
computed the maximal temperature $T_{max}$, above which the
internal space decompactifies, as well as the temperature $T_*$,
that is reached after the decay of the heaviest moduli. The
natural constraint $T_*<T_{max}$ implies a lower bound on the
allowed values of the internal volume ${\cal V}$. We found that
this restriction rules out a significant range of values
corresponding to smaller volumes of the order $\mathcal{V}\sim
10^{4}l_s^6$, which lead to standard GUT theories. Instead, the
bound favours values of the order $\mathcal{V}\sim 10^{15}l_s^6$,
which lead to TeV-scale supersymmetry desirable for solving the
hierarchy problem. Moreover, our result favours low-energy
inflationary scenarios with density perturbations generated by a
field, which is not the inflaton. In such a scenario, one could
achieve both inflation and TeV-scale supersymmetry, although
gravity waves would not be observable. Finally, we posed a
two-fold challenge for the solution of the cosmological moduli
problem. First, we showed that the heavy moduli decay before they
can begin to dominate the energy density of the Universe. Hence
they are not able to dilute any unwanted relics. And second, we
argued that, in order to obtain thermal inflation in the closed
string moduli sector, one needs to go beyond the present effective
field theory description.

The overall aim of my future work is to try to reach the goal of
building a true model where the moduli stabilisation and the
Standard Model building problems are solved simultaneously. At the
moment there is no example in the literature where these two
issues are successfully combined together. Given that this thesis
described a solid moduli stabilisation mechanism, I would like to
focus mainly on the model building part. However, this should be
done always with the help of a global perspective, since it will
be very important to check if the solutions to these two issues
are effectively decoupled, and so can be consistently studied
separately.

My general plan would be to produce, as I was mentioning before, a
comprehensive model where there is a mathematically rigorous
description of the compact Calabi-Yau background and both the
moduli stabilisation and the up-lifting procedure is well under
control. In addition, there is a localised Standard Model-like
construction with chiral matter via $D$-brane constructions, the
main hierarchies in Nature are explained and a clear spectrum of
soft supersymmetry breaking masses is derived. The model should,
at the same time, give rise to interesting cosmology and
astrophysics, being able to describe the inflationary and
reheating era, with also good dark matter candidates.

Finally two major experiments in particle physics and cosmology
are going to be performed, since PLANCK has just been launched in
May 2009 and the LHC is going to start operation rather soon.
Therefore my plan to focus on model building embedded in the
robust moduli stabilisation mechanism developed in this thesis,
becomes even more important. In fact, it is only with the help of
a detailed model building that one is able to derive from string
theory as many as possible theoretical predictions, that could be
put to experimental test via LHC or PLANCK. This is an absolutely
fundamental task because string theory was born about 40 years ago
and it has not been put to experimental test yet. I really look
forward to knowing the outcomes of these two main experiments in
particle physics and cosmology, being ready to orient my research
activity according to whatever signals of new physics could come
out.

A more detailed plan for future lines of work could be summarised
with the help of the following broad points:
\begin{itemize}
\item I would like to focus on an attempt to find explicit Calabi-Yau
realisations of LVS via hypersurfaces embedded in toric varieties.
This would be important both to render the models described in
this thesis, mathematically more rigorous and to explicitly show
their existence. Moreover, this work is extremely important since
it would set the basis for the first step towards the answer of a
fundamental question in string phenomenology: the realisation of a
local Standard Model-like construction within a \textit{compact}
Calabi-Yau with all moduli stabilised.

\item Having completed this project, I will be in possession of a
systematic study of moduli stabilisation for large classes of
compact Calabi-Yau manifolds with a well-defined mathematical
description. At this point, the natural thing to do, would be to
try to combine these results with the Standard Model-like
constructions which are, at the moment, available only for
non-compact Calabi-Yaus.

\item I will also try to address two main issues in string
cosmology. The first one is the complicated task of finding a
potential whose scale is able to give rise to inflation and
TeV-scale supersymmetry at the same time. The second one is the
fact that there are only a few string inflationary models in which
inflation is driven by the inflaton but the main contribution to
the generation of density perturbations comes from another field.
Both of these issues could be addressed at the same time by
improving the Fibre Inflation model described in this thesis. In
fact, one could build a curvaton scenario where the perturbations
are not generated by the inflaton, but by another modulus which
plays the r\^{o}le of the curvaton \cite{curvaton}. The scale of
the potential could be set such that TeV-scale supersymmetry is
achieved and the model would predict large non-gaussianities in
the CMB spectrum \cite{NGcurvaton}. This is very important to make
contact with experiments since the PLANCK satellite is probably
going to give an observational answer to the fundamental question
of the existence of large non-gaussianities in the CMB.

\item I would also try to explore if the inclusion of string loop
corrections to the K\"{a}hler potential allows me, as it suggests,
to perform a stringy derivation of the famous ADD scenario
\cite{add} in the case of a K3-fibered Calabi-Yau. In fact, the
moduli could be fixed in such a way to obtain an highly asymmetric
shape of the six extra-dimensions where two of them are larger
than the other four \cite{cliff}. Thus one obtains a
six-dimensional effective field theory where the radius of the two
extra dimensions could be of the order $0.1$ mm without violating
the present bounds on Newton's gravity law \cite{TestGravity}. The
exploration of the phenomenological implications of these
TeV-scale stringy scenarios could be very interesting in
connection with LHC data.
\end{itemize}

\begin{appendix}

\chapter[Appendix]{}

\section{Proof of the LARGE Volume Claim}
\label{Appendix A}

Let us now present a comprehensive argument in favour of the LARGE
Volume Claim which establishes the existence of LARGE Volumes in
IIB string compactifications.

\subsection{Proof for $N_{small}=1$}
\label{Appendix A1}

\textbf{Proof.} (LARGE Volume Claim for $N_{small}=1$) Let us
start from the scalar potential (\ref{scalar}) which we now
rewrite as:
\begin{equation}
V=\delta V_{(np1)}+\delta V_{(np2)}+\delta V_{\left( \alpha'
\right) },
\end{equation}
and perform the large volume limit as described in (\ref{limit})
with $N_{small}=1$ corresponding to $\tau_{1}$. In this limit
$\delta V_{\left( \alpha'\right) }$ behaves as:
\begin{equation}
\delta V_{\left( \alpha' \right) }\underset{\mathcal{V}\rightarrow
\infty }{ \longrightarrow }+\frac{3\hat{\xi}
}{4\mathcal{V}^{3}}e^{K_{cs}}\left\vert W\right\vert
^{2}+\mathcal{O}\left( \frac{1}{\mathcal{V}^{4}}\right) .
\label{Alfa}
\end{equation}
We also point out that:
\begin{equation}
e^{K}\underset{\mathcal{V}\rightarrow \infty }{\longrightarrow
}\frac{ e^{K_{cs}}}{\mathcal{V}^{2}}+\mathcal{O}\left(
\frac{1}{\mathcal{V}^{3}} \right) .  \label{Eallak}
\end{equation}
Let us now study $\delta V_{(np1)}$ which reduces to:
\begin{equation}
\delta V_{(np1)}\underset{\mathcal{V}\rightarrow \infty
}{\longrightarrow} e^{K}K_{11}^{-1}a_{1}^{2}\left\vert
A_{1}\right\vert^{2}e^{-a_{1}\left(T_{1}+\bar{T}_{1}\right)
}=\frac{K_{11}^{-1}}{\mathcal{V}^{2}}a_{1}^{2}\left\vert
A_{1}\right\vert^{2}e^{-2 a_{1}\tau_{1}}. \label{Imp}
\end{equation}
Switching to the study of $\delta V_{(np2)}$, we find that:
\begin{equation}
\delta V_{(np2)}\underset{\mathcal{V}\rightarrow \infty
}{\longrightarrow } -e^{K}\sum\limits_{k=1}^{h_{1,1}}K_{1k}^{-1}
\left[ \left( a_{1}A_{1}e^{-a_{1}\tau
_{1}}e^{-ia_{1}b_{1}}\bar{W}\partial _{ \bar{T}_{k}}K\right)
+\left( a_{1}\bar{A}_{1}e^{-a_{1}\tau
_{1}}e^{+ia_{1}b_{1}}W\partial _{T_{k}}K\right) \right] ,
\label{Jjj}
\end{equation}
where we have used the fact that $K_{1k}^{-1}=K_{k1}^{-1}$.
Equation (\ref{Jjj}) can be rewritten as:
\begin{eqnarray}
&&\delta V_{(np2)}\underset{\mathcal{V}\rightarrow \infty
}{\longrightarrow}-e^{K}\sum\limits_{k=1}^{h_{1,1}}K_{1k}^{-1}\left(
\partial _{T_{k}}K\right)a_{1} e^{-a_{1}\tau _{1}}\left[ \left(
A_{1}\bar{W}e^{-ia_{1}b_{1}}\right) +\left(
\bar{A}_{1}We^{+ia_{1}b_{1}}\right) \right]  \notag \\
&=&\left(
X_{1}e^{+ia_{1}b_{1}}+\bar{X}_{1}e^{-ia_{1}b_{1}}\right),
\end{eqnarray}
where:
\begin{equation}
X_{1}\equiv -e^{K}K_{1k}^{-1}\left( \partial _{T_{k}}K\right)
a_{1}\bar{A}_{1}W e^{-a_{1}\tau _{1}}.  \label{Def1}
\end{equation}
We note that for a general Calabi-Yau, the following relation
holds:
\begin{equation}
K_{1k}^{-1}\left( \partial _{T_{k}}K\right) =-2\tau _{1},
\end{equation}
and thus the definition (\ref{Def1}) can be simplified to:
\begin{equation}
X_{1}\equiv 2e^{K}a_{1}\tau _{1}\left\vert A_{1}\right\vert e^{-i
\vartheta_{1}}\left\vert W \right\vert e^{i \vartheta_{W}}
e^{-a_{1}\tau _{1}}=\left\vert X_{1}\right\vert e^{i(\vartheta
_{W}-\vartheta_{1})}. \label{DDEF1}
\end{equation}
Therefore:
\begin{equation}
\delta V_{(np2)}\underset{\mathcal{V}\rightarrow \infty
}{\longrightarrow } \left\vert X_{1}\right\vert \left( e^{+i\left(
\vartheta_{W}-\vartheta _{1}+a_{1}b_{1}\right) }+e^{-i\left(
\vartheta_{W}-\vartheta _{1}+a_{1}b_{1}\right) }\right)
=2\left\vert X_{1}\right\vert \cos \left( \vartheta_{W}-\vartheta
_{1}+a_{1}b_{1}\right).
\end{equation}
$\delta V_{(np2)}$ is a scalar function of the axion $b_{1}$
whereas $\vartheta_{1}$ and $\vartheta_{W}$ are to be considered
just as parameters. In order to find a minimum for $\delta
V_{(np2)}$ let us set its first derivative to zero:
\begin{equation}
\frac{\partial \left(\delta V_{(np2)}\right)}{\partial
b_{1}}=-2a_{1} \left\vert X_{1}\right\vert \sin
(\vartheta_{W}-\vartheta _{1}+a_{1}b_{1})=0. \label{Ggg}
\end{equation}
The solution of (\ref{Ggg}) is given by:
\begin{equation}
a_{1}b_{1}=p_{1}\pi +\vartheta _{1}-\vartheta_{W} ,\textrm{ \
}p_{1}\in \mathbb{Z}. \label{VVEV}
\end{equation}
We have still to check the sign of the second derivative evaluated
at $b_{1}$ as given in (\ref{VVEV}) and require it to be positive:
\begin{equation}
\frac{\partial^{2} \left(\delta V_{(np2)}\right)}{\partial
b_{1}^{2}} =-2 a_{1}^{2} \left\vert X_{1}\right\vert \cos
(\vartheta_{W}-\vartheta _{1}+a_{1}b_{1})>0  \Longleftrightarrow
p_{1}\in 2\mathbb{Z}+1. \label{secder}
\end{equation}
Thus we realise that at the minimum:
\begin{equation}
\delta V_{(np2)} =-2 \left\vert X_{1}\right\vert=-2 \left\vert
W\right\vert \left\vert A_{1}\right\vert a_{1}\tau _{1}
\frac{e^{-a_{1}\tau_{1}}}{\mathcal{V}^{2}}.  \label{Fff}
\end{equation}
We notice that the phases of $W$ and $A_{1}$ do not enter into
$\delta V_{(np2)}$ once the axion has been properly minimised and
so, without loss of generality, we can consider $W$ and $A_{1}\in
\mathbb{R} ^{+}$ from now on.

We may now study the full potential by combining equations
(\ref{Alfa}), (\ref{Imp}) and (\ref{Fff}):
\begin{equation}
V \simeq \frac{K^{-1}_{11}}{\mathcal{V}^{2}} A_{1}^{2}a_{1}^{2}
e^{-2a_{1}\tau_{1}}-\frac{W_{0}}{\mathcal{V}^{2}}A_{1}a_{1}\tau
_{1}e^{-a_{1}\tau_{1}}+\frac{\hat{\xi}
}{\mathcal{V}^{3}}W_{0}^{2}, \label{final potenzial}
\end{equation}
where we have substituted $W$ with its tree-level expectation
value $W_{0}$ because the non-perturbative corrections are always
subleading by a power of $\mathcal{V}$. Moreover, we have dropped
all the factors since they are superfluous for our reasoning.

We would like to emphasise that we know that the first term in
(\ref{final potenzial}) is indeed positive. In fact it comes from:
\begin{equation}
K_{11}^{-1}(\partial _{1}W)(\partial _{1}\bar{W}),
\end{equation}
and we know that the K\"{a}hler matrix is positive definite since
it gives rise to the kinetic terms. Moreover, as we have just
seen, the second term in (\ref{final potenzial}) comes from the
axion minimisation as so is definitely negative. Only the sign of
$\delta V_{\left( \alpha'\right) }$ is in principle unknown, but
the condition $h_{2,1}(X)>h_{1,1}(X)$ ensures that it is positive.
This condition will turn out to be crucial in showing that the
volume direction has indeed a minimum at exponentially large
volume.

We need now to study the form of $K_{11}^{-1}$. For a general
Calabi-Yau, the inverse K\"{a}hler matrix with $\alpha'$
corrections included, reads \cite{bobkov}:
\begin{equation}
K_{ij}^{-1}=-\frac{2}{9}\left(2\mathcal{V}+\hat{\xi}\right)k_{ijk}t^{k}+
\frac{4\mathcal{V}-\hat{\xi}}{\mathcal{V}-\hat{\xi}}\tau _{i}\tau
_{j}, \label{inversaalfa}
\end{equation}
which at large volume becomes:
\begin{equation}
K_{ij}^{-1}=-\frac{4}{9}\mathcal{V}k_{ijk}t^{k}+4\tau _{i}\tau
_{j}+\left( \textrm{terms subleading in }\mathcal{V}\right).
\label{inversa}
\end{equation}
Hence we can classify the behaviour of $K_{11}^{-1}$ depending on
the volume dependence of the quantity $k_{11j}t^{j}$ and find 4
different cases:

\begin{enumerate}
\item $k_{11j}t^{j}=0\textrm{ \ or \ }k_{11j}t^{j}\simeq\frac
{\tau _{1}^{1/2+3\alpha/2}}{\mathcal{V}^{\alpha }},$ $\alpha \geq
1$ $\Longrightarrow K_{11}^{-1}\simeq \tau _{1}^{2},$

\item $k_{11j}t^{j}=\frac{\tau _{1}^{1/2+3\alpha/2}}{\mathcal{V}^{\alpha }},$
$0<\alpha<1$ $\Longrightarrow K_{11}^{-1}\simeq\mathcal{V}^{\alpha
}\tau _{1}^{2-3\alpha /2},$ $0<\alpha <1,$

\item $k_{11j}t^{j}\simeq\sqrt{\tau _{1}}$ $\Longrightarrow
K_{11}^{-1}\simeq \mathcal{V}\sqrt{\tau _{1}},$

\item $k_{11j}t^{j}\simeq\mathcal{V}^{\alpha }\tau _{1}^{1/2-3\alpha/2},$
$\alpha >0$ $\Longrightarrow K_{11}^{-1}\simeq\mathcal{V}^{\alpha
}\tau _{1}^{2-3\alpha /2},$ $\alpha >1.$
\end{enumerate}

One could wonder why we are setting the conditions of the Claim on
the elements of the inverse K\"{a}hler matrix and not on the
intersection numbers or the form of the overall volume of the
Calabi-Yau from which it would be easier to understand their
topological meaning. The reason is that it is the inverse
K\"{a}hler matrix which enters directly with the superpotential
into the form of the scalar potential which is the one that
determines the physics.

Moreover, the Claim applies if the superpotential has the
expression (\ref{explicit}), but in this case we can still make
linear field redefinitions that will not change $W$, corresponding
to proper changes of basis, of the form:
\begin{equation}
\left\{
\begin{array}{c}
\tau _{j}\longrightarrow \tau _{j}'=\tau _{j},\textrm{ }\forall
j=1,...,N_{small}, \\
\tau _{j}\longrightarrow \tau
_{j}'=\tau_{j}+g_{1}(\tau_{i}),\textrm{ }\forall
j=N_{small}+1,...,h_{1,1}(X),
\end{array}
\right. \label{ChangeBasis}
\end{equation}
where $g_{1}(\tau_{i})$, $i=1,...,N_{small},$ is an homogeneous
function of degree 1. This means that the small 4-cycles will stay
small and the large ones will just be perturbed by the small ones.
We are therefore in the same situation and the physics should not
change. We conclude that the inverse K\"{a}hler matrix should not
change but both the intersection numbers and the form of the
volume can indeed vary. In fact, for an arbitrary Calabi-Yau, the
elements of the inverse K\"{a}hler matrix are given by:
\begin{equation}
K_{ij}^{-1}=-\frac{4}{9}\mathcal{V}k_{ijk}t^{k}+4\tau _{i}\tau
_{j}, \label{inversaAlfa}
\end{equation}
and so we see that in order to keep the form of $K^{-1}_{ij}$
unaltered, the quantity $(k_{ijk}t^{k})$ has not to vary, but the
intersection numbers $k_{ijk}$ can indeed change. This is the main
reason why we need to put our conditions on the $K^{-1}_{ij}$.

Let us illustrate this statement in the explicit example of the
orientifold of the Calabi-Yau threefold
$\mathbb{C}P^{4}_{[1,1,1,6,9]}(18)$ whose volume in terms of
2-cycle volumes is given by:
\begin{equation}
\mathcal{V}=6\left(t_{5}^{3}+t_{4}^{3}\right).
\end{equation}
The corresponding 4-cycle volumes look like:
\begin{equation}
\left\{
\begin{array}{c}
\tau _{4}=\frac{\partial \mathcal{V}}{\partial t_{4}}=18t_{4}^{2}, \\
\tau _{5}=\frac{\partial \mathcal{V}}{\partial t_{5}}=18t_{5}^{2},
\end{array}
\right. \textrm{ \ }\Longleftrightarrow \textrm{ \ }\left\{
\begin{array}{c}
t_{4}=-\frac{\sqrt{\tau _{4}}}{3\sqrt{2}}, \\
t_{5}=+\frac{\sqrt{\tau _{5}}}{3\sqrt{2}},
\end{array}
\right.
\end{equation}
and the volume in terms of the 4-cycles is:
\begin{equation}
\mathcal{V}=\frac{1}{9
\sqrt{2}}\left(\tau_{5}^{3/2}-\tau_{4}^{3/2}\right).
\label{initial form}
\end{equation}
Finally the superpotential reads:
\begin{equation}
W=W_{0}+A_{4}e^{-a_{4}T_{4}}+A_{5}e^{-a_{5}T_{5}}.
\end{equation}
It exists a well defined large volume limit when the 4-cycle
$\tau_{4}$ is kept small and $\tau_{5}$ is sent to infinity. In
this case the superpotential can be approximated as:
\begin{equation}
W\simeq W_{0}+A_{4}e^{-a_{4}T_{4}}. \label{form of W}
\end{equation}
We can now perform the following field redefinition:
\begin{equation}
\left\{
\begin{array}{c}
\tau _{4}\longrightarrow \tau _{4}'=\tau _{4}, \\
\tau _{5}\longrightarrow \tau _{5}'=\tau _{5}+\tau_{4},
\end{array}
\right. \label{redefy}
\end{equation}
which will not change the form of $W$ (\ref{form of W}). However
now the volume reads:
\begin{equation}
\mathcal{V}'=\frac{1}{9
\sqrt{2}}\left((\tau_{5}'-\tau_{4}')^{3/2}-\tau_{4}'^{3/2}\right)\simeq
\frac{1}{9
\sqrt{2}}\left(\tau_{5}'^{3/2}-\tau_{4}'\sqrt{\tau_{5}'}-\tau_{4}'^{3/2}\right),
\label{volumeprime}
\end{equation}
which is clearly different from the initial form (\ref{initial
form}). This means that also the intersection numbers are
different. However the elements of the inverse K\"{a}hler matrix
do not change. In particular we are interested in
$K^{-1}_{44}\simeq \mathcal{V}\sqrt{\tau_{4}}$ in this case as
$\tau_{4}$ is the small cycle. Its form stays unchanged since
$K'^{-1}_{44}\simeq \mathcal{V'}\sqrt{\tau_{4}'}$. From
(\ref{inversaAlfa}), this implies that:
\begin{equation}
\sqrt{\tau_{4}'}=(k_{44k}'t'^{k})=k_{444}'t'_{4}+k_{445}'t'_{5},
\end{equation}
and one would tend to say that $k_{445}'$ has to be zero but we
know from (\ref{volumeprime}) that this is definitely not the
case. This means that the field redefinition (\ref{redefy}) will
have the corresponding redefinition of the 2-cycle volumes which
will produce $t_{4}'$ and $t_{5}'$ that are both large 2-cycles
but such that the combination
$\left(k_{444}'t'_{4}+k_{445}'t'_{5}\right)$ stays small. This is
the reason why the form of the inverse K\"{a}hler matrix is left
invariant while the intersection numbers do vary. This can be
rephrased by saying that if $\tau_{j}$ is a small 4-cycle, in
general the corresponding $t_{j}$ has not to be a small 2-cycle
and viceversa. This is clear without the need to perform any field
redefinition in the case of the Calabi-Yau K3 fibration described
by the degree 12 hypersurface in $\mathbb{C}P^{4}_{[1,1,2,2,6]}$
whose overall volume in terms of 2-cycle volumes is:
\begin{equation}
\mathcal{V}=t_{1}t_{2}^{2}+\frac{2}{3}t_{2}^{3},
\end{equation}
giving relations between the 2- and 4-cycle volumes: \bea
\label{tay} \tau _{1}=t_{2}^{2}, & \qquad &
\tau_{2}=2t_{2}\left(t_{1}+t_{2}\right),
\nonumber \\
t_{2}=\sqrt{\tau _{1}}, & \qquad & t_{1}=\frac{\tau _{2}-2\tau
_{1}}{2\sqrt{\tau _{1}}}, \label{viceversa} \eea that allow us to
write:
\begin{equation}
\mathcal{V}=\frac{1}{2}\sqrt{\tau _{1}}\left( \tau
_{2}-\frac{2}{3}\tau _{1}\right) .  \label{volu11226}
\end{equation}
Looking at (\ref{volu11226}) we see that the large volume limit
can be performed keeping $\tau_{1}$ small and taking $\tau_{2}$
large. Nonetheless, as it is clear from (\ref{viceversa}), $t_{1}$
is big whereas $t_{2}$ is small. Therefore it is impossible to
impose that the quantity $k_{jji}t^{i}$ does not introduce any
volume dependence by requiring that some intersection numbers have
to vanish.

Going back to the proof of the LARGE Volume Claim for
$N_{small}=1$, let us assume that we are in case (3), so that
(\ref{final potenzial}) becomes:
\begin{equation}
V \simeq \frac{\sqrt{\tau_{1}}}{\mathcal{V}} A_{1}^{2}a_{1}^{2}
e^{-2a_{1}\tau_{1}} -\frac{W_{0}}{\mathcal{V}^{2}}A_{1}a_{1}\tau
_{1} e^{-a_{1}\tau_{1}} +\frac{\hat{\xi}
}{\mathcal{V}^{3}}W_{0}^{2}, \label{Finale potenzial}
\end{equation}
and when we take the decompactification limit given by:
\begin{equation}
\mathcal{V}\rightarrow \infty \ \ \textrm{with}\ \ e^{a_{1}\tau
_{1}}=\frac{\mathcal{V}}{W_{0}}, \label{VEry}
\end{equation}
all the terms in (\ref{Finale potenzial}) have the same volume
dependence:
\begin{equation}
V \simeq \frac{W_{0}^{2}}{\mathcal{V}^{3}}\left[ \left(A_{1}a_{1}
-\sqrt{\tau _{1}}\right)A_{1}a_{1}\sqrt{\tau _{1}}+\hat{\xi}
\right]. \label{Jkj}
\end{equation}
We can finally express the scaling behaviour of (\ref{Jkj}) as:
\begin{equation}
V\simeq \frac{W_{0}^{2}}{\mathcal{V}^{3}}\left( C_{1} \sqrt{\ln
\mathcal{V}}-C_{2} \ln \mathcal{V}+\hat{\xi} \right),
\label{Chiave}
\end{equation}
where $C_{1}$ and $C_{2}$ are positive constants of order 1 for
natural values of the parameter $A_{1}\simeq 1$. We conclude that
at large volume, the dominant term in (\ref{Chiave}) is the second
one and the scalar potential approaches zero from below. It is now
straightforward to argue that there must exist an exponentially
large volume AdS minimum.

In fact, at smaller volumes the dominant term in the potential
(\ref{Chiave}) is either the first or the third term, depending on
the exact value of the constants. Both are positive as we have
explained above. Thus at smaller volumes the potential is
positive, and so since it must go to zero at infinity from below,
there must exist a local AdS minimum along the direction in
K\"{a}hler moduli space where the volume changes.

One could argue that if at smaller volumes the dominant term in
(\ref{Chiave}) is the first one, then there is no need to require
$h_{2,1}(X)>h_{1,1}(X)$. In reality this is wrong, because $\xi
<0$ could still ruin the presence of the large volume minimum. In
fact we can rewrite the full scalar potential (\ref{Finale
potenzial}) as:
\begin{equation}
V=\frac{\lambda }{\mathcal{V}}\sqrt{\tau _{1}}e^{-2a_{1}\tau
_{1}}-\frac{\mu }{\mathcal{V}^{2}}\tau _{1}e^{-a_{1}\tau
_{1}}+\frac{\hat{\xi} W_{0}^{2}}{\mathcal{V}^{3}},  \label{Asfk}
\end{equation}
where $\lambda $, $\mu $ and $\nu $ are positive constants
depending on the exact details of this model. We can integrate out
$\tau _{1}$, so ending up with just a potential for $\mathcal{V}$.
Under the requirement $a_{1}\tau _{1}\gg 1$, $\partial V /
\partial \tau _{1}=0$ gives:
\begin{equation}
e^{-a_{1}\tau _{1}}=\frac{\mu }{2}\frac{\sqrt{\tau _{1}}}{\lambda
\mathcal{V}},
\end{equation}
which substituted back in (\ref{Asfk}) yields:
\begin{equation}
V=-\frac{1}{2}\frac{\mu ^{2}}{\lambda }\frac{\tau
_{1}^{3/2}}{\mathcal{V}^{3}}+\frac{\hat{\xi}
W_{0}^{2}}{\mathcal{V}^{3}}\sim \frac{-\left( \ln \mathcal{V}
\right) ^{3/2}+\hat{\xi} W_{0}^{2}}{\mathcal{V}^{3}}
\end{equation}
and is straightforward to see that we need $\hat{\xi} >0$ even
though the dominant term at small volumes in (\ref{Asfk}) is the
first one.

It remains to show that the scalar potential has also a minimum in
the other direction of the moduli space. In order to do that, let
us fix the Calabi-Yau volume and see what happens if we vary the
small K\"{a}hler modulus along that surface. Then as one
approaches the walls of the K\"{a}hler cone the positive first
term in (\ref{Finale potenzial}) dominates since it has the fewest
powers of volume in the denominator and the exponential
contributions of the modulus that is becoming small cannot be
neglected. Thus at large overall volume, we expect the potential
to grow in the positive direction towards the walls of the
K\"{a}hler cone.

On the other hand, when the small K\"{a}hler modulus becomes
bigger then the dominant term in (\ref{Finale potenzial}) is the
positive $\delta V_{(\alpha')}$ due to the exponential
suppressions in the other two terms. Given that the potential is
negative along the special direction in the moduli space that we
have identified and eventually raises to be positive or to vanish
in the other direction, we are sure to have an AdS exponentially
large volume minimum.

Since $V\sim \mathcal{O}(1/\mathcal{V}^{3})$ at the minimum, while
$-3e^{K}\left\vert W\right\vert ^{2}\sim
\mathcal{O}(1/\mathcal{V}^{2})$, it is clear that this minimum is
non-supersymmetric. We can heuristically see why the minimum we
are arguing for can be at exponentially large volume. The naive
measure of its location is the value of the volume at which the
negative term in (\ref{Chiave}) becomes dominant. As this occurs
only when $(\ln \mathcal{V})$ is large, we expect to find the
vacuum at large values of $(\ln \mathcal{V})$.

In reality the way in which we have taken the limit (\ref{VEry}),
tells us how the volume will scale, even though this can very well
not be the correct location of the minimum:
\begin{equation}
\mathcal{V}\sim W_{0} e^{a_{1}\tau _{1}}. \label{Hyu}
\end{equation}
Looking at (\ref{Hyu}) we realise that $W_{0}$ cannot be too
small, otherwise we would get a small volume minimum merging with
the KKLT one and our derivation would not make sense anymore.
However $W_{0}$ is multiplying an exponential, which means that in
order to destroy the large volume minimum $W_{0}$ has to be really
small.

Furthermore, we stress that there is no need to require
$h_{2,1}(X) \gg h_{1,1}(X)$ instead of just
$h_{2,1}(X)>h_{1,1}(X)$. In fact, in this proof we have used
$\hat{\xi}$ instead of $\xi$, so obscuring the presence of any
factors of $g_{s}$ but, as it is written explicitly in
(\ref{explicit}), in Einstein frame $\hat{\xi}$ is equivalent to
$\xi / g_{s}^{3/2}$. Therefore if we just have
$h_{2,1}(X)>h_{1,1}(X)$ then we can still adjust $g_{s}$ to make
sure that the AdS minimum is indeed at large volume.

We are now able to understand what happens if $K^{-1}_{11}$ is not
in case (3). For example, when it is in case (4) then the first
term in (\ref{final potenzial}) beats all the other ones and along
the direction (\ref{VEry}) the scalar potential either presents a
runaway or has no minimum at large volume depending on the exact
value of $\alpha$.

Moreover if $K^{-1}_{11}$ is in case (1) or (2) then the first
term in (\ref{final potenzial}) is subleading with respect to the
other two and at leading order in the volume, the scalar potential
looks like:
\begin{equation}
V \simeq -\frac{W_{0}}{\mathcal{V}^{2}}A_{1}a_{1}\tau
_{1}e^{-a_{1}\tau_{1}}+\frac{\hat{\xi}
}{\mathcal{V}^{3}}W_{0}^{2}. \label{POt}
\end{equation}
The minimisation equation for $\tau_{1}$, $\partial V/\partial
\tau_{1}=0$, admits the only possible solution $a_{1}\tau_{1}=1$,
that has to be discarded since we need $a_{1}\tau_{1}\gg 1$ in
order to avoid higher instanton corrections.

Finally, let us argue in favour of the last statement of the LARGE
Volume Claim. At the end of all our derivation we realised that
the small K\"{a}hler modulus $\tau_{1}$ plus a particular
combination which is the overall volume are stabilised. Therefore
we have in general $(N_{small}+1)$ fixed K\"{a}hler moduli and is
straightforward to see that if we have just one big K\"{a}hler
modulus then it will be fixed, whereas if we have more than one
big K\"{a}hler moduli, only one of them will be fixed and the
others will give rise to exactly $(h_{1,1}(X)-N_{small}-1)$ flat
directions. This is because they do not appear in the
non-perturbative corrections to the superpotential due to the
limit (\ref{limit}). This terminates our proof of the LARGE Volume
Claim for $N_{small}$=1.

\subsection{Proof for $N_{small}>1$}
\label{Appendix A2}

\textbf{Proof.} (LARGE Volume Claim for $N_{small}>1$) When
$N_{small}>1$ the situation is more involved due to the presence
of cross terms. However $\delta V_{\left( \alpha'\right) }$ has
still the form (\ref{Alfa}). Without loss of generality, we shall
focus on the case with $N_{small}=2$ K\"{a}hler moduli, which we
will call $\tau_{1}$ and $\tau_{2}$. $\delta V_{(np1)}$
generalises to:
\begin{eqnarray}
&&\delta V_{(np1)}\underset{\mathcal{V}\rightarrow \infty
}{\longrightarrow}
e^{K}\sum\limits_{j,k=1}^{2}K_{jk}^{-1}a_{j}A_{j}a_{k}\bar{A}
_{k}e^{-\left( a_{j}T_{j}+a_{k}\bar{T}_{k}\right) } \label{imp} \\
&=&e^{K}\left\{
\sum\limits_{j=1}^{2}K_{jj}^{-1}a_{j}^{2}\left\vert
A_{j}\right\vert ^{2}e^{-2a_{j}\tau
_{j}}+K_{12}^{-1}a_{1}A_{1}a_{2}\bar{A}_{2}e^{-\left( a_{1}\tau
_{1}+a_{2}\tau _{2}\right) }e^{i\left(
a_{2}b_{2}-a_{1}b_{1}\right) }\right\}.  \nonumber
\end{eqnarray}
In order to consider separately the axion-dependent part of
$\delta V_{(np1)}$, we write:
\begin{equation}
\delta V_{(np1)}=\delta V_{(np1)}^{real}+\delta V_{(np1)}^{AX}.
\end{equation}
Switching to the study of $V_{np2}$, we find that:
\begin{equation}
\delta V_{(np2)}\underset{\mathcal{V}\rightarrow \infty
}{\longrightarrow }
-e^{K}\sum\limits_{k=1}^{h_{1,1}}\sum\limits_{j=1}^{2}K_{jk}^{-1}
\left[ \left( a_{j}A_{j}e^{-a_{j}\tau
_{j}}e^{-ia_{j}b_{j}}\bar{W}\partial _{ \bar{T}_{k}}K\right)
+\left( a_{j}\bar{A}_{j}e^{-a_{j}\tau
_{j}}e^{+ia_{j}b_{j}}W\partial _{T_{k}}K\right) \right] ,
\label{JJjh}
\end{equation}
where we have used the fact that $K_{jk}^{-1}=K_{kj}^{-1}$.
Equation (\ref{JJjh}) can be rewritten as:
\begin{eqnarray}
\delta V_{(np2)}\underset{\mathcal{V}\rightarrow \infty
}{\longrightarrow}&-&e^{K}\sum\limits_{k=1}^{h_{1,1}}\sum
\limits_{j=1}^{2}K_{jk}^{-1}\left( \partial _{T_{k}}K\right)a_{j}
e^{-a_{j}\tau _{j}}\left[ \left(
A_{j}\bar{W}e^{-ia_{j}b_{j}}\right) +\left(
\bar{A}_{j}We^{+ia_{j}b_{j}}\right) \right]  \notag \\
&=&\sum\limits_{j=1}^{2}\left(
X_{j}e^{+ia_{j}b_{j}}+\bar{X}_{j}e^{-ia_{j}b_{j}}\right) ,
\end{eqnarray}
where:
\begin{equation}
X_{j}\equiv -e^{K}K_{jk}^{-1}\left( \partial _{T_{k}}K\right)
a_{j}\bar{A}_{j}W e^{-a_{j}\tau _{j}}.  \label{defin1}
\end{equation}
We note that for a general Calabi-Yau, the following relation
holds:
\begin{equation}
K_{jk}^{-1}\left( \partial _{T_{k}}K\right) =-2\tau _{j},
\end{equation}
and thus the definition (\ref{defin1}) can be simplified to:
\begin{equation}
X_{j}\equiv 2e^{K}a_{j}\tau _{j}\left\vert A_{j}\right\vert e^{-i
\vartheta_{j}}\left\vert W \right\vert e^{i \vartheta_{W}}
e^{-a_{j}\tau _{j}}=\left\vert X_{j}\right\vert e^{i(\vartheta
_{W}-\vartheta_{j})}. \label{DEFin1}
\end{equation}
Therefore:
\begin{equation}
\delta V_{(np2)}\underset{\mathcal{V}\rightarrow \infty
}{\longrightarrow } \sum\limits_{j=1}^{2}\left\vert
X_{j}\right\vert \left( e^{+i\left( \vartheta_{W}-\vartheta
_{j}+a_{j}b_{j}\right) }+e^{-i\left( \vartheta_{W}-\vartheta
_{j}+a_{j}b_{j}\right) }\right).
\end{equation}
Let us now reconsider $\delta V_{(np1)}^{AX}$, which we had set
aside for a moment. It can be rewritten as:
\begin{equation}
\delta V_{(np1)}^{AX}=e^{K} K_{12}^{-1}a_{1}a_{2}e^{-\left(
a_{1}\tau _{1}+a_{2}\tau _{2}\right) }\left(
A_{1}\bar{A}_{2}e^{i\left( a_{2}b_{2}-a_{1}b_{1}\right)
}+A_{2}\bar{A}_{1}e^{-i\left( a_{2}b_{2}-a_{1}b_{1}\right)
}\right),
\end{equation}
and finally as:
\begin{equation}
\delta V_{(np1)}^{AX}=Y_{12}e^{i\left(
a_{2}b_{2}-a_{1}b_{1}\right) }+\bar{Y}_{12}e^{-i\left(
a_{2}b_{2}-a_{1}b_{1}\right)},
\end{equation}
where:
\begin{equation}
Y_{12}\equiv e^{K}K_{12}^{-1}a_{1}a_{2}A_{1}\bar{A}_{2}e^{-\left(
a_{1}\tau _{1}+a_{2}\tau _{2}\right) }=\left\vert
Y_{12}\right\vert e^{i(\vartheta _{1}-\vartheta_{2})}.
\label{DEFin2}
\end{equation}
Therefore:
\begin{equation}
\delta V_{(np1)}^{AX}=\left\vert Y_{12}\right\vert \left(
e^{i\left( \vartheta
_{1}-\vartheta_{2}+a_{2}b_{2}-a_{1}b_{1}\right) }+e^{-i\left(
\vartheta _{1}-\vartheta_{2}+a_{2}b_{2}-a_{1}b_{1}\right)}\right).
\end{equation}
Thus, the full axion-dependent part of the scalar potential
$V_{AX}$ looks like:
\begin{eqnarray}
V_{AX}&=&\delta V_{(np2)}+\delta V_{(np1)}^{AX} \\
&=&2\sum\limits_{j=1}^{2}\left\vert X_{j}\right\vert \cos \left(
\vartheta_{W}-\vartheta _{j}+a_{j}b_{j}\right) +2\left\vert
Y_{12}\right\vert \cos \left( \vartheta
_{1}-\vartheta_{2}+a_{2}b_{2}-a_{1}b_{1}\right).  \notag
\end{eqnarray}

\subsubsection{Axion stabilisation}

$V_{AX}$ is a scalar function of the axions $b_{1}$ and $b_{2}$
whereas $\vartheta_{1}$, $\vartheta_{2}$ and $\vartheta_{W}$ are
to be considered just as parameters. In order to find a minimum
for $V_{AX}$ let us set its gradient to zero:
\begin{equation}
\left\{
\begin{array}{c}
\partial V_{AX}/\partial b_{1}=0\Longleftrightarrow\left\vert X_{1}\right\vert \sin
(\vartheta _{W}-\vartheta _{1}+a_{1}b_{1})=+\left\vert
Y_{12}\right\vert \sin \left( \vartheta
_{1}-\vartheta _{2}+a_{2}b_{2}-a_{1}b_{1}\right) ,\textrm{ } \\
\partial V_{AX}/\partial b_{2}=0\Longleftrightarrow\textrm{ \ }\left\vert X_{2}\right\vert \sin
(\vartheta _{W}-\vartheta _{2}+a_{2}b_{2})=-\left\vert
Y_{12}\right\vert \sin \left( \vartheta _{1}-\vartheta
_{2}+a_{2}b_{2}-a_{1}b_{1}\right) ,\textrm{ \ }
\end{array}
\right.   \label{GGgui}
\end{equation}
The solution of (\ref{GGgui}) is given by:
\begin{equation}
\left\{
\begin{array}{c}
\psi _{1}\equiv \left( \vartheta _{W}-\vartheta
_{1}+a_{1}b_{1}\right)
=p_{1}\pi ,\textrm{ \ }p_{1}\in \mathbb{Z}, \\
\psi _{2}\equiv \left( \vartheta _{W}-\vartheta
_{2}+a_{2}b_{2}\right) =p_{2}\pi ,\textrm{ \ }p_{2}\in \mathbb{Z},
\end{array}
\right.   \label{VEVov}
\end{equation}
and:
\begin{equation}
\psi_{12}\equiv \left( \vartheta
_{1}-\vartheta_{2}+a_{2}b_{2}-a_{1}b_{1}\right) =p_{12}\pi
,\textrm{ \ } p_{12}\in \mathbb{Z}. \label{ADjust}
\end{equation}
From (\ref{VEVov}) equation (\ref{ADjust}) requires
$p_{12}=p_{2}-p_{1}$. Let us summarise the points where the
gradient of the axion potential is zero in the following table:
\begin{equation}
\begin{tabular}{|c|c|c|c|c|}
\hline & (a) & (b) & (c) & (d) \\ \hline $\cos \psi _{1}$ & +1 &
-1 & +1 & -1 \\ \hline $\cos \psi _{2}$ & +1 & -1 & -1 & +1 \\
\hline $\cos \psi _{12}$ & +1 & +1 & -1 & -1 \\ \hline
\end{tabular}
\label{TABLE}
\end{equation}
We notice that the phases of $W$, $A_{1}$ and $A_{2}$ will not
enter into $\delta V_{(np2)}$ once the axions have been properly
minimised and so, without loss of generality, we can consider $W$,
$A_{1}$ and $A_{2}\in \mathbb{R} ^{+}$ from now on.

We have still to check the Hessian matrix evaluated at $b_{1}$ and
$b_{2}$ as given in (\ref{VEVov}) and require it to be positive
definite. Its diagonal elements are given by:
\begin{equation}
\left\{
\begin{array}{c}
\partial ^{2}V_{AX}/\partial b_{1}^{2}=-2a_{1}^{2}\left(
\left\vert X_{1}\right\vert \cos \psi _{1}+\left\vert
Y_{12}\right\vert \cos \psi _{12}
\right) , \\
\partial ^{2}V_{AX}/\partial b_{2}^{2}=-2a_{2}^{2}\left(
\left\vert X_{2}\right\vert \cos \psi _{2}+\left\vert
Y_{12}\right\vert \cos \psi _{12} \right),
\end{array}
\right.   \label{HEssian1}
\end{equation}
whereas the non-diagonal ones read:
\begin{equation}
\frac{\partial^{2} V_{AX}}{\partial b_{2} \partial b_{1}}
=\frac{\partial^{2} V_{AX}}{\partial b_{1} \partial b_{2}}=2 a_{1}
a_{2} \left\vert Y_{12}\right\vert \cos \psi_{12}.
\label{HEssian2}
\end{equation}
We can diagonalise the Hessian $\mathcal{H}$ to the identity by
decomposing it $a$ $la$ Choleski into
$\mathcal{H}=\mathcal{U}^{T}\mathbb{I}\mathcal{U}$, where the
elements of the upper triangular matrix $\mathcal{U}$ are given by
the following recursive relations:
\begin{equation}
\mathcal{U}_{11}^{2}=-2a_{1}^{2}\left(\left\vert X_{1}\right\vert
\cos\psi_{1}+\left\vert Y_{12}\right\vert \cos \psi_{12}\right),
\label{UU11}
\end{equation}
\begin{equation}
\mathcal{U}_{12}=\frac{2a_{1}a_{2}\left\vert Y_{12}\right\vert
\cos\psi_{12}}{\mathcal{U}_{11}}, \label{UU12}
\end{equation}
\begin{equation}
\mathcal{U}_{22}^{2}=-\mathcal{U}_{12}^{2}-2a_{2}^{2} \left(
\left\vert X_{2}\right\vert \cos\psi_{2} +\left\vert
Y_{12}\right\vert \cos\psi_{12}\right), \label{UU22}
\end{equation}
with $\mathcal{U}_{21}=0$. Determining if the Hessian is positive
definite is equal to checking that $\mathcal{U}$ is a real matrix.
Looking at (\ref{UU12}) we realise that $\mathcal{U}_{12}$ is
automatically real if $\mathcal{U}_{11}$ is real. Hence we have to
make sure just that both $\mathcal{U}_{11}^{2}>0$ and
$\mathcal{U}_{22}^{2}>0$. When we analyse the cases listed in
table (\ref{TABLE}) we realise that:

\begin{enumerate}
\item[(a)] can never be a minimum since $\textit{ }\mathcal{U}_{11}^{2}<0$;
in reality it turns out to be always a maximum,

\item[(b)] is a minimum only if $\left\vert X_{1}\right\vert >\left\vert
Y_{12}\right\vert $ and $\left\vert X_{1}\right\vert \left\vert
X_{2}\right\vert >\left\vert Y_{12}\right\vert \left( \left\vert
X_{1}\right\vert +\left\vert X_{2}\right\vert \right) ,$

\item[(c)] is a minimum only if $\left\vert Y_{12}\right\vert >\left\vert
X_{1}\right\vert $ and $\left\vert X_{2}\right\vert \left\vert
Y_{12}\right\vert >\left\vert X_{1}\right\vert \left( \left\vert
X_{2}\right\vert +\left\vert Y_{12}\right\vert \right) ,$

\item[(d)] is a minimum only if  $\left\vert Y_{12}\right\vert \left\vert
X_{1}\right\vert >\left\vert X_{2}\right\vert \left( \left\vert
X_{1}\right\vert +\left\vert Y_{12}\right\vert \right) ,$
\end{enumerate}
where, according to the definitions (\ref{DEFin1}) and
(\ref{DEFin2}), we have:
\begin{equation}
\left\{
\begin{array}{c}
\left\vert X_{1}\right\vert =2\left\vert A_{1}\right\vert
a_{1}\tau _{1}\left\vert W \right\vert \frac{e^{-a_{1}\tau
_{1}}}{\mathcal{V}^{2}}, \\
\left\vert X_{2}\right\vert =2\left\vert A_{2}\right\vert
a_{2}\tau _{2}\left\vert W \right\vert \frac{e^{-a_{2}\tau
_{2}}}{\mathcal{V}^{2}}, \\
\left\vert Y_{12}\right\vert =K_{12}^{-1}\left\vert
A_{1}\right\vert a_{1}\left\vert A_{2}\right\vert a_{2}\frac{
e^{-a_{1}\tau _{1}}e^{-a_{2}\tau _{2}}}{\mathcal{V}^{2}}.
\end{array}
\right. \label{ineq}
\end{equation}
In order to study the cases (b), (c) and (d), it is therefore
crucial to know the order of magnitude of the two exponentials
$e^{-a_{1}\tau _{1}}$ and $e^{-a_{2}\tau _{2}}$ given by their
scaling behaviour in the volume. This depends on the direction we
are looking at to find the minimum in the large volume limit which
can be performed in three different ways:
\begin{eqnarray}
I)\textrm{ }\mathcal{V} &\sim &e^{\gamma a_{1}\tau _{1}}\sim
e^{\gamma a_{2}\tau _{2}},\textrm{ \ }\gamma \in \mathbb{R}^{+}
\notag \\
II)\textrm{ }\mathcal{V} &\sim &e^{\beta a_{1}\tau _{1}}\sim
e^{\gamma a_{2}\tau _{2}},\textrm{ \ }\beta <\gamma,
\textrm{ \ }\gamma,\beta \in \mathbb{R}^{+},  \label{DIrections} \\
III)\mathcal{V} &\sim &e^{\beta a_{1}\tau _{1}}\sim e^{\gamma
a_{2}\tau _{2}},\textrm{ \ }\beta >\gamma\textrm{ \ }\gamma,\beta
\in \mathbb{R}^{+}.  \notag
\end{eqnarray}
Finally we need also to know the form of $K^{-1}_{12}$. We can
classify its behaviour according to the volume dependence of the
quantity $k_{12j}t^{j}$ and find 4 different cases:

\begin{enumerate}
\item $k_{12j}t^{j}=0\textrm{ \ or \ }k_{12j}t^{j}=\frac{f
(\tau _{1},\tau_{2})}{\mathcal{V}^{\alpha }},$ $\alpha \geq 1$
$\Longrightarrow K_{12}^{-1}=\tau _{1}\tau _{2};$

\item $k_{12j}t^{j}=\frac{g_{\gamma }(\tau _{1},\tau_{2})}{\mathcal{V}^{\alpha
}},$ $0<\alpha<1$, $g$ homogeneous function of degree $\gamma
=\frac{1+3\alpha}{2}$ $\Longrightarrow
K_{12}^{-1}=\mathcal{V}^{\alpha }g_{2-3\alpha /2}(\tau _{1},\tau
_{2}),$ \ $0<\alpha <1;$

\item $k_{12j}t^{j}=f_{1/2}(\tau _{1},\tau_{2}),$ $f$
homogeneous function of degree $1/2$ $\Longrightarrow
K_{12}^{-1}=\mathcal{V}f_{1/2}(\tau _{1},\tau _{2});$

\item $k_{12j}t^{j}=\mathcal{V}^{\alpha }h_{\beta }(\tau _{1},\tau_{2}),$
$\alpha >0$, $h$ homogeneous function of degree $\beta =
\frac{1-3\alpha}{2}$ $\Longrightarrow
K_{12}^{-1}=\mathcal{V}^{\alpha }h_{2-3\alpha /2}(\tau _{1},\tau
_{2}),$ \ $\alpha >1.$
\end{enumerate}
Let us now focus on the axion minimisation by analysing each of
these 4 cases in full detail. For each case we will have to study
if the inequalities (b), (c) and (d) admit a solution for any of
the three possible ways to take the large volume limit as
expressed in (\ref{DIrections}). We will always consider natural
values of the parameters $\left\vert A_{1} \right\vert \simeq
\left\vert A_{2} \right\vert \simeq \left\vert W \right\vert\simeq
1$.

From (I) of (\ref{DIrections}), we can immediately realise that,
regardless of the form of $K^{-1}_{12}$, at large volume
$\left\vert X_{1}\right\vert$ and $\left\vert X_{2}\right\vert$
have the same scaling with the volume and so we can denote both of
them as $\left\vert X \right\vert$. It is then straightforward to
see that both the second (c)-condition and the (d)-condition can
never be satisfied. In fact they take the form:
\begin{equation}
\left\vert X \right\vert \left\vert Y_{12} \right\vert >
\left\vert X \right\vert \left\vert Y_{12} \right\vert +
\left\vert X \right\vert^{2},
\end{equation}
which is manifestly an absurd. This implies that neither (c) nor
(d) can be a minimum along the direction (I) for any value of
$K^{-1}_{12}$. A further analysis reveals that the points (c) and
(d) can never be maxima so since we proved that they cannot be
minima, they are forced to be saddle points. We do not present the
details of this analysis here since it is not important for our
reasoning. On the other hand, the first (b)-condition is
automatically satisfied if the second one is true since it reduces
to:
\begin{equation}
\left\vert X \right\vert^{2} > 2 \left\vert X \right\vert
\left\vert Y_{12} \right\vert\textit{ \
}\Longleftrightarrow\textit{ \ }\left\vert X \right\vert > 2
\left\vert Y_{12} \right\vert. \label{disug}
\end{equation}

From (II) of (\ref{DIrections}), we also notice that, regardless
of the form of $K^{-1}_{12}$, at large volume $\left\vert
X_{1}\right\vert<\left\vert X_{2}\right\vert$ since
$\beta<\gamma$. It is then straightforward to see that in this
situation the (d)-condition can never be satisfied. This implies
that (d) is always a saddle point along the direction (II) for any
value of $K^{-1}_{12}$.

Furthermore (III) of (\ref{DIrections}) implies that, regardless
of the form of $K^{-1}_{12}$, at large volume $\left\vert
X_{1}\right\vert>\left\vert X_{2}\right\vert$ as $\beta>\gamma$.
Then we immediately see that the second (c)-condition can never be
satisfied. Therefore (c) is always a saddle point along the
direction (III) for any value of $K^{-1}_{12}$.

\bigskip

\textbf{Case (1):} $K^{-1}_{12}\simeq \tau_{1}\tau_{2}$

\begin{enumerate}
\item[$\bullet$] direction (I)

The volume dependence of the parameters (\ref{ineq}) is:
\begin{equation}
\left\{
\begin{array}{l}
\left\vert X_{1} \right\vert \simeq \left\vert X_{2} \right\vert
\simeq \mathcal{V}^{-(2+1/\gamma)}\ln\mathcal{V}, \\
\left\vert Y_{12} \right\vert \simeq
\mathcal{V}^{-(2+2/\gamma)}(\ln\mathcal{V})^{2}.
\end{array}
\right.  \label{ABsol}
\end{equation}
Looking at (\ref{ABsol}), we realise that at large volume
$\left\vert X_{1}\right\vert > 2\left\vert Y_{12} \right\vert$.
Therefore the second (b)-condition (\ref{disug}) is satisfied and
(b) is a minimum of the axion potential.

\item[$\bullet$] direction (II)

The parameters (\ref{ineq}) now read:
\begin{equation}
\left\{
\begin{array}{l}
\left\vert X_{1} \right\vert \simeq
\mathcal{V}^{-(2+1/\beta)}\ln\mathcal{V}, \\
\left\vert X_{2} \right\vert \simeq
\mathcal{V}^{-(2+1/\gamma)}\ln\mathcal{V}, \\
\left\vert Y_{12} \right\vert \simeq
\mathcal{V}^{-(2+1/\beta+1/\gamma)}(\ln\mathcal{V})^{2}.
\end{array}
\right.  \label{dabs}
\end{equation}
Looking at (\ref{dabs}), we realise that at large volume
$\left\vert X_{1}\right\vert > \left\vert Y_{12} \right\vert$,
which implies that (c) is a saddle point. Thus the first
(b)-condition is satisfied and the second becomes:
\begin{equation}
\frac{(\ln \mathcal{V})^{2}}{\mathcal{V}^{4+1/\beta +1/\gamma
}}>\frac{(\ln \mathcal{V})^{2}}{\mathcal{V}^{2+1/\beta +1/\gamma
}}\left( \frac{\ln \mathcal{V}}{\mathcal{V}^{2+1/\beta
}}+\frac{\ln \mathcal{V}}{\mathcal{V}^{2+1/\gamma}}\right)
\underset{1/\beta >1/\gamma }{\simeq }\frac{(\ln
\mathcal{V})^{3}}{\mathcal{V}^{4+1/\beta +2/\gamma}},
\end{equation}
which is true at large volume for values of $\gamma>\beta>0$ not
extremely big. Thus (b) is a minimum of the axion potential.

\item[$\bullet$] direction (III)

The parameters (\ref{ineq}) take the same form as (\ref{dabs}) but
now with $\beta>\gamma$. We have still $\left\vert
X_{1}\right\vert > \left\vert Y_{12} \right\vert$, which implies
that the first (b)-condition is satisfied. The second looks like:
\begin{equation}
\frac{(\ln \mathcal{V})^{2}}{\mathcal{V}^{4+1/\beta +1/\gamma
}}>\frac{(\ln \mathcal{V})^{2}}{\mathcal{V}^{2+1/\beta +1/\gamma
}}\left( \frac{\ln \mathcal{V}}{\mathcal{V}^{2+1/\beta
}}+\frac{\ln \mathcal{V}}{\mathcal{V}^{2+1/\gamma}}\right)
\underset{1/\beta<1/\gamma }{\simeq }\frac{(\ln
\mathcal{V})^{3}}{\mathcal{V}^{4+2/\beta +1/\gamma}},
\end{equation}
which is true at large volume. Thus (b) is a minimum of the axion
potential. On the contrary the simplified (d)-condition reads:
\begin{equation}
\frac{\ln\mathcal{V}}{\mathcal{V}^{1/\beta}}>1
+\frac{\ln\mathcal{V}}{\mathcal{V}^{1/\gamma}}
\end{equation}
which at large volume is clearly false for values of
$\beta>\gamma>0$ not extremely big. It follows that in this case
(d) is a saddle point.
\end{enumerate}

Let us summarise the results found in case (1) in the following
table:
\begin{center}
\begin{tabular}{|c|c|c|c|}
\hline & (I) & (II) & (III) \\ \hline (a) & max & max & max \\
\hline (b) & min & min & min \\ \hline (c) & saddle & saddle &
saddle \\ \hline (d) & saddle & saddle & saddle \\ \hline
\end{tabular}
\end{center}

\newpage

\textbf{Case (2):} $K_{12}^{-1}=\mathcal{V}^{\alpha }g_{2-3\alpha
/2}(\tau _{1},\tau _{2}),$ \ $0<\alpha <1$

\begin{enumerate}
\item[$\bullet$] direction (I)

The volume dependence of the parameters (\ref{ineq}) now reads:
\begin{equation}
\left\{
\begin{array}{l}
\left\vert X_{1} \right\vert \simeq \left\vert X_{2} \right\vert
\simeq \mathcal{V}^{-(2+1/\gamma)}\ln\mathcal{V}, \\
\left\vert Y_{12} \right\vert \simeq
\mathcal{V}^{-(2+2/\gamma-\alpha)}(\ln\mathcal{V})^{2-3\alpha/2}.
\end{array}
\right.  \label{tABsol}
\end{equation}
Substituting the expressions (\ref{tABsol}) in (\ref{disug}) we
find:
\begin{equation}
1>2
\frac{(\ln\mathcal{V})^{1-3\alpha/2}}{\mathcal{V}^{1/\gamma-\alpha}},
\label{IO}
\end{equation}
which at large volume is true if $\alpha<(1/\gamma)$, false if
$\alpha< (1/\gamma)$ or $\alpha=(1/\gamma)\leq 2/3$. On the
contrary, for $2/3<\alpha=(1/\gamma)<1$ the minimum is present.
Thus (b) can be a minimum of the axion potential.

\item[$\bullet$] direction (II)

The parameters (\ref{ineq}) now read:
\begin{equation}
\left\{
\begin{array}{l}
\left\vert X_{1} \right\vert \simeq
\mathcal{V}^{-(2+1/\beta)}\ln\mathcal{V}, \\
\left\vert X_{2} \right\vert \simeq
\mathcal{V}^{-(2+1/\gamma)}\ln\mathcal{V}, \\
\left\vert Y_{12} \right\vert \simeq
\mathcal{V}^{-(2+1/\beta+1/\gamma-\alpha)}(\ln\mathcal{V})^{2-3\alpha/2}.
\end{array}
\right.  \label{ddabso}
\end{equation}
The first (b)-condition becomes:
\begin{equation}
1>\frac{(\ln
\mathcal{V})^{1-3\alpha/2}}{\mathcal{V}^{1/\gamma-\alpha}},
\label{question}
\end{equation}
that is satisfied if either $\alpha<(1/\gamma)$ or
$2/3<\alpha=(1/\gamma)<1$. Otherwise (\ref{question}) is false
unless $\alpha=(1/\gamma)=2/3$ in which case we cannot conclude
anything just looking at the volume dependence. However the second
(b)-condition reads:
\begin{equation}
1>(\ln{\mathcal{V}})^{1-3\alpha/2}\left(\frac{1}
{\mathcal{V}^{1/\beta-\alpha}}+\frac{1}{\mathcal{V}^{1/\gamma-\alpha}}\right),
\label{SECONDB}
\end{equation}
which is definitely true at large volume if $\alpha<(1/\gamma)$ or
$2/3<\alpha=(1/\gamma)<1$. On the contrary, in the case
$\alpha=(1/\gamma)=2/3\Leftrightarrow (1/\beta-2/3)>0$,
(\ref{SECONDB}) becomes:
\begin{equation}
1>1+\frac{1} {\mathcal{V}^{1/\beta-2/3}},
\end{equation}
which is clearly impossible. Thus (b) can be a minimum of the
axion potential. We need now just to study the case (c) for which
the first inequality is:
\begin{equation}
1<\frac{(\ln
\mathcal{V})^{1-3\alpha/2}}{\mathcal{V}^{1/\gamma-\alpha}},
\label{Question}
\end{equation}
that is satisfied if either $(1/\gamma)<\alpha<1$ or
$\alpha=(1/\gamma)<2/3$. Otherwise (\ref{Question}) is false
unless $\alpha=(1/\gamma)=2/3$ in which case we cannot conclude
anything just looking at the volume dependence. However the second
(c)-condition can be simplified to give:
\begin{equation}
\frac{(\ln
\mathcal{V})^{1-3\alpha/2}}{\mathcal{V}^{1/\gamma-\alpha}}>
1+\frac{(\ln
\mathcal{V})^{1-3\alpha/2}}{\mathcal{V}^{1/\beta-\alpha}},
\label{llook}
\end{equation}
which is clearly satisfied at large volume if either
$(1/\gamma)<\alpha<1$ or $\alpha=(1/\gamma)<2/3$. On the contrary
when $\alpha=(1/\gamma)=2/3\Leftrightarrow (1/\beta-2/3)>0$,
(\ref{llook}) becomes:
\begin{equation}
1> 1+\frac{1}{\mathcal{V}^{1/\beta-2/3}},
\end{equation}
which is clearly false.

\item[$\bullet$] direction (III)

The parameters (\ref{ineq}) assume the same form as (\ref{ddabso})
but now with $\beta>\gamma$. Following lines of reasoning similar
to those used for direction (II), we get the results summarised in
the following table along with all the others for case (2).
\end{enumerate}

\begin{center}
\begin{tabular}{|c|c|c|c|}
\hline & (I) & (II) & (III) \\ \hline (a) & max & max & max \\
\hline
(b) & $%
\begin{array}{c}
\alpha <1/\gamma \text{ min} \\
2/3<\alpha =1/\gamma <1\text{ min} \\
\alpha =1/\gamma \leq 2/3\text{ saddle} \\
\alpha >1/\gamma \text{ saddle}
\end{array}%
$ & $%
\begin{array}{c}
\alpha <1/\gamma \text{ min} \\
2/3<\alpha =1/\gamma <1\text{ min} \\
\alpha =1/\gamma \leq 2/3\text{ saddle} \\
\alpha >1/\gamma \text{ saddle}%
\end{array}%
$ & $%
\begin{array}{c}
\alpha <1/\beta \text{ min} \\
2/3<\alpha =1/\beta <1\text{ min} \\
\alpha =1/\beta \leq 2/3\text{ saddle} \\
\alpha >1/\beta \text{ saddle}%
\end{array}%
$ \\ \hline
(c) & saddle & $%
\begin{array}{c}
\alpha <1/\gamma \text{ saddle} \\
2/3\leq \alpha =1/\gamma <1\text{ saddle} \\
\alpha =1/\gamma <2/3\text{ min} \\
1/\gamma <\alpha <1\text{ min}%
\end{array}
$ & saddle \\ \hline (d) & saddle & saddle & $
\begin{array}{c}
\alpha <1/\beta \text{ saddle} \\
2/3\leq \alpha =1/\beta <1\text{ saddle} \\
\alpha =1/\beta <2/3\text{ min} \\
\alpha >1/\beta \text{ min}
\end{array}
$ \\ \hline
\end{tabular}
\end{center}

\textbf{Case (3):} $K_{12}^{-1}=\mathcal{V}f_{1/2}(\tau _{1},\tau
_{2})$

\begin{enumerate}
\item[$\bullet$] direction (I)

The volume dependence of the parameters (\ref{ineq}) now looks
like:
\begin{equation}
\left\{
\begin{array}{l}
\left\vert X_{1} \right\vert \simeq \left\vert X_{2} \right\vert
\simeq \mathcal{V}^{-(2+1/\gamma)}\ln\mathcal{V}, \\
\left\vert Y_{12} \right\vert \simeq
\mathcal{V}^{-(1+2/\gamma)}\sqrt{\ln\mathcal{V}}.
\end{array}
\right.  \label{ABSOl}
\end{equation}
Substituting the expressions (\ref{ABSOl}) in (\ref{disug}) we
find:
\begin{equation}
\frac{\ln\mathcal{V}}{\mathcal{V}^{2+1/\gamma}}>2
\frac{\sqrt{\ln\mathcal{V}}}{\mathcal{V}^{1+2/\gamma}},
\end{equation}
which at large volume is true if $\gamma\leq 1$, false if
$\gamma>1$. Thus (b) can be a minimum of the axion potential.

\item[$\bullet$] direction (II)

The parameters (\ref{ineq}) now read:
\begin{equation}
\left\{
\begin{array}{l}
\left\vert X_{1} \right\vert \simeq
\mathcal{V}^{-(2+1/\beta)}\ln\mathcal{V}, \\
\left\vert X_{2} \right\vert \simeq
\mathcal{V}^{-(2+1/\gamma)}\ln\mathcal{V}, \\
\left\vert Y_{12} \right\vert \simeq
\mathcal{V}^{-(1+1/\beta+1/\gamma)}\sqrt{\ln\mathcal{V}}.
\end{array}
\right.  \label{dabS}
\end{equation}
Looking at (\ref{dabS}), we realise that the first (b)-condition
becomes:
\begin{equation}
\frac{\sqrt{\ln
\mathcal{V}}}{\mathcal{V}}>\frac{1}{\mathcal{V}^{1/\gamma}},
\label{firstB}
\end{equation}
which is satisfied only if $\gamma\leq 1$. Viceversa the first
(c)-condition is satisfied only for $\gamma>1$. Let us check now
the validity of the second (b)-condition which reads:
\begin{equation}
\frac{\sqrt{\ln
\mathcal{V}}}{\mathcal{V}}>\frac{1}{\mathcal{V}^{1/\beta}}
+\frac{1}{\mathcal{V}^{1/\gamma}}, \label{secondB}
\end{equation}
which, at large volume, is automatically true if $\gamma\leq 1$.
The second (c)-condition is also correctly satisfied for
$\gamma>1$ since it reads:
\begin{equation}
\frac{1}{\mathcal{V}^{1/\gamma}}>\frac{\sqrt{\ln
\mathcal{V}}}{\mathcal{V}}+\frac{1}{\mathcal{V}^{1/\beta}}.
\label{segno}
\end{equation}
Thus both (b) and (c) can be a minimum of the axion potential.

\item[$\bullet$] direction (III)

The parameters (\ref{ineq}) assume the same form as (\ref{dabS})
but now with $\beta>\gamma$. The inequality corresponding to the
(d)-condition reads:
\begin{equation}
\frac{1}{\mathcal{V}^{1/\beta}}>\frac{1}{\mathcal{V}^{1/\gamma}}+
\frac{\sqrt{\ln\mathcal{V}}}{\mathcal{V}},
\end{equation}
and becomes true if $\beta>1$. Moreover, the first (b)-condition
(\ref{firstB}) is again satisfied for $\gamma\leq 1$. On the other
hand, the second looks like (\ref{secondB}) and now, at large
volume, is true only if $\beta\leq 1$, which implies correctly
$\gamma\leq 1$ since in this case $\gamma<\beta$. It follows that
both (b) and (d) can be a minimum.
\end{enumerate}
\newpage
Let us summarise the results found in case (3) in the following
table:
\begin{center}
\begin{tabular}{|c|c|c|c|}
\hline & (I) & (II) & (III) \\ \hline (a) & max & max & max \\
\hline (b) & $\left\{
\begin{array}{c}
0<\gamma \leq 1\text{ min,} \\
\gamma >1\text{ saddle.}
\end{array}
\right. $ & $\left\{
\begin{array}{c}
0<\gamma \leq 1\text{ min,} \\
\gamma >1\text{ saddle.}
\end{array}
\right. $ & $\left\{
\begin{array}{c}
0<\gamma <\beta \leq 1\text{ min,} \\
\beta>1 \text{ saddle.} \\
\end{array}
\right. $ \\ \hline (c) & saddle & $\left\{
\begin{array}{c}
0<\gamma \leq 1\text{ saddle,} \\
\gamma >1\text{ min.}
\end{array}
\right. $ & saddle \\ \hline (d) & saddle & saddle & $\left\{
\begin{array}{c}
0<\gamma<\beta \leq 1\text{ saddle,} \\
\beta>1\text{ min.}
\end{array}
\right. $ \\ \hline
\end{tabular}
\end{center}

\bigskip

\textbf{Case (4):} $K_{12}^{-1}=\mathcal{V}^{\alpha }h_{2-3\alpha
/2}(\tau _{1},\tau _{2}),$ \ $\alpha >1$

\begin{enumerate}
\item[$\bullet$] direction (I)

The volume dependence of the parameters (\ref{ineq}) is given
again by (\ref{tABsol}) and (\ref{disug}) takes the same form as
the inequality (\ref{IO}) which at large volume is true if
$\alpha<1/\gamma$, false if $\alpha>1/\gamma$. The situation
$\alpha=1/\gamma$ is more involved and (\ref{IO}) simplifies to:
\begin{equation}
1>2(\ln\mathcal{V})^{1-3\alpha/2},
\end{equation}
which gives a positive result if $\alpha>2/3$. This is definitely
true in our case where $\alpha>1$. Thus (b) can be a minimum of
the axion potential.

\item[$\bullet$] direction (II)

The parameters (\ref{ineq}) now take the same form given in
(\ref{ddabso}). It follows then that the first (b)-condition
$\left\vert X_{1}\right\vert > \left\vert Y_{12} \right\vert$
looks like (\ref{question}) and is verified for $1/\gamma\geq
\alpha$. The second (b)-condition looks like (\ref{SECONDB}) which
at large volume is correctly true for $1/\gamma\geq \alpha$. Thus
(b) is a minimum of the axion potential. On the contrary the first
(c)-condition implies $1/\gamma<\alpha$, whereas the second is
similar to the inequality (\ref{llook}) which is again clearly
true for $1/\gamma<\alpha$. It follows that in this case (c) can
also be a minimum.

\item[$\bullet$] direction (III)

The parameters (\ref{ineq}) assume the same form as (\ref{ddabso})
but now with $\beta>\gamma$. The (d)-condition looks like:
\begin{equation}
\frac{(\ln{\mathcal{V}})^{1-3\alpha/2}}{\mathcal{V}^{1/\beta-\alpha}}>1
+\frac{(\ln{\mathcal{V}})^{1-3\alpha/2}}{\mathcal{V}^{1/\gamma-\alpha}},
\end{equation}
and is verified only if $1/\beta<\alpha$. On the other hand, the
first (b)-condition is again given by (\ref{question}) and so is
still solved for $1/\gamma\geq \alpha$. The second (b)-condition
looks like (\ref{SECONDB}) but now at large volume it is satisfied
for $1/\beta\geq\alpha$. It follows that in this case both (b) and
(d) can be a minimum of the axion potential.
\end{enumerate}

Let us summarise the results found in case (4) in the following
table:
\begin{center}
\begin{tabular}{|c|c|c|c|}
\hline & (I) & (II) & (III) \\ \hline (a) & max & max & max \\
\hline
(b) & $%
\begin{array}{c}
1<\alpha \leq 1/\gamma \text{ min,} \\
\alpha >1/\gamma \text{ saddle.}%
\end{array}%
$ & $%
\begin{array}{c}
1<\alpha \leq 1/\gamma \text{ min,} \\
\alpha >1/\gamma \text{ saddle.}%
\end{array}%
$ & $%
\begin{array}{c}
1<\alpha \leq 1/\beta <1/\gamma \text{ min,} \\
1/\beta <\alpha \text{ saddle.} \\
\end{array}%
$ \\ \hline
(c) & saddle & $%
\begin{array}{c}
1<\alpha \leq 1/\gamma \text{ saddle,} \\
\alpha >1/\gamma \text{ min.}%
\end{array}%
$ & saddle \\ \hline
(d) & saddle & saddle & $%
\begin{array}{c}
\alpha \leq 1/\beta \text{ saddle,} \\
\alpha >1/\beta \text{ min.}%
\end{array}%
$ \\ \hline
\end{tabular}
\end{center}
\bigskip

\subsubsection{K\"{a}hler moduli stabilisation}

After this long analysis of the axion minimisation, let us now
focus again step by step on the four cases according to the
different possible values of $K^{-1}_{12}$. In each case, we shall
fix the axions at their possible VEVs and then study the
K\"{a}hler moduli stabilisation depending on the particular form
of $K^{-1}_{11}$ and $K^{-1}_{22}$.

However, before focusing on each particular case, let us point out
some general features. At the axion minimum we will have:
\begin{equation}
\left\langle V_{AX}\right\rangle =2\left(\pm\left\vert
Y_{12}\right\vert \pm\left\vert X_{1}\right\vert \pm\left\vert
X_{2}\right\vert\right), \label{assione}
\end{equation}
where the "$\pm$" signs depend on the specific locus of the
minimum, that is (a) or (b) or (c), as specified in (\ref{TABLE}).
Now to write (\ref{assione}) explicitly, recall (\ref{ineq}) and
get:
\begin{equation}
\delta V_{(np2)}+\delta
V_{(np1)}^{AX}=\frac{2}{\mathcal{V}^{2}}\left\{
W\sum\limits_{j=1}^{2}\left(\pm 2 a_{j}\tau
_{j}e^{-a_{j}\tau_{j}}\right) \pm K^{-1}_{12}
a_{1}a_{2}e^{-(a_{1}\tau_{1}+a_{2}\tau_{2})}
 \right\}, \label{fffor}
\end{equation}
where we have set $A_{1}=A_{2}=1$. We may now study the full
potential by combining equations (\ref{Alfa}), (\ref{imp}) and
(\ref{fffor}):
\begin{eqnarray}
V &\sim &\frac{1}{\mathcal{V}^{2}}\left[
\sum\limits_{j=1}^{2}K^{-1}_{jj} a_{j}^{2} e^{-2a_{j}\tau_{j}}\pm
2K^{-1}_{12}a_{1}a_{2}e^{-(a_{1}\tau_{1}+a_{2}\tau_{2})}\right]  \notag \\
&&+4\frac{W_{0}}{\mathcal{V}^{2}}\sum\limits_{j=1}^{2}\left(\pm
a_{j}\tau _{j}e^{-a_{j}\tau_{j}}\right)+\frac{3}{4}\frac{\hat{\xi}
}{\mathcal{V}^{3}}W_{0}^{2}, \label{finalPotential}
\end{eqnarray}
where we have substituted $W$ with its tree-level expectation
value $W_{0}$ because the non-perturbative corrections are always
subleading by a power of $\mathcal{V}$.

When we take the generic large volume limit $\mathcal{V}\sim
e^{\beta a_{1}\tau_{1}}\sim e^{\gamma a_{2}\tau_{2}}$, with
$\beta$ and $\gamma \in \mathbb{R}$ without any particular
relation among them to take into account all the possible limits
(\ref{DIrections}), (\ref{finalPotential}) has the following
volume scaling:
\begin{equation}
V \sim \frac{K^{-1}_{11}}{\mathcal{V}^{2+2/\beta}}
+\frac{K^{-1}_{22}}{\mathcal{V}^{2+2/\gamma}}
\pm\frac{K^{-1}_{12}}{\mathcal{V}^{2+1/\beta+1/\gamma}}
\pm\frac{\tau _{1}}{\mathcal{V}^{2+1/\beta}}\pm\frac{\tau
_{2}}{\mathcal{V}^{2+1/\gamma}}+\frac{1}{\mathcal{V}^{3}}.
\label{semPlifica}
\end{equation}
Now given that we are already aware of the volume scaling of
$K^{-1}_{12}$, which was our starting point to stabilise the
axions, in order to understand what are the leading terms in
(\ref{semPlifica}), we need to know only the form of
$K^{-1}_{jj}$, $j=1,2$. We can classify its behaviour according to
the volume dependence of the quantity $k_{jjk}t^{k}$ and find 4
different cases as we did for $K^{-1}_{12}$:

\begin{enumerate}
\item $k_{jjk}t^{k}=0\text{ \ or \ }k_{jjk}t^{k}=\frac{f
(\tau _{1},\tau_{2})}{\mathcal{V}^{\alpha }},$ $\alpha \geq 1$
$\Longrightarrow K_{jj}^{-1}=\tau _{j}^{2};$

\item $k_{jjk}t^{k}=\frac{g_{\gamma }(\tau _{1},\tau_{2})}{\mathcal{V}^{\alpha
}},$ $0<\alpha<1$, $g$ homogeneous function of degree $\gamma
=\frac{1+3\alpha}{2}$ $\Longrightarrow
K_{jj}^{-1}=\mathcal{V}^{\alpha }g_{2-3\alpha /2}(\tau _{1},\tau
_{2}),$ \ $0<\alpha <1;$

\item $k_{jjk}t^{k}=f_{1/2}(\tau _{1},\tau_{2}),$ $f$
homogeneous function of degree $1/2$ $\Longrightarrow
K_{jj}^{-1}=\mathcal{V}f_{1/2}(\tau _{1},\tau _{2});$

\item $k_{jjk}t^{k}=\mathcal{V}^{\alpha }h_{\beta }(\tau _{1},\tau_{2}),$
$\alpha >0$, $h$ homogeneous function of degree $\beta =
\frac{1-3\alpha}{2}$ $\Longrightarrow
K_{jj}^{-1}=\mathcal{V}^{\alpha }h_{2-3\alpha /2}(\tau _{1},\tau
_{2}),$ \ $\alpha >1.$
\end{enumerate}
Before focusing on the K\"{a}hler moduli minimisation by analysing
all these 4 cases in full detail for each direction
(\ref{DIrections}), we stress that we can already show in general
that some situations do not lead to any LARGE Volume minimum.

For example, let us assume that the elements of $K^{-1}$ are such
that the dominant terms in (\ref{semPlifica}) are:
\begin{equation}
V \sim \frac{K^{-1}_{11}}{\mathcal{V}^{2+2/\beta}}-\frac{\tau
_{1}}{\mathcal{V}^{2+1/\beta}}-\frac{\tau
_{2}}{\mathcal{V}^{2+1/\gamma}}+\frac{1}{\mathcal{V}^{3}},
\label{SemPlifica}
\end{equation}
with $\beta=\gamma=1$ and
$K^{-1}_{11}=\mathcal{V}f_{1/2}(\tau_{1},\tau_{2})$. Therefore the
potential (\ref{finalPotential}) with $W_{0}=1$ looks like:
\begin{equation}
V \sim \frac{f_{1/2}(\tau_{1},\tau_{2}) a_{1}^{2}
e^{-2a_{1}\tau_{1}}}{\mathcal{V}} -\frac{4
a_{1}\tau_{1}e^{-a_{1}\tau_{1}}}{\mathcal{V}^{2}} -\frac{4
a_{2}\tau_{2}e^{-a_{2}\tau_{2}}}{\mathcal{V}^{2}}
+\frac{3}{4}\frac{\hat{\xi}}{\mathcal{V}^{3}}.
\label{FinalPotential}
\end{equation}
Thus from $\frac{\partial V}{\partial \tau_{1}}=0$ we find:
\begin{equation}
\mathcal{V}=\frac{4\tau_{1}}{\left(2a_{1}f_{1/2}-\frac{\partial
f_{1/2}}{\partial \tau_{1}}\right)}e^{a_{1}\tau_{1}},
\label{Etichetta1}
\end{equation}
whereas $\frac{\partial V}{\partial \tau_{2}}=0$ gives:
\begin{equation}
\mathcal{V}=- 4
\left(\frac{a_{2}}{a_{1}}\right)^{2}\frac{\tau_{2}}{\frac{\partial
f_{1/2}}{\partial \tau_{2}}}\frac{e^{2
a_{1}\tau_{1}}}{e^{a_{2}\tau_{2}}}. \label{Etichetta2}
\end{equation}
Now since we have $\beta=\gamma=1$, from the form of the large
volume limit (I) of (\ref{DIrections}), we infer that the minimum
should be located at $a_{1}\tau_{1}\simeq a_{2}\tau_{2}$. Making
this substitution and combining (\ref{Etichetta1}) with
(\ref{Etichetta2}), we end up with the following equation:
\begin{equation}
\frac{\partial f_{1/2}}{\partial
\tau_{2}}=\frac{a_{2}}{a_{1}}\frac{\partial f_{1/2}}{\partial
\tau_{1}}-2a_{2}f_{1/2}. \label{eor}
\end{equation}
Now using the homogeneity property of $f_{1/2}$, that is
$\tau_{1}\frac{\partial f_{1/2}}{\partial
\tau_{1}}+\tau_{2}\frac{\partial f_{1/2}}{\partial
\tau_{2}}=\frac{1}{2}f_{1/2}$, (\ref{eor}) takes the form:
\begin{equation}
\frac{\partial f_{1/2}}{\partial
\tau_{2}}=a_{2}\left(\frac{1}{4a_{2}\tau_{2}}-1\right)f_{1/2}.
\label{Eor}
\end{equation}
We can solve the previous differential equation getting
$f_{1/2}=\tau_{2}^{1/4}e^{-a_{2}\tau_{2}}$, which is not an
homogeneous function of degree $1/2$. Thus we deduce that this
case gives no LVS.

Another case in which we can show explicitly that no LARGE Volume
minimum is present, is the one where the dominant terms in
(\ref{semPlifica}) read:
\begin{equation}
V \sim \frac{K^{-1}_{11}}{\mathcal{V}^{2+2/\beta}}
+\frac{K^{-1}_{22}}{\mathcal{V}^{2+2/\gamma}}+\frac{1}{\mathcal{V}^{3}}.
\label{sSEMPLIl}
\end{equation}
with $\beta\leq\gamma$. The fact that all the three terms in
(\ref{sSEMPLIl}) are strictly positive leads us to conclude that
there would definitely be no LARGE Volume minimum in the volume
direction once we integrate out the small moduli. In fact,
(\ref{sSEMPLIl}) would take the generic form:
\begin{equation}
V\sim\frac{a(\ln{\mathcal{V}})^{b}+c}{\mathcal{V}^{3}},\text{ \
}c>0,
\end{equation}
which can be easily seen to have a minimum only if $a<0$ and
$b>0$.

We illustrate now a further case in which it is possible to show
explicitly that no LVS is present. The leading terms in
(\ref{semPlifica}) read:
\begin{equation}
V \sim \frac{K^{-1}_{11}}{\mathcal{V}^{2+2/\beta}}
+\frac{K^{-1}_{22}}{\mathcal{V}^{2+2/\gamma}}-\frac{\tau_{2}}{\mathcal{V}^{2+1/\gamma}},
\label{Se}
\end{equation}
with $\beta<\gamma$ and $\gamma>1$ to be able to neglect the
$\alpha'$ corrections that scale as $\mathcal{V}^{-3}$. The
necessary but not sufficient condition to fix the small K\"{a}hler
moduli is $K^{-1}_{11}=\mathcal{V}^{\delta}\tau_{1}^{2-3\delta/2}$
and
$K^{-1}_{22}=\mathcal{V}^{\eta}f_{2-3\eta/2}(\tau_{1},\tau_{2})$
with $\delta=2/\beta-1/\gamma$ and $\eta=1/\gamma$. We can now
prove that it is never possible to stabilise $a_{1}\tau_{1}\gg 1$.
In fact, the relevant part of the scalar potential
(\ref{finalPotential}) would read:
\begin{equation}
V \simeq \frac{a_{1}^{2}\tau_{1}^{2-3\delta/2}
e^{-2a_{1}\tau_{1}}}{\mathcal{V}^{2-\delta}}+\frac{a_{2}^{2}
f_{2-3\eta/2}(\tau_{1},\tau_{2})
 e^{-2a_{2}\tau_{2}}}{\mathcal{V}^{2-\eta}}-\frac{4
a_{2}\tau_{2}e^{-a_{2}\tau_{2}}}{\mathcal{V}^{2}}. \label{Flo}
\end{equation}
Now the equation $\frac{\partial V}{\partial \tau_{1}}=0$, admits
a solution of the form:
\begin{equation}
\mathcal{V}^{\eta-\delta}=\frac{2a_{1}^{3}
\tau_{1}^{2-3\delta/2}}{a_{2}^{2}\frac{\partial
f_{2-3\eta/2}}{\partial
\tau_{1}}}e^{2(a_{2}\tau_{2}-a_{1}\tau_{1})}, \label{Ppp}
\end{equation}
whereas $\frac{\partial V}{\partial \tau_{2}}=0$ gives:
\begin{equation}
\mathcal{V}^{\eta}=\frac{2
\tau_{2}}{a_{2}f_{2-3\eta/2}}e^{a_{2}\tau_{2}}. \label{Pp}
\end{equation}
The third minimisation equation $\frac{\partial V}{\partial
\mathcal{V}}=0$ looks like:
\begin{equation}
(\eta-2)a_{2}^{2}f_{2-3\eta/2}\mathcal{V}^{\eta}e^{-2a_{2}\tau_{2}}
+(\delta-2)a_{1}^{2}\tau_{1}^{2-3\delta/2}\mathcal{V}^{\delta}e^{-2a_{1}\tau_{1}}
+8a_{2}\tau_{2}e^{-a_{2}\tau_{2}}=0, \label{PPp}
\end{equation}
and substituting the results (\ref{Pp}) and (\ref{Ppp}), we
obtain:
\begin{equation}
2(\eta-2)+\frac{(\delta-2)}{a_{1}f_{2-3\eta/2}}\frac{\partial
f_{2-3\eta/2}}{\partial \tau_{1}}+8=0. \label{PPP}
\end{equation}
Solving the differential equation (\ref{PPP}), we realise that
$f_{2-3\eta/2}$ has an exponential behaviour in $\tau_{1}$ which
is in clear contrast with the requirement that it has to be
homogeneous. Following arguments very similar to this one it can
be seen that, as in the case with just one small modulus, the
presence of the $\alpha'$ corrections is crucial to find a LARGE
Volume minimum. In fact if we omit them, either it is impossible
to fix the small moduli large enough to ignore higher instanton
corrections or, once we integrate them out, we are left with a
run-away in the volume direction.

Lastly, we describe the final case in which it is possible to
prove the absence of a LARGE Volume vacuum. The leading terms in
(\ref{semPlifica}) are given by:
\begin{equation}
V \sim \frac{K^{-1}_{11}}{\mathcal{V}^{2+2/\beta}}
+\frac{K^{-1}_{22}}{\mathcal{V}^{4}}-\frac{\tau_{2}}
{\mathcal{V}^{3}}+\frac{1}{\mathcal{V}^{3}}, \label{USe}
\end{equation}
with $\beta<1$ and the axion minimum along the direction
$\mathcal{V}\sim e^{\beta a_{1}\tau_{1}}\sim e^{a_{2}\tau_{2}}$.
The necessary but not sufficient condition to fix the small
K\"{a}hler moduli is
$K^{-1}_{11}\simeq\mathcal{V}^{\delta}\tau_{1}^{2-3\delta/2}$ with
$\delta=2/\beta-1$, and
$K^{-1}_{22}=\mathcal{V}f_{1/2}(\tau_{1},\tau_{2})$. The relevant
part of the scalar potential (\ref{finalPotential}) takes the
form:
\begin{equation}
V \simeq \frac{a_{1}^{2}\tau_{1}^{2-3\delta/2}
e^{-2a_{1}\tau_{1}}}{\mathcal{V}^{2-\delta}}+\frac{a_{2}^{2}
f_{1/2}(\tau_{1},\tau_{2})
 e^{-2a_{2}\tau_{2}}}{\mathcal{V}}-\frac{4
a_{2}\tau_{2}e^{-a_{2}\tau_{2}}}{\mathcal{V}^{2}}
+\frac{3}{4}\frac{\hat{\xi}}{\mathcal{V}^{3}}. \label{UFlo}
\end{equation}
Now the equation $\frac{\partial V}{\partial \tau_{1}}=0$, admits
a solution of the form:
\begin{equation}
a_{1}^{2}\tau_{1}^{2-3\delta/2}\mathcal{V}^{\delta}e^{-2a_{1}\tau_{1}}=
\frac{a_{2}^{2}}{2a_{1}}\frac{\partial
f_{1/2}}{\partial\tau_{1}}\mathcal{V}e^{-2a_{2}\tau_{2}},
\label{UPp}
\end{equation}
whereas $\frac{\partial V}{\partial \tau_{2}}=0$ gives:
\begin{equation}
a_{2}\tau_{2}e^{-a_{2}\tau_{2}}=\frac{a_{2}^{2}}{2}f_{1/2}\mathcal{V}e^{-2a_{2}\tau_{2}}.
\label{UPpp}
\end{equation}
The third minimisation equation $\frac{\partial V}{\partial
\mathcal{V}}=0$ corresponds to:
\begin{equation}
(\delta-2)a_{1}^{2}\tau_{1}^{2-3\delta/2}\mathcal{V}^{\delta}e^{-2a_{1}\tau_{1}}
-a_{2}^{2}f_{1/2}\mathcal{V}e^{-2a_{2}\tau_{2}}-8a_{2}\tau_{2}e^{-a_{2}\tau_{2}}=
\frac{9}{4}\frac{\hat{\xi}}{\mathcal{V}}, \label{UPPp}
\end{equation}
and substituting the results (\ref{UPp}) and (\ref{UPpp}), we
obtain:
\begin{equation}
4a_{2}^2\mathcal{V}^{2}\left[\frac{(\delta-2)}{2
a_{1}}\frac{\partial f_{1/2}}{\partial\tau_{1}} -5 f_{1/2}\right]=
9\hat{\xi}e^{2a_{2}\tau_{2}}. \label{UPPPl}
\end{equation}
Now writing $f_{1/2}(\tau_{1},\tau_{2})=F(\frac{\beta
a_{1}}{a_{2}})\sqrt{\tau_{1}}$ for appropriate function $F$,
(\ref{UPPPl}) becomes:
\begin{equation}
\frac{a_{2}^2}{a_{1}\sqrt{\tau_{1}}}\mathcal{V}^{2}F\left(\frac{\beta
a_{1}}{a_{2}}\right)\left(\delta-2-20 a_{1}\tau_{1}\right)=
9\hat{\xi}e^{2a_{2}\tau_{2}}. \label{UQmuU}
\end{equation}
Given that a trustable minimum requires $a_{1}\tau_{1}\gg 1$, the
LHS of (\ref{UQmuU}) is negative while the RHS is definitively
positive and so this case does not allow us to find any LVS.

The general path that we shall follow to derive the conditions
which guarantee that we have enough terms with the correct volume
scaling to stabilise all the moduli at exponentially large volume,
is the following one. We learnt from the proof of the LARGE Volume
Claim for the case with just one small modulus $\tau_1$, that we
need to have two terms in the scalar potential with the same
volume scaling that depend on $\tau_1$ so that it can be
stabilised rather large in order to be able to neglect higher
instanton corrections. Then if we integrate out $\tau_1$, we have
to be left with at least two terms that depend on the overall
volume and have the same volume scaling. Lastly in order to find
the exponentially large volume minimum, the leading term at large
volume has to be negative. As we have seen before, the same
arguments apply here. Thus we shall first work out the conditions
to be able to fix both $a_{1}\tau_{1}\gg 1$ and $a_{2}\tau_{2}\gg
1$ by having at least two terms in the potential with a dependence
on these moduli and the same volume scaling. Then, we shall
imagine to integrate out these moduli, and derive the conditions
to be left with at least two terms dependent on $\mathcal{V}$ with
the leading one which is negative.

\newpage

\textbf{Case (1):} $K^{-1}_{12}\simeq \tau_{1}\tau_{2}$

\bigskip
The previous analysis tells us that, regardless of the particular
direction considered, the axion minimum is always in the case (b).
Thus we realise that at the minimum:
\begin{equation}
\left\langle V_{AX}\right\rangle =2\left(\left\vert
Y_{12}\right\vert -\left\vert X_{1}\right\vert-\left\vert
X_{2}\right\vert\right). \label{axione}
\end{equation}
Now recalling that in case (1) $K^{-1}_{12}\simeq
\tau_{1}\tau_{2}$, (\ref{finalPotential}) takes the form:
\begin{eqnarray}
V &\sim &\frac{1}{\mathcal{V}^{2}}\left[
\sum\limits_{j=1}^{2}K^{-1}_{jj} a_{j}^{2}
e^{-2a_{j}\tau_{j}}+2\tau_{1}\tau_{2}
a_{1}a_{2}e^{-(a_{1}\tau_{1}+a_{2}\tau_{2})}\right]  \notag \\
&&-4\frac{W_{0}}{\mathcal{V}^{2}}\sum\limits_{j=1}^{2}a_{j}\tau
_{j}e^{-a_{j}\tau_{j}}+\frac{3}{4}\frac{\hat{\xi}
}{\mathcal{V}^{3}}W_{0}^{2}. \label{finalpotential}
\end{eqnarray}
We shall now study the behaviour of (\ref{finalpotential}) by
taking the large volume limit along each direction
(\ref{DIrections}) and then considering all the possible forms of
$K^{-1}_{jj}$, $j=1,2$. When we take the large volume limit (I) of
(\ref{DIrections}), (\ref{finalpotential}) has the following
volume scaling:
\begin{equation}
V \sim \frac{K^{-1}_{11}}{\mathcal{V}^{2+2/\gamma}}
+\frac{K^{-1}_{22}}{\mathcal{V}^{2+2/\gamma}}
+\frac{\tau_{1}\tau_{2}}{\mathcal{V}^{2+2/\gamma}} -\frac{\tau
_{1}}{\mathcal{V}^{2+1/\gamma}}-\frac{\tau
_{2}}{\mathcal{V}^{2+1/\gamma}}+\frac{1}{\mathcal{V}^{3}}.
\label{semplifica}
\end{equation}
The third term in (\ref{semplifica}) is subleading with respect to
the fourth and the fifth. Thus it can be neglected:
\begin{equation}
V \sim \frac{K^{-1}_{11}}{\mathcal{V}^{2+2/\gamma}}
+\frac{K^{-1}_{22}}{\mathcal{V}^{2+2/\gamma}}-\frac{\tau
_{1}}{\mathcal{V}^{2+1/\gamma}}-\frac{\tau
_{2}}{\mathcal{V}^{2+1/\gamma}}+\frac{1}{\mathcal{V}^{3}}.
\label{Semplifica}
\end{equation}
We have seen that the presence of the $\alpha'$ corrections is
crucial to find the exponentially large volume minimum. Therefore
the fact that the third and the fourth terms in (\ref{Semplifica})
have to scale as $\mathcal{V}^{-3}$ tells us that $\gamma=1$.
\begin{equation}
V \sim \frac{K^{-1}_{11}}{\mathcal{V}^{4}}
+\frac{K^{-1}_{22}}{\mathcal{V}^{4}}-\frac{\tau
_{1}}{\mathcal{V}^{3}}-\frac{\tau
_{2}}{\mathcal{V}^{3}}+\frac{1}{\mathcal{V}^{3}}.
\label{SEmplifica}
\end{equation}
At this point it is straightforward to realise that if either
$K^{-1}_{11}$ or $K^{-1}_{22}$ were in case (4), then we would
have a run-away behaviour of the volume direction. Similarly the
situation with $K^{-1}_{11}$ and $K^{-1}_{22}$ either in case (1)
or (2) is not giving a LARGE Volume minimum since the first two
terms in (\ref{SEmplifica}) should be neglected without then the
possibility to stabilise $\tau_{1}$ and $\tau_{2}$ large. What
happens if either $K^{-1}_{11}$ or $K^{-1}_{22}$ is in case (3)
and the other one is either in case (1) or (2)? We do not find any
minimum. In fact, let us say that $K^{-1}_{11}$ is in case (3) and
$K^{-1}_{22}$ in case (1) or (2): then the second term in
(\ref{SEmplifica}) can be neglected. If we want to have still some
hope to stabilise $\tau_{2}$ large, then $K^{-1}_{11}$ should
better depend also on $\tau_{2}$: $K^{-1}_{11}\simeq
f_{1/2}(\tau_{1},\tau_{2})\mathcal{V}$. However this case has been
studied explicitly to show that it leads to an absurd. Thus only
if both $K^{-1}_{11}$ and $K^{-1}_{22}$ is in case (3) we can have
a LARGE Volume minimum.

On the contrary, if we took either the large volume limit (II) or
(III) in (\ref{DIrections}), (\ref{finalpotential}) would scale
as:
\begin{equation}
V \sim \frac{K^{-1}_{11}}{\mathcal{V}^{2+2/\beta}}
+\frac{K^{-1}_{22}}{\mathcal{V}^{2+2/\gamma}}
+\frac{\tau_{1}\tau_{2}}{\mathcal{V}^{2+1/\beta+1/\gamma}}
-\frac{\tau _{1}}{\mathcal{V}^{2+1/\beta}}-\frac{\tau
_{2}}{\mathcal{V}^{2+1/\gamma}}+\frac{1}{\mathcal{V}^{3}}.
\label{semplific}
\end{equation}
Let us focus on the direction (II) where $1/\beta>1/\gamma$. The
third and the fourth term in (\ref{semplific}) at large volume are
subdominant to the fifth and therefore they can be ignored:
\begin{equation}
V \sim \frac{K^{-1}_{11}}{\mathcal{V}^{2+2/\beta}}
+\frac{K^{-1}_{22}}{\mathcal{V}^{2+2/\gamma}}-\frac{\tau
_{2}}{\mathcal{V}^{2+1/\gamma}}+\frac{1}{\mathcal{V}^{3}},
\label{Semplific}
\end{equation}
If $1/\gamma>1$, then the $\mathcal{V}^{-3}$ term would be the
dominant one producing a run-away in the volume direction. Thus we
impose $1/\gamma\leq 1$. However we have already showed that the
situation with $1/\gamma< 1$ gives no LVS and so we deduce that we
need $1/\gamma=1$. Then we realise that the only possible
situation in which we can hope to fix $\tau_{2}$ large is when
either the first or the second term in (\ref{Semplific}) scales as
$\mathcal{V}^{-3}$. Now if the second term involving $K^{-1}_{22}$
were subleading with respect to the fourth term in
(\ref{Semplific}), then the first one should scale as
$\mathcal{V}^{-3}$. However at that point, knowing that
$K^{-1}_{11}$ will introduce a dependence on $\tau_{1}$, we would
not be able to stabilise $\tau_{1}$ large. Hence $K^{-1}_{22}$ has
to be in case (3): $K_{22}^{-1}=\mathcal{V}f_{1/2}(\tau
_{1},\tau_{2})$.

Now we have two different situations according to the fact that
$f_{1/2}$ indeed depends on both $\tau_{1}$ and $\tau_{2}$ or only
on $\tau_{2}$. The first possibility has already been studied with
the final conclusion that it produces no LVS. On the other hand,
when $K^{-1}_{22}$ depends only on $\tau_{2}$, i.e.
$K_{22}^{-1}\simeq\mathcal{V}\sqrt{\tau_{2}}$, we have that the
overall volume and $\tau_{2}$ are both stabilised by the interplay
of the second, the third and the fourth term in (\ref{Semplific}).
The first term is now subleading and can be used to fix $\tau_{1}$
if we write $K^{-1}_{11}\simeq
\mathcal{V}^{\alpha}\tau_{1}^{2-3\alpha/2}$ and then impose
$1/\beta=\alpha$ in order to make it scale as the fourth term in
(\ref{semplific}).

We point out that these results apply also to the direction (III)
where $1/\gamma>1/\beta$, if we swap $\gamma$ with $\beta$ and
$\tau_{1}$ with $\tau_{2}$. Let us finally summarise in the table
below what we have found for this case.

\newpage

\textbf{Case (1): $K^{-1}_{12}\simeq\tau_{1}\tau_{2}$}

\begin{center}
\begin{tabular}{|c|c|c|c|c|}
\hline $K_{11}^{-1}$ & $K_{22}^{-1}$ & (I), (b) & (II), (b) &
(III), (b) \\ \hline 1 & 1 & NO & NO & NO \\ \hline 1 & 2 & NO &
NO & NO \\ \hline 1 & 3 & NO & NO & NO \\ \hline 1 & 4 & NO & NO &
NO \\ \hline 2 & 1 & NO & NO & NO \\ \hline 2 & 2 & NO & NO & NO
\\ \hline 2 & 3 & NO & NO & NO \\ \hline 2 & 4 & NO & NO & NO \\
\hline 3 & 1 & NO & NO & NO \\ \hline 3 & 2 & NO & NO & NO \\
\hline 3 & 3 & OK, $\gamma =1$ & NO & NO \\ \hline 3 & 4 & NO & NO
& OK, $\beta =1$, ($\ast \ast $) \\ \hline 4 & 1 & NO & NO & NO \\
\hline 4 & 2 & NO & NO & NO \\ \hline 4 & 3 & NO & OK, $\gamma
=1$, ($\ast $) & NO \\ \hline 4 & 4 & NO & NO & NO \\ \hline
\end{tabular}
\end{center}
($\ast $) $ \ K_{22}^{-1}\simeq \mathcal{V}\sqrt{\tau _{2}}$, $%
K_{11}^{-1}\simeq \mathcal{V}^{\alpha }\tau _{1}^{2-3\alpha /2}$ with $\frac{%
1}{\beta }=\alpha $
\newline
($\ast \ast $) $K_{11}^{-1}\simeq \mathcal{V}\sqrt{\tau _{1}}$, $%
K_{22}^{-1}\simeq \mathcal{V}^{\alpha }\tau _{2}^{2-3\alpha /2}$ with $\frac{%
1}{\gamma }=\alpha $
\bigskip

\textbf{Case (2):} $K^{-1}_{12}\simeq
\mathcal{V}^{\alpha}g_{2-3\alpha/2}(\tau_{1},\tau_{2})$,
$0<\alpha<1$

\bigskip
This case is more involved than the previous one since, depending
on the direction chosen for the large volume limit and the exact
value of the parameter $\alpha$, the axion minimum can not only be
in case (b), but also in (c) and (d). Let us start by considering
each case in detail:
\begin{enumerate}

\item[$\bullet$] axion minimum at (c) along direction (II) for
$\alpha=1/\gamma<2/3$ or $1/\gamma<\alpha<1$

We can easily conclude that no LVS is present given that, looking
at the general volume scaling of the scalar potential
(\ref{semPlifica}), we can notice that the fifth term would be
dominant with respect to the last one since we have always
$1/\gamma<1$. Thus the $\alpha'$ correction would be negligible at
large volume, so producing no LVS.

\item[$\bullet$] axion minimum at (d) along direction (III) for
$\alpha=1/\beta<2/3$ or $1/\beta<\alpha<1$

This situation looks like the previous one if we swap $\gamma$
with $\beta$ and $\tau_{1}$ with $\tau_{2}$, therefore we conclude
that no LARGE Volume minimum will be present.

\item[$\bullet$] axion minimum at (b) along direction (II) for
$\alpha<1/\gamma$ or $2/3<\alpha=1/\gamma<1$

First of all we realise that the situation with
$2/3<\alpha=1/\gamma<1$ does not give rise to any LVS because the
leading order $\alpha'$ correction would be negligible at large
volume. We shall therefore focus on the case $\alpha<1/\gamma$.
The scalar potential (\ref{finalPotential}) takes the form:
\begin{eqnarray}
V &\sim &\frac{1}{\mathcal{V}^{2}}\left[
\sum\limits_{j=1}^{2}K^{-1}_{jj} a_{j}^{2} e^{-2a_{j}\tau_{j}}+2
a_{1}a_{2}g_{2-3\alpha/2}(\tau_{1},\tau_{2})\mathcal{V}^{\alpha}
e^{-(a_{1}\tau_{1}+a_{2}\tau_{2})}\right]  \notag \\
&&-4\frac{W_{0}}{\mathcal{V}^{2}}\left(a_{1}\tau
_{1}e^{-a_{1}\tau_{1}}+a_{2}\tau
_{2}e^{-a_{2}\tau_{2}}\right)+\frac{3}{4}\frac{\hat{\xi}
}{\mathcal{V}^{3}}W_{0}^{2}. \label{finalpotentialnn}
\end{eqnarray}
When we take the large volume limit (II) of (\ref{DIrections}),
(\ref{finalpotentialnn}) has the following volume scaling:
\begin{equation}
V \sim \frac{K^{-1}_{11}}{\mathcal{V}^{2+2/\beta}}
+\frac{K^{-1}_{22}}{\mathcal{V}^{2+2/\gamma}}
+\frac{g_{2-3\alpha/2}(\tau_{1},\tau_{2})}
{\mathcal{V}^{2+1/\beta+1/\gamma-\alpha}} -\frac{\tau
_{1}}{\mathcal{V}^{2+1/\beta}}-\frac{\tau
_{2}}{\mathcal{V}^{2+1/\gamma}}+\frac{1}{\mathcal{V}^{3}}.
\label{semplificann}
\end{equation}
Setting $1/\gamma=1$ and recalling that in this direction
$1/\beta>1/\gamma$, the dominant terms in (\ref{semplificann})
become:
\begin{equation}
V \sim \frac{K^{-1}_{11}}{\mathcal{V}^{2+2/\beta}}
+\frac{K^{-1}_{22}}{\mathcal{V}^{4}}-\frac{\tau
_{2}}{\mathcal{V}^{3}}+\frac{1}{\mathcal{V}^{3}}. \label{SEMPL}
\end{equation}
Now by noticing that equation (\ref{SEMPL}) has the same form of
(\ref{Semplific}) if we set $1/\gamma=1$, we can just repeat the
same consideration made before and obtain that $K^{-1}_{22}$ has
to be in case (3). Moreover if $K^{-1}_{22}$ depends on both
$\tau_{1}$ and $\tau_{2}$, then there is no LARGE Volume minimum,
but when $K^{-1}_{22}$ depends only on $\tau_{2}$, the first term
in (\ref{SEMPL}) is negligible at large volume and can be used to
fix $\tau_{1}$ if we make it compete with the fourth term in
(\ref{semplificann}) by writing $K^{-1}_{11}\simeq
\mathcal{V}^{\delta}\tau_{1}^{2-3\delta/2}$ and then imposing
$1/\beta=\delta$.

\item[$\bullet$] axion minimum at (b) along direction (III) for
$\alpha<1/\beta$ or $2/3<\alpha=1/\beta<1$

This situation looks like the previous one if we swap $\gamma$
with $\beta$ and $\tau_{1}$ with $\tau_{2}$, therefore we do not
need to discuss this case.

\item[$\bullet$] axion minimum at (b) along direction (I) for
$\alpha<1/\gamma$ or $2/3<\alpha=1/\gamma<1$

In this situation the full scalar potential still looks like
(\ref{finalpotentialnn}), but the volume scaling behaviour of its
terms now reads:
\begin{equation}
V \sim \frac{K^{-1}_{11}}{\mathcal{V}^{2+2/\gamma}}
+\frac{K^{-1}_{22}}{\mathcal{V}^{2+2/\gamma}}
+\frac{g_{2-3\alpha/2}(\tau_{1},\tau_{2})}
{\mathcal{V}^{2+2/\gamma-\alpha}} -\frac{\tau
_{1}}{\mathcal{V}^{2+1/\gamma}}-\frac{\tau
_{2}}{\mathcal{V}^{2+1/\gamma}}+\frac{1}{\mathcal{V}^{3}}.
\label{semplificanno}
\end{equation}
For $2/3<\alpha=1/\gamma<1$ the last term in (\ref{semplificanno})
would be subdominant with respect to the fifth one, but we know
that its presence is crucial to find a minimum and so we can
conclude that this case admits no minimum. On the other hand for
$\alpha<1/\gamma$, setting $1/\gamma=1$, the dominant terms in
(\ref{semplificanno}) become:
\begin{equation}
V \sim \frac{K^{-1}_{11}}{\mathcal{V}^{4}}
+\frac{K^{-1}_{22}}{\mathcal{V}^{4}}-\frac{\tau
_{1}}{\mathcal{V}^{3}}-\frac{\tau
_{2}}{\mathcal{V}^{3}}+\frac{1}{\mathcal{V}^{3}}. \label{SEMPLI}
\end{equation}
We immediately realise that (\ref{SEMPLI}) is absolutely similar
to (\ref{SEmplifica}). We can therefore repeat exactly the same
analysis and conclude that only if both $K^{-1}_{11}$ and
$K^{-1}_{22}$ is in case (3) we can have a LARGE Volume minimum.
Let us finally summarise in the table below what we have found for
this case.
\end{enumerate}

\textbf{Case (2):
$K^{-1}_{12}=\mathcal{V}^{\alpha}g_{2-3\alpha/2}(\tau_{1},\tau_{2})$,
$0<\alpha<1$}

\begin{center}
\begin{tabular}{|c|c|c|c|c|c|c|}
\hline $K_{11}^{-1}$ & $K_{22}^{-1}$ & $\underset{\alpha \leq
1/\gamma }{(I),(b),}$ & $\underset{\alpha \leq 1/\gamma
}{(II),(b),}$  & $\underset{\alpha \geq 1/\gamma }{(II),(c),}$  &
$\underset{\alpha \leq 1/\beta }{(III),(b),}$ & $\underset{\alpha
\geq 1/\beta }{(III),(d),}$  \\ \hline 1 & 1 &
NO & NO & NO & NO & NO \\ \hline 1 & 2 & NO & NO & NO & NO & NO \\
\hline 1 & 3 & NO & NO & NO & NO & NO \\ \hline 1 & 4 & NO & NO &
NO & NO & NO \\ \hline 2 & 1 & NO & NO & NO & NO & NO \\ \hline 2
& 2 & NO & NO & NO & NO & NO \\ \hline 2 & 3 & NO & NO & NO & NO &
NO \\ \hline 2 & 4 & NO & NO & NO & NO & NO \\ \hline 3 & 1 & NO &
NO & NO & NO & NO \\ \hline 3 & 2 & NO & NO & NO & NO & NO \\
\hline 3 & 3 & OK, $\gamma =1$ & NO & NO & NO & NO \\ \hline 3 & 4
& NO & NO & NO & OK, $\beta=1$ ($\ast\ast$) & NO \\ \hline 4 & 1 &
NO & NO & NO & NO & NO
\\ \hline 4 & 2 & NO & NO & NO & NO & NO \\ \hline 4 & 3 & NO &
OK, $\gamma=1$ ($\ast$) & NO & NO & NO \\ \hline 4 & 4 & NO & NO &
NO & NO & NO
\\
\hline
\end{tabular}
\end{center}
($\ast $) $\ K_{22}^{-1}\simeq \mathcal{V}\sqrt{\tau _{2}}$, $%
K_{11}^{-1}\simeq \mathcal{V}^{\delta }\tau _{1}^{2-3\delta /2}$
with $\frac{1}{\beta }=\delta $
\newline
($\ast \ast $) $K_{11}^{-1}\simeq \mathcal{V}\sqrt{\tau _{1}}$, $%
K_{22}^{-1}\simeq \mathcal{V}^{\delta }\tau _{2}^{2-3\delta /2}$
with $\frac{1}{\gamma }=\delta $
\bigskip

\textbf{Case (3):} $K^{-1}_{12}=
\mathcal{V}f_{1/2}(\tau_{1},\tau_{2})$

\bigskip
We shall now consider each particular situation according to the
different possible locations of the axion minimum:
\begin{enumerate}

\item[$\bullet$] axion minimum at (c) along direction (II) for
$\gamma>1$

Thus at the minimum:
\begin{equation}
\left\langle V_{AX}\right\rangle =2\left(\left\vert
X_{1}\right\vert-\left\vert Y_{12}\right\vert-\left\vert
X_{2}\right\vert\right). \label{axioneno}
\end{equation}
Therefore the full potential (\ref{finalPotential}) looks like:
\begin{eqnarray}
V &\sim &\frac{1}{\mathcal{V}^{2}}\left[
\sum\limits_{j=1}^{2}K^{-1}_{jj} a_{j}^{2} e^{-2a_{j}\tau_{j}}-2
a_{1}a_{2}f_{1/2}(\tau_{1},\tau_{2})\mathcal{V}
e^{-(a_{1}\tau_{1}+a_{2}\tau_{2})}\right]  \notag \\
&&+4\frac{W_{0}}{\mathcal{V}^{2}}\left(a_{1}\tau
_{1}e^{-a_{1}\tau_{1}}-a_{2}\tau
_{2}e^{-a_{2}\tau_{2}}\right)+\frac{3}{4}\frac{\hat{\xi}
}{\mathcal{V}^{3}}W_{0}^{2}. \label{finalpotentialno}
\end{eqnarray}
When we take the large volume limit (II) of (\ref{DIrections}),
(\ref{finalpotentialno}) has the following volume scaling:
\begin{equation}
V \sim \frac{K^{-1}_{11}}{\mathcal{V}^{2+2/\beta}}
+\frac{K^{-1}_{22}}{\mathcal{V}^{2+2/\gamma}}
-\frac{f_{1/2}(\tau_{1},\tau_{2})}
{\mathcal{V}^{1+1/\beta+1/\gamma}} +\frac{\tau
_{1}}{\mathcal{V}^{2+1/\beta}}-\frac{\tau
_{2}}{\mathcal{V}^{2+1/\gamma}}+\frac{1}{\mathcal{V}^{3}}.
\label{semplificano}
\end{equation}
Given that $1/\gamma<1$, the leading order $\alpha'$ correction
would be subleading in a large volume limit. However, we know that
its presence is crucial to find the minimum and so we conclude
that this case does not present any new LARGE Volume vacuum.

\item[$\bullet$] axion minimum at (d) along direction (III) for
$\beta>1$

This situation looks like the previous one if we swap $\gamma$
with $\beta$ and $\tau_{1}$ with $\tau_{2}$, therefore we conclude
that no LARGE Volume minimum is present.

\item[$\bullet$] axion minimum at (b) along direction (I) for
$0<\gamma\leq 1$

Thus at the minimum:
\begin{equation}
\left\langle V_{AX}\right\rangle =2\left(\left\vert
Y_{12}\right\vert-\left\vert X_{1}\right\vert-\left\vert
X_{2}\right\vert\right),
\end{equation}
and so the full potential (\ref{finalPotential}) becomes (setting
$W_{0}=1$):
\begin{eqnarray}
V &\sim &\frac{1}{\mathcal{V}^{2}}\left[
\sum\limits_{j=1}^{2}K^{-1}_{jj} a_{j}^{2} e^{-2a_{j}\tau_{j}}+2
a_{1}a_{2}f_{1/2}(\tau_{1},\tau_{2})\mathcal{V}
e^{-(a_{1}\tau_{1}+a_{2}\tau_{2})}\right]  \notag \\
&&-\frac{4}{\mathcal{V}^{2}}\left(a_{1}\tau
_{1}e^{-a_{1}\tau_{1}}+a_{2}\tau
_{2}e^{-a_{2}\tau_{2}}\right)+\frac{3}{4}\frac{\hat{\xi}
}{\mathcal{V}^{3}}. \label{finalpOtentialnnl}
\end{eqnarray}
When we take the large volume limit (I) of (\ref{DIrections}),
(\ref{finalpOtentialnnl}) has the following volume scaling:
\begin{equation}
V \sim \frac{K^{-1}_{11}}{\mathcal{V}^{2+2/\gamma}}
+\frac{K^{-1}_{22}}{\mathcal{V}^{2+2/\gamma}}
+\frac{f_{1/2}(\tau_{1},\tau_{2})} {\mathcal{V}^{1+2/\gamma}}
-\frac{\tau _{1}}{\mathcal{V}^{2+1/\gamma}}-\frac{\tau
_{2}}{\mathcal{V}^{2+1/\gamma}}+\frac{1}{\mathcal{V}^{3}}.
\label{semplificannolP}
\end{equation}
In the case where $1/\gamma>1$, the leading part of
(\ref{semplificannolP}) at large volume takes the form:
\begin{equation}
V \sim \frac{K^{-1}_{11}}{\mathcal{V}^{2+2/\gamma}}
+\frac{K^{-1}_{22}}{\mathcal{V}^{2+2/\gamma}}+\frac{1}{\mathcal{V}^{3}}.
\label{SEMPLIlh}
\end{equation}
We have already checked explicitly that this situation does not
present any LARGE Volume minimum, therefore we shall focus only on
the case $1/\gamma=1$, for which the volume scaling
(\ref{semplificannolP}) becomes:
\begin{equation}
V \sim \frac{K^{-1}_{11}}{\mathcal{V}^{4}}
+\frac{K^{-1}_{22}}{\mathcal{V}^{4}}
+\frac{f_{1/2}(\tau_{1},\tau_{2})} {\mathcal{V}^{3}} -\frac{\tau
_{1}}{\mathcal{V}^{3}}-\frac{\tau
_{2}}{\mathcal{V}^{3}}+\frac{1}{\mathcal{V}^{3}}. \label{simple}
\end{equation}
We immediately realise that as soon as either $K^{-1}_{11}$ or
$K^{-1}_{22}$ is in case (4) then the $\alpha'$ corrections would
be subleading in a large volume expansion. Thus we can reject this
possibility. Then we are left with three different situations.
Firstly when both $K^{-1}_{11}$ and $K^{-1}_{22}$ are subdominant
with respect to the last three terms in (\ref{simple}) (i.e. each
$K^{-1}_{jj}$ with $j=1,2$ is either in case (1) or (2)) the
scalar potential (\ref{finalpOtentialnnl}) takes the form:
\begin{equation}
V=\frac{2}
{\mathcal{V}}a_{1}a_{2}f_{1/2}(\tau_{1},\tau_{2})e^{-(a_{1}\tau_{1}+a_{2}\tau_{2})}
-\frac{4}{\mathcal{V}^{2}}\left(a_{1}\tau_{1}
e^{-a_{1}\tau_{1}}+a_{2}\tau_{2}e^{-a_{2}\tau_{2}}\right)+\frac{3}{4}\frac{\hat{\xi}
}{\mathcal{V}^{3}}. \label{Sp}
\end{equation}
Since we have to find a minimum such that $a_{j}\tau_{j}\gg 1$ for
$j=1,2$, we can work at leading order in a
$\frac{1}{a_{1}\tau_{1}}$ and $\frac{1}{a_{2}\tau_{2}}$ expansion
and obtain that $\frac{\partial^{2}V}{\partial \tau_{1}^{2}}\simeq
0$ due to the presence of just one exponential in $\tau_{1}$ in
both the first and the second term in (\ref{Sp}). In fact, if we
are interested in the dependence of $V$ on just $\tau_{1}$,
(\ref{Sp}) can be rewritten as:
\begin{equation}
V=c_{1}e^{-a_{1}\tau_{1}}\left(a_{2}f_{1/2}(\tau_{1},\tau_{2})e^{-a_{2}\tau_{2}}
-\frac{2\tau_{1}}{\mathcal{V}}\right) +c_{2}\equiv
c_{1}e^{-a_{1}\tau_{1}}g(\tau_{1})+c_{2}, \label{Spp}
\end{equation}
where $c_{1}$ and $c_{2}$ are constants and $g(\tau_{1})$ is the
sum of two homogeneous functions in $\tau_{1}$. Therefore at
leading order in a $\frac{1}{a_{1}\tau_{1}}$ expansion, we get:
\begin{equation}
\frac{\partial V}{\partial \tau_{1}}\simeq
-a_{1}c_{1}e^{-a_{1}\tau_{1}}g(\tau_{1})=0\Leftrightarrow
g(\tau_{1})=0,
\end{equation}
which implies:
\begin{equation}
\left. \frac{\partial ^{2}V}{\partial \tau _{1}^{2}}\right\vert
_{\min }\simeq \left. a_{1}^{2}c_{1}e^{-a_{1}\tau _{1}}g(\tau
_{1})\right\vert_{\min }=0.
\end{equation}
Similarly we have $\frac{\partial^{2}V}{\partial
\tau_{2}^{2}}\simeq 0$, whereas:
\begin{equation}
\left. \frac{\partial ^{2}V}{\partial \tau
_{1}\partial\tau_{2}}\right\vert_{\min }\simeq \left.
2a_{1}^{2}a_{2}^{2}f_{1/2}e^{-a_{1}\tau
_{1}}\frac{e^{-a_{2}\tau_{2}}}{\mathcal{V}}\right\vert_{\min
}\equiv c_{3}>0.
\end{equation}
Therefore considering $\mathcal{V}$ constant, the Hessian matrix
will look like:
\begin{equation}
\mathcal{H}\simeq
\begin{pmatrix}
0 & c_{3} \\
c_{3} & 0
\end{pmatrix}
\Longrightarrow \det \mathcal{H}=-c_{3}^{2}<0,
\end{equation}
so implying that we can never have a minimum.

Secondly we have to contemplate the possibility that only one of
the first two terms in (\ref{simple}) is competing with the last
three ones while the other is negligible. However even this case
does not yield any new LVS due the asymmetry of the dependence of
the scalar potential in $\tau_{1}$ and $\tau_{2}$ that does not
allow us to stabilise the small K\"{a}hler moduli large enough. In
fact, let us assume for example that $K^{-1}_{11}$ is in case (1)
or (2) and hence is negligible, whereas
$K^{-1}_{22}\simeq\mathcal{V}\sqrt{\tau_{2}}$. Consequently we
obtain:
\begin{eqnarray}
\frac{\partial V}{\partial \tau _{1}} &=&0\Longleftrightarrow
\mathcal{V}=
\frac{2\tau _{1}}{a_{2}f_{1/2}}e^{a_{2}\tau _{2}}, \label{qaz1} \\
\frac{\partial V}{\partial \tau _{2}} &=&0\Longleftrightarrow
\mathcal{V}= \frac{2\tau _{2}}{a_{1}f_{1/2}e^{-a_{1}\tau
_{1}}+a_{2}\sqrt{\tau _{2}} e^{-a_{2}\tau _{2}}}. \label{qaz2}
\end{eqnarray}
Now combining (\ref{qaz1}) with (\ref{qaz2}) we find:
\begin{equation}
a_{1}\tau _{1}f_{1/2}e^{a_{2}\tau_{2}-a_{1}\tau_{1}}+a_{2}\tau_{1}
\sqrt{\tau_{2}}=a_{2}\tau_{2}f_{1/2},
\end{equation}
which evaluated along the direction $a_{1}\tau_{1}\simeq
a_{2}\tau_{2}$ where the axion minimum is located, becomes:
\begin{equation}
a_{1}\tau _{1}f_{1/2}+a_{2}\tau _{1}\sqrt{\tau _{2}}\simeq
a_{1}\tau _{1}f_{1/2}\Longleftrightarrow \sqrt{a_{1}a_{2}}\tau
_{1}^{3/2}\simeq 0,
\end{equation}
which is the negative result we mentioned above. Lastly when both
$K^{-1}_{11}$ and $K^{-1}_{22}$ is in case (3) all the terms in
(\ref{simple}) have the same volume scaling. It can be seen that,
regardless of the form of $f_{1/2}$, the LARGE Volume minimum is
always present.

\item[$\bullet$] axion minimum at (b) along direction (II) for
$0<\gamma\leq 1$

In this situation the full scalar potential still looks like
(\ref{finalpOtentialnnl}), but the volume scaling behaviour of its
terms now reads:
\begin{equation}
V \sim \frac{K^{-1}_{11}}{\mathcal{V}^{2+2/\beta}}
+\frac{K^{-1}_{22}}{\mathcal{V}^{2+2/\gamma}}
+\frac{f_{1/2}(\tau_{1},\tau_{2})}
{\mathcal{V}^{1+1/\beta+1/\gamma}} -\frac{\tau
_{1}}{\mathcal{V}^{2+1/\beta}}-\frac{\tau
_{2}}{\mathcal{V}^{2+1/\gamma}}+\frac{1}{\mathcal{V}^{3}}.
\label{semplIficannl}
\end{equation}
Given that along the direction (II) $1/\beta>1/\gamma$, and the
axion minimum is present for $1/\gamma\geq 1$, the dominant terms
in (\ref{semplIficannl}) become:
\begin{equation}
V \sim \frac{K^{-1}_{11}}{\mathcal{V}^{2+2/\beta}}
+\frac{K^{-1}_{22}}{\mathcal{V}^{4}}-\frac{\tau
_{2}}{\mathcal{V}^{3}}+\frac{1}{\mathcal{V}^{3}}, \label{qwa}
\end{equation}
where we have already set $\gamma=1$ because, as we argued before,
this is the only possible situation when we can hope to find a
LARGE Volume minimum. We notice now that whenever the first two
terms in (\ref{qwa}) are negligible at large volume, then we have
only one term in $V$ dependent on $\tau_{2}$ and so we can never
obtain a minimum at $a_{2}\tau_{2}\gg 1$. This happens if
$K^{-1}_{22}$ is either in case (1) or in case (2) and
$K^{-1}_{11}$ is in case (1), (2), (3) or even in case (4) if its
term in (\ref{qwa}) still goes like $\mathcal{V}^{\alpha}$,
$\alpha>3$. Moreover if $K^{-1}_{22}$ is in case (4) then it would
beat the last two terms in (\ref{qwa}) so giving no LVS.

On the other hand, when only the first term in (\ref{qwa}) is
negligible at large volume, we have the possibility to find a new
LVS if $K^{-1}_{22}$ is in case (3) and does not depend on
$\tau_{1}$, that is $K^{-1}_{22}\simeq
\mathcal{V}\sqrt{\tau_{2}}$. In fact, at leading order in a large
volume expansion the scalar potential looks like the one we
studied for the case with just one small modulus and we know that
the corresponding LARGE Volume minimum would be present. However
we have still to fix $\tau_{1}$. If $K^{-1}_{11}$ is in case (1),
(2) or (3) then the first term in (\ref{semplIficannl}) is always
subleading with respect to the third and the fourth one and
therefore it can be neglected. We can now focus on the third and
fourth term in (\ref{semplIficannl}) which have the same volume
scaling. Then combining the solution of $\frac{\partial
V}{\partial \tau_{1}}=0$ and $\frac{\partial V}{\partial
\tau_{2}}=0$ we end up with $\tau_{1}=\tau_{2}$ which is correct
if we choose $\beta a_{1}=a_{2}$, $\beta<1$. However we have still
to check the sign of $\frac{\partial^{2} V}{\partial
\tau_{1}^{2}}$ which turns out to be positive only if, writing
$f_{1/2}(\tau_{1},\tau_{2})\sim\tau_{1}^{\alpha}\tau_{2}^{1/2-\alpha}$
for arbitrary $\alpha$, we have $\alpha<1$.

The situation when $K^{-1}_{22}\simeq \mathcal{V}\sqrt{\tau_{2}}$
and $K^{-1}_{11}$ is in case (4) needs to be studied more
carefully. Writing
$K^{-1}_{11}\simeq\mathcal{V}^{\delta}\tau_{1}^{2-3\delta/2}$,
$\delta>1$, if $\delta>\frac{2}{\beta}+1$ then the first term in
(\ref{qwa}) beats all the other ones so giving no LVS. If
$\delta=\frac{2}{\beta}+1$, then the first term in (\ref{qwa})
scales as the other ones but we have already shown that this is
not an interesting situation. The only way to get a LVS is to
impose $\delta\leq \frac{1}{\beta}$ to make the first term in
(\ref{semplIficannl}) scale as the third and fourth term or to
render it subdominant with respect to them.

Finally if $K^{-1}$ is in case (4) and $K^{-1}_{22}$ is either in
case (1) or (2) then we don't find any new LVS since the second
term in (\ref{qwa}) would be subleading with respect to the other
ones so leaving just one term, the first one, which depends on
$\tau_{1}$.

\item[$\bullet$] axion minimum at (b) along direction (III) for
$0<\gamma<\beta\leq 1$

This situation looks like the previous one if we swap $\gamma$
with $\beta$ and $\tau_{1}$ with $\tau_{2}$, therefore we do not
need to discuss this case. Let us finally summarise in the table
below what we have found for this case.
\end{enumerate}

\newpage

\textbf{Case (3):
$K^{-1}_{12}=\mathcal{V}f_{1/2}(\tau_{1},\tau_{2})$}
\begin{center}
\begin{tabular}{|c|c|c|c|c|c|c|}
\hline
$K_{11}^{-1}$ & $K_{22}^{-1}$ & $\underset{0<\gamma \leq 1}{(I),(b),}$ & $%
\underset{0<\gamma \leq 1}{(II),(b),}$ & $\underset{\gamma
>1}{(II),(c),}$ &
$\underset{0<\gamma <\beta \leq 1}{(III),(b),}$ & $\underset{\gamma \geq 1}{%
(III),(d),}$ \\ \hline 1 & 1 & NO & NO & NO & NO & NO \\ \hline 1
& 2 & NO & NO & NO & NO & NO \\ \hline 1 & 3 & NO & OK, ($\ast $)
& NO & NO & NO \\ \hline 1 & 4 & NO & NO & NO & NO & NO \\ \hline
2 & 1 & NO & NO & NO & NO & NO \\ \hline 2 & 2 & NO & NO & NO & NO
& NO \\ \hline 2 & 3 & NO & OK, ($\ast $) & NO & NO & NO \\ \hline
2 & 4 & NO & NO & NO & NO & NO \\ \hline 3 & 1 & NO & NO & NO &
OK, ($\star $) & NO \\ \hline 3 & 2 & NO & NO & NO & OK, ($\star
$) & NO \\ \hline 3 & 3 & OK, $\gamma =1$ & OK, ($\ast $) & NO &
OK, ($\star $) & NO \\ \hline 3 & 4 & NO & NO & NO & OK, ($\star
\star $) & NO \\ \hline 4 & 1 & NO & NO & NO & NO & NO \\ \hline 4
& 2 & NO & NO & NO & NO & NO \\ \hline 4 & 3 & NO & OK, ($\ast
\ast $) & NO & NO & NO \\ \hline 4 & 4 & NO & NO & NO & NO & NO \\
\hline
\end{tabular}
\end{center}

($\ast $) $\beta <\gamma=1$, $K_{22}^{-1}\simeq
\mathcal{V}\sqrt{\tau _{2}}$,
$f_{1/2}\sim\tau_{1}^{\alpha}\tau_{2}^{1/2-\alpha}$ with
$\alpha<1$

\bigskip
($\ast\ast$) $\beta<\gamma =1$, $K_{22}^{-1}\simeq
\mathcal{V}\sqrt{\tau_{2}}$,
$K^{-1}_{11}\simeq\mathcal{V}^{\delta}\tau_{1}^{2-3\delta/2}$ and
$f_{1/2}\sim\tau_{1}^{\alpha}\tau_{2}^{1/2-\alpha}$ with
$\alpha<1$ for $1<\delta<\frac{1}{\beta}$ and $\forall\alpha$ for
$\delta=\frac{1}{\beta}$

\bigskip
($\star $) $\gamma <\beta =1$, $K_{11}^{-1}\simeq\mathcal{V}
\sqrt{\tau_{1}}$,
$f_{1/2}\sim\tau_{2}^{\alpha}\tau_{1}^{1/2-\alpha}$ with
$\alpha<1$

\bigskip
($\star\star$) $\gamma <\beta =1$, $K_{11}^{-1}\simeq
\mathcal{V}\sqrt{\tau_{1}}$,
$K^{-1}_{22}\simeq\mathcal{V}^{\delta}\tau_{2}^{2-3\delta/2}$ and
$f_{1/2}\sim\tau_{2}^{\alpha}\tau_{1}^{1/2-\alpha}$ with
$\alpha<1$ for $1<\delta<\frac{1}{\gamma}$ and $\forall\alpha$ for
$\delta=\frac{1}{\gamma}$

\bigskip

\textbf{Case (4):} $K^{-1}_{12}\simeq
\mathcal{V}^{\alpha}h_{2-3\alpha/2}(\tau_{1},\tau_{2})$,
$\alpha>1$

\bigskip
Let us focus on each particular situation according to the
different possible positions of the axion minimum:
\begin{enumerate}

\item[$\bullet$] axion minimum at (c) along direction (II) for
$\alpha>1/\gamma$

Thus at the minimum:
\begin{equation}
\left\langle V_{AX}\right\rangle =2\left(\left\vert
X_{1}\right\vert-\left\vert Y_{12}\right\vert-\left\vert
X_{2}\right\vert\right). \label{axionenl}
\end{equation}
Therefore the full scalar potential (\ref{finalPotential}) reads:
\begin{eqnarray}
V &\sim &\frac{1}{\mathcal{V}^{2}}\left[
\sum\limits_{j=1}^{2}K^{-1}_{jj} a_{j}^{2} e^{-2a_{j}\tau_{j}}-2
a_{1}a_{2}h_{2-3\alpha/2}(\tau_{1},\tau_{2})\mathcal{V}^{\alpha}
e^{-(a_{1}\tau_{1}+a_{2}\tau_{2})}\right]  \notag \\
&&+4\frac{W_{0}}{\mathcal{V}^{2}}\left(a_{1}\tau
_{1}e^{-a_{1}\tau_{1}}-a_{2}\tau
_{2}e^{-a_{2}\tau_{2}}\right)+\frac{3}{4}\frac{\hat{\xi}
}{\mathcal{V}^{3}}W_{0}^{2}. \label{finalpotentialnl}
\end{eqnarray}
When we take the large volume limit (II) of (\ref{DIrections}),
(\ref{finalpotentialnl}) has the following volume scaling:
\begin{equation}
V \sim \frac{K^{-1}_{11}}{\mathcal{V}^{2+2/\beta}}
+\frac{K^{-1}_{22}}{\mathcal{V}^{2+2/\gamma}}
-\frac{h_{2-3\alpha/2}(\tau_{1},\tau_{2})}
{\mathcal{V}^{2+1/\beta+1/\gamma-\alpha}} +\frac{\tau
_{1}}{\mathcal{V}^{2+1/\beta}}-\frac{\tau
_{2}}{\mathcal{V}^{2+1/\gamma}}+\frac{1}{\mathcal{V}^{3}}.
\label{semplificanl}
\end{equation}
Recalling that in this direction $1/\beta>1/\gamma$, the dominant
terms in (\ref{semplificanl}) are:
\begin{equation}
V \sim \frac{K^{-1}_{11}}{\mathcal{V}^{2+2/\beta}}
+\frac{K^{-1}_{22}}{\mathcal{V}^{2+2/\gamma}}
-\frac{h_{2-3\alpha/2}(\tau_{1},\tau_{2})}
{\mathcal{V}^{2+1/\beta+1/\gamma-\alpha}}-\frac{\tau
_{2}}{\mathcal{V}^{2+1/\gamma}}+\frac{1}{\mathcal{V}^{3}}.
\label{eqproof}
\end{equation}
We know that the presence of the last term in (\ref{eqproof}) is
crucial to find the LARGE Volume minimum, so in order not to make
it subleading with respect to the fourth one, we need to have
$1/\gamma\geq 1$. If $1/\gamma=1$, the new volume scaling looks
like:
\begin{equation}
V \sim \frac{K^{-1}_{11}}{\mathcal{V}^{2+2/\beta}}
+\frac{K^{-1}_{22}}{\mathcal{V}^{4}}
-\frac{h_{2-3\alpha/2}(\tau_{1},\tau_{2})}
{\mathcal{V}^{3+1/\beta-\alpha}}-\frac{\tau
_{2}}{\mathcal{V}^{3}}+\frac{1}{\mathcal{V}^{3}}, \label{eqprooff}
\end{equation}
with $1/\beta\geq \alpha$ to keep the last term in
(\ref{eqprooff}). Now setting $1/\beta=\alpha>1$, (\ref{eqprooff})
reduces to:
\begin{equation}
V \sim \frac{K^{-1}_{11}}{\mathcal{V}^{2+2\alpha}}
+\frac{K^{-1}_{22}}{\mathcal{V}^{4}}
-\frac{h_{2-3\alpha/2}(\tau_{1},\tau_{2})}
{\mathcal{V}^{3}}-\frac{\tau
_{2}}{\mathcal{V}^{3}}+\frac{1}{\mathcal{V}^{3}}.
\label{eqproofff}
\end{equation}
By studying the expression (\ref{eqproofff}), we realise that
there are only three possible situations in which the necessary
but not sufficient conditions to stabilise $a_{j}\tau_{j}\gg 1$,
$j=1,2$, and not to neglect the leading order $\alpha'$
corrections, can be satisfied. The first one is when $K^{-1}_{22}$
is in case (3) and depends also on $\tau_{1}$:
$K^{-1}_{22}=\mathcal{V}f_{1/2}(\tau_{1},\tau_{2})$. In addition,
the first term in (\ref{eqproofff}) is subleading with respect to
the other ones given that $K^{-1}_{11}$ is in case (1) or (2) or
(3) or even in case (4) but still being subleading. However we can
again show that this case does not lead to any new LVS by noticing
that $\partial V/\partial \tau_{1}=0$ combined with $\partial
V/\partial \tau_{2}=0$ gives rise to a differential equation for
$f_{1/2}$ whose solution is not homogeneous.

The second situations takes place when the second term in
(\ref{eqproofff}) is subleading with respect to the others. This
occurs when $K^{-1}_{22}$ is either in case (1) or (2) and
$K^{-1}_{11}$ is in case (4). Moreover if
$K^{-1}_{11}\simeq\mathcal{V}^{\delta}\tau_{1}^{2-3\delta/2}$ we
have to impose $\delta=2\alpha-1$. However this case would not
work because the minimisation equation $\frac{\partial V}{\partial
\tau_{2}}=0$ would produce a negative volume:
\begin{equation}
\mathcal{V}^{\alpha}=-\frac{2\tau_{2}}{a_{1}h_{2-3\alpha/2}}e^{a_{1}\tau_{1}}.
\end{equation}
Finally we have to contemplate the possibility that all the terms
in (\ref{eqproofff}) have the same volume scaling. This can happen
only if $K^{-1}_{22}\simeq\mathcal{V}\sqrt{\tau_{2}}$ and
$K^{-1}_{22}\simeq\mathcal{V}^{\delta}\tau_{1}^{2-3\delta/2}$ with
$\delta=(2\alpha-1)>1$. Even this case can be seen to produce no
LVS. In fact, it is possible to integrate out the overall volume
from one of the usual minimisation equations, so being left with
two equations in $\tau_{1}$ and $\tau_{2}$. Then given that we
know that we are looking for a minimum located at $\beta
a_{1}\tau_{1}\simeq a_{2}\tau_{2}$, making this substitution, we
end up with two equations in just $\tau_{1}$ which can be seen to
disagree.

On the other hand, for $1/\beta>\alpha$, we would be left with:
\begin{equation}
V \sim \frac{K^{-1}_{11}}{\mathcal{V}^{2+2/\beta}}
+\frac{K^{-1}_{22}}{\mathcal{V}^{4}}-\frac{\tau
_{2}}{\mathcal{V}^{3}}+\frac{1}{\mathcal{V}^{3}}. \label{Eeq}
\end{equation}
Now by noticing that equation (\ref{Eeq}) has the same form of
(\ref{Semplific}) if we set $1/\gamma=1$, we can just repeat the
same consideration made before and obtain that $K^{-1}_{22}$ has
to be in case (3). Moreover if $K^{-1}_{22}$ depends on both
$\tau_{1}$ and $\tau_{2}$, then we have no LARGE Volume minimum.
On the other hand, when $K^{-1}_{22}$ depends only on $\tau_{2}$,
the first term in (\ref{Eeq}) is now negligible at large volume
and can be used to fix $\tau_{1}$ if we make it compete with the
third term in (\ref{eqprooff}) by writing $K^{-1}_{11}\simeq
\mathcal{V}^{\delta}\tau_{1}^{2-3\delta/2}$ and then imposing
$1/\beta+\alpha=1+\delta$.

On the contrary, if $1/\gamma>1$, (\ref{eqproof}) takes the form:
\begin{equation}
V \sim \frac{K^{-1}_{11}}{\mathcal{V}^{2+2/\beta}}
+\frac{K^{-1}_{22}}{\mathcal{V}^{2+2/\gamma}}
-\frac{h_{2-3\alpha/2}(\tau_{1},\tau_{2})}
{\mathcal{V}^{2+1/\beta+1/\gamma-\alpha}}+\frac{1}{\mathcal{V}^{3}}.
\label{EQproof}
\end{equation}
Now for $(1/\beta+1/\gamma-\alpha)>1$, (\ref{EQproof}) at leading
order in a large volume expansion, reduces to:
\begin{equation}
V \sim \frac{K^{-1}_{11}}{\mathcal{V}^{2+2/\beta}}
+\frac{K^{-1}_{22}}{\mathcal{V}^{2+2/\gamma}}
+\frac{1}{\mathcal{V}^{3}},
\end{equation}
which has already been proved to produce no LVS. On the other
hand, for $(1/\beta+1/\gamma-\alpha)<1$, the $\alpha'$ correction
would be negligible at large volume, so forcing us to impose
$(1/\beta+1/\gamma-\alpha)=1$ and obtain:
\begin{equation}
V \sim \frac{K^{-1}_{11}}{\mathcal{V}^{2+2/\beta}}
+\frac{K^{-1}_{22}}{\mathcal{V}^{2+2/\gamma}}
-\frac{h_{2-3\alpha/2}(\tau_{1},\tau_{2})}
{\mathcal{V}^{3}}+\frac{1}{\mathcal{V}^{3}}. \label{EQpf}
\end{equation}
However we can show explicitly that there is no LARGE Volume
minimum. In fact, setting $K^{-1}_{11}\simeq
\mathcal{V}^{\eta}\tau_{1}^{2-3\eta/2}$ with
$\eta=\frac{2}{\beta}-1$, and $K^{-1}_{22}\simeq
\mathcal{V}^{\delta}\tau_{2}^{2-3\delta/2}$ with
$\delta=\frac{2}{\gamma}-1$, and then substituting the solutions
of $\partial V/\partial \tau_{1}=0$ and $\partial V/\partial
\tau_{2}=0$ in $\partial V/\partial \mathcal{V}=0$, one finds that
it is never possible to fix the small K\"{a}hler moduli large
enough to be able to neglect the higher order instanton
contributions to $W$. If $h_{2-3\alpha/2}$ did not depend on both
$\tau_{1}$ and $\tau_{2}$ but just on one of them, this negative
result would not be altered as the term involving
$h_{2-3\alpha/2}$ depends always on both the two small moduli via
the two exponentials $e^{a_{1}\tau_{1}}e^{a_{2}\tau_{2}}$(see
(\ref{finalpotentialnl})).

\item[$\bullet$] axion minimum at (d) along direction (III) for
$\alpha>1/\beta$

This situation looks like the previous one if we swap $\gamma$
with $\beta$ and $\tau_{1}$ with $\tau_{2}$, therefore we do not
need to discuss this case.

\item[$\bullet$] axion minimum at (b) along direction (I) for
$1<\alpha\leq 1/\gamma$

Thus at the minimum:
\begin{equation}
\left\langle V_{AX}\right\rangle =2\left(\left\vert
Y_{12}\right\vert-\left\vert X_{1}\right\vert-\left\vert
X_{2}\right\vert\right), \label{axionennl}
\end{equation}
and so the full potential (\ref{finalPotential}) becomes:
\begin{eqnarray}
V &\sim &\frac{1}{\mathcal{V}^{2}}\left[
\sum\limits_{j=1}^{2}K^{-1}_{jj} a_{j}^{2} e^{-2a_{j}\tau_{j}}+2
a_{1}a_{2}h_{2-3\alpha/2}(\tau_{1},\tau_{2})\mathcal{V}^{\alpha}
e^{-(a_{1}\tau_{1}+a_{2}\tau_{2})}\right]  \notag \\
&&-4\frac{W_{0}}{\mathcal{V}^{2}}\left(a_{1}\tau
_{1}e^{-a_{1}\tau_{1}}+a_{2}\tau
_{2}e^{-a_{2}\tau_{2}}\right)+\frac{3}{4}\frac{\hat{\xi}
}{\mathcal{V}^{3}}W_{0}^{2}. \label{finalpotentialnnl}
\end{eqnarray}
When we take the large volume limit (I) of (\ref{DIrections}),
(\ref{finalpotentialnnl}) has the following volume scaling:
\begin{equation}
V \sim \frac{K^{-1}_{11}}{\mathcal{V}^{2+2/\gamma}}
+\frac{K^{-1}_{22}}{\mathcal{V}^{2+2/\gamma}}
+\frac{h_{2-3\alpha/2}(\tau_{1},\tau_{2})}
{\mathcal{V}^{2+2/\gamma-\alpha}} -\frac{\tau
_{1}}{\mathcal{V}^{2+1/\gamma}}-\frac{\tau
_{2}}{\mathcal{V}^{2+1/\gamma}}+\frac{1}{\mathcal{V}^{3}}.
\label{semplificannol}
\end{equation}
Due to the fact that $1<\alpha\leq 1/\gamma$, the third, the
fourth and the fifth term in (\ref{semplificannol}) are suppressed
with respect to the remaining ones by an appropriate power of the
volume and so we can neglect them. Thus the leading part of
(\ref{semplificannol}) takes the form:
\begin{equation}
V \sim \frac{K^{-1}_{11}}{\mathcal{V}^{2+2/\gamma}}
+\frac{K^{-1}_{22}}{\mathcal{V}^{2+2/\gamma}}+\frac{1}{\mathcal{V}^{3}}.
\label{SEMPLIl}
\end{equation}
We have already checked explicitly that this situation does not
present any LARGE Volume minimum.

\item[$\bullet$] axion minimum at (b) along direction (II) for
$1<\alpha\leq 1/\gamma$

In this situation the full scalar potential still looks like
(\ref{finalpotentialnnl}), but the volume scaling behaviour of its
terms now reads:
\begin{equation}
V \sim \frac{K^{-1}_{11}}{\mathcal{V}^{2+2/\beta}}
+\frac{K^{-1}_{22}}{\mathcal{V}^{2+2/\gamma}}
+\frac{h_{2-3\alpha/2}(\tau_{1},\tau_{2})}
{\mathcal{V}^{2+1/\beta+1/\gamma-\alpha}} -\frac{\tau
_{1}}{\mathcal{V}^{2+1/\beta}}-\frac{\tau
_{2}}{\mathcal{V}^{2+1/\gamma}}+\frac{1}{\mathcal{V}^{3}}.
\label{semplificannl}
\end{equation}
Given that along the direction (II) $1/\beta>1/\gamma$, and the
axion minimum is present for $1<\alpha\leq 1/\gamma$, the dominant
terms in (\ref{semplificannl}) become:
\begin{equation}
V \sim \frac{K^{-1}_{11}}{\mathcal{V}^{2+2/\beta}}
+\frac{K^{-1}_{22}}{\mathcal{V}^{2+2/\gamma}}+\frac{1}{\mathcal{V}^{3}}.
\label{SEMPLl}
\end{equation}
Thus we conclude that this case does not show any LVS, as we have
already showed.

\item[$\bullet$] axion minimum at (b) along direction (III) for
$1<\alpha\leq 1/\beta<1/\gamma$

This situation looks like the previous one if we swap $\gamma$
with $\beta$ and $\tau_{1}$ with $\tau_{2}$, therefore we do not
need to discuss this case. Let us finally summarise in the table
below what we have found for this case.
\end{enumerate}

\textbf{Case (4):
$K^{-1}_{12}=\mathcal{V}^{\alpha}h_{2-3\alpha/2}(\tau_{1},\tau_{2})$,
$\alpha>1$}
\begin{center}
\begin{tabular}{|c|c|c|c|c|c|c|}
\hline $K_{11}^{-1}$ & $K_{22}^{-1}$ & $\underset{1<\alpha \leq
1/\gamma }{(I),(b),} $ & $\underset{1<\alpha \leq 1/\gamma
}{(II),(b),}$ & $\underset{\alpha
> 1/\gamma }{(II),(c),}$ & $\underset{1<\alpha <1/\beta <1/\gamma }{(III),(b),%
}$ & $\underset{\alpha >1/\beta }{(III),(d),}$ \\ \hline 1 & 1 &
NO & NO & NO & NO & NO \\ \hline 1 & 2 & NO & NO & NO & NO & NO \\
\hline 1 & 3 & NO & NO & NO & NO & NO \\ \hline 1 & 4 & NO & NO &
NO & NO & NO \\ \hline 2 & 1 & NO & NO & NO & NO & NO \\ \hline 2
& 2 & NO & NO & NO & NO & NO \\ \hline 2 & 3 & NO & NO & NO & NO &
NO \\ \hline 2 & 4 & NO & NO & NO & NO & NO \\ \hline 3 & 1 & NO &
NO & NO & NO & NO \\ \hline 3 & 2 & NO & NO & NO & NO & NO \\
\hline 3 & 3 & NO & NO & NO & NO & NO \\ \hline 3 & 4 & NO & NO &
NO & NO & OK, $\beta =1$ ($\ast \ast $) \\ \hline 4 & 1 & NO & NO
& NO & NO & NO \\ \hline 4 & 2 & NO & NO & NO & NO & NO \\ \hline
4 & 3 & NO & NO & OK, $\gamma =1$ ($\ast $) & NO & NO \\ \hline 4
& 4 & NO & NO & NO & NO & NO \\ \hline
\end{tabular}
\end{center}
($\ast $) $\ K_{22}^{-1}\simeq \mathcal{V}\sqrt{\tau _{2}}$, $%
K_{11}^{-1}\simeq \mathcal{V}^{\delta }\tau _{1}^{2-3\delta /2}$
with $\frac{1}{\beta}+\alpha=1+\delta$, $\frac{1}{\beta}>\alpha$
\newline
($\ast \ast $) $K_{11}^{-1}\simeq \mathcal{V}\sqrt{\tau _{1}}$, $%
K_{22}^{-1}\simeq \mathcal{V}^{\delta }\tau _{2}^{2-3\delta /2}$
with $\frac{1}{\gamma}+\alpha=1+\delta$, $\frac{1}{\gamma}>\alpha$
\bigskip

Therefore we realise that the positive results represent cases
where all the $N_{small}$ small K\"{a}hler moduli plus a
particular combination, which is the overall volume, are
stabilised. It is then straightforward to see that at this stage
there will be $(h_{1,1}(X)-N_{small}-1)$ flat directions. This
terminates our proof of the LARGE Volume Claim.

\subsection{General picture}
\label{genAnal}

We shall try now to draw some conclusions from the previous LARGE
Volume Claim. This can be done by noticing that it is possible to
understand the topological meaning of two of the four cases for
the form of the elements of the inverse K\"{a}hler matrix.

Let us focus on the K\"{a}hler modulus $\tau_{1}$. From the
general expression of the inverse K\"{a}hler matrix for an
arbitrary Calabi-Yau (\ref{inversaAlfa}), we deduce that in this
case, dropping all the coefficients:
\begin{equation}
K_{11}^{-1}\simeq-\mathcal{V}k_{11i}t^{i}+\tau _{1}^{2}.
\label{inversa11}
\end{equation}
Case (1) states that $K^{-1}_{11}\simeq \tau_{1}^2$, therefore the
quantity $k_{11i}t^{i}$ has to vanish. This is definitely true if
$k_{11i}=0, \forall i=1,...,h_{1,1}(X)$, that is if the volume is
linear in $t_{1}$, the 2-cycle volume corresponding to $\tau_{1}$.
This is the definition of a three-fold with a K3 fibration
structure over the base $t_{1}$ \cite{oguiso}. Thus we realise
that Calabi-Yau K3 fibrations correspond to case (1). More
precisely if the three-fold is a single fibration only
$K^{-1}_{11}$ will be in case (1) but not $K^{-1}_{22}$. On the
contrary, double K3 fibrations will have both $K^{-1}_{11}$ and
$K^{-1}_{22}$ in case (1). Thus we have proved that:

\begin{quotation}
\textbf{Claim 1}
\textit{$K^{-1}_{11}\sim\tau_{1}^{2}\Leftrightarrow\tau_{1}$ is a
K3 fiber over the base $t_{1}$.}
\end{quotation}

One could wonder whether this reasoning is correct being worried
about possible field redefinitions since we showed that they can
change the intersection numbers. However this argument is indeed
correct because, as we have explained above, when one restricts
himself to changes of basis which do not alter the form of the
superpotential (\ref{form of W}), the form of the elements of the
inverse K\"{a}hler matrix do not change as the physics depends
only on them and we know that it should not be modified by changes
of basis. Therefore it suffices to calculate $K^{-1}_{11}$ in one
frame where the geometrical interpretation is clear.

The same procedure can be followed to prove that
$K^{-1}_{11}\sim\mathcal{V}\sqrt{\tau_{1}}$ if and only if
$\tau_{1}$ is a blow-up mode resolving a point-like singularity.
The blow-up of a singularity at a point $P$ is obtained by
removing the point $P$ and replacing it with a projective space
like $\mathbb{C}P^{1}$. This procedure introduces an extra
divisor, called \textit{exceptional}, with the corresponding extra
K\"{a}hler modulus that is what we call a blow-up mode. An
exceptional divisor $D_1$ is such that it has only its triple
self-intersection number non-vanishing \cite{fulton}:
\begin{equation}
D_1\cdot D_i\cdot D_j\neq 0\textit{ \ only \ if \ }i=j=1.
\end{equation}
\newline
\newline
\begin{figure}[ht]
\begin{center}
\epsfig{file=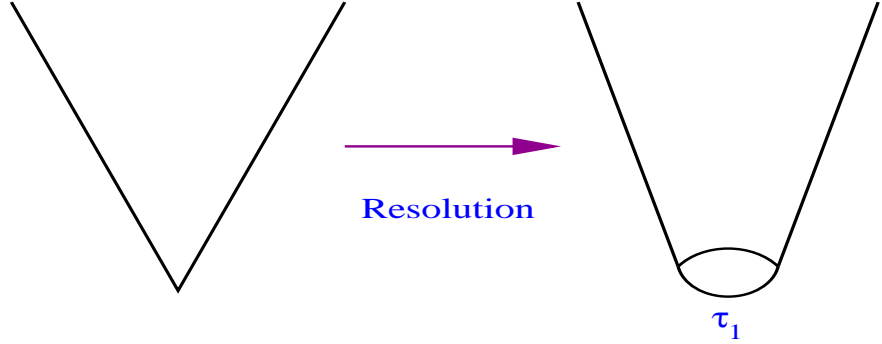, height=45mm,width=115mm}
\caption{Blow-up cycle $\tau_{1}$ resolving a point-like
singularity.}
\end{center}
\end{figure}

Therefore if $\tau_{1}$ is a blow-up then, in a suitable basis, we
can always write the volume as:
\begin{equation}
\mathcal{V}=f_{3/2}(\tau_{j})-\tau_{1}^{3/2},\text{ \ }j\neq 1
\label{blow-up}
\end{equation}
where $f_{3/2}(\tau_{j})$ is an homogeneous function of degree
3/2. It is then clear that these blow-up modes are purely
\textit{local} effects, since the change in the volume of the
Calabi-Yau as the blow-up cycle is collapsed, only goes as the
volume of the cycle, with no dependence on the overall volume. In
fact, a change of $\tau_{1}$, $\delta\tau_{1}$, would generate a
change of the total volume of the form:
\begin{equation}
\delta \mathcal{V}=\frac{\partial \mathcal{V}}{\partial \tau
_{1}}\delta \tau _{1}=-\frac{3}{2}\sqrt{\tau _{1}}\delta \tau
_{1}. \label{point}
\end{equation}
Let us approximate the volume (\ref{blow-up}) as
$\mathcal{V}\simeq f_{3/2}(\tau_{j})$ and calculate the K\"{a}hler
matrix:
\bigskip
\begin{equation}
\frac{\partial K}{\partial \tau _{1}}\simeq \frac{\sqrt{\tau
_{1}}}{\mathcal{ V}}\text{ \ }\Longrightarrow \text{ \ }\left\{
\begin{array}{c}
K_{11}=\frac{\partial ^{2}K}{\partial \tau _{1}^{2}}\simeq
\frac{1}{\sqrt{\tau _{1}}
\mathcal{V}}, \\
K_{1j}=\frac{\partial ^{2}K}{\partial \tau _{j}\partial \tau
_{1}}\simeq -\frac{ \sqrt{\tau
_{1}}}{\mathcal{V}^{2}}\frac{\partial \mathcal{V}}{\partial \tau
_{j}}\simeq o\left( \frac{1}{\mathcal{V}^{5/3}}\right) ,\text{
}j\neq 1.
\end{array}
\right. \label{argument}
\end{equation}
It turns out that $\frac{K_{1j}}{K_{11}}\simeq
o\left(\frac{1}{\mathcal{V}^{2/3}}\right)\ll 1$, $j\neq 1$ and so
we can immediately deduce the leading order term in the `11'
element of the inverse K\"{a}hler matrix by simply taking the
inverse of $K_{11}$, which gives case (3) or more explicitly
$K_{11}^{-1}\simeq\mathcal{V}\sqrt{\tau_{1}}$. But does
$K^{-1}_{11}\sim\mathcal{V}\sqrt{\tau_{1}}$ imply a form of the
volume as $\mathcal{V}=f_{3/2}(\tau_{j})-\tau_{1}^{3/2}$ with
$j\neq 1$? We shall prove now that this is indeed the case. Let us
focus on $N_{small}=1$ without loss of generality. Then:
\begin{equation}
K_{ij}\equiv\frac{\partial^{2} K}{\partial
\tau_{i}\partial\tau_{j}}=\frac{2}{\mathcal{V}}\left(\frac{1}{\mathcal{V}}\frac{\partial
\mathcal{V}}{\partial \tau_{i}}\frac{\partial
\mathcal{V}}{\partial \tau_{j}}
-\frac{\partial^{2}\mathcal{V}}{\partial\tau_{i}\partial
\tau_{j}}\right),\text{ \ \ for \ }i,j=1,2. \label{IFFF}
\end{equation}
We can then invert the K\"{a}hler matrix and find (by denoting
$\frac{\partial\mathcal{V}}{\partial\tau_{j}}\equiv\mathcal{V}_{j}$
and similarly for the second derivative):
\begin{equation}
K^{-1}_{11}=\frac{\mathcal{V}}{2}\left[\frac{\frac{1}{\mathcal{V}}\mathcal{V}_{2}^{2}-\mathcal{V}_{22}}{\left(\frac{1}{
\mathcal{V}}\mathcal{V}_{1}^{2}-\mathcal{V}_{11}\right)\left(\frac{1}{
\mathcal{V}}\mathcal{V}_{2}^{2}-\mathcal{V}_{22}\right)-\left(\frac{1}{
\mathcal{V}}\mathcal{V}_{1}\mathcal{V}_{2}-\mathcal{V}_{12}\right)^{2}}\right].
\end{equation}
Now if we impose that at leading order
$K^{-1}_{11}=c_{1}\mathcal{V}\sqrt{\tau_{1}}$ with
$c_{1}\in\mathbb{R}\smallsetminus\{0\}$, we get that at leading
order:
\begin{equation}
\left[\frac{\frac{1}{\mathcal{V}}\mathcal{V}_{2}^{2}-\mathcal{V}_{22}}{\left(\frac{1}{
\mathcal{V}}\mathcal{V}_{1}^{2}-\mathcal{V}_{11}\right)\left(\frac{1}{
\mathcal{V}}\mathcal{V}_{2}^{2}-\mathcal{V}_{22}\right)-\left(\frac{1}{
\mathcal{V}}\mathcal{V}_{1}\mathcal{V}_{2}-\mathcal{V}_{12}\right)^{2}}\right]=2c_{1}\sqrt{\tau_{1}}.
\label{IFF}
\end{equation}
Now using the homogeneity property of the volume in terms of the
4-cycle moduli,
$\tau_{1}\mathcal{V}_{1}+\tau_{2}\mathcal{V}_{2}=\frac{3}{2}\mathcal{V}$,
we derive:
\begin{equation}
\mathcal{V}_{2}=\frac{3\mathcal{V}}{2\tau
_{2}}-\frac{\mathcal{V}_{1}\tau _{1}}{ \tau _{2}},\text{ \ \ \ \
}\mathcal{V}_{12}=\frac{\mathcal{V}_{1}}{2\tau _{2}
}-\frac{\mathcal{V}_{11}\tau _{1}}{\tau _{2}},\text{ \ \ \ \
}\mathcal{V} _{22}=\frac{3\mathcal{V}-4\mathcal{V}_{1}\tau
_{1}+4\mathcal{V}_{11}\tau _{1}^{2}}{4\tau _{2}^{2}}. \label{IFF1}
\end{equation}
Now plugging (\ref{IFF1}) back in (\ref{IFF}), we find that at
leading order:
\begin{equation}
3\mathcal{V}\left(1+2c_{1}\sqrt{\tau_{1}}\mathcal{V}_{11}\right)
-2\mathcal{V}_{1}\left(2\tau_{1}+c_{1}\sqrt{\tau_{1}}\mathcal{V}_{1}\right)=0.
\label{IFF2}
\end{equation}
We can now write the general form of $\mathcal{V}_{1}$ as
$\mathcal{V}_{1}=c_{2}\mathcal{V}^{\alpha}\tau_{1}^{\frac{(1-3\alpha)}{2}}$
with $c_{2}\in\mathbb{R}\smallsetminus\{0\}$ and $\alpha\leq 1$.
It follows then that:
\begin{equation}
\mathcal{V}_{11}
=c_{2}\frac{\partial}{\partial\tau_{1}}\left(\mathcal{V}^{\alpha}
\tau_{1}^{\frac{(1-3\alpha)}{2}}\right)=\alpha
c_{2}^{2}\mathcal{V}^{2\alpha-1}\tau_{1}^{1-3\alpha}+c_{2}\frac{(1-3\alpha)}
{2}\mathcal{V}^{\alpha}\tau_{1}^{-\frac{(3\alpha+1)}{2}},
\label{IFF7}
\end{equation}
which for $\alpha<1$, $\alpha\neq 1/3$, at leading order reduces
to:
\begin{equation}
\mathcal{V}_{11}=c_{2}\frac{(1-3\alpha)}
{2}\mathcal{V}^{\alpha}\tau_{1}^{-\frac{(3\alpha+1)}{2}},
\label{IFF8}
\end{equation}
while for $\alpha=1/3$ reads:
\begin{equation}
\mathcal{V}_{11}=\frac{c_{2}^{2}}{3}\mathcal{V}^{-1/3},
\label{IFF80}
\end{equation}
 and for $\alpha=1$ becomes:
\begin{equation}
\mathcal{V}_{11}=c_{2}\frac{\mathcal{V}}{\tau_{1}^{2}}(c_{2}-1).
\label{IFF9}
\end{equation}
Now if $1/2<\alpha\leq 1$, then (\ref{IFF2}) at leading order
looks like:
\begin{equation}
3\mathcal{V}_{11}
=c_{2}^{2}\mathcal{V}^{2\alpha-1}\tau_{1}^{(1-3\alpha)}.
\label{IFF4}
\end{equation}
For $1/2<\alpha< 1$, using (\ref{IFF8}), (\ref{IFF4}) gives:
\begin{equation}
3c_{2}\frac{(1-3\alpha)}
{2}\mathcal{V}^{\alpha}\tau_{1}^{-\frac{(3\alpha+1)}{2}}
=c_{2}^{2}\mathcal{V}^{2\alpha-1}\tau_{1}^{(1-3\alpha)},
\label{IFF10}
\end{equation}
which at leading order reduces to:
\begin{equation}
3c_{2}\frac{(1-3\alpha)}
{2}\mathcal{V}^{\alpha}\tau_{1}^{-\frac{(3\alpha+1)}{2}}=0,
\label{IFF11}
\end{equation}
with the solution $\alpha=1/3$ that is in contradiction with the
fact that we are considering $\alpha>1/2$. For $\alpha=1$, using
(\ref{IFF9}), (\ref{IFF4}) becomes:
\begin{equation}
3(c_{2}-1) =c_{2}, \label{IFF12}
\end{equation}
but the solution $c_{2}=3/2\Rightarrow
\tau_{1}\mathcal{V}_{1}=\frac{3}{2}\mathcal{V}$, using the
homogeneity property of the volume,
$\tau_{1}\mathcal{V}_{1}+\tau_{2}\mathcal{V}_{2}=\frac{3}{2}\mathcal{V}$,
would imply $\mathcal{V}_{2}=0$ and so we have to reject it. On
the contrary if $\alpha= 1/2$, then (\ref{IFF2}) at leading order
takes the form:
\begin{equation}
6c_{1}\sqrt{\tau_{1}}\mathcal{V}_{11} =(2c_{1}c_{2}^{2}-3),
\label{IFF5}
\end{equation}
and by means of (\ref{IFF8}), this expression at leading order
becomes:
\begin{equation}
c_{1}c_{2}\mathcal{V}^{1/2}\tau_{1}^{-\frac{3}{4}} =0,
\label{IFF15}
\end{equation}
that clearly admits no possible solution. Finally if $\alpha<1/2$,
then (\ref{IFF2}) at leading order becomes:
\begin{equation}
3\mathcal{V}\left(1+2c_{1}\sqrt{\tau_{1}}\mathcal{V}_{11}\right)=0.
\label{IFF16}
\end{equation}
Due to (\ref{IFF8}), (\ref{IFF16}) for $\alpha\neq 1/3$ reads:
\begin{equation}
3\mathcal{V}\left(1+c_{1}c_{2}(1-3\alpha)
\mathcal{V}^{\alpha}\tau_{1}^{-\frac{3\alpha}{2}}\right)=0,
\label{IFF13}
\end{equation}
whereas using (\ref{IFF80}), (\ref{IFF16}) for $\alpha=1/3$ takes
the form:
\begin{equation}
0=3\mathcal{V}\left(1+2c_{1}\sqrt{\tau_{1}}\frac{c_{2}^{2}}{3}\mathcal{V}^{-1/3}\right)\simeq
3\mathcal{V}, \label{IFF160}
\end{equation}
which is impossible to solve. Now focusing on (\ref{IFF13}), if
$\alpha<0$ we do not find any solution, whereas if $0<\alpha<1/2$,
(\ref{IFF13}) reduces to:
\begin{equation}
3c_{1}c_{2}(1-3\alpha)
\mathcal{V}^{\alpha+1}\tau_{1}^{-\frac{3\alpha}{2}}=0,
\label{IFF130}
\end{equation}
which is solved by $\alpha=1/3$ that is in disagreement with the
fact that we are considering $\alpha\neq 1/3$. Lastly for
$\alpha=0$, (\ref{IFF13}) looks like:
\begin{equation}
3\mathcal{V}\left(1+c_{1}c_{2}\right)=0,
\end{equation}
which admits a solution of the form $c_{2}=-\frac{1}{c_{1}}$.
Therefore we have $\mathcal{V}_{1}=-\frac{\sqrt{\tau_{1}}}{c_{1}}$
and $\mathcal{V}_{11}=-\frac{1}{2c_{1}\sqrt{\tau_{1}}}$ that imply
an overall volume of the form
$\mathcal{V}=\lambda_{2}\tau_{2}^{3/2}-\lambda_{1}\tau_{1}^{3/2}$.
It is easy now to generalise this result for $N_{small}>1$ given
that we have shown that:

\begin{quotation}
\textbf{Claim 2} \textit{$K_{11}^{-1}\sim \mathcal{V}\sqrt{\tau
_{1}}\Leftrightarrow \mathcal{V}=f_{3/2}(\tau _{j})-\tau
_{1}^{3/2}$ with $j\neq 1$ $\Leftrightarrow $ $\tau _{1}$ is the
only blow-up mode resolving a point-like singularity.}
\end{quotation}

We point out also that the stressing that the blow-up has to
resolve a point-like singularity is exactly related to the fact
that it has to be a purely local effect. In fact, the resolution
of a hyperplane or line-like singularity would \textit{not} be a
local effect, even though it would still enable us to take a
sensible large volume limit by sending $\tau_{b}$ large and
keeping $\tau_{1}$ small. In this case, it is plausible to expect
an expression for the overall volume of the form:
\begin{equation}
\mathcal{V}=\tau_{b}^{3/2}-\tau_{b}\tau_{1}^{1/2}-\tau_{1}^{3/2}.
\label{linea}
\end{equation}
If we approximate the volume as $\mathcal{V}\simeq
\tau_{b}^{3/2}$, we can see that the change of $\mathcal{V}$ with
the increase of the cycle size $\tau_{1}$ does not depend on
powers of $\tau_{1}$ alone as in (\ref{point}) but it looks like:
\begin{equation}
\delta \mathcal{V}=\frac{\partial \mathcal{V}}{\partial \tau
_{1}}\delta \tau _{1}\simeq-\frac{1}{2}\frac{\tau_{b}}{\sqrt{\tau
_{1}}}\delta \tau _{1}\simeq
-\frac{1}{2}\frac{\mathcal{V}^{2/3}}{\sqrt{\tau _{1}}}\delta \tau
_{1}.
\end{equation}
Moreover the case (\ref{linea}) gives rise to an inverse
K\"{a}hler metric of the form $K_{11}^{-1}\simeq
\mathcal{V}^{1/3}\tau_{1}^{3/2}$ which does not satisfy the
condition of the LARGE Volume Claim exactly because $\tau_{1}$ is
not resolving a point-like singularity.

Let us show now that if we have $N_{small}=2$ with one small
modulus $\tau_{2}$ which is a local blow-up mode then the cross
term $K^{-1}_{12}$ has to be in case (1):
$K^{-1}_{12}\sim\tau_{1}\tau_{2}$. Without loss of generality we
can consider just one large modulus $\tau_{3}$ and so the volume
will look like
$\mathcal{V}=f_{3/2}(\tau_{3},\tau_{1})-\tau_{2}^{3/2}$. The
computation of the K\"{a}hler metric gives an expression like
(\ref{IFFF}) but now for $i,j=1,3$ with in addition:
\begin{equation}
K_{22}=\frac{3}{2\mathcal{V}\sqrt{\tau_{2}}},\text{ \ \ \
}K_{2j}=-\frac{3\sqrt{\tau_{2}}}{\mathcal{V}^{2}}\mathcal{V}_{j},\text{
\ with }j=1,3,
\end{equation}
and the `12' element of the inverse K\"{a}hler metric in full
generality reads:
\begin{equation}
K^{-1}_{12}=\frac{\tau_{2}\mathcal{V}^{2}\left(\mathcal{V}_{1}\mathcal{V}_{13}
-\mathcal{V}_{1}\mathcal{V}_{33}\right)}
{\left(3\tau_{2}^{3/2}-\mathcal{V}\right)\left(\mathcal{V}_{11}\mathcal{V}_{3}^{2}
+\mathcal{V}_{33}\mathcal{V}_{1}^{2}
-2\mathcal{V}_{1}\mathcal{V}_{3}\mathcal{V}_{13}\right)-\mathcal{V}^{2}\left(\mathcal{V}_{13}^{2}
-\mathcal{V}_{11}\mathcal{V}_{33}\right)}. \label{K-112}
\end{equation}
Using again the homogeneity property of the volume, we can find
the following relations:
\begin{gather*}
\mathcal{V}_{3}=\frac{1}{\tau _{3}}\left[ \frac{3}{2}\left(
\mathcal{V}+\tau _{2}^{3/2}\right) -\tau
_{1}\mathcal{V}_{1}\right] ,\text{ \ \ \ \ \ \ \ \ \ \
}\mathcal{V}_{13}=\frac{1}{\tau _{3}}\left[
\frac{\mathcal{V}_{1}}{2}
-\tau _{1}\mathcal{V}_{11}\right] , \\
\mathcal{V}_{33}=\frac{1}{4\tau _{3}^{2}}\left[ 3\left(
\mathcal{V}+\tau _{2}^{3/2}\right) +4\tau _{1}\left( \tau
_{1}\mathcal{V}_{11}-\mathcal{V} _{1}\right) \right] ,
\end{gather*}
which substituted back in (\ref{K-112}) give the final result:
\begin{equation}
K^{-1}_{12}=\frac{2\mathcal{V}^{2}\tau_{1}\tau_{2}}{2\mathcal{V}^{2}+6\mathcal{V}\tau_{2}^{3/2}-9\tau_{2}^{3}}\simeq
\tau_{1}\tau_{2}.
\end{equation}
Similarly one can check the correctness of Claim 2 by finding at
leading order $K^{-1}_{22}\sim\mathcal{V}\sqrt{\tau_{2}}$. Let us
summarise this result in the following:

\begin{quotation}
\textbf{Claim 3} \textit{If $N_{small}=2$ and $K_{22}^{-1}\sim
\mathcal{V}\sqrt{\tau_{2}}\Rightarrow K_{12}^{-1}\sim
\tau_{1}\tau_{2}$.}
\end{quotation}

We immediately realise that this Claim rules out the possible
LARGE Volume minima along the directions (II) and (III) for the
case (2), (3) and (4). However following arguments similar to the
ones presented to prove Claim 2, one can show that if
$K_{22}^{-1}\sim \mathcal{V}\sqrt{\tau_{2}}$ and so
$K_{12}^{-1}\sim \tau_{1}\tau_{2}$, $K_{11}^{-1}$ can never be in
case (4). Hence also the new would-be LVS along the directions
(II) and (III) for the case (1) have to be rejected because
mathematically inconsistent. Claim 3 also implies that the LVS
along the direction (I) for case (2) and (3) is viable only if
$K^{-1}_{jj}\sim\mathcal{V}h^{(j)}_{1/2}(\tau_{1},\tau_{2})$ with
$\frac{\partial^{2}h^{(j)}}{\partial\tau_{1}\partial\tau_{2}}\neq
0$ $\forall j=1,2$. In fact if
$\frac{\partial^{2}h^{(j)}}{\partial\tau_{1}\partial\tau_{2}}$
were vanishing, then Claim 3 would imply $K^{-1}_{12}$ in case (1)
and not (2) or (3).

In reality we understand these two cases better by realising that
we can go further in our connection of the topological features of
the Calabi-Yau with the elements of $K^{-1}$. In fact, one could
wonder what happens when a singularity is not resolved by just one
blow-up cycle but by several independent local blow-ups. A
concrete example where this happens, is the resolution of the
singularity at the origin of the quotient $\mathbb{C}^2/G$, where
$G$ is a finite subgroup of $SU(2)$ acting linearly on
$\mathbb{C}^2$. This resolution replaces the singularity by
several $\mathbb{C}P^1$'s which correspond to new K\"{a}hler
moduli whose number is determined by the group $G$. For example,
if $G=\mathbb{Z}_n$, one gets $n-1$ such $\mathbb{C}P^1$'s which
play the r\^{o}le of the simple roots of the Lie algebra $A_{n-1}
= su(n)$. After resolving the singularity of $\mathbb{C}^2/G$, one
obtains an example of an ALE space \cite{vafa}.
\newline
\begin{figure}[ht]
\begin{center}
\epsfig{file=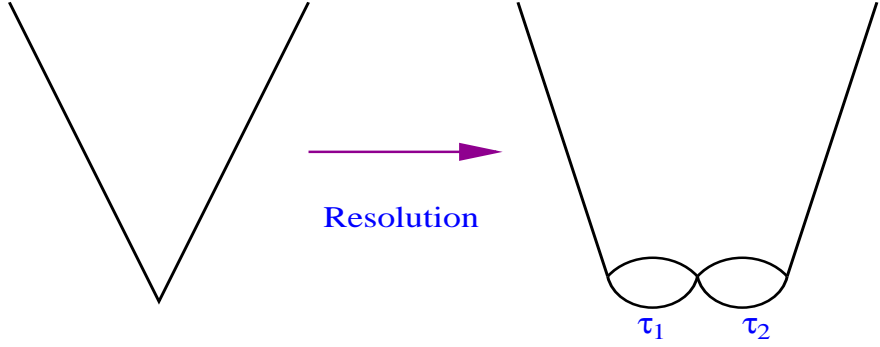, height=45mm,width=115mm}
\caption{Resolution by two independent blow-ups $\tau_{1}$ and
$\tau_{2}$.}
\end{center}
\end{figure}

Focusing on the case $N_{small}=2$, in a suitable basis, the
overall volume will take the general form:
\begin{equation}
\mathcal{V}=\tau_{b}^{3/2}-g_{3/2}(\tau_{1},\tau_{2}),
\label{blow-ups}
\end{equation}
with $g_{3/2}(\tau_{1},\tau_{2})\neq
\tau_{1}^{3/2}+\tau_{2}^{3/2}$ since in that special case
$\tau_{1}$ and $\tau_{2}$ would be blow-up cycles resolving two
different point-like singularities. If the total volume is given
by (\ref{blow-ups}) then the scaling with the volume of the
elements of the K\"{a}hler metric is (denoting $\partial
g_{3/2}/\partial \tau_{j}\equiv f_{j}$ and similarly for the
second derivative):
\begin{eqnarray*}
K_{bb} &\sim &\frac{1}{\mathcal{V}\sqrt{\tau _{b}}}+\frac{\tau
_{b}}{ \mathcal{V}^{2}}\sim \frac{1}{\mathcal{V}^{4/3}},\text{ \ \
\ \ }K_{12}\sim
\frac{g_{1}g_{2}}{\mathcal{V}^{2}}+\frac{g_{12}}{\mathcal{V}}\sim
\frac{1}{
\mathcal{V}},\text{\ \ } \\
K_{jb} &\sim &\frac{\sqrt{\tau _{b}}}{\mathcal{V}^{2}}g_{j}\sim
\frac{1}{ \mathcal{V}^{5/3}},\text{ \ \ \ \ \ \ }K_{jj}\sim
\frac{g_{j}^{2}}{\mathcal{V}^{2}}+\frac{g_{jj}}{\mathcal{V}}\sim
\frac{1}{\mathcal{V}},\text{ \ \ }j=1,2,
\end{eqnarray*}
therefore producing:
\begin{equation}
K_{ij}\sim \left(
\begin{tabular}{ccc}
\cline{1-2} \multicolumn{1}{|c}{$\mathcal{V}^{-1}$} &
$\mathcal{V}^{-1}$ &
\multicolumn{1}{|c}{$\mathcal{V}^{-5/3}$} \\
\multicolumn{1}{|c}{$\mathcal{V}^{-1}$} & $\mathcal{V}^{-1}$ &
\multicolumn{1}{|c}{$\mathcal{V}^{-5/3}$} \\ \cline{1-2}
$\mathcal{V}^{-5/3}$ & $\mathcal{V}^{-5/3}$ & $\mathcal{V}^{-4/3}$
\end{tabular}
\right) \label{DirK}
\end{equation}
where we have highlighted with a box the submatrix with the
leading powers of the volume. We have just to invert this
submatrix to get $K^{-1}_{jj}$ for $j=1,2$ and $K^{-1}_{12}$ which
turn out to be given by:
\begin{eqnarray}
K_{11}^{-1} &\sim &\mathcal{V}\left( \frac{g_{22}}{2\Delta
}\right) = \mathcal{V}h_{1/2}^{(1)}(\tau _{1},\tau _{2}),\text{ \
\ \ }K_{22}^{-1}\sim \mathcal{V}\left( \frac{g_{11}}{2\Delta
}\right) =\mathcal{V}
h_{1/2}^{(2)}(\tau _{1},\tau _{2}),  \notag \\
K_{12}^{-1} &\sim &\mathcal{V}\left( \frac{g_{12}}{2\Delta
}\right) = \mathcal{V}f_{1/2}(\tau _{1},\tau _{2}),\text{ \ \ \ \
where }\Delta \equiv g_{11}g_{22}-g_{12}^{2}.  \label{espa}
\end{eqnarray}
Following arguments similar to the ones used to prove Claim 2 we
can also show that starting from
$K^{-1}_{11}\sim\mathcal{V}h^{(1)}_{1/2}(\tau_{1},\tau_{2})$ with
$h^{(1)}$ really dependent on both the small moduli, the form of
the volume has to be (\ref{blow-ups}). A good intuition for this
result is that by setting $\tau_{1}=\tau_{2}$, this is the only
way to recover Claim 2. Therefore we have shown that:

\begin{quotation}
\textbf{Claim 4} \textit{$K_{11}^{-1}\sim
\mathcal{V}h^{(1)}_{1/2}(\tau_{1},\tau_{2})$ and $K_{22}^{-1}\sim
\mathcal{V}h^{(2)}_{1/2}(\tau_{1},\tau_{2}) \Leftrightarrow
\mathcal{V}=f_{3/2}(\tau_{j})-g_{3/2}(\tau_{1},\tau_{2})$ with
$j\neq 1,2$ $\Leftrightarrow $ $\tau_{1}$ and $\tau_{2}$ are two
independent blow-up modes resolving the same point-like
singularity,}
\end{quotation}

along with:

\begin{quotation}
\textbf{Claim 5} \textit{If $K_{11}^{-1}\sim
\mathcal{V}h^{(1)}_{1/2}(\tau_{1},\tau_{2})$ and $K_{22}^{-1}\sim
\mathcal{V}h^{(2)}_{1/2}(\tau_{1},\tau_{2})\Rightarrow
K_{12}^{-1}\sim \mathcal{V}f_{1/2}(\tau_{1},\tau_{2})$.}
\end{quotation}

In generality we can conclude that whenever $K^{-1}_{jj}\simeq
\mathcal{V} h^{(j)}_{1/2}(\tau_{1},\tau_{2},...,\tau_{N_{small}})$
then $\tau_{j}$ is a blow-up resolving a point-like singularity.
Moreover, if $h^{(j)}_{1/2}$ depends only on $\tau_{j}$, then
$\tau_{j}$ will be the only blow-up cycle resolving the
singularity and $K^{-1}_{ij}\sim\tau_{i}\tau_{j}$ $\forall i\neq
j=1,...,N_{small}$; on the contrary, if $h^{(j)}_{1/2}$ depends on
several 4-cycle moduli, say $\tau_{j}$ and $\tau_{k}$ for $j\neq
k$, the singularity is resolved by all those independent 4-cycles
with $K^{-1}_{jk}\sim\mathcal{V}f_{1/2}(\tau_{j},\tau_{k})$ and
$K^{-1}_{il}\sim\tau_{i}\tau_{j}$ $\forall i\neq
l=1,...,N_{small}$ for $l=j,k$.

These considerations imply that the would-be LVS along direction
(I) for case (2) is mathematically inconsistent. Thus for
$N_{small}=2$ we are left with just two cases that give rise to a
LARGE Volume minimum located at $\mathcal{V}\sim
e^{a_{1}\tau_{1}}\sim e^{a_{2}\tau_{2}}$:

\begin{enumerate}
\item $K^{-1}_{12}\sim \tau_{1}\tau_{2}$,
$K^{-1}_{jj}\sim\mathcal{V}\sqrt{\tau_{j}}$ $\forall j=1,2$ where
$\tau_{1}$ and $\tau_{2}$ are local blow-up modes resolving two
different point-like singularities;

\item $K^{-1}_{12}\sim \mathcal{V}f_{1/2}(\tau_{1},\tau_{2})$,
$K^{-1}_{jj}\sim\mathcal{V}h^{(j)}_{1/2}(\tau_{1},\tau_{2})$
$\forall j=1,2$ where $\tau_{1}$ and $\tau_{2}$ are two
independent blow-up modes resolving the same point-like
singularity.
\end{enumerate}

The only difference between these two cases is that the first one
works always whereas the second one gives a LVS only if, writing
the volume as
$\mathcal{V}=f_{3/2}(\tau_{j})-g_{3/2}(\tau_{1},\tau_{2})$ with
$j\neq 1,2$, the homogeneous function $g_{3/2}$ is symmetric in
$\tau_{1}$ and $\tau_{2}$. This can be seen easily by comparing
the solution of the two minimisation equations $\frac{\partial
V}{\partial \tau_{1}}=0$ and $\frac{\partial V}{\partial
\tau_{2}}=0$, then substituting the solution we are looking for,
that is $a_{1}\tau_{1}\simeq a_{2}\tau_{2}$, recalling
(\ref{espa}) and lastly finding that we do not get a contradiction
only if $g_{3/2}$ is symmetric.

\chapter[Appendix]{}

\section{Higher order corrections to the inflationary potential}
\label{Appendix A}

In this appendix we derive explicitly the leading corrections to
the fixed-volume approximation, which give rise to higher order
operators. We show that these operators do not introduce an $\eta$
problem since they are suppressed by inverse powers of the overall
volume.

\subsection{Derivation of the $\tau_1$ dependent shift of
  $\langle\mathcal{V}\rangle$}

\bigskip We start from the very general scalar potential
(\ref{PPpot1}):
\begin{equation}
 V=\left[-\mu_{4}(\ln
 \left(c\mathcal{V}\right))^{3/2}+\mu_{3}\right]
 \frac{W_{0}^{2}}{\mathcal{V}^{3}}+\frac{\delta
 _{up}}{\mathcal{V}^{2}}+\left( \frac{A}{\tau
 _{1}^{2}}-\frac{B}{\mathcal{V}\sqrt{\tau _{1}}}\right)
 \frac{W_{0}^{2}}{\mathcal{V}^{2}}, \label{app:pot1}
\end{equation}
where we have set $C=0$ since the loop corrections proportional to
$C$ turn out to be numerically small in the cases of interest,
both to finding the minimum in $\tau_{1}$ and to the inflationary
region. We now minimise this potential to obtain $\langle \mc{V}
\rangle$, first turning off the loop potential to investigate how
the uplifting term changes the minimum for $\mc{V}$. We follow
this by a perturbative study of the additional $\tau_1$-dependence
generated by the loop corrections: $\left\langle
\mathcal{V}\right\rangle =\mathcal{V}_{0} + \delta
\mathcal{V}(\tau _{1})$.

\subsubsection*{Uplifting only}

In the absence of loop corrections the potential reads:
\begin{equation}
 V=\left[ -\mu_{4}(\ln
 \left(c\mathcal{V}\right))^{3/2}+\mu_{3}\right]
 \frac{W_{0}^{2}}{\mathcal{V}^{3}}+\frac{\delta
 _{up}}{\mathcal{V}^{2}}, \label{app:Pot1}
\end{equation}
where the up-lifting term is chosen to ensure:
\begin{equation}
 \langle V \rangle=\left[-\mu_{4}(\ln
 \left(c\mathcal{V}_0\right))^{3/2}+\mu_{3}\right]
 \frac{W_{0}^{2}}{\mathcal{V}_0^{3}}+\frac{\delta
 _{up}}{\mathcal{V}_0^{2}}=0, \label{app:POt1}
\end{equation}
and so:
\begin{equation}
 \delta_{up}=\left[\mu_{4}(\ln
 \left(c\mathcal{V}_0\right))^{3/2}-\mu_{3}\right]
 \frac{W_{0}^{2}}{\mathcal{V}_0}\,. \label{app:up-lifting}
\end{equation}
Here $\mc{V}_0$ satisfies $\left.\partial
V/\partial\mc{V}\right|_{\mc{V}_0} = 0$, and so must solve:
\begin{equation}
 \frac{4\delta_{up}\mathcal{V}_0}{\mu_{4}W_0^2}+\frac{6\mu_{3}}{\mu_{4}}+3
 (\ln\left(c\mathcal{V}_0\right))^{1/2}-6(\ln\left(c\mathcal{V}_0\right))^{3/2}=0\,.
 \label{app:citala}
\end{equation}

This is most simply analysed once it is rewritten as:
\begin{equation}
 \psi+ p\, (\ln\left(c\mathcal{V}_0\right))^{1/2}
 -(\ln\left(c\mathcal{V}_0\right))^{3/2}=0,
\label{app:lunga}
\end{equation}
with:
\begin{equation}
 \psi :=\frac{\mu_3}{\mu_4}=\frac{\xi}{2\,\alpha\gamma}
 \left(\frac{a_3}{g_s}\right)^{3/2} \,,
\end{equation}
and the parameter $p$ takes the value $p = \frac32$ if we evaluate
$\delta_{up}$ using (\ref{app:up-lifting}), or $p = \frac12$ if we
take $\delta_{up} = 0$. Tracking the dependence on $p$ therefore
allows us to understand the sensitivity of the result to the
presence of the uplifting term.

The exact solution of (\ref{app:lunga}) is:
\begin{equation}
 \ln \left( c\mathcal{V}_{0}\right) = \frac{\left[12p +\left( 108
 \psi + 12 \sqrt{81 \psi^2 - 12 p^{3}} \right)^{2/3} \right]^2} {36
 \left(108 \psi + 12 \sqrt{81 \psi^{2}-12 p^{3}} \right)^{2/3}} \,,
 \label{app:prima}
\end{equation}
which approaches the $p$-independent result:
\begin{equation}
 \ln \left( c \mathcal{V}_0 \right) \simeq \psi^{2/3} =a_3
 \left( \frac{\hat\xi}{2\, \alpha \gamma} \right)^{2/3} \,,
\end{equation}
when $\psi \gg 1$, in agreement with eq. (\ref{app:tau3vsV})
together with expression (\ref{xx}) for $\tau_3$. This shows that
we may expect the uplifting corrections to $\mc{V}_0$ to be small
when $\psi \gg 1$.

\subsubsection*{Including loop corrections}

The potential now is given by (\ref{app:pot1}) and so the presence
of the loops will generate a $\tau_1$ dependent shift of
$\mathcal{V}$ such that:
\begin{equation}
 \mathcal{V}=\mathcal{V}_0+\delta \mathcal{V}(\tau_{1}),\text{ \
 with \ }\delta \mathcal{V}(\tau_{1})\ll \mathcal{V}_0\text{ \
 }\forall \tau_1.
\end{equation}
In order to calculate $\delta\mathcal{V}(\tau_1)$ at leading
order, let us solve the minimisation equation for the volume
taking into account that now that we have turned on the string
loops, we need to replace $\delta_{up}$ by $\delta_{up}' =
\delta_{up} + \mu_{up}$, where $\mu_{up}$ is the constant needed
to cancel the contribution of the loops to the cosmological
constant.
\begin{equation}
 \frac{\partial V}{\partial \mathcal{V}}=0\Longleftrightarrow
 \frac{4A\mathcal{V}}{\mu_4\tau_1^2}-\frac{6B}{\mu_4\sqrt{\tau_1}}
 +4\frac{\delta_{up}'\mathcal{V}}{\mu_{4}W_0^2}+6\frac{\mu_{3}}{\mu_{4}}+3
 (\ln\left(c\mathcal{V}\right))^{1/2}-6(\ln\left(c\mathcal{V}\right))^{3/2}=0.
\end{equation}
We notice that the logarithm in the previous expression can be
expanded as follows:
\begin{equation}
 \ln \left( c\mathcal{V}\right)=\ln \left(
 c\mathcal{V}_{0}\right)+\frac{\delta\mathcal{V}(\tau_1)}{\mathcal{V}_0},
\end{equation}
and by means of another Taylor series and the result
(\ref{app:citala}), we are left with:
\begin{gather}
 \left( 4\frac{\delta
 _{up}\mathcal{V}_{0}}{\mu_{4}W_{0}^{2}}+4\frac{\mu
 _{up}\mathcal{V}_{0}}{\mu_{4}W_{0}^{2}}+\frac{4A\mathcal{V}_{0}}{%
 \mu_{4}\tau _{1}^{2}}+\frac{3}{2}\left( \ln \left( c\mathcal{V}%
 _{0}\right) \right) ^{-1/2}-9\left( \ln \left(
 c\mathcal{V}_{0}\right) \right) ^{1/2}\right) \frac{\delta
 \mathcal{V}(\tau _{1})}{\mathcal{V}_{0}}=
 \notag \\
 -\frac{4A\mathcal{V}_{0}}{\mu_{4}\tau _{1}^{2}}+\frac{6B}{\mu_{4}%
 \sqrt{\tau _{1}}}-4\frac{\mu
 _{up}\mathcal{V}_{0}}{\mu_{4}W_{0}^{2}}.
\end{gather}
Now recalling the expression (\ref{app:up-lifting}) for
$\delta_{up}$ combined with (\ref{app:lunga}), we obtain:
\begin{equation}
 \frac{\delta \mathcal{V}(\tau _{1})}{\mathcal{V}_{0}}=\frac{\left(
 \frac{4A \mathcal{V}_0}{\mu_{4}\tau
 _{1}^{2}}-\frac{6B}{\mu_{4}\sqrt{\tau _{1} }}+4\frac{\mu
 _{up}\mathcal{V}_{0}}{\mu_{4}W_{0}^{2}}\right) }{\left(3\left( \ln
 \left( c\mathcal{V}_{0}\right) \right) ^{1/2}-\frac{3}{2}\left(
 \ln \left( c\mathcal{V}_{0}\right) \right) ^{-1/2}-\frac{4A
 \mathcal{V}_0}{\mu_{4}\tau _{1}^{2}}-4\frac{\mu _{up}\mathcal{V
 }_{0}}{\mu_{4}W_{0}^{2}}\right)}. \label{th}
\end{equation}
We can still expand the denominator in (\ref{th}) and working at
leading order we end up with:
\begin{equation}
 \frac{\delta \mathcal{V}(\tau _{1})}{\mathcal{V}_{0}}=\frac{\left(
 \frac{4A \mathcal{V}_0}{\mu_{4}\tau
 _{1}^{2}}-\frac{6B}{\mu_{4}\sqrt{\tau _{1} }}+4\frac{\mu
 _{up}\mathcal{V}_{0}}{\mu_{4}W_{0}^{2}}\right) }{\left(3\left( \ln
 \left( c\mathcal{V}_{0}\right) \right) ^{1/2}-\frac{3}{2}\left(
 \ln \left( c\mathcal{V}_{0}\right) \right) ^{-1/2}\right)}.
\label{tho}
\end{equation}
We have now all the ingredients to work out the canonical
normalisation.

\subsection{Canonical normalisation}

As we have seen in the previous subsection of this appendix,
$\mathcal{V}$ and $\tau_{3}$ will both have a $\tau_1$ dependent
shift of the form:
\begin{eqnarray}
\mathcal{V} &=&\mathcal{V}_{0}+\delta\mathcal{V}(\tau_{1}), \\
\tau_{3}
&=&\frac{\ln\left(c\mathcal{V}_0\right)}{a_3}+\frac{\delta\mathcal{V}(\tau_1)}{a_3\mathcal{V}_0},
\end{eqnarray}
which will cause $\partial_{\mu}\mathcal{V}$ and
$\partial_{\mu}\tau_3$ not to vanish when we study the canonical
normalisation of the inflaton field $\tau_1$ setting both
$\mathcal{V}$ and $\tau_{3}$ at its $\tau_1$ dependent minimum.
Thus we have:
\begin{eqnarray}
\partial_{\mu }\mathcal{V} &=&\frac{\partial\left(\delta\mathcal{V}(\tau_1)\right)}{\partial\tau_1} \partial _{\mu }\tau _{1},
\label{1}
\\
\partial _{\mu }\tau _{3} &=&\frac{1}{a_3\mathcal{V}_0}
\frac{\partial\left(\delta\mathcal{V}(\tau_1)\right)}{\partial\tau_1}
\partial _{\mu }\tau _{1}. \label{2}
\end{eqnarray}
The non canonical kinetic terms look like:
\begin{align}
-\mathcal{L}_{kin}& =\frac{1}{4}\frac{\partial ^{2}K}{\partial
\tau _{i}\partial \tau _{j}}\partial _{\mu }\tau _{i}\partial
^{\mu }\tau _{j}
\notag \\
& =\frac{3}{8\tau _{1}^{2}}\left( 1-\frac{2\alpha \gamma
}{3}\frac{\tau _{3}^{3/2}}{\mathcal{V}}\right) \partial _{\mu
}\tau _{1}\partial ^{\mu }\tau _{1}-\frac{1}{2\mathcal{V}\tau
_{1}}\left( 1-\alpha \gamma \frac{\tau
_{3}^{3/2}}{\mathcal{V}}\right) \partial _{\mu }\tau _{1}\partial
^{\mu }
\mathcal{V}  \notag \\
& +\frac{1}{2\mathcal{V}^{2}}\partial _{\mu }\mathcal{V}\partial
^{\mu } \mathcal{V}-\frac{3\alpha \gamma }{2}\frac{\sqrt{\tau
_{3}}}{\mathcal{V}^{2}}
\partial _{\mu }\tau _{3}\partial ^{\mu }\mathcal{V}+\frac{3\alpha \gamma }{8
}\frac{1}{\mathcal{V}\sqrt{\tau _{3}}}\partial _{\mu }\tau
_{3}\partial
^{\mu }\tau _{3}  \notag \\
& \simeq \frac{3}{8\tau _{1}^{2}}\partial _{\mu }\tau _{1}\partial
^{\mu }\tau _{1}-\frac{1}{2\mathcal{V}\tau _{1}}\partial _{\mu
}\tau _{1}\partial ^{\mu
}\mathcal{V}+\frac{1}{2\mathcal{V}^{2}}\partial _{\mu }\mathcal{V}
\partial ^{\mu }\mathcal{V}  \notag \\
& \qquad \qquad -\frac{3\alpha \gamma }{2}\frac{\sqrt{\tau
_{3}}}{\mathcal{V} ^{2}}\partial _{\mu }\tau _{3}\partial ^{\mu
}\mathcal{V}+\frac{3\alpha \gamma
}{8}\frac{1}{\mathcal{V}\sqrt{\tau _{3}}}\partial _{\mu }\tau
_{3}\partial ^{\mu }\tau _{3}.  \label{LKin}
\end{align}
Now using (\ref{1}) and (\ref{2}), we can derive the leading order
correction to the canonical normalisation in the constant volume
approximation:
\begin{equation}
-\mathcal{L}_{kin}=\frac{3}{8\tau _{1}^{2}}\left[
1-\frac{4\tau_1}{3}\frac{\partial}{\partial\tau_1}\left(\frac{\delta\mathcal{V}(\tau_1)}{\mathcal{V}_0}\right)
\right] \partial _{\mu }\tau _{1}\partial ^{\mu }\tau
_{1}=\frac{1}{2}\partial _{\mu }\varphi
\partial ^{\mu }\varphi,
\end{equation}
where $\varphi$ is the canonically normalised inflaton. Writing
$\varphi =g(\tau _{1})$ we deduce the following differential
equation:
\begin{equation}
\frac{\partial g(\tau_1)}{\partial
\tau_1}=\frac{\sqrt{3}}{2\tau_1}\sqrt{1-\frac{4\tau_1}{3}\frac{\partial}
{\partial\tau_1}\left(\frac{\delta\mathcal{V}(\tau_1)}{\mathcal{V}_0}\right)},
\end{equation}
which, after expanding the square root, admits the straightforward
solution:
\begin{equation}
\varphi =\frac{\sqrt{3}}{2}\ln \tau _{1}-\frac{1}{\sqrt{3}}\left(
\frac{ \delta \mathcal{V}(\tau _{1})}{\mathcal{V}_{0}}\right)
=\frac{\sqrt{3}}{2} \ln \tau _{1}\left[ 1-\frac{2}{3\ln \tau
_{1}}\left( \frac{\delta \mathcal{V} (\tau
_{1})}{\mathcal{V}_{0}}\right) \right],  \label{4}
\end{equation}
where the leading order term reproduces what we had in the main
text. We still need to invert this relation to get $\tau_1$ as a
function of $\varphi$ and then plug this result back in the
potential. We can write this function as:
\begin{equation}
\tau _{1}=e^{2\varphi/\sqrt{3}}\left(1+h(\varphi)\right),
\label{44}
\end{equation}
where $h(\varphi)\ll 1$.

At this point we can substitute (\ref{44}) in (\ref{4}) and by
means of a Taylor expansion, derive an equation for $h(\varphi)$:
\begin{gather}
\varphi =\varphi -\varphi \left[ \frac{2}{3\ln \tau _{1}}\left(
\frac{\delta \mathcal{V}(\tau _{1})}{\mathcal{V}_{0}}\right)
\right] _{\tau
_{1}=e^{2\varphi /\sqrt{3}}}+\frac{\sqrt{3}}{2}h(\varphi )+...  \notag \\
\Longrightarrow \text{ \ }h(\varphi )\text{ }=\frac{2}{3}\text{\
}\left. \left( \frac{\delta \mathcal{V}(\tau
_{1})}{\mathcal{V}_{0}}\right) \right\vert _{\tau _{1}=e^{2\varphi
/\sqrt{3}}},
\end{gather}
where we have imposed that the two first order corrections cancel
in order to get the correct inverse function. Therefore the final
canonical normalisation of $\tau_1$ which goes beyond the constant
volume approximation reads:
\begin{equation}
\tau _{1}=e^{2\varphi /\sqrt{3}}\left[ 1+\frac{2}{3}\left. \left(
\frac{ \delta \mathcal{V}(\tau _{1})}{\mathcal{V}_{0}}\right)
\right\vert _{\tau _{1}=e^{2\varphi /\sqrt{3}}}\right].
\label{444}
\end{equation}

\subsection{Leading correction to the inflationary slow roll}

In order to derive the full final inflationary potential at
leading order, we have now to substitute
$\mathcal{V}=\mathcal{V}_0+\delta\mathcal{V}(\tau_1)$ in
(\ref{app:pot1}) to obtain a function of just $\tau_{1}$. After
two subsequent Taylor expansions, the potential reads:
\begin{equation}
V=\left[-\mu_{4}(\ln \left(
c\mathcal{V}_{0}\right))^{3/2}\left(1+\frac{3\delta\mathcal{V}(\tau_1)}{2\mathcal{V}_0\ln
\left( c\mathcal{V}_{0}\right)}\right)+\mu_{3}+\frac{\delta
_{up}'\left(\mathcal{V}_0+\delta
\mathcal{V}(\tau_{1})\right)}{W_0^{2}}+ \frac{A\mathcal{V}}{\tau
_{1}^{2}}-\frac{B}{\sqrt{\tau _{1}}}\right]
\frac{W_{0}^{2}}{\mathcal{V}^{3}}. \notag
\end{equation}
Recalling the expression (\ref{app:up-lifting}) for $\delta_{up}$,
the leading contribution of the non-perturbative and $\alpha'$ bit
of the scalar potential cancels against the up-lifting term and we
are left with the expansion of $\delta V_{(np)}+\delta
V_{(\alpha')}+\delta V_{(up)}$ plus the loops:
\begin{equation}
V=\left[ -\frac{3\mu_{4}}{2}(\ln \left( c\mathcal{V}_{0}\right)
)^{1/2}\frac{\delta \mathcal{V}(\tau
_{1})}{\mathcal{V}_{0}}+\frac{\delta _{up}\delta \mathcal{V}(\tau
_{1})}{W_{0}^{2}}+\frac{\mu _{up}\mathcal{V}}{W_{0}^{2}}
+\frac{A\mathcal{V} }{\tau _{1}^{2}}-\frac{B}{\sqrt{\tau
_{1}}}\right] \frac{W_{0}^{2}}{\mathcal{V}^{3}}.
\label{interessante}
\end{equation}

It is now very interesting to notice in the previous expression
that the leading order expansion of the non-perturbative and
$\alpha'$ bit of the potential cancels against the expansion of
the up-lifting term. In fact from (\ref{interessante}), we have
that:
\begin{equation}
\delta V_{(np)}+\delta V_{(\alpha')}=-\frac{3\mu_{4}}{2}(\ln
\left( c\mathcal{V}_{0}\right) )^{1/2}\frac{\delta
\mathcal{V}(\tau _{1})}{\mathcal{V}_{0}}
\frac{W_{0}^{2}}{\mathcal{V}^{3}},
\end{equation}
along with:
\begin{equation}
\delta V_{(up)}=\frac{\delta _{up}\delta \mathcal{V}(\tau
_{1})}{\mathcal{V}^{3}}=\frac{3\mu_{4}}{2}(\ln \left(
c\mathcal{V}_{0}\right) )^{1/2}\frac{\delta \mathcal{V}(\tau
_{1})}{\mathcal{V}_{0}} \frac{W_{0}^{2}}{\mathcal{V}^{3}},
\end{equation}
where the last equality follows from (\ref{app:up-lifting}) and
(\ref{app:lunga}). This result was expected since we fine tuned
$\delta V_{(up)}$ to cancel $\delta V_{(np)}+\delta V_{(\alpha')}$
at $\mathcal{V}=\mathcal{V}_0$ and then we have applied the same
shift $\mathcal{V}=\mathcal{V}_0+\delta\mathcal{V}(\tau_{1})$ to
both of them, so clearly still obtaining a cancellation.

Thus we get the following \textit{exact} result for the
inflationary potential:
\begin{equation}
V_{inf}=\left[\frac{\mu _{up}\mathcal{V}}{W_{0}^{2}}
+\frac{A\mathcal{V} }{\tau _{1}^{2}}-\frac{B}{\sqrt{\tau
_{1}}}\right] \frac{W_{0}^{2}}{\mathcal{V}^{3}}.
\end{equation}
It is now possible to work out the form of $\mu_{up}$. The minimum
for $\tau_1$ lies at:
\begin{equation}
\langle\tau_1\rangle=\left(\frac{4A}{B}\mathcal{V}\right)^{2/3},
\end{equation}
and so by imposing $\langle V_{inf}\rangle=0$ we find:
\begin{equation}
\mu_{up}=\frac{3}{A^{1/3}}\left(\frac{B}{4}\right)^{4/3}\frac{W_0^2}{\mathcal{V}^{4/3}}.
\end{equation}
We can now expand again $\mathcal{V}$ around $\mathcal{V}_0$ and
obtain:
\begin{equation}
\mu_{up}=\frac{3}{A^{1/3}}\left(\frac{B}{4}\right)^{4/3}\frac{W_0^2}{\mathcal{V}_0^{4/3}}
\left(1-\frac{4}{3}\frac{\delta\mathcal{V}(\tau_1)}{\mathcal{V}_0}\right),
\label{muy}
\end{equation}
along with:
\begin{equation}
V_{inf}=V^{(0)}+\delta V,
\end{equation}
where:
\begin{equation}
V^{(0)}=\left[\frac{3}{A^{1/3}}\left(\frac{B}{4}\right)^{4/3}\frac{1}{\mathcal{V}_0^{1/3}}
+\frac{A\mathcal{V}_0 }{\tau _{1}^{2}}-\frac{B}{\sqrt{\tau
_{1}}}\right] \frac{W_{0}^{2}}{\mathcal{V}_0^{3}},
\end{equation}
is the inflationary potential derived in the main text in the
approximation that the volume is $\tau_1$-independent during the
inflationary slow roll, and:
\begin{equation}
\delta
V=\left(\frac{\delta\mathcal{V}(\tau_1)}{\mathcal{V}_0}\right)
\left[-\frac{10}{A^{1/3}}\left(\frac{B}{4}\right)^{4/3}\frac{1}{\mathcal{V}_0^{1/3}}
-2\frac{A\mathcal{V}_0 }{\tau _{1}^{2}}+3\frac{B}{\sqrt{\tau
_{1}}}\right] \frac{W_{0}^{2}}{\mathcal{V}_0^{3}},
\end{equation}
is the leading order correction to that approximation.

Now that we have an expression for the up-lifting term $\mu_{up}$
given by (\ref{muy}), we are able to write down explicitly the
form of the shift of $\mathcal{V}$ due to $\tau_1$ (\ref{tho}) at
leading order:
\begin{equation}
\frac{\delta \mathcal{V}(\tau _{1})}{\mathcal{V}_{0}}=\frac{\left(
\frac{4A \mathcal{V}_0}{\mu_{4}\tau
_{1}^{2}}-\frac{6B}{\mu_{4}\sqrt{\tau _{1}
}}+\frac{3B^{4/3}}{\mu_{4}\left(4A\right)^{1/3}}\frac{1}{\mathcal{V}_0^{1/3}}\right)
}{\left(3\left( \ln \left( c\mathcal{V}_{0}\right) \right)
^{1/2}-\frac{3}{2}\left( \ln \left( c\mathcal{V}_{0}\right)
\right) ^{-1/2}\right)}. \label{thor}
\end{equation}
Notice that the other possible source of correction to $V^{(0)}$
is the modification of the canonical normalisation of $\tau_1$ due
to $\delta\mathcal{V}(\tau_1)$ given by (\ref{444}). Let us
therefore evaluate now the contribution coming from this further
correction. Working just at leading order, we have to substitute
(\ref{444}) in $V^{(0)}$ and then expand obtaining:
\begin{equation}
V^{(0)}=V_{inf}^{(0)}+\delta V^{(0)},
\end{equation}
whereas we can just substitute $\tau_1=e^{2\varphi/\sqrt{3}}$ in
$\delta V$ since an expansion of this term would be subdominant.
At the end, we find that:
\begin{equation}
V_{inf}=V_{inf}^{(0)}+\delta V_{inf},
\end{equation}
where:
\begin{equation}
V_{inf}^{(0)}=\left[ \frac{3}{A^{1/3}}\left( \frac{B}{4}\right)
^{4/3}
\frac{1}{\mathcal{V}_{0}^{1/3}}+A\mathcal{V}_{0}e^{-4\varphi
/\sqrt{3} }-Be^{-\varphi /\sqrt{3}}\right]
\frac{W_{0}^{2}}{\mathcal{V}_{0}^{3}}, \label{55}
\end{equation}
is the canonically normalised inflationary potential used in the
main text in the constant volume approximation. Moreover, $\delta
V^{(0)}$ and $\delta V$ turn out to have the same volume scaling
and so their sum will give the full final leading order correction
to $V_{inf}^{(0)}$:
\begin{gather}
\delta V_{inf}=\delta V^{(0)}+\delta V,  \notag \\
\delta V_{inf}=-\frac{10}{3}\left. \left( \frac{\delta
\mathcal{V}(\tau _{1}) }{\mathcal{V}_{0}}\right) \right\vert
_{\tau _{1}=e^{2\varphi /\sqrt{3}}} \left[ \frac{3}{A^{1/3}}\left(
\frac{B}{4}\right) ^{4/3}\frac{1}{\mathcal{V}
_{0}^{1/3}}+A\mathcal{V}_{0}e^{-4\varphi /\sqrt{3}}-Be^{-\varphi
/\sqrt{3}} \right] \frac{W_{0}^{2}}{\mathcal{V}_{0}^{3}}.
\label{45}
\end{gather}

Comparing (\ref{55}) with (\ref{45}), we notice the interesting
relation:
\begin{equation}
\delta V_{inf}=-\frac{10}{3}\left. \left( \frac{\delta
\mathcal{V}(\tau _{1}) }{\mathcal{V}_{0}}\right) \right\vert
_{\tau _{1}=e^{2\varphi /\sqrt{3}}}V_{inf}^{(0)},
\end{equation}
which implies:
\begin{equation}
V_{inf}=V_{inf}^{(0)}\left[1-\frac{10}{3}\left. \left(
\frac{\delta \mathcal{V}(\tau _{1}) }{\mathcal{V}_{0}}\right)
\right\vert _{\tau _{1}=e^{2\varphi /\sqrt{3}}}\right].
\label{VeryGood}
\end{equation}
This last relation shows a special instance of the general
mechanism discussed in the main text of how this model avoids the
$\eta$-problems that normally plague inflationary potentials. In
particular, the corrections from the one loop potential is seen to
enter in the volume-suppressed combination
$\delta\mathcal{V}/\mathcal{V}_0\ll 1$, ensuring that their
contribution to the inflationary parameters $\varepsilon$ and
$\eta$ is negligible.

\section{Loop effects at high fibre}
\label{Appendix B}

In this section we investigate in more detail what happens at the
string loop corrections when the K3 fibre gets larger and larger
and simultaneously the $\mathbb{C}P^1$ base approaches the
singular limit $t_1\to 0$. One's physical intuition is that loop
corrections should signal the approach to this singular point. In
fact, we show here that the Kaluza-Klein loop correction in
$\tau_2$ is an expansion in inverse powers of $\tau_2$ which goes
to zero when $\tau_1\to\infty\Leftrightarrow t_1\to 0$, as can be
deduced from (\ref{tsoi}). Therefore the presence of the
singularity is signaled by the blowing-up of these corrections. We
then estimate the value $\tau_1^*$ below which perturbation theory
still makes sense and so we can trust our approximation in which
we consider only the first term in the 1-loop expansion of $\delta
V^{KK}_{\tau_2,1-loop}$ and we neglect all the other terms of the
expansion along with higher loop effects. However it will turn out
that, still in a region where $\tau_1<\tau_1^*$, $\delta
V^{KK}_{\tau_2}$, corresponding to the positive exponential in
$V$, already starts to dominate the potential and stops inflation.

Let us now explain the previous claims more in detail. Looking at
the expressions (\ref{LOOPoi}) for all the possible 1-loop
corrections to $V$, we immediately realise that both $\delta
V^{KK}_{(g_s),\tau_1}$ and $\delta V^{W}_{(g_s),\tau_1\tau_2}$
goes to zero when the K3 fibre diverges since
$t^*=\sqrt{\lambda_1\tau_1}$. Therefore these terms are not
dangerous at all. Notice that there is no correction at 1-loop of
the form $1/\left(t_1\mathcal{V}^3\right)$ because, just looking
at the scaling behaviour of that term, we realise that it should
be a correction due to the exchange of winding strings at the
intersection of two stacks of $D7$-branes given by $t_1$, but the
topology of the K3 fibration is such that there are no 4-cycles
which intersect in $t_1$, and so these corrections are absent.

However the sign at 1-loop that there is a singularity when
$\tau_1\to\infty\Leftrightarrow t_1\to 0$, is that $\delta
V^{KK}_{(g_s),\tau_2}$ blows-up. In fact, following the systematic
study of the behaviour of string loop corrections performed in
chapter 5, the contribution of $\delta K^{KK}_{\tau_2,1-loop}$ at
the level of the scalar potential is given by the following
expansion:
\begin{eqnarray}
\delta V_{\tau_2,1-loop}^{KK} &=&\sum\limits_{p=1}^{\infty }\left(
\alpha _{p}g_s^{p}\left( \mathcal{C}_{2}^{KK}\right)
^{p}\frac{\partial^{p}\left(K_{0}\right) }{\partial \tau
_{2}^{p}}\right) \frac{W_{0}^{2}}{\mathcal{V}
^{2}}\text{ }  \notag \\
\text{\ \ with }\alpha _{p} &=&0\text{ }\Longleftrightarrow p=1.
\label{App:V at 1-loop}
\end{eqnarray}
The vanishing coefficients of the first contribution is the
`extended no-scale structure.' Hence we obtain an expansion in
inverse powers of $\tau_2$:
\begin{eqnarray}
\delta V_{\tau _{2},1-loop}^{KK} &=&\left[ \alpha _{2}\left(
\frac{\rho }{ \tau _{2}}\right) ^{2}+\alpha _{3}\left( \frac{\rho
}{\tau _{2}}\right)
^{3}+...\right] \frac{W_{0}^{2}}{\mathcal{V}^{2}}  \notag \\
\text{with }\rho  &\equiv &g_{s}C_{2}^{KK}\ll 1\text{ and }\alpha
_{i}\sim \mathcal{O}(1)\text{ }\forall i.  \label{App:ewe}
\end{eqnarray}
We can then see that, since from (\ref{tsoi}) when
$\tau_1\to\infty\Leftrightarrow t_1\to 0$, $\tau_2\to 0$, all the
terms in the expansion (\ref{App:ewe}) diverge and perturbation
theory breaks down. Thus the region where the expansion
(\ref{App:ewe}) is under control is given by:
\begin{equation}
\frac{\rho}{\tau_2}\leq 2\cdot 10^{-2}\textit{ \
}\Leftrightarrow\textit{ \
}\frac{\mathcal{V}}{\alpha\sqrt{\tau_1}}\geq 50 g_s
C^{KK}_2\textit{ \ }\Leftrightarrow\textit{ \
}\tau_1\leq\sigma_1\mathcal{V}^2\text{ \ with \
}\sigma_1\equiv\left(50 \alpha g_s C^{KK}_2\right)^{-2}.
\label{constr1}
\end{equation}
We need still to evaluate what happens at two and higher loop
level. The behaviour of the 1-loop corrections was under rather
good control since it was conjectured from a generalisation of an
exact toroidal calculation \cite{bhp} and it was tested by a low
energy interpretation in chapter 5. However there is no exact
2-loop calculation for the toroidal case which we could try to
generalise to an arbitrary Calabi-Yau. Thus the best we can do, is
to constrain the scaling behaviour of the 2-loop corrections from
a low energy interpretation. A naive scaling analysis following
the lines of chapter 5, suggests that:
\begin{equation}
\frac{\partial^{2}\left( \delta K_{\tau_2,2-loops}^{KK}\right)
}{\partial \tau_2^{2}}\sim \frac{g_s}{16\pi ^{2}}\frac{1}{\tau_2
}\frac{\partial ^{2}\left( \delta K_{\tau_2,1-loop}^{KK}\right)
}{\partial \tau_2^{2}}, \label{App:2 loops}
\end{equation}
and so $\delta K_{\tau_2,2-loops}^{KK}$ is an homogeneous function
of degree $n=-4$ in the 2-cycle moduli, exactly as $\delta
K_{\tau_1\tau_2,1-loop}^{W}$. Given that:
\begin{equation}
\frac{\partial\left(\delta
K_{\tau_2,1-loop}^{KK}\right)}{\partial\tau_2}=-g_sC_{2}^{KK}\frac{\partial^2
\left(K_{tree}\right)}{\partial\tau_2^2},
\end{equation}
equation (\ref{App:2 loops}) takes the form:
\begin{equation}
\frac{\partial ^{2}\left( \delta K_{\tau_2,2-loops}^{KK}\right)
}{\partial \tau_2^{2}}\sim -\frac{g_s^2C_{2}^{KK}}{16\pi
^{2}}\frac{1}{\tau_2}\frac{\partial^3
\left(K_{tree}\right)}{\partial\tau_2^3}. \label{App:2loops}
\end{equation}
The previous relation and the homogeneity of the K\"{a}hler
metric, produce then the following guess for the Kaluza-Klein
corrections at 2 loops:
\begin{equation}
\delta K_{\tau_2,2-loops}^{KK}\sim -\frac{g_s^2C_{2}^{KK}}{16\pi
^{2}}\frac{\partial^2 \left(K_{tree}\right)}{\partial\tau_2^2},
\end{equation}
that at the level of the scalar potential would translate into:
\begin{equation}
\delta
V_{\tau_2,2-loops}^{KK}=\frac{g_s^{2}C_2^{KK}}{8\pi^2}\left[\frac{1}{\tau_{2}^{2}}
+\mathcal{O}\left(\frac{1}{\tau_2^3}\right)
\right]\frac{W_0^2}{\mathcal{V}^2}. \label{ewef}
\end{equation}
We notice that (\ref{ewef}) has the same behaviour of
(\ref{App:ewe}) apart from the suppression factor $\left( 8\pi
^{2}\mathcal{C}^{KK}_2\right) ^{-1}\sim \mathcal{O}(10^{-2})$.
This is not surprising since the leading contribution of $\delta
K_{\tau_2,1-loop}^{KK}$ in $V$ is zero due to the extended
no-scale but the leading contribution of $\delta
K_{\tau_2,2-loops}^{KK}$ in $V$ is non-vanishing. Thus we conclude
that in the region $\tau_1\ll\sigma_1\mathcal{V}^2$ both higher
terms in the 1-loop expansion (\ref{App:ewe}) and higher loop
corrections (\ref{ewef}) are subleading with respect to the first
term in (\ref{App:ewe}) which we considered in the study of the
inflationary potential.

However, writing everything in terms of the canonically normalised
inflaton field $\hat{\varphi}$ expanded around the minimum, the
first term in (\ref{App:ewe}) turns into the positive exponential
which, as we have seen in section \ref{SlowRoll}, destroys the
slow-roll conditions when it starts to dominate the potential at
$\hat{\varphi}_{max}=12.4$ for $R=2.3\cdot 10^{-6}$. In this point
$\delta V^{W}_{(g_s),\tau_1\tau_2}$ is not yet completely
subleading with respect to $\delta V^{KK}_{(g_s),\tau_2}$ and so
the slow-roll conditions are still satisfied. The form of this
bound in terms of $\tau_1$ can be estimated as follows:
\begin{equation}
\tau _{1}=\left\langle \tau _{1}\right\rangle e^{2\hat{\varphi}
_{\max }/\sqrt{3} }=\left( \frac{4A}{B}\right) ^{2/3}y_{max
}^{2}\mathcal{V}^{2/3}\mathit{\ } \Leftrightarrow \mathit{\ }\tau
_{1}\leq \sigma _{2}\mathcal{V}^{2/3}\text{ \ with \ }\sigma
_{2}\equiv 4.2\cdot 10^{6}\left( \frac{A}{B}\right) ^{2/3}.
\label{constr2}
\end{equation}

\begin{figure}[ht]
\begin{center}
\epsfig{file=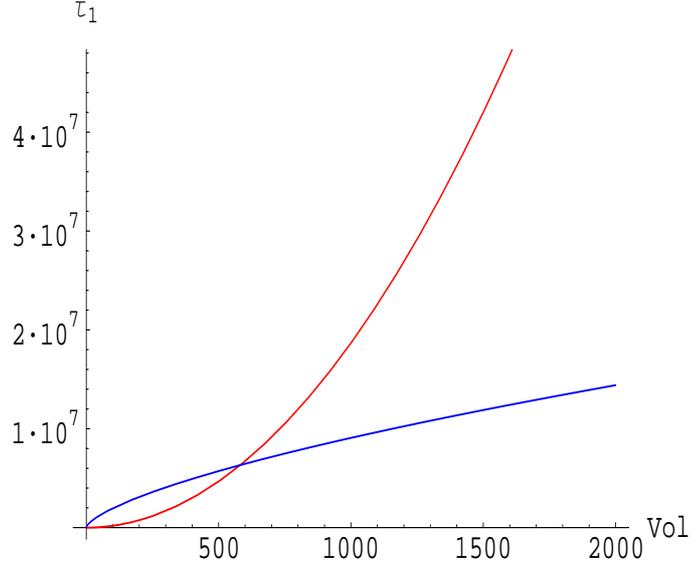, height=80mm,width=90mm} \caption{Plots of
the constraints $\tau_1^{max}=\sigma_1\mathcal{V}^2$ (red curve)
and $\tau_1^{max}=\sigma_2\mathcal{V}^{2/3}$ (blue curve) in the
($\tau_1$, $\mathcal{V}$) plane for the case SV2.}
   \label{Fig:walls}
\end{center}
\end{figure}

Let us now compare the bound (\ref{constr1}) with (\ref{constr2})
to check which is the most stringent one that constrains the field
region available for inflation. The value of the volume at which
the two bounds are equal is:
\begin{equation}
\mathcal{V}_*=\left(\frac{\sigma_2}{\sigma_1}\right)^{3/4}=1.65\cdot
10^{7}\alpha g_s^{5/2}\frac{C^{KK}_1
(C^{KK}_2)^{3/2}}{(C^W_{12})^{1/2}}.
\end{equation}
Using the natural choice of parameter values made in the main text
(for SV2 for example), $\mathcal{V}_*=582$, and so, since we
always deal with much larger values of the overall volume, we
conclude that the most stringent constraint is (\ref{constr2}) as
can be seen from Figure \ref{Fig:walls}.

Therefore the final situation is that, when the K3 fibre gets
larger, $\delta V^{KK}_{(g_s),\tau_2}$ starts dominating the
potential and ruining inflation well before one approaches the
singular limit in which the perturbative expansion breaks down and
these corrections blow-up to infinity.

\chapter[Appendix]{}

\section{Moduli couplings}
\label{App:ModuliCouplings}

We shall now assume that the MSSM is built via magnetised
$D7$-branes wrapping an internal 4-cycle within the framework of
four dimensional $N=1$ supergravity. The full Lagrangian of the
system can be obtained by expanding the superpotential $W$, the
K\"{a}hler potential $K$ and the gauge kinetic functions $f_i$ as
a power series in the matter fields \cite{SUGRAdecomp}:
\begin{eqnarray}
W &=&W_{mod}(\varphi )+\mu (\varphi
)H_{u}H_{d}+\frac{Y_{ijk}(\varphi
)}{6}C^{i}C^{j}C^{k}+..., \\
K &=&K_{mod}(\varphi ,\bar{\varphi} )+\tilde{K}_{i\bar{j}}(\varphi
,\bar{\varphi} )C^{i}C^{\bar{j}}+
\left[ Z(\varphi ,\bar{\varphi})H_{u}H_{d}+h.c.\right] +..., \\
f_{i} &=&\frac{T_{MSSM}}{4\pi }+h_{i}(F)S. \label{gaugeKinfunc}
\end{eqnarray}
In the previous expressions, $\varphi$ denotes globally all the
moduli fields, and $W_{mod}$ and $K_{mod}$ are the superpotential
and the K\"{a}hler potential for the moduli. $H_u$ and $H_d$ are
the two Higgs doublets of the MSSM, and the $C$'s denote
collectively all the matter fields. In the expression for the
gauge kinetic function (\ref{gaugeKinfunc}), $T_{MSSM}$ is the
modulus related to the 4-cycle wrapped by the MSSM $D7$-branes,
and $h_i(F)$ are 1-loop topological functions of the world-volume
fluxes $F$ on different branes (the index $i$ runs over the three
MSSM gauge group factors). Finally the moduli scaling of the
K\"{a}hler potential for matter fields
$\tilde{K}_{i\bar{j}}(\varphi ,\bar{\varphi})$ and $Z(\varphi
,\bar{\varphi})$, for LVS with the small cycle $\tau_s$ supporting
the MSSM, has been derived in \cite{ccq} and looks like:
\begin{equation}
\tilde{K}_{i\bar{j}}(\varphi
,\bar{\varphi})\sim\frac{\tau_s^{1/3}}{\tau_b}k_{i\bar{j}}(U)\text{
\ \ and \ \ }Z(\varphi
,\bar{\varphi})\sim\frac{\tau_s^{1/3}}{\tau_b}z(U).
\label{chiralMetric}
\end{equation}

\subsection{Moduli couplings to ordinary particles}
\label{Sec:ModCouplOrdPart}

We now review the derivation of the moduli couplings to gauge
bosons, matter particles and Higgs fields for high temperatures
$T>M_{EW}$. In this case all the gauge bosons and matter fermions
are massless.

\bigskip

$\bullet \textbf{ Couplings to Gauge Bosons}$

\bigskip

The coupling of the gauge bosons $X$ to the moduli arise from the
moduli dependence of the gauge kinetic function
(\ref{gaugeKinfunc}). We shall assume that the MSSM $D7$-branes
are wrapping the small cycle\footnote{The large cycle would yield
an unrealistically small gauge coupling: $g^2\sim
\langle\tau_b\rangle^{-1}\sim 10^{-10}$.}, and so we identify
$T_{MSSM}\equiv T_{s}$. We also recall that the gauge couplings of
the different MSSM gauge groups are given by the real part of the
gauge kinetic function, and that one obtains different values by
turning on different fluxes. Thus the coupling of $\tau_s$ with
the gauge bosons is the same for $U(1)$, $SU(2)$ and $SU(3)$. We
now focus on the $U(1)$ factor without loss of generality. The
kinetic terms read (neglecting the uninteresting flux dependent
part):
\begin{equation}
\mathcal{L}_{gauge}=-\frac{\tau_s}{M_P}F_{\mu\nu}F^{\mu\nu}.
\end{equation}
We then expand $\tau_s$ around its minimum and go to the
canonically normalised field strength $G_{\mu\nu}$ defined as:
\begin{equation}
G_{\mu\nu}=\sqrt{\langle\tau_s\rangle}F_{\mu\nu}, \label{redef}
\end{equation}
and obtain:
\begin{equation}
\mathcal{L}_{gauge}=-G_{\mu\nu}G^{\mu\nu}-\frac{\delta\tau_s}
{M_P\langle\tau_s\rangle}G_{\mu\nu}G^{\mu\nu}.
\label{FmunuFmunu}
\end{equation}
Now by means of (\ref{small}) we end up with the following
\textit{dimensionful} couplings:
\begin{eqnarray}
\mathcal{L}_{\chi XX}&\sim&\left(\frac{1}
{M_P\ln{\mathcal{V}}}\right)\chi G_{\mu\nu}G^{\mu\nu}, \\
\mathcal{L}_{\Phi XX}&\sim&\left(\frac{\sqrt{\mathcal{V}}}
{M_P}\right)\Phi G_{\mu\nu}G^{\mu\nu}. \label{ImpModuliCoupl}
\end{eqnarray}


$\bullet \textbf{ Couplings to matter fermions}$

\bigskip

The terms of the supergravity Lagrangian which are relevant to
compute the order of magnitude of the moduli couplings to an
ordinary matter fermion $\psi$
are its kinetic and mass terms:\footnote{Instead of the usual
2-component spinorial notation, we are using here the more
convenient 4-component spinorial notation.}
\begin{equation}
\mathcal{L}=\tilde{K}_{\bar{\psi}\psi}\bar{\psi}i\gamma^{\mu}\partial_{\mu}\psi+
e^{K/2}\lambda H\bar{\psi}\psi,
\end{equation}
where $H$ is the corresponding Higgs field (either $H_u$ or
$H_d$). The moduli scaling of $\tilde{K}_{\bar{\psi}\psi}$ is
given in (\ref{chiralMetric}), whereas $e^{K/2}=\mathcal{V}^{-1}$.
Expanding the moduli and the Higgs around their VEVs, we obtain:
\begin{eqnarray}
\mathcal{L} &=&\frac{\langle \tau _{s}\rangle ^{1/3}}{\langle \tau
_{b}\rangle }\left( 1+\frac{1}{3}\frac{\delta \tau _{s}}{\langle
\tau _{s}\rangle }-\frac{\delta \tau _{b}}{\langle \tau
_{b}\rangle }+...\right)
\bar{\psi}i\gamma^{\mu }\partial _{\mu }\psi   \notag \\
&&+\frac{1}{\langle \tau _{b}\rangle ^{3/2}}\left(
1-\frac{3}{2}\frac{\delta \tau _{b}}{\langle \tau _{b}\rangle
}+...\right) \lambda \left( \langle H\rangle +\delta H\right)
\bar{\psi}\psi .
\end{eqnarray}
We now canonically normalise the $\psi$ kinetic terms
($\psi\to\psi_c$) and rearrange the previous expression as:
\begin{eqnarray}
\mathcal{L} &=&\bar{\psi}_{c}\left(i\gamma^{\mu }\partial _{\mu
}+m_{\psi }\right) \psi _{c}+\left( \frac{1}{3}\frac{\delta \tau
_{s}}{ \langle \tau _{s}\rangle }-\frac{\delta \tau _{b}}{\langle
\tau _{b}\rangle } \right) \bar{\psi}_{c}\left( i\gamma^{\mu
}\partial _{\mu }+m_{\psi
}\right) \psi _{c}  \notag \\
&&-\left( \frac{1}{3}\frac{\delta \tau _{s}}{\langle \tau
_{s}\rangle }+ \frac{1}{2}\frac{\delta \tau _{b}}{\langle \tau
_{b}\rangle }\right) m_{\psi }\bar{\psi}_{c}\psi
_{c}+\mathcal{L}_{\delta H},  \label{sec}
\end{eqnarray}
where:
\begin{equation}
m_{\psi}\equiv\frac{\lambda\langle
H\rangle}{\langle\tau_s\rangle^{1/3}\langle\tau_b\rangle^{1/2}},
\end{equation}
and:
\begin{equation}
\mathcal{L}_{\delta
H}=\left(\frac{\lambda}{\langle\tau_b\rangle^{1/2}\langle\tau_s\rangle^{1/3}}\right)\delta
H\bar{\psi}_c\psi_c-\left(\frac{3\lambda}{2\langle\tau_b\rangle^{3/2}\langle\tau_s\rangle^{1/3}}
\right)\delta \tau_b\delta H \bar{\psi}_c\psi_c. \label{secc}
\end{equation}
The second term of (\ref{sec}) does not contribute to the moduli
interactions since Feynman amplitudes vanish for on-shell final
states satisfying the equations of motion. Writing everything in
terms of $\Phi$ and $\chi$, we end up with the following
\textit{dimensionless} couplings:
\begin{eqnarray}
\mathcal{L}_{\chi\bar{\psi}_c\psi_c}&\sim&\left(\frac{m_{\psi}}{M_P}\right)\chi\bar{\psi}_c\psi_c, \label{ein}\\
\mathcal{L}_{\Phi\bar{\psi}_c\psi_c}&\sim&\left(\frac{m_{\psi}\sqrt{\mathcal{V}}}{M_P}\right)
\Phi\bar{\psi}_c\psi_c. \label{zwei}
\end{eqnarray}
Moreover the first term in the Higgs Lagrangian (\ref{secc}) gives
rise to the usual Higgs-fermion-fermion coupling, whereas the
second term yields a modulus-Higgs-fermion-fermion vertex with
strength:
\begin{eqnarray}
\mathcal{L}_{\delta H\bar{\psi}_c\psi_c}&\sim&
\left(\frac{1}{\mathcal{V}^{1/3}}\right)\delta H\bar{\psi}_c\psi_c, \\
\mathcal{L}_{\chi\delta H\bar{\psi}_c\psi_c}&\sim&
\left(\frac{1}{M_P\mathcal{V}^{1/3}}\right)\chi\delta H\bar{\psi}_c\psi_c, \\
\mathcal{L}_{\Phi\delta
H\bar{\psi}_c\psi_c}&\sim&\left(\frac{1}{M_P\mathcal{V}^{5/6}}\right)
\Phi\delta H\bar{\psi}_c\psi_c.
\end{eqnarray}
We notice that for $T>M_{EW}$ the fermions are massless since
$\langle H\rangle=0$, and so the two direct moduli couplings to
ordinary matter particles (\ref{ein}) and (\ref{zwei}) are absent.
\bigskip

$\bullet \textbf{ Couplings to Higgs Fields}$

\bigskip

The form of the un-normalised kinetic and mass terms for the Higgs
from the supergravity Lagrangian, reads:
\begin{equation}
\mathcal{L}_{Higgs}=\tilde{K}_{\bar{H}H}\partial_{\mu}H
\partial^{\mu}\bar{H}
-\tilde{K}_{\bar{H}H} \left(\hat{\mu}^2+m_0^2\right)H\bar{H},
\label{LagraHiggs}
\end{equation}
where $H$ denotes a Higgs field (either $H_u$ or $H_d$), and
$\hat{\mu}$ and $m_0$ are the canonically normalised
supersymmetric $\mu$-term and SUSY breaking scalar mass
respectively. Their volume dependence, in the dilute flux limit,
is \cite{SoftSUSY}:
\begin{equation}
|\hat{\mu}|\sim m_0\sim\frac{M_P}{\mathcal{V}\ln\mathcal{V}}.
\label{mu}
\end{equation}
In addition to (\ref{LagraHiggs}), there is also a mixing term of
the form:
\begin{equation}
\mathcal{L}_{Higgs\text{ \ }mix}=Z\left(\partial_{\mu}H_d
\partial^{\mu}H_u+\partial_{\mu}\bar{H}_d
\partial^{\mu}\bar{H}_u\right)-\tilde{K}_{\bar{H}H}B\hat{\mu}\left(H_d
H_u+\bar{H}_d \bar{H}_u\right), \label{LagraHiggsMix}
\end{equation}
with:
\begin{equation}
B\hat{\mu}\sim m_0^2. \label{Bmu}
\end{equation}
However given that we are interested only in the leading order
volume scaling of the Higgs coupling to the moduli, we can neglect
the $\mathcal{O}(1)$ mixing of the \textit{up} and \textit{down}
components, and focus on the simple Lagrangian:
\begin{eqnarray}
\mathcal{L}_{Higgs}&=&\tilde{K}_{\bar{H}H}\left(\partial_{\mu}H
\partial^{\mu}\bar{H}
-\frac{M_P^2}{\left(\mathcal{V}\ln\mathcal{V}\right)^2}H\bar{H}\right) \\
&=&-\frac{1}{2}\tilde{K}_{\bar{H}H}\left[ \bar{H}\left( \square
+\frac{M_P^2}{\left( \mathcal{V}\ln \mathcal{V}\right)
^{2}}\right) H+H\left( \square +\frac{M_P^2}{\left( \mathcal{V}\ln
\mathcal{V} \right) ^{2}}\right) \bar{H}\right]. \notag
\label{LagrangianHiggs}
\end{eqnarray}
We now expand $\tilde{K}_{\bar{H}H}$ and
$\left(\mathcal{V}\ln\mathcal{V}\right)^{-2}$ and get:
\begin{eqnarray}
\mathcal{L}_{Higgs} &\simeq &-\frac{1}{2}K_{0}\left(
1+\frac{1}{3}\frac{ \delta \tau _{s}}{\langle \tau _{s}\rangle
}-\frac{\delta \tau _{b}}{\langle \tau _{b}\rangle }\right) \left[
\bar{H}\left( \square +m_{H}^{2}\left( 1-3\frac{\delta \tau
_{b}}{\langle \tau _{b}\rangle }
\right) \right) H+\right.   \notag \\
&&\left. H\left( \square +m_{H}^{2}\left( 1-3\frac{\delta \tau
_{b}}{\langle \tau _{b}\rangle }\right) \right) \bar{H} \right],
\end{eqnarray}
where
$K_0=\langle\tau_s\rangle^{1/3}\langle\mathcal{V}\rangle^{-2/3}$
and the Higgs mass is given by:
\begin{equation}
m_{H}\simeq\frac{M_P}{\langle\mathcal{V}\rangle\ln\langle
\mathcal{V}\rangle}.
\end{equation}
Now canonically normalising the scalar kinetic terms $H\rightarrow
H_c=\sqrt{K_0}H$, we obtain:
\begin{eqnarray}
\mathcal{L}_{Higgs} &=&-\frac{1}{2}\left[ \bar{H}_{c}\left(
\square
+m_{H}^{2}\right) H_{c}+H_{c}\left( \square +m_{H}^{2}\right) \bar{H}_{c}%
\right]   \notag \\
&&-\frac{1}{2}\left( \frac{1}{3}\frac{\delta \tau _{s}}{\langle
\tau
_{s}\rangle }-\frac{\delta \tau _{b}}{\langle \tau _{b}\rangle }\right) %
\left[ \bar{H}_{c}\left( \square +m_{H}^{2}\right)
H_{c}+H_{c}\left( \square
+m_{H}^{2}\right) \bar{H}_{c}\right]   \notag \\
&&+3\frac{\delta \tau _{b}}{\langle \tau _{b}\rangle }m_{H}^{2}\bar{H}%
_{c}H_{c}.
\end{eqnarray}
The second term in the previous expression does not contribute to
scattering amplitudes since Feynman amplitudes vanish for final
states satisfying the equations of motion. Thus the
\textit{dimensionful} moduli couplings to Higgs fields arise only
from the third term once we express $\delta\tau_b$ in terms of
$\Phi$ and $\chi$ using (\ref{big}). The final result is:
\begin{eqnarray}
\mathcal{L}_{\Phi \bar{H}_{c}H_{c}} &\sim &\left(
\frac{m_{H}^{2}}{M_{P} \sqrt{\mathcal{V}}}\right) \Phi
\bar{H}_{c}H_{c}\sim \left( \frac{M_{P}}{ \mathcal{V}^{5/2}(\ln
{\mathcal{V}})^{2}}\right) \Phi \bar{H}_{c}H_{c},
\label{Higgs1} \\
\mathcal{L}_{\chi \bar{H}_{c}H_{c}} &\sim &\left(
\frac{m_{H}^{2}}{M_{P}} \right) \chi \bar{H}_{c}H_{c}\sim \left(
\frac{M_{P}}{\mathcal{V}^{2}(\ln { \mathcal{V}})^{2}}\right) \chi
\bar{H}_{c}H_{c}.  \label{Higgs2}
\end{eqnarray}

\subsection{Moduli couplings to supersymmetric particles}
\label{Sec:ModCouplSUSY}

We shall now work out the moduli couplings to gauginos, SUSY
scalars and Higgsinos. Given that we are interested in thermal
corrections at high temperatures, we shall focus on $T>M_{EW}$.
Thus we can neglect the mixing of Higgsinos with gauginos into
charginos and neutralinos, which takes place at lower energies due
to EW symmetry breaking.

\bigskip

$\bullet \textbf{ Couplings to Gauginos}$

\bigskip

The relevant part of the supergravity Lagrangian involving the
gaugino kinetic terms and their soft masses looks like:
\begin{equation}
\mathcal{L}_{gaugino}\simeq
\frac{\tau_s}{M_P}\bar{\lambda'}i\bar{\sigma}^{\mu}\partial_{\mu}\lambda'+\frac{F^s}{2}\left(\lambda'\lambda'+h.c.\right),
\label{Ploj}
\end{equation}
where in the limit of dilute world-volume fluxes on the
$D7$-brane, the gaugino mass is given by
$M_{1/2}=\frac{F^{s}}{2\tau_s}$ \cite{SoftSUSY}. Now if the small
modulus supporting the MSSM is stabilised via non-perturbative
corrections, then the corresponding $F$-term scales as:
\begin{equation}
F^{s}\simeq\frac{\tau_s}{\mathcal{V}\ln\mathcal{V}}. \label{F}
\end{equation}
Notice that the suppression factor
$\ln\mathcal{V}\sim\ln(M_P/m_{3/2})$ in (\ref{F}) could be absent
in the case of perturbative stabilisation of the MSSM cycle
discussed in chapter 6. Let us expand $\tau_s$ around its VEV and
get:
\begin{eqnarray}
\mathcal{L}_{gaugino}&\simeq&
\langle\tau_s\rangle\left[\bar{\lambda'}i\bar{\sigma}^{\mu}\partial_{\mu}\lambda'+\frac{1}{2}\frac{M_P}{\mathcal{V}\ln\mathcal{V}}
\left(\lambda'\lambda'+h.c.\right)\right] \notag \\
&+&\frac{\delta\tau_s}{M_P}
\left[\bar{\lambda'}i\bar{\sigma}^{\mu}\partial_{\mu}\lambda'+\frac{M_P}{\langle\mathcal{V}\rangle\ln\langle\mathcal{V}\rangle}
\frac{\left(\lambda'\lambda'+h.c.\right)}{2}\right].
\label{labello}
\end{eqnarray}
We need now to expand also $\tau_b$ around its VEV in the first
term of (\ref{labello}):
\begin{equation}
\frac{1}{\mathcal{V}\ln\mathcal{V}}\simeq\frac{1}{\tau_b^{3/2}\ln\mathcal{V}}
\simeq\frac{1}{\langle\mathcal{V}\rangle\ln\langle\mathcal{V}\rangle}
\left(1-\frac{3}{2}\frac{\delta\tau_b}{\langle\tau_b\rangle}+...\right),
\end{equation}
and canonically normalise the gaugino kinetic terms
$\lambda'\rightarrow\lambda=\sqrt{\langle\tau_s\rangle}\lambda'$.
At the end we obtain:
\begin{eqnarray}
\mathcal{L}_{gaugino}&\simeq&
\bar{\lambda}i\bar{\sigma}^{\mu}\partial_{\mu}\lambda+\frac{M_P}
{\langle\mathcal{V}\rangle\ln\langle\mathcal{V}\rangle}
\frac{\left(\lambda\lambda+h.c.\right)}{2} \notag \\
&+& \frac{\left(\lambda\lambda+h.c.\right)}
{2\langle\mathcal{V}\rangle\ln\langle\mathcal{V}\rangle}\left(\frac{\delta\tau_s}
{\langle\tau_s\rangle}
-\frac{3}{2}\frac{\delta\tau_b}{\langle\tau_b\rangle}\right)
+\frac{\delta\tau_s}{\langle\tau_s\rangle M_P
}\bar{\lambda}i\bar{\sigma}^{\mu}\partial_{\mu}\lambda.
\label{labelloo}
\end{eqnarray}
From (\ref{labelloo}) we can immediately read off the gaugino
mass:
\begin{equation}
M_{1/2}\simeq\frac{M_P}{\langle\mathcal{V}\rangle\ln\langle\mathcal{V}\rangle}
\simeq\frac{F^s}{\tau_s}\sim\frac{m_{3/2}}{\ln\left(M_P/m_{3/2}\right)}.
\end{equation}
Let us now rewrite (\ref{labelloo}) as:
\begin{equation}
\mathcal{L}_{gaugino}\simeq
\left(1+\frac{\delta\tau_s}{\langle\tau_s\rangle M_P}\right)\left[
\bar{\lambda}i\bar{\sigma}^{\mu}\partial_{\mu}\lambda
+\frac{M_{1/2}}{2}\left(\lambda\lambda+h.c.\right)\right]
-\frac{3}{4}\frac{\delta\tau_b}{\langle\tau_b\rangle}\frac{M_{1/2}}{M_P}
\left(\lambda\lambda+h.c.\right). \label{labellooo}
\end{equation}
We shall now focus only on the third term in (\ref{labellooo})
since the second term does not contribute to decay rates. In fact,
Feynman amplitudes with on-shell final states that satisfy the
equations of motion, are vanishing. Using (\ref{big}), we finally
obtain the following \textit{dimensionless} couplings:
\begin{eqnarray}
\mathcal{L}_{\Phi \lambda\lambda } &\sim &\left(
\frac{M_{1/2}}{M_{P} \sqrt{\mathcal{V}}}\right) \Phi
\lambda\lambda \sim \left( \frac{1}{ \mathcal{V}^{3/2}\ln
{\mathcal{V}}}\right) \Phi \lambda\lambda ,
\label{rr} \\
\mathcal{L}_{\chi \lambda\lambda } &\sim &\left(
\frac{M_{1/2}}{M_{P}} \right) \chi \lambda\lambda \sim \left(
\frac{1}{\mathcal{V}\ln \mathcal{V}}\right) \chi \lambda\lambda .
\label{rrr}
\end{eqnarray}


$\bullet \textbf{ Couplings to SUSY Scalars}$

\bigskip

The form of the un-normalised kinetic and soft mass terms for SUSY
scalars from the supergravity Lagrangian, reads:
\begin{equation}
\mathcal{L}_{scalars}=\tilde{K}_{\alpha\bar{\beta}}\partial_{\mu}C^{\alpha}
\partial^{\mu}\bar{C}^{\bar{\beta}}
-\frac{\tilde{K}_{\alpha\bar{\beta}}}{\left(\mathcal{V}\ln\mathcal{V}\right)^2}
C^{\alpha}\bar{C}^{\bar{\beta}}. \label{Lagra}
\end{equation}
Assuming diagonal K\"{a}hler metric for matter fields:
\begin{equation}
\tilde{K}_{\alpha\bar{\beta}}=\tilde{K}_{\alpha}\delta_{\alpha\bar{\beta}},
\end{equation}
the initial Lagrangian (\ref{Lagra}) simplifies to:
\begin{eqnarray}
\mathcal{L}_{scalars} &=&\tilde{K}_{\alpha }\left( \partial _{\mu
}C^{\alpha }\partial ^{\mu }\bar{C}^{\bar{\alpha}}-\frac{1}{\left(
\mathcal{V}\ln \mathcal{V}\right) ^{2}}C^{\alpha
}\bar{C}^{\bar{\alpha}}\right) \\
&=&-\frac{1}{2}\tilde{K}_{\alpha }\left[
\bar{C}^{\bar{\alpha}}\left( \square +\frac{1}{\left(
\mathcal{V}\ln \mathcal{V}\right) ^{2}}\right) C^{\alpha
}+C^{\alpha }\left( \square +\frac{1}{\left( \mathcal{V}\ln
\mathcal{V} \right) ^{2}}\right) \bar{C}^{\bar{\alpha}}\right].
\nonumber \label{Lag}
\end{eqnarray}
We note that (\ref{Lag}) is similar to the Higgs Lagrangian
(\ref{LagrangianHiggs}). This is not surprising since for
temperatures $T>M_{EW}$, the Higgs behaves effectively as a SUSY
scalar with mass of the order the scalar soft mass: $m_{H}\sim
m_0$. Thus we can read off immediately the \textit{dimensionful}
moduli couplings to the canonically normalised SUSY scalars
$\varphi$ from (\ref{Higgs1}) and (\ref{Higgs2}):
\begin{eqnarray}
\mathcal{L}_{\Phi \bar{\varphi}\varphi} &\sim &\left(
\frac{m_{0}^{2}}{M_{P} \sqrt{\mathcal{V}}}\right) \Phi
\bar{\varphi}\varphi\sim \left( \frac{M_{P}}{
\mathcal{V}^{5/2}(\ln
{\mathcal{V}})^{2}}\right) \Phi \bar{\varphi}\varphi, \\
\mathcal{L}_{\chi \bar{\varphi}\varphi} &\sim &\left(
\frac{m_{0}^{2}}{M_{P}} \right) \chi \bar{\varphi}\varphi\sim
\left( \frac{M_{P}}{\mathcal{V}^{2}(\ln {
\mathcal{V}})^{2}}\right) \chi \bar{\varphi}\varphi.
\end{eqnarray}

\bigskip

$\bullet \textbf{ Couplings to Higgsinos}$

\bigskip

The relevant part of the supergravity Lagrangian involving the
Higgsino kinetic terms and their supersymmetric masses looks like:
\begin{equation}
\mathcal{L}_{Higgsino}\simeq
\tilde{K}_{\bar{\tilde{H}}\tilde{H}}\left[
\bar{\tilde{H}}_ui\bar{\sigma}^{\mu}\partial_{\mu}\tilde{H}_u
+\bar{\tilde{H}}_di\bar{\sigma}^{\mu}\partial_{\mu}\tilde{H}_d
+\hat{\mu}\left(\tilde{H}_u\tilde{H}_d+h.c.\right)\right].
\end{equation}
After diagonalising the supersymmetric Higgsino mass term, we end
up with a usual Lagrangian of the form:
\begin{equation}
\mathcal{L}_{Higgsino}\simeq
\tilde{K}_{\bar{\tilde{H}}\tilde{H}}\left[
\bar{\tilde{H}}i\bar{\sigma}^{\mu}\partial_{\mu}\tilde{H}+\hat{\mu}\left(\tilde{H}\tilde{H}+h.c.\right)\right],
\end{equation}
where $\tilde{H}$ denotes collectively both the Higgsino mass
eigenstates, which are the result of a mixing between the
\textit{up} and \textit{down} gauge eigenstates. We recall also
that since we are focusing on temperatures above the EWSB scale,
we do not have to deal with any mixing between Higgsinos and
gauginos to give neutralinos and charginos. Expanding the
K\"{a}hler metric (\ref{chiralMetric}) and the $\mu$-term
(\ref{mu}), we obtain:
\begin{equation}
\mathcal{L}_{Higgsino} \simeq K_{0}\left( 1+\frac{1}{3}\frac{
\delta \tau _{s}}{\langle \tau _{s}\rangle }-\frac{\delta \tau
_{b}}{\langle \tau _{b}\rangle }\right)
\left[\bar{\tilde{H}}i\bar{\sigma}^{\mu}\partial_{\mu}\tilde{H}+
\frac{m_{\tilde{H}}}{2}\left( 1-\frac{3}{2}\frac{\delta \tau
_{b}}{\langle \tau _{b}\rangle } \right)
\left(\tilde{H}\tilde{H}+h.c.\right)\right],
\end{equation}
where
$K_0=\langle\tau_s\rangle^{1/3}\langle\mathcal{V}\rangle^{-2/3}$
and the physical Higgsino mass is of the same order of magnitude
of the soft SUSY masses:
\begin{equation}
m_{\tilde{H}}\simeq\frac{M_P}{\langle\mathcal{V}\rangle\ln\langle
\mathcal{V}\rangle}\simeq M_{1/2}.
\end{equation}
Now canonically normalising the scalar kinetic terms
$\tilde{H}\rightarrow\tilde{H}_c=\sqrt{K_0}\tilde{H}$, we end up
with:
\begin{eqnarray}
\mathcal{L}_{Higgsino} &=&\left(1+\frac{1}{3}\frac{ \delta \tau
_{s}}{\langle \tau _{s}\rangle }-\frac{\delta \tau _{b}}{\langle
\tau _{b}\rangle }\right)\left[
\bar{\tilde{H}}_{c}i\bar{\sigma}^{\mu }\partial _{\mu
}\tilde{H}_{c}+\frac{m_{\tilde{H}}}{2}\left(\tilde{H}_{c}\tilde{H}_c+h.c.\right) \right] \notag \\
&&-\frac{3}{4}\frac{\delta \tau _{b}}{\langle \tau _{b}\rangle}
m_{ \tilde{H}}\left(\tilde{H}_c\tilde{H}_c+h.c.\right).
\label{secco}
\end{eqnarray}
Writing everything in terms of $\Phi$ and $\chi$, from the third
term of (\ref{secco}), we obtain the following
\textit{dimensionless} couplings:
\begin{eqnarray}
\mathcal{L}_{\chi\tilde{H}_c\tilde{H}_c}&\sim&\left(\frac{m_{\tilde{H}}}{M_P}\right)
\chi\tilde{H}_c\tilde{H}_c
\sim\left(\frac{1}{\mathcal{V}\ln\mathcal{V}}\right)
\chi\tilde{H}_c\tilde{H}_c, \\
\mathcal{L}_{\Phi\tilde{H}_c\tilde{H}_c}&\sim&\left(\frac{m_{\tilde{H}}}{M_P\sqrt{\mathcal{V}}}\right)
\Phi\tilde{H}_c\tilde{H}_c\sim\left(\frac{1}{\mathcal{V}^{3/2}\ln\mathcal{V}}\right)
\Phi\tilde{H}_c\tilde{H}_c.
\end{eqnarray}

\subsection{Moduli self couplings}
\label{SELF}

In this section we shall investigate if moduli reach thermal
equilibrium among themselves. In order to understand this issue,
we need to compute the moduli self interactions, which can be
obtained by first expanding the moduli fields around their VEV:
\begin{equation}
\tau_{i}=\langle\tau_i\rangle+\delta\tau_i,
\end{equation}
and then by expanding the potential around the LARGE Volume vacuum
as follows:
\begin{equation}
V=V(\langle \tau _{s}\rangle ,\langle \tau _{b}\rangle
)+\frac{1}{2}\left. \frac{\partial ^{2}V}{\partial \tau
_{i}\partial \tau _{j}}\right\vert _{\min }\delta \tau _{i}\delta
\tau _{j}+\frac{1}{3!}\left. \frac{\partial ^{3}V}{\partial \tau
_{i}\partial \tau _{j}\partial \tau _{k}}\right\vert _{\min
}\delta \tau _{i}\delta \tau _{j}\delta \tau _{k}+....
\label{expandi}
\end{equation}
We then concentrate on the trilinear terms which can be read off
from the third term of (\ref{expandi}). We neglect the
$\mathcal{O}(\delta\tau_i^4)$ terms since the strength of their
couplings will be subleading with respect to the
$\mathcal{O}(\delta\tau_i^3)$ terms since one has to take a
further derivative which produces a suppression factor. Taking the
third derivatives and then expressing these self-interactions in
terms of the canonically normalised fields:
\begin{eqnarray*}
\delta \tau _{b} &\sim &\mathcal{O}\left( \mathcal{V}^{1/6}\right)
\Phi +
\mathcal{O}\left( \mathcal{V}^{2/3}\right) \chi , \\
\delta \tau _{s} &\sim &\mathcal{O}\left( \mathcal{V}^{1/2}\right)
\Phi + \mathcal{O}\left(1\right) \chi,
\end{eqnarray*}
we end up with the following Lagrangian terms at leading order in
a large volume expansion:
\begin{eqnarray}
\mathcal{L}_{\Phi ^{3}} &\simeq
&\frac{M_{P}}{\mathcal{V}^{3/2}}\Phi ^{3}, \text{ \ \ \ \ \ \ \ \
\ }\mathcal{L}_{\Phi ^{2}\chi}\simeq \frac{M_{P}}{
\mathcal{V}^{2}}\chi \Phi ^{2}, \\
\mathcal{L}_{\chi^{2}\Phi} &\simeq
&\frac{M_{P}}{\mathcal{V}^{5/2}}\Phi \chi ^{2},\text{ \ \ \ \ \ \
\ \ \ }\mathcal{L}_{\chi ^{3}}\simeq \frac{M_{P}
}{\mathcal{V}^{3}}\chi ^{3}. \label{coupl}
\end{eqnarray}

\end{appendix}

\clearpage
\bibliography{thesisRefs}
\bibliographystyle{utphys}

\end{document}